\documentclass[12pt, twoside]{article}
\usepackage{latexsym}
\usepackage{amsmath, amsfonts, amscd, amssymb}
\usepackage{amsthm}
\usepackage{fancybox, eufrak, euscript, oldgerm, graphics}
\usepackage{pst-all}        
\usepackage{pst-poly}     
\usepackage{multido}      

\newcommand{\aconn}{\mathcal{A}}

\newcommand{\ter}{\mathfrak{t}}

\newcommand{\aut}{{\mathcal{A}}ut}

\newcommand{\conn}{\mathcal{D}}

\newcommand{\cons}{\mathbf{C}}

\newcommand{\curv}{R}
\newcommand{\ds}{\mathcal{S}}
\newcommand{\dss}{\mathfrak{S}}
\newcommand{\ric}{{\mathcal{R}}}
\newcommand{\smric}{^{\infty}\ric}
\newcommand{\field}{\mathfrak{F}}

\newcommand{\ricci}{\EuScript{R}}

\newcommand{\fa}{{\mathfrak{B}}}
\newcommand{\sfa}{{\mathcal{B}}}

\newcommand{\gelsp}{\mathfrak{M}}

\newcommand{\eh}{\mathfrak{E}\mathfrak{H}}

\newcommand{\gauge}{\mathcal{U}}

\newcommand{\Hom}{{\mathcal{H}}om}
\newcommand{\kd}{\text{\texttt{d}}}

\newcommand{\mink}{{\mathcal{M}}}
\newcommand{\inv}{\overleftarrow{\EuScript{P}}}
\newcommand{\invros}{\overleftarrow{\mathcal{N}}}
\newcommand{\inveinst}{\overleftarrow{\EuScript{E}}}
\newcommand{\invmod}{\overleftarrow{\EuScript{M}}}
\newcommand{\invs}{\overleftarrow{\EuScript{S}}}
\newcommand{\invsconn}{\overleftarrow{\mathfrak{A}}}
\newcommand{\invel}{\overleftarrow{\mathfrak{E}}}
\newcommand{\inveh}{\overleftarrow{\EuScript{EH}}}
\newcommand{\invcq}{\overleftarrow{\EuScript{Z}}}
\newcommand{\invtriad}{\stackrel{\rightleftarrows}{\EuScript{T}}}
\newcommand{\finv}{\overleftarrow{{\EuScript{G}}}}
\newcommand{\diromg}{\overrightarrow{\mathfrak{R}}}

\newcommand{\grouv}{{\mathcal{G}}}

\newcommand{\man}{{\mathcal{M}}an}
\newcommand{\modl}{\mathbf{\mathcal{E}}}
\newcommand{\morph}{\EuScript{F}}
\newcommand{\modll}{\mathbf{\mathcal{E}}^{\uparrow}}
\newcommand{\natf}{\EuScript{N}}

\newcommand{\omg}{\Omega}
\newcommand{\Omg}{\mathbf{\Omega}}
\newcommand{\Qaus}{{\vec{\Omg}}}

\newcommand{\cont}{\mathcal{C}^{0}}
\newcommand{\smooth}{\mathcal{C}^{\infty}}
\newcommand{\anal}{\mathcal{C}^{\omega}}
\newcommand{\sconn}{\textsf{A}}
\newcommand{\striad}{\large\texttt{T}_{\infty}}
\newcommand{\reg}{\succ_{reg}}
\newcommand{\sref}{\prec_{sref}}
\newcommand{\tref}{\prec_{tref}}
\newcommand{\ssmooth}{\EuScript{C}^{\infty}}

\newcommand{\struc}{\mathbf{A}}
\newcommand{\sstruc}{\mathbb{A}}

\newcommand{\triad}{{\mathfrak{T}}}
\newcommand{\ctriad}{{\mathfrak{D}}\triad}

\newcommand{\vol}{\varpi}
\newcommand{\wee}{\,{\scriptstyle\wedge}\,}
\newcommand{\bull}{{\scriptstyle\bullet}}

\newcommand{\com}{\mathbb{C}}
\newcommand{\mapto}{\longrightarrow}

\newcommand{\Z}{\mathbb{Z}}
\newcommand{\R}{\mathbb{R}}

\newcommand{\U}{\mathfrak{U}}
\newcommand{\hole}{{\mathcal{O}}}

\makeatletter

\def\diagram{\m@th\leftwidth=\z@ \rightwidth=\z@ \topheight=\z@
\botheight=\z@ \setbox\@picbox\hbox\bgroup}

\def\enddiagram{\egroup\wd\@picbox\rightwidth\unitlength
\ht\@picbox\topheight\unitlength \dp\@picbox\botheight\unitlength
\hskip\leftwidth\unitlength\box\@picbox}

\def\bfig{\begin{diagram}}
\def\efig{\end{diagram}}
\newcount\wideness \newcount\leftwidth \newcount\rightwidth
\newcount\highness \newcount\topheight \newcount\botheight

\def\ratchet#1#2{\ifnum#1<#2 \global #1=#2 \fi}

\def\putbox(#1,#2)#3{%
\horsize{\wideness}{#3} \divide\wideness by 2 {\advance\wideness
by #1 \ratchet{\rightwidth}{\wideness}} {\advance\wideness by -#1
\ratchet{\leftwidth}{\wideness}} \vertsize{\highness}{#3}
\divide\highness by 2 {\advance\highness by #2
\ratchet{\topheight}{\highness}} {\advance\highness by -#2
\ratchet{\botheight}{\highness}} \put(#1,#2){\makebox(0,0){$#3$}}}

\def\putlbox(#1,#2)#3{%
\horsize{\wideness}{#3} {\advance\wideness by #1
\ratchet{\rightwidth}{\wideness}} {\ratchet{\leftwidth}{-#1}}
\vertsize{\highness}{#3} \divide\highness by 2 {\advance\highness
by #2 \ratchet{\topheight}{\highness}} {\advance\highness by -#2
\ratchet{\botheight}{\highness}}
\put(#1,#2){\makebox(0,0)[l]{$#3$}}}

\def\putrbox(#1,#2)#3{%
\horsize{\wideness}{#3} {\ratchet{\rightwidth}{#1}}
{\advance\wideness by -#1 \ratchet{\leftwidth}{\wideness}}
\vertsize{\highness}{#3} \divide\highness by 2 {\advance\highness
by #2 \ratchet{\topheight}{\highness}} {\advance\highness by -#2
\ratchet{\botheight}{\highness}}
\put(#1,#2){\makebox(0,0)[r]{$#3$}}}

\def\adjust[#1]{} 

\newcount \coefa
\newcount \coefb
\newcount \coefc
\newcount\tempcounta
\newcount\tempcountb
\newcount\tempcountc
\newcount\tempcountd
\newcount\xext
\newcount\yext
\newcount\xoff
\newcount\yoff
\newcount\gap%
\newcount\arrowtypea
\newcount\arrowtypeb
\newcount\arrowtypec
\newcount\arrowtyped
\newcount\arrowtypee
\newcount\height
\newcount\width
\newcount\xpos
\newcount\ypos
\newcount\run
\newcount\rise
\newcount\arrowlength
\newcount\halflength
\newcount\arrowtype
\newdimen\tempdimen
\newdimen\xlen
\newdimen\ylen
\newsavebox{\tempboxa}%
\newsavebox{\tempboxb}%
\newsavebox{\tempboxc}%

\newdimen\w@dth

\def\setw@dth#1#2{\setbox\z@\hbox{\m@th$#1$}\w@dth=\wd\z@
\setbox\@ne\hbox{\m@th$#2$}\ifnum\w@dth<\wd\@ne \w@dth=\wd\@ne \fi
\advance\w@dth by 1.2em}

\def\t@^#1_#2{\allowbreak\def\n@one{#1}\def\n@two{#2}\mathrel
{\setw@dth{#1}{#2} \mathop{\hbox to
\w@dth{\rightarrowfill}}\limits \ifx\n@one\empty\else
^{\box\z@}\fi \ifx\n@two\empty\else _{\box\@ne}\fi}}
\def\t@@^#1{\@ifnextchar_{\t@^{#1}}{\t@^{#1}_{}}}
\def\to{\@ifnextchar^{\t@@}{\t@@^{}}}

\def\t@left^#1_#2{\def\n@one{#1}\def\n@two{#2}\mathrel{\setw@dth{#1}{#2}
\mathop{\hbox to \w@dth{\leftarrowfill}}\limits
\ifx\n@one\empty\else ^{\box\z@}\fi \ifx\n@two\empty\else
_{\box\@ne}\fi}}
\def\t@@left^#1{\@ifnextchar_{\t@left^{#1}}{\t@left^{#1}_{}}}
\def\toleft{\@ifnextchar^{\t@@left}{\t@@left^{}}}

\def\two@^#1_#2{\allowbreak
\def\n@one{#1}\def\n@two{#2}\mathrel{\setw@dth{#1}{#2}
\mathop{\vcenter{\lineskip\z@\baselineskip\z@
                 \hbox to \w@dth{\rightarrowfill}%
                 \hbox to \w@dth{\rightarrowfill}}%
       }\limits
\ifx\n@one\empty\else ^{\box\z@}\fi \ifx\n@two\empty\else
_{\box\@ne}\fi}}
\def\tw@@^#1{\@ifnextchar _{\two@^{#1}}{\two@^{#1}_{}}}
\def\two{\@ifnextchar ^{\tw@@}{\tw@@^{}}}

\def\tofr@^#1_#2{\def\n@one{#1}\def\n@two{#2}\mathrel{\setw@dth{#1}{#2}
\mathop{\vcenter{\hbox to \w@dth{\rightarrowfill}\kern-1.7ex
                 \hbox to \w@dth{\leftarrowfill}}%
       }\limits
\ifx\n@one\empty\else ^{\box\z@}\fi \ifx\n@two\empty\else
_{\box\@ne}\fi}}
\def\t@fr@^#1{\@ifnextchar_ {\tofr@^{#1}}{\tofr@^{#1}_{}}}
\def\tofro{\@ifnextchar^ {\t@fr@}{\t@fr@^{}}}

\def\mon{\mathop{\m@th\hbox to
      14.6\P@{\lasyb\char'51\hskip-2.1\P@$\arrext$\hss
$\mathord\rightarrow$}}\limits} 
\def\leftmono{\mathrel{\m@th\hbox to
14.6\P@{$\mathord\leftarrow$\hss$\arrext$\hskip-2.1\P@\lasyb\char'50%
}}\limits} 
\mathchardef\arrext="0200       

\def\settypes(#1,#2,#3){\arrowtypea#1 \arrowtypeb#2 \arrowtypec#3}
\def\settoheight#1#2{\setbox\@tempboxa\hbox{#2}#1\ht\@tempboxa\relax}%
\def\settodepth#1#2{\setbox\@tempboxa\hbox{#2}#1\dp\@tempboxa\relax}%
\def\settokens`#1`#2`#3`#4`{%
     \def\tokena{#1}\def\tokenb{#2}\def\tokenc{#3}\def\tokend{#4}}
\def\setsqparms[#1`#2`#3`#4;#5`#6]{%
\arrowtypea #1 \arrowtypeb #2 \arrowtypec #3 \arrowtyped #4 \width
#5 \height #6 }
\def\setpos(#1,#2){\xpos=#1 \ypos#2}

\def\settriparms[#1`#2`#3;#4]{\settripairparms[#1`#2`#3`1`1;#4]}%

\def\settripairparms[#1`#2`#3`#4`#5;#6]{%
\arrowtypea #1 \arrowtypeb #2 \arrowtypec #3 \arrowtyped #4
\arrowtypee #5 \width #6 \height #6 }

\def\resetparms{\settripairparms[1`1`1`1`1;500]\width 500}

\resetparms

\def\mvector(#1,#2)#3{
\put(0,0){\vector(#1,#2){#3}}%
\put(0,0){\vector(#1,#2){26}}%
}
\def\evector(#1,#2)#3{{
\arrowlength #3
\put(0,0){\vector(#1,#2){\arrowlength}}%
\advance \arrowlength by-30
\put(0,0){\vector(#1,#2){\arrowlength}}%
}}

\def\horsize#1#2{%
\settowidth{\tempdimen}{$#2$}%
#1=\tempdimen \divide #1 by\unitlength }

\def\vertsize#1#2{%
\settoheight{\tempdimen}{$#2$}%
#1=\tempdimen
\settodepth{\tempdimen}{$#2$}%
\advance #1 by\tempdimen \divide #1 by\unitlength }

\def\putvector(#1,#2)(#3,#4)#5#6{{%
\ifnum3<\arrowtype \putdashvector(#1,#2)(#3,#4)#5\arrowtype \else
\ifnum\arrowtype<-3 \putdashvector(#1,#2)(#3,#4)#5\arrowtype \else
\xpos=#1 \ypos=#2 \run=#3 \rise=#4 \arrowlength=#5 \ifnum
\arrowtype<0
    \ifnum \run=0
        \advance \ypos by-\arrowlength
    \else
        \tempcounta \arrowlength
        \multiply \tempcounta by\rise
        \divide \tempcounta by\run
        \ifnum\run>0
            \advance \xpos by\arrowlength
            \advance \ypos by\tempcounta
        \else
            \advance \xpos by-\arrowlength
            \advance \ypos by-\tempcounta
        \fi
    \fi
    \multiply \arrowtype by-1
    \multiply \rise by-1
    \multiply \run by-1
\fi \ifcase \arrowtype
\or \put(\xpos,\ypos){\vector(\run,\rise){\arrowlength}}%
\or \put(\xpos,\ypos){\mvector(\run,\rise)\arrowlength}%
\or \put(\xpos,\ypos){\evector(\run,\rise){\arrowlength}}%
\fi\fi\fi }}

\def\putsplitvector(#1,#2)#3#4{
\xpos #1 \ypos #2 \arrowtype #4 \halflength #3 \arrowlength #3
\gap 140 \advance \halflength by-\gap \divide \halflength by2
\ifnum\arrowtype>0
   \ifcase \arrowtype
   \or \put(\xpos,\ypos){\line(0,-1){\halflength}}%
       \advance\ypos by-\halflength
       \advance\ypos by-\gap
       \put(\xpos,\ypos){\vector(0,-1){\halflength}}%
   \or \put(\xpos,\ypos){\line(0,-1)\halflength}%
       \put(\xpos,\ypos){\vector(0,-1)3}%
       \advance\ypos by-\halflength
       \advance\ypos by-\gap
       \put(\xpos,\ypos){\vector(0,-1){\halflength}}%
   \or \put(\xpos,\ypos){\line(0,-1)\halflength}%
       \advance\ypos by-\halflength
       \advance\ypos by-\gap
       \put(\xpos,\ypos){\evector(0,-1){\halflength}}%
   \fi
\else \arrowtype=-\arrowtype
   \ifcase\arrowtype
   \or \advance \ypos by-\arrowlength
       \put(\xpos,\ypos){\line(0,1){\halflength}}%
       \advance\ypos by\halflength
       \advance\ypos by\gap
       \put(\xpos,\ypos){\vector(0,1){\halflength}}%
   \or \advance \ypos by-\arrowlength
       \put(\xpos,\ypos){\line(0,1)\halflength}%
       \put(\xpos,\ypos){\vector(0,1)3}%
       \advance\ypos by\halflength
       \advance\ypos by\gap
       \put(\xpos,\ypos){\vector(0,1){\halflength}}%
   \or \advance \ypos by-\arrowlength
       \put(\xpos,\ypos){\line(0,1)\halflength}%
       \advance\ypos by\halflength
       \advance\ypos by\gap
       \put(\xpos,\ypos){\evector(0,1){\halflength}}%
   \fi
\fi }

\def\putmorphism(#1)(#2,#3)[#4`#5`#6]#7#8#9{{%
\run #2 \rise #3 \ifnum\rise=0
  \puthmorphism(#1)[#4`#5`#6]{#7}{#8}#9%
\else\ifnum\run=0
  \putvmorphism(#1)[#4`#5`#6]{#7}{#8}#9%
\else
\setpos(#1)%
\arrowlength #7 \arrowtype #8 \ifnum\run=0 \else\ifnum\rise=0
\else \ifnum\run>0
    \coefa=1
\else
   \coefa=-1
\fi \ifnum\arrowtype>0
   \coefb=0
   \coefc=-1
\else
   \coefb=\coefa
   \coefc=1
   \arrowtype=-\arrowtype
\fi \width=2 \multiply \width by\run \divide \width by\rise \ifnum
\width<0  \width=-\width\fi \advance\width by60 \if l#9
\width=-\width\fi
\putbox(\xpos,\ypos){#4}
{\multiply \coefa by\arrowlength
\advance\xpos by\coefa \multiply \coefa by\rise \divide \coefa
by\run \advance \ypos by\coefa
\putbox(\xpos,\ypos){#5} }%
{\multiply \coefa by\arrowlength
\divide \coefa by2 \advance \xpos by\coefa \advance \xpos by\width
\multiply \coefa by\rise \divide \coefa by\run \advance \ypos
by\coefa
\if l#9%
   \putrbox(\xpos,\ypos){#6}%
\else\if r#9%
   \putlbox(\xpos,\ypos){#6}%
\fi\fi }%
{\multiply \rise by-\coefc
\multiply \run by-\coefc \multiply \coefb by\arrowlength \advance
\xpos by\coefb \multiply \coefb by\rise \divide \coefb by\run
\advance \ypos by\coefb \multiply \coefc by70 \advance \ypos
by\coefc \multiply \coefc by\run \divide \coefc by\rise \advance
\xpos by\coefc \multiply \coefa by140 \multiply \coefa by\run
\divide \coefa by\rise \advance \arrowlength by\coefa
\ifcase\arrowtype
\or \put(\xpos,\ypos){\vector(\run,\rise){\arrowlength}}%
\or \put(\xpos,\ypos){\mvector(\run,\rise){\arrowlength}}%
\or \put(\xpos,\ypos){\evector(\run,\rise){\arrowlength}}%
\fi}\fi\fi\fi\fi}}

\newcount\numbdashes \newcount\lengthdash \newcount\increment

\def\howmanydashes{
\numbdashes=\arrowlength \lengthdash=40 \divide\numbdashes by
\lengthdash \lengthdash=\arrowlength \divide\lengthdash by
\numbdashes
\increment=\lengthdash \multiply\lengthdash by 3
\divide\lengthdash by 5 }

\def\putdashvector(#1)(#2,#3)#4#5{%
\ifnum#3=0 \putdashhvector(#1){#4}#5 \else \ifnum#2=0
\putdashvvector(#1){#4}#5\fi\fi}

\def\putdashhvector(#1,#2)#3#4{{%
\arrowlength=#3 \howmanydashes
\multiput(#1,#2)(\increment,0){\numbdashes}%
{\vrule height .4pt width \lengthdash\unitlength} \arrowtype=#4
\xpos=#1 \ifnum\arrowtype<0 \advance\arrowtype by 7 \fi
\ifcase\arrowtype \or \advance\xpos by 10
    \put(\xpos,#2){\vector(-1,0){\lengthdash}}
    \advance\xpos by 40
    \put(\xpos,#2){\vector(-1,0){\lengthdash}}
\or \advance \xpos by 10
    \put(\xpos,#2){\vector(-1,0){\lengthdash}}
    \advance\xpos by  \arrowlength
    \advance\xpos by  -50
    \put(\xpos,#2){\vector(-1,0){\lengthdash}}
\or \advance\xpos by 10
    \put(\xpos,#2){\vector(-1,0){\lengthdash}}
\or \advance\xpos by \arrowlength
    \advance\xpos by -\lengthdash
    \put(\xpos,#2){\vector(1,0){\lengthdash}}
\or {\advance\xpos by 10
    \put(\xpos,#2){\vector(1,0){\lengthdash}}}
    \advance\xpos by \arrowlength
    \advance\xpos by -\lengthdash
    \put(\xpos,#2){\vector(1,0){\lengthdash}}
\or \advance\xpos by \arrowlength
    \advance\xpos by -\lengthdash
    \put(\xpos,#2){\vector(1,0){\lengthdash}}
    \advance\xpos by -40
    \put(\xpos,#2){\vector(1,0){\lengthdash}}
   \fi
}}

\def\putdashvvector(#1,#2)#3#4{{%
\arrowlength=#3 \howmanydashes \ypos=#2 \advance\ypos by
-\arrowlength
\multiput(#1,#2)(0,\increment){\numbdashes}%
    {\vrule width .4pt height \lengthdash\unitlength}
\arrowtype=#4 \ypos=#2 \ifnum\arrowtype<0 \advance\arrowtype by 7
\fi \ifcase\arrowtype \or \advance\ypos by \arrowlength
\advance\ypos by -40
    \put(#1,\ypos){\vector(0,1){\lengthdash}}
    \advance\ypos by -40
    \put(#1,\ypos){\vector(0,1){\lengthdash}}
\or \advance\ypos by 10
    \put(#1,\ypos){\vector(0,1){\lengthdash}}
    \advance\ypos by \arrowlength \advance\ypos by -40
    \put(#1,\ypos){\vector(0,1){\lengthdash}}
\or \advance\ypos by \arrowlength \advance\ypos by -40
    \put(#1,\ypos){\vector(0,1){\lengthdash}}
\or \advance\ypos by 10
    \put(#1,\ypos){\vector(0,-1){\lengthdash}}
\or \advance\ypos by 10
    \put(#1,\ypos){\vector(0,-1){\lengthdash}}
    \advance\ypos by \arrowlength \advance\ypos by -40
    \put(#1,\ypos){\vector(0,-1){\lengthdash}}
\or \advance\ypos by 10
    \put(#1,\ypos){\vector(0,-1){\lengthdash}}
    \advance\ypos by 40
    \put(#1,\ypos){\vector(0,-1){\lengthdash}}
\fi }}

\def\puthmorphism(#1,#2)[#3`#4`#5]#6#7#8{{%
\xpos #1 \ypos #2 \width #6 \arrowlength #6 \arrowtype=#7
\putbox(\xpos,\ypos){#3\vphantom{#4}}%
{\advance \xpos by\arrowlength
\putbox(\xpos,\ypos){\vphantom{#3}#4}}%
\horsize{\tempcounta}{#3}%
\horsize{\tempcountb}{#4}%
\divide \tempcounta by2 \divide \tempcountb by2 \advance
\tempcounta by30 \advance \tempcountb by30 \advance \xpos
by\tempcounta \advance \arrowlength by-\tempcounta \advance
\arrowlength by-\tempcountb
\putvector(\xpos,\ypos)(1,0)\arrowlength\arrowtype \divide
\arrowlength by2 \advance \xpos by\arrowlength
\vertsize{\tempcounta}{#5}%
\divide\tempcounta by2 \advance \tempcounta by20
\if a#8 %
   \advance \ypos by\tempcounta
   \putbox(\xpos,\ypos){#5}%
\else
   \advance \ypos by-\tempcounta
   \putbox(\xpos,\ypos){#5}%
\fi}}

\def\putvmorphism(#1,#2)[#3`#4`#5]#6#7#8{{%
\xpos #1 \ypos #2 \arrowlength #6 \arrowtype #7
\settowidth{\xlen}{$#5$}%
\putbox(\xpos,\ypos){#3}%
{\advance \ypos by-\arrowlength
\putbox(\xpos,\ypos){#4}}%
{\advance\arrowlength by-140 \advance \ypos by-70 \ifdim\xlen>0pt
   \if m#8%
      \putsplitvector(\xpos,\ypos)\arrowlength\arrowtype
   \else
   \putvector(\xpos,\ypos)(0,-1)\arrowlength\arrowtype
   \fi
\else
   \putvector(\xpos,\ypos)(0,-1)\arrowlength\arrowtype
\fi}%
\ifdim\xlen>0pt
   \divide \arrowlength by2
   \advance\ypos by-\arrowlength
   \if l#8%
      \advance \xpos by-40
      \putrbox(\xpos,\ypos){#5}%
   \else\if r#8%
      \advance \xpos by40
      \putlbox(\xpos,\ypos){#5}%
   \else
      \putbox(\xpos,\ypos){#5}%
   \fi\fi
\fi }}

\def\putsquarep<#1>(#2)[#3;#4`#5`#6`#7]{{%
\setsqparms[#1]%
\setpos(#2)%
\settokens`#3`%
\puthmorphism(\xpos,\ypos)[\tokenc`\tokend`{#7}]{\width}{\arrowtyped}b%
\advance\ypos by \height
\puthmorphism(\xpos,\ypos)[\tokena`\tokenb`{#4}]{\width}{\arrowtypea}a%
\putvmorphism(\xpos,\ypos)[``{#5}]{\height}{\arrowtypeb}l%
\advance\xpos by \width
\putvmorphism(\xpos,\ypos)[``{#6}]{\height}{\arrowtypec}r%
}}

\def\putsquare{\@ifnextchar <{\putsquarep}{\putsquarep%
   <\arrowtypea`\arrowtypeb`\arrowtypec`\arrowtyped;\width`\height>}}
\def\square{\@ifnextchar< {\squarep}{\squarep
   <\arrowtypea`\arrowtypeb`\arrowtypec`\arrowtyped;\width`\height>}}
\def\squarep<#1>[#2`#3`#4`#5;#6`#7`#8`#9]{{
\setsqparms[#1]
\diagram
\putsquarep<\arrowtypea`\arrowtypeb`\arrowtypec`
\arrowtyped;\width`\height>
(0,0)[#2`#3`#4`{#5};#6`#7`#8`{#9}]
\enddiagram
}}                                                 
\def\putptrianglep<#1>(#2,#3)[#4`#5`#6;#7`#8`#9]{{%
\settriparms[#1]%
\xpos=#2 \ypos=#3 \advance\ypos by \height
\puthmorphism(\xpos,\ypos)[#4`#5`{#7}]{\height}{\arrowtypea}a%
\putvmorphism(\xpos,\ypos)[`#6`{#8}]{\height}{\arrowtypeb}l%
\advance\xpos by\height
\putmorphism(\xpos,\ypos)(-1,-1)[``{#9}]{\height}{\arrowtypec}r%
}}

\def\putptriangle{\@ifnextchar <{\putptrianglep}{\putptrianglep
   <\arrowtypea`\arrowtypeb`\arrowtypec;\height>}}
\def\ptriangle{\@ifnextchar <{\ptrianglep}{\ptrianglep
   <\arrowtypea`\arrowtypeb`\arrowtypec;\height>}}
\def\ptrianglep<#1>[#2`#3`#4;#5`#6`#7]{{
\settriparms[#1]
\diagram
\putptrianglep<\arrowtypea`\arrowtypeb`
\arrowtypec;\height>
(0,0)[#2`#3`#4;#5`#6`{#7}]
\enddiagram
}}                                            

\def\putqtrianglep<#1>(#2,#3)[#4`#5`#6;#7`#8`#9]{{%
\settriparms[#1]%
\xpos=#2 \ypos=#3 \advance\ypos by\height
\puthmorphism(\xpos,\ypos)[#4`#5`{#7}]{\height}{\arrowtypea}a%
\putmorphism(\xpos,\ypos)(1,-1)[``{#8}]{\height}{\arrowtypeb}l%
\advance\xpos by\height
\putvmorphism(\xpos,\ypos)[`#6`{#9}]{\height}{\arrowtypec}r%
}}

\def\putqtriangle{\@ifnextchar <{\putqtrianglep}{\putqtrianglep
   <\arrowtypea`\arrowtypeb`\arrowtypec;\height>}}
\def\qtriangle{\@ifnextchar <{\qtrianglep}{\qtrianglep
   <\arrowtypea`\arrowtypeb`\arrowtypec;\height>}}
\def\qtrianglep<#1>[#2`#3`#4;#5`#6`#7]{{
\settriparms[#1]
\width=\height                                
\diagram
\putqtrianglep<\arrowtypea`\arrowtypeb`
\arrowtypec;\height>
(0,0)[#2`#3`#4;#5`#6`{#7}]
\enddiagram
}}

\def\putdtrianglep<#1>(#2,#3)[#4`#5`#6;#7`#8`#9]{{%
\settriparms[#1]%
\xpos=#2 \ypos=#3
\puthmorphism(\xpos,\ypos)[#5`#6`{#9}]{\height}{\arrowtypec}b%
\advance\xpos by \height \advance\ypos by\height
\putmorphism(\xpos,\ypos)(-1,-1)[``{#7}]{\height}{\arrowtypea}l%
\putvmorphism(\xpos,\ypos)[#4``{#8}]{\height}{\arrowtypeb}r%
}}

\def\putdtriangle{\@ifnextchar <{\putdtrianglep}{\putdtrianglep
   <\arrowtypea`\arrowtypeb`\arrowtypec;\height>}}
\def\dtriangle{\@ifnextchar <{\dtrianglep}{\dtrianglep
   <\arrowtypea`\arrowtypeb`\arrowtypec;\height>}}
\def\dtrianglep<#1>[#2`#3`#4;#5`#6`#7]{{
\settriparms[#1]
\width=\height                                
\diagram
\putdtrianglep<\arrowtypea`\arrowtypeb`
\arrowtypec;\height>
(0,0)[#2`#3`#4;#5`#6`{#7}]
\enddiagram
}}

\def\putbtrianglep<#1>(#2,#3)[#4`#5`#6;#7`#8`#9]{{%
\settriparms[#1]%
\xpos=#2 \ypos=#3
\puthmorphism(\xpos,\ypos)[#5`#6`{#9}]{\height}{\arrowtypec}b%
\advance\ypos by\height
\putmorphism(\xpos,\ypos)(1,-1)[``{#8}]{\height}{\arrowtypeb}r%
\putvmorphism(\xpos,\ypos)[#4``{#7}]{\height}{\arrowtypea}l%
}}

\def\putbtriangle{\@ifnextchar <{\putbtrianglep}{\putbtrianglep
   <\arrowtypea`\arrowtypeb`\arrowtypec;\height>}}
\def\btriangle{\@ifnextchar <{\btrianglep}{\btrianglep
   <\arrowtypea`\arrowtypeb`\arrowtypec;\height>}}
\def\btrianglep<#1>[#2`#3`#4;#5`#6`#7]{{
\settriparms[#1]
\width=\height                               
\diagram
\putbtrianglep<\arrowtypea`\arrowtypeb`
\arrowtypec;\height>
(0,0)[#2`#3`#4;#5`#6`{#7}]
\enddiagram
}}

\def\putAtrianglep<#1>(#2,#3)[#4`#5`#6;#7`#8`#9]{{%
\settriparms[#1]%
\xpos=#2 \ypos=#3 {\multiply \height by2
\puthmorphism(\xpos,\ypos)[#5`#6`{#9}]{\height}{\arrowtypec}b}%
\advance\xpos by\height \advance\ypos by\height
\putmorphism(\xpos,\ypos)(-1,-1)[#4``{#7}]{\height}{\arrowtypea}l%
\putmorphism(\xpos,\ypos)(1,-1)[``{#8}]{\height}{\arrowtypeb}r%
}}

\def\putAtriangle{\@ifnextchar <{\putAtrianglep}{\putAtrianglep
   <\arrowtypea`\arrowtypeb`\arrowtypec;\height>}}
\def\Atriangle{\@ifnextchar <{\Atrianglep}{\Atrianglep
   <\arrowtypea`\arrowtypeb`\arrowtypec;\height>}}
\def\Atrianglep<#1>[#2`#3`#4;#5`#6`#7]{{
\settriparms[#1]
\width=\height                                     
\diagram
\putAtrianglep<\arrowtypea`\arrowtypeb`
\arrowtypec;\height>
(0,0)[#2`#3`#4;#5`#6`{#7}]
\enddiagram
}}

\def\putAtrianglepairp<#1>(#2)[#3;#4`#5`#6`#7`#8]{{%
\settripairparms[#1]%
\setpos(#2)%
\settokens`#3`%
\puthmorphism(\xpos,\ypos)[\tokenb`\tokenc`{#7}]{\height}{\arrowtyped}b%
\advance\xpos by\height
\puthmorphism(\xpos,\ypos)[\phantom{\tokenc}`\tokend`{#8}]%
{\height}{\arrowtypee}b%
\advance\ypos by\height
\putmorphism(\xpos,\ypos)(-1,-1)[\tokena``{#4}]{\height}{\arrowtypea}l%
\putvmorphism(\xpos,\ypos)[``{#5}]{\height}{\arrowtypeb}m%
\putmorphism(\xpos,\ypos)(1,-1)[``{#6}]{\height}{\arrowtypec}r%
}}

\def\putAtrianglepair{\@ifnextchar <{\putAtrianglepairp}{\putAtrianglepairp%
   <\arrowtypea`\arrowtypeb`\arrowtypec`\arrowtyped`\arrowtypee;\height>}}
\def\Atrianglepair{\@ifnextchar <{\Atrianglepairp}{\Atrianglepairp%
   <\arrowtypea`\arrowtypeb`\arrowtypec`\arrowtyped`\arrowtypee;\height>}}

\def\Atrianglepairp<#1>[#2;#3`#4`#5`#6`#7]{{
\settripairparms[#1]
\settokens`#2`
\width=\height                                
\diagram
\putAtrianglepairp                            
<\arrowtypea`\arrowtypeb`\arrowtypec`
\arrowtyped`\arrowtypee;\height>
(0,0)[{#2};#3`#4`#5`#6`{#7}]
\enddiagram
}}

\def\putVtrianglep<#1>(#2,#3)[#4`#5`#6;#7`#8`#9]{{%
\settriparms[#1]%
\xpos=#2 \ypos=#3 \advance\ypos by\height {\multiply\height by2
\puthmorphism(\xpos,\ypos)[#4`#5`{#7}]{\height}{\arrowtypea}a}%
\putmorphism(\xpos,\ypos)(1,-1)[`#6`{#8}]{\height}{\arrowtypeb}l%
\advance\xpos by\height \advance\xpos by\height
\putmorphism(\xpos,\ypos)(-1,-1)[``{#9}]{\height}{\arrowtypec}r%
}}

\def\putVtriangle{\@ifnextchar <{\putVtrianglep}{\putVtrianglep
   <\arrowtypea`\arrowtypeb`\arrowtypec;\height>}}
\def\Vtriangle{\@ifnextchar <{\Vtrianglep}{\Vtrianglep
   <\arrowtypea`\arrowtypeb`\arrowtypec;\height>}}
\def\Vtrianglep<#1>[#2`#3`#4;#5`#6`#7]{{
\settriparms[#1]
\width=\height                                 
\diagram
\putVtrianglep<\arrowtypea`\arrowtypeb`
\arrowtypec;\height>
(0,0)[#2`#3`#4;#5`#6`{#7}]
\enddiagram
}}

\def\putVtrianglepairp<#1>(#2)[#3;#4`#5`#6`#7`#8]{{
\settripairparms[#1]%
\setpos(#2)%
\settokens`#3`%
\advance\ypos by\height
\putmorphism(\xpos,\ypos)(1,-1)[`\tokend`{#6}]{\height}{\arrowtypec}l%
\puthmorphism(\xpos,\ypos)[\tokena`\tokenb`{#4}]{\height}{\arrowtypea}a%
\advance\xpos by\height
\puthmorphism(\xpos,\ypos)[\phantom{\tokenb}`\tokenc`{#5}]%
{\height}{\arrowtypeb}a%
\putvmorphism(\xpos,\ypos)[``{#7}]{\height}{\arrowtyped}m%
\advance\xpos by\height
\putmorphism(\xpos,\ypos)(-1,-1)[``{#8}]{\height}{\arrowtypee}r%
}}

\def\putVtrianglepair{\@ifnextchar <{\putVtrianglepairp}{\putVtrianglepairp%
    <\arrowtypea`\arrowtypeb`\arrowtypec`\arrowtyped`\arrowtypee;\height>}}
\def\Vtrianglepair{\@ifnextchar <{\Vtrianglepairp}{\Vtrianglepairp%
    <\arrowtypea`\arrowtypeb`\arrowtypec`\arrowtyped`\arrowtypee;\height>}}
\def\Vtrianglepairp<#1>[#2;#3`#4`#5`#6`#7]{{
\settripairparms[#1]
\settokens`#2`
\diagram
\putVtrianglepairp                             
<\arrowtypea`\arrowtypeb`\arrowtypec`
\arrowtyped`\arrowtypee;\height>
(0,0)[{#2};#3`#4`#5`#6`{#7}]
\enddiagram
}}

\def\putCtrianglep<#1>(#2,#3)[#4`#5`#6;#7`#8`#9]{{%
\settriparms[#1]%
\xpos=#2 \ypos=#3 \advance\ypos by\height
\putmorphism(\xpos,\ypos)(1,-1)[``{#9}]{\height}{\arrowtypec}l%
\advance\xpos by\height \advance\ypos by\height
\putmorphism(\xpos,\ypos)(-1,-1)[#4`#5`{#7}]{\height}{\arrowtypea}l%
{\multiply\height by 2
\putvmorphism(\xpos,\ypos)[`#6`{#8}]{\height}{\arrowtypeb}r}%
}}

\def\putCtriangle{\@ifnextchar <{\putCtrianglep}{\putCtrianglep
    <\arrowtypea`\arrowtypeb`\arrowtypec;\height>}}
\def\Ctriangle{\@ifnextchar <{\Ctrianglep}{\Ctrianglep
    <\arrowtypea`\arrowtypeb`\arrowtypec;\height>}}
\def\Ctrianglep<#1>[#2`#3`#4;#5`#6`#7]{{
\settriparms[#1]
\width=\height                               
\diagram
\putCtrianglep<\arrowtypea`\arrowtypeb`
\arrowtypec;\height>
(0,0)[#2`#3`#4;#5`#6`{#7}]
\enddiagram
}}                                           
\def\putDtrianglep<#1>(#2,#3)[#4`#5`#6;#7`#8`#9]{{%
\settriparms[#1]%
\xpos=#2 \ypos=#3 \advance\xpos by\height \advance\ypos by\height
\putmorphism(\xpos,\ypos)(-1,-1)[``{#9}]{\height}{\arrowtypec}r%
\advance\xpos by-\height \advance\ypos by\height
\putmorphism(\xpos,\ypos)(1,-1)[`#5`{#8}]{\height}{\arrowtypeb}r%
{\multiply\height by 2
\putvmorphism(\xpos,\ypos)[#4`#6`{#7}]{\height}{\arrowtypea}l}%
}}

\def\putDtriangle{\@ifnextchar <{\putDtrianglep}{\putDtrianglep
    <\arrowtypea`\arrowtypeb`\arrowtypec;\height>}}
\def\Dtriangle{\@ifnextchar <{\Dtrianglep}{\Dtrianglep
   <\arrowtypea`\arrowtypeb`\arrowtypec;\height>}}
\def\Dtrianglep<#1>[#2`#3`#4;#5`#6`#7]{{
\settriparms[#1]
\width=\height                              
\diagram
\putDtrianglep<\arrowtypea`\arrowtypeb`
\arrowtypec;\height>
(0,0)[#2`#3`#4;#5`#6`{#7}]
\enddiagram
}}                                          
\def\setrecparms[#1`#2]{\width=#1 \height=#2}%

\def\recursep<#1`#2>[#3;#4`#5`#6`#7`#8]{{\m@th
\width=#1 \height=#2 \settokens`#3`
\settowidth{\tempdimen}{$\tokena$} \ifdim\tempdimen=0pt
  \savebox{\tempboxa}{\hbox{$\tokenb$}}%
  \savebox{\tempboxb}{\hbox{$\tokend$}}%
  \savebox{\tempboxc}{\hbox{$#6$}}%
\else
  \savebox{\tempboxa}{\hbox{$\hbox{$\tokena$}\times\hbox{$\tokenb$}$}}%
  \savebox{\tempboxb}{\hbox{$\hbox{$\tokena$}\times\hbox{$\tokend$}$}}%
  \savebox{\tempboxc}{\hbox{$\hbox{$\tokena$}\times\hbox{$#6$}$}}%
\fi \ypos=\height \divide\ypos by 2 \xpos=\ypos \advance\xpos by
\width \bfig
\putCtrianglep<-1`1`1;\ypos>(0,0)[`\tokenc`;#5`#6`{#7}]%
\puthmorphism(\ypos,0)[\tokend`\usebox{\tempboxb}`{#8}]{\width}{-1}b%
\puthmorphism(\ypos,\height)[\tokenb`\usebox{\tempboxa}`{#4}]{\width}{-1}a%
\advance\ypos by \width
\putvmorphism(\ypos,\height)[``\usebox{\tempboxc}]{\height}1r%
\efig }}

\def\recurse{\@ifnextchar <{\recursep}{\recursep<\width`\height>}}

\def\puttwohmorphisms(#1,#2)[#3`#4;#5`#6]#7#8#9{{%
\puthmorphism(#1,#2)[#3`#4`]{#7}0a \ypos=#2 \advance\ypos by 20
\puthmorphism(#1,\ypos)[\phantom{#3}`\phantom{#4}`#5]{#7}{#8}a
\advance\ypos by -40
\puthmorphism(#1,\ypos)[\phantom{#3}`\phantom{#4}`#6]{#7}{#9}b }}

\def\puttwovmorphisms(#1,#2)[#3`#4;#5`#6]#7#8#9{{%
\putvmorphism(#1,#2)[#3`#4`]{#7}0a \xpos=#1 \advance\xpos by -20
\putvmorphism(\xpos,#2)[\phantom{#3}`\phantom{#4}`#5]{#7}{#8}l
\advance\xpos by 40
\putvmorphism(\xpos,#2)[\phantom{#3}`\phantom{#4}`#6]{#7}{#9}r }}

\def\puthcoequalizer(#1)[#2`#3`#4;#5`#6`#7]#8#9{{%
\setpos(#1)%
\puttwohmorphisms(\xpos,\ypos)[#2`#3;#5`#6]{#8}11%
\advance\xpos by #8
\puthmorphism(\xpos,\ypos)[\phantom{#3}`#4`#7]{#8}1{#9} }}

\def\putvcoequalizer(#1)[#2`#3`#4;#5`#6`#7]#8#9{{%
\setpos(#1)%
\puttwovmorphisms(\xpos,\ypos)[#2`#3;#5`#6]{#8}11%
\advance\ypos by -#8
\putvmorphism(\xpos,\ypos)[\phantom{#3}`#4`#7]{#8}1{#9} }}

\def\putthreehmorphisms(#1)[#2`#3;#4`#5`#6]#7(#8)#9{{%
\setpos(#1) \settypes(#8)
\if a#9 %
     \vertsize{\tempcounta}{#5}%
     \vertsize{\tempcountb}{#6}%
     \ifnum \tempcounta<\tempcountb \tempcounta=\tempcountb \fi
\else
     \vertsize{\tempcounta}{#4}%
     \vertsize{\tempcountb}{#5}%
     \ifnum \tempcounta<\tempcountb \tempcounta=\tempcountb \fi
\fi \advance \tempcounta by 60
\puthmorphism(\xpos,\ypos)[#2`#3`#5]{#7}{\arrowtypeb}{#9}
\advance\ypos by \tempcounta
\puthmorphism(\xpos,\ypos)[\phantom{#2}`\phantom{#3}`#4]{#7}{\arrowtypea}{#9}
\advance\ypos by -\tempcounta \advance\ypos by -\tempcounta
\puthmorphism(\xpos,\ypos)[\phantom{#2}`\phantom{#3}`#6]{#7}{\arrowtypec}{#9}
}}

\def\setarrowtoks[#1`#2`#3`#4`#5`#6]{%
\def\toka{#1}
\def\tokb{#2}
\def\tokc{#3}
\def\tokd{#4}
\def\toke{#5}
\def\tokf{#6}
}
\def\hex{\@ifnextchar <{\hexp}{\hexp<1000`400>}}
\def\hexp<#1`#2>[#3`#4`#5`#6`#7`#8;#9]{%
\setarrowtoks[#9] \yext=#2 \advance \yext by #2 \xext=#1
\advance\xext by \yext \bfig
\putCtriangle<-1`0`1;#2>(0,0)[`#5`;\tokb``\tokd] \xext=#1 \yext=#2
\advance \yext by #2
\putsquare<1`0`0`1;\xext`\yext>(#2,0)[#3`#4`#7`#8;\toka```\tokf]
\advance \xext by #2
\putDtriangle<0`1`-1;#2>(\xext,0)[`#6`;`\tokc`\toke] \efig }
\makeatother

\title{\bf\LARGE $\smooth$-Smooth Singularities Exposed:\\ Chimeras of the
Differential Spacetime Manifold}

\author{Anastasios Mallios\thanks{Algebra and Geometry Section,
Department of Mathematics, University of Athens,
Panepistimioupolis 157 84, Athens, Greece; e-mail:
amallios@cc.uoa.gr} and Ioannis Raptis\thanks{Theoretical Physics
Group, Blackett Laboratory, Imperial College of Science,
Technology and Medicine, Prince Consort Road, South Kensington,
London SW7 2BZ, UK; e-mail: i.raptis@ic.ac.uk} $^{*}$\thanks{To
whom correspondence about this paper should be directed.}}

\date{}

\begin{document}

{\small
\begin{quotation}

\noindent ``{What a curious attitude scientists have: `{\em We
still don't know that; but it is knowable and it is only a
question of time till we know it}'! As if that went without
saying...}''---\cite{witt2}

\end{quotation}

\vskip 0.3in

\centerline{$<\bullet>$$<\bullet>$$<\bullet>$$<\bullet>$$<\bullet>$$<\bullet>$$<\bullet>$}

\vskip 0.3in

\begin{quotation}

\noindent ``{Physics does not {\em explain} anything. It simply
{\em describes} cases of concomitance.}''---\cite{witt6}

\end{quotation}

\vskip 0.3in

\centerline{$<\bullet>$$<\bullet>$$<\bullet>$$<\bullet>$$<\bullet>$$<\bullet>$$<\bullet>$}

\vskip 0.3in

\begin{quotation}

\noindent ``{People who are constantly asking `{\em why}' are like
tourists, who stand in front of a building, reading B\"adeker, and
through reading about the history of the building's construction
{\it etc etc} are prevented from {\em seeing}
it.}''---\cite{witt2}

\end{quotation}

\vskip 0.3in

\centerline{$<\bullet>$$<\bullet>$$<\bullet>$$<\bullet>$$<\bullet>$$<\bullet>$$<\bullet>$}

\vskip 0.3in

\begin{quotation}

\noindent ``{Tolstoy: the meaning (importance) of something lies
in its being something that everyone can understand. That is both
true and false. What makes the object hard to understand---if it's
significant, important---is not that you have to be instructed in
abstruse matters in order to understand it, but the antithesis
between understanding the object and what people {\em want} to
see. Because of this precisely what is more obvious may be what is
most difficult to understand. It is is not a difficulty for the
intellect but one for the will that has to be
overcome.}''---\cite{witt2}

\end{quotation}

\vskip 0.3in

\centerline{$<\bullet>$$<\bullet>$$<\bullet>$$<\bullet>$$<\bullet>$$<\bullet>$$<\bullet>$}

\vskip 0.3in

\begin{quotation}

\noindent ``{How hard it is for me to {\em see what is in front of
my eyes}.}''---\cite{witt1}

\end{quotation}

\vskip 0.3in

\centerline{$<\bullet>$$<\bullet>$$<\bullet>$$<\bullet>$$<\bullet>$$<\bullet>$$<\bullet>$}

\vskip 0.3in

\centerline{\underline{\large\bf Dedication}}

\vskip 0.1in

\begin{quotation}

\noindent The present work is wholeheartedly dedicated to all
theoretical/mathematical physics' researchers who ``{\em risk
their own lives, so that they may never be heard of again}'' and,
against all odds and with wax-plugged ears {\it contra} the
Sirens' song of their contemporary research trends, take the trip
into Feynman's ``{\em wild blue yonder to see if they can figure
it out}''...\footnote{Here we have partially and loosely quoted
Feynman's words in \cite{feyn0} about the importance of a
`non-fashionable', perhaps `iconoclastic', research pursuit of
Quantum Gravity: ``{\em ...It is very important that we do not all
follow the same fashion...It's necessary to increase the amount of
variety...and the only way to do this is to implore you few guys
to take a risk with your lives that you will not be heard of
again, and go off in the wild blue yonder to see if you can figure
it out...}''.}

\end{quotation}}

\newpage

{\catcode`\ =13\global\let =\ \catcode`\^^M=13
\gdef^^M{\par\noindent}}
\def\verbatim{\tt
\catcode`\^^M=13 \catcode`\ =13 \catcode`\\=12 \catcode`\{=12
\catcode`\}=12 \catcode`\_=12 \catcode`\^=12 \catcode`\&=12
\catcode`\~=12 \catcode`\#=12 \catcode`\%=12 \catcode`\$=12
\catcode`|=0 }

\maketitle

\pagestyle{myheadings}\markboth{\centerline {\small {\sc
{Anastasios Mallios and Ioannis Raptis}}}}{\centerline
{\footnotesize {\sc {$\smooth$-Smooth Singularities Exposed:
Chimeras of the Differential Spacetime Manifold}}}}

\pagenumbering{arabic}

\begin{abstract}

\noindent The glaringly serious conflict between the principle of
general covariance of General Relativity (GR) and the existence of
$\smooth$-smooth singularities assailing the differential
spacetime manifold on which the classical relativistic field
theory of gravity vitally depends, is resolved by using the basic
manifold independent and, {\it in extenso}, Calculus-free
concepts, techniques and results of Abstract Differential Geometry
(ADG). As a physical toy model to illustrate these ideas, the
ADG-theoretic resolution of both the exterior, but more
importantly, of the inner, Schwarzschild singularities of the
gravitational field of a point particle is presented, with the
resolution of the latter being carried out entirely by
finitistic-algebraic and sheaf-theoretic means, and in two
different ways. First, by regarding it as a localized, `static'
point-singularity, we apply Sorkin's finitary topological poset
discretization scheme in its Gel'fand dual representation in terms
of `discrete' differential incidence algebras
\cite{rapzap1,rapzap2} and the finitary spacetime sheaves thereof
\cite{rap2}. Then we exercise the ADG machinery on those sheaves
in the manner of \cite{malrap1,malrap2,malrap3} to show that the
vacuum Einstein equations still hold over the classically
offensive {\it locus} occupied by the point-mass both at the
`discrete' level of the finitary sheaves and at the `classical'
continuum (inverse and direct) limit of infinite topological
refinement of (a projective and inductive system of) the said
sheaves. On these grounds alone we infer that the essentially
algebraic differential geometric mechanism of ADG is in no way
impeded by the presence of singularities on a geometrical base
spacetime, be it a `discretum' or a `continuum'---a result which
goes to show that our ADG theoresis of gravity and its
singularities is genuinely background spacetime (`continuous' or
`discrete') independent. The second way in which we resolve the
interior Schwarzschild singularity is more straightforward, but
closely akin to the first. We carry it out by regarding the inner
singularity as a non-localized, time-extended, distributional
spacetime foam dense singularity in the sense of
\cite{malros2,malros3}---essentially, by smearing the original
point-singularity to a family of dense singularities extending
along the `wrist-watch' coordinate time-axis $\R$ of the
point-particle. Then, again we show that the vacuum Einstein
equations hold over the uncountable, densely singular {\it loci}
in the point-mass' time-line $\R$ when sheaves of Rosinger's
differential spacetime foam algebras of generalized functions
(distributions) are used as structure sheaves of generalized
coefficients or `coordinates' and the ADG-theoretic mechanism is
applied to them---as it were, to `engulf' or `absorb' them, but
still retain the said essentially algebraic differential geometric
mechanism in their very presence---in the manner of \cite{mall3}.
{\it In toto}, we intuit that Nature---{\it ie}, the laws of
physics---have no singularities, let alone that they `break down'
in any (differential geometric) sense at their {\it loci}, or that
the infinities that are normally associated with them have any
physical significance or are necessarily unavoidable as they
misleadingly appear to be from the manifold perspective, but that
it is precisely our conventional, $\smooth$-smooth manifold based
way of doing differential geometry, and in terms of which we have
hitherto formulated those physical laws as differential equations
relating $\smooth$-smooth---in the usual Calculus-theoretic,
manifold based notion of differentiability or smoothness---fields,
that `stumbles and falters' on singularities and their unphysical
infinities. It goes without saying that the traditional Calculus
or Analysis-based technology of dealing with ({\it ie}, try to
define and systematically study) $\smooth$-singularities always
within the confines of the mathematical framework of Classical
Differential Geometry (CDG), such as analytic inextensibility and
the associated notion of (smooth) geodesic incompleteness, as well
as the construction of various `topological' boundaries to an
otherwise `regular' spacetime manifold on which singularities are
then destined to be `asymptotically' or `marginally' situated, is
completely evaded since ADG does not employ at all a base
geometrical differential manifold to support its concepts and
constructions. In the process of this total singularity evasion we
provide a new, purely algebraic, Leibnizian-Kleinian expression of
the principle of general covariance, which is manifestly smooth
background spacetime manifold free as it does not involve at all
the latter's diffeomorphism `structure' group as in the classical,
manifold based theory. Rather, it concerns solely the automorphism
group of the vector and algebra sheaves involved, which, in turn,
from a geometric (pre)quantization vantage, are the state spaces
of the particles (`quanta') of the fields ({\it viz.} connections)
acting categorically, as sheaf morphisms, on those (associated)
vector sheaves' sections. On the one hand, this is a functorial
(with respect to the structure sheaf $\struc$ of generalized
coordinates or arithmetics) expression of the principle of general
covariance of GR since the differential equation of Einstein
representing the law of gravity in ADG involves the (Ricci)
curvature (`field strength') of the gravitational field
(:connection), which curvature is an $\struc$-sheaf morphism (an
$\otimes_{\struc}$-tensor). At the same time, we argue that this
is an autonomous, `self-referential' conception of general
covariance which concerns only the fields and their particle
quanta `in themselves'---what we call here `field-particle
solipsism', without reference at all to an external, underlying
spacetime manifold, and which we here coin `synvariance'. At the
heart of synvariance lies a radical revision---in fact, an
inversion---of the notions of (gravitational) kinematics and
dynamics to the effect that it enables us to argue, in striking
contrast to the traditional process of the construction of GR (in
point of fact, of any physical theory constructed so far), that
`dynamics comes before kinematics'. From a categorical
perspective, as befits the aforesaid functorial expression of
gravitational dynamics that ADG enables us to maintain, we argue
that a natural transformation-type of principle underlies the
notion of synvariance, which we coin the Principle of Algebraic
Relativity of Differentiability (PARD). In turn, PARD may be
philosophically interpreted as an abstract, generalized version of
Einstein's principle of physical reality, here coined the
Principle of Field Realism (PFR)---a principle which goes hand in
hand with the field solipsism mentioned earlier. Subsequently,
various implications that such a total ADG-theoretic evasion or
bypass of singularities and of the differential spacetime manifold
carrying them could have for our seemingly never ending attempts
to arrive at a genuinely quantum and inherently finite theory of
gravity are discussed in some detail. For instance, apart from
completely circumventing various caustic issues that are supposed
to trouble the (persistently differential manifold based) theory
at the quantum level of description of gravity ({\it eg}, in the
canonical or the path integral approach to quantum or `quantized'
GR on a smooth spacetime continuum) such as the inner
product/functional integral measure problem as well as the
so-called problem of time, by the notion of synvariance and the
autonomous conception of field dynamics that ADG enables us to
posit and actually practice algebraically (:sheaf-theoretically),
we can, already at the classical level of description of gravity,
evade, unscathed, the whole of Einstein's hole argument---a
`no-go' argument in GR originally proposed in order to put to the
test and ultimately `shoot down' the principle of general
covariance, when the latter is implemented via the diffeomorphism
group of the underlying smooth spacetime manifold. Based on a
generalized interpretation of the hole argument by Stachel, we
argue that its total ADG-assisted bypass is virtually equivalent
to the aforesaid priority of dynamics over kinematics that ADG
allows us to maintain. Concerning quantum gravity in particular,
we argue that the purely algebraic notion of field ({\it viz.}
$\struc$-connection $\conn$, for a suitably chosen $\struc$) in
ADG, and the gravitational dynamics (the differential equations of
Einstein) that it defines, is in a subtle sense `already quantum'
or `quantized-by-itself' hence in no need of either a process of
quantization, or conversely, of a correspondence principle (or
classical limit theory). We coin this feature of ADG-based
gravity, `third', or a more mouthful, `field (without an external
spacetime, whether a continuum or a discretum) self-quantization'.
Third quantization, as opposed to second quantization---let alone
first quantization, which is non-existent in ADG---makes us
question altogether the physical existence of a fundamental
space-time scale in Nature, like the Planck length-time is
supposed to be in the conventional (and persistently spacetime
continuum based in one way or another) approaches to quantum
gravity. In the end, we present a physico-philosophical critique
of the quite unsuccessful way we try to apply differential
geometric ideas in the quantum deep, as well as numerous potent
arguments we have gathered so far for the possibility of doing
ADG-theoretically field theory entirely by categorico-algebraic
and finitistic means, by referring directly and solely to the
(algebraic relations between the) fields themselves without at all
the `mediation' or interference in our constructions, or even
intervention in our calculations ({\it ie}, in our Calculus) of a
pointed background differential spacetime manifold in the guise of
$\smooth$-smooth coordinates. In the light of this critique, the
idea is entertained of potentially `marrying' Stachel's `two
Einsteins' \cite{stachel1}---one, the nowadays more popular facet
of Einstein's post GR work, advocating a unitary, `continuous'
field theory on the spacetime continuum (while at the same time
apparently maintaining a polemic stance against the quantum, which
he thought that a suitably completed and potentially
singularity-free field theory could actually `explain away'), the
other, arguably a less popular and currently much overlooked
aspect of Einstein's ideas, propounding a purely algebraic and
finitistic-combinatorial physics on a `fundamental discretum'. All
in all, the present `paper-book' may be viewed as a significant
extension of the trilogy \cite{malrap1,malrap2,malrap3} to a
tetralogy so as to include ADG's promising physical application
towards evading the $\smooth$-smooth gravitational singularities
of the differential spacetime manifold of GR, as well as to
explore the potential technical and conceptual consequences that
such an evasion has for both classical and quantum gravity
research. This work's recent forerunner, of a more modest size, is
the second author's paper \cite{rap5}, which the reader might like
to have a look at first as a `warmup reading'.

\vskip 0.1in



\end{abstract}

\newpage

\tableofcontents

\newpage

\setlength{\textwidth}{21.0cm} 

\oddsidemargin=-.8cm \evensidemargin=-1cm \headsep=.8cm

\setlength{\topskip}{0pt}  
\setlength{\textheight}{23.0cm} 
\setlength{\footskip}{-2cm}

\setlength{\topmargin}{0pt}

\section{Foreword}

The present work encapsulates (admittedly, in quite a large
`capsule' which, paying tribute to Laurent Schwartz, may as well
be coined `{\em paper-book}') more than 3-years' efforts of
exploring the potential application of ideas, concepts, techniques
and results of {\em Abstract Differential Geometry} (ADG) to {\em
Classical} and {\em Quantum Gravity} (QG), as well as to (quantum)
Yang-Mills (gauge) theories of matter. Primarily, it focuses on
showing how ADG may be employed to evade completely the {\em
singularities} and their associated unphysical infinities
assailing the {\em smooth background spacetime manifold} and, {\it
in extenso}, the {\em Classical Differential Geometry} (CDG) based
{\em General Relativity} (GR), as well as to anticipate various
conceptual (philosophical) and technical implications that such a
total singularity-bypass may have for a plethora of currently
`caustic' topics in QG research. The present treatment of gravity
by ADG-theoretic means is suitably called `{\em ADG-gravity}'.

\paragraph{On the physics side.} Much in the same way that, in the pure
mathematical subject of {\em Differential Geometry} (DG), ADG has
shown us how to do DG {\em purely algebraically}
(:sheaf-theoretically), based solely on the notion of an {\em
algebraic connection} $\conn$ ({\it viz.} generalized differential
operator) and {\em manifestly without depending ourselves on a
base differential} (:$\smooth$-smooth) {\em locally Euclidean
space} (:manifold), in its application to gravity ADG enables us
to formulate the latter as a {\em pure gauge theory} ({\it ie}, a
theoresis of gravity based only on the gravitational connection
field $\conn$), in a {\em genuinely background spacetime manifold
independent} fashion, and with algebraic `quantum traits' built
into the formalism from the very start ({\it ie}, without any need
of a formal procedure of quantization of the classical
relativistic field theory---GR---let alone of the base spacetime
continuum underlying it). `{\em Background independence}' in
particular, is a `hot' issue on which various current approaches
to QG---most notably, the {\em Loop Quantum Gravity} (LQG)
approach to (canonical) {\em Quantum General Relativity}
(QGR)---hinge. However, in contradistinction to these approaches
where `background independence' means `{\em background metric
independence}' while a base differential manifold is still present
in the theory galore, in `ADG-gravity' not only {\em a metric is
not employed at all} (as the sole dynamical variable in the theory
is the gravitational gauge connection field $\conn$), but also
{\em no background spacetime manifold is present at all}. In fact,
a background geometrical spacetime, whether a `classical
continuum' or a `quantal discretum', plays absolutely no role
whatsoever in ADG-gravity---{\it ie}, it plays no role in
formulating the gravitational dynamics, as {\em differential
equations} proper, in terms of the gravitational connection field
$\conn$ `in-itself'. {\it In summa}, {\em ADG-gravity is a
fundamentally background spacetimeless, `pure gauge'\footnote{The
(dynamical) formalism underlying ADG-gravity, in contradistinction
to the original metric-based formulation due to Einstein (second
order formalism) and to the more recent Palatini-Ashtekar one
which is based on both the (smooth) connection and tetrad
variables on a smooth manifold (first order formalism), is coined
{\em half-order formalism}', since the only dynamical variable
involved is the gravitational gauge connection field $\conn$ and
{\it a fortiori} no base differential manifold is present in the
theory. Especially due to the latter, ADG-gravity is called a
`{\em gauge theory of the third kind}', to distinguish it from the
usual gauge theories of the first (global) and second (local) kind
which, since they are based on an external (background) spacetime
continuum, necessarily draw a distinction between external
(spacetime) and internal (gauge) symmetries. All `symmetries' in
ADG-gravity are `internal'---{\it ie}, of the ADG-gravitational
field `in-itself'. (See remarks below about the novel terminology
that comes along with ADG-gravity.)} and `innately
quantum'\footnote{Due to the quantum concepts and associated
formalism which are built from the very start into ({\it ie},
almost `by definition' of) the ADG-gravitational field,
ADG-gravity is called `{\em third quantized}' or `{\em third
quantum}' (field) theory, if anything to distinguish it from the
usual first and second quantized (non-relativistic quantum
particle/finite degrees of freedom and relativistic quantum
field/infinite degrees of freedom, respectively) theories (of
matter) in which again a background space-time continuum is
invariably involved in one way or another. (Again, see remarks
below about the novel terminology used in ADG-gravity.)} field
theoresis of gravity}.

In turn, this fundamental base spacetimelessness of ADG-gravity
and its focusing solely on the gravitational field itself, helps
us shed light on, or even `resolve' and `bypass' completely and
directly ({\it ie}, in a `cutting of the Gordian knot' sort of
way), certain crucial problems encountered in both classical and
quantum gravity research: from the problem of the $\smooth$-smooth
spacetime singularities and Einstein's hole argument in the
manifold based GR, to the so-called {\em inner product/quantum
measure problem} in both the canonical and the covariant (path
integral) approaches to QG, as well as the {\em problem of time}
in (canonical) QGR and {\em the general problem of viewing
(vacuum) Einstein gravity as a gauge theory proper}, especially in
the quantum domain (QG) as a quantum gauge theory. And it must be
emphasized at this point that in all of the aforesaid problems the
{\em spacetime diffeomorphism `symmetry' group} $\mathrm{Diff}(M)$
of the underlying differential manifold $M$---the automorphism
group of the smooth spacetime background traditionally used to
implement mathematically the {\em Principle of General Covariance}
(PGC) of GR, is involved in one way or another. Thus, since
ADG-gravity is background spacetime manifold independent (hence
{\it a fortiori} no $\mathrm{Diff}(M)$ is involved in the theory),
none of these problems are encountered; as it were, they {\it ab
initio} become `{\em non-problems}' in formulating gravity
(`classically' or `quantally')\footnote{This signals that, as a
matter of fact, the traditional epithet-distinctions `classical'
and `quantum' (usually put in front of the noun `gravity') lose
their significance in ADG-gravity.} in the ADG-framework.

Concerning $\smooth$-smooth spacetime singularities in particular,
which are arguably `innate' in the spacetime manifold---{\it ie},
they are `inherent' in the {\em structure sheaf} $\smooth_{M}$ of
algebras of $\smooth$-smooth (`coordinate') functions on $M$
(which sheaf, in turn, {\em defines} $M$ as a differential
manifold proper in the first place by Gel'fand duality), their
total evasion in ADG-gravity brings to mind a tetrad of verses
from Constantinos Cavafis' poem `{\itshape Ithaca}' \cite{cav}:

\vskip 0.2in

\centerline{``\small\em ...Laestrygonians and Cyclopes, angry
Poseidon,}

\centerline{\small\em Such obstacles you will never encounter in
your way,}

\centerline{\small\em As long as you do not carry them in your
soul,}

\centerline{{\small\em As long as your soul does not raise them
before you...''}}

\vskip 0.2in

\noindent the basic `moral' being here (metaphorically speaking of
course) that {\em one will not encounter gravitational
singularities}, at least as insuperable differential geometric
obstacles (or even more formidably, as `{\em breakdown points}' of
Einstein's field equations---the way singularities are commonly
viewed by physicists nowadays) in the {\it aufbau} of one's theory
of gravity, {\em as long as one does not model} (in one's theory)
{\em spacetime after a differential manifold}. Accordingly, {\it
mutatis mutandis} for the other background $M$ and, {\it in
extenso}, $\mathrm{Diff}(M)$-related classical and quantum gravity
problems mentioned above.

\paragraph{On the mathematics side.} If anything, by ADG-gravity we can
reinstate the status of DG in QG research, a status and import of
the manifold based CDG that has been questioned in the past by
many workers in the field, with the most severe and direct critic
being (to our knowledge) Chris Isham, who claimed fairly recently
in \cite{ish}:\footnote{This quotation appears as (Q8.?) in the
main text and has also been used in the past, in the introduction
to \cite{malrap3}. Here, emphasis is ours.}

\begin{quotation}
\noindent ``{\small{\em ...at the Planck-length scale,
differential geometry is simply incompatible with quantum
theory}...{\small [so that]} {\small\em one will not be able to
use differential geometry in the true quantum-gravity
theory...}}''
\end{quotation}

\paragraph{On the philosophy side.} Since ADG, as a mathematical
framework, enables us to do DG {\em entirely algebraically} (in a
way reminiscent of the {\em relational} fashion in which Leibniz
envisioned the development of his `Geometric Calculus'---`{\it Ars
Combinatoria}' and `{\it Calculus Ratiocinator}') by referring
directly and solely to the `geometrical objects' (:the physical
fields) that `live' on `space(time)' without depending at all on
that external (to the fields themselves), `ambient' geometrical
space(time manifold) for its concepts and constructions,
philosophically speaking ADG-gravity may be viewed as a `{\em
purely realist}' theory---supported by what we coin the `{\em
Principle of Field Realism}' (PFR); or even for more effect, the
`{\em Principle of Field Solipsism}'. The PFR can be accommodated
by the ADG-framework due to the fact that the ADG-gravitational
field $\conn$ is a `{\em dynamically autonomous}' entity, in no
need of an external spacetime (manifold) for its `dynamical
subsistence' ({\it ie}, the dynamical Einstein equations that
$\conn$ defines via its curvature $\curv(\conn)$, although still
modelled after differential equations proper, do not depend at all
on a geometrical background $M$, as {\em they represent the
dynamics of the field `in-itself'}). More noteworthy is our
maintaining this realistic `attitude' even in the quantum domain,
in spite of the usual operationalistic (algebraic),
instrumentalist, `external (to the quantum system) observer
dependent' conception of physical reality that quantum theory is
traditionally supposed to entail.

This is also in contrast to the usual manifold based GR, which is
supported by an `operationalist' or `instrumentalist' philosophy
(and associated interpretation) according to which the components
of the smooth metric tensor field $g_{\mu\nu}$ on the differential
manifold $M$ not only represent the gravitational field
potentials, but also the (local) {\em chronogeometry} of $M$, as
they engage into the (infinitesimal proper time) line element
$ds^{2}$.\footnote{In the usual CDG-based (pseudo-)Riemannian
geometry supporting GR, the affine gravito-inertial connection
`inherits' the chronogeometrical interpretation by being required
to be compatible with the metric (metric or torsionless
connection)---a condition which is only optional in ADG-gravity,
and in a way it is {\em reversed} since now {\em the metric}
(which is an optional, externally prescribed structure in
ADG-gravity, imposed by the external to the gravitational field
`observer' or `measurer') {\em is made to be compatible with the
fundamental connection field, not the other way round}.} As a
result of this interpretation, the gravitational field is supposed
to encode all the information about our tampering with (`probing'
or `measuring') it with our (local) geometrical `spacetime gauges'
(:equicalibrated rods and synchronized clocks), with the results
of these (local) measurements being organized into the aforesaid
coordinate structure sheaf $\smooth_{M}$. Of course, GR is able to
sustain an analogue of the PFR via the PGC, yet its dependence on
a background geometrical differential manifold is vital: {\em how
else can one represent the gravitational Einstein equations as
differential equations proper in the first place?}

Unfortunately, a `self-referential' {\em vicious circle} is
lurking here: the standard perception nowadays of `genuine' or
`real' ({\it ie}, not coordinate) gravitational singularities is
as {\it loci} in the spacetime continuum where the Einstein field
law, viewed as a (partial) differential equation, {\em breaks
down}. In other words, the very structure sheaf $\smooth_{M}$ (or
equivalently, the base differential manifold $M$) that enables one
to set up (the dynamical law of) GR differential geometrically
({\it ie}, as a differential equation proper), is pregnant to GR's
own `destruction' in the form of singularities. In this subtle
(differential geometric) sense we view the Wheeler-Bergmann
`Popperian virtue' of GR, according to which the latter `{\em
carries in its belly the seeds of its own destruction}', its
`self-falsification' so to speak. From an ADG-theoretic vantage,
it is not that the gravitational field (and the law that it
defines/obeys) breaks down at a singularity, but simply that GR is
formulated within the manifold based CDG-theoretic framework,
which is out of its depth on the face of singularities (let alone
{\it contra} deeper and conceptually/interpretationally more
complicated QG issues). Plainly then, from the ADG-vantage
singularities are a `fault' of the mathematics---a glaring `proof'
that the mathematics ({\it ie}, the CDG used in GR) is inadequate
or `inappropriate', {\em not} of the physics (dynamical
gravitational field law). Alas, in the manifold based GR the
mathematics is so intimately enmeshed and entwined with the
physics that it misleads one into thinking that `{\em physical
spacetime}' is a (differential) manifold, when in
contradistinction, from the viewpoint of ADG, if there is any
`spacetime' at all it is `inherent' in the dynamical fields that
comprise `it'. This then is the main `aphorism' in the present
work: {\em in ADG-gravity, all is field} and no externally
prescribed spacetime is involved at all, the epitome of the
aforesaid `{\em field solipsism}'.

Moreover, when it comes to the quantum domain, where an
operational, `observer dependent realism' reigns supreme, it is
plain that the geometrical base spacetime continuum glaringly
`miscarries'. For one thing, there is supposed to be a minimal
space-time scale (the so-called Planck length-time
$\ell_{P}\approx 10^{-35}m$--$t_{P}\approx 10^{-40}s$) below which
one cannot localize (`measure') the gravitational field with
infinite accuracy without creating a singularity (concealed beyond
the horizon of a so-called `black hole'). This then seems to
suggest that below $\ell_{P}$ the smooth spacetime continuum
should give way to a `reticular-quantal and inherently cut-off' or
regularized structure, with the inevitable loss of one's
differential geometric privileges in the QG regime (see Isham
quotation above). From the background spacetimeless (whether a
continuum or a discretum) ADG-theoretic vantage, this is hard to
swallow. Similarly to Einstein's explicit dissatisfaction with
spacetime singularities in GR,\footnote{See the first quotation
(Q2.1) in section 2.} we cannot accept that there is a spacetime
scale above which the field law holds, but below which it
apparently breaks down or that it should be radically
modified---especially {\it vis-\`a-vis} ADG-gravity where no
external (to the fields), background spacetime (whether
`continuous' or `discrete') is involved at all. This is a `{\em
paraphysical antinomy}' of the very term `{\em physical law}' (pun
intended!) and its supposed universality.

All in all, it is evident by the foregoing that in the present
work we do not shy away from addressing conceptual (philosophical)
issues, especially {\it vis-\`a-vis} ADG-gravity's potential QG
import, for after all, as 't Hooft recently put it in
\cite{thooft}:

\begin{quotation}

\noindent ``{\small\em ...The problems of quantum gravity are much
more than purely technical ones. They touch upon very essential
philosophical issues\footnote{Our emphasis.}...}''

\end{quotation}

\paragraph{The gist of this `paper-book'.} {\it Vis-\`a-vis}
so-called `applied mathematics', `mathematical physics', or even
`mathematical methods in physics', it is fair to say that
twentieth century (`modern') theoretical physics was largely
dominated by applications of {\em differential geometry}. The
great interest of theoretical physicists in differential geometry
may be attributed to their idea of modelling physical laws after
{\em differential equations} if anything in order to implement
their primitive notion of `{\em infinitesimal locality}' or `{\em
differential local causality}',\footnote{The primitive intuition
that events causally affect others in their `infinitesimal
neighborhood'.} and arguably {\em differential geometry was
developed for (having a comprehensive theory of as well as for
solving) differential equations}.\footnote{Much in the same way
that {\em algebraic} geometry was originally developed for dealing
comprehensively with {\em algebraic} equations.} This of course is
the `classical' theory of differential geometry (CDG) which
vitally relies in one way or another on the notion of a {\em
smooth background space}: from the finite-dimensional (locally)
Euclidean spacetime (manifold) of the special, but more
importantly, the general theory of relativity in which the
(pseudo-)Riemannian geometry employed is grounded, to the
infinite-dimensional complete Euclidean (Hilbert) spaces modelling
the quantum configuration/phase (state) spaces used in quantum
mechanics and handled by functional analytic/operator-theoretic
means.\footnote{Let it be noted here that 19th century physics too
was essentially dominated by Newton's Differential Calculus---the
spry grandparent of nowadays modern differential geometry (CDG):
from the classical particle mechanics of Lagrange and Hamilton, to
the classical field theory of Faraday and Maxwell. It is fair to
say that the iconoclastic {\em physical} theories of the last
century---namely, GR and QM---introduced new, ground-breaking {\em
physical} concepts, but essentially relied on Newton's background
continuous space(time) dependent Differential Calculus (albeit, in
the more sophisticated language and technotropy of the manifold
based CDG) for their mathematical concepts, techniques, and more
importantly, for their {\em calculations}.} {\it A fortiori}, the
subsequent unison of quantum theory with SR to a QFT of matter and
the concomitant realization that matter forces are in fact {\em
gauge} forces, gave impetus for further development of
differential geometric ideas, concepts and techniques in
theoretical physics. To appreciate this, one has simply to recall
the boom in applications of the mathematics of fiber bundles to
gauge theory, to the extent that Michael Atiyah, the celebrated
mathematician, coined in \cite{atiyah} gauges theories as:

\vskip 0.1in

\centerline{``{\small ...{\em Physical} theories of a {\em
geometrical} character\footnote{Our emphasis.}...}''}

\vskip 0.1in

\noindent while one is justifiably tempted to add the epithet
`{\em differential}' to the word `{\em geometrical}' in the
quotation above.

Yet there still comes Isham's quotation above to `haunt' any
approach to QG that uses CDG-ideas. That is to say, there is this
`nagging paradox' that while CDG has enjoyed enormous success in
being applied to classical mechanics and field theory ({\it eg},
electrodynamics and GR), as well as to QM and QFT ({\it eg},
quantum gauge theories of matter), when one sets out to marry GR
with quantum theory to a quantum theory of gravity, the CDG of
smooth manifolds appears to be of little help (if any at all!) and
out of its depth. We understand this simply on the fact that the
main culprit for all the `pathologies' encountered in either GR
(in the form of singularities) or QFT (field-theoretic
infinities)---arguably, {\em the} `anomalies' that make us
question in the first place the CDG of smooth manifolds in the QG
realm---is {\em our} assumption of (physical) spacetime as a
differential manifold. Here we have {\em the} example {\it par
excellence} of the proverb `{\em throw away the baby together with
the bath-water}' in the sense that if the manifold, which causes
all these `unpleasant unphysicalities', will have to go in the
realm where GR is envisioned to be united with quantum theory (the
Planck regime), so will {\em differential geometry} as a whole.
For after all, {\em so far the only way we know how to do
differential geometry is by basing ourselves in one way or another
on a smooth manifold, which in physics is interpreted either as
the spacetime continuum or as the configuration space of a
physical system---classical or quantum}.\footnote{And let it be
noted here that fields are normally thought of as physical systems
with an uncountably infinite number of degrees of freedom, thus a
mathematical continuum such as a manifold appears to be tailor-cut
for modelling either the spacetime on which these fields
dynamically propagate and interact, or their corresponding state
spaces.} It is plain that if we have such a `CDG-monopoly', if the
manifold goes `bankrupt' in the quantum deep, inevitably so does
CDG which is vitally dependent on it.

It is precisely this issue we wish to challenge in the present
work, as it were to counter and ultimately evade Isham's `no-go'
of differential geometry in QG:\footnote{And let it be stressed
here that the {\em DG} that Isham was referring to was the CDG on
smooth manifolds. A recent exchange of the second author with
Chris about this quotation received back the counter-remark that
``{\em I may have been wrong}''. This we understand as implying
{\em not} that CDG on manifolds could be of import to QG research
after all (for this would be a sort of regress to old concepts and
dated technology so to speak), but that other new {\em theoretical
frameworks} for doing DG, such as Connes' Noncommutative
Differential Geometry (NDG) \cite{connes} which has enjoyed
numerous applications in the Standard Model and quantum spacetime
and gravity in the past decade \cite{connes1,connes2,connes3}, as
well as the category (topos) based Synthetic Differential Geometry
(SDG) of Kock and Lawvere \cite{kock,laven}, could be of great
value to the QG quest. We are aware that Chris is particularly
interested in the possibility of applying SDG-ideas to the quantum
structure of spacetime and gravity \cite{buttish4,ish3}. We too
are very keen on exploring in the immediate future various close
affinities between ADG, NDG and SDG, as well as, hopefully, to
unite forces at the QG research front (see 8.3.1).} we are going
to show and argue that since ADG is manifestly base manifold-free,
hence also ADG-gravity genuinely background spacetime manifold
independent, {\em one can still do differential geometry in the
realm of QG}. For one thing, as noted above, bypassing directly
the manifold with its `inherent' singularities and associated
unphysical infinities, while at the same time retaining most (if
not {\em all}!) the differential geometric panoply (mechanism) of
CDG in its manifest absence, as well as formulating gravity as a
`pure gauge theory', could prove to benefit tremendously both
classical and QG research.

In the present work we shed the weighty burden of the smooth
manifold and the inertia that its physical interpretation as
`spacetime' presents to the theoretical/mathematical physicist,
and we wish to travel light in Feynman's `wild blue yonder' QG
regime.\footnote{Hopefully not at the cost that we will never be
heard of again...} However, in order to appreciate how difficult
it may turn out to be for one to overcome the said `{\em
background spacetime manifold inertia}' and associated `{\em
CDG-conservatism and monopoly}',\footnote{Thus also how easy and
likely it is for one to go off to the QG `blue yonder' and never
be heard of again indeed!} one may recall some recent remarks of
Isham and Butterfield from \cite{ish9}, where the issue is raised
of what structures (other than the CDG-supporting, smooth, and as
`spacetime' interpreted continuum) to consider in QG research, as
well as how can the familiar structure and notion of the spacetime
continuum (of GR) emerge (or be recovered) from those `deeper'
structures:\footnote{In the excerpt below, all emphasis is ours
due to its importance.}

\begin{quotation}

\noindent ``{\small\em ...The usual tools of mathematical physics
depend so strongly on the real-number continuum, and its
generalizations (from elementary calculus `upwards' to manifolds
and beyond), that it is probably even harder to guess what
non-continuum structure is needed by such radical approaches, than
to guess what novel structures of dimension, metric etc. are
needed by the more conservative approaches that retain manifolds.
Indeed, there is a more general point: space and time are such
crucial categories for thinking about, and describing, the
empirical world, that it is bound to be ferociously difficult to
understand their emerging, or even some aspects of them emerging,
from `something else'...}''

\end{quotation}

\noindent On the other hand, Einstein's words from \cite{einst2}
immediately spring to mind here:\footnote{This quotation also
occurs in the main text (Q2.10).}

\begin{quotation}
\noindent ``{\small Time and space are modes by which {\em
we}\footnote{Our emphasis.} think, not conditions in which we
live.}'',
\end{quotation}

\noindent but more importantly, his urging us to question and
scrutinize familiar, well established concepts (like for instance
that of the spacetime continuum) in our physics
(re)searches:\footnote{What we find truly remarkable in this
quotation is that Einstein's words came only one year after the
development of GR, in which the concept of the base geometrical
spacetime continuum (and the CDG-based pseudo-Riemannian geometry
on it) triumphed. This quotation also occurs in the main text
(Q7.27), as well as in the conclusion of our last joint paper
\cite{malrap3}.}

\begin{quotation}
``{\small\em ...Concepts {\rm [like the spacetime
manifold]}\footnote{Our addition for making our point here
clearer.} which have proved useful for ordering things easily
assume so great an authority over us, that we forget their
terrestrial origin and accept them as unalterable facts. They then
become labelled as `conceptual necessities', `a priori
situations', etc. The road of scientific progress is frequently
blocked for long periods by such errors. It is therefore not just
an idle game to exercise our ability to analyse familiar concepts,
and to demonstrate the conditions on which their justification and
usefulness depend, and the way in which these developed, little by
little\footnote{Our emphasis throughout.}...}'' \cite{einst7}
\end{quotation}

\noindent come to accompany, comfort and inspire our endeavors
here. All in all, let the present paper-book mark the beginning of
the end of the epopee of the smooth background (spacetime)
manifold based (and thus largely CDG-dominated)
theoretical/mathematical physics research, and QG in
particular.\footnote{We are pleasurably indebted to John Stachel
for timely pointing out to us his belief that QG research in the
new millennium will focus primarily on formulating the theory in a
background independent fashion, to the extent that any current or
future approach to QG shall be ultimately `judged' on the degree
that it has achieved a genuinely background independent
formulation (Stachel in private communication with the second
author at Imperial College, Fall 2004). And let us further add
here our opinion that background independence should go all the
way---{\it ie}, that we should not content ourselves only with
background {\em metric} independence as it is `fashionable'
nowadays, but also look for background {\em smooth spacetime
manifold} independent scenaria, like the one ADG-gravity will
offer herein.}

However, in spite of all that `background geometrical spacetime
manifold inertia, habit and indolence' as well as our principal
aim to overcome them herein, and apart from the fact that quantum
gauge ({\it ie}, electrodynamics and non-abelian Yang-Mills)
theories and QG is the focal issue in the present treatise {\it
vis-\`a-vis} {\em physical applications} of ADG, one should not
lose touch with {\em the} philosophical essence of ADG as a
mathematical framework for doing differential geometry in a wider,
broader and more thorough sense. The (philosophical) gist of ADG
is primarily an `{\em epoptic}' one. In this respect, let us first
recall Wittgenstein's opening remarks in his posthumously
published book ``{\itshape Culture and Value}'' \cite{witt1}:

\begin{quotation}

\noindent ``{\small ...Our civilization typically constructs. Its
activity is to construct a more and more complicated structure.
And even clarity is only a means to this end and not an end in
itself. {\em For me on the contrary clarity, transparency, is an
end in itself. I am not interested in erecting a building but in
having the foundations of possible buildings transparently before
me}\footnote{Our emphasis. From another translation of ``{\itshape
Culture and Value}'' \cite{witt2} a year later than \cite{witt1},
we encounter another nice version of the same excerpt: ``{\em
...Our civilization is characterized by the word `progress'.
Progress is its form rather than making progress is one of its
features. Typically it constructs.  It is occupied with building
an ever more complicated structure. And even clarity is sought
only as a means to this end, not as an end in itself. For me on
the contrary clarity, perspicuity are valuable in themselves. I am
not interested in constructing a building, so much as in having a
perspicuous view of the foundations of possible buildings...}''.
The reader can choose the one (s)he prefers.}...}''

\end{quotation}

\noindent and slightly modify them to suit the general spirit of
the present work regarding, under the prism of ADG, the theory and
(physical) applications of differential geometry `at large':

\begin{quotation}

\noindent In this work we are not as much interested in tackling
this or that particular problem nowadays encountered in classical
and quantum gravity research as well as in quantum Yang-Mills
theories, let alone to come up with a grand structural scheme---a
{\it panacea} so to speak---for dealing with such problems, as to
attain a clear, `{\em epoptic}', bird's-eye-view as it were, of
possible applications of fundamental differential geometric
ideas---ones that are {\it ab initio} free from any commitment to
an {\it a priori} posited `background space(time manifold)
structure---to modern theoretical physics, as well as to
anticipate and explore the potent physical implications of such a
fundamental non-commitment (`background independence'). {\em How
far can we go in modern theoretical physics with a base
(spacetime) manifoldless differential geometry?}---that's {\em
the} basic question we would like to ask in the light of ADG the
modern theoretical physicist, and in particular the QG worker, who
is still willing, persistently in spite of Isham's `differential
geometric pessimism' in the quantum deep quoted above, to use
differential geometric ideas and technotropy in her QG research.
Let's hope it's a long way...

\end{quotation}

\noindent All in all, in many, closely intertwined,
cross-fertilizing and mutually affecting levels---physical,
mathematical and philosophical---our principal aim in this
paper-book is, once again emulating in a metaphorical way the
latter, post-Tractarian Wittgenstein's \cite{witt3,witt4,witt5}
remark that:

\begin{quotation}

\noindent ``{\small ...[The principal aim of my work]\footnote{Or
of philosophy in general. Our addition.} {\em is to show the fly
out of the fly-bottle}\footnote{Our emphasis. In a nutshell, the
meaning here is that philosophers trap themselves into
philosophical (pseudo-)problems by wrong `use' and associated
misinterpretations of (everyday colloquial) language, so that by
clarifying language---as it were, by casting the said problems in
their `natural language habitat' (by the way, as such `habitats',
Wittgenstein had introduced the relational and autonomous notion
of `language games')---the philosopher is led out of the trap that
he himself set up in the first place to tackle those problems by
means of language.}...}'',

\end{quotation}

\noindent the main aim of our work is

\begin{quotation}

\noindent {to `free' the mathematician, physicist and philosopher
of physics (who wishes to apply differential geometry to
theoretical physics) from the confines and shackles of the
background manifold that anyway she assumed up-front (and she
trapped herself into!) in order to do (and interpret!) field
theory differential geometrically, which `pseudo-physical'
spacetime continuum in turn creates all the problems, in the guise
of singularities and unphysical infinities, that she encounters in
both the classical and the quantum (field-theoretic)
domain.\footnote{The idea here is that our standard ({\it ie},
CDG-based) use of differential geometric concepts and
constructions via a base differential manifold is `wrong'
(`unnatural' or `unphysical')---the `wrongness' being exemplified
by the physically inadmissible singularities and infinities one
encounters in applying the manifold-based CDG-ideas to both
classical and quantum field theory. ADG studies differential
geometry in its `natural habitat'---that delimited by the
`geometrical objects' (:physical fields) (in-)themselves, without
the mediation (in the guise of smooth coordinates) of a smooth
background manifold, which is thus of no physical significance in
(physical applications of) the theory.} Our venture here is kind
of `therapeutic': as it were it aims to dispel the chimerical
`surrounding spacetime nimbus' and the mesmerizing magic that a
background geometrical (spacetime) manifold exercises on the
modern theoretical physicist, which in turn misleads her into
thinking that problems such as singularities and infinities that
she encounters when she models physical laws differential
geometrically via a base $M$ are actually {\em physical} problems,
when in fact they are only shortcomings and anomalies of the {\em
mathematical} framework ({\it ie}, CDG) that she employs in the
first place. The `ADG-therapy' prescribed here involves {\em
algebra} (:`relational differential geometry'), and, arguably,
{\em there is no infinity in algebra}, for infinities arise only
in our geometrical base space(time) manifold-mediated Analysis
({\it ie}, CDG or Differential Calculus).}

\end{quotation}

\noindent In the same line of thought we are tempted to recall
Evariste Galois' words in \cite{galois}:

\vskip 0.2in

\centerline{``{\small\em Les calcules sont
impracticables}'',\footnote{``{\em Calculations are
impractical}''.}}

\vskip 0.2in

\noindent and to modify them in our ADG-context to the following:

\begin{quotation}
\noindent The usual Differential Calculus (CDG), insofar as it is
effectuated---and so far, let it be stressed, it has actually been
effectuated(!)---via a base differential (spacetime) manifold, is
of little import in the QG deep.\footnote{Actually, it is of {\em
great} import, but of things of the `wrong' kind, such as
singularities, infinities and a host of other differential
geometric anomalies and pathologies, which cumulatively give one
the (false, in our opinion) impression that `DG miscarries with
quantum theory, and especially, in the realm of QG' (see again the
Isham quotation above).}
\end{quotation}

\paragraph{On the terminology side.} The novel perspective on gravity that ADG enables us to entertain
is inevitably accompanied by {\em new terminology}. We have thus
not refrained from engaging into vigorous `{\em lexiplastic
activity}', so that the present paper-book abounds with new terms
for novel concepts hitherto not encountered in the standard
theoretical physics' jargon and literature, such as `{\em gauge
theory of the third kind}', `{\em third quantization}', `{\em
synvariance}' and `{\em autodynamics}', to name a few. In this
respect, we align ourselves with Wallace Stevens' words in
\cite{stevens}:\footnote{This quote also appears in the main text,
in (Q3.?).}

\begin{quotation}
\noindent ``{\small\em ...Progress in any aspect is a movement
through changes in terminology...}''
\end{quotation}

\noindent with the `changes in terminology' in our case being not
just superficial (formal) `nominal' ones introduced  as it were
for `flash, effect and decor', but necessary ones coming from {\em
a significant change in basic theoretical framework for viewing
and actually doing DG}: from the usual geometrical manifold based
one (CDG), to the background manifoldless and purely algebraic
(:sheaf-theoretic) one of ADG. All the new terms and concepts are
defined and explained in a `{\em Glossary for ADG-gravity}'
appended at the end.

\paragraph{On the textual side.} The reader will have already noticed
that we are also not frugal in providing a plethora of {\em
footnotes} and {\em quotations}, as well as a long list of {\em
references} at the end. The purpose of footnotes is principally
explanatory, when we do not want to go into lengthy digressions
within the main text. We do hope that they prove to be more
helpful to the reader than distracting. The purpose of quotations
is sometimes to give a historical background and a `motivational
alibi' for the ideas presented---namely, that what is being said
has also been anticipated ({\it ie}, it is in accord with) similar
thoughts that have been expressed in the past by great thinkers or
specialist workers in the research field of QG. That is to say,
quotations essentially come to remind the reader that `{\em we are
standing on the shoulders of giants}', or at least that `{\em we
are not alone in this venture}', and enhance the `historical
continuity' of what is being said. However, other times quotations
are given as `counterpoints' to ({\it ie}, they provide a
contrasting platform against) the points being raised and
discussed. The ultimate hope of the present authors is that this
dialectical `{\em thesis-antithesis}' service of quotations will
prepare the ground for a fruitful `{\em synthesis}' of the new
theoretical paradigm for gravity (classical and quantum) that
ADG-gravity is in our opinion pregnant to. Finally, although to
the best of our ability and knowledge the long list of references
at the end is (intended to be) complete, we are certain that
important (even classic!) works of various people on a host of
subjects have been overlooked and unfortunately omitted. This was
done inadvertently and we apologize in advance.

\paragraph{Acknowledgments.} There are numerous people that we would like to thank wholeheartedly for fruitful
exchanges, encouragement, support and `colleagueal friendship'
during the nearly three-year writing of this paper-book: the names
{\em Keith Bowden, Jan de Graaf, David Finkelstein, Jim
Glazebrook, Achim Jung, Goro Kato, Ralph Kopperman, Jim Lambek,
Chris Mulvey, Prakash Panangaden, Steve Selesnick, Rafael Sorkin,
Freddy Van Oystaeyen} and {\em Roman Zapatrin} immediately spring
to mind. We also thank the participants in the first `{\em Glafka
2004: Iconoclastic Approaches to Quantum Gravity}' conference this
past mid-June in Athens (Greece) for their critical remarks and
discussions on our work, especially {\em Charis Anastopoulos,
Abhay Ashtekar, David Finkelstein, Jim Hartle, Joe Henson, Roman
Jackiw} and {\em Chris Mulvey}. The second author (IR) is
particularly indebted to {\em Chris Isham} for illuminating
discussions, as well as for his endless supply of fresh,
imaginative, `unorthodox' ideas on both classical and quantum
gravity, for his subtle and incisive critique of the present work,
but most importantly, for his unceasing moral encouragement,
understanding and fatherly support during times of many a
conceptual and technical impasses. A series of timely e-mail
exchanges with {\em John Stachel} (kindly initiated by Ray
Sorkin), culminating in his communicating to us his latest paper
\cite{stachel2} in which he gives a brief account of his idea to
use sheaf theory in classical and quantum gravity, is pleasurably
acknowledged. The second author would also like to thank the
European Union for funding this work, initially via a generous
2-year Marie Curie postdoctoral research fellowship held at
Imperial College, London (UK), and subsequently via a European
Reintegration Grant (ERG 505432) currently held at the Algebra and
Geometry Section, Department of Mathematics, University of Athens
(Greece). Last but not least, he greatly appreciates financial
assistance from Qualco---a private IT company in Athens---in the
guise of a generous research assistantship in connection with the
aforesaid EU grant. The second author is grateful to {\em Dr
Orestis Tsakalotos} (Qualco's CEO) for his invaluable help in
making this assistantship happen.

On the {\it TeX}tual side, we would like to thank cordially {\em
Mrs Popi Mpolioti} (University of Athens) for providing us with
nice {\it LaTeX} graphics' packages and some beautiful categorical
(commutative) diagram macros of her own `design', and also {\em
Roman Zapatrin} (University of St-Petersburg) for giving us
suitable {\it LaTeX} commands and simple tips for labelling and
referring to the plethora of quotations appearing in the main
text.

\newpage

\setlength{\textwidth}{12.4cm} 

\oddsidemargin=-.8cm \evensidemargin=-1cm \headsep=.8cm

\setlength{\topskip}{0pt}  
\setlength{\textheight}{18.7cm} 
\setlength{\footskip}{-2cm}

\setlength{\topmargin}{0pt}

\section{Introductory Remarks: General Covariance Versus Singularities}

That General Relativity (GR) predicts the existence of
singularities---{\it loci} in the smooth spacetime continuum where
Einstein's gravitational field equations do not hold or even
`break down'---is by now a virtually undisputed fact. By some
physicists this has been regarded as a positive feature of the
classical relativistic field theory of gravity---as if, in a
`Popperian falsifiability' sense, GR is a sound theory because in
a way it carries, as it were from within its conceptual and
technical edifice, its own limitations and, ultimately, it
foreshadows its own downfall, its self-destruction so to say. For
others, including Einstein, singularities were clearly unphysical,
pathological, at least anomalous, and certainly problematic,
features of the theory that should be somehow excluded from it; in
\cite{einst3} he notes for example:\footnote{Throughout the
present paper, quotations will be left-labelled by `Q*.$^\star$'
(the first ordinal `*' corresponding to the section number, and
the second, `$^\star$', to the number of the quotation in that
section), while numerous remarks of ours that we would like to
highlight and `retro-refer' to at later parts of the paper, will
be similarly marked by `R*.$^\star$'.}

\bigskip \noindent (Q2.1)\hskip 0.9in
\begin{minipage}{11cm}
\noindent ``{\small ...A field theory is not yet completely
determined by the system of field equations. Should one admit the
appearance of singularities?...{\small\em It is my opinion that
singularities must be excluded. It does not seem reasonable to me
to introduce into a continuum theory points (or lines {\it etc.})
for which the field equations do not hold}\footnote{Our
emphasis.}...}''
\end{minipage}

\vskip 0.1in

\noindent Bergmann, for instance, described GR's
`autocatastrophic' prediction of singularities as follows:

\bigskip \noindent (Q2.2)\hskip 0.9in
\begin{minipage}{11cm}
\noindent ``{\small ...What about shortcomings of general
relativity?...perfectly reasonable conditions at one time may lead
to field singularities at another. {\small\em That looks as if
general relativity carries within its conceptual belly the seeds
of its own destruction}\footnote{Our emphasis.}...}'' \cite{berg0}
\end{minipage}

\vskip 0.1in

\noindent while, more recently, Ashtekar too described
singularities and their associated infinities as an `intrinsic
fault' of GR:

\bigskip \noindent (Q2.3)\hskip 0.9in
\begin{minipage}{11cm}
\noindent ``{\small ...Although general relativity has been
accurately verified on macroscopic scales, it has an internal
blemish: it permits, and even predicts, the occurrence of singular
configurations in which physical quantities become infinite...}''
\cite{ash0}
\end{minipage}

\vskip 0.1in

The philosophical debate about whether singularities are
`positive' or `negative' traits of GR aside, we believe that very
few physicists would actually doubt that the principal pathogenic
gene in GR's conceptual genome---one that is the main culprit for
the existence of singularities---is the primitive conception, and
as a result, the basic assumption, of {\em modelling spacetime
after a differential} ({\it ie}, a $\smooth$-smooth) {\em
manifold}. Bergmann again, for instance, immediately after he
expressed his negative view about singularities above, added:

\bigskip \noindent (Q2.4)\hskip 0.9in
\begin{minipage}{11cm}
\noindent ``{\small ...Thus, let me begin by saying that {\em all
unitary field theories that I know have been based on one
topological model, that of a manifold}. They differ on the kind of
structures that they superimpose on that basic framework...}''
\end{minipage}

\vskip 0.1in

\noindent while Joshi admitted rather categorimatically right at
the very beginning of \cite{joshi} that:

\bigskip \noindent (Q2.5)\hskip 0.9in
\begin{minipage}{11cm}
\noindent ``{\small It is generally accepted now that any
reasonable classical theory of gravitation must admit
singularities where the curvatures grow unbounded and the usual
laws of physics breakdown. {\em Within the classical framework,
the physical universe is modelled by a spacetime manifold}...}''
\end{minipage}

\vskip 0.1in

Indeed, granted that a smooth manifold supplies one with the usual
(`classical' or `standard') differential geometric structure and
mechanism one needs in order to represent the laws of physics by
{\em differential equations} relating the various relevant smooth
physical quantities (fields), the coarse and intuitive picture one
has of singularities is as `regions' or `locations' in the
spacetime continuum where the `differentiability properties'
(smoothness) of physical quantities and, as a result, the
dynamical relations---the laws of Nature modelled after
differential equations---in which the latter
participate,\footnote{Thus, in a broad sense, by
`differentiability' we understand here the mathematical
representation of `dynamical variability' in physics---that
physical quantities (can) change. Equivalently, for us,
`measurable dynamical attributes' (commonly known as `{\em
observables}') and `differentiable quantities' (or `{\em
differentiables}' \cite{malrap3}) are effectively synonymous terms
standing for {\em dynamically variable properties of physical
systems}.} {\em break down} in one way or another, while at the
same time the very (smooth) fields that partake in those laws
become {\em unphysically} and mathematically {\em unmanageably}
infinite.

In fact, that singularities in GR (which, as it is well known,
represents its sole dynamical variable, the gravitational field,
by a smooth metric on a differential spacetime manifold) signal a
breakdown of differentiability, that is to say, that they
essentially mark the ineffectiveness of the entire edifice on
which Classical Differential Geometry (CDG), the so-called
Calculus on Manifolds, rests---{\it ie}, the $\smooth$-smooth
manifold---is built into their very definition. Clarke, for
example, remarks in \cite{clarke3}:\footnote{In this quotation the
words in in square brackets are our own additions for continuity,
completeness and clarity. We will return to comment in more detail
about this definition of smooth gravitational singularities in the
next section.}

\bigskip \noindent (Q2.6)\hskip 0.9in
\begin{minipage}{11cm}
\noindent ``{\small ...Thus the definition of a singularity
depends on the definition of an [analytic] extension of [the]
space-time [manifold], and so the question of what counts as a
singularity depends on what sort of extension is allowed. We call
a boundary point [of a smooth manifold] a class $C^{k}$
geometrical singularity if there is no [analytic] extension with a
$C^{k}$ metric that removes it; {\em i.e. if it is associated with
a breakdown of differentiability of the metric at the $C^{k}$
level}\footnote{Our emphasis.}...}''
\end{minipage}

\vskip 0.1in

This prompts us to emphasize here that it is the basic contention
of the present paper that

\bigskip \noindent (R2.1)\hskip 0.9in
\begin{minipage}{11cm}
\noindent behind both the {\em unphysicality} (`physical
inadmissibility') and the {\em unmanageability} (`mathematical
inadequacy and ineffectiveness in the handling') of singularities
and the infinities that they are associated with, lies the main
culprit for it all: {\em the}---in fact, {\em our}---{\em
representation of spacetime as a differential manifold and, as a
result, the CDG-framework and the various smooth constructions
(structures) within that framework that the manifold supports}.
\end{minipage}

\subsection{About Unphysicality of Singularities}

One of the main reasons for singularities and the infinities that
they are associated with is that the manifold picture of spacetime
allows one, even if just in theory, to pack an uncountable
infinity of events into a finite spacetime volume. We have no
actual physical experience of an infinity of events since we
invariably record a finite number of them (`field values') during
experiments of finite duration (`temporal extension') conducted in
laboratories of finite size (`spatial extension'). Einstein
himself, especially in view of the discrete or finitistic actions
of quanta, was sceptical about the infinities assailing the
geometrical spacetime continuum and, as a result, his continuous
field theory of gravity based on it, for as he remarked upon
concluding \cite{einst3}:\footnote{Einstein's quotation below is
the last paragraph of the last appendix D of ``{\itshape The
Meaning of Relativity}''. The complete paragraph (quotation) is
given subsequently in (Q2.?).}

\bigskip \noindent (Q2.7)\hskip 0.9in
\begin{minipage}{11cm}
\noindent ``{\small ...One can give good reasons why reality
cannot at all be represented by a continuous field. {\small\em
From the quantum phenomena it appears to follow with certainty
that a finite system of finite energy can be completely described
by a finite set of numbers}.\footnote{Our emphasis.} This does not
seem to be in accordance with a continuum theory, and must lead to
an attempt to {\small\em find a purely algebraic theory for the
description of reality}\footnote{Our emphasis.}...}''
\end{minipage}

\vskip 0.1in

\noindent Thus, simply on pragmatic or `experientially realistic'
grounds, this alone should suffice as a motivation to try to
somehow `discretize', `algebraicize' and, as a result, `quantize'
the geometrical spacetime manifold \cite{rapzap1,rapzap2} and,
{\it in extenso}, gravity \cite{malrap1,malrap2,malrap3}. Indeed,
as one can witness in \cite{malrap1}, Finkelstein's telling words
below were originally our principal (physical) motivation for
initiating the application of the algebraico-categorical and sheaf
theory based ADG to a finitistic, causal and quantal description
of spacetime and (vacuum Einstein-Lorentzian) gravity, which was
the subject matter of the aforementioned trilogy
\cite{malrap1,malrap2,malrap3}:

\bigskip \noindent (Q2.8)\hskip 0.9in
\begin{minipage}{11cm}
\noindent ``{\small ...The locality principle seems to catch
something fundamental about nature... Having learned that the
world need not be Euclidean in the large, the next tenable
position is that it must at least be Euclidean in the small, a
manifold. The idea of infinitesimal locality presupposes that the
world is a manifold. {\small\em But the infinities of the manifold
(the number of events per unit volume, for example) give rise to
the terrible infinities of classical field theory and to the
weaker but still pestilential ones of quantum field
theory}.\footnote{Our emphasis.} The manifold postulate freezes
local topological degrees of freedom which are numerous enough to
account for all the degrees of freedom we actually observe.

The next bridgehead is a dynamical topology, in which {\small\em
even the local topological structure is not constant but
variable}.\footnote{Our emphasis.} The problem of enumerating all
topologies of infinitely many points is so absurdly unmanageable
and unphysical that dynamical topology virtually forces us to a
more atomistic conception of causality and space-time than the
continuous manifold...}'' \cite{df5}

\end{minipage}

\vskip 0.1in

\subsubsection{The general covariance--versus--smooth singularities clash}

`Experiential realism' or `pragmatism' aside, we maintain that
{\it prima facie} there is an even stronger discord between
singularities and one of the conceptual pillars on which GR is
founded: the {\em Principle of General Covariance} (PGC). Here, to
express this fundamental disagreement, and also in order to
prepare the reader for our ADG-theoretic musings in the sequel, we
shall first abide by the following `heuristic', albeit intuitively
clear, version of the PGC which, in turn, is cast as a generalized
statement of the very Principle of Relativity (PR)
\cite{einst3,einst9}:\footnote{The generalized expression of the
PR in (R2.?), the fundamental non-objectivity of quantum theory
aside for the time being, is intimately related to Einstein's
conception of (objective) ``{\em physical reality}'', as follows:
``{\small ...Physics is an attempt conceptually to {\em\small
grasp reality as something that is considered to be independent of
its being observed}. In this sense one speaks of `{\em\small
physical reality}'...}'' \cite{einst9} (our emphasis).}

\bigskip \noindent (R2.2)\hskip 0.9in
\begin{minipage}{11cm}
\noindent the laws of Physics---here in particular, the law of
gravity---are independent of our measurements, {\it ie}, of (the
numerical results of) our observations of the physical fields that
take part in them.
\end{minipage}

\vskip 0.1in

\noindent This, together with the following identification that we
would like to emphasize in the present paper, namely that

\bigskip \noindent (R2.3)\hskip 0.9in
\begin{minipage}{11cm}
\noindent {\em a differential manifold $M$ is nothing but the
algebra $\sstruc=\smooth(M)$ of differentiable} ({\it ie}, smooth
coordinate) {\em functions on}/(of) {\em it}(s points),
\end{minipage}

\vskip 0.1in

\noindent which vital interdependence (or equivalence) we may
loosely cast as

\begin{equation}\label{eq1}
M\rightleftarrows\smooth(M),
\end{equation}

\noindent enables us to arrive swiftly at the usual (mathematical)
expression for the PGC by the following syllogism:

\begin{enumerate}

\item `Operationally' (algebraically) speaking, (the results of)
our acts of measurement and localization of smooth fields on the
points of the differential spacetime manifold $M$ are organized
into the algebra of (real-valued) smooth coordinates
$^{(\R)}\smooth(M)$\footnote{From now on we will omit the
pre-superscript `$\R$' from $\smooth(M)$ as we will implicitly
assume that our measurements `yield' (or simply that we record)
real numbers, unless of course we explicitly declare up-front in
the sequel another set of numbers as our recorded (generalized)
`measurement' or `coordinatization values'.} of those points
\cite{malrap3,mall8},\footnote{More precisely, such measurement
and point-event localization acts are conveniently grouped into
{\em sheaves} $\struc=\smooth_{M}$ of abelian topological algebras
$\sstruc$ (over the manifold $M$) such as $\smooth(M)$
\cite{mall8}, something that foreshadows our sheaf-theoretic, ADG
based endeavors in the sequel (in this respect, see also the
trilogy \cite{malrap1,malrap2,malrap3}).}

\item But the smooth fields involved in the physical laws---for
gravity in particular, as originally formulated in GR, the ten in
principle arbitrarily ({\it ie}, to any order) differentiable
gravitational potentials comprising the smooth metric tensor
$g_{\mu\nu}$, which engages into Einstein's equations---are
nothing else but {\em $\otimes_{\struc}$-tensors}\footnote{That
is, they are tensors whose components, expressed in local open
coordinate patches $U$ of $M$ ($U\subset M$), are elements of
$\smooth(U)$---smooth functions on $U$ (or equivalently, in
sheaf-theoretic terms, they are elements of
$\Gamma(U,\struc=\smooth_{M})$---local sections of the structure
sheaf $\smooth_{M}$ of smooth coordinates of the points of $M$).}
\cite{mall1,mall2}; {\it ergo},

\item Categorically speaking, the differential equations (physical
laws) in which these fields partake are independent of
(`invariant' so to speak, and the relevant fields are `covariant',
with respect to)\footnote{Technically speaking, `invariance'
usually refers to the Lagrangian (action) from which the laws
derive from the variation of the field variables, while
`covariance' pertains directly to the dynamical laws themselves
and to the field quantities involved in them \cite{auyang}.} the
group $\mathrm{Aut}(M)$ of automorphisms of $M$---the differential
structure `symmetry group' of `active point-transformations' of
the pointed base differential spacetime manifold on which these
fields are soldered (localized) and from which they actually
derive their `differentiability properties' ({\it ie}, they
actually qualify as being {\em smooth} in the classical sense of
this word, so that in the first place they can participate into
those laws, which are differential equations proper).\footnote{In
this sense, $\mathrm{Aut}(M)$ may be thought of as the manifold's
`auto-transformation group' respecting the various fields'
(classical) differentiability ($\smooth$-smoothness) property.}

\item Since, again categorically speaking,\footnote{That is to
say, implicitly working in the category $\man$ of
(finite-dimensional) $\smooth$-smooth manifolds.}
$\mathrm{Aut}(M)\equiv\mathrm{Diff}(M)$, we have effectively
recovered the standard (mathematical) expression of the PGC of GR.
In other words, the usual expression of the PGC maintaining that
the gravitational law of physics is a differential equation
involving tensorial quantities relative to arbitrary coordinate
transformations (which thus makes these quantities independent of
the local reference frames to which they are referred to and being
`measured'),\footnote{And one should recall that, since $M$ is by
definition locally Euclidean ({\it ie}, locally its model is the
Cartesian space $\R^{4}$ coming equipped with the usual
differential structure), one identifies the `structure group' of
GR with $GL(4,\R)$---the group of general (smooth) coordinate
transformations.} assumes here a precise {\em functorial form with
respect to our smooth point-localizations or `measurements' (of
the gravitational field) in} $\sstruc\equiv\smooth(M)$: the law of
gravity is a differential equation\footnote{As we will see in the
sequel, in ADG-theoretic terms even differential equations are of
an essentially categorical character as they are equations between
{\em sheaf morphisms}, the main sheaf morphism being the
gravitational connection $\conn$---the generalized (`localized' or
`gauged') differential operator $\partial$
\cite{mall3,malrap1,mall9,malrap2,malrap3}. It must also be noted
that what appears as the basic variable in the equations
themselves is the curvature of the connection which is an
$\struc$-morphism (tensor), thus our (classical) smooth
`observations'/localizations of the gravitational field (strength)
in $\struc\equiv\smooth_{M}$ respect it.} involving smooth
quantities obtained from our basic measurements
(`coordinatizations') in $\smooth(M)$ by the action of the {\em
homological tensor product functor}
$\otimes_{\struc}$\footnote{Again, it is tacitly assumed that we
are working within the category $\man$ of (finite-dimensional)
differential manifolds $M$---and more particularly, in the
category of sheaves $\struc\equiv\smooth_{M}$ of abelian
topological algebras $\sstruc\equiv\smooth(M)$ of (real or
complex-valued) smooth functions over them.} \cite{mall1,mall2} on
them.

\end{enumerate}

\noindent A concise categorical distillation of the syllogism 1--4
above, thus effectively also of the PGC of the manifold based GR,
is the following:

\bigskip \noindent (R2.4)\hskip 0.9in
\begin{minipage}{11cm}
\noindent The gravitational law (Einstein's equations) is {\em
functorial} relative to our smooth `measurements' and
`localizations' of the relevant fields---processes which are
represented algebraically by $\sstruc$, or in a sheaf-theoretic
(local) sense, by (local) sections of the sheaf
$\struc$.\footnote{In the sequel, where we will observe in
ADG-theoretic terms that the basic gravitational variable (field)
is the connection, rather than the metric, and in view of the fact
that the actual `observable' quantity---{\it ie}, the `measurable'
(geometrical) dynamical quantity engaging into the Einstein
equations---is actually the curvature of the connection, which is
an $\struc$-morphism, the `identification' above of the PGC with
`{\em functoriality of dynamics relative to our measurements in}
$\struc$' will become even more transparent and quite natural.}
\end{minipage}

\vskip 0.1in

An immediate `corollary' of the above expression of the PGC is
that {\em the law of gravity `permeates' or `sees through' the
base smooth spacetime manifold insofar as the latter is, in an
operational sense, identified with our operations of
coordinatization of its point events, which in turn are
conveniently grouped into the algebra $\smooth(M)$} (\ref{eq1})
(or equivalently and more `locally speaking', into the sheaf
$\smooth_{M}$ thereof over $M$).\footnote{So, in the following
sense the aforesaid expression of the PGC is closely akin to a
generalized sort of the PR \cite{einst3}: {\em the laws of Physis
are independent of the coordinates we lay out to chart (or in a
Cartesian sense, to `label'---{\it ie}, ascribe numerical values
to or `arithmetize') and localize the relevant fields on spacetime
points}.} Parenthetically we would like to point out that, of
course, Einstein himself had clearly intuited that the laws of
Nature should be unaffected by {\em our} (arguably subjective and
arbitrary) choices of coordinates. Characteristically, he remarks
in the `{\itshape Time, Space, and Gravitation}' article 12 in
\cite{einst10} about the essential motivation for his transition
from SR (inertial frames) to GR (arbitrary coordinate frames)
according to the PR:

\bigskip \noindent (Q2.9)\hskip 0.9in
\begin{minipage}{11cm}
\noindent ``{\small ...Must the independence of physical laws with
regard to a system of coordinates be limited to systems of
coordinates in uniform movement of translation with regard to one
another? {\small\em What has nature to do with the coordinate
systems that we propose and with their motions?}\footnote{Our
emphasis.} Although it may be necessary for our descriptions of
nature to employ systems of coordinates that we have selected
arbitrarily, the choice should not be limited in any way so far as
their state of motion is concerned\footnote{Or perhaps better
expressed, (the said arbitrary choice of any particular system of)
coordinates should not affect in any way the dynamical equations
(laws) of motion of the fields in focus.}...}''
\end{minipage}

\vskip 0.1in

\noindent Thus, in this sense, {\em GR may be regarded as a
manifestly background $M$-independent theory, as its dynamical
laws involving the gravitational field (Einstein equations) `see
through' the $\smooth$-charted spacetime and its smooth
`self-transmutations' in} $\mathrm{Aut}(M)\equiv\mathrm{Diff}(M)$
\cite{mall3,mall9}. If in turn we abide to a Wheelerian-type of
principle holding that

\bigskip \noindent (R2.5)\hskip 0.9in
\begin{minipage}{11cm}
\centerline{no theory is a {\em physical} theory unless it is a
{\em dynamical} theory,}
\end{minipage}

\vskip 0.1in

\noindent the above expression of the PGC, always in the classical
(CDG) framework of our (differential geometric) conception and
primitive assumption of spacetime as a smooth continuum,
vindicates Einstein's

\bigskip \noindent (Q2.10)\hskip 0.9in
\begin{minipage}{11cm}
\noindent ``{\small\em Time and space are modes by which we think,
not conditions in which we live}'' \cite{einst2},
\end{minipage}

\vskip 0.1in

\noindent since the differential spacetime manifold does not
actively participate in the gravitational dynamics, thus it is not
a physical, `organic', `living' condition on which, so to speak,
the laws of Nature vitally depend.

Parenthetically, it must be noted here that, already back in the
1920s, Einstein had conceived of the spacetime continuum as an
unphysical, ether-like substance \cite{einst1,malrap3}, ``{\em
acting but not being acted upon}'' \cite{einst3}. To be sure,
however, the smooth spacetime continuum, although it does not
participate itself into the dynamical equations for gravity, it
provides us with the {\em vital `differential conditions' or
`properties'}, or even better, the {\em essential differential
geometric mechanism}, for representing these equations as {\em
differential equations}: the structural and calculational
apparatus of CDG (Calculus). This was again noted by Einstein in
\cite{einst1} in the form of `{\em differential or infinitesimal
locality}' (Q2.?)---the basic assumption that field actions
connect ({\it ie}, dynamically evolve and interact between)
`infinitesimally contiguous' or `infinitely proximate' events in
the smooth spacetime continuum, something that in turn qualifies
(defines!) fields as being smooth (differentiable to an arbitrary
order) and, concomitantly, the dynamical equations in which they
participate as being {\em differential equations} proper (albeit
in the classical sense `afforded' and `furnished' by the
geometrical base differential spacetime manifold $M$). In
connection with these remarks, we would also like to quote
Finkelstein from \cite{df1}:

\bigskip \noindent (Q2.11)\hskip 0.9in
\begin{minipage}{11cm}
\noindent ``{\small ...Soon after the physics of light led
Einstein and others to special relativity, the physics of gravity
led Einstein to general relativity. The heart of relativity is
Einstein's locality principle:

{\small\em The law of nature is local and
causal.}\footnote{Finkelstein's emphasis.}

Here `local' means that the dynamical law relates any event $e$
only to events in its infinitesimal neighborhood. `Causal' means
that these other events are inside the light cone of $e$.

That is, {\em Einstein assumed that the law of nature was a
differential equation},\footnote{Emphasis is ours.} and that its
characteristics (the surfaces across which there could be jumps in
a solution) were null cones...}''
\end{minipage}

\vskip 0.1in

\noindent To return to the `illusive' nature of spacetime, even
more succinctly, Eddington pointed out, remarkably as early as
1916(!) \cite{eddington1}, that the PR essentially entails the
physical meaninglessness and artificiality of (the) base spacetime
(continuum of GR), relegating it merely to {\em our own mental
fiction}---or equivalently, Einstein's ``{\em mode by which we
think}'' in (Q2.?) above---a virtual background scaffolding, a
surrogate host (employed for the localization) of the relevant
fields, that does not play any role in the actual physical
dynamics:

\bigskip \noindent (Q2.12)\hskip 0.9in
\begin{minipage}{11cm}
\noindent ``{\small According to the principle of relativity in
its most extended sense, {\small\em the space and time of physics
are merely a scaffolding in which for our own convenience we
locate the observable phenomena of Nature. Phenomena are
conditioned by other phenomena according to certain laws, but not
by the space-time scaffolding, which does not exist outside our
brains}.\footnote{Our emphasis. Especially in the case of our
sheaf-theoretic ADG-endeavors in the sequel, we will see how
pertinent and prophetic(!) these remarks by Eddington are. Indeed,
if there is any value at all to a background spacetime---be it a
continuum or even a discretum as we shall see---it is as an
external `parameter' space, a virtual scaffolding assumed for our
convenience, for the (sheaf-theoretic) localization (gauging) and
concomitant dynamical variation of the actual `geometrical
objects' of Nature---the physical fields, but itself playing
absolutely no (structural) role in that dynamics. The
dynamics---{\it ie}, the laws of physics---`in itself' is
essentially algebraic---{\it ie}, it is defined by the algebraic
relations between the said fields ({\it viz}. connections) which
are themselves `{\em inherently algebraic and autonomous
entities}' and in no need of an external, background space(time)
for their subsistence, efficacy, or operativeness.}}''
\end{minipage}

\vskip 0.1in

In the light of these remarks about the `virtual', `imaginary' or
`mental' reality of the spacetime continuum, that is to say, if
one abides by the aforesaid version of the PGC, the existence of
singularities---{\it ie}, {\it \`a la} Einstein (Q2.?), {\em the
existence in a theory} which, like GR, is based on the smooth
spacetime manifold, {\em of points (or lines {\it etc.}) for which
the field equations do not hold}, or even more graphically {\it
\`a la} Joshi (Q2.?), {\em the existence of loci in the spacetime
continuum where the laws of physics break down}---comes straight
into conflict with it ({\it ie}, the PGC), for singularities
entail a `negative dependence' (even more impressively, a {\em
breakdown}!) of the law of gravity on {\em our own}
coordinatizations or measurements of the smooth physical fields
localized on $M$'s point events, or, what amounts to the same via
the identification (\ref{eq1}), on {\em our own}, and physically
arbitrary(!), mathematical modelling of physical `spacetime' by a
differential manifold $M$.\footnote{For recall again Einstein's
words in (Q2.?): ``{\em What has nature {\rm [{\it ie}, the
dynamical laws of physics]} to do with the coordinate systems that
\underline{we propose}?}''.}

\bigskip \noindent (R2.6)\hskip 0.9in
\begin{minipage}{11cm}
\noindent All in all, {\em the law of gravity appears to break
down at singularities which are due to our modelling of spacetime
after a differential manifold $M$, which in turn is tautosemous
with our smooth coordinatizations of its points in} $\smooth(M)$
(\ref{eq1}): stated thus, how {\em contingent and inextricably
dependent} on our mathematical model of spacetime as a
$\smooth$-smooth continuum---ultimately, on our own `smooth
measurements or chartings' of fields represented by
$\struc=\smooth_{M}$---does the physical law of gravity appear to
be?\footnote{This apparent contingency of the gravitational law of
Nature, especially in the classical domain of GR ({\it ie}, where
the `observer dependence' and concomitant `non-objectivity' of
physical reality that quantum theory brought about is not supposed
to play any role at all), on our (mathematical) model of
spacetime, we find hard to accept in the present paper. For one
thing, it comes straight into conflict with Einstein's PR in
footnote 5, as well as with our version of the PGC of GR in (R2.?)
and its categorical (functorial) expression in (R2.?). In fact,
this (theoretically) unacceptable feature of singularities and
{\it in extenso} of our $\smooth$-smooth manifold picture of
spacetime hosting them was our principal motivation for writing
this paper in the first place.}

\end{minipage}

\vskip 0.1in

To our knowledge, it was Einstein first in \cite{einst3}, and
subsequently Geroch in \cite{geroch}, who---the first implicitly
as in (Q2.?), the second more explicitly---found the PGC of GR
(much in the way it was expressed above) and singularities in
glaringly serious `physical disagreement'. In fact, in the very
first sentence of the abstract of \cite{geroch}, the author notes:

\bigskip \noindent (Q2.13)\hskip 0.9in
\begin{minipage}{11cm}
\noindent ``{\small The general covariance of relativity theory
creates serious difficulties in formulating a suitable definition
of a singularity in this theory.}''
\end{minipage}

\vskip 0.1in

\noindent To recapitulate, general covariance and singularities do
not seem to go hand in hand; moreover, this discord is so strong
that it apparently makes a clear-cut definition of singularities
in the manifold based GR a very subtle, elusive and difficult task
both technically (formally or `mathematically') \cite{clarke4},
but most importantly, conceptually (semantically or
`interpretationally').

In a nutshell, it is fair to say that the singularities-vs-PGC
tension is due to the following (mathematical) `vicious circle'
\cite{rap5}: on the one hand the $\smooth$-smooth structure of the
spacetime manifold $M$ appears to be a necessary assumption in
order to represent the dynamical law of gravity by a {\em
differential equation} proper, as well as to implement the PGC of
GR via $\mathrm{Diff}(M)$, while on the other, and as we will
argue extensively in the sequel, singularities are essentially
built into $\smooth(M)$---the algebra that {\em defines} $M$ as a
differential manifold in the first place (\ref{eq1}), and are
(differential) geometrically perceived precisely as {\it loci} in
$M$ where that smooth structure `breaks down', `becomes
anomalous', or at best `misbehaves' in one way or another.

\subsubsection{Begging the question: `real' or `virtual'
singularities?}

Having mentioned the conceptual and technical dissonance between
the PGC and singularities in the theoretical framework of the
differential manifold based CDG, it must be noted at this point
that it appears to be unanimously accepted now that there are two
kinds of singularities in GR, which might be coined `{\em
virtual}' and `{\em real}'. As virtual are characterized the
so-called {\em coordinate singularities}, the `canonical' example
being the exterior Schwarzschild one at a distance $r=2m$
(horizon) from the gravitating point particle of mass $m$
\cite{eddington3,df}. Coordinate singularities are not regarded as
being `real' ({\it ie}, physically significant), since they are
simply due to the fact that the physicist has laid down
inappropriate coordinate patches---she has chosen an inappropriate
coordinate frame of reference---in order to chart the spacetime
continuum and are not intrinsic physical anomalies of that
spacetime structure supporting the gravitational
field.\footnote{In the case of the exterior Schwarzschild
singularity for example, passing from spherical Schwarzschild
coordinates to the `logarithmic' ones of the so-called
Eddington-Finkelstein frame allows one to convert it to a
unidirectional membrane, thus show that it is not really a
singularity, but just a `{\em coordinate artifact}'
\cite{eddington3,df}. We will return to it in more detail in
section 5.} Geroch, for instance, declares up-front in
\cite{geroch}:

\bigskip \noindent (Q2.14)\hskip 0.9in
\begin{minipage}{13cm}
\noindent ``{\small ...We shall not be concerned with so-called
``coordinate singularities''. This term refers to a spacetime
which has been expressed in an improper coordinate system...
{\small\em The presence or absence of a coordinate singularity is
not a property of the spacetime itself, but rather of the
physicist who has chosen the coordinates by which the spacetime is
described}\footnote{Our emphasis.}...}''
\end{minipage}

\vskip 0.1in

\noindent Parenthetically, we could mention here, in contrast to
Geroch's remarks in (Q2.?) above, but in a similar manner of
expression in order to emphasize this contradistinction, ADG's
basic thesis regarding singularities, which will be amply
supported subsequently by the results of the present paper:

\bigskip \noindent (R2.7)\hskip 0.5in
\begin{minipage}{12cm}
\noindent In ADG, we are not troubled at all, as far as the
essentially algebraico-categorical differential geometric
mechanism is concerned, by any kind of `geometrical spacetime
singularity', and {\em the distinction of the latter into `real'
and `coordinate' ones loses its meaning in the theory}.
Singularities refer to a field which has been coordinatized by,
`measured' in, or simply referred to, an `improper' coordinate
system---the term `improper' pertaining to a structure sheaf
$\struc$ of coefficient functions or coordinates that host
`singularities' of some kind\footnote{Indeed, in the classical
case, $\struc\equiv\smooth_{M}$, and it is the smooth functions
that carry within them the seeds of the singularities and the
differential geometric anomalies of the spacetime manifold and of
the CDG (Calculus) based on it.}---yet, singularities that do not
affect or impede the said `innately algebraic' mechanism, let
alone cause it to `break down' and, concomitantly, the law---{\it
ie}, the differential equation which the field obeys (in fact,
{\em defines}), to stop being in force ({\it ie}, not to hold over
the `singular {\it loci}')...{\em The presence or absence of a
singularity is not a property of the field itself, which anyway
physically exists independently of what we perceive and measure
({\it ie}, coordinatize) as being `spacetime'},\footnote{See
Einstein's PR in footnotes 5 and 15, Eddington's remarks in
(Q2.?), as well as our categorical (functorial) expressions of the
PGC of GR in (R2.?) and (R2.?).} {\em but rather of the physicist
who has chosen the coordinates by which this field is described
({\it ie}, `measured', `coordinatized' or `numerically perceived',
thus soldered on or coordinated to, and placed within, the said
ambient `differential geometric spacetime
framework'\footnote{Indeed again, the singular spacetime manifold
of the classical theory corresponding to $\smooth_{M}$ that one
assumes for coordinate structure functions of its point-events
(\ref{eq1})---`events' here being understood as the (numerical)
results of our field-probings and measurements, our `Cartesian
arithmetizations' of (our numerical ascriptions to) the physical
fields, in accordance with Einstein's concluding sentence in the
quotation of footnote 15. To be sure, as we are going to argue
extensively in 7.5.5 motivated by Stachel's penetrating analysis
into the deeper meaning of Einstein's hole argument and the
significance of the PGC of GR, the points of $M$ do not have {\it
a priori} the right to be called `{\em spacetime events}'---`{\it
a priori}' here meaning `{\em kinematically fixed---before the
dynamical field of gravity is determined}' (as a solution of the
field equations). We wish to thank John Stachel for bringing this
subtle and (as we shall see later) important interpretational
matter to our attention in a timely e-communication. However,
keeping in mind the pending remarks in 7.5.5 about `physically
interpreting the points of a manifold as spacetime events proper',
we may abuse language below and refer to $M$'s points as events.})
that, after all \underline{we} impose on the field in order to
represent mathematically---as a differential equation---and study
the law that it obeys\footnote{As we will see in the context of
ADG below, it is more accurate to say `the law that the field {\em
defines}', not `the law that it {\em obeys}', although we may use
the two verbs interchangeably in the future.})}.
\end{minipage}

\vskip 0.1in

\noindent while Clarke in \cite{clarke3} further highlights the
supposedly fundamental distinction between virtual and true
singularities:\footnote{Again, in the quotation below, the words
in square brackets are our own additions for completeness and
clarity of exposition.}

\bigskip \noindent (Q2.15)\hskip 0.9in
\begin{minipage}{11cm}
\noindent ``{\small ...[Based on the definition of singularities
as boundary points beyond which the smooth causal geodesics in the
spacetime manifold cannot be analytically extended or continued,]
for the ideal endpoint to a curve to be called a singularity, as
opposed to a regular boundary-point, it must be the case that
there is no extension of the space time in which the curve in
question can be continued: {\small\em if there were such an
extension the purported singularity would be regarded as analogous
to a `coordinate singularity' such as the Schwarzschild horizon
written in Schwarzschild coordinates}\footnote{Our
emphasis.}...}''
\end{minipage}

\vskip 0.1in

\noindent It is plain from the above that `real' singularities are
supposed to represent true spacetime pathologies, uncircumventable
simply by (smooth) coordinate transformations and insuperable in
general by the technical $\smooth$-means of CDG obstructions to
the smooth spacetime structure. Such singularities, as Clarke
defined in (Q2.?) before, are {\it loci} where the differential
structure of the spacetime manifold and, concomitantly, the
physical laws that it supports as a smooth base space, manifestly
break down.\footnote{In the case of the Schwarzschild solution,
the fact is that the inner $r=0$ singularity cannot be, similarly
to the exterior one, `transformed away' by a coordinate change,
thus it shows us that {\em the classical theory---based on the
spacetime manifold---is out of its depth when trying to calculate
the gravitational field right at its point mass source}
\cite{eddington3,df}. It is widely accepted nowadays that only a
quantum theory of gravity may be able to describe the
gravitational field on its source, right at its `true singularity'
\cite{pen5,perry,modesto,husain}. We will return to such `real'
singularities and to the associated question whether the `true'
quantum gravity will be able to `heal' or ultimately to remove
singularities in sections 3 and 6, respectively.}

The last remarks bring us to the mathematical unmanageability of
$\smooth$-smooth manifolds and the CDG based on them in coping
with gravitational singularities, which in turn makes the latter
be viewed (always of course from within the CDG-framework) as {\em
incurable classical differential geometric diseases}. We will
argue, based on some glaring both conceptual (physical) and
technical (mathematical) `oxymorons' and in view of our
identification in (\ref{eq1}), that the notion of singularity
creates various `chimeras', various `false impressions' within the
framework of the CDG-based GR. One of these chimeras is the
distinction between virtual and real singularities, and we shall
argue that this distinction is in effect `begging the question'
of, and to a large extent obscures, what `truly' is a singularity
in GR \cite{geroch} as well as of what is its actual mathematical
`origin'. The points to be raised will allow us then to introduce
rather naturally ADG's view on singularities, as well as to
present its basic concepts and methods for evading them altogether
\cite{malros1,mall3,malros2,mall9,malros3,mall7,mall11}.

\subsection{About Mathematical Unmanageability}

On the face of it, the discussion above is pregnant to a
physico-mathematical oxymoron. On the one hand we have our basic
contention that the differential manifold {\em is} the algebra of
smooth `coordinate' functions on it (\ref{eq1}), and on the other
the nowadays general consensus that {\em real singularities are
not just coordinate ones}. One way out of this impasse is to
declare, by `geometrical fiat' as it were, (real) singularities as
being {\em ideal boundary points to the smooth spacetime manifold
which, `in themselves', are not parts of the set $M$ of spacetime
events proper, loci not lying in the `bulk interior manifold' of
the actual physical point-events where the gravitational field
equations hold, which in turn are deemed to be regarded as being
regular (non-singular)}. Clarke, for example, makes it clear from
the beginning that:\footnote{Again, in the quotation below, the
words in square brackets are our own additions for completeness
and clarity. The idea of viewing singularities as {\em ideal}
boundary points in spacetime---sites that are inaccessible (in
`finite time') by smooth (causal) paths followed by physical
particles---was explicitly pitched in \cite{geroch2} although it
was intuitively implicit already four years earlier in
\cite{geroch} in the guise of the idea of `{\em (causal) geodesic
incompleteness}' (for more details, the reader should wait until
the next section).}

\bigskip \noindent (Q2.16)\hskip 0.9in
\begin{minipage}{11cm}
\noindent ``{\small ...With each incomplete inextendible curve we
associate a{\rm [n ideal]} boundary point: a point added on to
{\rm [the]} space-time {\rm [manifold $M$]} mathematically, but
not forming part of the physical space-time... The aim of the
construction of boundary points is to build a mathematical model
based on a set $\overline{M}=M\cup\partial M$, where $\partial M$
is the set of boundary points and $M$ is the set of real physical
points in space-time, with $\overline{M}$ given a topology which
makes it the closure of $M$ (so that it is sensible to speak of
$\partial M$ as a boundary)...}'' \cite{clarke3}
\end{minipage}

\vskip 0.1in

\noindent A synthesis of the quotes (Q2.?), (Q2.?) and (Q2.?) lead
us to infer that, according to the definitions above, as {\em
physical} (regular) spacetime events are regarded only the {\it
loci} in the bulk (interior) $M$ of $\overline{M}$ where the
gravitational potentials are {\em differentiable} (smooth)
functions and the law of gravity {\em holds} ({\it ie}, it can be
expressed) as a differential equation in the realm of the smooth
$M$ and its CDG. This is well in line with our regarding smooth
singularities as `{\em obstructions to (classical)
differentiability}', always in the context of the CDG-theoretical
framework.

On the other hand, such an admittedly forced or {\it ad hoc}
geometrical declaration of singularities as `marginal points at
the edge of physical spacetime', having only mathematical purpose
and utility, but being almost devoid of physical significance,
appears to come straight into conflict with Geroch's distinction
between real and coordinate singularities in (Q2.?), because for
him, {\em true singularities, unlike virtual ones, are purported
to be properties of the (physical) spacetime itself and not merely
the result of improper spacetime descriptions (coordinates) used
by the physicist}.

\subsubsection{Enter ADG}

We contend that there is no need for such elaborate, artificial
physico-mathematical distinctions (and apparently deep conceptual
asymphonies!) between coordinate (virtual) and real (true)
singularities as long as one seriously abides by (\ref{eq1}),
namely, by the mathematical fact (in fact, {\em definition}!) that
the $\smooth$-smooth spacetime manifold $M$ is nothing else but
the algebra of its differentiable (coordinate) functions. The
latter, in turn, provides the foundations for the {\em aufbau} of
the entire theory of CDG which fares miserably upon trying to cope
with singularities---singularities of the very smooth functions
(and, {\it in extenso}, of the higher order smooth tensor fields
built out of them by (anti/symmetric) iterations of the
homological functor $\otimes_{\smooth_{M}}$ \cite{mall2}) that
define $M$ as a differential manifold. Stated in a positive way,

\bigskip \noindent (R2.8)\hskip 0.9in
\begin{minipage}{11cm}
\noindent all the singularities of the smooth spacetime manifold
$M$ are `coordinate' ones in the sense that they are anomalies of
certain elements in the function algebra $\smooth(M)$ that defines
it in the first place as a {\em differential space}\footnote{That
is, a space endowed with a differential structure.} proper. At the
same time, it is {\em our assumption} of modelling spacetime after
a $\smooth$-manifold (which automatically forces us to adopt the
usual concepts and technical tools of the CDG based on
$\smooth(M)$), that `stumbles and falters' (thus also the
differential equations for which CDG was created in the first
place\footnote{See subsection 7.6.} break down) on singularities.
Hence, singularities are shortcomings of our own theory (model) of
spacetime (the differential manifold) and of the way that we
mathematically represent the laws of Nature, as differential
equations, within the differential geometric framework (CDG)
supported by that model. {\em In summa}, {\em one cannot think of
singularities independently of the differential spacetime manifold
$M$; they are differential geometric diseases built into
$\smooth(M)$ and expressing the `inherent limitations' of CDG,
since the latter vitally depends on $M$, which in turn is
tautosemous with $\smooth(M)$ or, sheaf-theoretically, with the
structure sheaf $\smooth_{M}$ thereof}.
\end{minipage}

\vskip 0.1in

\noindent From this perspective it appears quite
inappropriate---not to say `{\em hubristic}'---to ascribe
singularities to Nature Herself ({\it ie}, give them a physical
significance), since they are problems inherent in {\em our}
spacetime model ($M$) and theory (CDG) of Her ({\it ie}, our
regarding of physical space(time) as being smooth or, differential
geometrically speaking, `regular') (R2.?). In other words, as
already briefly alluded to in (R2.?), we maintain that

\bigskip \noindent (R2.9)\hskip 0.9in
\begin{minipage}{11cm}
\noindent {\em Nature} ({\it ie}, the physical laws and the
physical fields participating in them) {\em has no singularities,
but it is our smooth model of Her} ({\it ie}, of what we
`axiomatically' accept {\it a priori} as being `physical
space(time)') {\em that is of limited applicability and validity}.
\end{minipage}

\vskip 0.1in

\noindent Of course, we tacitly abide by the `principle' that {\em
whenever there seems to be an asymphony between physics and the
mathematics one uses to describe and represent the physical
processes, situations and phenomena involved in that physics, one
should always question and change the maths} and never interpret
the dissonance as a `problem' of Physis, or even attempt to
uncover some supposedly hidden physical essence and significance
in those mathematical shortcomings and `anomalies'---in our own
`misrepresentations' of physical phenomena so to
speak.\footnote{To be sure, singularities (and their associated
infinities) may indeed be thought of as trying to tell us that
there is something fundamentally wrong with the (mathematical)
means that we employ to describe Nature (here, the manifold based
CDG and the description of physical laws---in particular, the law
of gravity---as differential equations within the theoretical
framework of Calculus), but surely not that there is a limitation
of the physical fields and of the laws that they obey/define at
their {\it loci} (Q2.?). We could summarize this subtle
distinction in the following: {\em it is not that the
gravitational field (law) breaks down at a singularity, but rather
that the manifold and, {\it in extenso}, the CDG-based GR does
so.}} For our main example above about the conflict between the
PGC and smooth singularities, we find it unreasonable, to say the
least, to doubt the first (which, after all, is more or less a
physical principle!\footnote{See the generalized Principle of
Relativity in (R2.?) above and its associated `definition' of
(objective) physical reality by Einstein in footnote 5.}) instead
of trying to question, find the main culprits for its disagreement
with the physics and, hopefully, change the second (which, anyway,
is merely a consequence of our mathematical, $\smooth$-smooth
manifold model of spacetime, and by no means Nature's `sacrosanct
quintessence').\footnote{This of course implies that if $M$, with
its intrinsic pathologies (singularities), has to go, so will the
standard mathematical representation by $\mathrm{Diff}(M)$ of the
PGC in the $M$-based GR. However, the PGC, as a basic {\em
physical} principle, should remain intact in the {\em physical}
theory; only its mathematical representation should change. This
is in line with what we said above, namely, that upon encountering
a problem, blame it on the mathematics, not the physics, and
consequently, try to change the former, not the latter.}

\noindent To summarize things:

\bigskip \noindent (R2.10)\hskip 0.6in
\begin{minipage}{13cm}
\noindent One cannot think of singularities apart from our
$\smooth$-smooth manifold model $M$ for spacetime, for they are
`intrinsic' to $\smooth(M)$, which defines $M$ in the first place
(\ref{eq1}). In this sense, all singularities are `coordinate',
`virtual' ones physically speaking, and at the same time {\em
mathematically `real'} ({\it ie}, they are mathematical
(arti)facts!), insofar as they express `innate' pathologies of $M$
and `inherent' limitations of the CDG that is based on it. {\em
Singularities are not physically real} ({\it ie}, they are not
Nature's own),\footnote{One could actually say that this is {\em
the} fundamental `{\em physical axiom}' (or intuition) of ours to
which the epithet `{\em chimeras}' in the title of the present
paper pertains. In connection with this basic intuition of ours,
see the remarks about our principal motivation for writing this
paper in the footnote of (R2.?).} but they simply express that it
is our Calculus---our mathematical Analysis based on the smooth
spacetime continuum---that is out of its depth when dealing with
certain physical situations, like for example when trying to
calculate the smooth gravitational field right at its point source
mass as in the Schwarzschild black hole scenario
\cite{eddington3,df} while at the same time maintain that this
offensive locus is part of the physical smooth spacetime manifold.
\end{minipage}

\vskip 0.1in

The contents of (R2.?) and (R2.?) may be thought of as being
`post-anticipations' of Ashtekar's telling remarks following
(Q2.?) above in which he comments on the singularities and their
associated infinities which result, in quite generic physical
situations in a smooth spacetime manifold, from the gravitational
collapse or accretion of a cloud of `cool, non-interacting matter'
(dust):

\bigskip \noindent (Q2.17)\hskip 0.9in
\begin{minipage}{11cm}
\noindent ``{\small ...Now, one believes that such infinities do
not actually occur in Nature, and their occurrence in a theory is
a signal that one is applying the theory beyond its domain of
validity...}'' \cite{ash0}
\end{minipage}

\vskip 0.1in

\noindent But which {\em theory} is Ashtekar referring to?---the
mathematical framework within which GR is cast ({\it ie}, CDG), or
to the actual dynamical law for the gravitational field ({\it ie},
the Einstein equations) which defines GR as a {\em physical}
theory? Our basic thesis here is that the theory referred to above
is the manifold-based CDG, not the field laws of GR.\footnote{As
noted earlier, in our opinion the `confusion' that is reflected in
the physicist's claim that GR, as a physical theory, breaks down
at singularities, is due to her identification of physical
spacetime with (the mathematical artifact corresponding to) a
locally Euclidean space $M$.} Indeed, at so-called `real'
singularities, CDG has reached the limit (in fact, a `dead end' as
far as physically meaningful constructions and calculations by
means of Calculus are concerned) of its applicability and
validity.\footnote{See 2.1 in the sequel.}

\subsubsection{ADG's kernel and leitmotif: `differentiability' is independent of smoothness}

Thus, confronted with CDG's unmanageability and ineffectiveness in
coping with singularities, and abiding by the aforesaid heuristic
`working principle' of `{\em changing the maths instead of blaming
it on the physics}', we would like to look for an alternative way
of doing differential geometry, a way that is not vitally
dependent---in fact, {\em not at all!}---on differential manifolds
with their `innate differential geometric diseases' in the guise
of the $\smooth$-smooth singularities.

Such a yearning has been already expressed by Einstein in two
ways, both of which have to do with his dissatisfaction with the
classical geometric continuum (manifold) picture of spacetime {\it
vis-\`a-vis} on the one hand the singularities that plague it, and
on the other, the `discrete' (finitistic) and algebraic
(relational) quantum paradigm.

The first way is rather `apologetic' and `confessional':

\bigskip \noindent (Q2.18)\hskip 0.9in
\begin{minipage}{11cm}
\noindent ``{\small ...Adhering to the continuum originates with
me not in a prejudice, but arises out of the fact that I have been
unable to think up anything {\em\small organic}\footnote{Our
emphasis. We will come back to comment further on this remark in
the last section (7.5).} to take its place...}'' \cite{einst2},
\end{minipage}

\vskip 0.1in

\noindent while the second is more suggestive and `wishful' (for
the mathematical structure and theory based on it that would
replace the classical one---CDG---on the geometrical spacetime
continuum):\footnote{Einstein's quotation below can be found in
\cite{stachel}. Again, the last sentence is written in {\em
emphasis script}.}

\bigskip \noindent (Q2.19)\hskip 0.9in
\begin{minipage}{11cm}
\noindent ``{\small ...The problem seems to me how one can
formulate statements about a discontinuum without calling upon a
continuum space-time as an aid; the latter should be banned from
theory as a supplementary construction not justified by the
essence of the problem---a construction which corresponds to
nothing real. {\em But we still lack the mathematical structure
unfortunately}\footnote{Our emphasis.}...}''
\end{minipage}

\vskip 0.1in

\noindent or even more suggestively:\footnote{This is a
one-sentence completion of the quotation (Q2.?) taken from
\cite{einst3}. The last sentence is written in {\em emphasis
script}.}

\bigskip \noindent (Q2.20)\hskip 0.9in
\begin{minipage}{11cm}
\noindent ``{\small ...One can give good reasons why reality
cannot at all be represented by a continuous field. From the
quantum phenomena it appears to follow with certainty that a
finite system of finite energy can be completely described by a
finite set of numbers (quantum numbers). This does not seem to be
in accordance with a continuum theory, and must lead to an attempt
to find a purely algebraic theory for the description of reality.
{\em But nobody knows how to obtain the basis of such a
theory}.}''
\end{minipage}

\vskip 0.1in

\noindent while, even his view above that a continuous field
theory founded on the geometrical spacetime continuum has many
shortcomings {\it contra} the algebraic quantum was essentially
based on the then (and even still today!) prominent lack of having
a (mathematical) theory---``{\em a method}'' in his own
words---that deals and handles effectively singularities (while,
optimally/preferably, still retaining the field-theoretic picture
and its differential geometric apparatus---{\it ie}, the laws in
which these fields participate can still be modelled after
differential equations \cite{stachel1}); as follows:

\bigskip \noindent (Q2.21)\hskip 0.9in
\begin{minipage}{11cm}
\noindent ``{\small ...Is it conceivable that a field
theory\footnote{Of course, Einstein was implicitly alluding to his
unitary field theory, which, according to his vision, could
hopefully `explain away' quantum phenomena. (Again, the reader
should refer to the last section for more discussion on this
point.)} permits one to understand the atomistic and quantum
structure of reality? Almost everybody will answer this question
with `no'. But I believe that at the present time nobody knows
anything reliable about it. {\em This is so because we cannot
judge in what manner and how strongly the exclusion of
singularities reduces the manifold of solutions. We do not possess
any method at all to derive systematically solutions that are free
of singularities}\footnote{Our emphasis.}...}'' \cite{einst3}
\end{minipage}

\vskip 0.1in

On the other hand, there is a recently developed theory, ADG
\cite{mall1,mall2,mall4}, that time and again has proven to be
suitable for evading completely the geometrical $\smooth$-smooth
differential manifold and its singularities
\cite{malros1,malros2,mall3,malros3,mall9,mall7,mall11}; moreover,
by its very algebraic (in fact, sheaf-theoretic and categorical)
character, it appears to be able to implement differential
geometrically quantum spacetime and gravity ideas in a directly
non-manifold based---{\it ie}, in a {\em manifestly background
spacetime manifold independent}---way\footnote{In this sense we
mean that one of the core ideas and central results of ADG, its
{\it leitmotif} so to speak, is that ``{\em differentiability is
independent of smoothness}'' (see concluding sentence-slogan in
\cite{malrap2}).}
\cite{mall5,malrap1,malrap2,malrap3,mall6,mall8}, while still
being able to do field theory by differential geometric means,
albeit, glaringly without $M$.\footnote{See concluding section.}
Indeed, {\em ADG offers us an entirely algebraic way of doing
differential geometry without at all the use of any Calculus},

\bigskip \noindent (R2.10)\hskip 0.9in
\begin{minipage}{11cm}
\noindent a way which, in a Leibnizian-Machian sense, deals
directly with the geometrical objects representing the physical
fields themselves and derives from their algebraic
interrelations---their `dynamical propagations and interactions'
so to speak---without its essential differential geometric
mechanism being dependent at all on or influenced by the
intervention or mediation of (smooth) coordinates, and {\it in
extenso} by (\ref{eq1}), on the base spacetime manifold $M$
\cite{mall7}.
\end{minipage}

\vskip 0.1in

\noindent And we call this relational-algebraic (and finitistic!
\cite{malrap1,malrap2,malrap3}) way of doing differential geometry
independently of a background space(time) `{\em Leibnizian}',
because Leibniz, in contradistinction to Newton, was searching for
{\em a combinatory-algebraic way of doing Calculus} (an `{\it ars
combinatoria cum calculus ratiocinator}'---a logico-combinatorial
`Geometric Calculus') {\em dealing directly with the geometrical
elements or objects of `space(time)'} without the presence, let
alone the intervention, of an ambient space(time) as such,
especially in the Cartesian guise of coordinates
\cite{leibniz1}.\footnote{We wish to highlight here the epithet
`{\em Geometric}' in front of `Calculus' in order to emphasize its
striking contrast to the conventional (Cartesian-Newtonian) `{\em
Analytic}' Calculus, thus anticipating our philosophical remarks
in 7.3 about the Euclidean {\it versus} the Cartesian conception
of (differential) geometry.} To further support this point, we
quote the Bourbakis \cite{bourb} maintaining that Leibniz wished
to

\bigskip \noindent (Q2.22)\hskip 0.9in
\begin{minipage}{11cm}
\noindent ``{\small Fonder un `calcul g\'eom\'etrique' {\em
op\'erant directement sur les \'el\'ements g\'eom\'etriques, sans
l'interm\'ediaire des coordonn\'ees}.}''\footnote{``Found a
`geometric calculus' {\em which operates directly on the
geometrical elements, without the mediation of coordinates}.'' Our
emphasis.}
\end{minipage}

\vskip 0.1in

In connection with the Leibnizian, relational ({\it ie},
algebraic) character of ADG, we read from the prologue to the
Russian edition of \cite{mall1} for instance:

\bigskip \noindent (Q2.23)\hskip 0.9in
\begin{minipage}{11cm}
\noindent ``{\small ...This special {\rm [unexpected]} help {\rm
[from ADG]} now, when the necessity has grown to study manifolds
with singularities and even to remove the underlying space (for
example, spacetime) {\em and proceed to a direct description of
the structures on this manifold}, may be important for many
branches of contemporary mathematical and theoretical
physics...}''\footnote{Our emphasis.}
\end{minipage}

\vskip 0.1in

In the same Leibnizian gist, such an essential background
spacetime manifold independence has been anticipated by Penrose,
albeit not in a differential geometric context proper like ours,
but in the infant developmental steps of his celebrated
combinatorial (relational-finitistic) approach to quantum
space(time and gravity) originally coined {\em spin-networks}
\cite{pen},\footnote{Spin-networks and their {\em spin-foam}
descendants are nowadays considered to be sound and promising
`discrete' approaches to non-perturbative (Lorentzian) quantum
gravity \cite{rovelli2,borissov,baez1,barrett,perez,rovelli}.}
much as follows:\footnote{Again, the part in the quotation below
written in emphasis script highlights the point we wish to make.}

\bigskip \noindent (Q2.24)\hskip 0.9in
\begin{minipage}{11cm}
\noindent ``{\small A reformulation is suggested in which
quantities normally requiring continuous coordinates for their
description are eliminated from primary consideration. In
particular, {\em since space and time have therefore to be
eliminated}, what might be called a form of Mach's principle must
be invoked: {\em a relationship of an object to some background
space should not be considered---only the relationships of objects
to each other can have significance}.\footnote{The reader should
refer to 7.5.8 for our ADG-theoretic generalization of Mach's
principle in the context of GR.}}'' \cite{pen-1}
\end{minipage}

\vskip 0.1in

Thus, in the same vain, having already applied ADG-theoretic ideas
to a finitistic, causal and quantal model for spacetime structure
and gravity \cite{rapzap1,malrap1,rapzap2,malrap2,malrap3}, we
focus here on ADG's `resolution' of $\smooth$-smooth
singularities.\footnote{We will see in the sequel that it is not
so accurate to say that ADG `{\em evades}' or `{\em resolves}'
singularities, as that it `{\em integrates}' or `{\em engulfs}',
or even `{\em absorbs}', them in the algebra (sheaf) $\struc$ of
`generalized coefficients' (coordinates)
\cite{mall3,rap5,mall7,mall11}. In fact, as we shall see later,
none of the aforesaid about singularities, the general
CDG-anomalies, their clash with the PGC of GR {\it etc}, could
stand on its own two feet had not we had the full fledged ADG
theory at our disposal. Thus, based on ADG, it is perhaps more
precise, considering what we actually do to the
$\smooth$-singularities of the classical theory (CDG), to say
`{\em dissolution}' (in $\struc$) rather than `{\em resolution}'
of singularities. In any case, `{\em singularity-resolution}' is a
term already preempted by methods of Algebraic Geometry ({\it eg},
singularity blow-up procedures) \cite{harts}. We do not wish to
confuse the established methods and practices of Algebraic
Geometry with the new, both technically and conceptually different
ones of ADG.}

In particular, the present paper may be regarded as a slightly
more technical and concrete physical applications oriented version
of \cite{mall7,mall11} while, in turn, it may also be thought of
as an extension of the trilogy \cite{malrap1,malrap2,malrap3} to a
tetralogy so as to include ADG's promising prospects concerning
$\smooth$-smooth spacetime singularities
\cite{malros1,mall3,malros2,malros3,mall11}. Thus, below we will
investigate the most general possible scenario for smooth
singularities and we wish to make plain precisely in what way ADG,
especially after its successful application in discretizing the
curved differential spacetime manifold of GR and in formulating
the vacuum Einstein equations, entirely algebraico-categorically,
on the finitary spacetime sheaves (finsheaves) of quantum causal
sets (qausets) in the aforesaid trilogy, manages to cope (in fact,
absorb, thus enable us to do explicit differential geometric
constructions and calculations) with them. This quest emanates
from the intriguing fact, repeatedly highlighted in our past
papers, that

\bigskip \noindent (R2.11)\hskip 0.9in
\begin{minipage}{11cm}
\noindent by ADG-theoretic means it has been shown that {\em the
laws of physics} ({\it eg}, vacuum Einstein gravity and free
Yang-Mills or gauge theories) {\em hold on spaces either hosting
the most general and unmanageable by $\smooth$-means
singularities} \cite{malros1,malros2,mall3,malros3,mall9}, {\em or
spaces looking `discrete' and structurally as remote as a space
can be from the featureless spacetime continuum of macroscopic
physics} and, moreover, these laws are expressed completely by
algebraic, categorical in essence, means \cite{malrap3} (though
still manifestly remaining {\em differential equations} proper!),
without depending on a background $\smooth$-smooth spacetime
manifold.
\end{minipage}

\vskip 0.1in

\noindent At this point it must be stressed that although the
ADG-based form of the (differential) equations (modelling the laws
of physics) remains effectively the same as in the classical
theory, their {\em physical interpretation} changes significantly.
In fact, while ADG uses only `conventionally' or `formally' the
terms and concepts ({\it eg}, connection, curvature {\it etc}) of
CDG, its essential abolition of $\smooth$-manifolds amounts to a
radically different {\em use} and {\em meaning} of those very same
terms. In a pragmatist sense, in ADG these differential geometric
concepts and techniques are {\em used} in a drastically different
way than in the $M$-dependent CDG, hence their {\em meaning}
(physical interpretation) is also significantly different.

{\it In toto}, it is our contention that ADG provides us with a
well geared and fine tuned, both conceptually and technically,
candidate for the ``{\em purely algebraic}'' \cite{einst3}, ``{\em
organic}'' \cite{einst2} theoretical framework that Einstein was
searching for motivated by the incurable headaches that
singularities and the quantum brought him when viewed (as he
viewed them indeed!) from the classical perspective of the smooth
spacetime continuum \cite{malrap3}.\footnote{The reader should go
to 8.? for a thorough discussion of our claims that, from a
differential geometric point of view (Q?.?), ADG could prove to be
the `right' (mathematical) framework in which to implement and
perhaps complete Einstein's unfulfilled unitary field theory
project.}

Below, after we provide a short summary of section 1,\footnote{In
this paper, each section closes with a summary of the section's
basic contents.} we give a brief outline of the contents of the
paper.

\subsection{Section's R\'esum\'e}

The second paragraph of a recent paper, \cite{vaas}, happens to be
tailor-cut for summarizing, by means of juxtaposition, this first,
introductory section, as well as for preparing the reader for
various important issues in both classical and quantum gravity
that we raise, tackle and discuss from an ADG-theoretic
perspective in the sections to follow:

\bigskip \noindent (Q2.25)\hskip 0.9in
\begin{minipage}{11cm}
\noindent ``{\small ...`{\em On the Planck scale there is a
precise, rich, and discrete structure}', says Ashtekar...The
Planck scale is the smallest possible length scale with units of
the order of $10^{-23}$ centimeters...At this scale, Einstein's
theory of general relativity fails. Its subject is the connection
between space, time, matter and energy. But on the Planck scale it
gives unreasonable values--absurd infinities and singularities. It
carries therefore--as the American physicist John Wheeler, who
knew Einstein personally, used to say--the seeds of its own
destruction.\footnote{Together with the Bergmann quote (Q2.2) from
\cite{berg0} about `singularities as auto-destruction {\it loci}
of GR', let us also mention here the Wheeler references
\cite{mtw,wheeler1}.} That means the theory indicates the
limitations of its own applicability. This is a restriction, but
at the same time also an advantage: physicists cannot avoid
looking for a better and more complete theory for the laws of
nature at this fundamental level. In other words: they need a
theory of quantum gravity in order to explain the behavior of
nature at all levels, from quarks to quasars...}''\footnote{Our
emphasis.}
\end{minipage}

\vskip 0.1in

\noindent Let is itemize the points raised in the quotation above
in order of importance for us as we will encounter them
subsequently in the present work:

\begin{enumerate}

\item First, it is maintained that {\em GR} fails---{\it ie}, it reaches the limit of its own
applicability, while its dynamical law (Einstein equations), which
traditionally is interpreted as linking spacetime geometry with
matter-energy-momentum and is represented as a {\em differential}
equation by CDG-means, reaches the threshold of its own
validity---{\it vis-\`a-vis} singularities and other unphysical
infinities. In what follows in the present paper-book we shall
elaborate in detail and repeatedly on the following subtle
distinction: it is not exactly that GR, as a {\em physical}
theory, breaks down at singularities and it is plagued by
infinities, but simply that {\em our} background differential
manifold model for spacetime, and the CDG-technology that comes
hand in hand with it, misbehave and miscarry in the quantum deep.
In other words, one should blame it on the
mathematics,\footnote{Philologically speaking, the seeds of GR's
destruction were planted in $M$, and `flowered' with the CDG and
GR, which are based on it.} not on the physics, so that one should
search for a `better' Calculus (or Analysis)---one that is not
vitally dependent on a smooth base manifold $M$ with its
`inherent' singularities and infinities, while also one that is
essentially algebraic in character in order to accord with the
quantum paradigm.

\item Second comes the supposition that there is a fundamental
space-time scale in Nature---the so-called Planck scale---below
which the spacetime continuum (:manifold) of the classical theory
(GR) gives way to something `discrete' (finitistic) and `quantal',
with the concomitant loss of one's differential geometric
privileges in the quantum deep. As we shall argue extensively in
the sequel in the light of ADG-gravity, ADG is quite indifferent
to the character of the background spacetime structure---whether
it is a classical continuum or a quantal discretum. ADG-gravity is
concerned exclusively and solely with the gravitational field
({\it viz.} connection) and the dynamics (Einstein equations,
still differential geometrically modelled after differential
equations) that it obeys (or better, {\em defines}), without
caring about the (mathematical) nature of the background, external
(to the gravitational field itself) spacetime. ADG-gravity is
fundamentally (:`by definition') spacetimeless, thus it is begging
the question in the theory on the one hand to posit (and impose!)
a fundamental spacetime length in Nature---a cut-off scale put in
by theoretical fiat in order to regularize and control the
infinite integrals arising in attempts to quantize gravity, and on
the other, to expect that a conceptually cogent and
calculationally finite QG will fundamentally hinge on (or even
result in) a quantization of spacetime itself.

\item And third, it follows from the above, GR is not a complete
theory exactly because its dynamical law (Einstein equations) that
define it as a {\em physical} theory proper stumbles and falters
on the singularities due to the underlying geometrical spacetime
manifold and, as a result, exactly because it is not
universal---{\it ie}, because of the presence of the minimum
spacetime length of Planck which is introduced and utilized
precisely in order to make sense of the Analytical nonsense ({\it
eg}, infinite expressions for physical quantities) that we obtain
when we formally apply the rules of QFT to the quantization of
gravity by explicitly (and forcedly!) retaining a background
continuum. The operative word here is `Analytical nonsense', and
the crux of the argument is that the `fault' lies with our
mathematical model and technical tools (manifold and CDG,
respectively) that we employ to describe the gravitational field
quantally, not with the gravitational field---regarded as a {\em
physical} entity---itself. All in all, our maths (CDG, Calculus or
Analysis) needs rectifying, completing and, ultimately,
fundamental revising, not the physical fields and the laws that
they define.

\end{enumerate}

\paragraph{Paper-book overview.} {\it Ex altis} viewed, and very briefly, the present paper unfolds as follows: in
the next section we revisit $\smooth$-smooth singularities, making
precise and plain the sense in which they cannot be thought of
independently of our differential manifold $M$ model for spacetime
and of the CDG that goes hand in hand with it. Since singularities
are `innate' in $M$, the basic observation is that it is quite
understandable that the manifold based CDG (Analysis) cannot cope
with them, thus we infer that it is high-time we found another
method of doing differential geometry---one that is manifestly
background smooth manifold independent, hence prepare the ground
for ADG. In section 4 we present GR under the prism of ADG by
basing ourselves on
\cite{mall1,mall2,mall3,malrap1,malrap2,malrap3,mall4}. We
highlight the sense in which the dynamical laws (Einstein
equations) for the gravitational field, which in our scheme is
represented by an {\em algebraic connection} and not by a smooth
metric as in the original formulation of GR by Einstein, are
manifestly smooth background spacetime $M$ independent, thus
further supporting our generalized version of the PGC in (R2.?).
In section 5 we explain the way in which ADG is purported to
`absorb' or `engulf', thus `dissolve', singularities in the
structure algebra sheaf $\struc$ of generalized coordinates or
`arithmetics', while leaving intact the essentially algebraic
differential geometrical mechanism according to which the physical
laws are expressed as differential equations proper. We thus infer
that smooth singularities are mathematical artifacts (`physically
virtual', `coordinate' singularities in the usual sense) and not
real physical `objects', let alone `incurable differential
geometric diseases' ({\it eg}, insuperable obstacles to
differentiability) in any sense, for {\em differentiability is
independent of smoothness}.\footnote{To be sure, singularities are
indeed incorrigible differential geometric anomalies, but only
when viewed from the classical perspective and tackled by the
$\smooth$-technical means of the usual Calculus on Manifolds (CDG)
which, ultimately, is due to our particular choice
$\struc\equiv\smooth_{M}$ for generalized arithmetics.} Even more
powerfully, we contend that from an ADG-theoretic perspective, the
so-called `real' singularities are as regular as the regular
points of the smooth spacetime manifold, with the latter locally
Euclidean realm being totally absent from our theory. In the
subsequent section, we present a concrete toy physical model in
which the (abstract) ideas of section 5 are put to work and come
to fruition, namely, the ADG-theoretic `absorption' or
`dissolution' of both the exterior ($r=2m$), but more importantly,
of the inner ($r=0$), Schwarzschild singularities, and in two
different ways. At the same time, and to illustrate the power and
effectiveness of ADG, we recall situations where ADG applies to GR
over (Euclidean and locally Euclidean) space(time)s much more
pathological and anomalous than Schwarzschild's---ones containing
an uncountable infinity of the {\em most non-linear and robust
singularities possible}---Rosinger's differential algebras of
generalized functions (non-linear distributions), the so-called
{\em spacetime foam dense singularities}
\cite{malros1,malros2,mall3,malros3}. Even more remarkably, these
uncountable singularities are seen to lie densely in the spacetime
manifold $M$'s `bulk', not just `sitting marginally', like virtual
mathematical points as it were, at its boundary, as the usual
smooth singularities are supposed to (in fact, {\em defined as}!)
\cite{clarke1,clarke3,clarke4}. In the penultimate section, mainly
based on our trilogy \cite{malrap1,malrap2,malrap3}, but also on
the numerous applications of ADG so far to field quantization
\cite{mall5,mall6,mall7,mall8}, we give various theoretical
reasons why ADG may be of significant import not only to classical
(GR), but also to {\em quantum gravity} (QG) proper. In
particular, the essentially base spacetimeless---whether this
background realm is assumed to be a continuum or a
discretum---formalism of ADG-gravity prompts us to question the
physical viability and significance of a fundamental spacetime
length in Nature, like the Planck length, which is usually evoked
in various essentially continuum based quantization schemes
(covariant or canonical) for gravity in order to `regularize' and
render finite various (Calculus based) expressions of physically
important quantities. We also question whether a genuinely quantum
theoresis of the gravitational field---like, we contend,
ADG-gravity is---should be concerned with a quantization of
spacetime itself, something that in Loop QG for example has very
recently been used to `resolve' the interior Schwarzschild
singularity \cite{modesto} and the (initial) cosmological ones
\cite{boj1,husain}. The paper, based on the general didactics of
ADG and their contrast against the glaring shortcomings of CDG in
dealing with gravitational singularities already at the classical
level, concludes with a general philosophical discussion of some
of the issues raised and treated in sections 2--7, being
particularly interested in discussing the extent to which
differential geometric ideas can fruitfully be applied to QG
research. The key point in this discussion is the subtle
distinction between the {\em Euclidean} (relational) way of doing
(differential) geometry---which involves directly and solely the
geometrical objects (in our context, the physical fields) {\it per
se} ({\it ie}, `in-themselves') without reference to an external,
ambient space(time), and the usual {\em Cartesian} (analytic) way
involving coordinates and, inevitably, the mediation of a
background geometrical space(time) (manifold). This distinction,
we contend guided by the central didagma of ADG,\footnote{Which,
we recall again, is that {\em one can do differential geometry
directly with the geometrical objects (fields) themselves without
the intervention of coordinates or, {\it in extenso}, of
space}---especially of the spacetime manifold employed by the CDG
to support the smooth gravitational field, which is identified
with the metric, in GR (R2.??, Q2.??).} lies at the heart of the
contrast between Leibniz's algebraic (relational) conception of
Calculus (Q2.?) and the more popular analytic-geometrical
(background space---intervening coordinates dependent) Newtonian
way \cite{mall7}. Based on this subtle distinction, we draw a
boundary between `{\em physical geometry}' and `{\em geometric
physics}' further inspired by some telling remarks of Peter
Bergmann in \cite{berg0}. We also suggest, again based on ADG, a
way of doing `continuous' field theory entirely by
categorico-algebraic and finitistic means manifestly independent
of a background spacetime continuum and its $\smooth$-smooth
singularities, thus potentially marrying the two seemingly
incompatible facets of Einstein \cite{einst1}. As a result, we
continue some points made in \cite{malrap3} and argue that ADG may
be a promising candidate for the ``{\em organic}'' \cite{einst2},
non-continuum based, ``{\em purely algebraic theory for the
description of reality}'' \cite{einst3} that ``{\em the other
Einstein}'' \cite{stachel1} was looking for in view of the
pathologies (singularities) of the geometric spacetime manifold of
GR and the `discontinuous' or `discrete' algebraic actions of
quanta.

\section{General Relativity: $\smooth$-Gravitational Singularities Revisited}

Our principal aim in this section is to lay bare and `beyond any
doubt' so to speak {\em the inextricable dependence of
singularities on the differential spacetime
manifold}---essentially, how one cannot think of the former apart
from the latter and in what way smooth singularities represent
`innate' shortcomings of CDG---anomalies that are built into
$\smooth(M)$---which therefore are uncircumventable by its
$\smooth$-manifold based concepts and analytic techniques. For
arguably, the anomalous and problematic character of singularities
in the manifold and CDG-based GR, and quite apart from the
physical meaninglessness of the infinities that are usually
associated with them, is reflected on the fact that so far there
has not been given any precise, `unambiguous' and unanimously
accepted definition of gravitational singularities, not least
because of the aforesaid clash between the PGC of GR (which is
mathematically implemented via the diffeomorphism group
$\mathrm{Diff}(M)$ of the smooth spacetime manifold $M$) and the
very existence of $\smooth$-singularities in GR (Q2.?)
\cite{geroch,clarke4,hawk0}. This will firmly support our
preliminary remarks in (R2.?) and prepare the reader for our
ADG-theoretic perspective on singularities in the sequel. To this
end, we first wish to review briefly and scrutinize to a certain
extent all kinds of (attempts at definitions of) singularities in
GR that have been proposed so far by different people in various
places like for example
\cite{clarke1,clarke2,clarke3,clarke4,geroch,geroch1,geroch2,hawk0,hawk1,hawk2,schmidt}.\footnote{This
we by no means claim to be a complete list of references to the
various works on gravitational singularities, but it is sufficient
for capturing pretty much all the basic `definitions' of smooth
singularities in GR that have been attempted hitherto. Especially
\cite{hawk0} and the more recent book \cite{clarke4} (and
references therein), provide one with a rather complete picture of
classical, $\smooth$-smooth spacetime singularities and their
study by Analytic (CDG) means.}

\subsection{All the Perceptions and `Definitions' (so Far) of $\smooth$-Smooth Singularities}

From the very early attempts at defining precisely what is a
singularity in GR, it has become clear that two central notions in
those definitions are:\footnote{Remarkably, Einstein, as early as
1918 and in the context of de Sitter's solution to GR's field
equations \cite{einst0}, had more or less figured out---even if he
did not express it in the technical terms we do today---the two
`properties' below that a {\it locus} in the spacetime continuum
must possess in order to qualify as a singularity proper.}

\begin{itemize}

\item the notion of {\em (analytic) inextensibility of the
spacetime manifold} $M$ past the offensive {\it
loci}---differential spacetime manifold extensions employed, as it
were, to `remove' singularities,\footnote{The importance of
(analytic) inextensibility as such was implicitly noted first by
Finkelstein (and then explicitly by Kruskal \cite{kruskal}) upon
encountering the interior Schwarzschild singularity in \cite{df}.}

\noindent and

\item the notion of {\em (causal) geodesic incompleteness}---the
inability so to speak of particles to `reach' singularities by
following smooth (causal) paths (geodesics) under the influence of
the gravitational field, something which in turn defines,
topologically speaking, the anomalous {\it loci} as points at the
boundary or `edge' $\partial M$ of the `physical, regular
spacetime bulk interior' $M$ (Q?.?), sites not properly belonging
to the latter.\footnote{The importance of (causal) geodesic
incompleteness was explicitly noted first in \cite{geroch}.
Subsequently, null geodesic incompleteness was the central
prediction of the celebrated singularity theorems of Hawking and
Penrose in \cite{hawk1}. However, as Clarke points out in
\cite{clarke4}, one need not consider only `free falling'
observers following causal {\em geodesics}, since other physically
admissible frames---ones with bounded acceleration for
example---may be able to reach the point-{\it loci} in question in
finite proper time, even though geodesic observers cannot. In
order to include the world-lines of such in principle arbitrarily
accelerated observers, curves more general than geodesics---ones
parameterized not by proper time, but by an arbitrary so-called
{\it (general) affine parameter} (see below)---must also be
included in the definition of incompleteness.}

\end{itemize}

\noindent In fact, an explicit combination of these two notions in
a concise, albeit informal and tentative definition of
singularities by Clarke in \cite{clarke3}, essentially reads as
follows:

\bigskip \noindent (R3.1)\hskip 0.9in
\begin{minipage}{11cm}
\noindent a singularity may be thought of as an ideal, boundary or
`end-point' of $M$---an unphysical {\it locus} not belonging to
the physical and otherwise regular spacetime bulk (interior of)
$M$---assigned to an incomplete inextensible curve.
\end{minipage}

\vskip 0.1in

\noindent while Hawking, based on these two notions, gives a
concise definition of a singular spacetime (manifold) in
\cite{hawk2}, as follows:

\bigskip \noindent (Q3.1)\hskip 0.9in
\begin{minipage}{11cm}
\noindent ``{\small\em A spacetime {\rm [manifold]} is singular if
it is timelike or null geodesically incomplete but cannot be
embedded in a larger spacetime {\rm [manifold]}.}''
\end{minipage}

\vskip 0.1in

\noindent Leaving the notion of ideal, boundary points of $M$ for
the more `rigorous' definition of singularities below,\footnote{To
the knowledge of these authors, such ideal `end-points' to causal
curves were first explicitly defined in \cite{geroch2}.} we first
make more precise the two notions of inextensibility and
incompleteness above.

\subsubsection{Analytic (or smooth) inextensibility of the (analytic or smooth) spacetime manifold $M$}

We initially discuss continuous extensibility of (curves in) $M$,
then we strengthen the notion of extensibility of $M$ so as to
include smooth and analytic extensions.\footnote{Recalling of
course that differentiability (smoothness) is stronger than
continuity, and analyticity still a bit stronger than smoothness.
In a familiar Calculus-theoretic or (Real) Analytic sense, and for
the ($\R$-valued) functions that are defined on a manifold $M$,
these two `strength relationships' are commonly expressed as
follows: `differentiability (of the functions involved) {\em
implies} continuity (of those functions)' (write:
`smoothness$>$continuity'), and `power (Taylor) series
expansibility (of the functions involved) {\em implies}
differentiability to any order (of those functions)' (write:
`analyticity$>$smoothness'). For the manifold $M$ itself, this
`order of strength' reads: `a topological manifold is weaker than
a smooth one, and a smooth one still weaker than a (real) analytic
one' (symbolically, write: ${}^{(\R)}\cont <{}^{(\R)}\smooth
<{}^{(\R)}\anal$). Plainly then, an analytic manifold is
automatically smooth, hence continuous as well. These last remarks
hint at something that will prove to be of great import in the
sequel, namely that, {\em structure-wise, whether a manifold $M$
is regarded as being topological, smooth or (real) analytic one
essentially depends on (in fact, it is defined by!) whether the
`structure coordinate' functions on it are (defined or assumed to
be) continuous, smooth, or (real) analytic ones, respectively}.
Ultimately, in a manifold-type of jargon: beginning with $M$ as a
structureless point-set, the epithets `{\em topological}', `{\em
smooth}' (or, anyway, of finite order $k$ of differentiability) or
`(real) analytic' that may be given to it depend on whether we
employ $\cont$-, $\smooth$- (or, anyway, $C^{k}$-), or
$\anal$-charts to cover or `coordinatize' it(s points). In turn,
these charts {\em define} structurally $M$ as a {\em topological},
{\em smooth} or {\em analytic} manifold, respectively. ({\bf
Note:} from now on, by an {\em analytic manifold} we mean a {\em
real analytic} one, which is usually denoted by ${}^{\R}\anal$, or
simply, omitting the pre-superscript `$\R$', by $\anal$. In fact,
and for all practical intents and purposes, in the sequel we will
not distinguish between a $\smooth$- and a $\anal$-manifold, in
spite of their slight technical difference mentioned above. That
is, in our ADG-theoretic musings below, we do not distinguish
between the terms `Calculus/Differential Geometry on Smooth
Manifolds' and `Analysis'. Indeed, here we regard the terms {\em
Differential Calculus} and {\em Analysis} as being synonymous.)}
In what follows, we make it precise and clear what the term `{\em
spacetime extension}'---when $M$ is taken to be a $\cont$-,
$\smooth$-, or $\anal$-manifold---actually refers to.

By a curve $\alpha$, with {\em (general) affine parameter} $t$, in
the manifold $M$, we mean, in general, a map $\alpha(t)$ from the
`clopen' interval $I_{\tau}=[t=0,t=\tau)\subset\R$ into $M$, that
is to say,

\begin{equation}\label{eq2}
\alpha:~t\in I_{\tau}\hookrightarrow M\ni\alpha(t)
\end{equation}

\noindent $\alpha$ may be regarded as a continuous, smooth or
analytic injection (embedding) of $I_{\tau}$ into $M$ depending on
whether the coordinates of its points in $M$ are $\cont$-,
$\smooth$-, or $\anal$-functions of $t$, which in turn is
conditional on whether $M$ itself is assumed to be a topological
($\cont$), smooth ($\smooth$) or analytic ($\anal$) manifold in
the first place.\footnote{See last footnote. Of course, $\R$ is
the archetypal (1-dimensional) $\anal$-manifold, carrying the
usual Euclidean differential (smooth) and topological structure.}

We first note that, by definition, the curve $\alpha$ has no
end-point (in $M$), since $t=\tau\not\in [0,\tau)$. Then, in the
case of continuous extension, $\alpha$ is said to be
(continuously) extensible if it is possible to {\em extend the map
continuously to an end-point $\alpha(\tau)$ in $M$}. Otherwise,
$\alpha$ is called {\em (continuously) inextensible}.

This definition of (continuous) inextensibility of (a curve in)
$M$ prompts us to make the following remarks which, in a way,
foreshadow our `global' stance against smooth and, {\it in
extenso}, analytic extensibility of a space(time) to be presented
in the sequel:

\bigskip \noindent (R3.2)\hskip 0.9in
\begin{minipage}{11cm}
\noindent since continuity is a question of topology, and since
$M$, regarded as a topological manifold, {\em is} the algebra
$\cont(M)$,\footnote{That is to say, $M$, initially taken to be a
structureless point-set, is covered by $\cont$-charts---{\it ie},
its points are `coordinatized' by $\cont$-functions (see
penultimate footnote).} one could say that the continuous
inextensibility of a curve $\alpha$ to a point $\alpha(\tau)$ in
$M$, means essentially that we have reached the limit of
applicability of the algebra $\cont(M)$ and, ultimately, the
representation of $M$ as a topological manifold. This however does
not mean that by changing the algebra of `continuous' functions on
$M$---thus, in effect, by changing the topology of
spacetime\footnote{For instance, one can change the algebra from
$\cont(M)$ to some other functional algebra which one may assume
up-front as being `continuous', now of course with respect to a
topology on $M$ (a topology that one is free to {\em define} on
$M$!) different from the usual (Euclidean) $\cont$-one that
defines it as a topological manifold.}---we cannot convert a
previously continuously inextensible curve in $M$ regarded as a
topological manifold, to one that is perfectly continuous at its
end-point. Plainly, it is {\em we} that assign a topology on
space(time) and, in the classical continuum case, {\em we} that
assume that spacetime is modelled after a $\cont$-manifold
precisely by choosing $\cont$-functions to label (`coordinatize')
its points.
\end{minipage}

\vskip 0.1in

\noindent The remarks in (R2.2) above apply then {\it mutatis
mutandis} to smooth and analytic extensions. For $\smooth$- and
$\anal$-extensions of $M$, we may convert them to the following
positive statement:

\bigskip \noindent (R3.3)\hskip 0.9in
\begin{minipage}{11cm}
\noindent The smooth or analytic inextensibility of a manifold $M$
past a certain {\it locus} (which is thus deemed to be identified
with a singularity subsequently) marks {\em our} inability to
cover (`coordinatize') the latter by $\smooth$- or $\anal$-charts
({\it ie}, label them with smooth or analytic `coordinates') thus,
concomitantly, apply the usual differential geometric (analytic)
concepts and techniques to it in the same way we do to the other
so-called `regular' (smoothly or analytically `chartable') {\it
loci} of $M$.
\end{minipage}

\subsubsection{$\smooth$-smooth singularities: theoretical `end-points'
of both Calculus and of the Calculus based GR}

\noindent In view of (R2.2) and (R2.3), the following `heuristic'
remark appears to be suitable here:

\bigskip \noindent (R3.4)\hskip 0.9in
\begin{minipage}{11cm}
\noindent {\it Vis-\`a-vis} smooth or analytic extensions, {\em
singularities represent `{\em mathematical limits}', `{\it
impasses}' or `{\em end-points}' to the applicability of CDG
(Analysis)},\footnote{Or, since GR depends essentially on CDG for
its mathematical formulation, which Calculus is in turn supported
by the smooth $M$, $\smooth$-singularities (with their associated
unphysical infinities) represent `{\em physical limit points}' to
the applicability of the classical spacetime continuum based
relativistic field theory of gravity at and beyond which the {\em
physical theory} itself (GR)---precisely because of the inadequacy
of the mathematical model for spacetime ($M$) and of the
differential geometric framework on which it rests (CDG)---is well
out of its depth (R2.?, Q2.??)---see also (R3.?) below. Plainly,
there apparently is an inevitable conflation (or even `confusion')
of the notions `{\em physical theory}' (GR) and `{\em mathematical
theory}' or `{\em mathematical framework}', since in GR we tend to
identify {\it a priori} {\em physical spacetime} with the
mathematical artifact {\em smooth manifold}, and, concomitantly,
identify the physical (dynamical) structures involved in the
relativistic field theory of gravity with those `afforded' by that
$\smooth$-background---the smooth fields on $M$ (in the case of
GR, $g_{\mu\nu}$). (The reader is advised to refer to section
7---and more in particular, to 7.4 and 7.5.5---for an
ADG-theoretic critique of this `{\em physical spacetime}$\equiv
M$' identification and {\it a priori} assumption in GR.)} or
simply, {\em they are {\it loci} where Calculus `breaks down' or
at best it is `ineffective'}, so to speak.
\end{minipage}

\vskip 0.1in

\noindent A preliminary physical corollary of (R2.3) and (R2.4) is
that

\bigskip \noindent (R3.5)\hskip 0.9in
\begin{minipage}{11cm}
\noindent we tend to identify `{\em physical}' spacetime events
with {\em regular points} (in $M$) on which the Einstein equations
(and their solutions) hold. Another way to say this, `physical'
spacetime events are precisely those points in the bulk (interior)
of $M$---those to which CDG concepts and techniques apply readily
({\it ie}, the differential equations of Einstein hold), without a
problem.\footnote{We thus put the word `physical' in single
quotation marks and write it in {\em emphasis script} just in
order to preliminarily catch the reader's attention about the
point made here, namely, that there is nothing really physically
real about the point-events of the smooth manifold (pun intended),
for as we will argue shortly, the bulk of $M$ (consisting of
regular points) is identified as being physical simply because we
can actually do CDG on it, and CDG on manifolds ({\it ie},
Calculus) is actually the only way we know how to do differential
geometry!---see (R2.6) next.} Formally, we write:
\end{minipage}

\begin{equation}\label{eq3}
\mathit{`physical'}~\mathrm{spacetime~events}\footnote{Under the
proviso that, physically speaking, the points of $M$ should not be
{\it a priori} interpreted as spacetime events proper---`{\it a
priori}' meaning here {\em before the gravitational field obeying
(or in ADG, defining) the gravitational dynamics (Einstein
equations) is specified} (again, see 7.5.5). Thus, after this
clarification and warning, from now on whenever we refer to the
points of the manifold $M$ as spacetime events without direct
allusion to the field equations, we will put the designation `{\it
sic!}' to remind the reader of this subtle interpretational
point.}\equiv\mathrm{regular~points~(in~the~interior)~of~}M
\end{equation}

The reader should note here the `negative' or `by exclusion' sense
in which singularities are defined. They are `negative' or
`$\smooth$-manifold excluding' and their definition is expressed,
in a way, by `negation' or `exclusion' as follows: it is as if
`real' singularities, in contradistinction to `virtual' or
`coordinate' ones (and, of course, `regular' spacetime points),
are precisely the ones to which we cannot further apply
CDG.\footnote{See 2.1.3 below.} In fact, this eliminative or
negative `definition-by-exclusion' of singularities is the very
essence of current CDG-based attitudes and approaches to them,
since it appears to be both technically and conceptually very
hard, if not impossible, to give a direct, concise and `positive'
definition of them within the confines of Calculus (Analysis). For
instance, below we quote the very first paragraph of the preface
to \cite{clarke4}:

\bigskip \noindent (Q3.2)\hskip 0.9in
\begin{minipage}{11cm}
\noindent ``{\small The central aim of this book is the
development of results and techniques needed to determine when it
is possible to extend a space-time through an `apparent
singularity' (meaning, a boundary-point associated with some sort
of incompleteness in the space-time). Having achieved this, we
shall obtain a characterisation of a `genuine singularity' as a
place where such an extension is not possible. {\small\em Thus we
are proceeding by elimination: rather than embarking on a direct
study of genuine singularities, we study extensions in order to
rule out all apparent singularities that are not genuine.} It will
turn out, roughly speaking, that the genuine singularities which
then remain are associated either with some sort of topological
obstruction of an extension, or with the unboundedness of the
Riemann tensor when its size is measured in a suitable norm.}''

\end{minipage}

\vskip 0.1in

Implicit here is the following `{\em $\smooth$-manifold and
CDG-conservative attitude}' (which, in fact, permeates throughout
the whole of the spacetime continuum based field physics, whether
classical or quantum!):

\bigskip \noindent (R3.6)\hskip 0.9in
\begin{minipage}{11cm}
\noindent {\bf $\mathbf{\smooth}$-conservatism and monopoly}.
Since Calculus on manifolds is the only way we know how to do
differential geometry and it has undoubtedly served us well in the
past for modelling mathematically relativistic gravitational
physics (GR) for example,\footnote{Not to mention the manifold
successful applications of the Minkowski spacetime
continuum---itself carrying the technical and conceptual panoply
of CDG---to the flat relativistic quantum theories of matter
fields (QFT); while, in the same context, one should not to forget
of course the numerous applications of CDG, in the guise of smooth
fiber bundle theory, to (quantum) Yang-Mills (gauge) theories. Of
course, these theories too are plagued by infinities coming from
the assumption of a base spacetime continuum $M$, infinities which
are not that different from the singularities of GR, although
admittedly the quantization procedure alleviates a bit their
robustness and strength (see section 6).} $\smooth$-smooth
gravitational singularities, which mark the ultimate
inapplicability or `breakdown' of classical differential geometric
concepts and techniques to GR, are distinguished from the `{\em
physical}' spacetime events ({\it sic}!) in the smooth $M$ ({\it
ie}, regular points or coordinate singularities) by being pushed
to the boundary of the regular $M$ (on which CDG still applies!).
Metaphorically speaking (R2.4), this `end of Calculus' at
singularities signifies that we cannot `calculate' ({\it ie},
apply differential geometric ideas and methods, and obtain
`sensible' numerical results---finite numbers) in the presence of
({\it ie}, in the vicinity of, let alone right at the spacetime
{\it loci} of) singularities. At the same time, the corresponding
CDG-based relativistic gravitational field physics (GR) is
assailed by physically unacceptable, because nonsensical,
infinities for many important $\smooth$-smooth
fields\footnote{That is, fields that are represented by
$\otimes_{\smooth(M)}$-tensors, such as the spacetime metric or
the Riemann curvature tensor.} (Q?.?, Q?.?, Q?.?).
\end{minipage}

\vskip 0.1in

The said `{\em smooth spacetime continuum conservatism}'---albeit,
above Planck scale\footnote{That is, still in the domain of
applicability of classical gravity (GR)---see 7.2 for an
ADG-theoretic critique of the Planck length scale and its apparent
physical importance (Q?.?).}---is nicely wrapped up in the
following quote by Hawking in the preamble to the first chapter of
\cite{hawk2}, just before he discusses singularities in GR and the
global (causal) structure of the classical spacetime continuum:

\bigskip \noindent (Q3.3)\hskip 0.9in
\begin{minipage}{11cm}
\noindent ``{\small ...Although there have been suggestions that
spacetime may have a discrete structure, {\em I see no reason to
abandon the continuum theories that have been so successful}.
General relativity is a beautiful theory that agrees with every
observation that has been made. {\em It may require modifications
on the Planck scale, but I don't think that will affect many of
the predictions that can be obtained from it}\footnote{Our
emphasis throughout.}...}''

\end{minipage}

\vskip 0.1in

\noindent with the principal `negative prediction' of GR being, of
course, its own `autocatastrophe': the existence of
$\smooth$-smooth spacetime singularities. We may thus discern in
(Q?.?) a rather positive view of continuum singularities quite in
contrast with Einstein's remarks in (Q?.?); furthermore, one may
infer a question (and possibly a doubt?) by Hawking whether QG
will ultimately remove singularities.\footnote{In anticipation of
(Q?.?).}

However, going against the grain (or trend!) of
`$\smooth$-conservatism', let us fuse into the following
`philological', albeit concise, statement the mathematical with
the physical spacetime continuum and CDG inapplicability remarks
in (R3.?--R3.?):

\bigskip \noindent (R3.7)\hskip 0.9in
\begin{minipage}{11cm}
\noindent Physically nonsensical infinities come from our
persistent trying to calculate (physically important quantities
modelled by $\smooth$-smooth fields) at---that is, our insistence
on applying Calculus to---singularities, which anyway come from
{\em our} assuming up-front the `structure algebra' $\smooth(M)$
(or its associated structure sheaf $\smooth_{M}$) to coordinatize
spacetime events ({\it sic}!) in the first place.
\end{minipage}

\vskip 0.1in

\subsubsection{(Causal geodesic) incompleteness}

Let us return now to our affinely parametrized curve $\alpha(t)$
in (\ref{eq2}). $\alpha$ is said to be incomplete if it has finite
(general) affine parameter $t$-length, by which we mean that

\begin{equation}\label{eq4}
\ell_{E}(\alpha)=\int_{0}^{\tau}\sqrt{\sum_{i=0}^{3}(x_{i}^{2})}~dt
<\infty ,
\end{equation}

\noindent where the subscript `$E$' denotes a {\it vierbein} that
is parallelly translated along $\alpha$ and with respect to which
the components $x_{i}$ of the tangent vector
$\dot{\alpha}:=\frac{d\alpha}{dt}$ of $\alpha$ are measured.

For {\em causal} (geodesic) curves $\alpha$,\footnote{By which we
mean curves whose tangent vector $\dot{\alpha}$ is timelike or
lightlike (null) everywhere on them.} incompleteness may be also
thought of as expressing that it takes finite (now proper) `length
of time' (duration) for free falling material particles to reach
the end-points of their evolution,\footnote{A tetrad field `$E$'
like the one in (\ref{eq2}) is now supposed to be comoving with
each and every one of such particles, thus defining a geodesic
frame adapted to ({\it ie}, with origin at) the particle.}
end-points which, as we shall see subsequently, are deemed to be
identified with singularities. All in all, and in a physical
sense, incompleteness is supposed to capture the apparently
contradictory quality of singularities as being `{\em finitely
asymptotic (gravito-focal)} {\it loci}'\footnote{`Finite
asymptoticness' pertaining to the fact that it takes finite proper
time (for geodesic frames adapted to free falling particles) or
affine parameter length (for general, arbitrary frames with
bounded acceleration) to reach (or `converge' to) the purported
singular {\it locus} which, however, itself does not belong to the
`physical', regular spacetime $M$ (see 2.1.4 next).} marking, in
the case of free falling matter, the end of the dynamical
histories of material particles propagating and accreting under
the focusing (attractive) action of the gravitational field.

\subsubsection{$\smooth$-Singularities: Mathematical artifact-points at the edge of
the `physical' (regular) spacetime manifold}

Having the notions of inextensibility and incompleteness at our
disposal, and following Clarke \cite{clarke3,clarke4}, we can then
proceed in two steps to define singularities more precisely:

\begin{itemize}

\item i) Such incomplete (causal) curves
(ICs)\footnote{Correspondingly, incomplete and inextensible curves
will be abbreviated by `IICs'.} define boundary points of the
spacetime manifold.

\item ii) Some of these `marginal' spacetime points are identified
with `true' singularities.

\end{itemize}

Starting formally, one may initially assume that the differential
spacetime manifold $M$ is, topologically speaking, {\em open and
bounded} (in the usual $\cont$-topology), with closure
$\overline{M}:=M\cup\partial M$, where $\partial M$ is the set of
boundary points of $M$. {\it En passant}, and in accordance with
(Q?.?) and i) above, we note that ICs define points on $\partial
M$. In particular, {\em boundary points represent equivalence
classes of ICs in} $M$, with the defining equivalence relation
being ``{\em ending at the same point}''
\cite{clarke3}.\footnote{Formally, for incomplete curves $c_{1}$
and $c_{2}$ ($c_{1},c_{2}\in$IC), we write `$c_{1}\ter\, c_{2}$'
for the aforesaid `{\em having the same terminal point}'
equivalence relation.} In other words, with each point $p$ on
$\partial M$ we associate a collection of ICs in $M$, namely, all
those that have $p$ as (common) terminus. Of course, this
definition of boundary points as $\ter$-classes of ICs in $M$ does
not give one an explicit expression for the relation `$\ter$';
consequently, it does not tell one how to actually construct
$\partial M$ as a boundary proper (at least in the topological
sense of the term `boundary').

Without going into any technical detail,\footnote{For a detailed
presentation, analysis and discussion of how to construct the
boundary of $M$, the reader is referred to \cite{clarke4}.} we
mention two approaches for deciding when two ICs terminate at the
same point on $\partial M$, thus effectively defining `$\ter$' and
pointing to a method for actually constructing the boundary of
$M$:

\begin{itemize}

\item $\alpha)$ The first, which is the physically more intuitive
of the two---here to be called the `{\em
c-method}',\footnote{`{\em c}' for `{\em causal}'.} essentially
rests on $\ter$-identifying two {\em future directed causal
curves}
$\overrightarrow{c}_{1},\overrightarrow{c}_{2}$\footnote{In
keeping with our previous notation in the trilogy
\cite{malrap1,malrap2,malrap3}, a right-pointing arrow over a
symbol corresponds to the adjective `{\em causal}' (which, of
course, may vary in meaning from context/model to context/model).}
when every point to the past of a point in, say,
$\overrightarrow{c}_{1}$, is also to the past of a point in
$\overrightarrow{c}_{2}$; write formally:

\begin{equation}\label{eq5}
\overrightarrow{c}_{1}\ter\,\overrightarrow{c}_{2}\Leftrightarrow
``Past(\overrightarrow{c}_{1})=:I^{-}(\overrightarrow{c}_{1})\equiv
I^{-}(\overrightarrow{c}_{2}):=Past(\overrightarrow{c}_{2})"\footnote{The
right hand side expression reading: ``{\em the past of
$\overrightarrow{c}_{1}$ coincides with the past of
$\overrightarrow{c}_{2}$}'' \cite{clarke3}. Similarly for past
directed causal curves (write $\overleftarrow{c}_{1}$ and
$\overleftarrow{c}_{2}$) and the identification of their futures
$I^{+}(\overleftarrow{c}_{1})$ and
$I^{+}(\overleftarrow{c}_{2})$.}
\end{equation}

\noindent Then, this particular definition of `$\ter$' leads to
the well known definition of {\em indecomposable past sets} which
figure prominently in the construction of the {\em conformal
boundary} ($c$-boundary) and associated `{\em null horizon}' in
\cite{geroch2}. The $c$-method of defining singular boundary
points---in particular, one that uses causal {\em
geodesics}---appears to be more `natural' for defining
singularities which form either from the gravitational accretion
of free falling non-interacting `cool matter' (dust) like for
example in the theoretical scenario for stellar collapse in
\cite{hawk1,schoen}, or in a cosmological setting for the
radiative focusing of black body emission \cite{hawk1,hawk0}. In
both cases, a geometrical configuration called `{\em trapped
surface}' forms---trapping as it were the aggregated matter and
confining the gravitationally focused radiation---which in turn
points, {\it vis-\`a-vis} the singularity theorems of Hawking and
Penrose \cite{hawk0,hawk1}, either to the existence of a boundary
point of $M$ or to a violation of causality within the physical
spacetime $M$.\footnote{The reader may wish to recall that the
singularity theorems of Hawking and Penrose were essentially based
on smooth causal (predominantly null) geodesics arguments as well
as on the following four rather generic assumptions about them,
$M$ and their relation to the smooth gravitational field strength
(smooth spacetime curvature): i) `{\em null geodesic focusing}':
the null geodesics from any point of $M$ are eventually focused
(or alternatively, the null geodesics from some closed $2$-surface
are converging), ii) `{\em Ricci curvature positivity}': for every
causal vector $V$, $V^{r}\ric_{rs}V^{s}>0$ holds, iii) `{\em
Riemann curvature non-degeneracy}': every causal geodesic
$\overrightarrow{\gamma}$ has a point where
$\dot{\overrightarrow{\gamma}}_{[r}R_{s]tu[v}\dot{\overrightarrow{\gamma}}_{w]}
\dot{\overrightarrow{\gamma}}^{t}\dot{\overrightarrow{\gamma}}^{u}\not=0$
holds, and finally, iv) `{\em causality protection condition}':
$M$ contains no closed timelike curves. For more technical details
and comments about the physical significance of the general
assumptions i)--iv), the reader is again advised to refer to
\cite{clarke4}.} And since causality is to be `protected' in any
(intuitively) physically reasonable model of relativistic
spacetime structure, (singular) boundary points to $M$ appear to
be inevitable consequences of the theory.\footnote{Roughly in this
sense one says that under quite general assumptions and conditions
GR predicts singularities.}

\item $\beta)$ The second method, which is the analytically
(mathematically) the more robust, elegant and fruitful of the
two---here to be called the `{\em b-method}' as it eventually
leads to the definition of the celebrated `{\em singularity
$b$-boundary}' of Schmidt \cite{schmidt,hawk0,clarke4}---involves
the construction of $\partial M$ not by working directly with ICs
in $M$, but rather with the frame bundle $FM$ over
$M$---essentially, by (parallelly) propagating frames along ICs in
$M$.\footnote{For a brief account of the construction of the
$b$-boundary, the reader is referred to \cite{clarke3}, but for
more details, either the original paper \cite{schmidt}, or again
the book \cite{clarke4}, should suffice.} The $b$-method appears
to be analytically more versatile and of greater calculational
import than the $c$-one, because it endows $\overline{M}$ with a
`natural' topology, inherited from $FM$,\footnote{That is to say,
a topology `projected down' from $\overline{FM}$---the Cauchy
completion of $FM$ with respect to Cauchy convergence (in the
metric topology of a suitably and rather naturally defined metric)
of sequences of frames tending to certain ones at the end-points
of ICs in $M$ \cite{schmidt,clarke3,clarke4}.} relative to which
it is (eventually) meaningful to talk about a neighborhood of a
singular point at $\partial M$, or calculate how the (Riemann)
curvature tensor (in a particular local frame) behaves (grows)
`asymptotically' as the frame (`local observer/particle')
approaches that singular {\it locus}.
\end{itemize}

On the other hand, leaving aside these admittedly ingenuous
technical attempts at defining $\partial M$---all of which are
manifestly based on the (analytic) concepts and tools of CDG,
which is in turn vitally dependent on the smooth spacetime
continuum $M$ in one way or another---we wish to emphasize here
that on the one hand,

\bigskip \noindent (R3.8)\hskip 0.9in
\begin{minipage}{11cm}
\noindent the points of $\partial M$ are {\it ab initio} declared
to be only mathematically adjoined to $M$, but devoid of any
physical significance in the sense that only the points of the
smooth bulk interior $M$ of $\overline{M}$ are regarded as being
{\em physical spacetime events}\footnote{{\it Sic!}} (Q?.?, R2.5)
\cite{clarke3}.
\end{minipage}

\vskip 0.1in

\noindent while on the other, that

\bigskip \noindent (R3.9)\hskip 0.9in
\begin{minipage}{11cm}
\noindent (smooth or analytic) inextensibility of $M$ past points
of $\partial M$ is precisely the criterion or `attribute' that
separates the latter into `regular' (`virtual' or `coordinate'
singularities included in $M$) and `singular' (`actual', `true' or
`real' singularities, excluded from $M$) (Q?.?, Q?.?, R2.5).
\end{minipage}

\vskip 0.1in

\noindent In 2.1.5 next, we make it clear, following Clarke
\cite{clarke3,clarke4}, which points on $\partial M$ qualify as
`actual' or `real', as opposed to `virtual' or `coordinate',
singularities, and we highlight for the reader that, so far, there
have been proposed three kinds of the former all of which depend
vitally on that `quintessential' property that qualifies them as
`true' singularities in the first place: {\em the (analytic or
smooth) inextensibility of $M$ past them}.

\subsubsection{Three types of `real' $\smooth$-smooth singularities: what underlies
them all}

There are three kinds of boundary points that are usually regarded
as being {\em `real' singularities}. As we will see below, `common
denominator' to all three types of singularities is the notion of
(analytic or smooth) inextensibility, while what distinguishes
them from one another is an auxiliary condition that must be
satisfied on top of our inability to extend $M$ analytically
beyond them.\footnote{Indeed, not actually of $M$, but of its
`$\struc$'; when one says `$M$', one invariably (albeit, more
often implicitly) refers to the `$\struc$' one employs in one's
differential geometric constructions and calculations!}

Thus, first returning to ii) of 3.1.4, we recall Clarke's words
from (Q?.?) maintaining that what distinguishes between `regular'
boundary points from `real' singularities is that while there is
an (analytic) extension of $M$ past the former, there is no such
`continuation' so as to include the latter into the `physical',
regular spacetime continuum (R2.8, R2.9). In a nutshell, and in
accord with our preliminary rationale in (R2.2, R2.3) and the
concomitant `heuristic remark' in (R2.4),

\bigskip \noindent (R3.10)\hskip 0.9in
\begin{minipage}{11cm}
\noindent `true' (and physically interesting or important)
singularities, as opposed to `virtual' or `coordinate' ones ({\it
ie}, regular boundary points of $M$), cannot be (smoothly or
analytically) coordinatized so as to be regarded as being
themselves `regular' differential geomerically or analytically
speaking; consequently, CDG (Analysis) cannot be applied to them:
the latter has simply reached its limit of applicability and
validity (Q?.?, Q?.?), or even more graphically and in a brute
sense, it (and the physical laws that it models differential
geometrically as differential equations) `breaks down'.
\end{minipage}

\vskip 0.1in

On the other hand, as briefly alluded to above, so far there have
been proposed three types of `truly' singular boundary points of
$M$, all of which have of course (analytic) inextensibility at
their basis, but at the same time are distinguished from one
another by an extra physico-mathematical condition that must be
satisfied in the vicinity of their {\it locus}. In order, they
are:

\begin{itemize}

\item $\alpha)$ {\em (Differential) geometric singularities}
(DGS): boundary points for which there is no $C^{k}$-differential
extension of (the metric on) $M$ so as to remove
them.\footnote{Following footnote ??, a $C^{k}$-differential
inextensibility of $M$ past one of its boundary points means that
there is no $C^{k}$-chart ({\it ie}, coordinates of finite order
`$k$' of differentiability) to cover it. Equivalently, there is no
(isometric) embedding of $M$, of (finite) differential order $k$
($1\leq k\leq\infty$), into a `larger' manifold $M^{'}$ (with
$g_{M}=g_{M^{'}}$) \cite{hawk0}.} DGSs are precisely the
singularities alluded to in (Q?.?) and, as Clarke points out
there, they explicitly mark a breakdown of differentiability (of
the metric) at the $C^{k}$-level. DGSs are direct (mathematical)
evidence of the ineffectiveness of CDG and the $\smooth$-smooth
manifolds supporting it in coping with singular points already at
a finite order of differentiability---and it must be noted that
the lower the order `$k$' of a $C^{k}$-DGS, the more robust the
latter is supposed to be\footnote{With continuous ({\it ie},
$\cont$-) DGSs  representing the most singular, topological
obstructions to (even defining!) fields on the spacetime
continuum. However, since the Einstein equations are second order
PDEs, the lack of even $C^{2}$-extensions, let alone $\cont$-ones,
past a {\it locus} in $M$, indicates the existence of a `true'
gravitational DGS \cite{hawk0}.} \cite{hawk0,clarke3,clarke4}.

\item $\beta)$ {\em (Various) energy singularities} (VES):
boundary points for which there is no (analytic) extension of $M$
that removes them satisfying at the same time various energy
conditions, the most prominent one being the {\em weak energy
condition} in the celebrated singularity theorems of Hawking and
Penrose \cite{hawk1,hawk0,clarke4}.

\item $\gamma)$ {\em (Solution) field singularities} (SFS):
boundary points for which there is no (analytic) extension of (the
metric on) $M$ that removes them and is a solution of the Einstein
field equations in question.\footnote{As Clarke mentions in
\cite{clarke3}, the Einstein equations referred to above may be
taken for example to be the Einstein-scalar or the Einstein-fluid
equations.} The important thing to mention here is that the term
{\em solution} to the field equations means {\em generalized} or
{\em smeared}, what is commonly known as {\em distributional},
solution \cite{clarke3,clarke4}.\footnote{As we will see in more
detail in section 5, in \cite{df} Finkelstein implicitly
characterized the interior Schwarzschild singularity as an SFS,
since he up-front assumed that the vacuum Einstein equations hold
on an analytic spacetime manifold $M$ (even though he did not
explicitly state his assumption of a possibly distributional
curvature solution to them).}

\end{itemize}

\paragraph{Loosely comparing the three species of `real' singularities.}
The order in which the three species of `true' singularities were
presented above was intended to depict their `conceptual and
structural/mathematical (differential geometric) strength'. To be
more precise, and from a mathematical standpoint, DGSs may be
regarded as being more `fundamental' than either VESs or SFSs as
they express a direct inapplicability of CDG, since `smoothness'
or `differentiability' (of the metric, whose components, as noted
earlier, represent the gravitational potentials in GR) is supposed
to break down at a finite level of differentiation in their
vicinity. Apart from the smooth or analytic inextensibility of $M$
past them, DGSs are manifest examples of the ineffectiveness of
the basic `structural pillars' supporting the entire edifice of
CDG\footnote{That is to say, in view of our remarks in section
1---and in particular, of the identification (\ref{eq1})---of the
`structure coordinate algebra (sheaf)' $\smooth(M)$
($\smooth_{M}$)!} in dealing with them. A DGS may be thought of as
being more basic than a singularity belonging to the other two
`species' in the sense that it is not just that a solution of
Einstein's field equations does not hold (or even uncontrollably
blows up!) at its {\it locus} (Q?.?), but that {\em
differentiability itself breaks down at it}, so that one cannot
even in principle set up a differential equation (to represent the
gravitational law) over it (Q?.?).

On the other hand, and now from a more physical perspective, SFSs
do not challenge directly the basic differential properties of the
spacetime manifold {\it per se}, since one is supposed to be able
to write `valid' differential (Einstein) equations on the smooth
manifold in focus; rather, the auxiliary physical condition
imposed in addition to the analytic inextensibility of $M$ is that
the Einstein equations do not admit even generalized
(distributional) gravitational field strength (curvature)
solutions.\footnote{For example, in ?? we will present the
ADG-theoretic resolution of the interior Schwarzschild singularity
using sheaves of Rosinger's algebras of generalized functions
(distributions) as coefficient structure sheaves---generalized
(smooth) functions hosting singularities on everywhere dense
subsets of finite-dimensional (locally) Euclidean spaces
(manifolds).} Finally, in both conceptual and structural strength,
VESs lie `in between' DGSs and SFSs, for as Clarke remarks in
\cite{clarke3},

\bigskip \noindent (Q3.3)\hskip 0.9in
\begin{minipage}{11cm}
\noindent ``{\small ...In the `singularity theorems' of Hawking
and Penrose an intermediate case {\rm [{\it ie}, in between what
we called DGSs and SFSs above]}\footnote{Our addition for
continuity with the preceding text, as well as for clarity.}
arises in which one does not impose full field equations but only
energy conditions. Correspondingly, we shall call a boundary point
where there is no extension satisfying the weak energy condition
(for example\footnote{For an analytical treatment of the role
played by the various energy conditions in the theory of and
theorems about spacetime singularities, the reader is once again
recommended to refer to either \cite{hawk0}, or to
\cite{clarke4}.}) {\em energy singularity}. One could also class
this situation with the field singularities,\footnote{The ones we
called SFSs above.} taking the view that the energy conditions
stand in for the field equations in those felicitous situations
where one can get away without using the full content of the field
equations...}''
\end{minipage}

\vskip 0.1in

\noindent Formally, we depict this `conceptual and structural
strength' order of `genuine' singularities as follows:

\begin{equation}\label{eq10}
\boxed{DGSs}>\boxed{VESs}>\boxed{SFSs}
\end{equation}

\noindent and further note that our ADG-theoretic musings in the
sequel will concentrate mainly on the (mathematically) most basic
and strongest kind of the three: the DGSs. Nevertheless, we will
also tackle SFSs ADG-theoretically, being especially motivated by
ADG's successful application in formulating Einstein's (vacuum)
gravitational equations over a space(time) $M$ `coordinatized' by
algebras of generalized functions (differential algebras of
non-linear `multifoam spacetime' distributions) having
singularities (of the most pathological and classically
unmanageable kind) everywhere densely in $M$, and not just on its
boundary \cite{malros1,malros2,mall3,malros3,mall9,mall7}.

It must be stressed however that from a physical perspective, but
always in the context of CDG, SFSs---in particular those
associated with specific differential equations of physical
interest, like for example the Einstein equations---may be viewed
as being physically more `relevant' and of greater importance (to
the physicist) than the `purely geometrical' DGSs. One way of
expressing this is how Clarke opens chapter 4 of \cite{clarke4}:

\bigskip \noindent (Q3.4)\hskip 0.9in
\begin{minipage}{11cm}
\noindent ``{\small Although many of our considerations will be
geometrical, treating space-time as a pseudo-Riemannian manifold
and asking whether or not this geometrical structure is breaking
down, it must always be remembered that we are working with a
physical theory, governed by particular physical equations for
fields and particles, and that it is the breakdown of the physics
that is primarily of interest. {\em The breakdown of geometry is
simply one possible manifestation of the breakdown of the
physics}...}''
\end{minipage}

\vskip 0.1in

On the other hand, and as we shall see subsequently from an
ADG-theoretic perspective, we will expressly disagree with certain
points in (Q?.?) above. For one thing, and this goes glaringly
against the grain of (R2.?) and (R2.?) which capture the very
essence of the present paper,\footnote{That is to say, our basic
contention that {\em Nature has no singularities, and in no way
physical laws are limited by them}.} as well as with Ashtekar's
remarks on the unphysical infinities associated with
$\smooth$-singularities in (Q?.?) and (Q?.?), ``{\em the breakdown
of} (differential) {\em geometry}''---the one based on ``{\em
treating space-time as a} (smooth) {\em pseudo-Riemannian
manifold}'' ({\it ie}, CDG)---cannot possibly mislead us into
thinking that the physics itself ({\it ie}, the laws of Nature,
represented by differential equations) breaks down at
singularities. Of course, this is so as long as one does not
confuse the actual {\em physical `spacetime' geometry}---the one
{\em defined} by the (dynamical) algebraic interrelations between
the `geometrical objects' themselves ({\it ie}, the physical
fields and their particles\footnote{See concluding
section.})---with {\em our} mathematical model of spacetime as a
differential manifold and, concomitantly, of the fields that
inhabit it as being, in the classical sense, $\smooth$-smooth. As
noted earlier, the shortcomings, inadequacies and
limitations---ultimately, the breakdown---of the CDG-theoretic
analysis of the latter should not be attributed to Nature, that
is, to the physical fields that actually comprise it. For surely,
as long as one understands by ``{\em the breakdown of physics}''
simply the fact that {\em the physical theory is out of its depth
when dealing with singularities} (R2.?), and if one tends to
`identify' the physical theory with the mathematical model for
spacetime (and the fields dynamically propagating and interacting
on it),\footnote{This `identification' may be justified as
follows: in a general Wheelerian sense, a physical theory `{\em
is}' the dynamical laws expressed within its mathematical
framework (R2.?). For the particular physical theory, GR, the
(mathematical) expression of the law of gravity is the Einstein
equations, which are (non-linear, hyperbolic, partial) {\em
differential equations} and which in turn vitally depend on our
{\it ab initio} modelling spacetime after a differential
manifold.} then the aforesaid breakdown of physics is just an
indication that {\em the smooth manifold model of
spacetime---ultimately, CDG---has reached the limits of its
applicability and validity} (Q?.?, R2.3, R2.4). Accordingly, the
said `misattribution' (of the differential geometric breakdown to
a breakdown of the physics itself) may be justified on the grounds
that {\em so far we have been unable to do differential geometry
without a base differential (spacetime) manifold}---what we
earlier coined `{\em the $\smooth$-smooth manifold conservative
attitude and its associated CDG-monopoly}' (R2.6, Q2.2).

At the same time, if the physically unacceptable `behavior' of
$\smooth$-singularities is supposed to be not just due to the
manifest breakdown of differentiability or smoothness {\it per se}
(as for example with DGSs), but that in the vicinity of their {\it
loci} certain physically important smooth fields `grow without
bound' (diverge) (Q?.?) and, ultimately, become infinite (Q?.?)
(as for example with SFSs which do not admit even `blown up' or
`smeared out' solutions), it behoves one, who still insists though
at confronting singularities with the CDG-theoretic panoply in a
$\smooth$-conservative manner, to search for the physical (field)
culprit that `misbehaves' at singularities. Thus, following
\cite{clarke4}, we mention {\it en passant} that, in the usual
$\smooth$-spacetime manifold context, one could attribute the
physically anomalous behavior at singularities to the following
three fields:

\begin{enumerate}

\item The smooth spacetime metric $g$ whose ten components, as
noted earlier, physically represent the gravitational
potentials.\footnote{For example, explicit allusion to the
`breakdown' of the metric, whose components $g_{\mu\nu}$ are of
finite differentiability class $C^{k}$, was involved in the
definition of DGSs above. Without any constraints on the order of
differentiability ({\it ie}, assuming a $\smooth$-inextensibility
of $M$), and as it was briefly alluded to in footnote ??, it is
plain that the components of $g_{\mu\nu}$ belong to
$\sstruc=\smooth(M)$, or locally, and in a sheaf-theoretic sense,
they are local sections of the `classical' structure sheaf of
smooth coordinates $\struc=\smooth_{M}$ of $M$ (write
$g_{\mu\nu}\in \Gamma(U,\struc),~U~\mathrm{open~in~}M$)
\cite{mall1,mall2,malrap2,malrap3}.}

\item The smooth affine (Levi-Civita) connection $\nabla$, which
is taken to be compatible with the metric in the usual Riemannian
geometry (Q3.1) \cite{hughston}, and which, gauge-theoretically
speaking, may be thought of as representing the gravitational
gauge potential field \cite{malrap3}.\footnote{Similarly to the
case of the metric, and after assuming that the smooth affine
connection is {\em metric}, or equivalently {\em torsionless}
({\it ie}, that it obeys the metric compatibility condition
$\nabla g=0$), one could also define in the context of CDG
`connection based DGSs'---that is, a $C^{k}$-inextensible $M$
admitting a connection whose Christoffel components
$\Gamma_{\mu\nu}^{\lambda}$ in a local frame are
$C^{k}$-differentiable functions. Again, with no differentiability
constraints imposed on the extensibility of $M$,
$\Gamma_{\mu\nu}^{\lambda}\in\Gamma(U,\struc\equiv\smooth_{M}),~U~\mathrm{open~subset~of~}M$.}

\item The smooth Riemann curvature tensor $\curv$ (or its partial
Ricci tensor $\ric$, and full Ricci scalar $\ricci$, contractions)
of the connection $\nabla$ (or equivalently, of the meric and its
first partial derivatives), which, again gauge-theoretically
speaking, may be thought of as representing the gravitational
gauge field strength \cite{malrap3}.\footnote{Similarly to the
metric and connection cases above, if one does not require
up-front that $M$ be inextensible at a particular finite order of
differentiation, the (local) components of $R$ (and as a result,
of $\ric$ and $\ricci$ as well) are elements of
$\Gamma(U,\struc),~U~\mathrm{open~in~}M$.}

\end{enumerate}

\vskip 0.1in

With \cite{clarke4} as compass, a general and brief comparison of
DGSs, VESs and SFSs, concentrating more on the relation between
the two `extreme' kinds ({\it ie}, DGSs and SFSs) and always in
the context of CDG, follows below:

\begin{itemize}

\item First, in view of the fact that the expression for $R$
involves the metric as well as its first and second derivatives,
it would {\it prima facie} appear that an SFS could be
automatically coined a $C^{2}$-DGS.\footnote{Due to this, many
people have maintained that in actual practice it is sufficient to
view GR as a theory `invariant' under $C^{2}$-diffeomorphisms, so
that the assumption of full smoothness ($\smooth$) ``{\em would be
regarded as an idealisation convenient for geometrical purposes
but of restricted validity}'' \cite{hawk0,clarke4}.}

\item Physically speaking, $\curv$ is usually held to be more
important than either $g$ or $\nabla$; consider for example:

\bigskip \noindent (Q3.5)\hskip 0.9in
\begin{minipage}{11cm}
\noindent ``{\small\em ...From some points of view, the Riemann
tensor is of more direct physical importance than either the
connection or the metric, in the sense that an unbounded Riemann
tensor would suggest unbounded tidal forces, while an unbounded
connection or metric component could be a purely coordinate
effect...}'' \cite{clarke4}
\end{minipage}

\vskip 0.1in

\noindent although at the same time we find Hawking in
\cite{hawk2}, in view of the unnaturalness of the method of
surgically removing singular regions out of the spacetime
manifold, tending more towards adopting the DGS-characterization
of singularities rather than as {\it loci} where the curvature
tensor diverges, as follows:

\bigskip \noindent (Q3.6)\hskip 0.9in
\begin{minipage}{11cm}
\noindent ``{\small\em ...One normally thinks of a spacetime
singularity as a region in which the curvature becomes unboundedly
large. However, the trouble with that as a definition is that one
could simply leave out the singular points and say that the
remaining manifold was the whole of spacetime. It is therefore
better to define spacetime as the maximal manifold on which the
metric is suitably smooth...\footnote{Then he goes on and defines
singularities as in (Q?.?), in terms of (smooth or analytic)
inextensibility and incompleteness (of causal geodesics).}}''
\end{minipage}

\vskip 0.1in

\noindent The same dissatisfaction with the physically `forced'
and {\it ad hoc} character of surgically excising singularities
out of the smooth and otherwise regular spacetime manifold was
expressed earlier in \cite{geroch}, who writes:

\bigskip \noindent (Q3.7)\hskip 0.9in
\begin{minipage}{11cm}
\noindent ``{\small ...We originally introduced geodesic
incompleteness because the concept appeared to give a precise
statement of [the following property]:\footnote{Our addition.
Geroch actually calls it `Property 1'.} {\em In a nonsingular
spacetime, one should like to be sure that `no regions have been
deleted from the spacetime manifold'}\footnote{Our
emphasis.}...}''
\end{minipage}

\vskip 0.1in

Arguably, such both Geroch's and Hawking's dissatisfaction with
removing from the spacetime manifold, as it were by `theoretical
fiat', singular {\it loci} is tautosemous with Einstein's
`disbelief' in (Q?.?).

\item On the other hand, in line with Hawking's reservations about
the relevance of the unboundedness of the curvature tensor in
defining singularities, it must also be stressed that ``{\em an
unbounded Riemann tensor does not in itself indicate a
singularity}'' \cite{clarke4}, while at the same time, the
boundedness of $R$ does not imply that the metric is
$C^{2}$-differentiable either.

\end{itemize}

Indeed, it appears that one could `go in circles and argue till
one is blue in the face' whether the breakdown of the
differentiability of the metric ({\it ie}, DGS) or whether the
unboundedness of the (even if of a distributional character)
curvature field solution of Einstein's equations ({\it ie}, SFS)
is the physically more meaningful definition of singularities
within the confines of the CDG-framework. On the other hand, since
at the bottom of both DGSs and SFSs lies the notion of analytic
spacetime manifold inextensibility, and in accord with Hawking's
lose `definition' of singularities in (Q?.?), we will content
ourselves with the following concise, almost `apofthegmatic' and
`aphoristic', conception of singularities.

\subsubsection{Calculus is dead!; long live a background manifoldless differential geometry: ADG-theoretic forebodings}

Singularities are {\it loci} where `analyticity', and in
particular, `differentiability' or `smoothness', in the classical
CDG- or Calculus-theoretic sense of these words, `breaks
down'---{\it ie}, it just does not work. In turn, this means that
at singularities, CDG simply ceases to be an effective
mathematical framework within which to analyze and do
gravitational physics, and concomitantly, a base $\smooth$-smooth
manifold altogether ceases to be a sound model of `physical
spacetime', since the law of gravity appears to `halt', stop being
in force, or even `break down', at them.

At the same time, by smooth (or analytic) inextensibility of the
spacetime manifold---the notion that as we saw above underlies all
the aforementioned definitions of $\smooth$-singularities---and
the concomitant relegation of the singular {\it loci} at the
boundary or edge of an otherwise regular `physical' spacetime
manifold $M$, we simply understand that the algebra $\smooth(M)$
(or its associated structure sheaf $\smooth_{M}$) cannot be
further employed so as to coordinatize those `differentially
pathological' points (DGSs). Indeed, supposedly, if that could be
done, from the $\smooth$-smooth perspective (CDG) the purported
singularities would be characterized as being merely `coordinate'
or `virtual' ones, of no physical significance. {\it In summa},
and in connection with (\ref{eq1}), it is precisely $\smooth(M)$
({\it ie}, $M$ itself!) that `inherently' carries the seeds for
those `differential geometric pathologies', thus, insofar as
$\smooth_{M}$ (or the differential manifold $M$ itself) and CDG is
employed to deal with singularities, the problem is insuperable
simply because, in a strong sense, it is
`self-referential'.\footnote{That is to say, the `germs' of
singularities are already built into $\smooth(M)$!} Another way to
say this, if one tries to change coordinate algebra (sheaf)

\begin{equation}\label{eq6}
\struc_{M}\mapto\struc^{'}_{M^{'}}
\end{equation}

\noindent so as to smoothly extend $M$ (to $M^{'}$) past the
singular {\it loci}, albeit, always staying within the classical
one $\smooth_{M}$ of $\smooth$-smooth functions, singularities
will persist ({\it ie}, $M$ will remain smoothly inextensible),
since they are `innate' to the very coordinate `structure algebra'
$\smooth(M)$ (or the classical structure sheaf
$\struc\equiv\smooth_{M}$).

On the other hand, it must have become now fairly transparent that
if one did not stay within $\smooth(M)$ ({\it ie}, within $M$) and
the classical differential geometric framework, but could change
altogether the structure algebra sheaf of coordinates
(arithmetics)---with a concomitant, of course, change of the
classical notion of `differentiability' or `smoothness'---while at
the same time one was able to retain the full conceptual and
technical panoply of CDG and still have in one's hands the
latter's full calculational power,\footnote{What we cumulatively
referred to before as the `inherent' differential geometric
mechanism or machinery of Calculus (CDG).} the differential
geometric pathologies (singularities) of the differential
spacetime manifold could be totally evaded. We may coin this
change `{\em differential geometric extensibility}', which means
of course that we enlarge or generalize (in fact, {\em abstract}!)
the CDG-mechanism in a way that is not vitally (in fact, not at
all) dependent on classical smoothness, which is secured by
$\smooth_{M}$ ({\it ie}, by the very nature of the base locally
Euclidean manifold $M$). Of course, in line with what was said
before, this `differential geometric extensibility' is an abstract
analogue of `coordinate (structure sheaf) change' and, {\it in
extenso}, the purported total singularity-evasion will reveal
that, in a deep sense, {\em all singularities are effectively
`coordinate singularities'}. This last point lies at the heart of
ADG and foreshadows our ADG-theoretic musings in the sequel.

All in all, in the smooth manifold context ({\it ie}, working
always within the coordinate structure sheaf $\smooth_{M}$), by
`{\em analytic extension}', and for all intents and purposes, one
essentially  means (\ref{eq6}), so that

\begin{itemize}

\item when one is able to include by such a procedure a singular
{\it locus} with the other regular points, the said singularity is
said to be a `virtual' or `coordinate' one, and classical
differentiability (CDG) is essentially retained, but

\item when no such extension past the offensive point is possible,
the {\it locus} in focus is regarded as a `real' or `true'
singularity, and classical smoothness apparently breaks
down---{\it ie}, CDG simply becomes inadequate for coping with the
singularity.

\end{itemize}

\noindent In contradistinction, as briefly noted above and as we
shall see subsequently in the light of ADG, (\ref{eq6}) above
acquires a new, completely different meaning (than the notion of
smooth extension in CDG) on two accounts:

\begin{itemize}

\item On the one hand, and more from a mathematical perspective,
not just remaining within the confines of differential manifolds
and the associated (both technical and conceptual)
$\smooth$-conservatism, (\ref{eq6}) can be interpreted not only as
a (smooth or analytic) coordinate change aimed at integrating
apparently problematic {\it loci} with the other regular points of
a smooth manifold $M$,\footnote{Analytic or smooth extension
perceived here effectively as an enlargement, an embedding of $M$
(`charted' by $\smooth_{M}$) into a `larger' manifold $M^{'}$
(`charted' by $\struc^{'}_{M^{'}}$) \cite{hawk0}, always remaining
though within the category of smooth manifolds
($\smooth$-conservatism).} but, changing (in fact, enlarging!)
categories (of differential spaces) themselves---as it were, to
change altogether {\em differential geometric framework}
(theory).\footnote{What we coined above `{\em differential
geometric extensibility}. In fact, as we shall see in the next
section (3.1.7), the category of smooth manifolds is a subcategory
of the category of differential triads---the abstract differential
(sheaf) spaces of ADG.} Indeed, (\ref{eq6}) may be taken to mean
not just a smooth coordinate change, but as a change from, say,
the usual (`classical') smooth coordinate functions (in
$\smooth_{X}$),\footnote{Structure functions that define (the base
topological space) $X$ as a differential manifold $M$, placing it
thus within the category of smooth manifolds.} to another
functional algebra structure sheaf\footnote{Of function-like
entities which the theorist declares (assumes) up-front as being
`differentiable'.} which might be significantly different from the
`classical' one $\smooth_{M}$ of algebras of infinitely
differentiable functions, provided of course the new structure
sheaf of generalized arithmetics furnishes us with a `viable'
differential geometric mechanism as versatile and potent as the
classical one of Calculus (CDG). On such changes of differential
geometric framework rests the so-coined {\em Principle of
Relativity of Differentiability}(PRD) that we shall encounter in
the sequel.\footnote{This principle has been already anticipated
in \cite{malrap3}.}

\item On the other hand, and more from a physical viewpoint, such
changes (`enlargements') of algebras of differentiable functions
(generalized coordinate or coefficient arithmetics), while still
retaining in full force and effect the (inherently algebraic)
differential geometric mechanism of Calculus (albeit, in the
manifest absence of base $\smooth$-manifolds!) \cite{malrap2}, may
prove to be of great physical import since what might have
appeared to be insuperable,  ({\it ie}, `analytically
inextendible'---something that cannot be included with the rest of
the regular interior points of $M$) `truly' singular {\it loci} in
a manifold ($\struc\equiv\smooth_{M}$), now from the novel
differential geometric perspective afforded by the new
$\struc^{'}$ may appear to be completely `regular' and in no way
inhibiting (let alone collapsing) the said (inherently algebraic)
`differential geometric mechanism' that $\struc^{'}$ also
possesses in full effect. Moreover, what might have appeared as
`genuinely' singular points (or functions) from the classical
vantage (of $\struc\equiv\smooth_{M}$) may now be integrated in
(absorbed or engulfed by) $\struc^{'}$, while at the same time the
physical law of gravity, which is still modelled after a
differential equation nourished and sustained by the mechanism
afforded by $\struc^{'}$, may hold in full force in the presence
of or at the {\it locus} of those very singularities. Thus the
latter sites in no way represent breakdown points of the law of
gravity (Einstein equations), and hence they are completely evaded
or bypassed (algebraically).

\end{itemize}

\subsection{CDG Cannot Cope with $\smooth$-Singularities:
$\struc\equiv\smooth_{M}(\equiv M)$ is the Culprit}

The crux of the foregoing presentation and discussion is on the
one hand that $\smooth$-singularities cannot be thought of
independently of $\smooth_{M}$ (in point of fact, of $M$ itself!)
as they are built into that classical structure sheaf of
differentiable functions, and on the other, that the manifold
based CDG breaks down at ({\it ie}, it proves to be grossly
inadequate for coping with) them. Hence, in this light we
vindicate what we have been arguing for in the past two sections,
namely, that

\bigskip \noindent (R3.11)\hskip 0.9in
\begin{minipage}{11cm}
\noindent it is not gravity {\it per se} ({\it ie}, the physical
law that the gravitational field obeys) that carries within it the
seeds for its own destruction (Q?.?, Q?.?, Q?.?), but rather that
CDG itself---which is geometrically effectuated (or expressed) via
the background spacetime manifold---hosts its `autocatastrophe'.
At the same time, GR appears to set its own limit of validity by
predicting singularities exactly because it relies on Calculus
({\it ie}, on base manifolds) for its (differential) geometrical
expression (modelling). {\it In toto}, in a philological,
self-referential sense, {\em it is not that `gravity collapses
under its own weight', but rather, that it is the manifold that
breaks down due to its own $\smooth$-smooth foundation;
singularities are not pathologies of the gravitational law
(field), but of the (differential) geometrical medium (background)
employed by CDG---the differential manifold $M$---in order to
express that law differential geometrically, as a differential
equation}. Thus, it is the mathematics to `blame', not the
physics.\footnote{Alas, unfortunately the maths gets intimately
entwined and `confused' with the physics that it is supposed to
model, since one assumes up-front that {\em physical spacetime} is
(modelled after) a smooth manifold supporting a smooth metric
satisfying the differential equations of Einstein.}
\end{minipage}

\vskip 0.1in

\noindent On the other hand, it is well appreciated by now that
the smooth coordinates labelling the points of $M$ do not have a
direct metrical meaning; hence, since the smooth spacetime metric
is the sole dynamical variable in GR as originally formulated by
Einstein, the coordinates play no role in the gravitational
dynamics, thus they have no physical significance either
(PGC).\footnote{This is in line with our earlier Wheelerian remark
that {\em no theory is a physical theory unless it is a dynamical
theory}. Dynamics is everything!} We turn now to discuss in more
detail precisely this issue.

\subsubsection{The metric insignificance of coordinates in GR}

Here we would like to recall that Einstein himself `struggled' for
a long time to understand the metric meaning of spacetime
coordinates---that is, to clarify what is the relation from the
perspective of gravitational dynamics between the smooth
coordinates (in $\smooth(M)$) of the manifold's points, which {\it
prima facie} appear to have only a kinematical significance, and
the sole dynamical variable in GR, the spacetime metric
$g_{\mu\nu}$ representing the ten gravitational potentials in the
original relativistic theory of gravity. This quest was
essentially motivated by his desire to give a precise
(mathematical) formulation of the PGC, given the satisfaction of
the Equivalence Principle (EP),\footnote{Which, coupled to the
assumption of (differential) {\em locality}, requires that the
spacetime manifold of GR is locally flat---{\it ie}, that locally
$g_{\mu\nu}$ reduces to the
$\eta_{\mu\nu}=\mathrm{diag}(-1,1,1,1)$ of the Minkowski space of
SR.} and it was eventually answered in the negative ({\it ie},
that in GR, the spacetime coordinates have no direct dynamical
meaning), as the following quotation from \cite{einst9} shows:

\bigskip \noindent (Q3.8)\hskip 0.9in
\begin{minipage}{11cm}
\noindent ``{\small ...We start with an empty, field-free space,
as it occurs---related to an inertial system---within the meaning
of the special theory of relativity, as the simplest of all
imaginable situations. If we now think of a noninertial system
introduced by assuming that the new system is uniformly
accelerated against the inertial system in one direction, then
there exists {\small\rm [according to the EP]}\footnote{Our
addition.} with reference to this system a static parallel
gravitational field. The reference system may be chosen to be
rigid, Euclidean in its three dimensional metric properties. But
the time in which the field appears as static is not measured by
equally constituted stationary clocks.} {\small\em From this
special example one can already recognize that the immediate
metric significance of coordinates is lost once one admits
nonlinear transformations of the coordinates.}\footnote{Our
emphasis.} {\small\em To do the latter is, however, obligatory if
one wants to do justice to the equality of gravitational and
inertial mass through the foundations of the
theory}\footnote{Einstein's own emphasis.} {\small\rm [{\it ie},
the EP]}...\footnote{Again, our addition.}

{\small\em If, then, one must give up the notion of assigning to
the coordinates an immediate metric {\rm [ and
operational!]}\footnote{Our addition.} meaning (differences of
coordinates $=$ measurable lengths, or times), one cannot but
treat as equivalent all coordinate systems that can be created by
continuous transformations of the coordinates. The general theory
of relativity, accordingly, proceeds from the following principle:
Natural laws are to be expressed by equations that are covariant
under the group of continuous coordinate
transformations\footnote{Our emphasis.}...}''
\end{minipage}

\vskip 0.1in

\noindent while, in an even more direct, straight-out manner,
Einstein justified the long time it took him to construct GR
starting from SR exactly on the grounds that he had to shed first
his `prejudice' that coordinates have direct metrical significance
\cite{einst2}:

\bigskip \noindent (Q3.9)\hskip 0.9in
\begin{minipage}{11cm}
\noindent ``{\small ...Why were another seven years required for
the construction of the general theory of relativity? {\em The
main reason lies in the fact that it is not so easy to free
oneself from the idea that coordinates must have an immediate
metrical meaning}...}''
\end{minipage}

\vskip 0.1in

Thus, in GR, the spacetime coordinates have no (dynamical)
metrical significance, and one would intuit that their role in the
theory is relegated merely to a `kinematical' one. Albeit, in the
sequel we will argue that their role is not just kinematical
either; it is a deeper one, having to do with the very classical
differential geometric (in fact, analytic) foundations of the
theory. We discuss this point now.

\subsubsection{$\smooth$-coordinates, `geometrical points' and intervening space(time) are sacrosanct in CDG and GR for
differentiability's sake}

We shal start with some remarks of Chern from \cite{chern1} and
\cite{chern}, respectively:

\bigskip \noindent (Q3.10)\hskip 0.9in
\begin{minipage}{11cm}
\noindent {\small\em ``...From the proliferation of coordinate
systems it is natural to have a theory of coordinates. General
coordinates need only the property that they can be identified
with points; i.e., there is a one-to-one correspondence between
points and their coordinates---their origin and meaning are
inessential...''} \cite{chern1}
\end{minipage}

\vskip 0.1in

\noindent culminating in a `definition' of `{\em geometric objects
or properties}':

\bigskip \noindent (Q3.11)\hskip 0.9in
\begin{minipage}{11cm}
\noindent {\small\em ``...A property is geometric, if it does not
deal directly with numbers or if it happens on a manifold, where
the coordinates themselves have no meaning...''} \cite{chern}
\end{minipage}

\vskip 0.1in

After having established in 2.2.1  that the smooth coordinates of
the differential spacetime manifold have no direct metrical
significance in GR, hence also since $g_{\mu\nu}$ is the sole
dynamical variable in that theory, that they play no dynamical
role either, one is tempted to ask what exactly is the `operative'
role played by the smooth coordinates of the spacetime manifold's
points in the relativistic field theory of gravity.

The answer we suggest is that, although the coordinates of the
$\smooth$-smooth spacetime manifold $M$, as encoded in
$\smooth(M)$, have no dynamical meaning in the theory, they play a
vital role in setting up the theory within a CDG-framework. That
is, since one wishes, as Einstein did for the sake of
locality,\footnote{See below.} to model physical laws---in
particular, the law of gravity---by {\em differential equations},
and since as it has been mentioned emphatically earlier, the
differential spacetime manifold $M$, as a pointed space, is
nothing else but $\smooth(M)$ (or the structure sheaf
$\smooth_{M}$ thereof) (\ref{eq1}),\footnote{Some clarifying
technical remarks are due here: underlying the `identification'
(better, equivalence) in (\ref{eq1}) is the general notion of {\em
Gel'fand duality}---whose usual application is the so-called
Gel'fand representation theory---according to which, $M$, as a
pointed set endowed with its usual topology and differential
structure, is the (real) Gel'fand spectrum $\gelsp$ of the
topological algebra $\smooth(M)$, or its sheaf $\smooth_{M}$
(write: $\gelsp(\smooth(M))$ or $\gelsp(\smooth_{M})$)
\cite{mall0,mall-1,mall1,malrap3}. This duality is supposed to be
captured by the amphidromous arrow in (\ref{eq1}).} the latter
algebra provides the very differentiability property for the
metric in GR\footnote{Again, whose components in a local frame are
elements of $\smooth(M)$ ({\it ie}, smooth functions) since it is
by definition a $\otimes_{\struc\equiv\smooth_{M}}$-tensor.} so
that it can partake into the dynamical {\em differential}
equations of Einstein in the first place.

Another way to say this is that the smooth coordinates (in
essence, $M$ itself) provide the {\em the vital `precondition' of
differentiability or `smoothness' for GR}---{\it ie}, so that one
can apply the methods and techniques of Calculus on Manifolds
(CDG) in the theory of gravity, so that in turn the dynamical
equations of motion of Einstein that {\em define} the latter (as a
physical theory) are {\em differential equations} proper. In a
metaphorical sense, while smooth coordinates ({\it ie}, the
manifold) are not active parts in the dynamics of GR, they are
crucial `initial conditions' for it, as the assumption of a smooth
spacetime manifold is posited (fixed or `postulated') up-front  by
the theorist in order to place the theory in the (familiar)
CDG-framework, thus represent physical laws (here, the Einstein
equations) by differential equations within the latter (and secure
locality as we will see soon).

These arguments then apply {\it mutatis mutandis} for the points
of the manifold, which can also be recovered by Gel'fand duality
from $\smooth(M)$ or $\smooth_{M}$.\footnote{One may wish to
recall that the points $p\in M$ are represented by maximal, or
Zariski prime (since $\smooth(M)$ is a regular topological
algebra), ideals in $\gelsp(\smooth_{M})$
\cite{mall0,mall-1,mall1,malrap3}; see also 4.2.3 for an extended
discussion of the application of Gel'fand duality to $\smooth(M)$
in order to recover the $\smooth$-smooth manifold $M$ as a pointed
differential continuum.} Once one fixes the algebra of coordinates
to $\smooth(M)$ (or correspondingly, the structure sheaf to
$\smooth_{M}$), one automatically obtains $M$ as a pointed
manifold with its `built in' topological and differential
structures.

All in all, from this viewpoint, points and the smooth coordinates
labelling them appear to be sacrosanct in GR, and, {\it in
extenso}, in any theory that wishes to represent the dynamical
laws that define it by differential equations. A corollary is, of
course, that CDG (Calculus on Manifolds) is vital for the
(mathematical) formulation of GR.

\paragraph{Smoothness at the service of locality.} To recapitulate, coordinates (and points) are neither dynamical nor
kinematical proper entities in GR. Of course, one wishes to secure
{\it ab initio} `smoothness' in one's theory in order to implement
the `classical' notion of locality (Q?.?), which in the CDG
(classical spacetime continuum) context can be expressed as
follows:

\bigskip \noindent (R3.12)\hskip 0.9in
\begin{minipage}{11cm}
\noindent physical processes, corresponding to dynamical (field)
actions, connect infinitesimally separated point-events---a
statement equivalent to the assumption that the physical laws of
dynamics should be (modelled after) differential equations
\cite{malrap1,malrap2,malrap3}.
\end{minipage}

\vskip 0.1in

\noindent As noted in the introductory section, Einstein, for
example, was well aware of the fact that the {\it a priori}
conception and assumption of spacetime as a continuum (manifold),
with the differentiability (smoothness) properties that the latter
offers in order to implement the `classical' notion of
(infinitesimal) locality, is the last relic of an ether-like,
absolute structural element in GR---a structure which, in a
philological sense, `acts, but is not acted upon'\footnote{`Acts,
but is not acted upon' here meaning, as explained above, that
differentiability ({\it ie}, the differential structure of the
spacetime continuum) is a vital precondition for setting up the
dynamics (as differential equations) in GR, but this dynamics does
not in turn affect it in any way, as `differentiability' is fixed
up-front by the theorist to the classical notion of
$\smooth$-smoothness of CDG---a notion which is part and parcel of
the algebra $\smooth(M)$ of coordinates labelling the manifold's
points.} \cite{einst5,einst1,malrap1,malrap3}.

\paragraph{How come not a `variable' differentiability?} In this
paragraph we would like to address the issue of still not
possessing (or even having intuited yet) in theoretical physics a
`variable' notion of differentiability or `smoothness'. The
following discussion will anticipate the ADG-theoretic `{\em
algebraic relativization of differentiability}', culminating in
the {\em Principle of Algebraic Relativity of Differentiability}
(PARD), to be discussed in detail in 3.2 subsequently.

Our initial considerations in this respect are motivated the
following tower of structures `postulated' in GR:\footnote{The
following diagram is borrowed almost unaltered from \cite{df5}. We
wish to thank David Finkelstein for allowing us to do so.}

{\fontsize{0.19in}{0.19in}
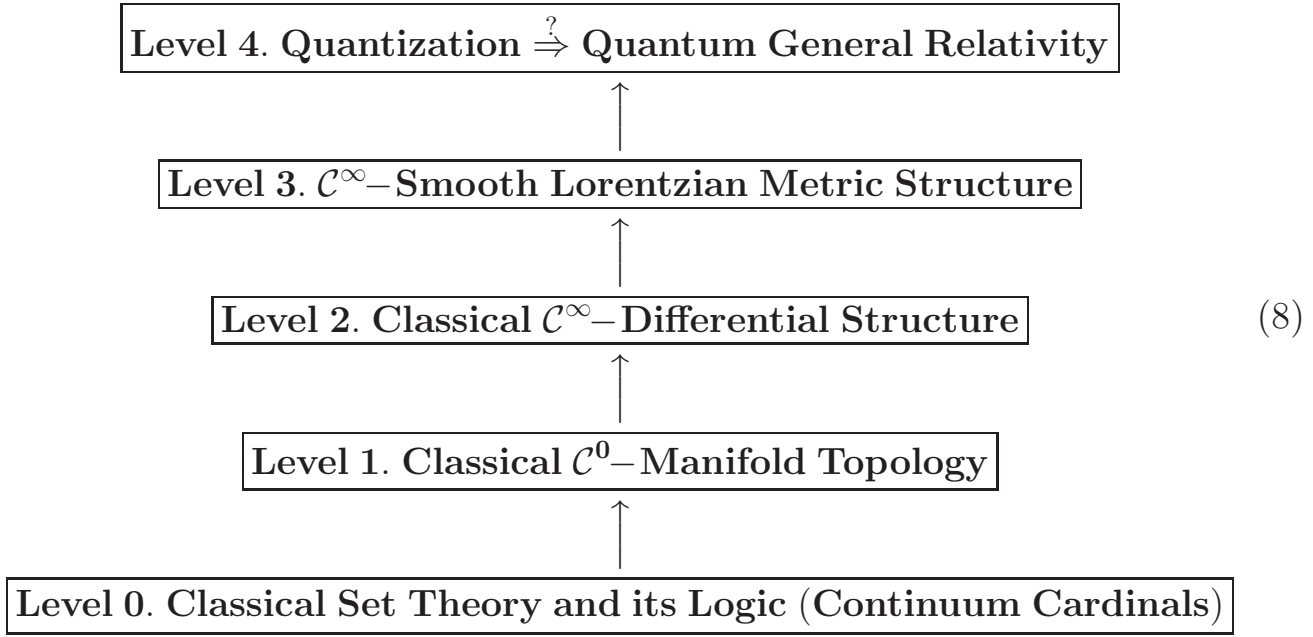
\begin{equation}\label{eq7}
\begin{CD}
\boxed{\mathbf{Level~4.~Quantization}\stackrel{\mathbf{?}}{\Rightarrow}\mathbf{Quantum~General~Relativity}}\\
@AAA\\
\boxed{\mathbf{Level~3.~\smooth\!\!-\! Smooth~Lorentzian~Metric~
Structure}}\\
@AAA\\
\boxed{\mathbf{Level~2.~Classical~\smooth\!\!-\! Differential~Structure}}\\
@AAA\\
\boxed{\mathbf{Level~1.~Classical~\cont\!\!-\! Manifold~Topology}}\\
@AAA\\
\boxed{\mathbf{Level~0.~Classical~Set~Theory~and~its~Logic~(Continuum~Cardinals)}}
\end{CD}
\end{equation}}

\noindent Finkelstein used this diagram in \cite{df5} to make the
following point: the attempts to arrive at a genuinely quantum
theoresis of spacetime and gravity by a direct quantization of the
smooth metric field on a pointed spacetime manifold, as for
example the various attempts in the context of the so-called
Quantum General Relativity (QGR)\footnote{That is, generally
speaking, the application of the usual continuum based QFT
formalism, canonically (via Hamiltonian methods) or covariantly
(via Lagrangian, action methods), to GR. Such an approach to QG,
especially via the Lagrangian, action based way, would in general
be favored by particle physicists who use predominantly
field-theoretic methods, rather than by general relativists who by
and large use classical differential geometric methods
\cite{feyn2,wein}, although after the advent of Ashtekar's new
(self-dual connection) variables approach to GR \cite{ash}, as
well as the loop variables approach to QG based on smooth
holonomies of Ashtekar's connections
\cite{rovsm,baez0,thiem2,thiem3,smolin}, differential geometric
methods (on the moduli spaces of those connections) have been
revived in the context of QGR
\cite{ashish,ashlew1,ashlew2,ash0,ash1,ash2}.} aspire to arrive at
QG, appear to be `superficial', `non-economical' and `contrived'
since only the `surface structure' of GR is involved ({\it ie},
the smooth metric at level 4 in (\ref{eq7})), but the deeper ones,
such as the differential and topological ones, let alone the
set-theoretic (logical) one at the bottom, are {\it a priori}
fixed to the ones of the classical picture---{\it ie}, the
classical spacetime continuum (manifold) regarded as a classical
(albeit uncountable) point-set. Accordingly, he favors a
`bottom-up' approach to quantum spacetime structure and dynamics
whereby, on {\it ab initio} manifestly finitistic and quantal
grounds, one should first attain a quantum relativistic picture of
set theory and its logic (with a `discrete' version of
relativistic causality still in force), and then build on it the
higher level spacetime structures such as the topological for
instance \cite{df2,df6,df3}, an ambitious endeavor which
eventually developed into a full-fledged Quantum Causal Net and,
ultimately, Quantum Relativity theory \cite{df1}. Finkelstein's
basic thesis is essentially that the deeper structures of GR, not
only the top level metric one, should be subjected to some kind of
`quantization' and `relativization', and altogether starting from
scratch---{\it ie}, from the logical one at level 0 in
(\ref{eq7}).

{\it En passant}, and in view of the essentially pointless
algebraico-categorical ADG-machinery supporting a background
spacetimeless ADG-field theory in general, and ADG-gravity in
particular, as it will be exposed in great detail in the coming
sections, it should be noted here that Finkelstein already in his
pioneering paper \cite{df0} (which laid the foundations for and
marked the commencement of his Quantum Set and Net Theory, as well
as his inspired Quantum Relativity Theory subsequently) had
emphasized the `unreasonableness' of the spacetime manifold
supported GR, as well as any attempt at directly quantizing its
`surface' (metric) structure in order to arrive at QGR (regarded
as a QFT on the classical spacetime continuum) and QG. As we read
below, what he set out to do in \cite{df0} was `antipodal' to the
tower of structures in (\ref{eq7}) above:

\bigskip \noindent (Q3.12)\hskip 0.9in
\begin{minipage}{11cm}
\noindent ``{\small ...This approach [{\it ie}, the `Space-Time
Code'],\footnote{Our addition for textual continuity.} which
attributes to space-time points an intricate internal structure,
seems upside down from the point of view of general relativity,
and general relativity seems upside down from here, seems too
complicated a theory of too simple a thing to be fundamental
rather than phenomenological. {\em What is too complicated about
general relativity is the delicate vertical structure of laws that
would have to be legislated on the first day of creation: set
theory, holding up topology, holding up differential manifolds,
holding up pseudometric geometry, with a precarious topping of
quantization.\footnote{See again (\ref{eq7}) above.} What is too
simple about general relativity is the spacetime point. It looks
as if a point might be an enormously complicated thing. Each
point, as Feynman once put it,\footnote{See quotation (Q7.54)
taken from \cite{feyn0} in section 7 and comments therein in the
light of our ADG-theoretic perspective on (gauge) field theory and
especially (quantum) gravity.} has to remember with precision the
value of indefinitely many fields describing indefinitely many
particles; has to have data inputs and outputs connected to
neighboring points; has to have a little arithmetic element to
satisfy the field equations; and all in all might just as well be
a complete computer}\footnote{Our emphasis.}... }''
\end{minipage}

\vskip 0.1in

\noindent Leaving Finkelstein's reticular `quantum set theory' at
the deepest, bottom logical level in (\ref{eq7}) aside for a
moment, it should be mentioned that, starting from Wheeler's
original foam conception of a `quantal', dynamically fluctuating
spacetime `microtopology' at sub-Planckian scales \cite{wheeler},
theoretical physicists have worked on various finitistic and
algebraic `quantization' scenarios for spacetime topology
\cite{ish0,ish,df3,rapzap1,rapzap2}, with Isham's recent
`quantizing on a general category' scenario \cite{ish5,ish6,ish7}
being the category-theoretic evolution of the `group quantization
of spacetime topology' ({\it \`a la} Mackey's imprimitivity) ideas
put forth initially in \cite{ish-1} and further elaborated
subsequently in \cite{ish}. At a more abstract level,
mathematicians too have worked on `quantal', noncommutative
conceptions of topology, with Mulvey's quantales in the context of
Gel'fand representation theory for non-abelian $C^{*}$-algebras
\cite{mul,mulpel1,mulpel2}, as well as Van Oystaeyen's
noncommutative topologies in the context of noncommutative
algebraic geometry \cite{voyst1,voyst2,voyst3,voyst4}, being the
two most prominent paradigms in our view.

At the next level of {\em differential} structure (level 2 in
(\ref{eq7}))---in which we are predominantly interested in the
present paper, while, as noted above, physicists have intuited
that both in the classical context of GR \cite{horowitz}, but more
importantly, in a genuinely quantum theoresis of spacetime
structure and dynamics \cite{wheeler}, one's theory should be able
to account for dynamical processes of spacetime topology change,
no one has intuited yet similar changes in the differential
structure of spacetime.\footnote{And this ellipsis is kind of
`short-sighted' in our opinion, since the differential is normally
regarded as a higher level structure than the topological
(\ref{eq7}). In other words, if `continuity' should be somehow
`quantized' and `relativized'---{\it ie}, be regarded as a
non-fixed, {\em variable structure} that partakes into dynamics
\cite{df1}, why not also `differentiability' or `smoothness'?}
Presumably, this is because the smooth manifold provides the sole
theoretical framework (CDG) we possess so far within which we know
how to differentiate, and the automatic, almost `reflex'-attitude
of the theorist is to assume and fix it up-front in her theory.
This `smooth manifold monopoly' and associated
`CDG-solipsism'---especially in theoretical physics where CDG
methods and applications reign supreme, with impressive
experimental success and support(!), from the microcosm of the
Standard Model (SM) to the macrocosm of GR---is the epitome of the
manifold conservative attitude criticized throughout this paper,
which attitude, in turn, as mentioned earlier, appears to be
mandatory if one wishes to represent the laws of physics by
differential equations proper.

{\it Prima facie}, one could readily counter the aforesaid almost
religious commitment to the (commutative)\footnote{Commutativity
here pertains to the fact that
$\smooth(M)^{\mathbb{K}}~(\mathbb{K}=\R ,\com)$ is an abelian
algebra.} spacetime manifold by bringing up the by now well
developed and quite popular Noncommutative (Differential) Geometry
(NDG) of Connes \cite{connes,kastler} already enjoying numerous
applications in high energy (SM) theoretical physics
\cite{connes2,connes3} and potentially in (quantum) spacetime
structure and gravity research \cite{connes1}.\footnote{It must be
noted at this point that NDG appears to be particularly `popular'
nowadays in the (super)string-theoretic approach to QG, for which
however we provide no direct references herein. The reader can go
via \cite{rovelli} to find references about how NDG appears to be
of import in current string theory research.}
However,\footnote{And this is meant as a mild critical remark
about NDG {\it vis-\`a-vis} ADG and what we are trying to
accomplish in the present paper.} NDG, {\it \`a la} Connes, may be
perceived as an attempt to `{\em quantize Calculus}' by functional
analytic means, {\em while the notion of manifold still survives
in it, fixed as it were at the background}, of course, with all
the classical differential geometric anomalies and diseases in the
guise of `singularities' that such a `manifold conservatism'
entails. On the other hand, the whole algebraic (categorical),
being sheaf-theoretic, machinery of ADG has been developed
independently of any notion of base differential manifold
\cite{mall1}. Being free of any smooth manifold dependent concept
and construction, ADG is able to cope with problems pertaining to
`singularities' still by applying methods of NLPDEs (`generalized
functions').\footnote{Indeed, for a similar {\em en passant}
critique of NDG on the face of ADG's successful application to the
so-called spacetime foam dense singularities associated with
Rosinger's algebraic theory of non-linear generalized functions
(distributions), the reader is referred to \cite{malros2}. We will
revisit in detail this application of ADG in subsection 4.2.} It
is also worth noting here the conceptual simplicity of the
machinery of ADG, as well as that, in general terms, resorting to
noncommutativity,

\begin{itemize}

\item ``{\small\em does not mean a loss of interest in, let alone,
the abandonment of commutative theories. Indeed, in many important
problems, the latter turns out to be both more simple and far more
effective}'' \cite{malros2}.\footnote{Recently for example, Jet
Nestruev ({\it ie}, Vinogradov {\it et al.}) \cite{vinogradov}
have been rather critical of noncommutative theories as regards
their conceptual and effective calculational import in the theory
of PDEs.}

\item Related to the point above, (differential geometric)
theories are noncommutative ``{\em only on their so-called
`global' level, while in the last instance of their detailed
calculations, that is, `locally',\footnote{And it should be
emphasized that `differentiability' is, by definition, a local
notion.} they reduce to commutative ones}'' \cite{malros2}.

\item The latter point, and from a quantum-theoretic perspective,
becomes even more important {\it vis\`a-vis} concrete calculations
from direct application of differential geometric ideas in physics
if one recalls Bohr's correspondence principle according to which,
the values of our measurements of the so-called `observables' of
quantum systems---the noncommuting in general $q$-numbers of
quantum mechanics---are `classical', commutative $c$-numbers
\cite{malros2,malrap1,malrap2,malrap3}.

\item Finally, even more incisive is Finkelstein's remark that
``{\small\em noncommutativity is by no means the quintessential
characteristic of `quantumness'; rather, the essence of the latter
is `relativization' and dynamical variability of physical
attributes}''.\footnote{David Finkelstein quoted roughly from an
old private communication with the second author. In other words,
`quantization' (if one would still want to use this term in a
theory such as Quantum Relativity, and also our approach
\cite{malrap3}, in which it is fundamentally assumed that ``{\em
all is quantum}'') is `(quantum) relativization' entailing
dynamical variability \cite{df1}.} Equivalently, {\em
noncommutativity does not necessarily imply
`quantumness'}.\footnote{For instance, think of the non-commuting
boosts in SR, which is a classical theory.}

\end{itemize}

After this short critical interlude about NDG, returning to
(\ref{eq7}) with its `differential manifold monopoly' and the
associated `CDG solipsism' at level 2, we would like to prepare
here the ground for (a) the ADG-theoretic `dynamicalization' of
coordinates and spacetime (points),\footnote{By the term
`dynamicalization' of coordinates and spacetime (points) we mean
the dynamical `consequence' of `spacetime'---in other words, that
spacetime is the result of the gravitational dynamics, not fixed
up-front to a manifold as in the classical theory, and that this
dynamical dependence is in turn due to a process of
`relativization of differential structure' as the latter is
effectively encoded in the structure sheaf $\struc$. The latter is
essentially the PARD alluded to before.} and (b) the associated
Principle of Algebraic Relativity of Differentiability (PARD),
both to be expounded in more detail in subsection 3.2.

\subsubsection{Anticipating the central field-axiomatic of ADG:
`field-solipsism' and the associated Principle of Field Realism
(PFR)}

At this point, one must highlight and admire the intuition,
imagination and resourcefulness of the physicist in attempts to
transcend the technical (mathematical) rigidity and monopoly of
the differential spacetime manifold. For example, ever since the
first `toddling' steps of GR, Einstein had intuited that the
point-events (and, {\it in extenso}, their coordinate labels) of
the spacetime continuum are the `results' of gravitational
dynamics---in a strong sense, {\it spacetime is `inherent' and
foreshadowed in the dynamical gravitational field itself}, and
that the (empty) spacetime continuum, devoid of (the dynamics of)
the gravitational field, is a physically meaningless concept and
structure. Characteristically, we quote:\footnote{The following
quotations can be also found in \cite{malrap3}, but due to their
importance we recast them here.}

\bigskip \noindent (Q3.13)\hskip 0.9in
\begin{minipage}{11cm}
\noindent ``{\small ...According to general relativity, the
concept of space detached from any physical content does not
exist. {\small\em The physical reality of space is represented by
a field whose components are continuous functions of four
independent variables---the coordinates of space and time. It is
just this particular kind of dependence that expresses the spatial
character of physical reality}.\footnote{Our emphasis.}}''
\cite{kostro}\footnote{Page 175 and reference therein.}
\end{minipage}

\vskip 0.1in

\noindent while

\bigskip \noindent (Q3.14)\hskip 0.9in
\begin{minipage}{11cm}
\noindent ``{\small ...If the laws of this field are in general
covariant, that is, are not dependent on a particular choice of
coordinate system, then the introduction of an independent
(absolute) space is no longer necessary. That which constitutes
the spatial character of reality is simply the four-dimensionality
of the field. {\small\em There is no `empty' space, that is, there
is no space without a field}.\footnote{Again, our emphasis.}}''
\cite{kostro}\footnote{Again, page 175 and reference therein.}
\end{minipage}

\vskip 0.1in

\noindent as well as

\bigskip \noindent (Q3.15)\hskip 0.9in
\begin{minipage}{11cm}
\noindent ``{\small\em ...No space and no portion of space can be
conceived of without gravitational potentials; for these give it
its metrical properties without which it is not thinkable at all.
The existence of the gravitational field is directly bound up with
the existence of space\footnote{Emphasis is ours.}...}''
\cite{einst5}
\end{minipage}

\vskip 0.1in

\noindent and

\bigskip \noindent (Q3.16)\hskip 0.9in
\begin{minipage}{11cm}
\noindent ``{\small\em ...according to the general theory of
relativity even empty space has physical qualities, which are
characterized mathematically by the components of the
gravitational potential.\footnote{That is, even in the vacuum
Einstein equations, it is the dynamical gravitational field that
gives spacetime its physical identity. We will return to elaborate
more on this in the final section when we discuss Einstein's hole
argument.}''} \cite{kostro}\footnote{Again, page 111 and reference
therein.}
\end{minipage}

\vskip 0.1in

\noindent moreover,

\bigskip \noindent (Q3.17)\hskip 0.9in
\begin{minipage}{11cm}
\noindent ``{\small Thus, once again `empty' space appears as
endowed with physical properties, {\it i.e.}, no longer as
physically empty, as seemed to be the case according to special
relativity. One can thus say that the ether is resurrected in the
general theory of relativity, though in a more sublimated form.}''
\cite{kostro}\footnote{Page 111 and reference therein.}
\end{minipage}

\vskip 0.1in

\noindent so that we come to the `apotheosis' of this `{\em no
gravitational field, no spacetime}' apofthegma of Einstein:

\bigskip \noindent (Q3.18)\hskip 0.9in
\begin{minipage}{11cm}
\noindent ``{\small\em ...There is no such thing as empty space,
{\it i.e.}, a space without field. Space-time does not claim
existence on its own, but only as a structural quality of the
field\footnote{Our emphasis.}...}'' \cite{einst8}
\end{minipage}

\vskip 0.1in

\noindent which can be augmented by his later remark that

\bigskip \noindent (Q3.19)\hskip 0.9in
\begin{minipage}{11cm}
\noindent ``{\small\em ...space has lost its independent physical
existence, becoming only a property of the field\footnote{Again,
our emphasis.}...}'' \cite{einst3}
\end{minipage}

\vskip 0.1in

Subsequently, Bergmann and Komar, working on (a possible extension
of) the diffeomorphism symmetries of GR, which implement the PGC
in that differential spacetime manifold based theory \cite{berg2},
were led to the conclusion that points (and their coordinates) are
not independent structural entities in GR, that they have no
actual physical meaning, but rather that in a sense they derive
their `identity' from the field equations
themselves:\footnote{Peter Bergmann's life-long work on the role
of $\mathrm{Diff}(M)$ in GR has repeatedly emphasized that points
(and coordinates) are not absolute, `sacrosanct', self-sustaining
entities in the theory, but rather that they are `consequences' of
the dynamical symmetries of the theory (David Finkelstein in
private e-correspondence).}

\bigskip \noindent (Q3.20)\hskip 0.9in
\begin{minipage}{11cm}
\noindent ``{\small ...Such transformations might point in a
direction in which {\em fields get unified under an invariance
group in which the space-time manifold no longer plays a
pre-eminent role. A world point would derive its identity from its
dynamic environment, or it might possess no identity at
all.}\footnote{Emphasis is ours.}}'' \cite{berg1}
\end{minipage}

\vskip 0.1in

\noindent While, the climax of this `dynamicalization' of
spacetime points and their coordinates appears to be the principal
aim in Stachel's latest work, as we read from \cite{stachel2}:

\bigskip \noindent (Q3.21)\hskip 0.9in
\begin{minipage}{11cm}
\noindent ``{\small ...I shall then emphasize the ways in which
the space-time structure of the general theory of relativity, by
virtue of its dynamization\footnote{Here we use the slightly
different term `dynamicalization' instead of Stachel's
`dynamization' in order to prepare the reader for a significantly
different ADG-theoretic stance about these issues, to be expounded
in the sequel ({\it cf.} last section when we discuss Einstein's
hole argument motivated Stachel's extensive work on it).} of the
chrono-geometry, differs radically from that of all previous
physical theories. I shall stress two fundamental differences...
{\em 2. the role of diffeomorphism invariance in precluding the
existence of any pre-assigned (kinematical) spatio-temporal
properties of the points of the manifold (even locally) that are
independent of the choice of a solution to the field equations (no
kinematics before dynamics).\footnote{The reader should wait for
our arguments about the ADG-theoretic reversal, from `{\em
kinematics-before-dynamics}', to `{\em
dynamics-before-kinematics}', to be given in 3.2.3. Note also the
comments about `differentiables'---properties (better, `initial
conditions') pertaining to the `differentiability attributes' of
the structures involved in the theory.} The physical points of
space-time thus play a secondary, derivative role in the theory,
and cannot be used in the formulation of physical questions within
the theory (they are part of the answer, not part of the
question)\footnote{Our emphasis.}...\footnote{From our point of
view, this is not quite; rather: {\em they are part of the
`initial (pre)conditions for differentiability'}, which will be
relativized ADG-theoretically below (again, see subsection 3.2).
`Differentiables' lie with the `observer'/`measurer', who, like
initial conditions, chooses what $\struc$ to employ, thus
differential geometrize her theory.}}}''
\end{minipage}

\vskip 0.1in

\noindent All this points at the following {\em `axiom' of
field-solipsism} in ADG:

\bigskip \noindent (R3.13)\hskip 0.9in
\begin{minipage}{11cm}
\noindent In ADG, the concept of (gravitational)
field---represented by the pair $(\modl ,\conn)$\footnote{This is
a `superfluous' representation. Strictly speaking, for us the
field (of gravity) is simply the connection $\conn$, but since the
latter is a map---{\it ie}, the $\cons$-linear sheaf morphism
$\modl\mapto\modl(\omg)\equiv\modl\otimes_{\struc}\omg$---we also
include for completeness its domain $\modl$ (or even its range
$\omg :=\modl^{*}$ \cite{malrap3}).}---is a fundamental, `ur',
primitive notion. {\em There is no background spacetime
structure;\footnote{Or rather, this background surrogate
`space(time)'---whether a continuum or a discretum---plays no role
whatsoever in the algebraic differential geometric mechanism
offered by ADG.} all there is and is of physical significance are
the `geometric objects'---{\it ie}, the fields themselves.} In an
Aristotelian \cite{arist,aristotle,popper}, but more explicitly,
Leibnizian sense, the fields, like Leibniz's monads
\cite{leibniz2,leibniz,leibniz1}, are autonomous, self-sustaining
entities, without being in need of an external, background
spacetime for their subsistence, but unlike monads, our fields are
not `windowless', as they interact with each other. Thus, {\em the
notion of field in ADG comes to replace that of the spacetime
manifold and its point-events in the classical theory of gravity
(GR). In a strong sense, the `events' in our theory are the fields
(in) themselves, and there is nothing else except fields and the
dynamics that they define as differential equations.}
\end{minipage}

\vskip 0.1in

\noindent All in all, and in contradistinction to Einstein's GR,
in which too ``{\small [In the theory of relativity,] \em the
field is an independent, not further reducible fundamental
concept...and the theory presupposes the independence of the field
concept}'' \cite{einst3}, our `solipsistic' field

\bigskip \noindent (R3.14)\hskip 0.9in
\begin{minipage}{11cm}
\noindent {\em does not depend at all on a background spacetime
continuum (manifold) for its support and (differential geometric)
subsistence.}\footnote{See concluding section.}
\end{minipage}

\noindent In \cite{malrap3} we argued that the aforesaid
`field-solipsism' sets the stage for the ADG-theoretic formulation
of a {\em genuinely `unitary' field theory---one dealing solely
with the fields themselves and which therefore is in no need of an
external (background) geometrical spacetime, whether the latter is
modelled after a continuum (manifold) or a
discretum}.\footnote{Again, see the concluding section 7 for
further discussion of the possibility that ADG, with its
fundamental field-solipsism, is an appropriate mathematical
framework in which to materialize Einstein's unitary field
vision.}

\subsection{Section's R\'esum\'e}

Although there is a plethora of issues raised in the past section
that we could summarize now, here is a short list of points made
that we would like the reader to keep in mind as the paper
unfolds:

\begin{enumerate}

\item $\smooth$-smooth singularities in GR are inherent in
$\smooth_{M}$; in effect, they are built into $M$ itself. One
cannot think of gravitational singularities apart from the
assumption (mathematical model) that (physical) spacetime is a
differential manifold.

\item Singularities signify a breakdown and end of the manifold
based CDG (Calculus or Analysis), not of the dynamical
gravitational field laws of GR, but since the latter is
inextricably tied to the base spacetime manifold, we are misled
into thinking that singularities are pathologies of the physical
theory itself ({\it ie}, of the gravitational dynamics---Einstein
differential equations) and not just `anomalies' of the
mathematical model for spacetime ($M$) and of the general
differential geometric framework fundamentally based on it (CDG).

\item From the ADG-theoretic perspective, all singularities are
virtual, `coordinate' ones, as they are part and parcel of the
coordinate structure sheaf $\struc$ one uses in the theory (in the
classical case, $\struc\equiv\smooth_{M}$).

\item Spacetime (manifold) points and their (smooth) coordinates are not
fundamental, that is to say, physical ({\it ie}, dynamical)
entities in GR. Traditionally, they ({\it ie}, the pointed
manifold and the smooth coordinates labelling its points) are
thought of as being part of the kinematics of GR. Here we regard
them as vital structural preconditions for setting up the physical
theory differential geometrically in the first place. At the same
time, if there is any `space(time) geometry' at all, then it is
inherent in $\struc$ that {\em we} assume in order to
(differential) geometrically represent the gravitational field
$\conn$.

\item What is normally regarded as being `irreducible' in the
smooth manifold and {\it in extenso} CDG-based GR, namely, the
manifold's points and the gravitational field $g$ defined on
them---totally known as `{\em spacetime events}', in ADG-gravity
is replaced by the `{it ur}'-notion of field, namely, a connection
$\conn$ on a vector sheaf $\modl$.\footnote{Equivalently, the
ADG-gravitational field is the pair $(\modl ,\conn)$
\cite{malrap3}.} Moreover, the causal nexus between events---which
in the standard GR is supposed to be a dynamical
variable\footnote{Since $g_{\mu\nu}$---the sole dynamical variable
in the original formulation of GR due to Einstein---defines
(locally) the Minkowski lightcone, which in turn delimits the
local causal relations between spacetime points (events).}---is
replaced in ADG-gravity by the dynamical law (vacuum Einstein
equations) that the (vacuum) ADG-gravitational field $\conn$
defines on $\modl$.\footnote{See (\ref{eqy23}) below.} As a
pun-type of metaphor, the dynamical gravitational connection field
$\conn$ represents in ADG-gravity the dynamical causal connection
\cite{malrap1}, and the primitive notion of causality is replaced
by {\em the} fundamental ADG-notion of (algebraic) connection.
{\em Connection $\conn$ is causality}.

\end{enumerate}

\vskip 0.1in

\section{GR from the ADG-Theoretic Point of View}

We start this section with the simple but fundamental observation
that in order to do differential geometry one needs to possess
some sort of {\em differential operator} $d$. There is no DG
proper without $d$. In turn, to be able to differentiate one needs
somehow to {\em form differences} ({\it ie}, possess some kind of
{\em linear structure}), as well as to {\em take limits} ({\it
ie}, possess some sort of {\em topology}). Thus, {\it prima
facie}, all one needs to have in hand in order to differentiate
({\it ie}, to define a differential $d$) is a type of {\em
topological vector space} structure. On top, if one further needs
one's derivative $d$ to obey a Leibnizian condition ({\it ie},
behave in a particular way regarding products of `functions'---the
sort of `animals' on which $d$ acts), one should possess a kind of
{\em topological algebra} structure, with $\smooth(M)$ (and the
coordinate structure sheaf $\smooth_{M}$) underlying differential
manifolds $M$ and on which the entire CDG edifice is based
(\ref{eq1}) being the archetypal example of a `non-normable'
topological algebra (structure sheaf). All in all,

\bigskip \noindent (R4.1)\hskip 0.9in
\begin{minipage}{11cm}
\noindent {\em `differentiability' is a topologico-algebraic
notion.}
\end{minipage}

\vskip 0.1in

\noindent {\bf Comment:} It is important to note here that ADG's
original aim was to formulate the central differential geometric
notion of {\em connection} ({\it viz.} `generalized differential
operator') {\em entirely algebraico-categorically}, in a way that
explicitly does not depend at all on a background geometrical
$\smooth$-smooth manifold (thus show that the classical case of a
smooth connection arises only when one chooses
$\smooth_{M}$---{\it ie}, a base differential manifold
(\ref{eq1})---for one's structure sheaf of generalized
`coordinates' or `coefficients').

\bigskip \noindent (R4.2)\hskip 0.9in
\begin{minipage}{11cm}
\noindent {\em Alas, one can get a differential $d$ (or a
generalized differential $\conn$), thus a generalized notion of
differentiability or `smoothness', from structure sheaves
glaringly different from $\smooth_{M}$---and more importantly,
from ones that may be teeming with singularities or ones that are
more suitable for addressing and tackling one's (physical) problem
at issue}.
\end{minipage}

\vskip 0.1in

\noindent What is important to stress here is that, in effect,
ADG's main aim is to `{\em algebraicize Calculus}'. That is to
say, in view of the topologico-algebraic character of $d$
mentioned above, in ADG there is the tendency to underplay and
`atrophize' the `spatial', topological (:geometric) aspect (and
dependence) of Analysis, and to pronounce its purely algebraic
aspect. Indeed, as we shall see below, the arbitrary base
topological space $X$ on which the algebra and differential module
sheaves of interest are soldered plays no role in the {\em
`inherently' algebraic} differential geometric mechanism of ADG,
which, as it were, derives from the stalk of the said
sheaves---{\it ie}, from the algebraic structures inhabiting the
relevant sheaf spaces. But it is high time now to expound the
basic tenets of ADG.

\subsection{Rudiments of Abstract Differential Geometry: The ABC of ADG}

In this section we present the basic concepts and mathematical
structures involved in ADG. We begin by giving several {\em
physico-mathematical} `shortcomings' and `philosophical blemishes'
of differential ({\it ie}, $\smooth$-smooth) manifolds and, {\it
in extenso}, of the CDG based on them, `inadequacies' which could
be regarded {\it a posteriori}\footnote{That is, now after ADG has
been developed, applied in both mathematics and physics, and
contrasted against CDG.} as `reasons' for developing ADG, as a
mathematical theory with potent physical applications (and being
supported by sound conceptual/philosophical foundations), in the
first place, although the principal {\it a priori} (mathematical)
motivation of the first author for constructing the theory will be
clearly highlighted and distinguished from those `lateral',
`by-the-way' and secondary reasons. The following exposition is
taken mainly from \cite{rap8,rap5}.

\subsubsection{Manifold reasons against the (spacetime) manifold: `lateral' mathematical and
physical motivations for developing ADG}

\paragraph{Mathematical `deficiencies' of manifolds.} As in \cite{rap8,rap5},
we follow mainly Papatriantafillou \cite{pap1,pap2,pap3,pap4} and
first list some mathematical shortcomings of (the category $\man$
of) finite-dimensional smooth manifolds:

\begin{enumerate}

\item An arbitrary subset of a (smooth) manifold is not a (smooth)
manifold. In categorical terms, $\man$ {\em has no canonical
subobjects}. For example, a curve with a corner or a cusp in a
differential manifold $M$ is not itself a smooth submanifold of
$M$.

\item The quotient space of a manifold by an (arbitrary)
equivalence relation is not a manifold. Consider for instance the
`discrete' $T_{0}$-poset obtained by a suitable equivalence
relation from a continuous manifold $M$ relative to a locally
finite open cover of the latter in \cite{sork0}.\footnote{This
example will be of special interest to us subsequently (section
5), when we `resolve' the inner Schwarzschild singularity by
finitistic-algebraic, ADG-based means.}

\item In general, $\man$ is not closed under projective (inverse)
or inductive (direct) limits.\footnote{In category-theoretic
jargon, these are known as categorical limits and colimits,
respectively.} In categorical terms, $\man$ {\em is not
bicomplete}. To be sure, in general, the projective limit of an
inverse system of finite-dimensional manifolds is not a manifold;
while, in certain cases, as in the theory of jets, where the
inverse limit exists, it is infinite dimensional.

\item $\man$ has no initial or final structures. For instance, one
cannot pull-back or push-out a smooth atlas by a continuous map.
As a consequence, there is no what one might coin `differential
geometry of topological spaces'.

\item In general, {\em there are no categorical products
and co-products in} $\man$.

\end{enumerate}

\paragraph{Physical shortcomings of manifolds.} We have already alluded
to the physical `anomalies' in the guise of singularities and
associated unphysical infinities of smooth fields on a spacetime
modelled after a $\smooth$-continuum. We briefly recapitulate them
below and add a few more physically problematic features of
differential manifolds:

\begin{enumerate}

\item In GR, singularities are due to our {\it a priori}
assumption of modelling spacetime after a pointed 4-dimensional
differential manifold $M$.

\item But then, anyway, there is no well defined notion of
singularities in GR, if anything, because one needs
$\mathrm{Diff}(M)$ to implement the PGC within the confines (and
limitations!) of a smooth spacetime framework (CDG). We may coin
this `{\em the Geroch singularities-versus-differential spacetime
manifold vicious circle effect}' \cite{geroch,rap5}.

\item Similarly, in Quantum Field Theory (QFT), the weaker, but
still troublesome unphysical infinities of the quantum fields of
matter may be ultimately attributed to the fact that Minkowski
space is, after all, a pointed $\smooth$-smooth manifold in which
one can pack infinite degrees of freedom (field modes) in a finite
spacetime volume.

\item In QG, string theory aside, the occurrence of a `natural'
length scale---the so-called Planck length/time
($\ell_{P}=\sqrt{\frac{G\hbar}{c^{3}}}\approx 10^{-33}cm$), and
dually, momentum/energy---could be interpreted as indicating that
the smooth continuum model for `macroscopic' spacetime becomes
inadequate for studying very energetic processes such as those
expected to underlie quantum gravitational effects. In section 6
we will comment more on the potential evasion of $\ell_{P}$ by the
background manifold-free approach to DG that ADG offers, as
regards potential ADG-applications to QG.

\item And finally, due to some robust problems that various smooth
manifold and CDG-based (canonical or covariant) approaches to QG
have, such as the {\em inner product problem} and {\em the problem
of time}, both of which essentially involve
$\mathrm{Diff}(M)$)---the `structure group' of the
$\smooth$-smooth spacetime manifold $M$ \cite{malrap3}---the
current `trend' or `tendencies' in QG research (especially in the
loop QG approach to QGR) is to {\em develop the theory in a
manifestly and genuinely background spacetime manifold independent
way} \cite{ash0,ash1,ash4,ash6,ash5,smolin,thiem2,thiem3}. Again,
we will return to comment more on these issues in section 6.

\end{enumerate}

\noindent {\it In summa}, {\it vis-\`a-vis} the envisioned physics
of the `quantum deep', where quantum theory and GR are united into
a conceptually sound and consistent (and moreover, a
calculationally finite!) QG, theoretical physicists have time and
again questioned and criticized explicitly the smooth spacetime
continuum and, inevitably, the whole edifice and technical panoply
of CDG that goes hand in hand with it.\footnote{See for example
the three quotations (Q?.?)--(Q?.?) by Einstein, Feynman and Isham
in section 7.}

\subsubsection{A brief prehistory of ADG culminating in ADG}

To account briefly about different ideas that preceded the
inception and development of ADG, we follow mainly \cite{mall1}.
Thus, {\it circa} 1970, some of the aforesaid differential
manifold `deficiencies' led to the theory of {\em differential
spaces} \cite{sikorski1,sikorski2,mostow} in which new classes of
smooth functions, generalizing those in $\smooth(M)$ defining a
smooth manifold $M$, were considered. In theoretical physics, this
theory was applied, by also using sheaf-theoretic means, to GR,
singularities, and to some extent in QG research, by the Polish
school of Heller {\it et al.}
\cite{hell0,hell1,hell2,hell3,hell4,hell,hell5}. It is fair to say
that the original impetus along this algebraic line of approach to
GR was given in Geroch's celebrated paper \cite{geroch4}.

In the late 80s--early 90s, and quite independently of the
aforementioned antecedents, the first author conceived of an
entirely algebraic (:sheaf-theoretic), axiomatic scheme for doing
differential geometry without at all the use of any Calculus, or
what amounts to the same, without using at all any base
differential (smooth) manifold. This scheme was subsequently
developed into a full-fledged theory, which was coined ADG or
`{\em Geometry of Vector Sheaves}'---the title of a recent
monograph \cite{mall1,mall2}.

In retrospect, one can argue that the central aim of ADG was
originally {\em to arrive at an entirely algebraic notion of
derivative ({\it viz.} `connection') without any commitment to a
background locally Euclidean space---or what amounts to the same,
without using at all any notion of Calculus} \cite{mall-3,mall-2}.
The basic observation in this direction was that {\em the geometry
of a $\smooth$-smooth manifold $M$ is actually deduced from the
algebraic structure of its structure sheaf $\smooth_{M}$, thus
relegating the local structure of $M$ and the functional character
of $\smooth_{M}$ to a secondary, auxiliary role}. Thus, the usual
CDG assumptions about the local structure of $M$, such as charts,
atlases {\it etc}, are replaced in ADG by assumptions on the
existence of an algebraic derivation on an arbitrary sheaf of
algebras\footnote{Algebras which in turn may be
`non-functional'---{\it ie}, non-function-like ({\it eg},
\cite{malrap1,malrap2,malrap3}), or even if they are functional,
possibly be far from smooth in the standard sense of $\smooth(M)$
({\it eg}, \cite{malros1,malros2,malros3}).} on some arbitrary
base topological space $X$, with the latter sheaf playing a role
analogous to the structure sheaf $\smooth_{M}$ of (germs of)
smooth functions on $M$ in the usual CDG.

The essence of that original intuition of the first author is that
{\em in principle any sheaf of algebras}, provided that they
furnish us with a suitable differential operator $d$ in order to
be able to do differential geometry in the first place, {\em may
be employed as the structure sheaf in ADG}---as it were, to
replace the `classical' one $\smooth_{M}$ defining a differential
manifold $M$ on which CDG is based. This brings us to the two key
notions in ADG.

\subsubsection{$\mathbf{K}$-algebraized spaces and their associated differential
triads: the heart and soul of ADG}

The two primitive, fundamental, {\it ur} as it were notions in ADG
are those of $\mathbf{K}$-{\em algebraized space} and {\em
differential triad}, which we now define.

Let $X$ be an arbitrary topological space, and $\struc_{X}$ a
sheaf of abelian, associative, unital $\mathbb{K}$-algebras
$\mathbb{A}$ ($\mathbb{K}=\R ,\com$)\footnote{In fact, one could
in principle use any number field as linear coefficients for the
elements of $\mathbb{A}$.} on it, called the {\em structure
sheaf}. The pair $(X,\struc_{X})$ is called a $\mathbf{K}$-{\em
algebraized space}, and it may be regarded as the (abstract)
differential geometric analogue of the notion of a {\em ringed
space} in algebraic geometry \cite{harts,shaf}.\footnote{The
reader should note that the constant sheaf $\cons\equiv\mathbf{K}$
of the real or complex numbers is embedded into $\struc_{X}$ ({\it
ie}, $\cons\stackrel{\subset}{\rightarrow}{\struc}_{X}$). It is
tacitly assumed that for every open set $U\subset X$, the algebra
$\struc(U)\equiv\Gamma(U,\struc)$ of continuous local sections of
$\struc_{X}$ is a unital, commutative and associative algebra
$\mathbb{A}$ over $\mathbb{K}$ (in fact, even globally it is
assumed that: $\mathbb{A}\simeq\struc(X)\equiv\Gamma(X,\struc)$!)
\cite{mall1}. Furthermore, for sheaf-cohomological purposes,
$\struc_{X}$ is usually assumed to be {\em fine} and {\em flabby}
\cite{mall1}. Fineness and flabbiness are two properties of
$\struc$ that will prove to be of great import later, in the next
section, when we discuss the application of ADG to spacetime foam
dense singularities \cite{malros1,malros2,mall3,malros3}. In the
sequel, when it is rather clear what the base topological space
$X$ is, we will omit it from $\struc_{X}$ and simply write
$\struc$.}

Now, a {\em differential triad} $\triad$ relative to
$(X,\struc_{X})$ is the triplet

\begin{equation}\label{eqy1}
\triad :=(\struc_{X} ,\partial ,\Omg^{1}(X))
\end{equation}

\noindent consisting of the structure sheaf $\struc_{X}$ involved
in the $\mathbf{K}$-algebraized space $(X,\struc_{X})$ above, a
sheaf $\Omg^{1}$ of (first-order differential)
$\mathbb{A}$-modules $\omg^{1}$ over $X$,\footnote{$\omg^{1}$ is
the ADG-theoretic analogue of the usual sheaves of germs of smooth
$1$-forms over a differential manifold $M$---{\it ie}, when
$\struc\equiv\smooth_{M}$.} and a derivation map $\partial$
defined as the following {\em $\cons\equiv\mathbf{K}$-sheaf
morphism}

\begin{equation}\label{eqy2}
\partial :~\struc\mapto\Omg^{1}
\end{equation}

\noindent which is $\mathbf{K}$-linear and satisfies the Leibniz
(product) rule

\begin{equation}\label{eqy3}
\partial(s\cdot t)=s\cdot\partial(t)+t\cdot\partial(s)
\end{equation}

\noindent for any local sections $s$ and $t$ of $\struc$ ({\it
ie}, $s,t\in\struc(U)$, with $U\subset X$ open).\footnote{One must
emphasize here the intimate link between $\mathbf{K}$-algebraized
spaces and differential triads as expressed by the following
result: ``{\em every $\mathbf{K}$-algebraized space defines a
triad $\triad$}'' \cite{mall1,pap2,pap3}. This is but a
preliminary indication of the purely algebraic (:sheaf-theoretic)
conception of `differentiability' (`smoothness') that ADG
supports, one that is significantly wider, more general and
versatile than the traditional $\smooth$-one associated with
differential manifolds (CDG), which is just a particular case when
one assumes $\struc\equiv\smooth_{X}$ ({\it ie}, $X\equiv M$).
Also, about a conventional-notational `ambiguity': so far in the
ADG-literature one encounters differential triads defined as in
(\ref{eqy1}) above ({\it eg}, see \cite{mall2}), or equivalently,
as triplets $(X,\struc ,\partial)$ ({\it eg}, see
\cite{malros2})---{\it ie}, in the first case one includes in the
definition both the `domain' ($\struc$) and the `range'
($\Omg^{1}$) of $\partial$ (\ref{eqy2}); while in the second,
while regards as important to include the base topological space
$X$ of the corresponding $\mathbf{K}$-algebraized space diad, and
exclude the target sheaf $\Omg^{1}$ of $\partial$. In what follows
we will use both definitions interchangeably.}

With the definition of the operator $\partial$ in $\triad$, we
come to the central definition in ADG: that of an
$\struc$-connection $\conn$. As noted before, the purely algebraic
and background manifold-free notion of connection in ADG is the
key concept in the theory.

\subsubsection{$\struc$-connections and their curvatures}

In ADG, the operator $\partial$ in (\ref{eqy2}) readily
generalizes to an $\struc$-connection $\conn$, as follows: given a
differential triad $\triad=(\struc,\partial ,\Omg^{1})$ as above,
let $\modl$ be an $\struc$-module sheaf on $X$.\footnote{For
$\modl$, regarded as a vector sheaf of rank $n$, one has by
definition the following $\struc|_{U}$-isomorphisms:
$\modl|_{U}=\struc^{n}|_{U}=(\struc|_{U})^{n}$ and, concomitantly,
the following equalities section-wise:
$\modl(U)=\struc^{n}(U)=\struc(U)^{n}$ (with $\struc^{n}$ the
$n$-fold Whitney sum of $\struc$ with itself). One says that
$\modl$ {\em is a locally free $\struc$-module of finite rank
$n$}---an appellation synonymous to {\em vector sheaf} in ADG. We
further assume that for the vector sheaf $\modl$ endowed with the
$\struc$-connection $\conn$ as above, the $\struc$-module sheaf
$\Omg^{1}$ in the given differential triad $\triad$ is its dual
({\it ie}, $\Omg^{1}=\modl^{*}\equiv{\Hom}_{\struc}(\modl
,\struc)$).} Then, one first defines $\conn$ as a map

\begin{equation}\label{eqy4}
\conn:~\modl\mapto\Omg(\modl)\equiv\modl\otimes_{\struc}\Omg\cong
\Omg\otimes_{\struc}\modl
\end{equation}

\noindent which is a $\cons$-linear morphism of the
$\mathbb{K}$-vector sheaves involved, and then one requires that
this map satisfies the following condition

\begin{equation}\label{eqy5}
\conn(\alpha \cdot s)=\alpha\cdot\conn(s)+s\otimes\partial(\alpha)
\end{equation}

\noindent for $\alpha\in\struc(U)$, $s\in\modl(U)$, and $U\subset
X$ open. The connection $\conn$, as defined above, is commonly
known as a {\em Koszul linear connection}.\footnote{For more about
the algebraic $\struc$-connection $\conn$ above, as for example
its local gauge-split form $\conn=d+\aconn$ as well as its affine
local gauge transformation properties, the reader is referred to
\cite{malrap3}, and of course to the exhaustive treatment
\cite{mall1,mall2}. Additionally, in case the reader would like to
learn more about the standard theory of Koszul connections from
the traditional geometrical viewpoint of smooth fiber bundles,
(s)he can refer to \cite{darling}.}

\paragraph{The curvature of $\conn$.} In order to define the curvature
$\curv(\conn)$ of the $\struc$-connection $\conn$, we first need
$\struc$-modules $\Omg^{2}$ of $2$-form-like entities. To this
end, one defines inductively $\struc$-modules of higher-order
`differential forms' by iteration of the completely antisymmetric
homological tensor product functor $\wedge_{\struc}$, as follows:
$\Omg^{0}:=\struc,~\Omg^{1}:=\struc\wedge_{\struc}\Omg
,~\Omg^{2}=\struc\wedge_{\struc}\Omg^{1}\wedge_{\struc}\Omg^{1},\cdots\Omg^{i}\equiv(\Omg^{1})^{i}:=
\wedge^{i}_{\struc}\Omg^{1}$. Then, one introduces the first-order
extension of $\partial\equiv d^{0}$ as a $\mathbf{K}$-linear
exterior differential operator $\kd\equiv d^{1}$ effecting sheaf
morphisms

\begin{equation}\label{eqy6}
\kd:~\Omg^{1}\mapto\Omg^{2}
\end{equation}

\noindent and obeying relative to $\partial$ the following
`cohomological nilpotency' condition\footnote{This nilpotency of
$d^{i}$ ($i\geq 0$) marks the starting point in ADG of abstract
(de Rham) sheaf-cohomology, which in turn relates to the exactness
of the following complex:
$\mathbf{0}(\equiv\mathbf{\Omega^{-2}})\stackrel{\imath\equiv
d^{-2}}{\mapto}\mathbf{C}(\equiv\mathbf{\Omega^{-1}})\stackrel{\epsilon\equiv
d^{-1}}{\mapto}\mathbf{\struc}(\equiv\mathbf{\Omega^{0}})$
$\stackrel{d^{0}\equiv\partial}{\mapto}\mathbf{\Omega^{1}}\stackrel{d^{1}\equiv\kd}{\mapto}
\mathbf{\Omega^{2}}\stackrel{d^{2}}{\mapto}\cdots\mathbf{\Omega^{n}}
\stackrel{d^{n}}{\mapto}\cdots$, with each
$d^{i}:~\Omg^{i}\mapto\Omg^{i+1}$ ($i\geq2$) being defined like
$\partial\equiv d^{0}$ and $\kd\equiv d^{1}$ before, {\it ie}, as
a sheaf morphism between the relevant $\struc$-modules of
higher-order ($i\geq2$) differential forms involved
\cite{mall1,mall2,malrap2}.}

\begin{equation}\label{eqy7}
\kd\circ\partial\equiv d^{1}\circ d^{0}\equiv d^{2}=0
\end{equation}

Then, in complete analogy to the `extension' of the connection
$\partial \equiv d^{0}$ in {\ref{eqy2}) to $\kd\equiv d^{1}$ in
(\ref{eqy6}) above, given a $\struc$-module $\modl$ endowed with
an $\struc$-connection $\conn$, one can define the first
prolongation of $\conn$ to be the following $\mathbf{K}$-linear
vector sheaf morphism

\begin{equation}\label{eqy8}
\conn^{1}:~\Omg^{1}(\modl)\mapto\Omg^{2}(\modl)
\end{equation}

\noindent satisfying section-wise relative to
$\conn(\equiv\conn^{0})$

\begin{equation}\label{eqy9}
\conn^{1}(s\otimes t):=s\otimes\kd t-t\wee\conn s,
\end{equation}

\noindent with $s\in\modl(U),t\in\Omg^{1}(U)$.

We thus come to define the curvature $\curv$ of an
$\struc$-connection $\conn$ by the following commutative diagram

\begin{equation}\label{eqy10}
\Vtriangle[\modl`\Omg^{1}(\modl)\equiv\modl\otimes_{\struc}\Omg^{1}`\Omg^{2}(\modl)\equiv\modl\otimes_{\struc}
\Omg^{2};\conn^{0}`\curv\equiv\conn^{1}\circ\conn^{0}`\conn^{1}]
\end{equation}

\noindent from which one reads directly that

\begin{equation}\label{eqy11}
\curv\equiv \curv(\conn):=\conn^{1}\circ\conn
\end{equation}

\noindent and from which it follows that
$\curv(\partial)=d^{2}=0$---{\it ie}, that $\partial$ {\em is a
flat $\struc$-connection}.\footnote{Again, for more about
$\curv(\conn)$ on a vector sheaf $\modl$ of rank $n$, as for
example its local gauge character as a $0$-cocycle of local
$n\times n$ matrices having for entries local sections of
$\Omg^{2}$, as well its covariant local gauge transformation
properties, the reader is referred to \cite{malrap3}, and of
course to the exhaustive treatment \cite{mall1,mall2}.}

For the sake of our exposition here, the property of the curvature
$\curv$ of an $\struc$-connection $\conn$ that we would like to
highlight, in contradistinction to the connection $\conn$ itself,
is its {\em geometrical character}, which can be expressed in a
concise manner ADG-theoretically as follows:

\bigskip \noindent (R4.3)\hskip 0.9in
\begin{minipage}{11cm}
\noindent {\em while $\curv$ is an $\struc$-morphism, $\conn$ is
only a constant sheaf $\mathbf{K}$-morphism} \cite{malrap3}.
\end{minipage}

\vskip 0.1in

In turn, (R2.?) above has the following `algebraico-geometrical
implications' first noted in \cite{malrap3}:\footnote{These
implications will be of great import in 3.2.2 and 3.2.3 where we
discuss the novel ADG-perspective on the PGC of GR as well as on
(gravitational) kinematics and dynamics. Moreover, they will play
a significant role when we interpret quantally (quantum
theoretically) the inherently algebraic formalism of ADG in
section 6.}

\bigskip \noindent (R4.4)\hskip 0.9in
\begin{minipage}{11cm}
\noindent {\em Since $\curv$ is an $\struc$-morphism, our `local
measurements', our gauge acts of coordinatization or localization
(of the fields involved), which are organized into the
`geometry-encoding algebraic apparatus' $\struc$ of ADG---the
sheaf of generalized coefficients or `arithmetics' in the theory,
respect it. Thus, $\curv$ is a geometrical object ({\it ie}, an
$\struc$-tensor) in our theory, while $\conn$, which is respected
only by the constant sheaf $\mathbf{K}$ but not by our (local)
measurements in $\struc$, is not a geometrical object and it
eludes our acts of trying to measure (localize) it. Rather,
$\conn$ is an algebraic entity not captured by our
`geometrical/measuring apparatus' (`space') which is effectively
encoded in $\struc$ (Gel'fand duality). This is reflected in the
affine or inhomogeneous (local gauge non-covariant) and the
homogeneous (local gauge covariant) transformation laws of $\conn$
and $\curv(\conn)$ under a local change of gauge (coordinates),
respectively} \cite{mall1,mall2,malrap3}.

\end{minipage}

\subsubsection{Manifold-free algebraic (pseudo)-Riemannian structures in ADG}

In complete analogy to the smooth (pseudo-)Riemannian structures
on the differential spacetime manifold of GR, one can define
ADG-theoretically, in an entirely algebraic (:sheaf-theoretic) and
manifestly base manifold-free way, the following structures:

\paragraph{The $\struc$-metric $\rho$.} By an $\struc$-valued
(pseudo)-Riemannian inner product $\rho$ on a vector sheaf $\modl$
we mean a {\em sheaf} morphism

\begin{equation}\label{eqy12}
\rho :~\modl\oplus\modl\mapto\struc
\end{equation}

\noindent which is {\em $\struc$-bilinear} between the
$\struc$-modules concerned, {\em symmetric indefinite} ({\it ie},
$\rho(s,t)=\rho(t,s),~s,t\in\modl(U)$ and of indefinite
signature), as well as iii) {\em strongly
non-degenerate}.\footnote{For more description of these three
defining properties of $\rho$, see \cite{malrap3}.} For any two
local sections $s$ and $t$ of $\modl$ in $\modl(U)$, $\rho$ is
implemented via the canonical isomorphism

\begin{equation}\label{eqy13}
\modl\stackrel{\tilde{\rho}}{\cong}\modl^{*}
\end{equation}

\noindent between $\modl$ and its dual $\modl^{*}$,\footnote{As
noted earlier, we assume that for the vector sheaf $\modl$ endowed
with the $\struc$-connection $\conn$, the sheaf $\Omg$ of
differential $1$-form-like entities in the given differential
triad $\triad$ is the dual of $\modl$ appearing in
(\ref{eqy13})---{\it ie},
$\Omg=\modl^{*}\equiv{\Hom}_{\struc}(\modl ,\struc)$.} in the
following way

\begin{equation}\label{eqy14}
\tilde{\rho}(s)(t):=\rho(s,t)
\end{equation}

\noindent with (\ref{eqy13}) being true up to an
$\struc$-isomorphism \cite{malrap3}.

\paragraph{Christoffel-Riemannian metric $\struc$-connections.} In complete analogy with the usual Christoffel theory,
we can define a {\em linear connection} $\nabla$ as follows

\begin{equation}\label{eqy15}
\nabla :~\modl\times\modl\mapto\modl
\end{equation}

\noindent acting on $\modl(U)$ as

\begin{equation}\label{eqy16}
\nabla(s,t)\equiv\nabla_{s}(t):=\conn(t)(s)
\end{equation}

Now, one says that {\em $\conn$ is a pseudo-Riemannian
$\struc$-connection} or that it is compatible with the indefinite
metric $g$ of the inner product $\rho$ in (\ref{eqy12}), whenever
it fulfills the following two conditions:

\begin{itemize}

\item {\em Riemannian symmetry}: $\nabla(s,t)-\nabla(t,s)=[s,t]$;
for $s,t\in\modl(U)$ and $[\, .\, ,\, .\,  ]$ the usual Lie
bracket (product).

\item {\em Ricci identity}:
$\partial(\rho(s,t))(u)=\rho(\nabla(u,s),t)+\rho(s,\nabla(u,t))$;
for $s,t,u\in\modl(U)$, as usual.

\end{itemize}

In particular, for a Lorentzian $\rho$ and its associated
$g$,\footnote{With respect to a {\em local (coordinate) gauge}
$e^{U}\equiv\{ U;~(e_{i})_{0\leq i\leq n-1}\}$ of the vector sheaf
$\modl$ of rank $n$,
$\rho(e_{i},e_{j})=g_{ij}=\mathrm{diag}(-1,+1,\cdots)$
\cite{mall1,mall2}.} an $\struc$-connection $\conn$ is said to be
compatible with the Lorentz $\struc$-inner product $\rho$ on
$\modl$\footnote{Such a metric connection is commonly known as
{\em Levi-Civita connection}.} when its associated Christoffel
$\nabla$ in (\ref{eqy15}) satisfies

\begin{equation}\label{eqy17}
\nabla\rho=0
\end{equation}

\noindent which, in turn, is equivalent to the following `{\em
horizontality}' condition for the canonical isomorphism
$\tilde{\rho}$ in (\ref{eqy13}) relative to the {\em connection}
$\conn_{\modl\otimes_{\struc}\modl^{*}}$ in the tensor product
vector sheaf $\Hom_{\struc}(\modl
,\modl^{*})=(\modl\otimes_{\struc}\modl)^{*}=\modl^{*}\otimes_{\struc}\modl^{*}$
induced by the $\struc$-connection $\conn$ on $\modl$

\begin{equation}\label{eqy18}
\conn_{\Hom_{\struc}(\modl ,\modl^{*})}(\tilde{\rho})=0
\end{equation}

\noindent It is worth reminding the reader who is familiar with
the usual theory that (\ref{eqy18}) above implies that the
Levi-Civita $\struc$-connection $\conn$ induced by the Lorentz
$\struc$-metric $\rho$ is {\em torsion-free} \cite{mall3}.

The manifold-free algebraic (pseudo)-Riemannian structures above
have enabled us to formulate the vacuum Einstein equations for
gravity as {\em differential equations proper}, but in the
prominent absence of a background differential spacetime
continuum. We briefly recapitulate this ADG-based formulation
next.

\subsubsection{ADG-vacuum Einstein
gravity without the base spacetime manifold}

The Calculus-free ADG-machinery has been applied to writing the
Einstein equations for the `pure' ({\it ie}, vacuum) gravitational
field over space(time)s that differ dramatically from the
classical, featureless locally Euclidean spacetime manifold, and
ranging from the `ultra-smooth' (and `ultra-singular'!)
\cite{malros1,malros2,mall3,malros3},\footnote{With \cite{mall3}
the key reference. We will encounter this `ultra-singular' case in
the next section and, regarding the evasion of the inner
Schwarzschild singularity, also in 5.2.3.} to the `discrete'
(`discontinuous') \cite{malrap1,malrap2,malrap3}.\footnote{With
\cite{malrap3} the key reference. Again, concerning the evasion of
the interior Schwarzschild singularity, we will encounter this
`discrete', finitary case in 5.2.2.}

However, in order to actually write down the ADG-theoretic version
of the vacuum Einstein equations, together with the
pseudo-Riemannian notions above we also need the {\em Ricci
curvature tensor} and its {\em scalar} contraction---structures
that are readily available in ADG again in a manifestly
manifoldless guise \cite{mall1,mall2,malrap3}.

Thus, given a (real) Lorentzian vector sheaf $\modll :=(\modl
,\rho)$ of rank $n$ equipped with a non-flat $\struc$-connection
$\conn$, one can define the following {\em Ricci curvature
operator} $\ric$ relative to a `local gauge' $U$ of $\modl$

\begin{equation}\label{eqy21}
\ric(\, .\, ,\,\! s)t\in (\modl nd\modl)(U)=M_{n}(\struc(U))
\end{equation}

\noindent for local sections $s$ and $t$ of $\modl$ in
$\modl(U)=\struc^{n}(U)=\struc(U)^{n}$. $\ric$ is an $\modl
nd\modl$-valued operator, thus called a `{\em curvature
endomorphism}'.

Since $\ric$ is matrix-valued with (local) matrix-entries in
$\struc(U)$, we can take its trace, thus define the following {\em
Ricci scalar curvature operator} $\ricci$

\begin{equation}\label{eqy22}
\ricci(s,t):=tr(\ric(\, .\, ,\,\! s)t)
\end{equation}

\noindent which is $\struc(U)$-valued.

With the ADG-version of the Ricci scalar in hand, we read from
\cite{mall3,malrap3} that the ADG-based {\em vacuum Einstein
equations} for Lorentzian gravity are:

\begin{equation}\label{eqy23}
\ricci(\modl)=0
\end{equation}

\noindent for a given choice of $\mathbf{K}$-algebraized space and
its associated differential triad---or essentially, for a given
choice of structure sheaf $\struc$ of generalized coefficients or
coordinates.\footnote{And it is precisely this {\em freedom to
choose} $\struc$---generalized coefficients or `coordinates' quite
different from the classical $\smooth$-smooth ones in
$\struc\equiv\smooth_{M}$---that ADG allows us, which qualifies
the title of the present sub-subsection `{\em base spacetime
manifoldless vacuum Einstein gravity}'.}

It is also important to mention here from \cite{mall3,malrap3}
that {\em the equations} (\ref{eqy23}) {\em can be obtained from
the variation of an $\struc$-valued Einstein-Hilbert action
functional} $\eh$ {\em on the affine space}
$\sconn_{\struc}(\modl)$ {\em of Lorentzian metric
$\struc$-connections on $\modl$, which are the sole dynamical
variables in the ADG-theoretic perspective on
gravity}.\footnote{This last remark is made in order to prepare
the reader for the fact, to be further corroborated in the sequel,
that {\em in ADG}---in contradistinction to the original CDG based
GR where the basic dynamical variable is supposed to be the smooth
spacetime metric---{\em the only gravitational dynamical variable
is the connection}, not the metric (see 3.3 next). That is to say,
{\em according to ADG, gravity is another gauge theory}, but
unlike the other spacetime manifold (and CDG) based gauge theories
of matter (electromagnetism, Yang-Mills theories), {\em a
background spacetime manifoldless one}.}

\subsubsection{The categorical imperative: the category $\ctriad$ of
differential triads, its properties and versatility compared to
the category $\man$ of smooth manifolds}

In the foregoing sections we have mentioned numerous times that
ADG is an {\em algebraico-categorical} scheme for doing
differential geometry. Its {\em algebraic} (:sheaf-theoretic)
character has already been amply manifested, thus here we reveal
in some detail its purely {\em category-theoretic}
features.\footnote{Of course, the categorically minded reader has
perhaps already noticed terms like {\em sheaf morphism} ({\it eg},
$\conn$ and $R(\conn)$) and {\em homological functor} ({\it eg},
sheafification and tensor product functor), which already indicate
that we have implicitly been working all along in a {\em category
of sheaves} of some kind.} We itemize this presentation into four
paragraphs, {\bf i--iv}, briefly borrowing the basic concepts,
constructions and results from \cite{pap4} ({\bf i}), \cite{pap3}
({\bf ii}), \cite{pap1} ({\bf iii}) and \cite{pap2} ({\bf iv}),
respectively.\footnote{The reader should refer to the reference
list at the back and note that the order in which the categorical
features of ADG are presented below does not accord with the
chronological order in which these were originally studied by
Papatriantafillou.} The last paragraph ({\bf iv}) in particular
will prove to be of great import later when we `resolve'
ADG-theoretically the interior Schwarzschild singularity by
finitistic-algebraic means---a culmination of our finitary
application of ADG to classical and quantum gravity in the past
trilogy \cite{malrap1,malrap2,malrap3}.

\paragraph{i. Morphisms of differential triads: generalized (abstract) differentiable maps.} The
first observation in the endeavor to explore and study categorical
features of ADG is that one could regard differential triads as
objects in some category---{\em the category of differential
triads}, which we will denote by $\ctriad$. This aim  could at
least be motivated on the following grounds: since a differential
triad purports to abstract and generalize (the differential
structure of) a differential manifold, and since the latter is an
object in the category $\man$ of (finite-dimensional) smooth
manifolds, having for arrows the usual arbitrarily differentiable
(smooth) maps between them, in the same way differential triads
could be viewed as objects in a category. Thus, the issue arises
of what are the arrows in $\ctriad$, a question that brings us to
the definition of {\em morphisms of differential triads} which we
borrow from \cite{pap4}.

Let $X$ and $Y$ be topological spaces, which are base spaces of
the $\mathbf{K}$-algebraized spaces $(X,\struc_{X})$ and
$(Y,\struc_{Y})$, respectively. In addition, let
$\triad_{X}=(\struc_{X},\partial_{X},\Omg_{X})$ and
$\triad_{Y}=(\struc_{Y},\partial_{Y},\Omg_{Y})$ be differential
triads over them. Then, a {\em morphism} $\morph$ between
$\triad_{X}$ and $\triad_{Y}$ is a triplet of maps
$\morph=(f,f_{A},f_{\Omg})$, such that:

\begin{enumerate}

\item the map $f:~X\mapto Y$ is {\em continuous};

\item the map $f_{\struc}:~\struc_{Y}\mapto f_{*}(\struc_{X})$ is
a {\em morphism of sheaves of} $\mathbb{K}${\em -algebras over}
$Y$ preserving the respective algebras' unit elements ({\it ie},
$f_{\struc}(1)=1$);\footnote{In the expression for $f_{\struc}$
above, $f_{*}$ is the {\em push-out} along the continuous $f$, a
map which carries each element of a differential triad into a like
element in the sense that, for any triad $\triad$,
$f_{*}(\triad):=(f_{*}(\struc),f_{*}(\partial),f_{*}(\Omg))$ is
also a differential triad---the one `induced' by $f$
\cite{pap4,pap3}; whence, term-wise for our triads $\triad_{X}$
and $\triad_{Y}$ above (and omitting the base topological space
subscripts):
$f_{*}(\struc):=(f_{*}(\struc)(U):=\struc_{f^{-1}(U)}),~(U\subseteq
Y~\mathrm{open})$ is a sheaf of unital, abelian, associative
$\mathbb{K}$-algebras over $Y$,
$f_{*}(\Omg):=(f_{*}(\Omg)(U):=\Omg_{f^{-1}(U)}),~(U\subseteq
Y~\mathrm{open})$ a sheaf of $f_{*}(\struc)$-modules (of 1st-order
differential form-like entities), and
$f_{*}(\partial):=(f_{*}(\partial)(U):=\partial_{f^{-1}(U)}),~(U\subseteq
Y~\mathrm{open})$ an induced $\mathbf{K}$-linear, Leibnizian sheaf
morphism \cite{pap4}.} and the following categorical diagram is
obeyed:

\begin{equation}\label{eqy19}
\bfig
\putsquare<1`-1`-1`1;700`700>(700,700)[X`Y`\struc_{X}`\struc_{Y};f```]
\putbtriangle<0`-1`0;700>(1400,700)[``~~~f_{*}(\struc_{X});`f_{\struc}`]
\putmorphism(1400,700)(1,0)[\phantom{TT'T'}`%
\phantom{TT'}`]{700}1a \efig
\end{equation}

\item the map $f_{\Omg}:~\Omg_{Y}\mapto f_{*}(\Omg_{X})$, as noted
in the last footnote, is a {\em morphism of sheaves of}
$\mathbb{K}${\em -vector spaces over} $Y$, with
$f_{\Omg}(\alpha\omega)=f_{\struc}(\alpha)f_{\Omg}(\omega),~\forall
\alpha\in\struc_{Y},~\omega\in\Omg_{Y}$; and finally,

\item with respect to the $\cons\equiv\mathbf{K}$-sheaf morphism ({\it viz.} flat
connection) $\partial$ in the respective triads, and as it has
also been alluded to in the last footnote, the following diagram
is commutative:

\begin{equation}\label{eqy20}
\bfig
\putsquare<1`1`1`1;700`700>(700,700)[\struc_{Y}`\Omg_{Y}`f_{*}(\struc_{X})`f_{*}(\Omg_{X});
\partial_{Y}`f_{\struc}`f_{\Omg}`f_{*}(\partial_{X})]
\efig
\end{equation}

\noindent which is readily read as $f_{\Omg}\circ
\partial_{Y}=f_{*}(\partial_{X})\circ f_{\struc}$.

\end{enumerate}

\noindent Thus, the category $\ctriad$ has $\triad$s as objects
and $\morph$s between them as arrows. Furthermore, we saw above in
the definition of a morphism $\morph$ of differential triads what
an important role a continuous map $f$ between their respective
base spaces plays via its push-out $f_{*}$. In fact, it has been
shown that the (essentially algebraic) differential (geometric)
mechanism encoded into a differential triad can be transferred
forward and backwards by any continuous map; moreover, the final
and initial structures obtained by such push-forward and pull-back
actions induced by $f$ enjoy certain {\em universal properties}
that qualify $f$ to a {\em differentiable map} \cite{pap3}. This
result opens up new possibilities of developing what one might
coin `{\em the differential geometry of topological spaces}'.
Below, we discuss briefly this `universality' of differential
triad morphisms and its potent consequences, especially in the
application of ADG to our finitary, causal and quantal approach to
Lorentzian gravity, and more particularly, in the evasion of the
inner Schwarzschild singularity by finitistic-algebraic means as
it will be entertained in 5.2.2 later on.

\paragraph{ii. Towards a `differential geometry of topological spaces'; Calculus
`reversed': continuity `implies' differentiability.} We begin with
some general observations. We noted earlier that the notion of
{\em sheaf} is essentially {\em topological}, while that of `{\em
differentiability}' essentially {\em topologico-algebraic} (R3.1),
and {\em local} (in the topological space involved). Thus, it is
not surprising that ADG manages to capture so many (if not all!)
the features of the usual Calculus on smooth manifolds by using
{\em sheaves of algebras}---as it were, the harmonious combination
and coherent intertwining of the local-topological character of
sheaves with the algebraic structure of the (functional) objects
localized sheaf-theoretically (in $\struc$) results in the
definition of a {\em differential} operator $d\equiv\partial$
({\it viz.} connection), which is {\em the} basic object-tool with
which one can actually do differential geometry. Furthermore, and
perhaps more importantly, all this is achieved in the prominent
absence of a base smooth manifold when, as a matter of fact ({\it
ie}, of fundamental assumption in ADG), all that is needed is an
in principle arbitrary topological space (to serve as a `{\em
surrogate localization-scaffolding}' for the said algebras), a
background stage which plays no role whatsoever in the said
inherently algebraic differential geometric mechanism of ADG. In
continuation of (R3.1), we may distill all this to the following
general, but fundamental, {\it motto} of ADG:

\bigskip \noindent (R4.5)\hskip 0.9in
\begin{minipage}{11cm}
\noindent {\em topological + }(sheaf-theoretically localized on
it) {\em algebraic structure = differential} (geometric) {\em
structure} (without any $\smooth$-smoothness $\equiv$ base
differential manifold $M\equiv\smooth(M)$).
\end{minipage}

\vskip 0.1in

\noindent In what follows, we simply corroborate further (R2.?)
above by following mainly \cite{pap3}.

The main result in \cite{pap3}, as summarized in its abstract, is
that

\bigskip \noindent (Q4.1)\hskip 0.9in
\begin{minipage}{11cm}
\noindent ``{\small the differential mechanism induced by a
differential triad is transferred backwards and forward by any
continuous map $f$. The initial and final structures thus obtained
satisfy appropriate universal conditions that turn the continuous
map $f$ into a differentiable map.}''
\end{minipage}

\vskip 0.1in

In a nutshell, given a continuous $f:~X\mapto Y$, with $X$ the
base space of a differential triad $\triad_{X}$, it has been shown
that $f$ pushes forward the (essentially algebraic) differential
mechanism of $\triad_{X}$, so that a new and unique differential
triad---one that satisfies a {\em universal mapping}
condition---is defined on $Y$, and thus $f$ becomes
differentiable. The relevant theorem,\footnote{Theorem {\bf 3.1}
in \cite{pap3}.} which uses some ideas already expressed in the
last footnote and paragraph ({\bf i.}), can be stated as
follows:\footnote{For the corresponding proof, the reader is
referred to \cite{pap3}.}

\bigskip \noindent (R4.6)\hskip 0.9in
\begin{minipage}{11cm}
\noindent {\small Let
$\triad_{X}=(\struc_{X},\partial_{X},\Omg_{X})~\in\ctriad_{X}$,\footnote{Plainly,
$\ctriad_{X}$ is the sub-category of $\triad$ consisting of all
differential triads and triad morphisms with common base
topological space $X$.} and $f:~X\mapto Y$ continuous. If $Y$
inherits
$f_{*}(\triad_{X}):=(f_{*}(\struc_{X}),f_{*}(\partial_{X}),f_{*}(\Omg_{X}))$
from the push-out $f_{*}$ of $f$, then there is a morphism of
differential triads
$\morph=(f,f_{\struc},f_{\Omg}):~\triad_{X}\mapto
f_{*}(\triad_{X})~ (\in\ctriad)$---{\it ie}, $f$ becomes {\em
differentiable}. Furthermore, the pushed-forward triad
$f_{*}(\triad_{X})$ satisfies the following universal
(composition) property: given a triad
$\triad_{Y}=(\struc_{Y},\partial_{Y},\Omg_{Y})\in\ctriad_{Y}$, as
well as a morphism
$\tilde{\morph}:=(f,\tilde{f}_{\struc},\tilde{f}_{\Omg}):~\triad_{X}\mapto\triad_{Y}$,
there is a unique morphism
$(id_{Y},g_{\struc},g_{\Omg}):~f_{*}(\triad_{X})\mapto\triad_{Y}$
such that: \vskip 0.1in
\centerline{$\tilde{\morph}=(id_{Y},g_{\struc},g_{\Omg})\circ\morph$}}
\end{minipage}

\vskip 0.1in

\noindent Accordingly, the `dual' (converse) scenario involving
$f$'s {\em pull-back action} $f^{*}$, when now the range of $f$ is
a differential triad $\triad_{Y}$ on $Y$ while $X$ ($f$'s domain)
is merely a topological space not being endowed {\it a priori}
with a differential (triad) structure, $f^{*}$ too is seen to
transfer (induce) backwards the differential mechanism encoded in
$\triad_{Y}$ on $X$, thus rendering it a {\em differential} (not
just a topological) space and in the process promoting $f$ to a
{\em differentiable} (not just a continuous) map
\cite{pap3}.\footnote{Although, as noted in \cite{pap3}, the proof
of this `dual', pull-back theorem is slightly more involved than
the push-out one.}

\paragraph{The `aftermath' of the maths.} The principal value of these push-forward and pull-back results
lies with what one might call `{\em induced differential
structures from topological structures}', or perhaps more
strikingly, `{\em the differential geometry of topological spaces
via a Calculus-reversal}'. Let us explain this issue a bit more.
We saw earlier when we discussed the tower of structures imposed
on (pseudo-)Riemannian manifolds (\ref{eq7}) that, from the
vantage of CDG, the topological (`continuous') structure on a
manifold is usually regarded as being more basic (deeper) than its
differential (`smooth') one, something that is already reflected
in the so-called {\em Fundamental Theorem of Calculus} that we
learn as undergraduates, namely, that

\vskip 0.1in

\centerline{\em `differentiability implies continuity'}

\vskip 0.1in

\noindent or, functionally speaking, that {\em a differentiable
function is continuous}.\footnote{In footnote ?? we formally cast
this `power relationship' between $C^{i}-$ $(i=1\ldots\infty)$ and
$\cont$-functions on $M$ as $C^{i}>\cont$.} Here, however, we
encounter exactly the reverse phenomenon, namely that,

\bigskip \noindent (R4.7)\hskip 0.9in
\begin{minipage}{11cm}
\noindent as long as one has sheaf-theoretically localized an
algebraic structure ($\struc$) on a topological space ($X$) so
that one is able to define a differential triad ($\triad_{X}$) on
the latter {\it \`a la} ADG, then a continuous map ($f$) from the
said topological space ($X$) to another topological space ($Y$)
endows the latter with a differential structure ($f_{*}$-induced
differential triad on $Y$), and in the process the map ($f$)
becomes `smooth' (differentiable). In a nutshell, and in a strong
sense, {\em in ADG, continuity implies differentiability}---{\it
ergo}, the aforementioned `Calculus-reversal'.
\end{minipage}

\vskip 0.1in

These results indicate the `superiority' and `versatility' of
$\ctriad$ (and {\it in extenso} of ADG) relative to $\man$ (and
{\it in extenso} of CDG), since, to begin with,\footnote{Shortly
we will give more `reasons' about the superiority of $\ctriad$
compared to $\man$---in effect, how ADG is able to resolve all the
deficiencies and shortcomings of the smooth manifold based CDG
mentioned earlier in 3.1.1.} the latter does not allow for such
{\em initial} (pulled back) and {\em final} (pushed forward) {\em
differential structures} in the sense that, and we quote {\it
verbatim} Papatriantafillou from the introduction of \cite{pap3},

\bigskip \noindent (Q4.2)\hskip 0.9in
\begin{minipage}{11cm}
\noindent ``{\small if $(X,{\mathcal{A}})$ is a smooth manifold,
$Y$ a topological space (resp. $X$ is a topological space and
$(Y,{\mathcal{B}})$ a smooth manifold) and $f:~X\mapto Y$ is
continuous, one cannot `push-out' the atlas ${\mathcal{A}}$ (resp.
pull-back the atlas ${\mathcal{B}}$), in order to make $Y$ (resp.
$X$) a smooth manifold and $f$ a differentiable
map.}''\footnote{This virtue of ADG compensates for item number 4
in the list of mathematical `deficiencies' of $\man$ and CDG given
in 3.1.1.}
\end{minipage}

\vskip 0.1in

\noindent Moreover, as also noted in \cite{pap3} as a corollary of
the theorem (R3.?) above,\footnote{The corollary below is written
in {\em emphasis}-typeface, as in the original paper \cite{pap3}.}

\bigskip \noindent (Q4.3)\hskip 0.9in
\begin{minipage}{11cm}
\noindent ``{\small\em Let $(\struc ,\partial, \Omg)$ be a
differential triad over $X$, and let $X$ be endowed with an {\rm
[in principle arbitrary]}\footnote{Our addition.} equivalence
relation $\sim$. Then the quotient space $\tilde{X}:=X/ \sim$ is
provided with a differential triad, so that the canonical {\rm
[projection]}\footnote{Our addition.} map $\pi :~X\mapto\tilde{X}$
is differentiable.}''
\end{minipage}

\vskip 0.1in

\noindent As the author of \cite{pap3} observes in connection with
this corollary, while the quotient space of a manifold $X$ by an
in principle arbitrary equivalence relation is not a manifold,
thanks to the results above, in $\ctriad$ the `quotient-fold'
$\tilde{X}$ itself acquires a differential structure
(triad).\footnote{This virtue of ADG resolves item number 2 in the
list of mathematical shortcomings of $\man$ and CDG in 3.1.1 and
it will be seen to figure prominently in the ADG-theoretic
finitary-algebraic `resolution' of the inner Schwarzschild
singularity in 5.2.2 \cite{rap5}.} As a characteristic example,
Papatriantafillou gives the differential triad that the orbifold
($G$-fold) or moduli space $X/G$ inherits from $X$ when a
topological group $G$ acts continuously on the
latter.\footnote{This example is very important {\it vis-\`a-vis}
our connection based ADG-theoretic approach to gravity since, as
noted earlier in this paper and extensively in \cite{malrap3}, the
relevant gravitational configuration space in our theoresis is the
affine space $\sconn_{\struc}(\modl)$ of $\struc$-connections on
$\modl$; moreover, since the PGC of GR is implemented in ADG not
via $\mathrm{Diff}(M)$ as in the usual $M$-based theory (GR), but
via the group (sheaf) $\aut(\modl)$ of automorphisms of the
gravitational field $(\modl ,\conn)$, the `effective' physical
configuration space is the orbifold
$\sconn_{\struc}(\modl)/\aut(\modl)$. Now, in the differential
manifold based approaches to canonical QGR such as loop QG, it has
become imperative to develop differential geometric ideas on the
moduli space of Ashtekar's spin-Lorentzian connections modulo
diffeomorphisms in $\mathrm{Diff}(M)$ \cite{ashlew2}. Analogously
here, ADG has been significantly developed on
$\sconn_{\struc}(\modl)$ \cite{mall1,mall2,malrap3,mall4};
therefore, Papatriantafillou's example above gives one a fleeting
glimpse at the fact that one can straightforwardly transfer the
(abstract) differential structure from $\sconn_{\struc}(\modl)$
onto the $G$-fold $\sconn_{\struc}(\modl)/\aut(\modl)$, thus be
able to further develop differential geometric ideas on the latter
\cite{mall1,mall2,malrap3,mall4}---and what's more, unlike in the
$M$-based loop QG approach, {\em all this is accomplished in the
manifest absence of a background differential spacetime
manifold}.}

\paragraph{iii. The category of differential triads: further properties.} Now that we
have discussed above some of the advantages and virtues of
(working with) $\ctriad$ relative to $\man$, in this paragraph and
the next we want to give briefly without proofs some more
properties of $\ctriad$ that $\man$ simply lacks---properties,
such as categorical bicompleteness ({\bf iv}), that not only will
prove to be of great import in our finitary-algebraic resolution
of the inner Schwarzschild singularity in the next section, but
also ones that could be used in the future to promote $\ctriad$
into an {\em elementary topos} (Lawvere-Tierney)
\cite{gold,macmo}---a structure which promises to enjoy great
applications, both mathematical and physical alike, as anticipated
in \cite{rap3,rap4,rap6} and is currently under intense
development \cite{rap7}, especially in exploring the (quantum)
logical underpinnings of our `finitary', ADG-theoretic approach to
Lorentzian QG \cite{malrap1,malrap2,malrap3}.

The first of those properties pertains to the existence of {\em
canonical subobjects} in $\ctriad$. Following \cite{pap1}, we note
that $\man$ simply lacks (canonical) subobjects, since an
arbitrary subset of a differential manifold is not a
manifold.\footnote{This pertains to item number 1 in the list of
mathematical deficiencies of $\man$ and CDG mentioned in 3.1.1.}
Again, just quoting Papatriantafillou from the abstract of
\cite{pap1},\footnote{Again, for proofs of all statements made
below, the reader is referred to the relevant papers cited.}

\bigskip \noindent (Q4.4)\hskip 0.9in
\begin{minipage}{11cm}
\noindent ``{\small We prove that every subset of the base space
of a differential triad determines a differential triad, which is
a subobject of the former, a property missing from the category of
manifolds.}''\footnote{It must be noted here that in the
aforementioned project to promote $\ctriad$ to a topos, the result
above ({\it ie}, that $\ctriad$ has subobjects) might prompt one
to look for the {\em subobject classifier} structure
$\mathbf{\Omega}$ that would define $\ctriad$ as a topos proper
\cite{gold,macmo}. Preliminary investigations \cite{rap7} seem to
indicate that in the same way that the `quintessential' paradigm
of a topos---the topos $\mathbf{\mathcal{S}h}_{X}$ of sheaves (of
structureless, `variable') sets (varying) over a topological space
$X$ ({\it eg}, a locale)---has a non-Boolean, Heyting algebra-type
of `generalized truth values object' $\mathbf{\Omega}$ (:a synonym
for the subobject classifier) and hence an intuitionistic internal
language (logic) \cite{lambek,macmo}, so $\ctriad$ too has a
similar subobject classifier and, as a result, a non-classical
(non-Boolean) kind of internal logic. This is glaring contrast to
the classical, Boolean logic `inherent' in the topos
$\mathbf{Set}$ of `constant' sets on which arguably the $M$-based
GR rests (after all, $M$ is a `classical' point-set), and whose
subobject classifier is the Boolean binary alternative
$\mathbf{\Omega}=\mathbf{2}=\{\top ,\bot\}$ of truth values (hence
the denomination `classical' given to $M$). In such a project the
main aim is to do all the GR constructions {\em internally},
within the topos $\ctriad$. But for more on this project, as well
as for its physical implications, the reader should wait for
\cite{rap7}.}
\end{minipage}

\vskip 0.1in

\noindent Furthermore, Papatriantafillou shows that $\ctriad$ has
finite products (and dually, co-products).

\paragraph{iv. Bicompleteness of $\ctriad$.} This paragraph is
about another property of $\ctriad$ that $\man$ does not possess,
namely, that it is {\em bicomplete}---{\it ie}, closed under {\em
projective} and {\em inductive limits}.\footnote{Projective limits
are also known as {\em inverse} or {\em categorical limits}, while
synonyms for inductive limits are {\em direct} or {\em categorical
colimits}. The notion of projective limit is categorically dual to
that of inductive limit, hence the prefix `{\em co-}' in front of
the latter.} That is to say, in $\ctriad$, projective (inverse)
systems of differential triads have differential triads as
projective (inverse) limits. The same holds dually for inductive
(direct) systems.\footnote{This pertains to item 3 in the list of
mathematical deficiencies of $\man$ and CDG mentioned in 3.1.1.}
As we did above, we quote directly the relevant result from the
abstract of \cite{pap2} (omitting the corresponding proofs), and
also from its introduction we remark the associated deficiency of
$\man$:

\bigskip \noindent (Q4.5)\hskip 0.9in
\begin{minipage}{11cm}
\noindent ``{\small We prove that in the category of differential
triads projective and inductive systems have limits\footnote{From
the abstract of \cite{pap2}.}... [That is,]\footnote{Our addition
for continuity of the text.} we prove that
$\mathcal{D}\mathcal{T}$\footnote{Papatriantafillou symbolizes the
category of differential triads by $\mathcal{D}\mathcal{T}$, we by
$\ctriad$.} is closed for the projective and inductive limits of
differential triads. To this end, we prove that a projective
(resp. inductive) system of differential triads over the same base
space has a limit in
$\mathcal{D}\mathcal{T}$.\footnote{Propositions 3.2 and 3.3 in
\cite{pap2}.} Then, for a projective (resp. inductive) system of
differential triads over a projective system of base
spaces,\footnote{This is precisely the result we will use in the
sequel (5.2.2) to `resolve' finitarily {\it \`a la} Sorkin
\cite{sork0} and by algebraic ADG-theoretic means the interior
Schwarzschild singularity \cite{rap5}.} we construct a new
projective (resp. inductive) system of differential triads over
the projective limit of the projective limit of the base spaces,
and we prove that the limit satisfies the universal property of
the projective (resp. inductive) limit for the initially given
family\footnote{Theorems 4.4 and 4.5 in \cite{pap2}.}...[These
results must be contrasted against the situation in $\man$
where]\footnote{Our addition for continuity of the text.} the
limit of a projective system of manifolds is not, in general, a
manifold. In some cases, as in the theory of jets, it is a
manifold, but it is infinite-dimensional. Thus, the category
[$\man$] of smooth, finite dimensional manifolds is not closed
under projective limits. The same is true for inductive limits of
manifolds...\footnote{From the introduction to \cite{pap2}.}}''
\end{minipage}

\noindent Thus, in contradistinction to $\man$, $\ctriad$ is
bicomplete ({\it ie}, both complete and co-complete). Also, in
connection with our topos remarks just before, (finite)
bicompleteness is another defining property for a category to
qualify as an elementary topos \cite{gold,macmo}.\footnote{{\it In
toto}, for a category to pass as an elementary topos {\it \`a la}
Lawvere-Tierney, it must: (a) have canonical subobjects and hence
a subobject classifier, (b) be (finitely) bicomplete, (c) have a
so-called initial (and a terminal) object, and finally, (d) have
an exponential structure \cite{gold,macmo}. The last property
means essentially that given any to objects $A$ and $B$ in the
category, one can form their exponential $A^{B}$ which represent
the collection of all arrows (morphisms or maps) from $B$ to $A$.
For the case of $\ctriad$ in particular, it can be shown that it
possesses such a structure, since for any two differential triads
$\triad$ and $\triad^{'}$ (over, say, the same topological space),
one can always group together, into the exponential
$\triad^{\triad^{'}}$, all the triad morphisms (`differentiable
maps') from $\triad^{'}$ to $\triad$.} Finally, as noted in
various footnotes above, concerning our finitary-algebraic,
ADG-theoretic resolution of the inner Schwarzschild singularity
later in 5.2.2, this bicompleteness of $\ctriad$ (in its finitary
analogue)\footnote{See next paragraph {\bf v}.} will prove to be
an invaluable result.

\paragraph{v. A `bonus' paragraph: the finitary analogue and the `Newtonian spark' of ADG.} The
three main results-properties of $\ctriad$ discussed above,
namely, the universal mapping property of triad morphisms, their
push-out and pull-back virtues {\it vis-\`a-vis} continuous maps
(between their underlying topological spaces), as well as
$\ctriad$'s bicompleteness, have been particularly useful and
fruitful in our ADG-based approach to a finitary, causal and
quantal version of Lorentzian gravity
\cite{malrap1,malrap2,malrap3}. In this last paragraph of the
present subsection we make it precise and clear exactly when and
in what way the results above were used, either explicitly or
implicitly, in the said approach.\footnote{Such an explicit and
direct presentation has not been given earlier in our trilogy
\cite{malrap1,malrap2,malrap3}, but have been partly exposed in
the latest paper by the second author \cite{rap5}.}

To briefly recapitulate things, our starting point in this
approach was Sorkin's finitary (:locally finite) replacements of
continuous manifolds \cite{sork0}.\footnote{The exposition below
will be encountered again later, in 4.3.2, but for slightly
different purposes.} In a nutshell, Sorkin considered locally
finite open coverings $\gauge_{i}$\footnote{The index-set $I=\{
i\}$ over which $i$ varies being a poset or net---the `topological
refinement net' (see below).} of an open and bounded region $X$ in
a continuous ({\it ie}, topological or $\cont$-) manifold $M$.
With respect to one such covering, Sorkin then grouped $X$'s
points into equivalence classes according to the following
equivalence relation:

\begin{equation}\label{eqxx1}
X\ni x\stackrel{\gauge_{i}}{\sim}y\in X\Leftrightarrow \Lambda
|_{\gauge_{i}}(x)=\Lambda |_{\gauge_{i}}(y)
\end{equation}

\noindent where $\Lambda |_{\gauge_{i}}(x):=\bigcap\{
U\in\gauge_{i}|~x\in U\}$---the `smallest' open set (in the
subtopology $\tau_{i}$ of $X$ generated by the open sets in
$\gauge_{i}$\footnote{$tau_{i}$ is generated by arbitrary unions
and finite intersections of the covering open sets in
$\gauge_{i}$.}) containing $x$, which we here coin `{\em Sorkin's
{\it ur}-cell of $x$ relative to $\gauge_{i}$}'.

The {\em quotient space} $X/\stackrel{\gauge_{i}}{\sim}=:P_{i}$,
consisting of $\stackrel{\gauge_{i}}{\sim}$-equivalence classes of
points in $X$,\footnote{That is, the `blown-up' points of $P_{i}$
are Sorkin's {\it ur}-cells of the original $X$'s points $x$.} was
then seen to be a $T_{0}$-topological space with the structure of
a (locally finite) partially ordered set (poset), and it was
pitched as ``{\em the finitary substitute of the continuous
topology of $X$}''.

The principal result in \cite{sork0}, and one that qualifies the
$P_{i}$s above as genuine `discrete' approximations of the
topological continuum, was that an inverse system (or net) $\inv$
of the $P_{i}$ was seen to yield at the projective limit of
infinite refinement of the said finitary posets\footnote{Roughly,
the act of topological refinement (or localization of $X$'s points
\cite{rap2,malrap1,malrap2}) is represented by the relation
$\gauge_{i}\preceq\gauge_{j}$ between the coverings involved,
which reads: $\gauge_{i}$ {\em is a finer covering than}
$\gauge_{j}$ (or with respect to the index-net $I$: $i\leq
j$)---which means in effect that $\gauge_{j}$ contains more and
`smaller' open sets than $\gauge_{i}$ (or equivalently, that the
subtopology $\tau_{j}$ of $X$ generated by the open sets in
$\gauge_{j}$ is {\em finer} than the corresponding $\tau_{i}$)
\cite{sork0}.} a space---call it $P_{\infty}$---that is
`essentially' topologically equivalent ({\it ie}, homeomorphic) to
the $\cont$-manifold $X$ that we started with.\footnote{A minor
detail here: the adverb `essentially' above pertains to the fact
that, actually, at the inverse limit of $\inv$ one does not get
back $X$ itself, but a (non-Hausdorff) space $P_{\infty}$ having
$X$ as a dense subset. However, one can recover $X$ from
$P_{\infty}$, by a procedure commonly known as {\em Hausdorff
reflection}, as the set of the latter's closed points
\cite{kopperman}.}

As highlighted in \cite{sork0}, the key result for setting up the
projective system $\inv$ is that continuous surjections
(corresponding to the canonical projection maps) from $X$ to the
$\stackrel{\gauge_{i}}{\sim}$-moduli $T_{0}$-spaces $P_{i}$ enjoy
a {\em universal mapping property} expressed by the diagram below:

\begin{equation}\label{eqxx2}
\qtriangle[X`P_{j}`P_{i};f_{j}`f_{i}`f_{ji}]
\end{equation}

\noindent That is, $f_{i}=f_{ji}\circ f_{j}$ ($i\leq j$ in $I$),
and reading that {\em the map (canonical projection) of $X$ onto
the finitary substitutes is universal among maps into
$T_{0}$-spaces}, with $f_{ji}$ the {\em unique} map---itself a
continuous surjection of $P_{j}$ onto $P_{i}$\footnote{Itself
corresponding to the aforesaid act of topological refinement (or
coarse-graining) $\gauge_{i}\preceq\gauge_{j}$ ($i\leq j$ in the
index-net $I$). Above, the epithet `continuous' for $f_{ji}$
pertains to the fact that one can assign a `natural'
topology---the so-called Sorkin lower-set topology---to the
$P_{i}$s, whereby an open set is of the form $\mathcal{O}(x):=\{
y\in P_{i}:~y\mapto x\}$, and where $\mapto$ is the partial order
relation in $P_{i}$ (with basic open sets involving the links or
covering---`immediate arrow'---relations in $P_{i}$). Plainly
then, $f_{ji}$ is a monotone (partial order-preserving) surjection
from $P_{j}$ to $P_{i}$, hence continuous with respect to the
Sorkin topology.}---mediating between the continuous projections
$f_{i}$ and $f_{j}$ of $X$ onto the $T_{0}$-posets $P_{i}$ and
$P_{j}$, respectively. With these canonical continuous projections
of $X$ onto the $P_{i}$s, the said inverse system of finitary
posets can be represented by $\inv :=(P_{i}, f_{ij})$; while
formally, the inverse limit result may be written as
$\underleftarrow{\lim}\inv\equiv\lim_{\infty\leftarrow
i}P_{i}\equiv P_{\infty}\stackrel{\mathrm{homeo}.}{\simeq}X$
(modulo Hausdorff reflection) \cite{malrap2,malrap3}.

At this point it must be noted however that Sorkin's
considerations in \cite{sork0} were purely topological, without
any allusion at all to the differential (smooth) structure of the
locally Euclidean $X$. The rich differential structure
`hybernating' in the finitary posets in Sorkin's scenario was
unveiled subsequently when Zapatrin, in collaboration with the
second author, passed to the Gel'fand-dual algebraic picture of
the $P_{i}$s above, involving the incidence (Rota) algebras
$\omg_{i}$ of those posets \cite{rapzap1,rapzap2}. Indeed,
recognizing from the beginning  that the $P_{i}$s may be viewed
homologically as (finitary) \v{C}ech-Alexandrov simplicial
complexes (nerves) ${\mathcal{K}_{i}}$ \cite{alex,cech}, their
associated incidence algebras\footnote{Write
$\omg_{i}({\mathcal{K}}_{i})$, or simply $\omg_{i}$ as above.}
were seen to be {\em $\Z_{+}$-graded discrete differential
algebras}\footnote{Or equivalently {\it \`{a}-la} Dimakis {\it et
al.}, {\em discrete differential manifolds}
\cite{dimu1,dimu2,dimu3}.}
$\omg_{i}=\bigoplus^{j\in\Z_{+}}\omg_{i}^{j}$,\footnote{With the
$j=0$-graded elements of $\omg_{i}$ in $\omg_{i}^{0}\equiv A_{i}$
constituting an abelian subalgebra of the noncommutative in
general $\omg_{i}$. In turn, ${\mathcal{R}}_{i}:=\bigoplus^{j\geq
1 }\omg_{i}^{j}$ was seen to be an $A_{i}$-module---a module of
discrete differential form-like entities \cite{rapzap1,rapzap2}.}
having a nilpotent (exterior) K\"{a}hler-Cartan differential $\kd$
as their basic differential operator,\footnote{$\kd$ was seen to
effect linear Leibnizian maps $\kd
:~\omg_{i}^{j}\mapto\omg_{i}^{j+1}$ between the linear subspaces
$\omg_{i}^{j}$ of $\omg_{i}$, raising their grade by $1$ in the
process.} which operator, in turn, can expressed via the
homological border (boundary) and coborder (coboundary) operators
of the ${\mathcal{K}}_{i}$s \cite{zap1}.

In fact, in \cite{rapzap1,rapzap2,zap1} it was shown that the
correspondence
$P_{i}~\mathrm{or}~{\mathcal{K}}_{i}:~\mapto\omg_{i}$ is {\em
functorial} in the sense that there is a (contravariant) functor
from the category $(\mathrm{finitary~posets},
\mathrm{monotone~maps})$\footnote{Monotone maps, or equivalently,
poset morphisms ({\it ie}, partial order-preserving maps) are
exactly the `continuous' ones---maps preserving the aforementioned
Sorkin $T_{0}$-topology of the $P_{i}$s \cite{malrap2,malrap3}.},
or equivalently, the category
$(\mathrm{finitary~simplicial~complexes},
\mathrm{simplicial~maps})$, to the category (incidence algebras,
algebra homomorphisms). Subsequently in \cite{rap2}, and then in
\cite{malrap1}, it was further recognized that the aforesaid
functorial correspondence is in fact an instance of a `discrete'
{\em sheafification functor}, in the sense that {\em the map
$P_{i}\mapto\omg(P_{i})$ is a local homeomorphism}---a {\em
sheaf}---as the `local topology' of the $P_{i}$s is carried over
to the `local topology' of their corresponding
$\omg_{i}$s.\footnote{That is, the generating or covering
relations (links) of the aforementioned Sorkin topology on the
$P_{i}$s was mapped by a procedure called {\em Gel'fand
spatialization} (itself an instance of {\em Gel'fand duality}) to
the generating relations of the so-called Rota topology on the
$\omg_{i}$s \cite{zap0,rapzap1,rapzap2}.} Thus, {\em finitary
spacetime sheaves} (finsheaves) of incidence algebras over
Sorkin's finitary substitutes of $\cont$-manifolds were born.

Then, with ADG in mind, it did not take long to see that the said
finsheaves define {\em finitary differential triads} (fintriads)
$\triad_{i}$ \cite{malrap2}.\footnote{$\triad_{i}$s are supposed
to live in the category $\ctriad_{i}$---the finitary analogue (in
fact, a subcategory) of $\ctriad$.} So now we are in a position to
expose how fruitful, in a finitary context, the results about
$\ctriad$ that we presented earlier are.

To begin with, the continuous surjections $f_{ji}$ in Sorkin's
system lift to {\em fintriad morphisms}---abstract {\em
differentiable} maps, not merely topological (continuous) ones.
Furthermore, the universal mapping property (\ref{eqxx2}) that
Sorkin's finitary poset morphisms enjoy translates directly to the
universal mapping condition that triads satisfy in (R3.?). {\it In
toto}, Sorkin's projective (inverse) system $\inv$ of
$T_{0}$-posets now translates to a projective system of fintriads.
But the following question arises now:

\vskip 0.1in

\centerline{How come the finitary differential triads $\triad_{i}$
in the first place?}

\vskip 0.1in

\noindent The answer to this question follows straight from the
aforementioned push-out result (Q?.?) and its quotient space
corollary (Q?.?), namely that,

\bigskip \noindent (R4.8)\hskip 0.9in
\begin{minipage}{11cm}
\noindent {\em the continuous canonical projection} $f_{i}$ {\em
from $X$ to the moduli space}
$X/\stackrel{\gauge_{i}}{\sim}=:P_{i}$ {\em in} (\ref{eqxx2}) {\em
induces (via its push-out $f_{i*}$) a differential triad on
$P_{i}$---the fintriad $\triad_{i}$}.\footnote{We tacitly assume
here that the locally Euclidean space (manifold) $X$ carries the
usual differential structure, which is encoded in the classical
differential triad
$\triad_{\infty}\equiv\triad_{X}=(\struc\equiv\smooth_{X},\partial
,\Omg)$ ($X$ a differential manifold).}
\end{minipage}

\vskip 0.1in

\noindent This is a {\em direct} way to account for the
differential structure that the finitary (originally taken to be
purely) topological posets inherit from the continuum, instead of
the {\em roundabout} way, via homological (simplicial) arguments,
that we have given hitherto.\footnote{Of course, the
(Cartan-K\"ahler) differential operator involved in the incidence
algebras of the Euclidean `triangulations' (nerve-simplices) of
$X$ is the same as the usual (de Rham) one of the smooth continuum
$X$, since the \v{C}ech-covering sets are `nice'
\cite{malrap2}---for, after all, the Sorkin {\it ur}-cells
(nerves) $\Lambda_{\gauge_{i}}(x)$ involved in (\ref{eqxx1}) are
nothing else but open subsets of the locally Euclidean $X$.}
Moreover, this direct account is an instance of what we call the
`{\em Newtonian spark}' conceptual paradigm of ADG, which we now
briefly explain:\footnote{We will return to discuss further this
paradigm, and in more general terms, in 7.6.1.}

\bigskip \noindent (R4.9)\hskip 0.9in
\begin{minipage}{11cm}
\noindent In an `ophelimistic' (or `opportunistic') sense, while
the differential involved in our application of ADG to the
finitary context comes from the original continuum $X$, that
manifold is not involved at all, as a background
space,\footnote{As is the case in CDG.} in the development of a
differential geometry {\it \`a la} ADG. Here in particular, the
essentially algebraic differential geometric mechanism (in effect,
the differential $d$) is abstracted (induced) from $X$ (via
$f_{i*}$) and used by the algebraic (:sheaf-theoretic) means of
ADG in a finitary setting, while at the same time $X$ is being
disposed of (here, replaced by the $P_{i}$s)---{\it ie}, it does
not play any role whatsoever in the {\it aufbau} of that
differential geometry in the reticular realm of the $P_{i}$s and
their associated $\omg_{i}$s.
\end{minipage}

\noindent As noted above, in general terms one may characterize
this ADG `attribute' as `{\em differential geometric forgetful
opportunism}', in the following sense: in ADG we do not care where
from ({\it ie}, from what kind of `space') $d$ comes, but once we
have got hold of and secured it, we develop with it all our
differential geometric {\it aufbau} purely algebraically
(:sheaf-theoretically), independently of that original
`space'---as it were, regardless of `the source of $d$'.
Philologically put, from the Newtonian spark ($d$) we start, by
ADG-means, the differential geometric fire, which then burns down
that initial `geometrical background' (`space'). Equivalently, to
parallel in a metaphorical way the Tractarian Wittgenstein's words
upon concluding \cite{witt},

\bigskip \noindent (Q4.6)\hskip 0.9in
\begin{minipage}{11cm}
\noindent ``{\small ...one must so to speak throw away the ladder,
after he has climbed up on it...}'' \cite{witt},
\end{minipage}

\vskip 0.1in

\noindent similarly here, {\em for doing differential geometry
{\it \`a la} ADG,  one must throw away the underlying space(time)
after she has gathered a differential from it}, for she does not
actually need it ({\it ie}, here, the base manifold $X$), and
thus, {\it a fortiori,} she avoids directly various (differential
geometric) anomalies and pathologies that are inherent in it ({\it
eg}, `singularities'). Once we have obtained the
differential---the `Newtonian spark'---from $X$, we throw the
latter away, and we build our differential geometric edifice
purely algebraically, in a Leibnizian way, without the background
manifold's presence or assistance via the mediation of its smooth
coordinate structure (sheaf) $\smooth_{X}$, which anyway carries
the differential geometric diseases of the classical theory (CDG).

Now, on with the translation of the inverse limit result in
\cite{sork0} to our finitary-algebraic ADG-theoretic setting
involving the fintriads. First we note that, by Gel'fand
duality,\footnote{Or equivalently, by the categorical
contravariant functor-duality between the categories of the
$P_{i}$s (and monotone maps) and their $\omg_{i}$ (and incidence
algebra homomorphisms) mentioned earlier.} since the $P_{i}$s
comprise the {\em projective} system $\inv$, their corresponding
incidence algebras $\omg_{i}$ constitute an {\em inductive} system
$\diromg$ \cite{malrap2,malrap3,rap5}.\footnote{In other words,
while the system $\inv$ of $T_{0}$-posets in Sorkin's scheme is an
{\em inverse} one, by Gel'fand duality (or the contravariant
functor in the respective categories noted above), the
corresponding system of incidence algebras is a {\em direct} one.
Moreover, since we are actually working with (fin){\em sheaves} of
incidence algebras (over Sorkin's $T_{0}$-posets), the inductive
system $\diromg$ of the latter and its inductive `continuum' limit
may be interpreted as the act of infinite localization of the
incidence algebras, defining in the process {\em stalks} of the
`\texttt{classical}' continuum sheaves which are inhabited by {\em
germs} of $\ssmooth$-\texttt{smooth} functions and differential
forms over them (see below explanation for the adjective
`\texttt{smooth}'). Indeed, as also briefly alluded to earlier,
the aforesaid contravariant functor may be thought of as a {\em
Gel'fand-type sheafification (sheaf-theoretic localization)
functor} \cite{mall-1,mall8}.} Then, the resulting fintriads that
the finsheaves of the $\omg_{i}$s define, can be organized into an
inverse/direct system $\invtriad$.\footnote{The joint epithet
`inverse/direct' to the system $\invtriad$ pertains precisely to
the categorical (contravariant functor), Gel'fand-type of duality
mentioned above: while the $P_{i}$s---the base spaces of the
$\triad_{i}$s---constitute an inverse system $\inv$ subject to a
projective limit procedure, their categorically, Gel'fand-dual
$\omg_{i}$s---inhabiting the stalks of the finsheaf spaces in the
$\triad_{i}$s---are organized into the direct system $\diromg$
which is subjected to an inductive procedure. Informally speaking,
in a chiral sense, $\gauge_{i}$-refinement for the base $P_{i}$s
goes `from-right-to-left', while for their associated $\omg_{i}$s
`from-left-to-right'.} Then, thanks to $\ctriad$'s
bicompleteness,\footnote{And especially Papatriantafillou's result
in \cite{pap2} giving {\em the direct limit differential triad of
a direct system of differential triads defined on a projective
system of base topological spaces} (Q?.?).} in the same way that
Sorkin's projective system $\inv$ `converges', at the inverse
limit of infinite refinement, to the original $\cont$-manifold
$X$,\footnote{Again, modulo Hausdorff reflection.} the
corresponding inverse system of fintriads yields (or perhaps
better, {\em defines}) at the projective/inductive limit of
maximal topological refinement of the underlying $P_{i}$s and, as
a result, of the $\omg_{i}$ inhabiting the stalks of the
finsheaves in the $\triad_{i}$s, an `infinitary' differential
triad, which we have fittingly coined in the past the
\texttt{smooth} (or even, `\texttt{classical}') continuum
differential triad $\striad:=(\ssmooth ,\partial ,\Omg)$
\cite{malrap2,malrap3,rap5}.\footnote{The denomination `{\em
classical continuum}' for $\striad$ comes from the physical
interpretation of Sorkin's projective limit for the $P_{i}$s (and
correspondingly, the continuum direct limit of their $\omg_{i}$s)
as Bohr's Correspondence Principle ({\it ie}, as a {\em classical
limit} process), because of the quantum interpretation that the
spaces that the $\omg_{i}$s represent enjoy
\cite{rapzap1,rapzap2}.} $\striad$ comes as close as possible (via
Sorkin's scheme) to the classical one
$\triad_{\infty}=(\smooth_{X},\partial ,\Omg_{X})$ supported by
the differential manifold $X$.\footnote{As also pointed out in
\cite{malrap3,rap5}, the expression `comes as close as possible to
$\triad_{\infty}$' pertains to the fact that, much in the same way
that one does not actually recover $X$ as the inverse limit space
of $\inv$, one also does not exactly get $\smooth_{X}$ and the
$\smooth(X)$-module sheaf $\Omg$ of (germs of) smooth differential
forms (over the differential manifold $X$'s points) at the direct
limit of (infinite sheaf-theoretic localization of) the
$\omg_{i}$s in $\diromg$. Rather, similarly to the fact that one
gets a `larger' inverse limit topological space $P_{\infty}$
having $X$ as a dense subset in Sorkin's scheme ({\it ie},
roughly, $P_{\infty}$ has more points than $X$) \cite{sork0}, one
anticipates $\diromg$ to yield at the inductive limit an abelian
(`topological') algebra $\aconn_{\infty}$ `larger' than
$\smooth(X)$ and consequently an $\aconn_{\infty}$-module
$\mathcal{R}_{\infty}$ of differential form-like entities `larger'
than the standard $\smooth$-one. In Zapatrin's words, when he was
working out continuum limits of incidence algebras of simplicial
complexes \cite{zap1,zap2}: ``{\em it is as if too many functions
and forms want to be smooth in the continuum limit}'' (Roman
Zapatrin in private e-mail correspondence with the second author).
One intuits that much in the same way that Hausdorff reflection
does away with the `extra points' of $P_{\infty}$ to recover the
$\cont$-manifold $X$ (as a dense subset of $P_{\infty}$), so by
getting rid of the extra functions and forms on $P_{\infty}$ from
$\omg_{\infty}$ ({\it eg}, by factoring it by a suitable
differential ideal \cite{zap1,zap2}), one should recover the usual
smooth functions and forms over the differential manifold $X$.
Anyhow, the important point here is that {\em one does indeed get
a continuum differential triad}, which however, only in order to
be formally distinguished from the classical $\smooth$-smooth one
$\triad_{\infty}$ to avoid any minor technical misunderstanding,
we might call `$\ssmooth$-\texttt{smooth}' and symbolize it by
$\striad$ \cite{malrap3,rap5}. On the other hand, after having
alerted the reader to this slight distinction between $\striad$
and $\triad_{\infty}$, in the sequel, for all intents and purposes
and in order to avoid proliferation of redundant symbols, we shall
abuse language and notation and assume that $\striad$ and
$\triad_{\infty}$ are `essentially isomorphic' ({\it ie},
effectively equivalent and indistinguishable) while both will be
generically referred to as {\em the classical continuum
differential triad} (CCDT), with the symbols $\triad_{\infty}$ and
$\striad$ used interchangeably.} As noted earlier, this
`classical' continuum limit result will play a crucial role in our
ADG-based finitary-algebraic `resolution' or `evasion' of the
inner Schwarzschild singularity (5.2.2).

In the next subsection we would like to digress a bit and discuss
a basic `aphorism' in ADG, namely, that all differential geometry
is essentially based on the structure sheaf $\struc$ of
generalized arithmetics or coordinates---the space of
`differentiable' functions that one may freely choose {\it a
priori} in order to base all her differential geometric
constructions. From the point of view of CDG, as repeatedly noted
earlier, the initial choice is mandatorily
$\struc\equiv\smooth_{X}$, which is tantamount to fixing a
background differential manifold ($X\equiv M$) on which then the
whole CDG-edifice is erected and vitally relies. On the other
hand, the main didagma of ADG is that one can use structure
sheaves glaringly different from the classical one $\smooth_{M}$
and still be able to develop all the traditionally continuum based
differential geometric constructions without that locally
Euclidean continuum being present in the background.

\subsection{All Differential Geometry Boils Down to $\struc$: (A)DG Begins with (and Ends in) $\struc$}

The fundamental motto of ADG can be diagrammatically expressed as
follows:

\begin{equation}\label{eqxx3}
\bfig
\Vtriangle[\struc`\conn(\equiv\partial)`\mathbf{DG};(c)`(a)`(b)]
\efig
\end{equation}

\noindent Let us explain the schema above, starting with arrow
$(b)$:

\begin{itemize}

\item $(b)$ All {\em Differential} Geometry (DG) is subsumed under the
(existence of a) differential (operator) $\conn$---the
connection.\footnote{Recall, the basic observation that initially
motivated the development of ADG is the recognition that the usual
differential $d\equiv\partial$ is a particular ({\it ie}, a flat)
instance of a connection (and vice versa: $\conn$ is a
generalized, an abstracted and prolongated differential $d$).} In
other words, {\em there is no} {\bf D}G {\em without a
connection}: as a symbolic pun, `$\conn$' stands for the letter
`D' in `DG', and we write:

\begin{equation}\label{eqxx4}
\mathrm{\underline{D}G}\longleftrightarrow
\underline{\conn}\mathrm{G}
\end{equation}

\noindent Arrow $(b)$ in (\ref{eqxx3}) above pictures this
$\conn$-input in DG.

\item $(a)$ Of course, $\conn$, being an operator (a map!), acts
`somewhere', on some `domain' or `action-space'---the {\em
representation} or {\em carrier space} of $\conn$, so to speak.
Insofar as the word `{\em geometry}' (G) enters into the term
D{\bf G}, it pertains to specifying the substrate---background as
it were---on which $\conn$ is soldered and operates. In ADG, that
domain---ultimately, the source\footnote{See explanation of arrow
$(c)$ below.}---of $\conn$ is provided by (specifying)
$\struc$.\footnote{And the reader should recall that in ADG the
domain of $\partial$ is $\struc$ (\ref{eqy2}), while of
$\partial$'s generalization $\conn$, it is $\modl$
(\ref{eqy4})---which vector sheaf anyway is (locally), by
definition, a finite power of $\struc$ ($\modl
|_{U}\equiv\modl(U)\simeq\struc^{n},~U~\mathrm{open~in}~X$).}
However, it must be stressed at this point that in ADG if any
`space' is involved at all, it is not assumed {\it a priori}, but
only evoked indirectly, via $\struc$.\footnote{For we have time
and again emphasized it here and in our previous trilogy
\cite{malrap1,malrap2,malrap3}, that if any `geometrical
space(time)' enters into our ADG-considerations at all, it enters
through $\struc$---in point of fact, {\em it is inherent in}
$\struc$ (Gel'fand duality/spectral theory).} For in any case, in
ADG $\conn$ is termed an `{\em algebraic connection}' (or
`$\struc$-{\em connection}' for short), emphasis being placed here
on the epithet `{\em algebraic}', as the notion of connection is
an entirely algebraic one, without any {\it a priori} geometrical
commitment to (or dependence on) an underlying `space(time)'---a
background stage or medium which enables us to represent and
interpret(!) $\conn$ geometrically. The intention here is to
clearly separate and distinguish up-front the purely algebraic
$\conn$ from its possible `{\em geometrical realization}' (or
representation, and concomitant geometrical interpretation)
accomplished by specifying (choosing) a specific $\struc$ (and the
space inherent in it).\footnote{In much the same way for example
that one should distinguish between an abstract algebraic
structure ({\it eg}, a group) and its representations ({\it eg},
matrix-operator realizations on some linear action/carrier space);
hence the jargon above. This geometrical representation of the
algebraic $\conn$, and the latter's geometrical interpretation (in
that space chosen), recalls the Newton--Leibniz `dispute' about
the notion and meaning of derivative ({\it viz.} differential $d$,
and in ADG, $\conn$) in the context of the infant developmental
steps of Calculus (CDG). Leibniz envisaged $d$ as a {\em
relational} (we would nowadays say, {\em algebraic}) entity
independent of a surrounding space and its mediating coordinates
{\it \`a la} Descartes. Newton on the other hand was more
Cartesian, as $d$ for him was heavily burdened by its geometrical
representation and interpretation in an ambient space. Recall that
for Newton the derivative of a function (at a point in its domain
space) is tautosemous to the slope of the tangent line to the
curve-graph of the function at the said point---a representation
and interpretation that `masks' in our view the purely algebraic,
Leibnizian essence of $d$. Alas, the manifold based CDG followed
Newton's steps, while ADG is markedly Leibnizian. For, as we also
noted in \cite{malrap1,malrap2}, if $d$ was an `inherently'
geometrical object, would it not be redundant and meaningless---in
fact, begging the question---to ask for its geometrical
interpretation? One could imagine for example the following
question being asked about a purely geometrical (`spatial') object
like the triangle: `{\em What is the geometrical interpretation of
the triangle?}'---plainly, the triangle, being `inherently' a
geometrical object, is in no need of a geometrical interpretation!
In contradistinction, it is a meaningful question to ask for the
geometrical realization and interpretation of the derivative,
which goes to show that $d$ is not an inherently geometrical
object; it is an algebraic entity. In the last section (7.3), we
will return to discuss further this all important distinction
between the Newtonian (what we prefer to call, {\em Cartesian})
and the Leibnizian (what we coin, {\em Euclidean}) conceptions
(and practices) of DG, by contrasting in the process CDG against
ADG.}

To recapitulate the joint meaning of arrows $(a)$ and $(b)$: {\em
all DG pertains to the activity (action) of some (generalized)
differential $\conn$ exercised (or carried out) on some `space'
(`geometry'), which in turn is inherent in some (freely chosen)
$\struc$---the `source' (or `representation') algebra of $\conn$}.
In turn, $\struc$, being the `geometrical source' of
$\conn$,\footnote{Again, see explanation of arrow $(c)$ in
(\ref{eqxx3}) below.} allows us to maintain the title of this
subsection, namely that,

\centerline{\em all DG boils down to $\struc$;}

\noindent under the important proviso that it is {\em us} that
choose the representation algebra $\struc$ (and, {\it in extenso},
the space inherent in it) for $\conn$.\footnote{Our maintaining
that the term `geometry' goes hand in hand with {\em our} choice
of $\struc$ becomes even more valid if one interprets (as we have
done throughout the aforesaid trilogy) the elements of $\struc$ as
{\em generalized coordinates} or {\em measurements} (the words
coefficients and arithmetics are synonyms) of the field $\conn$.
This is our `geometrical capturing'---our representing, measuring
and concomitant localizing in `space(time)' (in the `space'
inherent in $\struc$!)---of $\conn$, which anyway exists
independently of us (field realism; see below). For there is no
geometry without measurement, and no measurement without us---the
`observers' and `measurers' (`geometers') to carry it out. {\em
The (algebraic) field} {\it ie}, $\conn$) {\em is Nature's; the
geometrical representation of it} ({\it ie}, $\struc$) {\em ours}.
Moreover, as it was also emphasized in \cite{malrap2,malrap3}, in
a quantum-theoretic sense, the algebraic $\conn$ lies on the
quantum side of the quantum divide (the so-called Heisenberg {\it
schnitt}), while (the commutative) `geometric' $\struc$ on the
classical side. In this sense, `{\em geometrization (or
representation) is classicalization}' (Bohr's correspondence
principle). At the same time, {\em the geometrical representation
of $\conn$ via $\struc$ results in the geometrical interpretation
of the derivative}, as noted two footnotes above. In CDG, at least
in the usual Newtonian conception of Calculus, the essentially
algebraic character of the differential is in a sense `masked' by
the intervention of (representation) space (manifold) in the guise
of {\em our} choice of $\struc\equiv\smooth_{M}$ for structure
sheaf of coordinates, to the effect that one (misleadingly) tends
to view $\conn$ as a geometrical notion (see `triangle oxymoron'
two footnotes above). Of course, in the classical case (CDG), it
is precisely our {\it a priori} assumption of a background locally
Euclidean space (manifold) $M$ that on the one hand mandates
$\smooth_{M}$ for $\struc$ and then furnishes us with the usual
$d$ (and {\it in extenso}  with the standard smooth $\conn$). This
assumption masks the basic didactic point of ADG, namely that,
{\em the differential ({\it viz.} connection) may come from (the
assumption/choice of) structure sheaves $\struc$ totally different
from $\smooth_{M}$ and, as a result, without any {\it a priori}
commitment to a geometrical base differential manifold} (recall,
``{\em differentiability is independent of smoothness}''
\cite{malrap2}).}

In turn, as noted in the last footnote, all our (generalized)
measurements of the (physical) fields ({\it ie}, the
connections)---which measurement-records comprise our geometry,
and concomitantly, our own (`mental') fiction of an ambient
`space(time)' (Q?.?, Q?.?)---take values in $\struc$.\footnote{As
mentioned repeatedly in the foregoing trilogy
\cite{malrap1,malrap2,malrap3}, from an ADG-theoretic perspective,
the `geometrical objects', or physically speaking, the
`measurable'/`observable' entities in the theory, are basically
$\struc$-valued $\otimes_{\struc}$-tensors (or equivalently,
$\struc$-morphisms) built out of the fundamental fields ({\it
viz}. the connections $\conn$), such as the curvature $R(\conn)$
of the connection.} In this sense, {\em (A)DG not only begins with
$\struc$} (`representation-domain'), {\em but also it ends in
$\struc$} (`value-range').\footnote{Characteristically, recall
from \cite{malrap3} that in ADG the metric $\rho$---arguably, {\em
the} entity in terms of which one can meaningfully speak about
{\em geometry} proper---is a map (sheaf morphism) with domain
$\modl$ (in fact, $\modl\oplus\modl$), which locally reduces to a
power of $\struc$, and range $\struc$ again ({\it ie}, $\rho$ is
an {\em $\struc$-valued metric}). Thus $\rho$ (geometry) begins
(domain) and ends (range) with $\struc$.}

\item $(c)$ This arrow simply reflects the basic ADG maxim mentioned above, namely,
that {\em the differential} $d\equiv\partial$ ($\conn$) {\em comes
from algebra}.\footnote{Also being implicit here that in the
classical case of CDG where $\struc\equiv\smooth_{M}$, the
differential comes from our {\it a priori} assumption of a base
manifold---a conflation of $d$ or $\conn$ with the mediation of a
background `space' which masks the differential's essentially
algebraic character and misleads one into thinking on the one hand
that the notion of connection is geometrical, and on the other
that, in one way or another, all DG is supported by a locally
Euclidean background space---what we called earlier, the manifold
and {\it in extenso} CDG-conservatism and monopoly.}

\item One could compress all the $(a)$-$(c)$ discussion above into the following symbolic `pun-equivalence':

\begin{equation}\label{eqxx5}
\mathrm{\underline{AD}G}\longleftrightarrow\underline{\struc\conn}
\mathrm{G}\footnote{With a Kleinian replacement of $G$ (by the
principal group sheaf $\grouv\equiv\aut(\modl)$ of the field's
$(\modl ,\conn)$-automorphisms) to be accomplished in
sub-subsection 3.2.3 below, where we discuss an ADG-extended
(abstract or generalized) version of the PGC of the manifold and
CDG based GR.}
\end{equation}

\end{itemize}

\subsubsection{The ADG-theoretic Principle of Algebraic Relativity of Differentiability}

In the foregoing discussion we saw how `differentiability' in ADG,
freed from the topological shackles of Analysis (CDG), is a purely
algebraic notion. Moreover, the geometrical or `measurement'
freedom---{\em our} freedom---to choose generalized coordinate
structure sheaf $\struc$ (other than the standard one
$\smooth_{M}$ of CDG) mentioned in connection with arrow $(a)$ in
(\ref{eqxx3}) above, reflects what we call the {\em Principle of
Algebraic Relativity of Differentiability} (PARD) in ADG, namely,
that

\bigskip \noindent (R4.10)\hskip 0.9in
\begin{minipage}{11cm}
\noindent DG---or how the algebraic field $\conn$ is geometrically
captured (`measured') and expressed ({\it ie}, represented as
acting on some `space(time)-geometry', which is anyway inherent in
$\struc$)---varies for different choices of our arithmetics
$\struc$: in other words, {\em choose an $\struc_{1}$, and you get
a DG$_{1}$; choose another $\struc_{2}$, and you get a different
DG$_{2}$.}
\end{minipage}

\vskip 0.1in

\noindent The import of PARD in applications of ADG to physical
problems is significant. For example, concerning the problem of
singularities in the spacetime manifold (and, {\it in extenso},
the CDG) based GR---the main problem addressed in the present
paper, namely, changing arithmetics (`coordinates') from the
classical ones $\struc_{1}\equiv\smooth_{M}$ defining $M$ (and, in
effect, the CDG) which in turn are responsible for ({\it ie}, they
have built into them) the singularities of GR, to another
$\struc_{2}$ conveniently chosen so as to integrate, `absorb' or
`engulf' those singularities but still retain the essentially
algebraic differential geometric mechanism `supplied' by the
(gravitational) field $\conn$ (which anyway exists independently
of our geometry-defining measurements in the $\struc$ we choose to
geometrically represent it---field realism), {\em the problem of
singularities simply disappears}. It appears appropriate to
metaphorically `paraphrase' again, in a suitably modified way,
Wittgenstein from (an amalgamation of two quotations from)
\cite{witt} and \cite{witt1}, respectively,

\bigskip \noindent (R4.11)\hskip 0.9in
\begin{minipage}{11cm}
\noindent {\em the solution of the problem of singularities is
seen in the vanishing of this problem,\footnote{First quotation
from \cite{witt}: ``{\rm ...The solution of the problem of life is
seen in the {\em vanishing} of this problem...}'' (our emphasis).}
which is achieved simply by changing structure sheaf of
generalized coordinate-arithmetics---one's
observations-measurements (geometry); and more generally, by
changing the way (the theoretical framework within which) one
looks at DG as a whole and the so-called singularities troubling
the `conventional' way of looking at DG ({\it ie}, the CDG-way
effectuated via the background $M$, or equivalently, via
$\struc\equiv\smooth_{M}$!).}\footnote{Second, more extensive,
quotation from \cite{witt1}: ``{\rm ...The way to solve the
problem you see in life is to live in {\em a way that will make
what is problematic disappear}. The fact that life is problematic
shows that the shape of your life does not fit into life's mould.
So you must change the way you live and, once your life does fit
into the mould, {\em what is problematic will disappear}...Getting
hold of the difficulty {\rm [or problem]} {\em deep down} is what
is hard. {\em It {\rm [{\it ie}, the problem]} has to be pulled
out by the roots; and that involves our beginning to think about
these things in a new way}. The change is as decisive as, for
example, that from the alchemical to the chemical way of thinking.
The new way of thinking is what is so hard to establish, {\rm
[but]} {\em once the new way of thinking has been established, the
old problems vanish...For they go with our way of expressing
ourselves and, if we clothe ourselves in a new form of expression,
the old problems are discarded along with the old garment}...}''
(our emphasis).}

Indeed, ADG is a totally new way of looking at DG, so that what
appeared to be problematic from a CDG-standpoint ({\it viz.}
singularities), now completely disappears. Of course, the cause of
all these differential geometric anomalies and diseases---the base
differential manifold---has been ``{\em pulled out by the
roots}'', and the new, essentially algebraic, ``form of
expression'' of ADG results in the effective discarding of the
`old' CDG-problems of singularities, which ``{\em are discarded
along with the old garment}'' of CDG, namely, the `alchemical'
({\it ie}, physically non-real or `fiducial') background spacetime
continuum (Q?.?, Q?.?, Q?.?).
\end{minipage}

\vskip 0.1in

\noindent It is indeed as if what appeared to be problematic (and,
quite paradoxically, of physical value) from the (old/traditional)
viewpoint of the manifold based CDG ({\it ie}, the singularities),
under the new prism of the (background spacetime) manifoldless
ADG, it appears to be `poor', of little (if not at all!) physical
significance. We cannot refrain from quoting Wittgenstein
\cite{witt1} for the third time here:

\bigskip \noindent (Q4.7)\hskip 0.9in
\begin{minipage}{11cm}
\noindent ``{\small\em ...The solution of philosophical problems
can be compared with a gift in a fairy tale: in the magic castle
it appears enchanted and if you look at it outside in daylight it
is nothing but an ordinary bit of iron (or something of the
sort)\footnote{Our emphasis throughout.}...}''
\end{minipage}

\vskip 0.1in

\noindent and appropriately paraphrasing him to suit the title of
the present treatise:

\bigskip \noindent (R4.12)\hskip 0.9in
\begin{minipage}{11cm}
\noindent The solution of the problem of singularities can be
compared with the various chimerical creatures in mythology:
within the mythical realm of the (background spacetime) manifold
(based CDG) it appears formidable (indeed, out of reach!), but if
you look at it `outside'---in the base manifoldless environment of
ADG---it is nothing but part and parcel of the structure sheaf of
generalized arithmetics one chooses to employ (freely and at
will!) as generalized coordinates in the theory---{\it ie}, from
the ADG-vantage all singularities are `coordinate' ones (or
something of the sort), while in view of the
$\struc$-functoriality of the gravitational dynamics (Einstein
equations) in ADG-gravity, the said singularities are of no
physical significance.
\end{minipage}

\vskip 0.1in

\noindent In view of the three Wittgenstein quotations and
concomitant remarks above, as well as in the two footnotes
therein, we cannot resist quoting at this point Wallace Stevens
from \cite{stevens}:

\bigskip \noindent (Q4.8)\hskip 0.9in
\begin{minipage}{11cm}
\noindent ``{\small\em ...Progress in any aspect is a movement
through changes in terminology...}''\footnote{And in the case of
ADG, {\em through changes in theoretical framework---the main
change in ADG being to look at and actually do DG without a
background manifold}.}
\end{minipage}

\vskip 0.1in

\noindent {\it In summa}, the aforesaid generalized arithmetics'
or coordinates' change is from the classical one involving the
background differential manifold $\struc_{1}\equiv\smooth_{M}$
which supports CDG, to another $\struc_{2}$ (thus also a different
DG$_{2}$ altogether!), which is not only manifestly {\em not}
supported by and effectuated via a base geometrical manifold like
in CDG, but also it possibly incorporates the smooth singularities
(of $\struc_{1}$) in $\struc_{2}$ and at the same time leaves the
essentially algebraic differential geometric mechanism of the
$\conn$-fields intact and fully operative in their very
presence!.\footnote{See 5.2.2 and 5.2.3 for two explicit and
concrete physical examples of this ADG-theoretic application.}

It must be mentioned however that PARD is not a relativity
principle `proper'---{\it ie}, there is no {\em transformation
theory} linking the different $\struc$s
\cite{df1}.\footnote{Except of course for the generalized `general
coordinate transformations' within the same chosen $\struc$. For
example, for a chosen, fixed as it were, $\struc$ (and {\it in
extenso} $\modl$, which by definition is locally $\struc^{n}$),
the coordinate relativity/transformation group (supporting our
generalized version of the PGC of GR as we will see numerous times
in the sequel) is $\aut\modl$ (see 3.2.2 next). Accordingly, for
the background manifold $M$ and thus CDG-based GR, the said
transformation theory implementing the PGC involves
$\mathrm{Aut}M\equiv\mathrm{Diff}M$, it being tacitly assumed in
this case that one is working within the category $\man$ of
differential manifolds having $\struc\equiv\smooth_{M}$ as the
chosen structure sheaf of coordinates. However, in the next
sub-subsection we shall argue that, categorically speaking, PARD
may be expressed via {\em natural transformations}
\cite{maclane,macmo}. In other words, PARD is supported by some
kind of `{\em categorical transformation theory}'---a
`super'-transformation theory since we are not just talking about
generalized coordinate changes within the same (chosen) $\struc$
({\it eg}, $\struc$ chosen to be $\smooth_{M}$ and the relevant
category $\man$), but about changes between different $\struc$s
altogether---thus effectively `transcending' $\man$ and working
within the wider and more flexible (for doing differential
geometry) category $\ctriad$ of differential triads that we saw
earlier.} This is to be expected since the choice of $\struc$ (in
effect, of DG!) lies with the `external' (to the fields
themselves) `observer' or `experimenter', and it is {\em she} that
chooses how to `geometrize'\footnote{Here, synonyms to the word
`geometrize' are `represent', `coordinatize', `arithmetize', or
even `localize' and `particle represent' (as we will see from a
geometric prequantization viewpoint subsequently).} the field
$\conn$, while at the same time, as we have already elaborated on
in \cite{malrap3}, {\em such `generalized coordinate gauge
choices' lie with the (macroscopic) exo- or epi-system and not
with the (microscopic) endo-system}\footnote{Which in ADG are the
fields themselves (or perhaps better, `{\em in
themselves}).}---they are the experimenter's (measurer's or
geometer's) free choices \cite{df1}.

\subsubsection{The field, the whole field, and nothing but the field:
an ADG-generalized version of the PGC of the manifold based GR;
the case for Synvariance}

As we have already pointed out in the foregoing trilogy
\cite{malrap1,malrap2,malrap3},\footnote{Especially in
\cite{malrap3}.} as well as earlier in the present paper, and as
we will also elaborate on further in the sequel\footnote{See for
example our ADG-treatment of the Einstein hole argument in GR in
7.5.5.} since ADG is manifestly base differential manifold
independent, and as a result Calculus-free, it manages to bypass
without any difficulty various problems that GR, as well as some
attempts to quantize it by still retaining though a background
spacetime continuum for differentiability's sake,\footnote{What we
earlier coined `manifold (and CDG) conservatism'. We shall discuss
those problems later in this paper.} encounter.

Concerning GR, the first `problematic' issue that is completely
evaded is the mathematical representation of the PGC by
$\mathrm{Diff}(M)$---the (group of) automorphisms of the external
(to the gravitational field) smooth $M$. On the face of the
problems that this representation creates in trying to cope with
(actually, even to define!) singularities in GR as we discussed in
the first two sections, the said evasion is more than welcome.
However, we would still like to possess an ADG-theoretic analogue
of the PGC of the $M$ and CDG-based GR, even in the latter's
prominent absence. The analogy is straightforward and goes hand in
hand with the following {\em differential geometric
`correspondence principle'} with CDG that ADG allows us to draw:
since CDG can be `recovered' from (or even be regarded as a very
particular---and to that, `{\em singular}'!---instance of) ADG
when one chooses $\smooth_{X}$ for structure sheaf of coefficients
(`coordinates') in the theory---a choice which automatically
`converts' the underlying, {\it a priori} arbitrary, topological
space $X$ to a smooth manifold $M$, and since as noted before in
the classical theory $\mathrm{Diff}(M)\equiv\mathrm{Aut}M$, {\em
in ADG the abstract version of the PGC is mathematically
represented by the (principal) group sheaf $\grouv\equiv\aut\modl$
of automorphisms of the vector sheaf $\modl$ involved}.
Conversely, {\em $\modl$ is the representation (associated) sheaf
of the principal sheaf $\aut\modl$}
\cite{vas1,vas2,vas3,malrap3,vas4}---effectively, the action or
`carrier' (sheaf) space of the field $\conn$. Thus, in the {\em
Kleinian} sense of the word `geometry',\footnote{Roughly, that the
`geometry' of an object is completely described by its (group of)
symmetries.} the principal sheaf $\aut\modl$ of
self-transmutations (`auto-symmetries') of $\conn$---what we coin
`{\em the esoteric geometry of the field}' in the
sequel\footnote{See (\ref{eqxx7}) below. The epithet `{\em
esoteric}' here essentially means `{\em the auto-transformations
of the field in-itself, without recourse to a background spacetime
structure (and especially, a manifold!) supporting it}'.}---is
represented ({\it ie}, acts) on (the local sections of) $\modl$
which, from a geometric prequantization vantage, are the (local)
particle states of the (prequantized) field
\cite{mall1,mall2,mall5,mall6,malrap2,malrap3}.\footnote{For more
about the ADG-based geometric prequantization of gravity, see 6.1
below.}

\bigskip \noindent (R4.13)\hskip 0.9in
\begin{minipage}{11cm}
\noindent This is a completely {\em autonomous} conception and
representation of GC---one that is self-sustained by the field and
nothing else,\footnote{That is, by no structure other than the
field itself---as it were, a structure `external' to the field
(and its `innate' quanta/particles).} which we coin `{\em
synvariance}'.\footnote{The prefix `{\em syn-}' here is the Greek
correspondent preposition of the Latin prefix `{\em co-}'. We use
the new term `synvariance', because `covariance' is heavily
`loaded' with (and `burdened' by) connotations from the usual,
external spacetime continuum based GR, whereas ADG does not allude
to, let alone employ, at all a(n external to the connection
fields) base manifold in its constructions. Synvariance goes hand
in hand with the `{\em genuinely unitary}', `{\em pure gauge}'
conception and practice of field theory (especially of gravity)
that ADG allows us to maintain---a field theory developed with the
objects (fields) `in-themselves', without recourse to a background
spacetime structure \cite{malrap3}, and {\it a fortiori},
regardless of whether the latter is a continuum or a discretum.
See diagram (\ref{eqxx6}) and relevant discussion next.}
\end{minipage}

\vskip 0.1in

\noindent We can formally picture the aforesaid {\em differential
geometric `correspondence principle'} between ADG and CDG by the
following diagram:

\begin{equation}\label{eqxx6}
\bfig
\putsquare<1`1`1`1;900`900>(900,900)[\boxed{\mathrm{ADG}:\modl}`\boxed{\mathrm{CDG}:M}`\boxed{\aut\modl}`\boxed{\mathrm{Diff}(M)};
\struc\equiv\smooth_{X}`\mathrm{Synvariance}`\mathrm{PGC}`X\equiv
M] \efig
\end{equation}

\noindent However, in view of our remarks above about synvariance,
this analogy/correspondence, and the associated difference in
representation of the PGC, between ADG and CDG is far from
trivial, as we explain now:

\bigskip \noindent (R4.14)\hskip 0.9in
\begin{minipage}{11cm}
\noindent {\em In ADG, in striking contrast to CDG, no underlying
space(time manifold) is employed ({\it ie}, mediates) to
effectuate the essentially algebraic ({\it ie},
Leibnizian-`relational', between the connection fields
in-themselves) differential geometric mechanism. All DG is carried
out exclusively in the sheaf space $\modl$---in point of fact, in
terms of the (local) sections of $\modl$\footnote{For after all,
{\em a sheaf is its (local) sections} \cite{mall1,mall2}.}---with
sole resource the algebraic relations between the `geometrical
objects' (fields and their particle-quanta) inhabiting it. As a
result, in contradistinction to the manifold and CDG based GR, the
PGC in our ADG-theoretic perspective on gravity---{\it ie}, the
concept of synvariance---concerns solely and exclusively the
self-transmutations of the fields in-themselves, without an
allusion to or dependence on a background, external to those
fields, spacetime manifold.}
\end{minipage}

\vskip 0.1in

Characteristically, to pronounce the contrast between the manifold
and CDG-based GR thus also of the $\mathrm{Diff}(M)$-implemented
PGC, and the base space(time manifold) independent formulation of
gravity {\it \`a la} ADG with the concomitant autonomous
conception of field theory and the associated notion of
synvariance, we first quote Stachel from \cite{stachel1}:

\bigskip \noindent (Q4.9)\hskip 0.9in
\begin{minipage}{11cm}
\noindent ``{\small ...The general theory of relativity provides a
field-theoretical account of gravitation; moreover, {\em through
its incorporation of the principle of general covariance, it seems
to demand that any future physical theory that includes
gravitation be built on the foundation of the space-time
continuum...}}''\footnote{Our emphasis. We will return to comment
further on this point---that is to say, about doing field theory
ADG-theoretically, independently of a background spacetime
manifold, in subsection 7.5.}
\end{minipage}

\vskip 0.1in

\noindent and juxtapose it with the following remarks of the first
author in \cite{mall9}:

\bigskip \noindent (Q4.10)\hskip 0.9in
\begin{minipage}{11cm}
\noindent ``{\small ...We see that [in ADG] fundamental notions
(for example, connections) and relations of a similar nature, are
indeed independent of the base space of the sheaves involved. The
aforesaid {\em fundamental notions/relations are}, by definition,
{\em referred to the sheaf-spaces themselves}, thus, {\em not to
the corresponding base spaces}...}''
\end{minipage}

\vskip 0.1in

\noindent We take the foregoing discussion as an opportunity to
complete, in a Kleinian way, the third letter ($\mathrm{G}$) in
the symbolic
`$\mathrm{\underline{AD}G}\longleftrightarrow\underline{\struc\conn}
\mathrm{G}$'-pun expressed in (\ref{eqxx5}). We do this
pictorially by virtue of the following diagram\footnote{We wish to
thank Mrs Popi Mpolioti---the secretary of the Algebra and
Geometry section of the maths department of the University of
Athens (Greece)---for constructing using LaTeX and supplying us
with this diagram. This diagram appears in a slightly different
guise in the first author's latest book \cite{mall4}.}

{\small
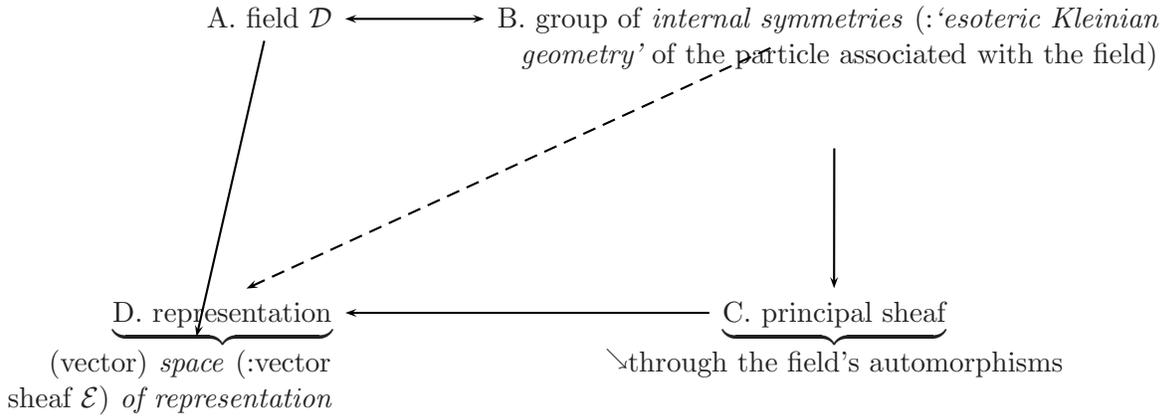
\[
\setlength{\arraycolsep}{1.1cm}%
\begin{array}{rcl}%
 \Rnode{a}{\text{A. field~$\conn$}}%
 &\Rnode{b}{\text{B. group of \emph{internal~symmetries}~\text{(:}\textit{`esoteric Kleinian }}}\\%
 \Rnode{c}{}
  & \Rnode{d}{\ \textit{geometry'}~\text{of~the~particle~associated~with~the~field)}}\\
 \Rnode{e}{}
  & \Rnode{f}{}\\
 \Rnode{g}{}
 & \Rnode{h}{}\\[2cm]%
 \Rnode{i}{\text{\qquad $\underbrace{\mathrm{D.~representation}}$}}
 & \Rnode{j}{\underbrace{\text{C. principal sheaf}}} \\
 \Rnode{k}{\text{(vector) \emph{space} (:\text{vector}}}
  & \Rnode{l}{\overset{\searrow}{} \text{through the field's automorphisms}}\\
 \Rnode{m}{\text{sheaf}~\modl)~\emph{of representation}}
 & \Rnode{n}{}
\end{array}
\everypsbox{\scriptstyle}%
\psset{nodesep=5pt,arrows=<->}%
\ncline{a}{b}\taput{}%
\psset{nodesep=5pt,arrows=->}%
\ncline{a}{k}\tlput{}%
\ncline{h}{j}\tlput{}
\ncline[linestyle=dashed]{b}{i}%
\psset{nodesep=5pt,arrows=<-}%
\ncline{i}{j}\tbput{}%
\]
}

\noindent which we readily explain now clockwise:

\begin{itemize}

\item A. In ADG, a field is the pair $(\modl ,\conn)$, with the (local) sections of $\modl$ representing,
from a geometric prequantization viewpoint, the (local)
quantum-particle states of the field.

\item B. In a Kleinian sense, this is the group of `internal' symmetries\footnote{In keeping with the notion of
synvariance, perhaps the epithet `internal' or `esoteric'(as
opposed to `external' or `exoteric', which is usually reserved for
the external symmetries of a background spacetime) is redundant
here as {\em there is no external, base spacetime (manifold)---a
realm independent of the fields themselves and the dynamics that
they define} (as differential equations, being themselves
connections). However, we will keep the epithet `internal' if
anything because it invokes ideas from gauge theory ({\it eg}, the
gauge symmetries of a system are usually referred to as {\em
internal symmetries} just to distinguish them from the external,
base spacetime manifold ones), and our ADG-theoretic perspective
on gravity has been coined `{\em pure gauge}' or `{\em genuinely
unitary}' field theory \cite{malrap3}. We will comment further on
ADG's genuinely unitary field theory, especially {\it vis-\`a-vis}
the evasion of smooth gravitational singularities and its
implications for QG, in the sequel (section 6).} of the particle
(states) associated with the field---in effect, these are the
self-transformations (technically speaking, the automorphisms) of
the field, living in $\mathrm{Aut}\modl$.

\item C. The said automorphisms are in fact organized into the principal (group) sheaf $\aut\modl$, which
is thought of as being represented by the (associated) vector
sheaf $\modl$ of the field---the carrier or representation
(vector) space of $\conn$, the space on which $\conn$
acts.\footnote{A space which is, by definition, locally isomorphic
to $\struc^{n}$, and constituting the `local geometry' of the
field ($\conn$)/particle ($\modl$). (Parenthetically, we see here
the traditionally quantum field-particle duality encoded in the
fundamental ADG-conception of field as a pair $(\modl ,\conn)$.
That is, {\em the ADG-notion of field is `self-dual'}. Later on,
in 6.2, we will make further remarks on its quantum-theoretic
significance.) Anyway, as a result of the local isomorphism
$\modl\stackrel{\mathrm{loc.}}{\simeq}\struc^{n}$, locally
$\aut\modl |_{U}=M_{n}(\struc(U))~(U~\mathrm{open~in}~X)$.}

\item D. This is the said representation sheaf $\modl$ associated
with the internal, `Kleinian auto-symmetries of the
particle-field' sheaf $\aut\modl$. Of course, since the field
$\conn$ {\em defines} the (vacuum Einstein gravitational) dynamics
on $\modl$ (\ref{eqy23}), one may think of the
`auto-transmutations' of the field in $\aut\modl$ as the
`invariances' (symmetries) of the said dynamics. In this sense
`synvariance' may be understood as Sorkin put it recently, in
\cite{sork7}, in the context of the `genealogical growth-dynamics'
of causets: ``{\em general covariance and becoming
coexist}''.\footnote{The term `becoming' here meaning, in analogy
to the causal sequential growth of causets in Sorkin's scheme, the
dynamical $\conn$-connection of local (`graviton') states
represented by the local sections of $\modl$ on which the field
$\conn$ acts.}

\end{itemize}

\noindent All in all, in view of the above, and in the Kleinian
sense of the word `geometry', we may complete the symbolic
pun-equivalence in (\ref{eqxx5}) to:

\begin{equation}\label{eqxx8}
ADG\longleftrightarrow \struc\conn\grouv
\end{equation}

In view of synvariance and the points A-D above, some remarks are
due now:

\begin{enumerate}

\item We first come to ask in an Einsteinian way, but rather
rhetorically: {\em What `space(time)' does the field} (or its
quanta-particles) {\em `see' or `feel' in the course of its
dynamics?}\footnote{And we call this kind of question Einsteinian,
because it recalls one of the primitive, gedanken-questions that
Einstein asked originally in formulating the theory of relativity.
In a slightly modified way, he pondered on the question: `{\em If
I was riding on a beam of light, what would I see---what would
spacetime look like?}'.} In view of synvariance, the answer is an
emphatic `{\em no space(time)!}'. If there is any {\em physical}
space(time) that $\conn$ `sees' or `feels' in its dynamical
self-propagation (autodynamics), it is the `{\em solution
space(time)}' of the law that $\conn$ obeys (in fact, that $\conn$
{\em defines!} as a differential equation) in the first place. In
a strong sense, {\em physical spacetime is inherent in the
$\conn$-dynamics, not an ambient `kinematical' realm that $\conn$
can see during its dynamical evolution---an ambient realm that can
host and at the same time delimit $\conn$'s possible moves
(kinematics)}.\footnote{The priority of dynamics over kinematics
will be discussed analytically in 3.2.5 next, and in connection
with Einstein's hole argument in 7.5.5.} In turn, that $\conn$
sees no {\it a priori} existing `space(time)' (like for instance a
manifold)---a {\em background geometry} (which, as we noted above,
is anyway inherent in $\struc$)---is reflected on the fact that
{\em the ADG-theoretic expression of Einstein's equations involves
the curvature $R(\conn)$ of the field ({\it viz.} field), which is
an $\otimes_{\struc}$-tensor, or better, an $\struc$-sheaf
morphism}. Thus, {\em the dynamics sees through our geometry
(:`spacetime') built into $\struc$, and is not affected at all by
it} (synvariance)--- {\it ie}, ultimately, $\struc$ {\em (:our
generalized measurements) plays absolutely no role in the
gravitational field dynamics} (:field realism). Our geometry (in
$\struc$) does not affect the field-dynamics.\footnote{This
`extreme' field realism does not preclude a quantum interpretation
for the `self-dual' ADG particle-field pair $(\modl ,\conn)$,
which we give in 6.2. It just puts into perspective the by now
standard `{\em external (to the quantum system) observer or
measurer dependence of physical reality}' that the standard
quantum theory appears to have forced on us.}

\item We now come to emphasize again that $\modl$, which by definition is
locally $\struc^{n}$, is the `representation sheaf' associated
with the principal group sheaf $\aut\modl$ of self-transmutations
(`esoteric symmetries') of the field. In turn, since from a
gemetric prequantization perspective $\modl$ is the `particle
representation sheaf' of the field, in the sense that local
sections of $\modl$ correspond to local ({\it ie}, localized in
`spacetime', or what amounts to the same, measured in $\struc(U)$)
quantum-particle states of the field, this picture justifies our
remark earlier that $\struc$ is the `geometric representation' of
the field---the `spati(otempor)al' ({\it ie}, localized in the
`space(time)' built into $\struc$), particle (quantum) aspect of
the field $\conn$ which is being captured exactly by the
$\struc$-valued local sections of $\modl |_{U}\simeq\struc^{n}(U)$
({\it ie}, vector $n$-tuples with entries in the arithmetics'
structure sheaf $\struc$ relative to a local gauge $U$ in $X$
\cite{mall1,mall2,malrap3}).

\item We come to stress again that since synvariance is modelled after
the self-transformations of the field in $\aut\modl$, the
ADG-theoretic conception of gravity is as a pure, genuinely
external spacetime-free, gauge field theory. All there is `out
there' is the (field-$\conn$ and particle-$\modl$ aspect of the)
field $(\modl ,\conn)$,\footnote{Again, for a discussion of the
self-quantum-dual (`self-complementary') particle-$\modl$ and
field-$\conn$ aspects of the ADG-field pair $(\modl ,\conn)$, the
reader should wait for 6.2.} and no spacetime (especially a
manifold) external to that field is involved at all. In turn, the
action of the principal $\aut\modl$ on its associated $\modl$
({\it ie}, on the local particle states of the field) makes the
field a `quantum fuzzy', `foamy' entity.

\item The last sentence above prompts to remark that {\em if any sort of
`noncommutative geometry' is involved at all in our scheme, that
is a `{\em noncommutative Kleinian geometry}' being (locally)
effectuated in our theory by}
$\aut\modl(U)=(M_{n}(\struc))^{\bull}(U)~(U~\mathrm{open~in}~X)$.\footnote{For,
interestingly enough, a rather canonical example of a non-abelian
$C^{*}$-algebra, which seems to crop-up frequently in Connes'
noncommutative geometry \cite{connes}, is $M_{n}(\cont(X))$---the
algebra of $n\times n$-matrices of continuous functions on a
locally compact space $X$, vanishing at infinity \cite{block}.
Thus, if any `noncommutative space(time) (geometry)' might creep
into our theory, that may as well be through the spectrum of
$M_{n}(\struc)$, in much the same way that, as noted earlier, a
(commutative) space(time) geometry is spectrally built into the
abelian structure sheaf $\struc$ (Gel'fand duality). It is also
interesting to note here that, in the general search for
noncommutative spaces, what figures prominently is the formulation
of a noncommutative version of the Gel'fand-Naimark representation
theory for commutative $C^{*}$-algebras---a search culminating in
noncommutative spaces known as $C^{*}$-quantales
\cite{mul,mulpel1,mulpel2}.}

\end{enumerate}

\subsubsection{The issue of the $\struc$-functoriality of the
gravitational dynamics and the PARD expressed categorically via
natural transformations}

Having in hand the remarks above about synvariance (the fields'
`{\em external spacetimeless auto-covariance}'), one might go a
bit further and claim that the `bottom-line' of ADG concerning the
expression of the (vacuum) Einstein equations in (\ref{eqy23}) is
that,

\bigskip \noindent (R4.15)\hskip 0.9in
\begin{minipage}{11cm}
\noindent The mathematical expression (:differential equation) for
the gravitational dynamics (in vacuo) is {\em functorial} with
respect to our generalized measurements (coordinates) in $\struc$.
Equivalently, the said expression involves mathematical quantities
(measurable dynamical variables commonly known as `observables')
that are $\otimes_{\struc}$-tensors\footnote{Again, with
$\otimes_{\struc}$ the homological tensor product {\em functor}.}
({\it alias}, `geometrical objects' which `by definition' our
generalized arithmetics, our measurements defining as it were our
`spacetime geometry' encoded in $\struc$), respect. Or perhaps
even better expressed in a sheaf-theoretic parlance, the said
`geometrical objects' are $\struc$-morphisms, the prime example
being the curvature $R(\conn)$ of the connection (field) $\conn$
involved in (\ref{eqy23}) \cite{mall10}. Effectively, this is the
content of the PGC of GR\footnote{As well as of the generalized
Principle of Relativity in (R2.?) before, and of Einstein's
`definition' of (objective) physical reality associated with the
latter (see footnote 5).} when viewed
ADG-theoretically.\footnote{The deeper physical meaning of this
functorial expression of the PGC of GR will be unveiled in 7.5.5
where we discuss {\it \`a la} Stachel the deeper significance of
Einstein's hole argument, albeit, from our ADG-theoretic
perspective.}
\end{minipage}

\vskip 0.1in

\noindent The remarks above about functoriality {\it vis-\`a-vis}
the PGC of GR and the `observables' involved in its
dynamics\footnote{Again, in our ADG-case, the Ricci curvature
involved in (\ref{eqy23}), which is what we call a `{\em
geometrical object}'---{\it ie}, {\em an $\struc$-sheaf morphism
or $\otimes_{\struc}$-tensor}---an object (better, map) which is
respected by our generalized measurements (arithmetics) in
$\struc$.} can be further supported by Baez's words taken from
\cite{baez5}:

\bigskip \noindent (Q4.11)\hskip 0.9in
\begin{minipage}{11cm}
\noindent ``{\small ...We can also express the principle of
general covariance and the principle of
gauge-invariance\footnote{Two notions that are not distinct from
each other (in fact, in a sense they are identical to each other!)
in our ADG-theoretic perspective on GR, since gravity according to
ADG is another gauge theory \cite{malrap3}---and what's more, a
`pure', `genuine', external (background) spacetime-less (or
spacetime-independent) gauge theory ({\it ie}, {\em a gauge theory
of the `third kind'}, as we shall see in 3.3 shortly).} most
precisely by saying that {\em observables are
functorial}.\footnote{Our emphasis. ADG has furthermore taught us
that it is important to ask {\em relative to what} ({\it ie}, to
what structure) {\em is something} (here the law of gravity) {\em
functorial?} ADG's answer is: {\em relative to our generalized
coordinates (`measurements') in} $\struc$.} So physicists should
regard functoriality as mathematical for `able to be defined
without reference to a particular choice of coordinate
system'\footnote{And what's more from an ADG-theoretic viewpoint,
the law of gravity is possible to be defined, as a differential
equation proper, without reference not only to a particular choice
of a system of smooth coordinates, but also to {\em any} smooth
coordinate frame with values in $\smooth_{M}$---{\it ie}, without
reference to smooth background spacetime manifold $M$.}...}''
\end{minipage}

\vskip 0.1in

\paragraph{Functoriality is not merely `freedom from coordinates'.}
There is a subtle point in the last sentence of the quotation
above and in the footnote that follows it that we would like to
discuss briefly here in order to avoid misunderstandings
concerning the significance of functoriality in the application of
ADG to gravity. The $\struc$-functoriality of ADG is not just
`coordinate independence' or `coordinate-free
definition/description' as one might be tempted to read Baez's
concluding remark above having in mind the so-called
coordinate-free descriptions of modern CDG. To be sure, in the
modern, coordinate-independent formulation of differential
geometry one is able to write the Einstein equations in a
coordinate-free manner---{\it ie}, without explicit commitment to
a particular reference system of (still though, $\smooth$-smooth!)
coordinates. However, in {\em actual calculations},
$\struc\equiv\smooth_{M}$ ({\it ie}, the background manifold $M$)
is invariably invoked, for after all, {\em the coordinate-free
modern differential geometry still is `CDG in disguise'}. For
otherwise how else could one, for instance, write down Einstein's
equations as {\em differential} equations proper?\footnote{Here is
again the differential manifold conservatism and monopoly.} The
fact is that $\smooth_{M}$ is always implicitly present there, and
when it is invoked to actually do CDG or Calculus ({\it ie}, in
actual, concrete calculations!), it brings along all the
differential geometric pathologies and anomalies ({\it eg},
singularities) that are inherent in the background manifold.

All this has already been anticipated and discussed in
\cite{mall7}. What we want to emphasize here is that the
$\struc$-functoriality of the gravitational dynamics formulated in
ADG-theoretic terms goes much deeper than a superficial `freedom
from coordinates'. Essentially, it means that {\em the dynamical
law of gravity is free from {\em any} coordinates---smooth
($\struc\equiv\smooth_{M}$) or other}, which in turn means, via
the interpretation that we have given to (the elements of)
$\struc$ as {\em our} generalized `measurements' of the
gravitational field and as {\em our} `spacetime (differential)
geometric' representation of it (which is inherent in $\struc$),
on the one hand that {\em the law of gravity}---{\it ie}, the
connection field $\conn$ defining it---{\em is `out there',
unaffected by our `measurement perturbations' and `geometrical
representations' (of it) in $\struc$} (PFR), and on the other,
that (\ref{eqy23}) {\em is a genuinely background spacetime
(manifold or not) independent description of gravitational
dynamics}\footnote{These two points will be further corroborated
shortly when we express the PARD in terms of natural
transformations.}---and what's more, {\em this still is a
differential geometric description proper}.

We will return to discuss the issue of `dynamical
$\struc$-functoriality' (as opposed to merely a `kinematical
freedom from coordinates' one)---especially in the light of first,
second, and what we call {\em third}, quantization---in 6.1.2, but
here we would like to express the PARD via the categorical notion
of {\em natural transformation}. The reader may already be aware
of the fact that the {\it raison d'\^{e}tre}, or even {\it de
faire}, of category theory is the notion of {\em natural
transformation}. For one may recall Saunders MacLane, one of the
founders of category theory, claiming in \cite{maclane} that

\bigskip \noindent (Q4.12)\hskip 0.9in
\begin{minipage}{11cm}
\noindent ``{\small ...Category theory was not invented to talk
about functors. It was invented to talk about {\em natural
transformations}\footnote{Our emphasis. The reader should note
that in earlier drafts of the paper we inappropriately used the
first-person, {\it ie}, `{\em I invented}', when, as a matter of
well known fact, category theory was the making of both Samuel
Eilenberg and Saunders MacLane. The second author apologizes for
the misquotation.}...}''
\end{minipage}

\vskip 0.1in

\noindent Now then, having established that the categorical notion
of {\em functor} plays a central role in the expression of
(vacuum) gravitational dynamics {\it \`a la} ADG as it ({\it ie},
functoriality with respect to the structure sheaf $\struc$)
effectively represents the PGC of GR in the more general and
abstract, as well as external spacetimeless, terms of ADG, we
further maintain that the even more important (at least according
to MacLane) notion of {\em natural transformation} has a `natural'
(pun intended!) physical realization in the PARD discussed above.
To explain this, recall first that, technically speaking, {\em a
natural transformation is a map $\natf$ between functors} (say for
example, functors $F$ and $G$ between two categories $C_{1}$ and
$C_{2}$) such that the following diagram commutes

\begin{equation}\label{eqxy1}
\bfig \putsquare<1`1`1`1;700`700>(700,700)[F(A)`F(B)`G(A)`G(B);
F(f)`\natf(A)`\natf(B)`G(f)] \efig
\end{equation}

\noindent with $A,B\in\mathrm{Obj}(C_{1})$,
$f\in\mathrm{Hom}(A,B)\in\mathrm{Arr}(C_{1})$,
$[F(A),F(B),G(A),G(B)]\in\mathrm{Obj}(C_{2})$, and
$[F(f)\in\mathrm{Hom}(F(A),F(B)),
G(f)\in\mathrm{Hom}(G(A),G(B))]\in\mathrm{Arr}(C_{2})$. {\it In
summa},

\begin{equation}\label{eqxyz1}
G(f)\circ\natf(A)=\natf(B)\circ F(f)
\end{equation}

\noindent Analogously, one may think of the aforementioned changes
of structure sheaves $\struc_{1}\mapto\struc_{2}$ involved in the
PARD as effecting natural transformation-type of changes between
the functorial expression of (vacuum) gravitational dynamics
(\ref{eqy23}) relative to $\struc_{1}$ and $\struc_{2}$,
respectively. Heuristically, one may cast this by the following
natural transformation-type of commutative diagram:

\begin{equation}\label{eqxy2}
\bfig
\putsquare<1`1`1`1;900`900>(900,900)[\struc_{1}(:\modl_{1}\stackrel{\mathrm{loc.}}{\simeq}\struc_{1}^{n})`\ricci(\modl_{1})=0
`\struc_{2}(:\modl_{2}\stackrel{\mathrm{loc.}}{\simeq}\struc_{2}^{n})`\ricci(\modl_{2})=0;
\otimes_{\struc_{1}}`\natf_{\struc}`\natf_{\conn}`\otimes_{\struc_{2}}]
\efig
\end{equation}

\noindent which we explain below:

\begin{enumerate}

\item First (upper left corner of the commutative square above),
one chooses $\struc_{1}$ to `catch' (solder/localize) and
(differential) geometrically represent (on
$\modl_{1}:\stackrel{\mathrm{loc.}}{\simeq}\struc^{n}$) the
gravitational field ({\it viz.} connection $\conn$).

\item Then (upper right corner), the gravitational dynamical law
that the field {\em defines} is expressed {\em functorially}, with
respect to the $\struc_{1}$ chosen, via the homological tensor
product functor $\otimes_{\struc_{1}}$---or what amounts to the
same, via the (Ricci) curvature tensor of the connection field,
which is an $\struc_{1}$-morphism: $\ricci(\modl_{1})=0$.

\item Then one considers a change $\natf_{\struc}:~\struc_{1}\mapto\struc_{2}$ of structure sheaf of
generalized arithmetics---{\it ie}, one assumes a different
$\struc_{2}$ ($\struc_{2}\not=\struc_{1}$) via which one can
(differential) geometrically represent the gravitational
connection field on another vector sheaf
$\modl_{2}:\stackrel{\mathrm{loc.}}{\simeq}\struc_{2}^{n}$, and
again express the gravitational law that the field obeys
$\struc_{\struc_{2}}$-{\em functorially} relative to the new
$\struc_{2}$, as: $\ricci(\modl_{2})=0$ (again, with
$\ricci(\modl_{2})$ an $\struc_{2}$-morphism).

\item Finally, the `natural transformation' $\natf_{\struc}$ of
structure sheaves is supposed to `induce' (or `lift' to) a
corresponding change $\natf_{\conn}$ between the
$\struc$-functorial expressions of the gravitational field law;
formally:
$\natf_{\conn}:~\ricci(\modl_{1})=0\mapto\ricci(\modl_{2})=0$.

\item {\it In toto}, heuristically, and in loose formal analogy to (\ref{eqxyz1}) above:

\begin{equation}\label{eqxyz2}
\otimes_{\struc_{1}}\circ\natf_{\struc}=\natf_{\conn}\circ\otimes_{\struc_{2}}
\end{equation}

\end{enumerate}

\noindent All in all, the PARD can be represented in ADG as a
`{\em natural transformation}' kind of change (`map') between the
$\otimes_{\struc}$-functorial expressions of the gravitational
field ($\conn$) dynamics (vacuum Einstein equations). The said
transformation $\natf$ is in fact a pair
$(\natf_{\struc},\natf_{\conn})$ of `adjoint' maps, with
$\natf_{\struc}$ the actual change of structure sheaves, and
$\natf_{\conn}$ the induced passage from the expression of
(vacuum) gravitational field ($\conn$) dynamics in the generalized
coordinates $\struc_{1}$, to the corresponding expression of the
{\em same} field\footnote{That is, same gravitational field
$\conn$. The PFR is implicit here (see next paragraph and
sub-subsection): the (gravitational) field $\conn$ (defining the
law of gravity as a differential equation) exists `out there'
independently of ({\it ie}, it remains unperturbed, `invariant'
with respect to) our various `geometrical representations' of it
when we employ different kinds of $\struc$ (and, in effect,
different associated/representation sheaves
$\modl\stackrel{\mathrm{ loc.}}{\simeq}\struc^{n}$).} dynamics in
$\struc_{2}$.\footnote{The expression (\ref{eqxyz2}) can be
formally stated as follows: {\em the structure sheaf morphism
(change) $\struc_{1}\mapto\struc_{2}$ `lifts' to a natural
transformation-type of map between the corresponding homological
tensor product functors $\otimes_{\struc_{1}}$ and
$\otimes_{\struc_{2}}$ respectively, with respect to which the
dynamics is expressed via the respective
$\otimes_{\struc}$-tensors $\ricci$}.} As noted earlier in
footnote ???, underlying PARD is this `{\em categorical
transformation theory}' expressed via natural transformation-type
of maps.

Of course, since, as we saw before, {\em the choice of generalized
arithmetics $\struc$ is ours and not Nature's own}, one should be
cautious and `critical' of the epithet `natural' in the term
`natural transformations' ({\it ie}, `natural' is not `Natural';
pun intended). On the other hand, the PARD, viewed in the manner
above as a natural transformation ($\natf_{\conn}$) between the
$\struc$-functorial gravitational dynamics for different choices
(changes) of $\struc$ ($\natf_{\struc}$), reveals something quite
natural indeed, namely, that {\em with respect to the
gravitational dynamics}---essentially, with respect to the
gravitational field $\conn$ defining that dynamics (as a
differential equation proper)---{\em all choices of
structure/coordinate sheaf $\struc$ are on an `equal footing', to
the effect that the dynamics itself is indifferent to and
unaffected by} (`invariant' so to speak, {\it ie},
$\struc$-functorial with respect to) {\em any such choice}. To
stress it once more, the gravitational field dynamics `sees
through' our coordinates, our generalized measurements
(:perturbing acts of localization) of the gravitational field
({\it viz.} connection) which are organized in $\struc$. Which
brings us to the deeper meaning of the PARD.

\subsubsection{What is being `relativized' and what remains
`invariant' by PARD?: the Principle of Field Realism (PFR)
abolishing the differential manifold as the last relic of an
ether-like substance in the manifold based GR}

Plainly, what is being relativized by the PARD is $\struc$---the
structure sheaf of our generalized coordinates (`field
measurements'), while since all DG boils down to $\struc$ as we
argued earlier, {\em what is in effect relativized is
`differentiability' itself}. That is to say, the {\it a priori}
(and in that sense, absolutely!) fixed `classical'
differentiability or smoothness that comes with assuming
$\struc\equiv\smooth_{M}$ for structure sheaf,\footnote{Or
equivalently, {\it a priori} fixing the base space(time) to a
differential manifold (\ref{eq1}) securing thus the
CDG-conservatism and monopoly.} is now relativized by allowing
other structure algebra sheaves, different from the classical one
of $\smooth_{M}$, be used in (A)DG.\footnote{Once again, ``{\em
differentiability is independent of smoothness}'' \cite{malrap2}.}
At the same time, as briefly noted above, what remains
`form-invariant' so to speak is the differential
equation-expression of the field dynamics, with the gravitational
connection field $\conn$, `underlying' the Ricci curvature
$\struc$-tensor, `remaining intact at the background'---being
unaffected by these changes of $\struc$. It follows that the
gravitational field $\conn$---and, {\it in extenso}, the
gravitational field law (vacuum Einstein equations) that it
defines---remains unaffected, `unperturbed' by our generalized
`measurements' (of it) in $\struc$. {\it In toto}, ADG allows us
not only to generalize the PGC of GR to synvariance by abolishing
the external, base spacetime continuum (manifold), but also, via
the natural transformation-type of representation of the PARD, to
generalize Einstein's PR\footnote{Again, see footnote 5.} in a way
that it suits the fundamental (gravitational) field solipsism
underlying synvariance, in the following way:

\bigskip \noindent (R4.16)\hskip 0.9in
\begin{minipage}{11cm}
\noindent \ovalbox{\bf PFR:} {\em the gravitational field}
($\conn$), at least in its differential geometric expression
(representation) via its curvature $\curv(\conn)$ in the dynamical
(differential) equation modelling the physical law of gravity in
ADG, {\em remains unaffected (`unperturbed')\footnote{And the
corresponding dynamical law is $\struc$-functorial.} no matter
what $\struc$ one employs to `geometrically represent', `measure',
`localize', or `arrest'/`freeze' it (in spacetime)}. In this sense
we say that at the base of the PARD lies the PFR.\footnote{Of
course, as we shall argue extensively in 6.2 subsequently, the
`genuinely unitary' particle-field pair $(\modl ,\conn)$ is indeed
`internally perturbed'---in a quantum-theoretic sense---by our
`geometrization' or `localization' actions in (each chosen)
$\struc$. Accordingly, the field itself is `{\em self-quantum}',
or even `{\em quantum foamy}', being (dynamically) self-transmuted
by $\aut\modl$, but {\em our} (differential) geometric expression
of the law that the field obeys (or defines as a differential
equation) is $\struc$-invariant (functorial) indeed.}
\end{minipage}

\vskip 0.1in

\subsubsection{Kinematics before dynamics, or vice
versa?; `dynamicalization' of coordinates and space(time):
space(time) structure (`geometry') is `inherent' in (is a
consequence of) the dynamical field (`algebra');
`$\conn\Longrightarrow R(\conn)$'}

The fundamental background spacetimelessness and `synvariance',
the $\struc$-functoriality, the PARD, and its deeper consequence,
the PFR (and field autonomy and solipsism), discussed above,
especially in view of our remarks about the role that the smooth
coordinates play in GR in 2.2.1 and 2.2.2 before, in a deep sense
force us to posit that in the ADG perspective on gravity (GR), and
contrary to the current theoretical physics `consensus, {\em
dynamics is prior to kinematics}. Let us explain this by first
arraying certain pertinent quotations that apparently point to the
opposite.

We read from \cite{einst9} that Einstein, in formulating GR, in a
clear and direct way posited what one would now call {\em the
priority of kinematics over dynamics}---the basic idea there being
that one should set up the kinematical variables {\em before}
prescribing the dynamical equations of motion that these variables
satisfy:

\bigskip \noindent (Q4.13)\hskip 0.9in
\begin{minipage}{11cm}
\noindent ``{\small ...The major question for anyone searching in
this field is this: Of which mathematical type are the variables
(functions of the coordinates) that permit the expression of the
physical properties of space (`structure')? {\em Only after
that:}\footnote{Our emphasis.} What equations are satisfied by
those variables?}''

\end{minipage}

\vskip 0.1in

\noindent Einstein then goes on and describes the transition from
kinematics to dynamics in GR as the theoretical procedure
involving first the kinematical structure of SR---{\it ie}, the
flat, (gravitational) field free Minkowski space, with its
`observable' (in his own words, ``{\em measurable, with real
physical meaning, quantity}'')
$ds^{2}=g_{\mu\nu}dx_{\mu}dx_{\nu}$---the infinitesimal distance
between spacetime events involving the quintessential
chronogeometric structure of Minkowski space: $g_{\mu\nu}\equiv
\eta_{\mu\nu}$. He then recognizes that the transition from SR to
GR essentially involved the `dynamicalization' of the metric
structure of SR---that is, of $g_{\mu\nu}$. {\it In summa}, and in
view of the quotation (Q?.?) above,

\bigskip \noindent (R4.17)\hskip 0.9in
\begin{minipage}{11cm}
\noindent Einstein's approach to GR (starting from SR) set the
paradigm (or at least the trend) for almost all physical theory
formation attempts in the sequel, namely, {\em by first delimiting
what can possibly happen}---the structure of the nowadays called
{\em kinematical space} of the theory---as well as the
to-be-dynamically-variable quantities in it, {\em and then what
actually happens}---the {\em dynamical equations} satisfied by
those variables; or in other words, the {\em dynamics} of the
theory. In this sense, it is fair to say that in every physical
theory-construction procedures nowadays, {\em kinematics comes
before dynamics}.\footnote{See (Q?.?) below where Sorkin in
\cite{sork1} `quotes' Taketani from \cite{taketani}.}

\end{minipage}

\vskip 0.1in

\noindent The description above is tautosemous to the by now well
established fact that, while SR is in a strong sense all about
(Minkowski) spacetime kinematics,\footnote{That is, the
chronogeometric structure of Minkowski space is fixed and purely
kinematical, not dynamical (in pictorial geometrical terms,
Minkowski space is flat).} GR is a `dynamicalization' of SR by
making the latter's main kinematical entity---the spacetime
manifold's metrical structure $g_{\mu\nu}$---a dynamical
variable.\footnote{From which it follows that the kinematical
space of GR, at least in its original formulation in terms of
$g_{\mu\nu}$, is the space of Lorentzian metrics (`superspace').}

More recently, this priority of kinematics over dynamics in
setting up a physical theory has been emphasized by Sorkin in
\cite{sork1}, where he describes `historically' the conceptual
development and construction of his `discrete' causet theoretical
scenario for QG. Sorkin, inspired by the writings of Taketani
\cite{taketani} on the various stages through which a physical
theory (such as Newtonian mechanics) passes during its upbringing,
notes characteristically:

\bigskip \noindent (Q4.14)\hskip 0.9in
\begin{minipage}{11cm}
\noindent \noindent ``{\small ...Equally important [in the
construction of causet theory]\footnote{Our addition for
completeness.} however, was the influence of M. Taketani's
writings, which convinced me that {\em there is nothing wrong with
taking a long time to understand a structure `kinematically'
before you have a real handle on its dynamics}...}''
\end{minipage}

\vskip 0.1in

\noindent And even more recently, and importantly for what we want
to highlight here in the context of the ADG-formulation of
gravity, Sorkin clearly delineates the kinematical structure of GR
\cite{sork2}:

\bigskip \noindent (Q4.15)\hskip 0.9in
\begin{minipage}{11cm}
\noindent \noindent ``{\small ...In the case of General
Relativity, the Kinematics comprises a {\em differentiable
manifold $M$} of dimension four, a {\em Lorentzian metric}
$g_{ab}$ on $M$, and a structure which, although it is closely
intertwined with the metric, I want to regard as distinct, namely
the {\em causal order-relation} $\prec$\footnote{All emphasis
above is ours.}...}''
\end{minipage}

\vskip 0.1in

\noindent In what follows in the present paper, we will see that
in the ADG-perspective on gravity, all the three `traditional'
kinematical structures of GR that Sorkin outlines above are
replaced by the single notion of {\em connection} (field) $\conn$,
and more importantly, that this notion effectively reverses the
roles of kinematics and dynamics in the sense that {\em in
ADG-gravity dynamics comes before kinematics}. To dwell a bit more
on these remarks without going into any detail for the time being,
first, as we have been arguing numerous times above, {\em in ADG
there is no base differential manifold}. Second, as we will see
soon, the metric is not {\em the} fundamental dynamical variable
in ADG-gravity, {\em the connection is}. And third, we shall see
that `causality', in a generalized (abstract) sense, is already
represented by $\conn$---{\it ie}, {\em the causal connection
(nexus) is already encoded in the gravitational
$\struc$-connection $\conn$} (pun intended). In addition, we will
see as a consequence of these replacements that even the
fundamental in GR notion of {\em spacetime events} is substituted
in ADG by the notion of the (autonomous) field $(\modl ,\conn)$.
The bottom-line is that in ADG the basic notion of
$\struc$-connection (field) is a dynamical one, since $\conn$ {\em
defines} (via its curvature) the dynamical (differential)
equations of Einstein; moreover, by virtue of the
$\struc$-functoriality of that dynamics, whatever `background
kinematical spacetime geometry' is inherent in $\struc$, it is
`transparent' to $\conn$---the field `sees through it' and remains
undisturbed by it.\footnote{The epitome here, especially {\it
vis-\`a-vis} singularities, is that no matter what singularities
are present in the structure sheaf $\struc$ that one chooses to
employ, these pose no obstacle or `breakdown threat' to the
dynamical law of gravity defined by the field $\conn$. The latter
`sees through' singularities and remains virtually unperturbed by
them (DGS beware!).}

Of course, in ADG we still retain the traditional (in physics)
Aristotelian categories of {\em possibility} (in the form of
kinematics) and {\em actuality} (in the form of dynamics) in that
we still possess a different, more abstract notion of
gravitational kinematics in ADG-gravity, since now the possibility
space is the affine space $\sconn_{\struc}(\modl)$ of
$\struc$-connections;\footnote{Or to be more precise, in view of
synvariance, the moduli space $\sconn_{\struc}(\modl)/\aut\modl$
\cite{malrap3}. Of course, it must be noted here that in the
current Ashtekar connection-based approaches to both classical and
quantum gravity, the moduli space of smooth connections modulo
$\mathrm{Diff}(M)$ is regarded as the relevant gravitational
kinematical space, but still this has obviously not been entirely
freed yet from the `geometrical shackles' of a background
differential manifold \cite{malrap3}.} furthermore, the said
(vacuum) Einstein equations can be derived from varying
(extremizing) the ADG-analogue of the Einstein-Hilbert action
$\eh$, which is an $\struc$-valued functional on the said abstract
kinematical space of `gauge equivalent' connections
\cite{malrap3}. Yet, because there is no fixed background
differential manifold in ADG-gravity, we are reluctant to give the
terms gravitational kinematics and dynamics the same meaning that
they have in the usual background $M$-dependent theory (whether
classical or quantum). For one thing, in view of the PFR the
dynamical gravitational field $\conn$ exists autonomously `out
there' and it is {\em us}---the external (to the field) observers
(experimenters or `measurers'/`geometers')---that bring along our
own $\struc$, with the kinematical-geometrical space(time) built
into it, in order to `geometrize' ({\it ie}, geometrically
represent) it. Arguably, kinematics lies with the (external)
experimenter (the `exosystem' \cite{df1}), who in the course of an
{\it a priori} planned experiment `constrains' or `dictates' to
the system its possible moves/paths (within the `geometrical
spacetime' framework inherent in the $\struc$ that she chooses
{\it ab initio}), not with the dynamical `system' itself (the
`endosystem' \cite{df1})---the algebraic $\struc$-connection
$\conn$ which {\it a priori} knows no (spacetime) geometry. Of
course, in view of the $\struc$-functoriality of dynamics, the
field itself (and the dynamical law that it defines) is in no way
actually constrained by the $\struc$ that the external
experimenter `imposes' on it---{\it ie}, once again, the
gravitational dynamics is in no way conditioned, let alone
impeded, by any ambient, external (background to the field)
kinematical/geometrical space(time) that {\em we}, the external
`observers', adopt to geometrically picture (realize) it. This is
the essence of the PFR in ADG-gravity.

But without carrying any further this discussion, since anyway we
are going to encounter again these issues in section 7 in
connection with the ADG's perspective on Einstein's hole argument,
let us summarize the traditional way of thinking about kinematics
and dynamics.

\paragraph{The traditional or `classical' analogy: `kinematics before
dynamics'.} We may summarize the above by the following
`quotient':

\[
\mathrm{Dynamical~histories}=\frac{\mathrm{Kinematical~histories}}{\mathrm{Equations~of~motion}}\Leftrightarrow
\frac{\mathrm{Possibility~space}}{\mathrm{Dynamical~law}}=\mathrm{Actuality~space}
\]

\noindent which may be formally and pseudo-technically read as
`{\em dynamics (what actually happens) is the `moduli space' of
the kinematical space (what can possibly happen) factored out by
the dynamical equations of motion}', which in turn presupposes
that the structure of the kinematical space has already been
charted before the dynamical equations cut through (or select
from) it the `orbits' (dynamical histories or tracks) that are
actually materialized---as it were, the geometrical footprints
left by the system as it dynamically evolves through its
`pre-existent' ({\it a priori} determined and fixed) kinematical
space.

Of course, regarding the notion of the geometrical background
spacetime kinematics in GR, again we witness Einstein's remarkable
insight into the fundamental concepts and workings of his
relativistic field theory of gravity (GR) in the following words
found in article 12 in \cite{einst10}, titled `{\itshape Time,
Space, and Gravitation}':

\bigskip \noindent (Q4.16)\hskip 0.9in
\begin{minipage}{11cm}
\noindent \noindent ``{\small\em ...In the generalized theory of
relativity, the doctrine of space and time kinematics is no longer
one of the absolute foundations of general physics. The
geometrical states of bodies and the rates of clocks depend in the
first place on their gravitational fields, which again are
produced by the material system concerned\footnote{Our emphasis
throughout.}...}''
\end{minipage}

\vskip 0.1in

\noindent Arguably, the quotation above hints at the fact that,
for Einstein too, the dynamical field (of gravity) is prior to an
{\it a priori} fixed `geometrical spacetime kinematics' (pun
intended), which should thus {\em not} be regarded as being prior
to dynamics and lying at the basis---as it were as an absolute,
up-front fixed foundational structure---in the construction of any
physical theory.\footnote{It is perhaps the aftermath of the
gedanken `{\em hole argument}' that Einstein had originally come
up with in order to `test' the validity of the PGC of GR, that led
him to review his theoresis of gravitational kinematics and
dynamics as (Q?.?) and (Q?.?) show. It is on the basis of this
argument, whose deeper meaning has been explored and explained in
a long series of works by John Stachel
\cite{stachel4,stachel3,stachel5,stachel0,stachel9,stachel6,stachel7,stachel2},
that we shall later further support our claim that in ADG-gravity
dynamics comes before kinematics (8.5.5, 8.5.6).}

Let us wrap up this traditional kinematics-before-dynamics
`principle' of theory-construction with the following words from
\cite{cao} corroborating what has been said above:

\bigskip \noindent (Q4.17)\hskip 0.9in
\begin{minipage}{11cm}
\noindent ``{\small ...{\em Generally speaking, a dynamical
theory, regardless of its being a description of fundamental
interactions or not, must \underline{presume} some geometry of
space for the formulation of its laws and interpretation. In fact
a choice of a geometry predetermines or summarizes its dynamical
foundations, namely, its causal and metric
structures}.\footnote{Our emphasis.} For example, in Newtonian (or
special relativistic) dynamics, Euclidean (or Minkowskian)
(chrono-) geometry with its affine structure, which is determined
by the kinematic symmetry group (Galileo or Lorentz group) as the
mathematical description of the kinematic structure of
space(time), determines or reflects the inertial law as its basic
dynamical law. {\em In these theories, the kinematic structures
have nothing to do with the dynamics. Thus dynamical laws are
invariant under the transformations of the kinematic symmetry
groups. This means that the kinematic symmetries impose some
restrictions on the form of the dynamical laws. {\em However, this
is not the case for general relativistic theories}.\footnote{Our
emphasis again. We will come back to comment on this important
remark from the vantage of Einstein's hole argument (as exposed by
Stachel) in the last section.} In these theories, {\em there is no
a priori kinematic structure of spacetime},\footnote{Something
that appears to contradict Sorkin's quotation (Q?.?) above. There
is no real contradiction if one waters down the `{\it a priori}'
above: a smooth manifold (and its diffeomorphism) group is indeed
part of the kinematical structure of GR if one considers on the
one hand the virtual (theoretical) constraint of requiring the
Einstein equations to be {\em differential} equations proper, and
on the other, the actual $\mathrm{Diff}(M)$-constraints when
(quantum) gravity is regarded as a constrained (quantum) gauge
theory.} and thus there is no kinematic symmetry and no
restriction on the form of the dynamical laws. That is, the
dynamical laws are valid in every conceivable four-dimensional
topological manifold, and thus have to be generally covariant. It
should be noted, therefore, that the restriction of GC upon the
form of dynamical laws in general relativistic theories is
different in nature from the restrictions on the forms of
dynamical laws imposed by kinematic symmetries in non-general
relativistic theories}\footnote{Again, our emphasis. Again, please
wait for our ADG-treatment of Einstein's hole argument in the last
section.}...}''
\end{minipage}

\vskip 0.1in

\paragraph{Turning the tables around ADG-theoretically: `dynamics prior to kinematics' from `field
solipsism'.} [RETURN]

\[
\begin{array}{c}
\mathrm{Algebraic~connection~is~the~cause~of~geometric~curvature}\Leftrightarrow\conn\Longrightarrow
\curv(\conn)\cr
\mathrm{The~dynamical~field~entails~geometry}\Leftrightarrow
\mathrm{Geometry~is~inherent~in~the~dynamical~field}\cr
\mathrm{Sheaf-cohomologically}: (\modl
,\conn)\Leftrightarrow(\conn ,\curv(\conn))
\end{array}
\]

\texttt{Perhaps mention here Stachel's ideas on the
`dynamicalization of coordinates' from \cite{stachel2}.}

\subsection{GR \`a la ADG: The Connection $\conn$ is Primary, the Metric $\mathbf{g}$ Secondary}

We commence this subsection with the basic observation, already
emphasized in \cite{malrap3}, that {\em in the ADG-picture of GR
the fundamental---in fact, the sole---dynamical variable is the
connection $\conn$}. This must be contrasted against the original,
(pseudo)-Riemannian geometry based formulation of GR by Einstein,
in which the gravitational potentials were identified with the ten
components of the metric tensor $g_{\mu\nu}$. Of course, it is
well known that soon after the original `metric formulation' of
GR, there were people, such as Eddington and Cartan for example,
who tried alternative scenarios for gravity in which the
fundamental dynamical variable was assumed to be the affine
connection instead of the metric. Weyl, for instance, in early
attempts to unite gravity with electromagnetism, thought of
weakening the Riemannian idea of an invariant infinitesimal
distance by replacing it with relative distance, in the sense that
only ratios of distances, not (`absolute') distance itself, should
be invariant, thus anticipating the (`local') gauge principle.

It must be emphasized here, however, that Einstein himself, on
many occasions, alluded to the physical significance of the notion
of (the Levi-Civita-Christoffel) connection in GR, and sometimes
he went as far as to almost regard it as more important than the
spacetime metric, as the following quotations show:

\bigskip \noindent (Q4.18)\hskip 0.9in
\begin{minipage}{11cm}
\noindent \noindent ``{\small ...The development just sketched of
the mathematical theories essential for the setting up of general
relativity\footnote{Here Einstein alludes to the work of Gauss
and, more importantly, Riemann, on the (differential) geometry
({\it eg}, the study of the curvature) of (high-dimensional)
metric spaces `in themselves'---{\it ie}, independently of whether
or not these spaces are embedded in higher-dimensional (Euclidean)
spaces.} had the result that {\em at first the Riemannian metric
was considered the fundamental concept on which the general theory
of relativity and thus the avoidance of the inertial system were
based. Later, however, Levi-Civita rightly pointed out that the
element of the theory that makes it possible to avoid the inertial
system is rather the infinitesimal displacement field
$\Gamma^{l}_{ik}$. The metric or the symmetric tensor field which
defines it is only indirectly connected with the avoidance of the
inertial system in so far as it determines a displacement
field}\footnote{Our emphasis.}...}'' \cite{einst3}
\end{minipage}

\vskip 0.1in

\noindent At the same time, it is well known that in the CDG-based
Riemannian geometry it is precisely the auxiliary requirement (or
condition) that the connection be compatible with the
metric\footnote{Which defines $\conn$ as a `{\em metric
connection}'. In turn, this metric compatibility condition on the
connection implies that the gravitational field is {\em
torsionless}.} which establishes an `equivalence' between the
original metric formulation of GR and the latter's `connection
picture', as we read for example from \cite{hughston}:

\bigskip \noindent (Q4.19)\hskip 0.9in
\begin{minipage}{11cm}
\noindent ``{\small ...In relativity theory it is convenient to
distinguish three levels or layers of geometry, with increasing
structure and complexity: (a) the underlying differentiable
manifold, (b) the connection, (c) the metric...For Riemannian
geometry (b) and (c) are made compatible, in a special sense, by
the requirement $\nabla_{a}g_{bc}=0$, with $\nabla_{a}$
torsion-free...{\em Remarkably, these conditions are sufficient to
determine $\nabla_{a}$ uniquely in terms of $g_{ab}$, and it is
this natural link between metric and connection that constitutes
the foundation of Riemannian geometry}\footnote{Our
emphasis.}...}''
\end{minipage}

\vskip 0.1in

\noindent Motivated by the quotation (Q?.?) above, we digress a
bit and note that the bottom `layer (a) of geometry', namely, the
differential manifold ($\smooth$-) structure, is left intact and
quite independent by this `identification' of the `connection
layer' (b) with the `metric layer' (c), which are treated on a par
({\it ie}, they are viewed as being equivalent by the torsion-free
connection condition). We may cast this CDG-based picture of
Riemannian geometry by the following {\large $\top$}-shaped `{\em
geometrical}' ({\it ie}, metric based) diagram:

{\fontsize{0.19in}{0.19in}
\begin{equation}\label{eq15}
\begin{CD}
\boxed{\mathbf{Connection~\nabla}}\longleftrightarrow\boxed{\mathbf{Metric~g_{\mu\nu}}}\\
\hskip 0.31in @A\mathrm{metric~compatibility}A\mathrm{of~the~connection}A\\
\boxed{\mathbf{Differential~manifold~structure}}
\end{CD}
\end{equation}}

\noindent In contradistinction, in ADG, in a deep sense this
{\large $\top$}-shaped diagram of the CDG and smooth
manifold-dependent structure of Riemannian {\em geometry} is
inverted ({\it ie}, it looks upside-down!) to the following
{\large $\perp$}-shaped {\em algebraic} ({\it ie},
$\struc$-connection based) one

{\fontsize{0.19in}{0.19in}
\begin{equation}\label{eq16}
\begin{CD}
\hskip 0.2in \boxed{\mathbf{\struc\!\!-\!\!metric~\mathbf{\rho}}}\\
\hskip 0.35in @A\mathrm{connection~compatibility}A\mathrm{of~the~metric}A\\
\boxed{\mathbf{Structure~sheaf
~\struc}}\dashleftarrow\dashrightarrow\boxed{\mathbf{Connection~\conn}}
\end{CD}
\end{equation}}

\noindent which we briefly explain below:

\begin{enumerate}

\item First we note that the differential manifold $M$ in
(\ref{eq15}) is replaced in (\ref{eq16}) by the structure sheaf
$\struc_{X}$ of `differential algebras' of generalized coordinates
(`arithmetics') over an arbitrary topological space $X$, which
itself carries no differential structure whatsoever {\it a
priori}.\footnote{Of course, as noted earlier, and in the
classical case, one recovers $M$ (in effect, one identifies it
with $X$) when $\struc_{X}\equiv\smooth_{M}$. Yet, the same can
still happen for suitably chosen `geometric topological algebra
sheaves' $\struc$ \cite{mall1,mall8}, $\smooth_{X}$ being of
course a very special case.} In (\ref{eq15}), and in accordance
with (\ref{eq7}), the connection and the metric structures are
imposed on the manifold's $\smooth$-smooth CDG-structure,
inheriting thus the usual epithet `{\em smooth}' in front, as in
`{\em smooth connection}' and `{\em smooth metric}'. By contrast,
in ADG, the differential structure, primarily represented by the
differential $\partial\equiv d^{0}$, `derives' from $\struc$
itself; moreover, more importantly, it was ADG that first
appreciated that the differential operator $\partial$ is nothing
else but a particular kind of {\em connection}---albeit, as
mentioned in 3.1, a flat one.\footnote{Then, as we saw in 3.1, one
generalizes $\partial$ ADG-theoretically to a `curved connection'
$\conn$.} The dashed bidirected arrow in (\ref{eq16}) just
signifies that in ADG the notion of connection ({\it viz.},
generalized differential $\partial\equiv d$) is an algebraic one.
In other words, the connections defined in ADG are `{\em algebraic
connections}'\footnote{Indeed, the term used by ADG to refer to
the $\conn$s defined and studied in it is {\em
$\struc$-connections}
\cite{mall-3,mall-2,mall1,mall2,vas1}.}---{\it ie}, inextricably
tied to (ultimately, coming from) the structure sheaf of
differential algebras $\struc$. This is another corroboration of
the title of 3.2, namely, that {\em all differential geometry
boils down to} $\struc$, for it is plain that {\em one cannot do
differential geometry without a differential operator $d$ (or more
generally, without a connection $\conn$), and the latter
originates with or has as its `source' (action domain)
$\struc$}.\footnote{For one thing, recall K\"{a}hler: essentially,
{\em every commutative unital ring comes equipped with a
differential} \cite{mall2,malrap2,malrap3}.} These remarks are
also in line with the brief historical account of the initial
development of ADG and its `precursors' we gave in 3.1 according
to which the first author's original aim upon formulating ADG was
{\em to arrive at an entirely algebraic notion of (generalized)
differential ({\it ie}, connection) by sheaf-theoretic means,
without the use of any underlying smooth manifold}, thus, in
effect, {\em to do differential geometry without the use of any
Calculus, or anyway, without any classical notion of smoothness in
the traditional $\smooth$-sense---that supported and furnished by
a locally Euclidean base space (manifold)}.

\item Secondly, we come to the fundamental difference between
ADG's perspective on GR and the CDG-based Riemannian geometry
underlying GR: as noted above, while in the latter the main
geometrical structure is the smooth spacetime metric $g_{\mu\nu}$
which is thus regarded as the basic dynamical variable in the
corresponding physical theory (GR), in the former the basic,
purely algebraic, structure is the connection $\conn$, which is
thus being regarded as the sole dynamical entity in the algebraic
({\it ie}, sheaf-theoretic) ADG-picture of GR
\cite{mall3,malrap3}.

This difference is partially reflected in the asymmetry between
the {\large $\perp$}- and the {\large $\top$}-shaped diagrams
above, as follows: {\em while in Riemannian geometry one makes the
connection compatible with the metric, with the latter held as
being more fundamental than the former, in ADG one goes the other
way around and requires that the $\struc$-metric be compatible
with the $\struc$-connection, which is thus assumed to be more
basic than the former}. {\it Prima facie}, this reversal might
appear to be only formal and rather superficial---as it were, it
might remind one of the circular question

\vskip 0.1in

\centerline{\em which came first, the egg or the chick?,}

\vskip 0.1in

\noindent but {\em conceptually} we believe that there is much
more to it. For recall that ADG-theoretically the $\struc$-metric
$\rho$ is an $\struc$-bilinear $\struc$-valued sheaf morphism.
Being an $\struc$-morphism, like the curvature $\curv(\conn)$ of
the connection $\conn$, $\rho$ {\em is a geometrical object} ({\it
ie}, a $\otimes_{\struc}$-tensor), as it respects our generalized
arithmetics in $\struc$ \cite{malrap3}. On the other hand, {\em
the connection $\conn$, being a $\cons\equiv\mathbf{K}$- but not
an $\struc$-morphism, is not a geometrical object}; rather, {\em
it is an algebraic notion which eludes purely geometrical
characterization} \cite{malrap3}.\footnote{As repeatedly
emphasized in \cite{malrap3}, from a gauge-theoretic viewpoint,
the non-geometrical character of the connection, as opposed to its
curvature, corresponds to the fact that, while the former
transforms inhomogeneously (affinely) under a gauge
transformation, the latter transforms homogeneously (tensorially).
In other words, the connection is not a tensor under a symmetry
(local gauge) transformation, hence it is not a geometrical object
proper.} In the context of GR, the physical importance of the
non-tensorial nature of the connection, as opposed to the
tensorial character of either the (connection-compatible) metric
or the (connection's Riemann) curvature (tensor) has been
emphasized by Stachel \cite{stachel4,stachel3}:\footnote{The
second author would like to thank cordially John Stachel for
communicating to him, quite timely with respect to the writing of
the present paper, the two papers \cite{stachel4,stachel3}.}

\bigskip \noindent (Q4.20)\hskip 0.9in
\begin{minipage}{11cm}
\noindent ``{\small ...Within a few years, Levi-Civita, Weyl and
Cartan generalized the Christoffel symbols to the concept of
affine connection. This concept served to make the relationship
between the mathematical representations of various physical
concepts much clearer. {\small\em Just because it is not a tensor
field, the connection field provides an adequate representation of
the gravitational-cum-inertial field required by Einstein's
interpretation of the equivalence principle. There is no (unique)
decomposition of the connection field into an inertial connection
plus a gravitational tensor field}\footnote{Our emphasis.}...}''
\cite{stachel3}
\end{minipage}

\vskip 0.1in

\noindent who continues, in support of the quotation above, by
quoting Einstein from a 1950 letter to Max Von Laue:

\bigskip \noindent (Q4.21)\hskip 0.9in
\begin{minipage}{11cm}
\noindent ``{\small\em ...what characterizes the existence of a
gravitational field from the empirical standpoint is the
non-vanishing of the {\rm\small [components of the affine
connection]}\footnote{Stachel's addition.}, not the non-vanishing
of {\rm\small [the components of the Riemann
tensor]}.\footnote{Again, Stachel's addition.} If one does not
think in such intuitive ({\it Anschaulich}) ways, one cannot grasp
why something like curvature should have anything at all to do
with gravitation. In any case, no rational person would have hit
upon anything otherwise. The key to the understanding of the
equality of the gravitational and inertial mass would have been
missing\footnote{Our emphasis throughout.}...}''
\end{minipage}

\vskip 0.1in

\noindent In fact, Stachel goes a step further and argues, in
retrospect of course, that had the notion of connection, in
contradistinction to the metric of Riemannian geometry, been
developed around when Einstein was laying the foundations of his
relativistic field theory of gravitation, he would have chosen
{\em it} instead of the metric as the `natural' mathematical
representation of the gravito-inertial field as mandated by the
EP. In this respect, we read from \cite{stachel11}:

\bigskip \noindent (Q4.22)\hskip 0.9in
\begin{minipage}{11cm}
\noindent ``{\small ...With hindsight, one can see that Einstein's
attempt to find the best way to implement mathematically the
physical insights about gravitation incorporated in the
equivalence principle was hampered significantly by the absence of
the appropriate mathematical concepts. {\em His insight, as he put
it a few years later, that gravitation and inertia are
`essentially the same'} (`{\it Wesensgleich}'), {\em cries out for
implementation by their incorporation into a single
gravito-inertial field, represented mathematically by a non-flat
affine connection on a four-dimensional manifold}.\footnote{Our
emphasis.} But the concept of such a connection was only developed
{\em after},\footnote{Stachel's emphasis.} and largely in response
to, the formulation of the general theory. {\em So Einstein had to
make do with what was available: Riemannian geometry and the
tensor calculus as developed by the turn of the century, i.e.,
based on the concept of the metric tensor, without a geometrical
interpretation of the covariant derivative. As I have suggested
elsewhere, this absence is largely responsible for the almost
three-year lapse between the end of Act I and the close of the
play}\footnote{Our emphasis. Let it be noted here that the first
act in the development of GR was, according to Stachel, the
inception by Einstein of the EP, while the third and final act is
the formulation of the correct ({\it ie}, generally covariant)
field equations for gravity.}...}''
\end{minipage}

\vskip 0.1in

\noindent Furthermore, we would like to believe with the advent of
ADG, the differential geometry (DG) that Einstein would have
chosen for GR, had such a mathematical framework been around
during the early, formative years of the latter theory, would not
have been the CDG ({\it ie}, manifold) based (pseudo-)Riemannian
geometry of $g_{\mu\nu}$, but a DG based primarily---if not {\em
solely} and {\em exclusively}(!)---on the notion of connection,
much like ADG is. Moreover, the additional virtue of ADG, in
contradistinction even to some of Stachel's remarks above, is that
its basic concept of (affine) connection does not depend at all on
a smooth geometrical base manifold---it is a purely algebraic
notion, without an {\it a priori} need of a geometrical
interpretation as a covariant derivative ({\it eg}, parallel
transporter of other structures/fields on the smooth spacetime
manifold).\footnote{A background manifold independence that, as we
argue throughout the present paper, not only proves to be
invaluable {\it vis-\`a-vis} the singularities of GR, but one that
is of potential import in QG research (see section 6).} In turn,
it is our contention that the dual role of the metric in
Einstein's original formulation of GR was a `misfortunate accident
of labor' in the {\it aufbau} of the theory, one that apparently
necessitates the geometrical spacetime manifold based
interpretation of the theory, and at the same time, always within
the confines of the CDG-based Riemannian geometry, it
unfortunately masked the purely algebraic character of the
connection (which was put on a par with $g_{\mu\nu}$). The said
dual role pertains of course to the well known fact that the
components of $g_{\mu\nu}$ stand both for the gravitational
potentials {\em and} for the entries in the infinitesimal line
element ($ds^{2}=g_{\mu\nu}dx^{\mu}dx^{\nu}$) of the smooth
spacetime manifold marking the {\em chronogeometry} and delimiting
the {\em local causal structure of spacetime} (lightcone soldered
at each point of $M$). In this respect, we quote again Stachel
from \cite{stachel3}:

\bigskip \noindent (Q4.23)\hskip 0.9in
\begin{minipage}{11cm}
\noindent ``{\small ...The main difficulty at this stage was to
grasp the dual nature of the metric tensor: it is both the
mathematical object which represents the space-time structure
(chronogeometry) and the set of `potentials', from which
representations of the gravitational field (Christoffel symbols)
and of the tidal `forces' (Riemann tensor) may be derived...}''
\end{minipage}

\vskip 0.1in

\noindent In a nutshell, our contention is that exactly because
the original formulation of GR (accidentally) happened to be based
on $g_{\mu\nu}$ rather than on $\conn$, we have been
misorientated, misdirected and ever since felt `obliged' to look
similarly for a geometrical (smooth) spacetime interpretation of
the latter when, {\em a priori}, it is a purely algebraic entity.
Moreover, we were misled to think that GR was a theory of the
structure of spacetime---a theory of `{\em
choro-chronogeometry}'---whereas, as we maintain herein and
throughout our trilogy \cite{malrap1,malrap2,malrap3}, it is a
theory solely about the (purely algebraic) {\em gravitational
field} ({\it viz.} connection), without any `smooth geometrical
background spacetime impurities' being {\it a priori} (and
externally)\footnote{Here we simply mean that it is the
experimenter (measurer or geometer), or at least the theoretician,
that `geometrizes' the gravitational connection field by bringing
along $\struc\equiv\smooth_{M}$.} added to that
picture.\footnote{Impurities which then prove to be the main
culprits for the singularities assailing the CDG and manifold
based GR, for all smooth gravitational singularities are, in one
way or another, inherent in $\smooth_{M}$.}

{\it In toto}, a combination of Einstein's quotations (Q?.?) and
(Q?.?) point to the plausible conjecture that, for him too, at
least intuitively ({\it Anschaulichkeit}), the notion of
connection is more suitable for representing the gravito-inertial
field than either the metric or the (Riemann) curvature;
furthermore, assisted by Stachel's remarks in (Q?.?), we are led
to infer that it is precisely due to the non-geometrical (purely
algebraic in ADG \cite{malrap3}) character of the connection, in
contradistinction to the tensorial character of either
$g_{\mu\nu}$ or $\curv(\conn)$, that $\conn$ may be viewed as {\em
the} appropriate structure to implement the `entire' or `unitary'
gravito-inertial field that the EP mandates in GR.\footnote{The
epithet `{\em unitary}' means here `{\em taken as an indivisible
unit, an integral whole---one that is not (physically
meaningfully) separable into its inertial and gravitational
components}'. In the penultimate section, we will relate this
notion of `unitary gravitational field' with the `unitary',
`purely gauge', or even `fully covariant' (synvariant) conception
of field theory propounded by ADG \cite{malrap3}, as well as
discuss the theoretical consequences that such a conception might
have in QG research, especially {\it vis-\`a-vis} the smooth
singularities of the classical theory (GR). (See also next
paragraph.)}

This is precisely how the algebraic connections $\conn$ are viewed
in ADG: as {\em integral wholes}---that is to say, as we have
repeatedly emphasized in our previous trilogy
\cite{malrap1,malrap2,malrap3}, while by `connection' theoretical
physicists usually refer only to the gauge potential part $\aconn$
of $\conn$,\footnote{Thus $\conn$, as a whole, is usually referred
to as the `{\em covariant derivative}', with all the `extraneous
geometrical baggage' that the latter term carries, such as being
geometrically interpreted as a parallel transporter of various
structures ({\it eg}, tangent vectors and other local higher-rank
tensors) along smooth curves in the underlying manifold.} we do
rather mean $\conn$ taken as an inseparable, integral whole. Of
course, locally ({\it ie}, in a particular local gauge $U$),
$\conn$ splits into the flat `{\em inertial connection}' (Q?.?)
$\partial$\footnote{And, it must be emphasized again, this is
precisely how $\partial$ is viewed ADG-theoretically: as {\em a
flat connection} \cite{mall1}. In this respect, it must be
recalled from \cite{malrap1} that the flat derivation $\partial$
is mathematically represented by the (smooth) vector fields in GR,
which in turn are local sections of the tangent bundle $TM$ to
$M$. And one must keep in mind that the EP forces the curved $M$
to be locally Minkowskian---{\it ie}, that the fibers of $TM$ are
isomorphic to (copies of) Minkowski space---so that its local
sections are Minkowski vectors, which comprise {\em inertial
frames} in the {\em basis} or {\em frame bundle} $FM$ associated
with $TM$.} and to the gravitational (gauge potential $1$-form)
$\aconn$, $\conn=\partial+\aconn$, but, as Stachel noted in (Q?.?)
above, in a manifestly non-geometrical way,\footnote{Here one
could also use the terms `non-tensorial', or perhaps even better,
`gauge non-covariant' way.} as $\aconn$ itself is not a
`geometrical object' {\it per se} ({\it ie}, an
$\otimes_{\struc}$-tensor or $\struc$-morphism in
ADG),\footnote{This is unlike either the metric $g$ or the
curvature $\curv$, which are {\em the} `geometrical objects'
($\otimes_{\struc}$-tensors or $\struc$-morphisms) {\it par
excellence} \cite{malrap3}.} because it transforms inhomogeneously
or affinely under (local) gauge transformations. As briefly noted
3 footnotes ago, the potential, both technical and conceptual,
import that the entirely algebraic (non-geometrical), non-smooth
({\it ie}, non-smooth base manifold dependent) and
`unitary'---what in \cite{malrap3} was coined `{\em fully
covariant}'\footnote{Which has been coined here `{\em synvariant}'
after the generalization in 3.2.2 of the PGC of GR in our
background spacetime manifold free ADG-scenario called `{\em
synvariance}'.}---ADG-perspective on gravitational connections
might have for current issues in QG research, especially in view
of the $\smooth$-smooth background spacetime singularities that
assail GR, will be discussed in more detail in section 6.

\item In this light, we are in a position to `rationalize' about
the {\em connection-compatibility of the metric} condition in ADG,
in contrast to the reverse {\em metric-compatibility of the
connection} in Riemannian geometry: in ADG, since the
gravitational algebraic connection $\conn$ is the basic dynamical
structure regarded as being on a par (\ref{eq16}) with the basic
structure sheaf $\struc$ to which, after all, {\em all
differential geometry boils down} (3.2), the $\struc$-metric that
might be imposed on this differential geometric foundation as a
secondary, non-fundamental (`derivative' or `contingent')
structure,\footnote{And it must be emphasized here that, since
$\struc$ ({\it ie}, `geometry' or `space(time)' in ADG) {\em is a
free choice of the theorist, then so is the $\struc$-metric
relative to it}! \cite{malrap3}} must presumably be such that it
`respects' the said basic structure---namely, the algebraic
connection $\conn$. By contrast, in the CDG-based Riemannian
geometry on which GR rests, at the bottom one assumes up-front a
fixed smooth base spacetime manifold (providing one with the usual
CDG-machinery), and relative to that `background absolute
structure' one regards as the fundamental gravitational variable
the smooth spacetime metric $g_{\mu\nu}$. Then one may adjoin to
this structure the Levi-Civita-Christoffel connection $\nabla$ in
a way that respects the metric, thus defining a torsionless,
metric connection. By this rationale one is able to view the
metric-compatibility of the connection condition in the
(pseudo)-Riemannian GR as {\em a physically non-fundamental,
contingent condition that is somehow linked to the macroscopic
spacetime continuum}.\footnote{Indeed, ``{\em Quantum spacetime
will turn out to be not distortion-free. The torsionless condition
on the covariant derivative is a classical {\rm (spacetime
continuum)} requirement}'' (David Finkelstein in private
e-communication with the second author 4 years ago).}

\end{enumerate}

Talking about the fundamental, non-geometrical character of the
gravitational connection, it is interesting to quote Ehlers from
\cite{ehlers}, where this virtue of $\conn$ is emphasized, but
shortly after it is remarkably `down-played':\footnote{Something
that is understandable however, since during the `pre-connection',
pre-Sen and Ashtekar \cite{sen,ash} period, theorists struggled to
view gravity as a (pure) gauge theory (let alone QG as a quantum
gauge theory) \cite{ivanenko}.}

\bigskip \noindent (Q4.24)\hskip 0.9in
\begin{minipage}{11cm}
\noindent ``{\small ...the tensor field $g$, besides determining
times, angles and lengths, also plays the part of a {\small\em
gravitational-inertial}\footnote{Ehlers' own emphasis. Compare it
with Stachel's denomination of $g_{\mu\nu}$ as the
gravito-inertial field.} potential...They {\rm [{\it ie}, the
Christoffel symbols $\Gamma^{a}_{bc}$, formed from $g_{ab}$ and
its first derivatives,]} are components of the Riemannian
connection associated with $g$. One may call this connection the
gravitational-inertial {\small\em field strength},\footnote{Again,
Ehlers' emphasis.} and the $\Gamma^{a}_{bc}$ its components
relative to $(x^{a})$. {\small\em This is an example of a
physically important object which is neither a tensor nor a
spinor. It is non-local in the sense that it does not have a
coordinate-independent `value' a spacetime point},\footnote{Our
emphasis this time.} in accordance with Einstein's elevator
argument showing that the gravitational field can be `transformed
away' at an event, like inertial forces in Newtonian theory...}''
\end{minipage}

\vskip 0.1in

\noindent However, as noted above, a few lines later we read:

\bigskip \noindent (Q4.25)\hskip 0.9in
\begin{minipage}{11cm}
\noindent ``{\small ...The formulation of the `field kinematics'
of GR in terms of principal bundles and their associated bundles
allows one to consider GR as a {\small\em gauge
theory},\footnote{Ehlers' emphasis.} and thus to compare its
structure---which differs considerably from Yang-Mills-type gauge
theories---with other such theories. (As far as I can see such
considerations have not led to a deeper understanding of GR as
such, nor n particular of the role of the Poincar\'e group in GR.
The physical role of translations in SR, as space-time
displacements, is taken over in GR by {\small\em
diffeomorphisms}\footnote{Ehlers' emphasis.} of spacetime. I at
least fail to see that the use of affine bundles with affine (in
Cartan's sense) connections changes this fact, nor does it help me
to appreciate it more deeply)\footnote{This `lack of appreciation'
for $\conn$ will be lifted by Feynman's words shortly.}...}''
\end{minipage}

\vskip 0.1in

After this digression, and following \cite{malrap3}, we would like
to make a few more remarks about our preference for the
connection, rather than the metric, formulation of gravity. Our
opting for $\conn$ instead of $g_{\mu\nu}$\footnote{In ADG, we
symbolize the metric by $\rho$, not $g$ \cite{malrap3}.}, apart
from the PFR which acknowledges the existence of the connection
field `out there' while it also recognizes that the `geometrical'
$\struc$ and {\it in extenso} the $\struc$-metric that goes hand
in hand with it is what {\em we} the external observers-{\it
cum}-`geometers' impose on the gravitational field in order to
`measure' it, may be regarded as being tautosemous to the
following fundamental statement:

\bigskip \noindent (R4.18)\hskip 0.9in
\begin{minipage}{11cm}
\noindent {\em From the ADG-theoretic point of view, gravity is
another gauge theory} \cite{mall4}.
\end{minipage}

\vskip 0.1in

To our knowledge, it was Feynman first, during his quantum
field-theoretic attempts to quantize the gravitational field, to
propose explicitly that gravity should be `properly' viewed as a
gauge theory, and, perhaps even more suggestively for our
endeavors herein, that the fact that the gravitational field was
originally identified with the spacetime metric was just a happy,
albeit `accidental', coincidence coming from the original
formulation of GR within the theoretical confines of the ``{\em
fancy-schmanzy}'' manifold-based Riemannian differential geometry
\cite{feyn2}. For its worth and relevance to the discussion above,
below we recall from \cite{malrap3} the excerpt from the
``{\itshape Quantum Gravity}'' forward to Feynman's ``{\em
Lectures on Gravitation}'' written by Brian Hatfield:

\bigskip \noindent (Q4.26)\hskip 0.9in
\begin{minipage}{11cm}
\noindent ``{\small ...Thus it is no surprise that Feynman would
recreate general relativity from a non-geometrical viewpoint. The
practical side of this approach is that one does not have to learn
some `fancy-schmanzy' (as he liked to call it) differential
geometry in order to study gravitational physics. (Instead, one
would just have to learn some quantum field theory.) However, when
the ultimate goal is to quantize gravity, Feynman felt that the
geometrical interpretation just stood in the way. From the field
theoretic viewpoint, one could avoid actually defining---up
front---the physical meaning of quantum geometry, fluctuating
topology, space-time foam, {\it etc.}, and instead look for the
geometrical meaning after quantization...Feynman certainly felt
that the geometrical interpretation is marvellous, `{\em but the
fact that a massless spin-$2$ field can be interpreted as a metric
was simply a coincidence that might be understood as representing
some kind of gauge invariance'}\footnote{Our emphasis of Feynman's
words as quoted by Hatfield.}...}''
\end{minipage}

\vskip 0.1in

\noindent Subsequently, and after the spin-connection formulation
of GR had been established earlier, principally by the work of
people like Cartan \cite{cartan,gosch}, Bergmann \cite{berg} and
Sen \cite{sen}, the connection-based picture for GR (and
potentially for QG!) found its `apotheosis' with the work of
Ashtekar on the new spin-Lorentzian ({\it ie},
$sl(2,\com)$-valued) connection variables for classical and
quantum gravity \cite{ash}. Currently, it is fair to say that the
epicenter of main approaches to QG (modulo string theory), whether
canonical (Hamiltonian) or covariant (Lagrangian), such as loop QG
and cosmology for example
\cite{rovsm,ash3,thiem2,thiem3,ash4,ash5,ash6,smolin},
concentrates mainly on arriving in one way or another at a quantum
version of the connection formulation of GR---a scheme which is
usually coined {\em Quantum General Relativity}. In other words,
{\em QG is tackled as a quantum gauge theory}, and as the main
quantum configuration arena---the relevant quantum configuration
space for QG---one takes the affine space of smooth
spin-Lorentzian connections, or more appropriately due to the
diffeomorphism symmetries of the underlying differential spacetime
manifold $M$ implementing the PGC in the classical theory (GR),
{\em the moduli space of $\mathrm{Diff}(M)$-equivalent smooth
connections}.

Of course, this persistent abiding by the background smooth
spacetime continuum of GR, even in the QG regime(!), creates all
sorts of both technical and conceptual problems coming mainly from
the infinite dimensional `gauge' group $\mathrm{Diff}(M)$ of
GR---the automorphism group of the $\smooth$-smooth base spacetime
manifold. These problems seriously inhibit the actual carrying out
of the (in some cases formal, in others, heuristic) quantization
procedure, whether canonical ({\it eg}, problem of time) or
covariant ({\it eg}, path integral measure problem), of the smooth
gravitational connection field on the classical background
spacetime continuum \cite{malrap3}. For this, many people have
insisted upon {\em formulating the elusive QG in a manifestly
background spacetime manifold independent way}
\cite{ash0,ash1,ash4,ash5,alvarez,ash6,smolin}.\footnote{In 6.3 we
will return to discuss in more detail the issue of `background
independence' in loop QG, how the latter has been recently used to
resolve the interior Schwarzschild singularity \cite{modesto} (in
the wider context of quantum resolution of black hole
singularities by Loop Quantum Gravity means \cite{husain}), and
how it compares with what we regard as a `{\em genuinely
background manifold independent}' formulation of gravity (whether
classical or quantum) that ADG enables us to accomplish.} However,
the basic dynamical entities involved, namely, the gravitational
connection variables, are still assumed to be {\em smooth}, and
the necessity has arisen to develop a differential
geometry---albeit, still manifestly along CDG-lines---on the
moduli space of connections \cite{ashlew2}, an endeavor again
motivated by loop QG \cite{ashish,ashlew1}. Thus, as critically
pointed out in \cite{malrap3}, the smooth spacetime manifold is
still lurking (even if just dormant or in an apparently `inert'
disguise) at the background. Moreover, since, as we argued
extensively earlier, the base spacetime $M$ is the culprit for the
singularities and their associated unphysical infinities in GR,
while at the same time even the smooth connection-based approaches
to both GR and QG appear essentially to retain $M$ at the
background for differential geometry's sake,\footnote{Thus, QG too
falling prey to the CDG and the associated differential
manifold-conservatism and monopoly that we observed in the first
two sections.}

\bigskip \noindent (R4.19)\hskip 0.9in
\begin{minipage}{11cm}
\noindent {\em what reasons are there for one to hope that a
quantum theory of gravity will ever resolve---{\rm or even more
drastically}, evade---the problem of singularities in the
classical theory (GR)?} In our view, as long as the approaches to
QG abide by the $\smooth$-smooth base spacetime manifold of GR and
the CDG-technology on it, the answer is an emphatic {\em None at
all; the singularities of GR will persist and will inevitably have
to be reckoned with in the quantum theory (QG)}!
\end{minipage}

\vskip 0.1in

\noindent We will return to address and elaborate further on this
crucial issue in section 6. Now, we would like to conclude this
discussion about the `dominance' (more physical significance) of
the connection over the spacetime metric by quoting from John
Stachel's paper ``{\itshape Eddington and Einstein}'' in
\cite{stachel6} a very pertinent passage indeed. As noted before,
Eddington, primarily motivated by Weyl's so-called `{\em gauge
theory of the first kind}' \cite{weyl1}, in some sense neglected
on purpose the fundamental theorem of Riemannian geometry which
identifies the metric with the connection on a smooth manifold,
and set out to formulate general relativity solely in terms of the
affine connection.\footnote{See also \cite{malrap3} for a little
discussion of Eddington's attempts.} The quotation below nicely
describes Eddington's `ahead of its time' effort:\footnote{Below,
all emphasis is ours.}

\bigskip \noindent (Q4.27)\hskip 0.9in
\begin{minipage}{11cm}
\noindent ``{\small ...The most important influence Eddington's
work had on Einstein was in connection with the question of a
unified theory. Einstein was searching for a physical theory that
would not only provide an {\small\em organic}\footnote{We
emphasize the word `organic' {\it vis-\`a-vis} Einstein's unitary
field theory, because we will return and comment on it extensively
in later parts of this paper (section 7, 7.5).} unification of the
gravitational and electromagnetic fields, but hopefully would also
provide a theory of matter, in particular an explanation of the
appearance of quantum effects. Eddington's motivation was rather
more mathematical and epistemological. In particular, he
interpreted his unified theory as providing an {\small\em
explanation} of gravitation, as opposed to the mere {\small\em
description} provided by  Einstein's 1915 theory. Before going
into this, I must say a few words about the mathematical side of
the theory. After 1915, it had been realized by several
mathematicians that metrical and affine structures on a manifold
were logically independent, even though there exists a unique
symmetric affine connection associated with each metric. Weyl was
the first to take advantage of the possibilities arising from this
distinction to set up a physical theory, in which an additional
element---a vector field---is needed, besides the metric field, in
order to determine the affine structure of space-time. He
identified this vector field physically with the electromagnetic
potentials.

Noting the very special nature of Weyl's generalization, Eddington
started by assuming that {\small\em there was no a priori
connection between the metric and initially arbitrary symmetric
connection}. He observed that the curvature or Riemann tensor and
its contraction, usually referred to today as the Ricci tensor,
may be formed from the affine connection without any metric. This
Ricci tensor, however, will in general not be symmetric, even
though the connection is. {\small\em What has this to do with
physics?} Eddington noted that an affine connection enables us to
compare tensors at neighboring points (in particular, to say when
two neighboring vectors are parallel). {\small\em He regarded the
possibility of such a comparison between quantities at neighboring
points in space-time as the minimum element necessary to do any
physics: ``For if there were no comparability of relations, even
the most closely adjacent, the continuum would be divested of even
the rudiments of structure and nothing in nature would resemble
anything else.}''\footnote{Quotation from \cite{eddington}.}

The purely affine theory discussed so far involves no metric, and
thus provides no basis for a comparison of lengths at different
points, even neighboring ones. Such a comparison is provided by
what Eddington, following Weyl's terminology, calls {\small\em
gauge}. {\small\em Thus, the pure affine theory is a
gauge-invariant theory...}}''
\end{minipage}

\vskip 0.1in

\subsubsection{Gravitational singularities: `idols' of the spacetime metric?
`Smooth misery' of the first and second kind.} Keeping in mind the
question posed in (R3.2) above, and now that we have noted that
there has recently been a sea-change in both classical and quantum
general relativity research---especially after Ashtekar proposed
the new spin-Lorentzian variables \cite{ash}---in {\em apparently}
viewing the connection, and not the metric, as the fundamental
gravitational dynamical variable,\footnote{The adverb `apparently'
will be justified below.} thus tending to view GR, {\it \`a la}
Feynman, more as a gauge theory and less as a physical application
of the CDG-based Riemannian geometry (Q?.?), {\em what is the
`true' origin and problematic role of singularities in both the
classical and the (potentially) quantum theory of gravity?}

{\it Prima facie}, this is a legitimate question that behooves the
working classical general relativist, for it is fair to say that,
at least originally, singularities in GR were conceived as {\it
loci} in the spacetime continuum where (some of the components of)
the solution-metric of Einstein's equations could not be
continuously defined---{\it eg}, the Schwarzschild metric, the
Friedmann metric, or even the de Sitter metric
\cite{clarke4}---thus they were intimately connected with the
original metric-formulation of the theory. Thus, by founding the
theory on the connection rather than the metric, one would be
tempted to enquire whether singularities could be evaded
altogether in the connection-based picture, {\em as if
gravitational singularities were due to our using the wrong
dynamical variables in the first place}---as if Einstein's
equations, in the original metric-formulation of GR, were
inappropriately expressed in terms of $g_{\mu\nu}$, which in some
sense `inherently' carries these pathologies.\footnote{In a
metaphorical sense, this could be regarded as the dynamical
analogue of the virtual, coordinate singularities we talked about
earlier.}

Any such hopes, however, are deemed to be short-lived. They
quickly get shattered by imposing the fundamental
metric-connection condition of the CDG-based Riemannian geometry,
which establishes the equivalence between $g_{\mu\nu}$ and
$\nabla$ (\ref{eq15})---that is, that the (smooth) connection is
completely and uniquely determined by the (smooth) metric (and its
partial derivatives) (Q?.?). For even if one does not wish to
impose up-front, as it were `by hand', the condition of
metric-compatibility of the connection by committing oneself to
the usual metric-picture of GR in which the (vacuum) Einstein
equations are obtained from varying the Einstein-Hilbert action

\begin{equation}
\int\ricci\vol
\end{equation}

\noindent with respect to the metric,\footnote{The so-called {\em
second-order formalism}.} but one rather wishes to resort to the
connection-picture {\it \`a la}, say, Palatini or Ashtekar
\cite{malrap3}, in which the fundamental variables are the
connection $\aconn$ and the vierbein $e$,\footnote{The so-called
{\em first-order formalism}.} one on the one hand obtains the
(vacuum) Einstein equations from varying the following Ashtekar
action functional

\begin{equation}
S_{ash}[\aconn^{(+)},e]=\frac{1}{2}\int_{M}\epsilon(e\wee
e\wee\curv^{(+)}(\aconn^{(+)}))
\end{equation}

\noindent with respect to the comoving 4-frame ({\it vierbein})
$e$ ({\it ie}, one establishes that the connection is Ricci-flat),
but on the other, by varying $S_{ash}$ with respect to the
(self-dual) connection $\aconn^{(+)}$,\footnote{Again here the
reader should note that we use the physicist's jargon according to
which by the term `connection' one means the gauge potential part
$\aconn$ of the connection $\conn$. $\aconn^{+}$ signifies {\em
self-dual connection}.} one recovers the Riemannian
metric-compatibility condition for the latter \cite{malrap3}.
Thus, while in the second-order formalism (metric-picture of GR)
the metric-compatibility condition of the connection is an
additional, apparently {\it ad hoc}, requirement (added to the
equations of motion for the metric), in the first-order formalism
(the vierbein/connection picture of GR) it is derived, loosely
speaking, as `the equation of motion of the
connection'.\footnote{Thus, with respect to dynamics, the
first-order (connection) formulation is like a second-order
(metric) picture `in disguise', since variation with respect to
$e$ yields the Einstein equations (like varying with respect to
$g_{\mu\nu}$ in the second-order picture), and the connection is
used only as an `auxiliary variable' to secure its own Riemannian
equivalence with the metric (\ref{eq15}). In other words, in the
first-order formalism the `real' gravitational variable ({\it ie},
the one upon variation of which one obtains the Einstein equations
for gravity) is the {\it vierbein}, not the connection. This is
not surprising considering that $e$ is the `flat (locally
Minkowskian/Lorentzian) square root of the metric':
$e_{\mu}^{a}\eta_{ab}e_{\nu}^{b}=g_{\mu\nu}$!} Moreover, this
observation, coupled to the fact that both the metric and the
connection in either formalism are {\em $\smooth$-smooth
structures} ({\it ie}, they are based on a background differential
spacetime manifold $M$ and their components in a local frame are
elements of $\smooth(M)$),\footnote{For, as we have emphasized
throughout the present paper, it is the smooth base spacetime
manifold that is the culprit for the gravitational singularities
in GR.} shatters the hopes of anyone who would like to see
singularities as being merely the artifacts of an inappropriately
expressed dynamics ({\it ie}, one expressed in terms of the metric
instead of the connection). For in any case, both the original
second- and the more recent first-order formalism appear to have
built-into them the Riemannian condition for the
metric-compatibility of the connection, while at the same time in
either formalism, in effect {\em the basic gravitational dynamical
variable is the metric}---explicitly in the second-order
formalism, and implicitly, in $g_{\mu\nu}$'s `$e$-disguise', in
the first-order one.

\paragraph{The manifold-free, `half-order', purely algebraic
connection-formalism of ADG.} In contradistinction to either the
first or the second-order formalism of GR, both of which are
intimately bound to the manifold and hence CDG-based Riemannian
geometry as we argued above, in ADG:

\begin{itemize}

\item First of all, {\em the sole gravitational dynamical variable
is the entire algebraic connection} $\conn$
\cite{malrap3},\footnote{`Entire' above means `$\conn$ as a whole,
not just its local frame-dependent gauge potential part $\aconn$'.
We will return to discuss the importance---especially for QG
research---of this inseparable, `holistic' conception of the
gravitational field in section 6.} and not the metric as in the
original second-order (metric) formulation of GR, or the {\it
vierbein}-cum-gauge potential part $\aconn$ of the connection in
the subsequent first-order (connection) one. {\em The ADG-picture
of gravity is a genuinely and exclusively connection-based one}.

\item Secondly, and more importantly regarding the
$\smooth$-smooth singularities of GR, our algebraic gravitational
connection field $\conn$, unlike $g_{\mu\nu}$ (2nd-order) or
$(e,\aconn)$ (1st-order), is not dependent at all on a background
differential manifold for its existence and sustenance. Since the
background smooth manifold $M$ is the culprit for the
singularities in GR,\footnote{That is, all singularities are
inherent in $\smooth_{M}$.} {\em the purely algebraic,
$\conn$-based description of gravity afforded by ADG can evade
smooth gravitational singularities altogether}, as we will see in
the sequel. For, in any case, in the context of the
(pseudo-)Riemannian geometry on a differential spacetime manifold
on which GR is founded, neither the smooth Levi-Civita connection,
nor the smooth Lorentzian $g_{\mu\nu}$, appear to be the
appropriate structures to study, in order to pin-point ({\it ie},
define precisely), gravitational singularities (Q?.?)
\cite{clarke4}.

\item {\it In summa}, since by the connection alone, and without
commitment to a background geometrical spacetime manifold, ADG
represents the gravitational field, we may coin the ADG-theoretic
formulation of GR {\em an entire, or unitary, algebraic field
`half-order' formalism}---{\em a `pure' gauge theory} without at
all a dependence on an external spacetime (in fact, whether the
latter is a continuum or a discretum).\footnote{We explain it
again: `unitary algebraic field' pertains to the fact that in the
ADG-theoretic formulation of GR, `the gravitational field is
represented solely by the algebraic connection $\conn$ (on an
appropriately defined vector sheaf---the `representation' sheaf of
$\conn$ associated with the field's `auto-symmetries' in
$\aut\modl$), without this representation being dependent in any
way on a smooth locally Euclidean base space(time)'
\cite{malrap3}, while the `half-order' epithet pertains to the
fact that half the variables of the first-order formalism are
involved; namely, only the connection---but the {\em entire
gravito-inertial connection} $\conn$, not just its gauge potential
part $\aconn$. This `entirety' or `non-separability' of the notion
of connection in our ADG-perspective on gravity is another aspect
of the `unitary' or `holistic' and (dynamically) autonomous field
character of the ADG-approach to gravity, and as it will be argued
in section 6, it might be of significant import to QG research.}
\end{itemize}

\medskip

\subsubsection{ADG-gravity: gauge encounters of the third kind}

We may subsume ADG-gravity under the heuristic term `{\em gauge
theory of the third kind}'. Let us explain more this denomination.
Historically, the terms {\em gauge (field) theory of the first and
second kind} pertain to physical theories with global and local
(over an external, background spacetime manifold) internal gauge
symmetries, respectively. The epithets `internal' and `external'
are traditionally used to distinguish between gauge and base
spacetime symmetries, respectively. As briefly noted in (Q?.?)
above, with the long abandoned and forgotten gauge theory of the
first kind (`global scale theory') one normally associates the
name Hermann Weyl \cite{weyl1}. As it is well known, Maxwellian
electrodynamics and the (flat) Yang-Mills theories of matter
interactions (weak and strong) are examples of (abelian and
non-abelian, respectively) gauge theories of the second
kind---so-called `{\em local gauge theories}', theories with
(gauge) symmetries localized on a base spacetime manifold $M$
(normally taken to be flat Minkowski space $\mink$). They both
emphasize the importance of the notion of connection ({\it viz}.
gauge field).\footnote{Usually, the said gauge fields are modelled
after connections on principal fiber bundles (over a base
manifold) having the local gauge (symmetry) groups as structure
groups in their fibers. As noted earlier, in the usual manifold
based gauge-theoretic jargon, by `gauge field' one normally means
the (local) gauge potential part $\aconn$ of the connection
$\conn$ \cite{malrap3}.}

With the advent of the connection formulation of GR \cite{ash},
people got to appreciate more the gauge-theoretic nature of the
gravitational force, although it is perhaps `wrong'\footnote{And
certainly misleading, as we will argue in the context of QG in
section 6.} to think of the $\mathrm{Diff}(M)$-invariances of the
(external spacetime manifold based) theory of gravity (GR) as
gravitational gauge symmetries proper \cite{wein}. For plainly,
technically speaking, the diffeomorphisms in $\mathrm{Diff}(M)$
are (by definition) the (external) `symmetries' (automorphisms) of
the (external) $\smooth$-smooth base spacetime manifold ({\it ie},
$\mathrm{Aut}(M)\equiv\mathrm{Diff}(M)$, with $M$ smooth), {\em
not elements of the automorphism group of a principal fiber
bundle}--the structure commonly used to represent the other three
gauge forces of nature. In other words, diffeomorphisms are
external not internal `invariances' of GR.\footnote{Indeed,
especially {\it vis-\`a-vis} the problem of (canonical/Hamiltonian
or covariant/Lagrangian) quantization of gravity, regarding
$\mathrm{Diff}(M)$ as gravity's gauge symmetry group proper leads
to a number of formidable problems, such as the so-called {\em
problem of time} and the {\em inner product/quantum measure
problem}. We will discuss them in more detail in section 6.}

In contradistinction, in the ADG-formulation of gravity (GR), no
external, background spacetime manifold is involved whatsoever and
the sole dynamical variable is the gravitational
$\struc$-connection part $\conn$ of the ADG-gravitational field
$(\modl ,\conn)$ \cite{malrap3}.\footnote{This is reflected in the
`{\em half-order}' formulation of GR {\it \`a la} ADG mentioned
earlier, whereby, the dynamical (vacuum gravitational) equations
of Einstein are obtained from varying the ADG-analogue of the
Einstein-Hilbert functional (on the space of connections) {\em
only with respect to} $\conn$ \cite{mall3,malrap3}.} Thus, for ADG
too, gravity is a gauge theory, albeit, a `{\em pure gauge
theory}', or equivalently, a `{\em gauge theory of the third
kind}', one in which a base spacetime continuum arena plays no
role at all and has no physical significance
whatsoever;\footnote{To be fair, the background topological space
$X$ in ADG plays only the role of a surrogate scaffolding for the
sheaf-theoretic localization of the vector and algebra sheaves
involved, on which then the gravitational connection fields are
then defined and act, but as stressed earlier, $X$ plays
absolutely no role in the gravitational dynamics (Einstein
equations), unlike $M$ and its $\mathrm{Diff}(M)$ in the usual
manifold based formulation of GR.} while moreover, the local gauge
`auto-symmetries' of the ADG-gravitational field $(\modl ,\conn)$
`in-itself' are indeed organized into the principal sheaf
$\aut\modl$.\footnote{Consequently, as noted earlier, `covariance'
is replaced by `synvariance'.}

{\it In toto}, ADG is concerned directly and solely with the
gravitational field `in-itself', without reference (or commitment)
to (let alone the mediation of) a locally Euclidean background
spacetime $M$ (in the guise of smooth coordinates in
$\smooth(M)$).\footnote{In fact, as we saw in this paper,
regardless of any background spacetime whatsoever (whether
continuous or discrete).} {\em ADG-gravity is fundamentally
background (external) spacetimeless}---{\it ie}, it is only
concerned with the gravitational field ({\it viz.} connection
$\conn$) itself. In turn, this is reflected in the fact that the
PGC, which in the standard $M$-based theoresis of gravity (GR) is
mathematically effectuated via $\mathrm{Diff}(M)$, in ADG is
modelled after $\aut\modl$---the (principal) group sheaf of
self-transformations (`auto-transmutations') of the gravitational
field $(\modl ,\conn)$, glaringly without reference to the base
space $X$.\footnote{It can be noted here that should one wish to
study the manifold based GR as a gauge theory in bundle-theoretic
terms, the `natural' bundle that could be associated with it is
the {\em frame bundle}, with `natural' (local) structure (Lie)
group (resp. algebra) of (local) gauge symmetries $GL(4,\R)$
(resp. $gl(4,\R)$)---the group of general frame (coordinate)
transformations. In the purely gauge-theoretic formulation of GR
{\it \`a la} ADG, the analogous statement is that the principal
(group) sheaf $\aut\modl$ of `local gauge auto-symmetries' of the
(gravitational) field $(\modl ,\conn)$ is (by definition) locally
isomorphic to $(M_{n}(\struc))^{\bull}$. On the other hand, in the
external manifold based GR, $\mathrm{Diff}(M)$ and $GL(4,\R)$ are
two entirely different `animals', often confused (and misused!) by
physicists and philosophers of science alike, with unfortunately
grave misconceptions and false directions hindering development in
QG research as we will argue in section 6 \cite{wein}.}

The last thing that should be noted before we conclude this
section is that the said conception of ADG-gravity as a gauge
theory of the third kind is closely related to our viewing our ADG
perspective on gravity as being `{\em already quantum}' or `{\em
self-quantum}' ({\it ie}, quantized by or in itself), a feature
which we subsume under the term {\em third-} or {\em
self-quantization}.\footnote{For instance, in \cite{weinstein} it
is mentioned that exactly because $\mathrm{Diff}(M)$ is not a
gauge group proper, one has trouble in defining observables in a
possible quantization scenario for vacuum Einstein gravity (in a
compact spacetime manifold) \cite{torre1,torre2}. In subsection
7.5, in connection with the ADG-theoretic resolution of the
Einstein hole argument, we will argue that the `third-gauged' and
`third-quantized' ADG-gravity enables us straightforwardly to view
vacuum Einstein gravity as an external spacetimeless, pure gauge
theory, with quantum traits already built into the fundamental
formalism.} We will elaborate further on third quantization in 6.1
and 6.2.

\subsection{Section's R\'esum\'e}

A brief summary of this section can be itemized as follows:

\begin{enumerate}

\item First we note that ADG is an algebraico-categorical way of
doing DG, without commitment to a background smooth manifold. From
a categorical viewpoint the category of differential triads
$\ctriad$ involved in ADG has significant advantages over the
category $\man$ of differential manifolds involved in CDG ({\it
eg}, it is bicomplete, it has canonical subobjects, products and
coproducts, and it behaves well under quotients---properties that
$\man$ simply lacks). These advantages may have significant
physical implications and applications ({\it eg}, applying ADG to
Sorkin's finitary topological spaces
\cite{malrap1,malrap2,malrap3}, or developing a topos-theoretic
perspective on such a finitary gravity \cite{rap7}).

\item From the ADG-theoretic viewpoint, all DG boils down to the
structure sheaf $\struc$ of generalized arithmetics (or
`differentiable coordinates') that one employs, as both $d$ and
its generalization, $\conn$, are essentially `derived' from it. In
a (quantum) physical sense, our (generalized) measurements (in
$\struc$) induce dynamical changes ($\conn$).

\item Further to the point above, as befits ADG's algebraico-categorical character, the
(vacuum) gravitational dynamics (Einstein's equations) is
functorial relative to $\struc$. This entails that the dynamics
remains `invariant' under our general coordinate transformations,
while since no external spacetime manifold $M$ is involved (and
hence no $\mathrm{Diff}(M)$ either), the PGC of the manifold and
CDG-based GR is replaced in ADG-gravity by $\aut\modl$ and the
principle of Synvariance (:dynamical auto-transmutations of the
field). Moreover, $\struc$-functoriality of the ADG-gravitational
dynamics is pregnant to a natural transformation-type of
`phenomenon' for the Einstein equation, which can be
(mathematically) interpreted on the one hand as the PARD, and on
the other, (physically) as the PFR.

\item Finally, since the whole ADG-gravity formalism is manifestly {\em not} supported by a
base locally Euclidean spacetime (:differential manifold), it
should be contrasted against either Einstein's original smooth
pseudo-Riemannian metric based second-order formalism, or the
first-order, metric-affine Palatini formalism, or even more
recently, Ashtekar's new variables scenario \cite{ash}, in which
both the {\em smooth} affine connection {\em and} the {\em smooth}
metric in the guise of the {\em smooth} {\it vierbein} field are
engaged. Fittingly, the `ADG-gravity' formalism can been coined
`{\em background smooth spacetime manifoldless, half-order, third
gauge formalism}',\footnote{And we shall see in the sequel, also
`{\em third quantum field}' (or `{\em third quantized field}', or
even `{\em self-quantum}') formalism.} and it appears to be a
promising candidate scenario for solving the long standing problem
of viewing (quantum) gravity as a (quantum) gauge theory proper
\cite{ivanenko,weinstein}.\footnote{See section 7 below.}

\end{enumerate}

\section{ADG-Theoretic `Algebraic Absorption' of Singularities}

There have been recent attempts to deal with the singularities of
GR in an algebraic and sheaf-theoretic way,\footnote{The algebraic
approach to Einstein gravity was pioneered by Geroch in
\cite{geroch4}.} by also resorting to some sort of `algebraically
generalized differential spaces'
\cite{hell0,hell1,hell2,hell3,hell4,hell,hell5}; albeit, all of
them still rely essentially on a background $\smooth$-smooth
spacetime manifold for `drawing' their differential (and
differentiable!) structures. One expected result of this manifold
dependence, even in those generalized algebraic approaches, which
also aspire to address structural issues in the QG regime, is that
the singular {\it loci} are situated at the boundary of the
spacetime manifold, them too stubbornly resisting (analytic)
extensibility thus exposing the insuperable weaknesses of the
classical theory on which GR is based (CDG) that we saw in section
2.

More recently, and quite explicitly, Stachel \cite{stachel2}, in
connection with the analytic inextensibility feature unambiguously
characterizing `real' spacetime singularities in the
Calculus-based GR, notes rather `pessimistically':\footnote{We
wish to thank Rafael Sorkin for communicating to us John Stachel's
latest paper \cite{stachel2}.}

\bigskip \noindent (Q5.1)\hskip 0.9in
\begin{minipage}{11cm}
\noindent ``{\small ...{\em sheaf theory}\footnote{Our emphasis.}
might be the appropriate mathematical tool to handle the problem
[of analytic manifold inextensibility {\it vis-\`a-vis}
$\smooth$-spacetime singularities]\footnote{Our addition for
clarity.} in general relativity. {\small\em As far as I know, no
one has followed up on this suggestion, and my own recent efforts
have been stymied by the circumstance that all treatments of sheaf
theory that I know assume an underlying manifold\footnote{Again,
our emphasis.}...}}''
\end{minipage}

\vskip 0.1in

All in all, the base differential spacetime manifold, with its
inherent $\smooth$-smooth singularities, still appears to persist
in all the algebraic, sheaf-theoretic approaches to GR so far---a
persistence that, taking into account the Einstein-Feynman-Isham
`no-go' of CDG in the quantum deep,\footnote{See quotes
(Q?.?--Q?.?) in the sequel.} appears to halt the potential
mathematical and physical applications of algebraic,
sheaf-theoretic ideas to differential geometry and QG respectively
already at the classical, $\smooth$-continuum level. {\em This is
not the case though in applications of ADG, since the latter
employs sheaf theory alright, but in a manifestly base manifold
free way}. Parenthetically we digress a bit and note that in
chapter 7, by way of contrast against the basic tenets of ADG, we
will comment on and criticize the inappropriateness of the way we
have so far applied differential geometric ideas, based
essentially on a geometrical locally Euclidean spacetime, in QG
research. This critique will lead us then to `define' the notion
of `{\em physical, Euclidean geometry}'---a notion that in our
view appears to be fruitful in QG research---in contradistinction
to the nowadays widely adopted and used conception of a `{\em
mathematical, Cartesian geometry}'---a conception that in the
context of {\em differential} geometry appears to be
`monopolized', with grave consequences for QG research, by the
(use of the base) differential manifold.

In the next subsection we argue on general grounds that ADG, which
uses sheaf theory in a way that Stachel would like ({\it ie},
without the presence of a base differential manifold), is able to
evade the entire spectrum of $\smooth$-gravitational
singularities: from the purely differential geometric DGSs, to the
generalized, distributional SFSs.

\subsection{Killing Two $\smooth$-Singular Birds, DGSs and SFSs, With One ADG Stone}

On general grounds, and as we shall witness shortly in connection
with both ways that we `resolve' the interior Schwarzschild
singularity by ADG-means, ADG, essentially due to its up-front
abolishing the background differential spacetime manifold $M$
carrying the $\smooth$-smooth singularities in its coordinate
structure sheaf $\struc\equiv\smooth_{M}$, evades all (and all
sorts of) spacetime singularities that have been hitherto
`classified' by the usual Analytic (CDG) means \cite{clarke4}:
from the DGSs, to the SFS, and of course the VESs `in between'
these two categories (\ref{eq10}). That is to say, by the way we
will evade ADG-theoretically the inner Schwarzschild singularity
in the next section, it will become transparent that:

\begin{enumerate}

\item To begin with, since no base spacetime manifold is
used at all in ADG, no issue of analytic extension of such a
background past the singular {\it locus} arises whatsoever, thus
the entire spectrum of $\smooth$-smooth singularities---from DGSs,
to SFS---is evaded `in one go'.

\item With respect to DGSs in particular, since to begin with
the $C^{k}$-metric ($k=0\ldots\infty$) is {\em not} the basic
dynamical gravitational field variable in ADG (but rather the
background manifoldless, algebraic $\struc$-connection $\conn$
is),\footnote{Again, see 3.3.} no issue of its $C^{k}$-extension
past the singular {\it locus} arises whatsoever, let alone, of
course, that the law of gravity---essentially, the dynamical
variable $g_{\mu\nu}$ entering the Einstein equations as
formulated traditionally---breaks down at the $C^{k}$-order of
differentiation.\footnote{Again, see $\alpha)$ in 2.1.5.}

\item Also, concerning distributional SFSs in particular, we will
witness in 5.2.3 that Einstein's equations hold over space(time)s
that are densely packed with `distributional singularities'---{\it
ie}, the gravitational field law is formulated and holds entirely
when Rosinger's differential algebras of generalized functions
(non-linear distributions), hosting dense singularities in
(locally) Euclidean space(time)s, are used as coefficient
functions (structure sheaves of generalized coordinates) in
ADG.\footnote{Again, see $\beta)$ in 2.1.5.}

\item Finally, it follows from (\ref{eq10}) that VESs too, lying in between
DGSs and SFSs, pose no problem to the application of ADG to vacuum
Einstein gravity.\footnote{Again, see $\gamma)$ in 2.1.5.}

\end{enumerate}

In order to prepare the reader for the `static', spatial
point-resolution, as well as for the `dynamic', temporal, extended
line-resolution, of the interior Schwarzschild singularity by
ADG-means in the next section, we recall below some basic results
from the application of ADG to the so-called spacetime foam dense
singularities in GR as originally treated in
\cite{malros1,malros2,malros3}.

\subsection{A Concrete Application of ADG: Uncountable `Spacetime Foam Dense
Singularities' in the Smooth Manifold's Bulk}

In this subsection we will present elements from the application
of ADG to Euclidean and finite-dimensional locally Euclidean
(manifold) space(time)s $X$ hosting the most numerous and
unmanageable by CDG means singularities: the so-called `spacetime
foam' dense singularities
\cite{malros1,malros2,mall3,malros3,mall9}.\footnote{Of these, the
central references are \cite{malros2,malros3}. It must be noted
here that a bit earlier, in \cite{malros1}, it was shown that ADG
can be successfully applied so as to incorporate singularities
situated on arbitrary closed nowhere dense subsets of Euclidean
and locally Euclidean (manifold) space(time)s, sets that can have
arbitrarily large Lebesgue measure \cite{oxtoby}. As we will see
below, the spacetime foam dense singularities in
\cite{malros2,malros3} are much more numerous (and unmanageable by
CDG-theoretic means) than the nowhere dense ones, albeit, the
entire spectrum of ADG still applies to them without a problem.}
In mathematics, these are singularities of {\em generalized
functions} (distributions) which have been used as coefficients in
and have been occurring as solutions of non-linear (both
hyperbolic and elliptic) partial differential equations (NLPDEs),
as originally treated entirely algebraically by Rosinger in
\cite{ros1,ros}.\footnote{Hereafter we will refer to Rosinger's
theory as {\em non-linear algebraic theory of generalized
functions}, or better, as {\em non-linear algebraic distribution
theory} (NLADT).} As briefly mentioned before, in physics,
interest in such singularities has arisen in the study of
`spacetime foam' structures in GR and QG, as studied primarily by
the Polish school of Heller {\it et al.} \cite{hell2,hell5}.

Arguably, {\em these are the most robust and numerous
singularities that have appeared so far in the theory of NLPDEs},
but two of their prominent features that we would like to
highlight here are:

\begin{itemize}

\item First, {\em their cardinality}. These are singularities on
arbitrary subsets of the underlying topological space(time) $X$.
In particular, they can be concentrated on {\em dense} subsets of
$X$,\footnote{That is why one generally refers to them as `{\em
dense singularities}'.} under the proviso that their complements,
consisting of non-singular (regular) points, are also dense in
$X$. In case $X$ is a Euclidean space or a finite-dimensional
manifold, {\em the cardinality of the set of singular points may
be larger than that of the regular ones}. For instance, when one
takes $X=\R$, the dense singular subsets of it may have the
cardinal of the continuum---{\it ie}, the singularities may be
situated on the irrational numbers, while the regular ones are
also dense but countable in $\R$ and situated on the
rationals.\footnote{In 5.2.3 we will exploit this application of
ADG to spacetime foam dense singularities and cast the interior
singularity of the Schwarzschild solution of the Einstein
equations as such an extended dense singularity extending along
the `wristwatch time-line' ($\R$) of a point particle of mass $m$
at rest.}

\item Second, {\em their situation in the manifold's bulk}. As it
is evident from the above, the dense singularities, apart from
their uncountable multiplicity, are not situated merely at the
boundary of the underlying topological space(time) (manifold), but
occupy `central' points in its interior. This is in striking
contrast to the usual theoresis of $\smooth$-smooth spacetime
singularities that we revisited in section 2
\cite{clarke3,clarke4}, which we may now coin `separated and
isolated', or `solitary', or even `{\em spacetime marginal}' for
effect.\footnote{This situation is also in contrast to the
`algebraically generalized differential spaces/spacetime foam'
approach to GR and QG of Heller {\it et al.}, as they too assume
that singularities---in fact, them too {\em nowhere dense
singularities} like the ones in \cite{malros1}---sit right at the
edge of the spacetime manifold (see
\cite{hell0,hell1,hell4,hell,hell5,hell3}, and especially
\cite{hell6}).}

\item And third, {\em the differential algebras of generalized
functions in Rosinger's NLADT contain both the usual algebra
$\smooth(X)$ of smooth functions and Schwartz's linear
distributions} \cite{malros2,malros3}. Furthermore, the NLADs,
either with nowhere dense, or even more prominently, with dense
singularities, have proven to be more versatile (and potentially
more useful in differential geometric applications) than the,
lately quite popular in the theory of NLPDEs and its applications
to various equations of mathematical physics, non-linear
generalized functions of Colombeau \cite{colombeau}.\footnote{See
\cite{malros2} for a discussion of the (differential geometric)
virtues of the NLADs compared to the Colombeau distributions.}

\end{itemize}

\subsubsection{Dense singularities in Euclidean and locally
Euclidean spaces}

We first note that in \cite{malros2} as the domain of definition
of Rosinger's NLADs one takes any nonempty open subset $X$ of a
finite-dimensional Euclidean space ($\R^{n}$), or more generally,
any non-void region of a space that is locally isomorphic to
$\R^{n}$---an $n$-manifold $M$. Then, various collections of
singularities in $X$ are described by giving a family $\ds$ of
subsets $\dss$ of $X$, with each $\dss$ corresponding to a set of
singularities of a certain given generalized function (NLAD).

\paragraph{Dense singularities are the most numerous ones.} Arguably, cardinality-wise, the largest collection $\ds$ of
such Euclidean or manifold singularity-sets $\dss\subset X$ that
one can encounter is

\begin{equation}\label{eqx-3}
\ds_{dns}(X):=\{\dss\subset
X:~X\setminus\dss~\mathrm{is~dense~in~}X\}
\end{equation}

\noindent with the subscript `{\it dns}' denoting `{\em dense}'.
It is plain that there are many such dense singularity-sets in
$X$; moreover, for some of them, the cardinality of $\dss$ may be
bigger than that of its complement $X\setminus\dss$, which in turn
means that the set of singular {\it loci} is not only dense in
$X$, but also larger than the set of $X$'s regular points.

Due to the maximum cardinality property of $\ds_{dns}$, the
various families $\ds$ of singularity-sets in $X$ considered in
\cite{malros2,malros3} are `less numerous' than $\ds_{dns}$---{\it
ie}, they are taken to satisfy the condition
$\ds\subseteq\ds_{dns}$.\footnote{For a more detailed treatment of
dense singularity-sets, the reader can refer to
\cite{ros2,ros3,ros}.} Characteristically, two such collections
that were considered in \cite{malros1,malros2,malros3} are

\begin{equation}\label{eqx-2}
\ds_{ndns}(X):=\{\dss\subset
X:~\dss~\mathrm{is~closed~and~nowhere~dense~in}~X\}
\end{equation}

\noindent and

\begin{equation}\label{eqx-1}
\ds_{B_{1}}(X):=\{\dss\subset
X:~\dss~\mathrm{is~of~first~Baire~category~in}~X\}
\end{equation}

\noindent with $\ds_{ndns}$ representing families of {\em nowhere
dense singularity-sets} (hence the subscript `{\it ndns}')
\cite{malros1}, while $\ds_{B_{1}}$ singularity-sets of the first
Baire category \cite{malros2,malros3}. In terms of their
cardinality, they are ordered by inclusion as follows:

\begin{equation}\label{eqx0}
\boxed{\ds_{ndns}}\subset\boxed{\ds_{B_{1}}}\subset\boxed{\ds_{dns}}
\end{equation}

\noindent with $\ds_{dns}$, as noted above, being the most
numerous (and the most problematic from the usual CDG-viewpoint)
ones, and the ones in which we are interested here.

\paragraph{Spacetime foam algebras and nets of dense singularities with asymptotically
vanishing differential ideals: a singularity analogue of Sorkin's
topological finitary posets `smeared point-quotient trick'.} In
this paragraph, and in view of the two-fold ADG-theoretic
resolution of the interior Schwarzschild singularity that we are
going to present in the sequel, by regarding it either as an
`extended' spacetime foam dense time-line singularity {\it \`a la}
Mallios and Rosinger \cite{malros2,malros3} (5.2.3), or as a
static spacetime `localized' point-singularity and then treating
it by Sorkin's `blow-up' finitary topological posets methods
\cite{sork0}, their corresponding differential incidence algebras
\cite{rapzap1,rapzap2} and the finitary spacetime sheaves thereof
\cite{rap2} in the manner of ADG \cite{malrap1,malrap2,malrap3}
(5.2.2), we present the basic ideas in the construction of
spacetime foam algebras of generalized functions as originally
developed in \cite{ros2,ros3,ros}. In the process, we will witness
fundamental theoretical affinities, conceptual similarities and
common technical `tricks' between Rosinger's and Sorkin's ideas,
especially when the latter are presented in an algebraic way as in
\cite{rapzap1,rapzap2}.\footnote{We will see that, in a strong
sense, the algebraic description of spacetime foam given in
\cite{rapzap1,rapzap2} is closely related, both conceptually and
technically, to Rosinger's spacetime foam algebras hosting dense
singularities, while in the end, the differential geometric
platform for unifying both is provided by the sheaf-theoretic
means of ADG. Under the prism of ADG, this is more than a nominal,
superficial analogy between the Raptis-Zapatrin algebraic
conception of spacetime foam and the Mallios-Rosinger spacetime
foam dense singularities, and this is what we wish to highlight by
the presentation herein.}

The main thing that the reader should notice in the brief
presentation below is that Rosinger, by the spacetime foam algebra
construction, manages to encode the information on various
singularity point-sets ({\it eg}, such as $\ds_{ndns}$, or even
more prominently, $\ds_{dns}$) {\em entirely algebraically} in
much the same way that the topological poset spacetime
discretizations where cast again {\em purely algebraically} in
\cite{rapzap1,rapzap2}. Then, the concomitant organization of the
respective algebras\footnote{Rosinger's spacetime foam algebras
may be coined a mouthful `{\em singularity spacetime foam
differential algebras}' (SSTFDAs), while Raptis and Zapatrin's,
`{\em discrete spacetime foam differential algebras}' (DSTFDAs).}
into sheaves, and after crucially realizing that both kinds of
algebras are {\em differential algebras},\footnote{Differential
algebras of generalized functions (distributions) in Rosinger's
case, and discrete differential algebras of generalized maps
(poset arrows) in the Raptis-Zapatrin case.} the application of
the manifold and Calculus-free sheaf-theoretic ADG-technology to
either enables one to do differential geometry (and apply it to
gravity!) over space(time)s that are far from being smooth (in the
usual $\smooth$-manifold sense)---that is to say, space(time)s
that are densely populated by singularities (of the most general
type of functions) in Rosinger's case
\cite{malros1,malros2,malros3,mall2,mall3}, and `discrete' or
`reticular' space(time)s that are far from being smooth continua
like the classical differential spacetime manifold $M$
\cite{malrap1,malrap2,malrap3}. So far, these are the two main
applications of ADG to mathematical physics---in particular, to
classical and quantum gravity research.

There are three basic steps in the construction of Rosinger's
spacetime foam algebras ({\it ie}, the SSTFDAs), steps that, as
noted above, bear close similarities with the algebraic approach
to spacetime foam ({\it ie}, the DSTFDAs) \cite{rapzap2} which was
originally based on Sorkin's finitary (locally finite) topological
poset substitutes of continuous manifolds \cite{sork0,rapzap1}:

\begin{enumerate}

\item {\bf `Coarse or fine graining of singularity-sets':} This
step consists on imposing the following conditions on any
collection $\ds$ of singularity-sets $\dss\subset X$:

\begin{equation}\label{eqx1}
\mathbf{Density~of~regular~points:}~\forall\dss\in\ds
:~X\setminus\dss~\mathrm{is~dense~in~X}
\end{equation}

\noindent and

\begin{equation}\label{eqx2}
\mathbf{`Coarse-graining'~of~singularity\!\!-\!sets:}~\forall\dss
,\dss^{'}\in\ds :~\exists\dss^{''}\in\ds
:~\dss\cup\dss^{'}\subseteq\dss^{''}
\end{equation}

\noindent We note that for such an arbitrary collection $\ds$ of
singularity-sets in $X$ one has the inclusion
$\ds\subseteq\ds_{dns}$; while plainly, both $\ds_{ndns}$ and
$\ds_{B_{1}}$ satisfy (\ref{eqx1}) and (\ref{eqx2}) above.

The condition (\ref{eqx2}) may be interpreted as signifying that
`{\em the singularity-set $\dss^{''}$ is the coarse-graining of
$\dss$ and $\dss^{'}$}';\footnote{In turn, $\dss^{''}$ may be
thought of as being refined (decomposed) into the `smaller' (more
localized!) singularity-sets $\dss$ and $\dss^{'}$. Technically
speaking, this condition is equivalent to positing that $\ds$ is
an `upwards directed' $\cup$-semilattice, which may be coined
`{\em the coarse-graining singularity semilattice in $X$}'.}
moreover, it may be abstracted to the following {\em algebraic
construction} involving not the singularity-subsets of $X$ {\it
per se}, but rather the $\smooth$-smooth functions that live on
that space(time): so let $L=(\Lambda ,\leq)$ be a right-directed
partially ordered set---a net.\footnote{As noted in
\cite{malros2,malros3}, {\it prima facie} $L$ may not be thought
of as being dependent on $\dss$, but in certain special situations
the two may be related \cite{ros2}.} Then consider the algebra
$(\smooth(X))^{\Lambda}$ of sequences of smooth functions on $X$
indexed by $\Lambda$,\footnote{That is, $(\smooth(X))^{\Lambda}=\{
(s_{\lambda}): ~\lambda\in\Lambda\}$.} which brings us to the next
basic step in the construction of spacetime foam algebras.

\item {\bf `Differential ideals covering singular points':} For
any singularity-set $\dss$ in an arbitrary singularity-family
$\ds$ in $X$, define the following {\em (differential) ideal} in
$(\smooth(X))^{\Lambda}$:

\begin{equation}\label{eqx3}
\begin{array}{c}
{\mathcal{I}}_{L,\dss}(X):=\{
(s_{\lambda})\in(\smooth(X))^{\Lambda}:\cr \forall x\in
X\setminus\dss
,~\exists\lambda\in\Lambda~\mathrm{such}~\mathrm{that},~\forall\mu
\in \Lambda ,~\mathrm{with}~\mu\geq\lambda ,~\mathrm{and}~\forall
p\in\mathbb{N}^{n}:\cr D^{p}s_{\mu}(x)=0\}
\end{array}
\end{equation}

\noindent with $D^{p}$ in the last line standing generically for
the smooth function ($p=0$) entries in $s_{\lambda}$ and their
(term-wise) partial derivations of arbitrary order ($\mathbb{N}\ni
p\geq1$). The last line is an {\em asymptotic vanishing
condition}, which, as noted in \cite{malros3}, secures
that\footnote{In the quotation below emphasis is ours, as usual.}

\bigskip \noindent (Q5.2)\hskip 0.9in
\begin{minipage}{11cm}
\noindent ``{\small ...the sequences of smooth functions in the
ideal ${\mathcal{I}}_{L,\dss}(X)$ will in a way cover with their
support the singularity set $\dss$, and at the same time vanish
outside it, together with all their partial derivatives. In this
way, the ideal  ${\mathcal{I}}_{L,\dss}(X)$ carries in an {\em
algebraic} manner the information on the singularity set $\dss$.
Therefore, a {\em quotient} in which the factorization is made
with such ideals may in certain ways {\em do away} with
singularities, and do so through {\em purely algebraic
means}...}''\footnote{In connection with the definition
(\ref{eqx3}) above, we also read from \cite{malros2} that the
assumption that $L$ is right-directed is actually used in proving
the additive closure of $\mathcal{I}_{L,\dss}$, namely, that
$\forall
s,s^{'}\in\mathcal{I}_{L,\dss},~s+s^{'}\in\mathcal{I}_{L,\dss}$,
which is effectively a defining property for an ideal (as a linear
subspace). This assumption will also be instrumental in exploring
{\em singularity refinement} properties of spacetime foam algebras
in the sequel. In connection with the last remark, we also note
from \cite{malros2,malros3} that a judicious choice of $L$ so that
the resulting ideals are non-trivial
($\mathcal{I}_{L,\dss}\not=\{0\}$) may result in
$\mathcal{I}_{L,\dss}$ being rather {\em large}---in fact, {\em
uncountable}. Such can be, for instance, the lattice of all
topologies on a $\cont$-manifold \cite{ish0}, and we will
encounter such an example shortly when we explore the close
similarities between the Mallios-Rosinger \cite{malros2,malros3}
and the Raptis-Zapatrin \cite{rapzap1,rapzap2} algebraic
approaches to spacetime foam under the unifying sheaf-theoretic
prism of ADG.}
\end{minipage}

\vskip 0.1in

\noindent The last sentence in (Q?.?) above brings us to the
third, {\em defining act} for spacetime foam algebras (with dense
singularities in $X$).

\item {\bf `Factoring out the singular loci'---the definition of
spacetime foam algebras:} First we note that, in case the
collection $\ds$ of singularity-sets consists of only one such set
$\dss$ ({\it ie}, $\ds=\{\dss\}$), albeit, one that is dense in
$X$ (and moreover in the aforesaid sense that the set of regular
points $X\setminus\dss$ is dense in $X$ (\ref{eqx-3}), which in
turn identifies $\ds$ with $\ds_{dns}$),\footnote{From now on we
will omit the subscript `$dns$' from $\ds$, since in what follows
we will be interested exclusively in dense singularity-sets in
$X$.} the {\em spacetime foam algebra} is defined by the following
quotient:

\begin{equation}\label{eqx4}
\fa_{L,\dss}(X):=(\smooth(X))^{\Lambda}/{\mathcal{I}}_{L,\dss}(X)
\end{equation}

\noindent At the same time, by observing that when $\ds$ consists
of more than one dense singularity-set $\dss$ the set

\begin{equation}\label{eqx5}
{\mathcal{I}}_{L,\ds}:=\bigcup_{\dss\in\ds}{\mathcal{I}}_{L,\dss}
\end{equation}

\noindent is also an ideal in $(\smooth(X))^{\Lambda}$, one
defines the {\em spacetime multi-foam algebras} in complete
analogy with (\ref{eqx4}) above, as follows:

\begin{equation}\label{eqx6}
\fa_{L,\ds}(X):=(\smooth(X))^{\Lambda}/{\mathcal{I}}_{L,\ds}(X)
\end{equation}

\noindent noting along the way that, when $\ds=\{\dss\}$,
$\fa_{L,\ds}(X)\equiv\fa_{L,\dss}(X)$---{\it ie}, spacetime
multi-foam and foam algebras coincide.\footnote{In line with
\cite{malros2,malros3}, we will call both spacetime foam and
multi-foam algebras simply {\em spacetime foam algebras}.}

\end{enumerate}

\subsubsection{Spacetime foam algebras as differential algebras of generalized
functions, their regularity, and the fine and flabby sheaves
thereof}

As it happens, the bottom-line of the preceding discussion is that
{\em the spacetime foam algebras may be regarded as differential
algebras of generalized functions}---`smeared', (non-linear)
distribution-type of functions that enlarge the usual smooth ones
in $\smooth(M)$, they contain the linear distributions of
Schwartz, while for differential geometry's sake (and from the
sheaf-theoretic perspective of ADG), they have certain structural
(sheaf-theoretic) advantages over Schwartz's linear spaces, but
more importantly, for example, over Colombeau's algebras of
generalized functions. Below, we would like to discuss briefly
these qualities of spacetime foam algebras.

To begin with, the main reason why special attention was paid in
\cite{malros2,malros3} on the collection $\ds_{dns}$ of dense
singularity-sets in $X$ is that if $X\setminus\dss\in\ds$, then

\begin{equation}\label{eqx7}
{\mathcal{I}}_{L,\dss}(X)\cap{\mathcal{X}}_{\Lambda
,\dss}^{\infty}(X)=\{ 0\}
\end{equation}

\noindent where ${\mathcal{X}}_{\Lambda ,\dss}^{\infty}(X)$ stands
for the {\em diagonal} of the power set
$(\smooth(X))^{\Lambda}$---that is to say,

\begin{equation}\label{eqx8}
\begin{array}{c}
{\mathcal{X}}_{\Lambda
,\dss}^{\infty}(X)=\{\chi(\phi):=(\phi_{\lambda}:~\lambda\in\Lambda)\}~\mathrm{such~that}\cr\forall
\lambda\in\Lambda:~\phi_{\lambda}=\phi
,~\mathrm{as}~\phi~\mathrm{ranges~over}~\smooth(X)
\end{array}
\end{equation}

\noindent This in turn entails the following algebra isomorphism:

\begin{equation}\label{eqx9}
\smooth(X)\ni\phi\mapsto\chi(\phi)\in{\mathcal{X}}_{\Lambda
,\dss}^{\infty}(X)
\end{equation}

\noindent {\it In summa}, the following {\em algebra embedding} is
secured by (\ref{eqx9}):

\begin{equation}\label{eqx10}
\smooth(X)\ni\phi\mapsto\chi(\phi)+{\mathcal{I}}_{L,\dss}(X)\in\fa_{L,\dss}(X)
\end{equation}

\noindent And in words,

\bigskip \noindent (R5.1)\hskip 0.9in
\begin{minipage}{11cm}
\noindent  the algebra $\smooth(X)$ of smooth functions on $X$ is
{\em embedded} (included) into the spacetime foam
algebras,\footnote{And, of course, to multi-foam algebras as one
obtains a similar embedding
$\smooth(X)\ni\phi\mapsto\chi(\phi)+{\mathcal{I}}_{L,\ds}(X)\in\fa_{L,\ds}(X)$
in case the collection $\ds$ consists of more than one dense
singularity-set $\dss$ in $X$.} which in turn means that the
latter are algebras of {\em generalized functions}
(distributions).\footnote{In fact, the spacetime foam algebras are
{\em unital}, with unit element
$\chi(\mathbf{1})+{\mathcal{I}}_{L,\dss~\mathrm{or}~\ds}$ (where
$\mathbf{1}$ is the constant unit-valued function on $X$).}
\end{minipage}

\vskip 0.1in

In addition, the aforementioned asymptotic vanishing condition
(\ref{eqx3}) entails the following `{\em closure}' of
$\fa_{L,\dss}(X)$ and $\fa_{L,\ds}(X)$ under (partial)
differentiation of arbitrary order:

\begin{equation}\label{eqx11}
\begin{array}{c}
D^{p}\fa_{L,\dss}(X)\subseteq\fa_{L,\dss}(X)\cr
D^{p}\fa_{L,\ds}(X)\subseteq\fa_{L,\ds}(X)
\end{array}
\end{equation}

\noindent which means that $\fa_{L,\dss}(X)$, and {\it in extenso}
$\fa_{L,\ds}(X)$, are {\em differential} algebras of generalized
functions.

We also read from \cite{malros3} that

\bigskip \noindent (Q5.3)\hskip 0.9in
\begin{minipage}{11cm}
\noindent ``{\small the {\rm [spacetime foam and]} multi-foam
algebras contain the Schwartz distributions, that is, {\rm
[similarly to (\ref{eqx10})]}, we have linear embeddings which
respect the arbitrary partial derivation of smooth functions,
namely $\conn^{'}(X)\subset\fa_{L,\dss}(X)$, for
$\dss\in\ds_{dns}(X)$ and
$\conn^{'}(X)\subset\fa_{L,\ds}(X)$.\footnote{With $\conn^{'}(X)$
denoting the linear (vector) space of the Schwartz
distributions.}}''
\end{minipage}

\vskip 0.1in

For more details about the issues to be presented in the following
two paragraphs, the reader is referred to \cite{malros2,malros3}.

\paragraph{A note on regularity properties of the spacetime foam generalized functions: quotient regularity with a
twist of Sorkin's finitary topological refinement;
`singularity-refinement'.}

In order to discuss briefly the regularity properties of the
non-linear spacetime foam distributions, we first note the
following set-theoretic inclusion of the aforesaid foam into
multi-foam differential ideals:

\begin{equation}\label{eqx12}
{\mathcal{I}}_{L,\dss}\subseteq{\mathcal{I}}_{L,\ds}
\end{equation}

\noindent an inclusion which can be algebraically translated into
the following {\em surjective algebra homomorphism} between the
corresponding foam and multi-foam algebras:

\begin{equation}\label{eqx13}
\fa_{L,\ds}(X)\ni v+{\mathcal{I}}_{L,\dss}(X)\mapsto
v+{\mathcal{I}}_{L,\ds}(X)\in\fa_{L,\ds}(X)
\end{equation}

Now, in the context of generalized functions, we read from
\cite{malros3} that (\ref{eqx13}) means that ``{\em the typical
generalized functions in $\fa_{L,\ds}(X)$ {\rm [(the multi-foam
distributions)]} are more regular than those in $\fa_{L,\dss}$
{\rm [(the foam distributions)]}}''.\footnote{We may formally
symbolize this `{\em regularity power-relation}' as
$\fa_{L,\ds}(X)\reg \fa_{L,\dss}(X)$.} In a way, this is to be
expected since, given two spaces $\mathcal{S}_{1}$ and
$\mathcal{S}_{2}$ of generalized functions, with, say,
$\smooth(X)\subset\mathcal{S}_{1}\subset\mathcal{S}_{2}$, the
elements of the {\em larger} space $\mathcal{S}_{2}$ normally
appear to be {\em less regular} than those of the {\em smaller}
one $\mathcal{S}_{1}$---a property that one may wish to coin {\em
subset regularity}.\footnote{Again, formally one may write
$\mathcal{S}_{1}\reg\mathcal{S}_{2}$. In connection with subset
regularity, a word of caution already noted in \cite{malros3}:
while subset regularity is satisfied, for example, by the
inclusions $\smooth(\R)\subset\mathcal{C}^{1}(\R)\subset\cont(\R)$
(when $X\equiv\R$), there is also the linear surjective mapping
$D:~\mathcal{C}^{1}(\R)\mapto\cont(\R)$, which by no means
suggests that continuous ($\cont$) functions (on $\R$) are more
regular than the singly differentiable ($C^{1}$) ones.}

A notion akin to subset regularity is, what is coined in
\cite{malros3}, {\em quotient regularity}.\footnote{It too to be
formally symbolized here by $\reg$.} This notion can be briefly
explained as follows. Again, let $\mathcal{S}_{1}$,
$\mathcal{S}_{2}$ be as above, and also assume that they are
obtained as quotient vector spaces (of some `big' linear space
$\mathcal{S}$) in the following way:

\begin{equation}\label{eqx14}
\mathcal{S}_{1}=\mathcal{S}/\mathcal{W},~\mathcal{S}_{2}=\mathcal{S}/\mathcal{V};~\mathrm{with}~
\mathcal{V}\subseteq\mathcal{W}
\end{equation}

\noindent Let also
$\mathcal{F}:~\mathcal{S}_{2}\mapto\mathcal{S}_{1}$ be the {\em
canonical} surjective linear mapping between them given point-wise
by

\begin{equation}\label{eqx15}
\mathcal{S}_{2}=\mathcal{S}/\mathcal{V}\ni f+\mathcal{V}\mapsto
f+\mathcal{W}\in\mathcal{S}/\mathcal{W}=\mathcal{S}_{1}
\end{equation}

\noindent Then, as noted in \cite{malros3}, one also has for the
two quotient spaces $\mathcal{S}_{1},~\mathcal{S}_{2}$ of
generalized functions the {\em quotient regularity power relation}

\begin{equation}\label{eqx16}
\mathcal{S}_{1}\reg\mathcal{S}_{2}
\end{equation}

\noindent reading that ``{\em the typical elements of
$\mathcal{S}_{1}$ are at least as regular as those of
$\mathcal{S}_{2}$}''. Thus, in the sense above, it is plain that

\bigskip \noindent (R5.2)\hskip 0.9in
\begin{minipage}{11cm}
\noindent {\em the typical spacetime multi-foam generalized
function is more regular than the typical foam one}.
\end{minipage}

\vskip 0.1in

\noindent In other words,

\begin{equation}\label{eqx17}
\fa_{L,\ds}\reg\fa_{L,\dss}
\end{equation}

\noindent in the sense of quotient regularity.\footnote{For,
plainly, $\mathcal{I}_{L,\dss}\subseteq\mathcal{I}_{L,\ds}$
(\ref{eqx12}), and $\mathcal{S}\equiv(\smooth(X))^{\Lambda}$.}

By the foregoing discussion it has become clear that {\em the more
dense singularity-sets $\dss$ a collection $\ds$ includes, the
more quotient regular the corresponding spacetime multi-foam
algebra $\fa_{L,\ds}$ is}. This prompts us to introduce the notion
of {\em singularity refinement}, a notion which will prove to be
very useful in comparing and exploring the close similarities
between the two algebraic approaches to spacetime foam---namely,
the Mallios-Rosinger approach in \cite{malros2,malros3} (SSTFDAs),
and the Raptis-Zapatrin approach in \cite{rapzap1,rapzap2}
(DSTFDAs), both of which, under the unifying perspective of ADG,
will be used in the next section to `resolve' the interior
Schwarzschild singularity.

The basic idea behind singularity refinement is that as one
employs `smaller and more numerous'\footnote{The expression
`smaller and more numerous' may be cumulatively coined `{\em
finer}', hence the term `{\em refinement}'. Shortly we will see
the close link between this notion of singularity refinement and
the one of {\em topological refinement} originally due to Sorkin
\cite{sork0}.} (dense) singularity-sets in order to cover the
(densely) singular point-{\it loci} of the underlying topological
space(time) $X$---a procedure which may be formally symbolized by
the {\em singularity refinement relation}

\begin{equation}\label{eqx18}
\ds\sref\ds^{'}
\end{equation}

\noindent between the corresponding (dense) singularity families
$\ds$---the respective differential ideals $\mathcal{I}_{L,\ds}$
and $\mathcal{I}_{L,\ds^{'}}$ in $(\smooth(X))^{\Lambda}$ are
(partially) ordered by inclusion according to (\ref{eqx12}) ({\it
ie}, $\mathcal{I}_{L,\ds}(X)\subseteq\mathcal{I}_{L,\ds^{'}}(X)$),
which in turn entails {\em surjective spacetime multi-foam algebra
homomorphisms} of the kind (\ref{eqx13})

\begin{equation}\label{eqx19}
\fa_{L,\ds}(X)\mapto\fa_{L,\ds^{'}}(X)\Leftrightarrow\fa_{L,\ds^{'}}(X)\reg\fa_{L,\ds}(X)
\end{equation}

\noindent In fact, (\ref{eqx19}) is equivalent to (\ref{eqx18}),
but simply expressed in algebraic terms ({\it ie}, in terms of the
corresponding multi-foam algebras instead of the dense singularity
point-subsets of $X$ themselves). {\it In toto}, and in view of
the expression {\ref{eqx18}) and the right hand side of
(\ref{eqx19}), we note that {\em singularity refinement and
quotient regularity are `order-covariant' notions}.\footnote{That
is to say, as one refines one's dense singularity-sets, the
corresponding spacetime multi-foam algebras become more regular,
since, after all, in the process one does away with ({\it ie}, one
factors out) more singular point-{\it loci} (the ones lying in the
respective $\mathcal{I}_{L,\ds}(X)$s).}

The foregoing discussion gives us a hint that we are actually
dealing with a kind of `{\em inverse system}'
$\invros=(\ds,\sref)$ of collections ($\ds$) of dense
singularity-sets ($\dss$) covering the densely singular point-{\it
loci} of $X$. In turn, $\invros$ may be interpreted as a {\em
singularity refinement net}. Now, having in hand the material
above, we are in a position to explore the aforesaid close, both
conceptually and technically, similarities between SSTFDAs and
DSTFDAs, both of which were ultimately developed and presented by
using the algebraic ({\it ie}, sheaf-theoretic) technology of ADG.

\subsubsection{The conceptual `leit motif' underlying both SSTFDAs
and DSTFDAs: from an ADG-theoretic and physical perspective, more
than a superficial, formal resemblance}

First of all, we state it up-front that, geometrically speaking,
the singularities and various other (differential) geometric
pathologies of the (smooth) manifold are arguably due to its
point-set character. For any of the points of $M$ can be the {\it
locus} of a singularity of some physically important (smooth)
field. Thus, the common `strategy' behind SSTFDAs and DSTFDAs is
to somehow downplay (or perhaps even evade) the pointed nature of
$M$, the former with differential geometrical aims\footnote{In
fact, with the theory of non-linear PDEs.} in mind, while the
latter with topological issues in mind. Below, we briefly bring
forth from 3.1.7 and recapitulate from \cite{sork0} the basic
steps in the construction of the so-called {\em finitary
topological spaces}, which are structures originally intended to
replace or `approximate' the topology of continuous (spacetime)
manifolds by finitistic means. In the process, we will see and
highlight the close ties between the two approaches, especially
when Sorkin's constructions are cast algebraically---the so-called
combinatory-algebraic description of spacetime foam
\cite{rapzap1,rapzap2}.\footnote{The reader should note that in
5.2.1, before we give a finitary-algebraic and ADG-based
resolution of the inner Schwarzschild singularity, we are going to
discuss further the virtues of DSTFDAs.} For technical details
about what is going to be mentioned below, the reader is referred
to \cite{sork0,rapzap1,rapzap2}.

\begin{enumerate}

\item {\bf Smearing points.} The first step in Sorkin's {\em
finitary topological replacements of a continuous (spacetime)
manifold}\footnote{Say, spacetime regarded as a topological ({\it
ie}, $\cont$-) manifold.} is {\em the replacement of the latter's
points by `coarse', open sets about them}. The basic intuition
underlying this substitution is that geometrical points represent
ideal elements in the theory, when, in fact, what we actually
determine by our `coarse' space(time) measurements are never
points (or `instances'), but rather `large' (`blown-up' or
`extended') regions (or time-intervals) about them. This
`smearing' of points by regions in the DSTFDAs theory is the
analogue in the SSTFDA theoresis of concentrating on distributions
rather than on the `geometrical point-functions'
$\smooth(M)$---{\it ie}, the main idea is to generalize (or
enlarge) the (class of) smooth (differentiable) functions on the
point-set $M$.

\item {\bf Factoring out points set-theoretically
(topologically).} As we saw in 3.1.7, after the said substitution,
Sorkin effectively does away with points by grouping them into
`{\em indistinguishability equivalence classes}'. Namely, he first
covers a bounded region $X\subset M$ with a {\em locally finite
(finitary) open covering}\footnote{Shortly we will see how this
assumption of `finitarity' (local finiteness) is also of great
import in the SSTFDAs theory.} $\gauge_{i}$, and then he defines
the following `indistinguishability\footnote{Relative to our
coarse observations or `measurements' $U\in\gauge_{i}$.}
equivalence relation' between $X$'s points:\footnote{This
expression is identical to (\ref{eqxx1}) given earlier. In fact,
the short exposition of Sorkin's scheme given below is the same as
the one given in the bonus paragraph in 3.1.7, but with different
interpretational emphasis.}

\begin{equation}\label{eqx20}
X\ni x\stackrel{\gauge_{i}}{\sim}y\in X\Leftrightarrow \Lambda
|_{\gauge_{i}}(x)=\Lambda |_{\gauge_{i}}(y)
\end{equation}

\noindent where $\Lambda |_{\gauge_{i}}(x):=\bigcap\{
U\in\gauge_{i}|~x\in U\}$ is the `smallest' open set in the
subtopology $\tau_{i}$ of $X$ generated by the open sets in
$\gauge_{i}$\footnote{That is, $\tau_{i}(X)$ is generated by
finite intersections of countable unions of open sets $U$ in
$\gauge_{i}$.} that contains $x$.

Then, it is shown that {\em the quotient space
$P_{i}:=X/\stackrel{\gauge_{i}}{\sim}$, consisting of
$\stackrel{\gauge_{i}}{\sim}$-equivalence classes of points in
$X$, is a partially ordered set (poset) with the structure of a
$T_{0}$-topological space}, and it is coined {\em the finitary
substitute of $X$}.

As noted in section 3, the $P_{i}$s can also be viewed as {\em
simplicial complexes} $\mathcal{K}_{i}$ when one regards each
poset as a family of abstract $k$-simplices ($k=1,2,3,\ldots$),
with each $k$-simplex being defined as a collection of $k+1$ open
sets $U_{\alpha}\in\gauge_{i}~(\alpha=0\ldots k)$ of non-trivial
common intersection; that is to say,

\begin{equation}\label{eqx21}
\{ U_0, \ldots , U_k\} \in \mathcal{K} \: \Leftrightarrow \: U_0
\cap U_1 \cap \ldots \cap U_k \neq \emptyset
\end{equation}

\noindent The $k$-simplices in $\mathcal{K}_{i}$ are then
partially ordered by the homological incidence relation `{\em is
the face of}' to the effect that $\mathcal{K}_{i}$ itself can be
viewed as a poset.\footnote{Indeed, one may recall that
$\mathcal{K}_{i}$ is sometimes referred to as the
Alexandrov-\v{C}ech `{\em nerve complex}' associated with the
locally finite covering $\gauge_{i}$ of $X$ \cite{alex,cech}. For
short, and for obvious reasons, we call it a `{\em finitary
complex}'.}

\item {\bf Factoring out points algebraically.} One can get a
clearer idea of the close affinities between Sorkin's work on
topological locally finite poset discretizations of
$\cont$-manifolds and the spacetime foam dense singularity
constructions, by passing to the {\em Gel'fand-dual algebraic
picture} of the finitary topological posets/simplicial complexes
above involving the so-called {\em incidence algebras} associated
with the said $P_{i}$s, or equivalently, with the
$\mathcal{K}_{i}$s \cite{zap0,rapzap1,rapzap2,zap1}.

Without going into any technical detail, as also mentioned earlier
when we were discussing the finitary differential triads, one
associates with Sorkin's $P_{i}$s associative, but in general
noncommutative, $\mathbb{K}$-algebras ($\mathbb{K}=\R$ or $\com$)
$\omg_{i}(P_{i})$s, which are called {\em the incidence algebras
of the finitary topological posets}. The main characteristic of
the $\omg_{i}$s is that they are {\em discrete differential
algebras of generalized maps (arrows)}
\cite{dimu1,dimu2,dimu4,dimu3}, which can be decomposed into a
direct sum of an abelian subalgebra
$\sstruc_{i}\equiv\omg^{p=0}_{i}$ of `coordinate', position-like
maps, and a $\Z_{+}$-graded $\sstruc_{i}$-bimodule
$\bigoplus_{p}\omg_{i}^{p\geq1}$ of discrete differential,
momentum-like maps comprising $p$-graded linear subspaces in
$\omg_{i}$. At the same time, there is a {\em linear,
$\sstruc_{i}$-Leibnizian and nilpotent Cartan-K\"ahler
differential operator} $\kd$ effecting maps
$\kd:~\omg^{p}_{i}\mapto\omg_{i}^{p+1}$.\footnote{Indeed, $\kd$
defines the $\omg_{i}$s as discrete {\em differential} algebras.
As noted in section 3, $\kd$ can be expressed in terms of the
homological boundary (border) and coboundary (coborder) operators
defined on the respective $\mathcal{K}_{i}$s \cite{zap1}.}

One then, by virtue of a procedure called {\em Gel'fand
spatialization} \cite{rapzap1,rapzap2}, recovers the finitary
poset's points (vertices) and the Sorkin graph-topology that the
arrows in its Hasse diagram define by going to the {\em primitive
spectrum} of $\omg_{i}$\footnote{Whose points, by definition, are
{\em primitive ideals} in $\omg_{i}$, which in turn may be
identified with the {\em (kernels of the) $\omg_{i}$'s irreducible
representations.}} and by assigning the so-called {\em Rota
topology} on it.\footnote{In fact, one first defines algebraically
the {\em generating relation} of the Rota topology (corresponding
to the {\em immediate arrows}, or {\em links}, or even {\em
covering relations} in the Hasse diagram of the $P_{i}$), and then
recovers the finitary poset topology as the {\em transitive
closure} of this generating relation.} Moreover, one can show that
the said ideals are {\em differential ideals} relative to the
Cartan-K\"ahler differential $\kd$ \cite{zap1}.

\item {\bf $X$'s points are `blown-up' or smeared, and then quotiented-out:
Gel'fand duality on $\smooth$-manifolds; forced `surgery' on
singular points.} The conceptual distillation of the steps 1--3
above is that first the geometrical points of the original
(topological) space(time) manifold $X$ were `blown-up', `smeared',
or `enlarged' by being replaced by the $U$s in $\gauge_{i}$ about
them,\footnote{In Sorkin's scheme in particular, by the minimal
open cells $\Lambda(X)|_{\gauge_{i}}$---what we coined Sorkin's
{\it ur}-cells (relative to $\gauge_{i}$) in 3.1.7.} to the effect
that the `points' of the resulting $P_{i}$ are `coarse'
equivalence classes of points. Moreover, by going to the algebraic
realm of the corresponding $\omg_{i}$s, on the one hand the points
(:$\stackrel{\gauge_{i}}{\sim}$-equivalence classes) of $P_{i}$
are replaced by (differential) ideals in its $\omg_{i}$, and on
the other, one uncovers differential geometric, not just
topological, information encoded in the graded discrete
differential incidence algebras.\footnote{Indeed, here we witness,
in a finitary setting, one of the basic tenets of ADG: that {\em
differentiability} (or differential structure) {\em comes from
algebraicity} (or algebraic structure). This is in the spirit of
Leibniz's relational (:algebraic) conception of differentiation
(derivative).}

The aforedescribed enlargement and the concomitant `modding-out'
(substitution) of the geometrical points of $X$ by ideals {\it \`a
la} Gel'fand may be viewed as a discrete analogue of the basic
result from the application of Gel'fand duality to
$\smooth$-smooth manifolds, namely, the reconstruction of a
$\smooth$-smooth manifold $M$ as the spectrum $\gelsp$ of its
topological algebra ${}^{(\R)}\smooth(M)$ of smooth ($\R$-valued)
functions \cite{mall0,mall-1}.\footnote{Gel'fand duality is the
basis for the equivalence between $M$ and $\smooth(M)$ (or
$\smooth_{M}$) noted in (\ref{eq1}).} To recall briefly this
construction, let $M$ be a differential manifold and $p$ one of
its points. One considers then the following collection of smooth
$\R$-valued functions on $M$

\begin{equation}\label{eqx22}
\mathcal{I}_{p}=\{\phi:~M\mapto\R
|~\phi(p)=0\}\subset{}^{\R}\smooth(M)
\end{equation}

\noindent It is easy to verify that $\mathcal{I}_{p}$ is a {\em
maximal ideal} in ${}^{(\R)}\smooth(M)$ and that the quotient of
the latter by the former is isomorphic to the reals:

\begin{equation}\label{eqx23}
{}^{\R}\smooth(M)/\mathcal{I}_{p}\simeq\R
\end{equation}

\noindent Accordingly, the set
$Spec[{}^{(\R)}\smooth(M)]$\footnote{As noted above, often
symbolized by $\gelsp[{}^{(\R)}\smooth(M)]$.} of all maximal
ideals $\mathcal{I}_{p}~(\forall p\in M)$ of ${}^{(\R)}\smooth(M)$
such that

\begin{equation}\label{eqx24}
\R\hookrightarrow{}^{\R}\smooth(M)\mapto{}^{\R}\smooth(M)/\mathcal{I}_{p}
\end{equation}

\noindent (within an isomorphism of the first term), is called
{\em the (real) (Gel'fand) spectrum of} ${}^{(\R)}\smooth(M)$.
Furthermore, if ${}^{(\R)}\smooth(M)$---regarded algebraic
geometrically as a commutative ring---is endowed with the
so-called Zariski topology, or equivalently, with the usual
Gel'fand topology \cite{harts},\footnote{That the Gel'fand and the
Zariski topology on $\gelsp[{}^{\R}\smooth(M)]$ are identical is
due to the fact that ${}^{(\R)}\smooth(M)$ is an abelian {\em
regular} topological algebra---{\it ie}, maximal ideals in
${}^{(\R)}\smooth(M)$ are also its prime ideals, so that the
maximal spectrum of ${}^{(\R)}\smooth(M)$ coincides with its prime
spectrum \cite{mall0}.} then the `point-wise' map

\begin{equation}\label{eqx25}
M\ni p\mapsto \mathcal{I}_{p}\in\gelsp[{}^{(\R)}\smooth(M)]
\end{equation}

\noindent proves to be a {\em homeomorphism} between the usual
$\cont$-topology of $M$ ({\it ie}, $M$ being regarded simply as a
topological manifold) and the Gel'fand (Zariski) topology on
$\gelsp[{}^{\R}\smooth(M)]$.\footnote{Of course,
$\gelsp[{}^{\R}\smooth(M)]$, initially regarded as a structureless
point-set (having for points the ideals $\mathcal{I}_{p}$),
inherits from the one-to-one and onto map in (\ref{eqx25}) not
only the usual locally Euclidean topological structure of $M$, but
also the differential ({\it ie}, the $\smooth$-smooth) one. It is
precisely this identification (both in topological as well as in
differential structure) of $M$ with $\gelsp[{}^{\R}\smooth(M)]$
that underlies the left-pointing arrow in the tautesis of $M$ with
$\smooth(M)$ in (\ref{eq1}), which in turn supports our claim in
section 1 that {\em a differential manifold $M$ is nothing else
but the algebra $\sstruc=\smooth(M)$ of $\smooth$-smooth functions
on (coordinates of) its points} (or equivalently, $M$ {\em is} the
structure sheaf $\struc=\smooth_{M}$).} All in all, the essential
idea of Gel'fand duality here is to substitute the underlying
space(time) continuum by the (algebras of) objects
(functions/fields) that live on it, and then recover it by a
suitable technique, which we have coined {\em Gel'fand
spatialization} in the past \cite{rapzap1,rapzap2,malrap3}.

Furthermore, {\it vis-\`a-vis} singularities, which, as noted
earlier, are arguably due to the pointed nature of $M$, the
factoring-out of $\mathcal{I}_{p}$ from ${}^{(\R)}\smooth(M)$ in
(\ref{eqx23}) (to yield $\R$---the reals, which, in turn, induce
the usual Euclidean topological and differential structure to the
quotient space), in a way factors out singularities completely
analogously to the differential ideal-smearings and `modding-outs'
involved in the definition of spacetime foam algebras in
(\ref{eqx6}). This is straightforward to see if for example one
considers briefly the interior Schwarzschild singularity
associated with the gravitational field of a point particle $p$ of
mass $m$.\footnote{We will come back to tackle this example from a
DSTFDA and a SSTFDA ADG-perspective in the next section. This is
just a sketchy discussion of the inner Schwarzschild singularity
here just to illustrate this potential doing away with
singularities by algebraic `ideal-smearing' and factoring-out
points of the $\smooth$-smooth spacetime continuum.} There, as it
is well known, the Schwarzschild solution to the vacuum Einstein
equations has a singularity right at the point mass $m$---that is
to say, when the differential spacetime manifold $M$ is charted by
a smooth coordinate system\footnote{In fact, by {\em any}
$\smooth$-smooth coordinate system!} with origin at the point
particle $p$, the Schwarzschild metric solution $g_{S}$ has a
singularity at distance $r=0$ from the particle, since the smooth
function $\phi(p)=r(p)(=\sqrt{x^{2}+y^{2}+z^{2}}),~\forall p\in M$
\footnote{In parenthesis above, is the distance function expressed
in the usual smooth cartesian coordinates. Also, the expression
`$\forall p\in M$' above may be read `{\em on any point $p$ of the
smooth spacetime manifold $M$ one decides to place the point-mass
$m$}'.} appears as a coefficient in the denominator in the
expression for $g_{S}$.\footnote{See (\ref{eqyy3}) in 5.1.} But
the smooth functions like $r(p)$ are precisely the ones that {\em
define} the maximal ideals $\mathcal{I}_{p}$ in
${}^{(\R)}\smooth(M)$ (\ref{eqx22})\footnote{Algebraic
geometrically speaking, the points of
$\gelsp[{}^{\R}\smooth(M)]$.} and, moreover, precisely the ones
that are factored out of $\smooth(M)$ in (\ref{eqx22}).

Of course, one cannot claim that the procedure of Gel'fand duality
exercised on the smooth manifold actually does away with
point-singularities, since, simply set-theoretically speaking, the
factoring out of the ideals from $\smooth(M)$ is tantamount to
removing by fiat---as it were, surgically dissecting in an {\it ad
hoc} and forced fashion---from $M$ the offensive points. In the
Schwarzschild case, this corresponds to removing by fiat the point
mass $m$ placed at $p$, leaving thus $M-\{ p\}$---a `{\em
punctured spacetime}'---as an effective (regular) manifold in the
theory. Equivalently, as we will see in more detail in 5.1, as an
effective, regular spacetime manifold in the Schwarzschild case,
one may regard the space $M-L_{t}$ corresponding to the original
(total) spacetime manifold $M$ less the singular continuous
`wristwatch time-line' $L_{t}:=\{ p\in M:~x_{i}(p)=0\}$ of the
point-mass placed at the origin of an (analytic) Cartesian
coordinate system $(x_{\mu})$ charting $M$
\cite{df}.\footnote{That is, in this case, the dissected effective
regular spacetime is $M-L_{x_{0}=t}~(t\in\R)$.} In either case,
this is a forced exclusion of singularities that {\it prima facie}
appears to accord with Einstein's dictum for excluding
singularities in (Q2.1), but which, from our ADG-theoretic
viewpoint, simply indicates both the inadequacy of the
$\smooth$-smooth (spacetime) manifold based CDG in describing the
(here gravitational) field and the law (differential equation)
that it obeys right at its singular source-{\it locus}. In
contradistinction,

\bigskip \noindent (R5.3)\hskip 0.9in
\begin{minipage}{11cm}
\noindent the factoring out of spacetime foam dense singularities
expressed in (\ref{eqx6}), and the concomitant application of the
differential manifold-free ADG to sheaves of spacetime foam
algebras over the (in principle) arbitrary base space(time) $X$,
in no way represents such a theoretically {\it ad hoc} surgical
exclusion of the singular {\it loci} from $X$; rather, {\em when
sheaves of the differential spacetime foam algebras are used as
structure sheaves $\struc$ of generalized coefficient functions in
the theory, one simply witnesses on the one hand the absorption or
integration of $\smooth$-smooth singularities in those generalized
`coordinate-arithmetics',\footnote{For after all, the algebra of
$\smooth$-smooth functions is manifestly embodied in the foam
algebras (\ref{eqx10}).} while on the other, that the entire,
essentially algebraic, differential geometric mechanism of the
theory still holds, and in no way breaks down, in their presence}
\cite{mall2,malros1,mall3,malros2,malros3}.\footnote{We will
briefly recall the results of the application of ADG to spacetime
foam dense singularities shortly, in 4.2.4 below.}
\end{minipage}

\end{enumerate}

\paragraph{Topology refinement versus singularity refinement: some
common grounds and close ties resting on (sheaf-theoretic)
localization.} In the foregoing we argued that the points of the
smooth spacetime continuum $M$ are the `bearers' or `{\em
carriers}' of its singularities.\footnote{Or equivalently, by
Gel'fand duality, the smooth coordinates of $M$'s points in
$\smooth(M)$, which themselves are the bearers of $M$'s points
(\ref{eqx25}), are the carriers of its singularities. In other
words, any of $M$'s points can serve as the host-{\it locus} of a
singularity of some physically important $\smooth$-smooth field,
which is just a $\otimes_{\sstruc\equiv\smooth(M)}$-tensor.}

In a topological context---and in particular in the aforementioned
$T_{0}$-poset discretizations of $\cont$-spacetime
manifolds---Sorkin too basically assumes that ``{\em the points of
the continuum are the carriers of its topology}'' \cite{sork0}.
Moreover, and in order for the $T_{0}$-poset substitutes to
qualify as genuine replacements of the continuum that are coarsely
approximating its continuous ($\cont$) topology, as we saw in
section 3 Sorkin provides an {\em inverse limit} (re)construction
of $X$, regarded as a point-set, from an inverse system (or net)
$\inv=\{ P_{i}\}$ of finitary $T_{0}$-posets. The basic notion
here is that of {\em topological refinement},\footnote{Which may
be also coined {\em covering refinement}.} which we may formally
symbolize as

\begin{equation}\label{eqx26}
\gauge_{i}\tref\gauge_{j}\Leftrightarrow\tau_{i}(X)\tref\tau_{j}(X)
\end{equation}

\noindent and read directly as follows:\footnote{The reader should
note the use of the same `precedence' or `order' relation
`$\prec$' for both ($t$)opological ($\tref$) and ($s$)ingularity
($\sref$) ($ref$)inement. Below, we will show the close links
between the two notions, which in turn reflect the intimate
technical and conceptual affinities between the algebraic DSTFDA
and SSTFDA approaches to spacetime foam respectively, especially
when viewed from the unifying algebraic (:sheaf-theoretic)
perspective of ADG.} {\em the subtopology $\tau_{i}$ of $X$
generated by the open sets of the locally finite open covering
$\gauge_{i}$ of $X$ is coarser than the subtopology $\tau_{j}(X)$
generated by $\gauge_{j}$}.\footnote{And {\it vice versa},
$\tau_{j}$ is {\em finer} than $\tau_{i}$. One may represent the
said inverse system $\inv$ of finitary posets by the pair
$(P_{i},\tref)$. In fact, we read from \cite{sork0}, it is this
(partial order) relation of topological refinement, which can be
also read simply as `{\em $\tau_{i}(X)$ is a subtopology of
$\tau_{j}(X)$}', that makes one think of the set $\{\tau_{i}(X)\}$
as a {\em net}---essentially, {\em a right-directed set of
elements} (like the $L=(\Lambda ,\leq)$ we encountered earlier in
defining spacetime foam algebras; also, `{\em right-directed}'
here meaning that $\forall\tau_{i},\tau_{j}$, or simply index-wise
$i,j$, in the net, $\exists k$ in the net, such that $i,j\tref
k$)---which in turn indexes the elements $P_{i}$ of the projective
system $\inv$. However, by abuse of terminology, and without
causing any confusion, in what follows we may use the terms
`inverse system' and `net' interchangeably for $\inv$.} As we
noted in section 3, the physical semantics of this inverse limit
procedure, as also Sorkin points out in \cite{sork0}, is that as
one employs more numerous and `smaller' or `finer' open sets in
order to cover (the points of) $X$---as it were, as one employs
higher (microscopic) power of resolution to locate (or
`localize'\footnote{That is, determine the {\it locus} of a point
in $X$. We will come back to the issue of (sheaf-theoretic)
localization shortly.}) and effectively separate (or distinguish
between) $X$'s points by using open sets about them,\footnote{This
for instance may be regarded as the essence of the, in this case,
$T_{0}$-axiom of separation of point-set topology, namely, that
the points of a set are the carriers of the open sets (about
them), which sets, in turn, {\em separate} them and at the same
time {\em define} $X$'s topology---{\it ie}, qualify the {\it a
priori} structureless point-set $X$ as a {\em topological space}
proper.} at the (ideal\footnote{And we use the epithet `{\em
ideal}', because the points of the continuum are operationally
non-pragmatic, non-realistic theoretical artifacts; while, the
operationally realistic entities---what we actually determine by
our acts of localization of `events'---are `large', `extended',
coarse regions about them \cite{sork0,sork1,rapzap1,rapzap2}.}) of
infinite refinement (or localization) of the finitary posets, one
obtains a space that is topologically indistinguishable ({\it ie},
effectively homeomorphic, modulo Hausdorff reflection
\cite{kopperman}) to the original topological space(time) manifold
$X$ that one started with. Formally, one writes:

\begin{equation}\label{eqx27}
\underleftarrow{\lim}\inv=\lim_{\infty\leftarrow
i}P_{i}=P_{\infty}\stackrel{\mathrm{homeo.}}{\simeq}X
\end{equation}

Now, a functional representation of topological refinement and of
the inverse system $\inv$ bears a close resemblance to how we
represented singularity refinement (and its `order-covariant'
notion of quotient regularity $\reg$) in (\ref{eqx19}), namely, we
read from \cite{sork0} that

\begin{equation}\label{eqx28}
\gauge_{i}\tref\gauge_{j}(\equiv\tau_{i}(X)\tref\tau_{j}(X))\Leftrightarrow
P_{j}\stackrel{f_{ij}}{\mapto}P_{i}
\end{equation}

\noindent which says that `$\gauge_{j}$ is topologically finer
than $\gauge_{i}$' is equivalent to `the existence of a {\em
continuous surjection} $f_{ij}$ between the corresponding
$T_{0}$-posets (\ref{eqxx2}).\footnote{Or equivalently,
order-theoretically speaking, an onto monotone map ({\it ie}, a
partial order preserving epimorphism).} In fact, in terms of such
continuous surjections, the inverse system can now be equivalently
written as the pair $\inv=(P_{i},f_{ij})$. This inverse system of
finitary $T_{0}$-posets and continuous epi-maps between them
enjoys a {\em universal mapping property} (\ref{eqxx2}), which in
turn guarantees an inverse limit topological space $P_{\infty}$
appearing in (\ref{eqx26})\footnote{It too a $T_{0}$ topological
space---{\it ie}, non-Hausdorff (not $T_{2}$).} as a continuous
surjection $f_{i\infty}:~P_{\infty}\mapto P_{i}$.

In a similar functional way, and by Gel'fand duality, topological
refinement corresponds to a {\em continuous surjective algebra
homomorphism $\hat{f}_{ij}$ between the finitary posets'
respective incidence algebras $\omg_{j}$ and $\omg_{i}$}:
$\hat{f}_{ij}:~\omg_{j}\mapto\omg_{i}$
\cite{rapzap1,rapzap2}.\footnote{Here {\em continuous} means {\em
respecting the aforementioned Rota topology on the corresponding
incidence algebras' primitive spectra}---a topology which derives
from the algebraic structure of the incidence algebras, which in
turn is preserved by the said homomorphisms. In fact, as also
noted in section 3, this correspondence
($(P_{i},f_{ij})\mapto(\omg_{i},\hat{f}_{ij})$) is {\em
functorial} between the respective categories (of finitary
posets/simplicial complexes--order preserving maps/simplicial maps
and their finitary incidence algebras/algebra homomorphisms)
\cite{rapzap1,rapzap2,zap1}.} Since {\em the $\omg_{i}$s are
categorically-dual to the $P_{i}$s} \cite{malrap2,malrap3,zap1},
it follows that the collection
$\diromg=\{\omg_{i}\}=(\omg_{i},\hat{f}_{ij})$ is an {\em
inductive system} under topological refinement. As such, and in
complete analogy to the inverse limit procedure exercised on
$\inv$ in (\ref{eqx27}), it possesses a {\em direct} or {\em
inductive limit} incidence algebra $\omg_{\infty}$, which we may
formally write as follows

\begin{equation}\label{eqx29}
\underrightarrow{\lim}\diromg=\lim_{i\rightarrow
\infty}\omg_{i}=\omg_{\infty}
\end{equation}

\noindent The relevance and utility of this inductive limit
process becomes even more transparent when one views, in the
sheaf-theoretic context of ADG, {\em finitary spacetime sheaves
(finsheaves) of incidence algebras}
\cite{rap2,malrap1,malrap2,malrap3} and one emphasizes the fact
that they define (finitary versions of) {\em differential triads}
(the fintriads we saw in section 3), which are the basic building
blocks of ADG \cite{mall1,mall2,malrap2,malrap3}.

\paragraph{Finsheaves.} To recall the relevant concepts and constructions briefly,
initially, the basic intuition behind \cite{rap2} was to do the
same to the sheaf ${}^{\R}\cont_{X}$ of (germs of) continuous
($\R$-valued) functions on the topological manifold $X$ as Sorkin
did for the base space $X$ itself in \cite{sork0}. Thus, {\em
emphasis was placed on the functions that `live' on space{time}
rather than on the background space(time)} {\it per se}, which is
traditionally the domain of definition of those
functions.\footnote{See the last section for a discussion of the
traditional `{\em geometric}' (`domain dependent') {\it versus}
the modern `{\em algebraic/categorical}' (`domain independent')
conception of functions.} The sheaf-theoretic analogues of the
finitary substitutes $P_{i}$ were then coined {\em finitary
spacetime sheaves}\footnote{And symbolized by $\mathcal{S}_{i}$ in
\cite{rap2}.} and were seen to be sheaves of continuous functions
over Sorkin's locally finite topological posets\footnote{Write
formally, $\mathcal{S}_{i}(P_{i})$.}---as it were, the finitary
substitutes of the aforementioned `continuum sheaf' $\cont_{X}$.
With our discussion above in mind, in a sheaf-theoretic context
the most important notion is that of {\em localization}, a notion
that is essentially tautosemous to the technical one of the
process of {\em sheafification} (of a presheaf)
\cite{mall8}.\footnote{In fact, we read from \cite{mall8} that
``{\em sheafification is localization}''.} Indeed, the very
definition of the {\em stalks} of a sheaf---the elementary,
`point-like' building blocks of the sheaf space, each erected over
each point $p$ of the base topological space $X$\footnote{For
recall that for a general sheaf $\mathcal{S}_{X}$, the sheaf space
is given by $\mathcal{S}=\bigcup^{\mathrm{dis}}_{p\in
X}\mathcal{S}_{p}=\bigoplus_{p\in X}$, where
$\bigcup^{\mathrm{dis}}$ denotes {\em disjoint union}
\cite{mall1}.}---involves in an essential way a {\em direct limit}
process exercised on an inductive system of open subsets of $X$
relative to which a presheaf has been first defined
\cite{mall1,rap2}.\footnote{This direct limit procedure may be
(topologically) interpreted as {\em fine graining}---the
resolution or analysis of the sheaf space into its `{\em ultra
local}' elements, its stalks (fibers). Thus, we can roughly say
here, in complete analogy to what we said earlier for the
topological space $X$ and its points, that {\em the topology of
the sheaf space is carried by its stalks}---the stalks which, in
turn, like the {\it a priori} thought of as being completely
disconnected points of the point-set $X$, inherit from
$\mathcal{S}$ the {\em discrete relative topology}. Of course,
this is more than a formal analogy (between the points of $X$ and
the stalks of $\mathcal{S}$) if one considers the very definition
of a sheaf $\mathcal{S}_{X}$ as a {\em local homeomorphism}
(between $X$ and $\mathcal{S}$) \cite{mall1,rap2}.} Moreover, from
a physical point of view,

\bigskip \noindent (R5.4)\hskip 0.9in
\begin{minipage}{11cm}
\noindent {\em localization\footnote{Roughly speaking, making the
objects/structures one is dealing with $U$-dependent ($U\subset
X$), with $U$ `varying' over the lattice of open subsets of $X$
which {\em defines} the latter as a topological space in the first
place.} may be thought of as the process of `relativization',
`gauging' and concomitant `dynamicalization'}\footnote{That is to
say, making the objects/structures one is dealing with {\em
dynamical variables} that are formally varying with respect to the
background topological space, which thus acts as a `virtual',
`surrogate' external `parameter space'
\cite{malrap1,malrap2,malrap3}. (The epithets `virtual' and
`surrogate' for the external `carrier' topological space $X$ will
be qualified shortly, after the ADG-theoretic notion of a
connection on a sheaf is introduced.)}; hence, so is
sheafification.\footnote{Let us furthermore note at this point
Finkelstein's motto in the context of his Quantum Relativity
theory that, in essence, ``{\em (quantum) relativization is
dynamicalization}''. We will return to address quantum (gravity)
issues in the sheaf-theoretic context of ADG in section 6.} In
other words, {\em a sheaf (of objects of any kind) by definition
entails the dynamical variability of those objects}.
\end{minipage}

\vskip 0.1in

With the remarks in (R4.4) in mind, and as noted in section 3, we
stress that the second author's original {\em mathematical} aim in
defining finsheaves as noted at the end of \cite{rap2}, was to
organize the discrete differential incidence algebras that we saw
earlier into sheaf-like structures so as to apply the first
author's ADG-theoretic concepts and differential geometric
machinery to a finitistic or combinatory-algebraic setting quite
remote from the realm of the $\smooth$-smooth spacetime continuum,
something that {\it prima facie} befits ADG's differential
calculus (CDG) and differential manifold-free character
\cite{mall1}. At the same time, the {\em physical} idea underlying
this sheaf-aggregation of incidence algebras was that, when the
locally finite posets were interpreted not as finitary topological
spaces proper, but as {\em causal sets} (causets)
\cite{bomb87,sork4,sork-1,sork1,sork2,sork3}, while their
associated incidence algebras as {\em quantum causal sets}
(qausets) \cite{rap1}, the said agglomeration of qausets into
(fin)sheaves would provide one with a natural setting in which to
study the `localization', `gauging', `curving' and the concomitant
dynamical variation of `discrete quantum causality'
\cite{malrap1}. All in all, finsheaves of qausets were intuited as
being `natural' models for developing a {\em `discrete' version of
Lorentzian QG}; moreover, one which is {\em manifestly background
spacetime manifold independent},\footnote{Yet still employing the
full differential geometric panoply of CDG; albeit, in a reticular
setting and entirely by algebraico-categorical means
\cite{malrap2}.} quite unlike the usual ({\it eg}, canonical or
covariant) $\smooth$-smooth spacetime manifold based approaches to
QG which still effectively employ a background differential
manifold \cite{malrap1,malrap2,malrap3}.

As also briefly alluded to in section 3, the five essential
observations (results) for actually setting up finsheaves of
incidence algebras (ultimately, with an eye towards applying ADG
to them) were that:

\begin{enumerate}

\item The correspondence finitary posets/simplicial complexes
$\mapto$ incidence algebras is {\em functorial} (Gel'fand duality
functor) \cite{rapzap1,rapzap2,zap1}.

\item In fact, first in \cite{malrap1} it was noted that the map
$P_{i}\mapto\omg_{i}$ is {\em by construction} (of the Rota
topology on $\omg_{i}$ via Gel'fand spatialization/duality
\cite{zap0,rapzap1,rapzap2}) a local homeomorphism---{\it ie}, a
sheaf \cite{mall1,rap2}.\footnote{From now on we write
$\Omg_{i}(P_{i})$, or simply $\Omg_{i}$, for finsheaves of
incidence algebras. When we wish to emphasize the qauset
interpretation of the incidence algebras dwelling in the stalks of
the said finsheaves, we will use the symbol $\Qaus_{i}$ for the
latter, as we did throughout the trilogy
\cite{malrap1,malrap2,malrap3}.}

\item Finsheaves of the {\em differential} incidence algebras
define {\em finitary differential triads} $\triad_{i}$
\cite{malrap2}.

\item Categorically speaking, and by Gel'fand duality, {\em
inverse limit} `classical continuum localizations' of the
$P_{i}$s\footnote{The epithet `{\em classical}' above comes from
\cite{rapzap1} when the continuum inverse limit of finitary posets
and, dually, of their associated incidence algebras, was
physically interpreted as {\em Bohr's correspondence principle or
limit}.} correspond to {\em direct limit} `classical continuum
localizations' of the corresponding incidence algebras and, {\it
in extenso}, of the fintriads $\triad_{i}$ that their finsheaves
define \cite{malrap2}.\footnote{As explained in section 3,
fintriads comprise in fact an inverse-{\em cum}-direct system
$\invtriad$. In 5.2.2, based precisely on Papatriantafillou's
inverse/inductive limit results of of differential triads as
applied to our fintriads, we will show how to totally evade the
inner Schwarzschild singularity---regarded as a localized,
`static' point-singularity---solely by
finitistic-algebraic/categorical means.}

\end{enumerate}

By now, and after this short digression on (fin)sheaf-theoretic
localizations (of the DSTFDAs), we hope that the close conceptual
and structural similarities between the SSTFDA and the DSTFDA
approach to spacetime foam have become more transparent. In the
next couple of paragraphs we bring the two approaches even closer
together by first exploiting, in a sheaf-theoretic context, the
issue of finitarity (local finiteness) and its import in the
SSTFDA scheme, as well as by giving heuristic arguments linking
topological and singularity refinement.

\paragraph{`Finitarity' (local finiteness): the condition underlying the fineness and
flabbiness of (structure) sheaves of multi-foam algebras.} Before
we proceed with the next paragraph, where we unite SSTFDAs and
DDSTFAs under the sheaf-theoretic roof of ADG, and before the
sub-subsection 4.2.4 below, where we outline the results of the
successful application of ADG to spacetime foam dense
singularities and the sheaves of spacetime foam algebras thereof,
we would like to dwell for a while on the issue of {\em
finitarity} or {\em local finiteness}. In the SSTFDAs context, the
assumption of local finiteness plays an important role in actually
showing that the sheaves of spacetime foam algebras---employed as
structure sheaves of generalized arithmetics or `coordinates' to
replace the `classical' one $\smooth_{X}$ of smooth functions on
the topological space $X$---are {\em fine} and {\em flabby}. In
turn, the flabbiness and fineness properties of those sheaves
secure the application of very basic sheaf-cohomological
constructions and associated applications of ADG---applications
which are of great importance in mathematical physics, as for
instance the construction of a {\em short exact exponential sheaf
sequence} and the consequences that this construction has for the
process of {\em geometric (pre)quantization} ({\it eg}, Weil's
integrality theorem) \cite{mall1,mall2,malrap2}. Putting it in a
negative way, in the case of Colombeau algebras for example, the
sheaves of which manifestly lack the flabbiness
property,\footnote{In a paragraph in 4.2.4 below, we will mention
that, actually, the failure of sheaves of Colombeau's non-linear
distributions \cite{colombeau} to be flabby is due to the
imposition of several {\em growth conditions} that these
generalized functions must satisfy at the vicinity of their
singularities. Moreover, as noted in \cite{malros3}, ``{\em this
lack of flabbiness of the Colombeau algebras is quite closely
related to a number of deficiencies}'' \cite{kaneko}. By contrast,
the non-linear spacetime foam distributions considered here do not
have to obey any such growth conditions around open neighborhoods
of their dense singularities in $X$, and one of the advantages of
that `growthlessness', {\it vis-\`a-vis} ADG and their potential
import in mathematical physics applications, is that the sheaves
thereof are indeed flabby.} a short exact exponential sheaf
sequence cannot be constructed; hence they would be of little
(abstract) differential geometric utility if one wished to use
them ADG-theoretically as structure sheaves of generalized
coefficients. So, let us see briefly how finitarity or local
finiteness appears as an important assumption in the theory of
spacetime foam algebras.

Once again we follow \cite{malros2,malros3}. As before, one
considers a collection $\ds$ of (dense) singularity-sets in a
Euclidean or locally Euclidean (manifold) space(time) $X$, the
elements $\dss\subset X$ of which are also assumed to satisfy the
two conditions (\ref{eqx1}) (`regularity density') and
(\ref{eqx2}) (`singularity coarse-graining') that we saw earlier.
Then, in order to show that the structure sheaf
$\struc_{X}\equiv\fa_{L,\ds,X}$ of spacetime foam algebras on $X$
is {\em fine} and {\em flabby}, one assumes that $\ds$ is {\em
locally finitely additive}. By the latter, and in complete analogy
to the locally finite open coverings of $X$ assumed in
\cite{sork0}, one roughly means that {\em every point $x$ of $X$
has an open neighborhood about it that non-trivially intersects a
finite number of the singularity-sets}.\footnote{Compare with
Sorkin's definition of a locally finite open covering $\gauge$ of
$X$ that we saw earlier: ``{\em for every point $x$ of $X$ there
is an open set about it that meets a finite number of the covering
sets in $\gauge$}''.} More precisely, one considers any sequence
$\dss_{k}$ ($k\in\mathbb{N}$) of singularity-sets in $\ds$ and
takes their union $\dss=\bigcup_{k}\dss_{k}$, which belongs to
$\ds$ by (\ref{eqx2}).\footnote{Write formally,
$\bigcup_{k}\dss_{k}=\dss\subseteq\ds$.} Then, for $U\subseteq X$
open, $\dss\cap U\in\ds|_{U}$, whenever

\begin{equation}\label{eqx30}
\begin{array}{c}
\forall x\in U, ~\exists V\subseteq
U~(V\mathrm{a~neighborhood~of}~x):\cr
\mathrm{the~set}~\{k\in\mathbb{N}:~\dss_{k}\cap
V\not=\emptyset\}~\mathrm{is~finite}
\end{array}
\end{equation}

\noindent where $\ds|_{U}$ denotes {\em the restriction of $\ds$
to $U$}. With respect to these restriction mappings, one then
proceeds and constructs a {\em complete presheaf}\footnote{Or
equivalently, a {\em sheaf} \cite{mall1}.} $\sfa$ of spacetime
foam algebras $\fa$ over the topological space $X$, which,
moreover, with the help of (\ref{eqx29}), proves to be fine and
flabby.\footnote{For more details about this construction, the
reader is referred to \cite{malros2,malros3}.}

\paragraph{Heuristic remarks about `continuum'
inverse/inductive limits under topological/singularity
refinement.} We begin with Sorkin's assumption that $X$ is a
relatively compact (open and bounded) region of the spacetime
manifold $M$, that it admits locally finite open covers
$\gauge_{i}$, and that it is inhabited by dense singularities
(whose set-theoretic complements in $X$ are also dense). In turn,
the open sets comprising the covers are densely packed with
singularities of `all sorts'. The basic physical idea behind
Sorkin's open coverings' refinement is that as one employs higher
and higher power of resolution (of $X$ into is points---``{\em the
carriers of its topology}'' \cite{sork0}) so as to localize them
with higher accuracy by using smaller and more numerous open sets
about them, at the inductive or projective limit of infinite
refinement (:infinite power of resolution and localization) one
effectively recovers (modulo Hausdorff reflection) the continuous
point-set topology of $X$. For a singular {\it locus} (:point) in
particular,...

\paragraph{Growthlessness suitable for ADG's `background spacetimelessness'.}
One could argue that the main reason why global topological (and
differential topological) tools and methods were extensively
developed and used from the 60s until the 80s in the analysis of
spacetime singularities---in particular, in endowing
$\overline{M}=M\cup\partial M$ with a certain topology while at
the same time relegating the singular {\it loci} of the spacetime
manifold $M$ `asymptotically' or `marginally', to its boundary
$\partial M$ (relative to the chosen topology)---was that one
wished to study and make precise (as it were, rigorously
`quantify' in an analytic way) for instance how physically
observable geometrical properties of the gravitational field, such
as the Riemann curvature tensor\footnote{Which represents
gravitational tidal forces.} as well as its Ricci-tensor ($\ric$)
and Ricci-scalar ($\ricci$) contractions, {\em grow} ({\it ie},
`converge' or `diverge') as one (say, a gravitated test particle
of mass $m$) approaches by a continuous (or more fittingly,
smooth) path the singular {\it locus} at the manifold's
boundary.\footnote{In a mathematical sense, one would like to
model the growth of the gravitational field strength $R$ on the
particle, as the latter approaches the singular point, after a
{\em continuous limit-convergence} relative to the chosen
topology.} This is clearly stated in the following passage from
Clarke's book \cite{clarke4}:\footnote{By the way, the following
quotation further corroborates our arguments in 2.1.2 (based on
(Q?.?)) about singularities, situated at the boundary $\partial M$
of the otherwise smooth and regular spacetime manifold `bulk' $M$,
as being sites where CDG (Calculus or Analysis) ends, or more
graphically, `breaks down'.}

\bigskip \noindent (Q5.4)\hskip 0.9in
\begin{minipage}{11cm}
\noindent ``{\small...We must now give the definition of the noun
`singularity'. The fundamental idea is that space-time itself (the
structure $(M,g)$) consists entirely of regular points at which
$g$ is well behaved [{\it ie}, regular],\footnote{Our addition.}
while singularities belong to a set $\partial M$ of additional
points---`ideal points'---added onto $M$. {\small\em We denote the
combined set $M\cup\partial M$ by $ClM$, the closure of $M$, and
define the topology of this set to be such that phrases like `a
continuous curve in $M$ ending at a singularity $p$ in $\partial
M$', or `the limit of $R$\footnote{Here, by `$R$' Clarke
symbolizes the Riemann curvature tensor.} as $x$ tends to a
singularity $p$ is...' all have meanings corresponding to one's
intuitive picture of what they ought to mean}\footnote{Our
emphasis.}...}''
\end{minipage}

\vskip 0.1in

\noindent Indeed, and from a more general perspective, in the
theory of (non-linear) PDEs where, apart from the linear
distribution theory of Schwartz and the related theory of
generalized functions {\it \`a la} Sobolev, the non-linear
Colombeau distributions have recently enjoyed much popularity,
(differential) {\em growth conditions} are invariably imposed in
the neighborhood of all those functions' singularities. For
instance, we witness in \cite{clarke4}, where an entire section is
devoted to Sobolev spaces and their manifold applications to the
analysis of spacetime singularities, a plethora of conditions that
the {\it \`a la} Sobolev functionally generalized $g$ and $R$ must
satisfy locally, {\it ie}, in the vicinity of a
singularity.\footnote{Such conditions are usually imposed in order
to control and classify the `differential blow-up' ({\it eg}, give
upper bounds to the order of differentiability) of the metric and
its Riemann curvature tensor in the proximity of a singularity
({\em Sobolev `differentiability' classes} \cite{clarke4}).} On
the other hand, as we read from \cite{malros2}, in the context of
the application of ADG to the spacetime foam dense singularities
of Rosinger's non-linear distributions, the imposition of {\em
such growth conditions manifestly lifts the flabbiness property of
the respective structure sheaves}. That is to say, if for example
one wished to employ sheaves of Colombeau algebras (instead of
Rosinger's) as structure sheaves of generalized arithmetics in the
manner of ADG, precisely due to the growth conditions near their
singularities that the Colombeau distributions are demanded to
satisfy, the said sheaves would simply fail to be flabby. In turn,
as noted before, exactly because of this shortcoming, the
non-flabby Colombeau structure sheaves do not allow for basic
differential geometric constructions such as that of a short exact
exponential sheaf sequence---a vital construction in the
application of ADG to geometric (pre)quantization (via Weil's
integrality) for example \cite{mall2,mall5,malrap2}.\footnote{We
will return to discuss in more detail the application of ADG to
geometric (pre)quantization of gravity in 6.1.} Thus, in glaring
contradistinction to the CDG-based analysis of spacetime
singularities in \cite{clarke4}, we read from \cite{malros2}:

\bigskip \noindent (Q5.5)\hskip 0.9in
\begin{minipage}{11cm}
\noindent ``{\small\em ...Furthermore, as in the case of the
nowhere-dense algebras and also in space-time foam algebras, no
kind of condition is asked on the generalized functions in the
neighborhood of their singularities\footnote{Emphasis is
ours.}...}''
\end{minipage}

\vskip 0.1in

\noindent Indeed, this {\em growthlessness} of Rosinger's
non-linear generalized functions appears to be tailor-cut for
ADG's fundamental {\em spacetimelessness}. That is to say, the
purely algebraic, background geometrical spacetime manifold-free
ADG-theoretic perspective on gravity (GR) and its singularities
({\it eg}, via Rosinger's entirely algebraic theoresis of
spacetime foam dense singularities) is not concerned at all with
the standard (differential) geometric-analytic question above
whether and in what manner the Riemann curvature tensor grows
without bound (`diverges') as a singularity (situated on $\partial
M$) is approached in a continuous manner by a (smooth or regular)
path that a particle follows in $M$. ADG totally bypasses, by
purely algebraic means, the CDG-based analysis of gravitational
singularities, while the geometrical manifold-based reasoning (and
picturization!) that goes hand in hand with that analysis is
simply rendered obsolete and is of no import in our theory. Of
course, all this is not surprising, because in a `{\em cutting the
Gordian knot}' fashion ADG evades all the problems that
$\smooth$-smooth singularities present to CDG (Calculus or
Analysis):

\bigskip \noindent (R5.5)\hskip 0.9in
\begin{minipage}{11cm}
\noindent {\em Since there is no background differential spacetime
manifold in ADG's perspective on gravity (GR), there are no
singularities either, and it is begging the question to try to
classify, let alone attempt to define, singularities by analytic,
CDG-theoretic means when they do not `exist' in the first
place.\footnote{Again, by `{\em do not exist}' we mean that
singularities, which are built into the structure sheaf
$\struc\equiv\smooth_{M}$ of the underlying differential manifold
$M$, do not perturb the slightest bit the inherently algebraic,
base manifold independent differential geometric mechanism of ADG
and therefore, unlike CDG, they are not perceived as {\it loci}
where differentiability breaks down or is limited in one way or
another as DGSs appear to indicate.}}
\end{minipage}

\vskip 0.1in

\subsubsection{Differential geometry \`a la ADG in the presence of the most numerous and `wildest' from the CDG
perspective singularities: the versatility of ADG}

As repeatedly mentioned in the previous sections, and from a
sheaf-theoretic perspective, the CDG of $\smooth$-smooth manifolds
$M$ is entirely captured by employing $\smooth_{M}$ as the
structure sheaf of coefficients (or coordinates labelling $M$'s
points). At the same time, the gist of ADG consists in showing
that one can actually do differential geometry by using structure
sheaves $\struc$ other than the classical one $\smooth_{M}$, of
course, as long as these generalized coefficient algebra sheaves
provide one with the essential(ly algebraic) `differential
mechanism'\footnote{In point of fact, with a linear, Leibnizian
differential operator $\partial$, which is the `canonical' example
of a flat $\struc$-connection in ADG \cite{mall1}.} by virtue of
which one actually does differential geometry, and which in the
classical theory (CDG) is furnished by $\smooth_{M}$ or, what
amounts to the same, by the locally Euclidean character of the
underlying space(time) $M$.\footnote{In other words, as repeatedly
noted in the previous sections, ADG has taught us that in order
for one to be able to do differential geometry, one does not have
to exclusively commit oneself to the structure sheaf $\smooth_{M}$
(or equivalently, to classical differential manifolds $M$), but
one can explore other structure sheaves of coefficients that may
possibly be {\em far from smooth}. To stress it once again, the
principal didactic of ADG is that ``{\em differentiability is
independent of $\smooth$-smoothness}'' \cite{malrap2}.} Thus, as
noted earlier, the first successful application of ADG was to do
differential geometry over space(time)s that are dense with
singularities of the most general, and `problematic' or
`anomalous' from the viewpoint of smooth Euclidean or locally
Euclidean spaces on which CDG vitally relies, kind by using,
instead of the $\struc\equiv\smooth_{M}$ of the classical theory
(CSDG), sheaves of Rosinger's spacetime foam algebras of
generalized functions as structure coefficients.

Indeed, one witnesses in a series of papers
\cite{malros1,malros2,malros3} fundamental differential geometric
constructions, normally being associated exclusively with the
presence of $\smooth$-smooth base space(time manifold)s, to apply,
completely unaltered and uninhibited in any way, on space(time)s
teeming with singularities of the classically most unmanageable
sort---the aforedescribed dense singularities encoded in the
(structure sheaves of) spacetime foam algebras of generalized
functions. A long list of results of this application of ADG
includes among other things:

\begin{enumerate}

\item Poincar\'e's lemma and, {\it in extenso}, de Rham's theorem
\cite{mall1,mall2,malros1,malros2,malros3,malrap2},

\item Maxwell's equations: \cite{mall1,mall2,mall4,mall9,mall4},

\item Yang-Mills equations \cite{mall1,mall2,mall3,mall9,malrap3,mall4},

\item Einstein's equations
\cite{mall3,malrap3,mall9,mall7,mall11,mall10,mall4},

\item Geometric Prequantization and field (second) quantization
\cite{mall2,mall5,mall6,malrap2,malrap3,mall4},

\item General remarks and philosophical implications on
(gravitational) singularities
\cite{mall3,mall7,mall11,mall10,mall4}.

\end{enumerate}

\subsection{Section's R\'esum\'e}

The epitome, `bottom line' as it were, of this section is that in
ADG, by absorbing into the structure sheaf $\struc$ of generalized
arithmetics (`differentiable coordinates') singularities of the
most unmanageable (at least from the CDG-vantage) kind, like
Rosinger's spacetime foam dense singularities of non-linear
generalized functions (distributions), one is able to develop the
entire differential geometric conceptual panoply and technical
constructions in their very presence, as if singularities were not
there. In this sense ADG bypasses or evades singularities, and
moreover, unlike the DGSs or the SFSs that we saw in section 3,
which are regarded as pathologies and differential geometric
anomalies (or `breakdown points' for the differential equations
corresponding to the CDG-implemented gravitational field law)
uncircumventable by the manifold-grounded CDG (Analytic or
Differential Calculus-based) means. {\it In summa}, ADG `sees
through' $\struc$;\footnote{That is, the gravitational field law
(\ref{eqy23}) in ADG-gravity is $\struc$-functorial
(Synvariance).} hence also through the singularities that are
built into it.

\section{A Concrete Toy Model and Playground Application of ADG: a Spatial Point-Localized Finitary-Algebraic and a
Temporal Line-Distributional Spacetime Foam Resolution of the
Interior Schwarzschild Singularity}

First of all, we state it up-front that in the context of the
usual CDG-analysis of $\smooth$-manifolds, the three most
`canonical' and familiar examples of `true' DGSs and, perhaps more
appropriately, SFSs (even though distributional solutions are not
normally associated {\it ab initio} with them), are those of the
Schwarzschild, the de Sitter, and the Friedmann solutions to the
(vacuum) Einstein equations \cite{hawk0,clarke4}. Thus, in both
cases the Ricci curvature scalar appears to grow without bound as
smooth causal curves, that material particles are supposed to
follow in the differential spacetime manifold, approach the
corresponding singular {\it loci}. Even more strikingly, (certain
components of) the respective `solution metrics' ({\it ie}, the
Schwarzschild, the de Sitter, and the Friedmann metric) appear not
to be able to be continuously defined on the singular points of
the locally $\R^{4}$ differential spacetime manifold $M$ carrying
them \cite{hawk0,clarke4}.

In this section we choose the interior singularity of the
Schwarzschild solution as a playground physical model to
illustrate its ADG-theoretic `absorption' or `dissolution'. At the
same time, having also in hand the exterior, so-called `virtual'
or `coordinate', Schwarzschild singularity and its successful
evasion by Finkelstein in \cite{df}, we will make it clear how
{\em both} the superable exterior {\em and} the supposedly
insuperable interior singularities\footnote{In point of fact, the
interior singularity is manifestly insuperable by CDG-theoretic
(analytic) means!} are, in a subtle ADG-theoretic sense, `{\em
virtual}' or {\em coordinate} ones, and that {\em in no way they
inhibit the `inherent', essentially algebraic ({\it ie},
sheaf-theoretic), differential geometric mechanism of ADG, which
still applies galore over them}. Thus, we will maintain that {\em
the $\smooth$-smooth singularities of GR in no way indicate a
halting or breakdown of differentiability, contrary to what CDG
has `forced' us to believe so far; while, we will also show how to
`absorb' the inner Schwarzschild singularity into
suitable\footnote{The epithet `suitable' here means `appropriate
to the physical problem at issue' (see below).} structure sheaves
$\struc$ of generalized arithmetics (coordinates) and still
possess at our disposal the vacuum Einstein equations in full
force over the classically offensive} ($r=0$) {\it locus}.

In more detail, below we will revisit, with an ADG-theoretic eye,
the anomalies of the Schwarzsc- hild solution to the vacuum
Einstein equations. To begin with, we will briefly recall from
\cite{df} how Finkelstein showed that the exterior Schwarzschild
singularity is not a `true', `genuine' singularity, but only a
`virtual', coordinate one. Essentially, he showed that when the
Schwarzschild spacetime manifold associated with a stationary
gravitating point-particle of mass $m$ is suitably recoordinatized
({\it ie}, in effect, {\em analytically extended})---from
(analytic) Schwarzschild coordinates, to (analytic)
Finkelstein-Eddington coordinates charting its point events---the
exterior singularity disappears and the $3$-spherical surface of
Schwarzschild radius ($r=2m$)---what is usually referred to as the
{\em Schwarzschild horizon}, appears to behave like a
unidirectional membrane---a temporally semi-permeable shell
allowing only future or past-directed causal signals to cross its
surface. Finkelstein then inferred that this `{\em gravitational
osmosis}' may be regarded as a classical model for distinguishing
(future propagating) particles from their time-reversed (past
moving) antiparticles, a phenomenon which, as he observes, is
essentially due to the non-linear nature of the gravitational
force field. We will then use Finkelstein's rationale to outline
key technical and conceptual features of our ADG-theoretic
approach to `resolving' $\smooth$-smooth spacetime singularities.

Of course, Finkelstein's work also pronounced, even if just
indirectly---as it were, by elimination---the unavoidability and
`unevadability' of the interior Schwarzschild singularity right at
the point-mass $m$ ($r=0$), namely, it made it clear that the
gravitational field becomes unmanageably infinite (in fact, that
it becomes undefinable, that the law that it obeys breaks down,
and that there is no further analytic extension of the smooth
spacetime manifold past it!) right at its point-source. From a
CDG-theoretic or Analytic point of view, the latter {\it locus} is
deemed to be characterized as a `real', `genuine' singularity
\cite{df,clarke4}, and one is tempted to infer (like most
physicists do infer!) that GR, as a physical theory, is out of its
depth when trying to describe the gravitational field and the law
that it obeys right at the point where its mass source is
situated.

Motivated by this apparently insuperable shortcoming of GR, {\em
we will apply the ADG-technology and show how to evade completely
the interior Schwarzschild singularity}, in two different, but
closely related both in their technical aspects and their
underlying philosophy, ways:

\begin{itemize}

\item (a) The internal Schwarzschild singularity will be resolved
by finitistic-algebraic means stemming from our previous work on
extending algebraically and by using the sheaf-theoretic means of
ADG Sorkin's {\em topological} discretizations of continuous
($\cont$) spacetime manifolds \cite{sork0} to the {\em
differential geometric} regime, thus arriving at a finitary
(locally finite), causal and quantal description of Lorentzian
gravity \cite{malrap1,malrap2,malrap3}. This is a natural
continuation and extension of the said trilogy to a tetralogy so
as to show, in a straightforward, and by virtue of a concrete
physical example, way, the sense in which ADG totally evades, in a
finitary-algebraic, sheaf-theoretic fashion the commonly regarded
as being `real', `true' or `genuine' spacetime singularities of GR
\cite{malros1,malros2,mall3,mall9,malros3,mall7}. In a sense, to
be contrasted against (b) below, this is a {\em `static-point
resolution'} of the inner Schwarzschild singularity, regarded as a
`{\em stationary or static point-singularity}' (of the DGS kind).
Physically speaking, its essence is that, since only a (locally)
finite number of `degrees of freedom' (spacetime point-events) are
involved (or `excited' by the gravitational field) , and apart
from the fact that the vacuum gravitational field equations will
be manifestly shown to hold over the interior
singularity,\footnote{That is, they do not actually breakdown in
the (differential) geometric sense of DGSs \cite{clarke4}.} it is
not actually the case, in contradistinction to the smooth
spacetime continuum based GR, that the gravitational field becomes
infinite (let alone undefined!) in the vicinity of (or right at)
the singularity. In fact, by suitable inverse and direct limit
techniques and results of ADG, we will show that in a suitable
`continuum limit' the vacuum Einstein equations still hold over
the interior singular {\it locus}; while, furthermore, we will
also argue that there is nothing to suggest that the continuum
gravitational field strength, represented by the Ricci scalar,
becomes unbounded (infinite) at the singular point. This last
argument will put into perspective the nowadays supposedly
mandatory transition from a continuous (manifold) to `discrete',
`cut-off' spacetime below Planck length in various QG approaches
\cite{sork0,bomb87}---a transition which is apparently
necessitated by a need to render finite, even if `by force of
hand', the gravitational path integral (regularization) or the
entropy of the horizon of a black hole which is supposed to
conceal in its core the singularity under focus \cite{sork1}. In
fact, these arguments will pave the way towards our questioning of
the `physicality' of the Planck length and, ultimately, of the
endeavor to quantize spacetime itself (section 6) in a theory,
such as ADG, which deals solely with the fields themselves and the
laws that they obey (differential equations), independently of a
background (base) spacetime structure, whether the latter is
assumed up-front to be a `classical continuum' (manifold) or a
`quantal discretum' (section 7).

\item (b) The second way in which the interior Schwarzschild
singularity will be evaded is by presenting the latter, not as a
localized, `isolated or solitary' as it were, point-singularity as
in (a) above, but rather as a {\em distributional} one of
SFS-type---a singularity continuously extending along the
point-particle's time-line and viewed from the perspective of the
spacetime foam dense singularities of Mallios and Rosinger
\cite{malros2,malros3}. Stemming from Finkelstein's work
\cite{df}, one realizes that the `effective' (analytic)
Schwarzschild manifold $X$ associated with a gravitating
point-particle is the total spacetime manifold $M$ minus the
singular continuous `wristwatch time-line' $L_{t}:=\{ p\in
M:~x_{i}(p)=0\}$ of the point-mass placed at the origin of an
(analytic) Cartesian coordinate system $(x_{\mu})$ charting $M$:
one writes $X=M-L_{t}$.\footnote{With $x_{0}=t$, $x_{1}=x$,
$x_{2}=y$ and $x_{3}=z$, as usual.} Without a loss of generality,
we then regard $L_{t}$ as a (locally) Euclidean space ({\it ie},
locally homeomorphic to $\R$), which is everywhere dense with
$\smooth$-singularities in the `spacetime foam' sense of
generalized functions (distributions) of
\cite{malros2,malros3},\footnote{For instance, we may think of
Schwarzschild-type of singularities situated on the irrational
points of the `time-line' $L_{t}$ of $m$.} thus apply the ADG
technology as in the latter papers\footnote{See subsection 4.2.}
so as to show that the vacuum Einstein equations still hold over
the entire $L_{t}$; hence in effect, over all $M$, without the
need of ``{\em reducing the manifold of solutions}''
(Q?.?),\footnote{See also next paragraph.} by `forced surgery' as
it were, to the effective $X$. This is a concrete example of the
application of ADG to GR by using a sheaf of Rosinger's algebras
of generalized functions as structure sheaf of generalized
arithmetics (coordinates), an idea originally entertained in the
context of GR and gravitational singularities in
\cite{mall3,mall9,mall7,mall11}.
\end{itemize}

The physical moral of both of these resolutions can be appreciated
in the light of two of Einstein's negative remarks about
singularities encountered in \cite{einst3}: ``{\small\em It does
not seem reasonable to me to introduce into a continuum theory
points (or lines {\it etc.}) for which the field equations do not
hold...}'' (Q2.1) and ``{\small\em ...we cannot judge in what
manner and how strongly the existence of singularities reduces the
manifold of solutions.}'' (Q?.?). Namely, with the two
ADG-theoretic evasions of the inner Schwarzschild singularity
anticipated above, we will vindicate Einstein by showing that
actually the (vacuum) gravitational field equations actually hold
over singular points\footnote{The localized static Schwarzschild
point-singularity in the resolution foreshadowed in (a) above.}
and lines\footnote{The extended `time-line' Schwarzschild
singularity in the resolution anticipated in (b) above.} in the
spacetime continuum $M$,\footnote{Let alone that the law that the
(vacuum) gravitational field obeys breaks down in any
(differential geometric) sense at the singular {\it loci}.}, and
in no way, the occurrence of either ({\it ie}, of singular points
or lines), reduces the manifold of solutions.\footnote{For
example, in case (a), the reduced spacetime, consisting of regular
points on which the vacuum Einstein equations hold, is a smooth
manifold $M$ punctured at the locus of the point mass; while in
case (b), the reduced manifold is the smooth regular space
$X=M-L_{t}$, with the singular wristwatch time-line of the
particle excised. In fact, any forced `surgical' procedure
removing by hand and in an {\it ad hoc} fashion the singular
points, lines or surfaces where the gravitational field equations
do not hold, will simply not (and actually does not!) do from an
ADG-theoretic viewpoint.} Quite on the contrary, we will maintain
in the sequel, it is not that GR, as a physical theory (defined by
local physical-dynamical laws---here, the vacuum Einstein
equations), breaks down, but rather, that the CDG mathematical
framework (and its associated base differential spacetime
manifold) within which we model GR differential
geometrically\footnote{That is, we model the physical law, that
{\em defines} GR as a dynamical theory of the gravitational field,
after a {\em differential} equation proper.} is of limited
applicability and validity {\it vis-\`a-vis} $smooth$-smooth
singularities, since, as noted above, the vacuum Einstein
equations still hold over the classically ({\it ie}, from the
manifold based CDG-viewpoint) offensive {\it loci}.

\subsection{Finkelstein's Resolution of the Exterior Schwarzschild
Singularity, its Physical Interpretation and its `Aftermath'
Viewed Under the Prism of ADG}

As it is well known, the Schwarzschild solution represents the
spherically symmetric gravitational field outside a massive,
spherically symmetric body of mass $m$. On grounds of physical
utility alone, our choice of this particular solution on which to
exercise our ADG-machinery and results may be justified on the
fact that experimentally all the differences between
non-relativistic (Newtonian) gravity and GR have been based on
predictions by this solution. Also, since comparison with
Newtonian gravity allows us to interpret the Schwarzschild
solution as the gravitational field (in empty spacetime) produced
by a point-mass source $m$ viewed from far away ({\it ie}, from
infinity) \cite{hawk0}, Finkelstein's original treatment of the
Schwarzschild gravitational field as being produced by a
point-mass in an otherwise vacuous spacetime manifold \cite{df}
appears to be a good starting choice.\footnote{The following
recollection of results about the Schwarzschild singularities can
be also found in the recent paper \cite{rap5}.}

So first, following Finkelstein in \cite{df}, one assumes that
spacetime is a smooth ($\smooth$) or even analytic ($\anal$)
manifold $X$,\footnote{In this paper we do not distinguish between
a $\smooth$- and a $\anal$-manifold (or for the same reason,
between CDG and Calculus or Analysis). From an ADG-theoretic
viewpoint, as noted earlier, a smooth manifold $X$ corresponds to
choosing $\smooth_{X}$ for structure sheaf, while an analytic one
has $\struc\equiv\anal_{X}$---the structure sheaf of coordinate
functions (of $X$'s points) each admitting analysis (expansion)
into power series. Admittedly, $\anal$- is a slightly stronger
assumption for a manifold than $\smooth$-, but this does not
change, let alone inhibit, the points we wish to make here about
the S-sing and its bypass in the light of ADG.} and then one
places at its `center' (interior) a point-mass $m$. The
`effective' spacetime manifold of this point-particle becomes $X$
minus the particle's `wristwatch' time-line $L_{t}:=\{ p\in
X:~x_{i}(p)=0,~(i=1,2,3,~t\equiv x_{0})$ (expressed in a Cartesian
coordinate system with $m$ situated at its origin); that is,

\begin{equation}\label{eqyy1}
X_{S}=X-L_{t}\footnote{The subscript `$S$' stands for
`($S$)chwarzschild'.}
\end{equation}

\noindent Then, one observes that $m$ is the source of a
gravitational field, represented by a smooth (or analytic)
spacetime metric $g_{\mu\nu}$, satisfying the vacuum Einstein
equations (\ref{eqy23}) which are cast here as follows

\begin{equation}\label{eqyy2}
\smric=0
\end{equation}

\noindent with the pre-superscript `$\infty$' indicating the
$\smooth$-smoothness of the base spacetime manifold $X$ (and
therefore also of the smooth Ricci tensor, which is a function of
the smooth $g_{\mu\nu}$ and its partial derivatives)\footnote{One,
like Finkelstein, could also use the pre-superscript `$\omega$' to
indicate an analytic $X$, metric and Ricci tensor. Of course, from
an ADG-perspective, all this essentially boils down to choosing
$\struc\equiv\smooth_{X}$ for structure sheaf.} on which the
partial differential equations above hold.

The Schwarzschild solution of the said equations is the
Schwarzschild metric $g_{\mu\nu}^{S}$ (expressed in
Cartesian-Schwarzschild coordinates), which in turn defines an
infinitesimal proper time interval, as follows:

\begin{equation}\label{eqyy3}
ds_{S}^{2}=(1-r_{S}^{-1})(dx_{S}^{0})^{2}-(1-r_{S}^{-1})^{-1}dr_{S}^{2}-(dx_{S}^{i}dx_{S}^{i}-dr_{S}^{2})
\end{equation}

\noindent expressed in `natural units' in which the so-called
Schwarzschild radius ($r=2m$) and the speed of light
($c=10^{8}m/s$) are equal to $1$.\footnote{Also, in (\ref{eqyy3})
above, $r_{S}=\sqrt{x_{S}^{i}x_{S}^{i}}$ and
$dr_{S}=r_{s}^{-1}x_{S}^{i}dx_{S}^{i}$. The more familiar ({\it
ie}, not in `natural units') expression for the Schwarzschild line
element in cartesian coordinates is
$(1-\frac{2m}{r})dt^{2}+dx^{2}+dy^{2}+dz^{2}+\frac{2m}{r(r-2m)}(xdx+ydy+zdz)^{2}$,
while in spherical-Schwarzschild coordinates (again not in natural
units), it reads
$-(1-\frac{2m}{r})dt^{2}+(1-\frac{2m}{r})^{-1}dr^{2}+r^{2}(d\theta^{2}+\sin^{2}\theta
d\phi^{2})$.}

Evidently, $g_{\mu\nu}^{S}$ has two singularities: one right at
the {\it locus} of the point-mass---the Cartesian origin ($r=0$),
and one at the Schwarzschild radius ($r=1$) delimiting a spacelike
$3$-dimensional unit-spherical shell in $X$, commonly known as the
Schwarzschild horizon. The two singularities are usually pitched
as the {\em interior} (inner) and {\em exterior} (outer)
Schwarzschild singularities, respectively.\footnote{The said
hypersurface is {\em the horizon of the Schwarzschild black hole},
which is supposed to have the inner singularity in its core, as it
were, `beyond the horizon', which screens it from the view of an
external observer.}

In \cite{df}, Finkelstein initially considered an analytic metric
$g_{\mu\nu}^{EF}$ on $X$, expressed in what is nowadays usually
called Eddington-Finkelstein coordinates (frame),\footnote{The
Eddington-Finkelstein frame consists of so-called {\em
logarithmic-null spherical coordinates} $(n^{\pm},r,\theta
,\phi)$, with the null coordinate $n^{\pm}$ being either {\em
advanced} $n^{+}:=t+r^{'}$ or {\em retarded} $n^{-}:=t-r^{'}$, and
$r^{'}$ defining a logarithmic radial coordinate
$r^{'}:=\int\frac{dr}{1-2mr^{-1}}=r+2m\log(r-2m)$.} defining the
following infinitesimal spacetime line element

\begin{equation}\label{eqyy4}
\begin{array}{c}
ds^{2}_{F}=(1-r_{F}^{-1})(dx_{F}^{0})^{2}+2r_{F}^{-1}dx_{F}^{0}dr_{F}-
(1+r_{F}^{-1})dr_{F}^{2}-(dx_{F}^{i}dx_{F}^{i}-dr_{F}^{2})=\cr
-(1-\frac{2m}{r})(dn^{\pm})^{2}\pm
2dn^{\pm}dr+r^{2}(d\theta^{2}+\sin^{2}\theta d\phi^{2})
\end{array}
\end{equation}

\noindent and he then showed that, for the region of $X$ outside
the Schwarzschild horizon ($r_{F}>1$), the following simple
`logarithmic time coordinate change' from the analytic Finkelstein
${}^{\omega}\struc_{F}={}^{\omega}(x^{\mu}_{F})$ coordinates to
the also analytic Schwarzschild ones
${}^{\omega}\struc_{S}={}^{\omega}(x^{\mu}_{S})$

\begin{equation}\label{eqyy5}
\begin{array}{c}
\underline{{}^{\omega}\struc_{F}\mapto{}^{\omega}\struc_{S}:}\cr
\cr x_{F}^{0}\mapto x_{S}^{0}=x_{F}^{0}+\mathrm{ln}(r_{F}-1)\cr
x_{F}^{i}\mapto x_{S}^{i}=x_{F}^{i}
\end{array}
\end{equation}

\noindent transforms the line element $ds^{2}_{F}$ (and the
associated $g_{\mu\nu}^{F}$) in (\ref{eqyy4}) to the Schwarzschild
one $ds_{S}^{2}$ (and its associated $g_{\mu\nu}^{S}$) in
(\ref{eqyy3}).

Conversely, he argued that since $\smric$ in (\ref{eqyy2}) is a
tensor with respect to the $^{\omega}\struc_{F}$ coordinates, the
vacuum Einstein equations hold in all $X$ (now coordinatized by
${}^{\omega}\struc_{F}$\footnote{Which we may just as well
symbolize by $X_{F}$.})---in particular, they hold on the horizon
unit-shell.

{\it In toto}, Finkelstein showed that the analytic coordinate
change

\begin{equation}\label{eqyy6}
X_{S}\equiv(X,{}^{\omega}\struc_{S})\mapto(X,{}^{\omega}\struc_{F})\equiv
X_{F}
\end{equation}

\noindent amounts to an {\em analytic extension} of $X_{S}$
(coordinatized by the Cartesian ${}^{\omega}\struc_{S}$ and
carrying the analytic $g_{\mu\nu}^{S}$ defining $ds_{S}^{2}$
above---which is singular at $r=1$), to $X_{F}$ (coordinatized by
the analytic ${}^{\omega}\struc_{F}$ and carrying the analytic
$g_{\mu\nu}^{F}$ defining $ds^{2}_{F}$, which is {\em not}
singular at the Schwarzschild radius!).

In fact, Finkelstein showed that the said analytic extension of
$X_{S}$ to $X_{F}$ can be carried out in two distinct
ways,\footnote{Depending of course on whether one chooses advanced
or retarded logarithmic-null coordinates.} each one being the
time-reversed picture of the other, which in turn means that the
$r=1$ horizon, far from being a site of singular {\it loci}, acts
as ``{\em a true unidirectional membrane}'' in the sense that
``{\em causal influences can pass through it only in one
direction}'' and, moreover, he gave a particle-antiparticle
interpretation of this gravitational time-asymmetry
\cite{df}.\footnote{The null (in the Finkelstein frame)
hypersurface at $r=1$ is also known as a {\em closed trapped
surface} \cite{hawk0}, which `traps' past- (resp. future-)
directed causal ({\it ie}, timelike or null) signals depending on
whether one chooses advanced (resp. retarded) Finkelstein
coordinates to chart the original manifold. Let it be also noted
here that it can be easily seen that inside Schwarzschild horizon
the original time and radial coordinates exchange roles.}

On the other hand, about the inner Schwarzschild singularity
Finkelstein concluded that {\em the theory} ({\it ie}, the
CDG-based GR) {\em is out of its depth} as there is no (analytic)
coordinate change that can remove it like the outer
one.\footnote{Indeed, the maximal analytic extension of $X$, which
is unable to include the $r=0$ {\it locus}, is the well-known one
constructed by Martin Kruskal \cite{kruskal,hawk0}.} In other
words, the interior singularity, right at the point-particle $m$,
is regarded as being a `genuine', `true' singularity of the
gravitational field, not removable (or `resolvable') by analytic
({\it ie}, CDG-theoretic) means
\cite{df,hawk0,clarke5,clarke4}.\footnote{Indeed, in the
$n^{+}$-picture, any future-directed causal curve crossing the
horizon can reach $r=0$ in finite affine parameter distance.
Moreover, it can be shown that as $r\mapto 0$ the Ricci scalar
curvature $\ricci$ in (\ref{eqyy2}) blows up as
$\frac{m^{2}}{r^{6}}$, while there is no further analytic
extension (in a $C^{2}$-, or even in a $C^{0}$-, fashion!) of the
Schwarzschild spacetime manifold across the $r=0$ {\it locus}.}

Characteristically, Rindler \cite{rindler} draws a sharp
distinction between `true', `genuine', `non-coordinate'
singularities, and `virtual', coordinate ones---while more
importantly for our ADG-based exposition here, he (implicitly)
regards the former as being {\em physically real}---by using both
the exterior and the interior Schwarzschild singularities as
`canonical' examples of the said distinction:

\bigskip \noindent (Q6.1)\hskip 0.9in
\begin{minipage}{11cm}
\noindent ``{\small ...For many years it was believed that there
was a real singularity at $r=2m$, in the sense that the local
physics would become unusual there...However, in 1933 Lema\^{i}tre
found that the [exterior]\footnote{Our addition.} Schwarzschild
singularity is not a {\em physical}\footnote{Rindler's emphasis.}
singularity at all, but merely a {\em
coordinate}\footnote{Rindler's emphasis.} singularity, i.e., one
entirely due to the choice of the coordinate system...Of course,
in general it will not be so obvious how to remove a coordinate
singularity, or even whether a given singularity is due to
coordinates or not. One way of deciding this last question is to
calculate the invariants of the curvature tensor and testing
whether these remain finite as the singularity is approached: {\em
if they do, the singularity is probably not a physical
one}\footnote{Our emphasis. Thus, implicitly, Rindler regards the
inner Schwarzschild singularity as being a {\em physical}
one---precisely what we want to challenge in the present paper (at
least in the way that Rindler puts it above, namely that: ``{\em
local physics becomes unusual at a real singularity}''---{\it ie},
that the differential geometrically expressed, as a differential
equation, law of gravity `breaks down' or `ceases to hold' at a
true singularity).}...}''
\end{minipage}

\vskip 0.1in

Which, having ADG in mind, brings us to the aftermath of
Finkelstein's resolution of the exterior Schwarzschild
singularity.

\paragraph{The moral of the story from an ADG-theoretic vantage.}
Finkelstein's resolution or evasion of the outer Schwarzschild
singularity, always within the confines of CDG, on the one hand
showed that the said singularity is not `genuine', `true' or
`physical', but merely a coordinate one---an indication that the
manifold (and the smooth gravitational metric field on it) was
expressed (charted) by the physicist in a `wrong' system of
coordinates; while on the other, it showed that the inner
singularity is a `true' or `physical' one, as it resists all
smooth coordinate changes (extensions past it) intended to include
it with the other regular points of the smooth $X$. This then set
the paradigm of how to go about, within the realm of Analysis, and
distinguish between `virtual' (coordinate, unphysical) and
`genuine' (physical) singularities that we discussed in section 2,
namely, the method (and subsequent trend) of analytic extension
and (causal) completion accompanied with the construction of
various singular topological and causal boundaries on which
`genuine' singularities are `asymptotically'
situated.\footnote{With the concomitant general classification of
`true' singularities into DGSs, VESs and SFSs, as we reviewed in
section 2.} {\it In toto}, Finkelstein's analysis of the
Schwarzschild singularities, when suitably abstracted
(generalized), points to a general Analytical (CDG-based) method
of dealing with (smooth) spacetime singularities
\cite{clarke4}---at least it provides one with a general method
for deciding when a singularity is a coordinate or a genuine one,
something that was `frustratedly' demanded by Eddington
\cite{eddington4} as the following quotation from \cite{clarke5}
shows:

\bigskip \noindent (Q6.2)\hskip 0.9in
\begin{minipage}{11cm}
\noindent ``{\small ...Eddington deplored the lack of a method to
distinguish between a real and a coordinate singularity: `{\em It
is impossible to know whether to blame the world-structure or the
coordinate system}'\footnote{Our emphasis of Eddington's
quotation.}...}''
\end{minipage}

\vskip 0.1in

\noindent On the face of all this, the serious conflict between
the PGC of the manifold based GR and the `definition' of (and
distinction between) coordinate and true gravitational
singularities with CDG-methods may be expressed thus: on the one
hand the dynamics---Einstein's equations---treats all coordinate
systems on an equal footing (being in fact indifferent with
respect to what frame the said equations are written down), while
on the other, a metric-solution to those equations possesses
singularities whose character ({\it ie}, whether they are
coordinate or `real' ones) actually depends on what coordinate
system is used.\footnote{More precisely, coordinate singularities
appear to be frame-dependent ({\it ie}, singular in one frame, but
non-singular in another), while genuine ones are in a negative
sense coordinate-independent (always within the CDG-theoretic
framework): no matter what $\smooth$-smooth system of coordinates
is used to chart the spacetime manifold, true singularities remain
singularities and cannot be removed or `gauged away' by any
coordinate (`gauge') change!} In other words,

\bigskip \noindent (R6.1)\hskip 0.9in
\begin{minipage}{11cm}
\noindent {\em how can the gravitational field equations, which
purport to be generally covariant, admit solutions that appear to
be `coordinate-sensitive'?}\footnote{In a metaphorical sense, the
singularities of GR may be perceived as analogues of the anomalies
of QFT. Anomalies are usually thought of as indicating the
`breakdown' (or lifting) of a classical symmetry of (the
Lagrangian defining the dynamics of) a classical theory {\em upon
quantization} \cite{jackiw}; whereas here, singularities may be
thought of as the `breakdown' of the $\mathrm{Diff}(M)$-`symmetry'
of the (classical) theory ({\it ie}, of the $M$-based Einstein
equations defining GR) {\em upon solution} (of those equations,
while still remaining in the classical domain of CDG where no
quantization is invoked). In this sense, singularities (especially
DGSs) may be coined `{\em differential geometric
solution-anomalies}'. A bit later on, in 5.3, we are going to
challenge, having the ADG-theoretic evasion of gravitational
singularities in hand, the traditional viewpoint that the concrete
{\em solutions} (solution-metrics) of the gravitational field
equations are the physically significant entities in GR. Further
later on, in section 6, we will challenge, again based on the
ADG-didactics learned from the evasion of the inner Schwarzschild
singularity, the nowadays well-established intuition (or
expectation!) that a quantization of GR (or of spacetime
structure) will remove singularities.}
\end{minipage}

\vskip 0.1in

From an ADG-vantage however, the Schwarzschild singularities'
story acquires a whole new meaning (interpretation) and teaches us
a very different moral. To begin with, since no background
manifold is involved at all in the theory, the CDG-based
techniques of analytic extensibility (of a manifold), smooth
(causal) geodesic completion (of a manifold) and the associated
construction of various (causal/topological) boundaries (to the
manifold) in order to accommodate true singularities altogether
lose their meaning and import.\footnote{For in ADG we do not do
Calculus `geometrically', in a Cartesian-Newtonian fashion---{\it
ie}, by `picturing' (or representing) differential geometrically
physical situations with the assistance (mediation in our
calculations) of a smooth background space(time) in the guise of
coordinates in $\smooth{M}$---but rather {\em purely
algebraically} (relationally), in a Leibnizian way. (See
concluding section.)} In particular, they are not used to
distinguish between coordinate and genuine singularities, hence
the latter distinction loses its significance in ADG-gravity.
Indeed, from an ADG-theoretic perspective all singularities are
virtual, coordinate ones as they all are inherent in the
particular structure sheaf $\smooth_{M}$ of generalized
arithmetics (coordinates) that uniquely characterizes (defines) a
differential manifold $M$ and {\it in extenso} the CDG that is
based on it. Moreover, Finkelstein's `method' of changing
coordinates in order to show that the exterior Schwarzschild
singularity is a virtual one, followed by the novel {\em physical
interpretation} (time-asymmetry) that he gives to the situation
now `pictured' in the new coordinate system (the Schwarzschild
horizon acts as a unidirectional membrane effecting a `causal
osmosis'), is abstracted by ADG to the following `method' of
dealing with (in fact, evading completely!) arbitrary
singularities of any kind:

\bigskip \noindent (R6.2)\hskip 0.9in
\begin{minipage}{11cm}
\noindent {\em Upon encountering a singularity (or in general, à
`non-smooth' situation), one should look for an `appropriate'
structure sheaf $\struc$ of generalized arithmetics
(coordinates)\footnote{As noted earlier, the epithet `appropriate'
to $\struc$ in the context of ADG means that the said (functional)
structure sheaf provides one with a differential $d$ to the extent
that the functions involved in $\struc$ can be branded
`differentiable', like in the classical case of CDG where
$\struc\equiv\smooth_{M}$.} that incorporates the (function that
misbehaves at) the said singularity (or more generally, a
structure sheaf that somehow represents faithfully the
`non-smooth' situation in hand), yet it is still able to furnish
one with the essentially algebraic differential geometric
mechanism one needs in order to apply differential geometric ideas
to the physical situation or problem that one wishes to tackle.
{\it In summa}, the basic general didagma here is that upon
encountering a differential geometric problem ({\it eg}, a
singularity or a non-smooth situation) one could evade it (or even
include it in one's now `generalized Calculus'!) by judiciously
changing (or enlarging) one's structure algebra sheaf $\struc$ of
`differentiable functions';\footnote{Within the confines of the
differential manifold based CDG, the procedure of analytic
extension does precisely this: {\em one tries to enlarge one's
algebra of differentiable functions so as to include the
`pathological' one having the singularity}.} moreover, in the new
differential geometric situation afforded by the new $\struc$,
one, like Finkelstein, should look for a novel {\em physical
interpretation} for how the physical situation `looks' now under
the prism of the new ``coordinatizations' (generalized
`observations' or `measurements').\footnote{Metaphorically
speaking, $\struc$ is like one's reading-glasses: if something
appears to be problematic ({\it eg}, blurry and unreadable), one
should change one's glasses, but by no means attribute his failure
to read to the text he is trying to read---to emphasize it again:
{\em singularities lie with $\struc$, which is ours; not with
Nature}.}}
\end{minipage}

\vskip 0.1in

\noindent Of course, the CDG-conservatism and monopoly ({\it ie},
the fact that so far the only way we knew how to do differential
geometry is, in one way or another, by involving a base manifold)
virtually `mandated' the distinction between true and coordinate
singularities as it identified the latter with {\it loci}
resisting Analysis---sites where the manifold based CDG ceases to
apply (and hence the law of gravity, modelled after a differential
equations, appears to `break down'---DGSs). In contradistinction,
in view of the discussion and (R5.?) above, {\em from the
ADG-viewpoint all singularities are in a deep sense coordinate
ones, being inherent in (or built into) $\struc$} (which in the
classical case is $\smooth_{M}$---{\it ie}, singularities are
innate anomalies of the background differential spacetime
manifold).

\paragraph{Gravitational `differential geometric
solution-anomalies': no sufficient reason there.} The question we
posed in (R5.?) above and the implicit reply we gave in its
footnote, namely, that singularities (or at least DGSs) may be
interpreted as `differential geometric solution-anomalies', with
our subsequent regarding all $\smooth$-smooth spacetime
singularities as virtual, coordinate ones from the perspective of
ADG-gravity, recalls Finkelstein's fundamental question upon
resolving the exterior, `virtual' (coordinate) Schwarzschild
singularity in \cite{df}:\footnote{In the excerpt below, emphasis
is ours; while in square brackets are our own additions for
clarity and completeness.}

\bigskip \noindent (Q6.3)\hskip 0.9in
\begin{minipage}{11cm}
\noindent ``{\small ...Thus it seems that $T$ [time] invariance
and general relativity are incompatible for a spherical point
particle: although the requirements $(abcde)$ [that Finkelstein
imposes in the beginning of \cite{df} on the gravitational field
and its background spacetime (manifold) symmetries] do not
distinguish between past and future, the only universe which obeys
them does. {\em How is it possible that causes which are symmetric
can have effects that are not? Such a violation of the principle
of sufficient reason must be attributed to the nonlinear nature of
gravitational theory}...\footnote{And then Finkelstein goes on to
anticipate a theoretical situation in which the nonlinear
character of gravity will `choose' one of the two alternative
time-asymmetric Schwarzschild universes, namely, the phenomenon of
spherical gravitational collapse of a spherical body ({\it eg}, a
star) past the critical point when its radius equals the
Schwarzschild radius.}}
\end{minipage}

\vskip 0.1in

\noindent In complete analogy, as we argued above, from the
CDG-viewpoint singularities too appear to violate the principle of
sufficient reason in the following sense: while the Einstein
equations (the `{\em causes}' in our case) are generally covariant
(no preferred coordinate system) and thus
`coordinate-insensitive', their solutions (the `{\em effects}' in
our case)\footnote{And it must be stressed here that from an
ADG-viewpoint this appellation of the gravitational field ({\it
viz.} algebraic $\struc$-connection), defining the field equations
as differential equations proper, as `{\em the cause}', while on
the other hand, of a solution to these equations as `{\em the
effect}', is not just a formal matter. For us, as we shall argue
in detail later on, the {\em algebraic} connection $\conn$
defining the gravitational field law is, abstractly speaking, `the
cause' of the `{\em solution-geometry}' ({\it ie}, `{\em solution
space}') of that law. In turn, the latter is the effect (result)
of that law---it is {\em the realm where the field (law) holds}.
This is another way to say what we mentioned earlier and we will
further discuss in the last section, namely, that `geometry' (in
the generalized sense of `solution space') is the effect of the
`primary cause'---the field ($\conn$) itself defining the
gravitational law (differential geometrically) as a differential
equation in the first place. {\it In toto}, as it has been argued
in 3.2.5 and will be further corroborated in section 7, {\em
dynamics (algebra) comes before kinematics (geometry), and the
former is the `cause' of the latter}.} appear to be
`coordinate-sensitive'. We then coined this
`$\mathrm{Diff}(M)$-breaking' phenomenon associated with the
appearance of singularities of exact solutions of the Einstein
equations `{\em gravitational differential geometric
solution-anomalies}'.\footnote{Arguably, this lies at the heart of
the glaring conflict between the PGC of GR and the existence od
$\smooth$-smooth spacetime singularities mentioned in section 1,
an asymphony that in turn makes a precise definition (by classical
Analytic means) of singularities a difficult, slippery issue
indeed \cite{geroch,clarke4}.} On the other hand, we argued above,
from the vantage of ADG-gravity there is no violation of
sufficient reason simply because in the first place our {\it a
priori} assumption of spacetime as a differential manifold is not
sufficiently reasonable (pun intended).\footnote{Thus, in
contradistinction to Finkelstein in \cite{df}, we did not
attribute the apparent violation of the principle of sufficient
reason to the gravitational field itself ({\it ie}, for example,
to its non-linear nature), but to the smooth background spacetime
manifold $M$ which carries in its coordinate structure sheaf
$\smooth_{M}$ singularities, now perceived as gravitational
`differential geometric solution-anomalies'. That is, in our case,
we `blame it' on the external {base} spacetime (manifold), and by
no means on the field itself (which anyway in ADG-gravity is not
the smooth metric, but the base manifold independent algebraic
$\struc$-connection $\conn$; see below).}

\subsubsection{A Finitary-Algebraic and ADG-based `Resolution',
`Evasion', `Dissolution', or `Engulfment', of the Interior
Schwarzschild Singularity}

In this central part of the present paper we provide arguments
that lead us straightforwardly to a finitary (:locally finite),
algebraic and ADG-based `resolution' (better, `total evasion' and
`dissolution') of the inner singularity of the Schwarzschild
solution of the vacuum Einstein equations for the smooth
gravitational field of a point-particle of mass $m$ situated at
the `center' of the differential spacetime manifold . We will
show, by finitistic-algebraic and ADG-based ({\it ie},
sheaf-theoretic and categorical) arguments,

\begin{itemize}

\item that this singularity is not actually a breakdown {\it
locus} of the differentiability (smoothness) of the gravitational
field \cite{clarke3,clarke4} (Q2.?), which anyway in our ADG
picture is not identified with a smooth metric, but with an
algebraico-categorically defined and manifestly background
manifold independent $\struc$-connection variable $\conn$
\cite{malrap1,malrap2,malrap3},

\item that there is no divergence of, no infinity associated with,
the Ricci tensor as one approaches $m$ \cite{hawk0,clarke4}
(Q2.?), since it is plain that, physically speaking, only a finite
number of events (or `degrees of freedom' of the gravitational
potential field---the aforesaid `finitary spin-Lorentzian
connection' \cite{malrap1,malrap2,malrap3}) are involved or
`excited' (at least locally---{\it ie}, in the `immediate
neighborhood' of the point mass source $m$),

\item that this singularity is `isolated'\footnote{Or better,
`solitary', as the term `isolated singularity' has a different, by
now rather standard, meaning in the usual GR and singularities'
terminology.} and `weak' compared to the classically ({\it ie}, by
CDG means) most numerous, unmanageable, robust, non-smooth and
non-pointed ({\it ie}, `smeared out' or distributional) spacetime
foam dense singularities we encountered in subsection 4.2, with
which ADG copes without a problem in the sense that they are not
at all impediments to its `inherent' algebraico-categorical
differential geometric mechanism \cite{malros2,malros3}, while the
(vacuum) Einstein equations, which are expressed via the
ADG-machinery, hold in full force over them
\cite{mall3,mall9,mall7}, and

\item that we give a `classical continuum limit' construction,
with a concomitant new and generalized, because abstract,
definition of `classical smoothness', which shows that the (new)
differential spacetime continuum vacuum Einstein equation still
holds over its point mass source $m$---which, in turn, in the
usual $\smooth$-smooth manifold case, is supposedly a `real' or
`true' spacetime singularity. Thus for example, there is neither a
need to subject $M$ to a physically {\it ad hoc} `surgery'
\cite{geroch,hawk2} in order to remove the offensive {\it locus}
$m$, nor to push the latter to the boundary of $M$, both of which
in turn point to the fact that the inner singularity, as far as
the `innate differential geometric mechanism' (of ADG) is
concerned, is neither ``{\em a point where the (gravitational)
field equations do not hold}'' (Q2.1) (let alone `break down'
(Q2.?) in any differential geometric sense of the DGSs), nor a
{\it locus} at the edge of spacetime beyond which $M$ cannot be
smoothly (or analytically) continued. Moreover, this bypass of the
inner Schwarzschild singularity will show that ADG effectively
offers us a direct, Gordian knot-cutting answer to Einstein's
quest to ``{\em judge in what manner and how strongly the
existence of singularities reduces the manifold of solutions}''
(Q2.?), namely, {\em not at all!}; while at the same time, it goes
a long way towards offering ``{\em a method to derive
systematically solutions that are free of singularities}'' (Q2.?),
as singularities are not impediments to, let alone breakdown
points of, the physical law of gravity, which holds galore in
their very presence.\footnote{To be precise, ADG does not offer a
method of deriving (new) solutions of the Einstein equations that
are singularity-free, but rather, a (differential geometric)
method of showing that the Einstein equations can be written, as
differential equations proper, in the presence of {\em any}
singularity, {\em as if singularities were not there}. Not only
gravity, but differential geometry in general, has been freed from
singularities!}

\end{itemize}

\noindent The basic theoretical background material for this
`finitary' and sheaf-theoretic resolution can be found in our
trilogy \cite{malrap1,malrap2,malrap3} and it has been briefly
reviewed in the previous sections of this paper. But first, as a
suitable preamble to the promised evasion, we wish to make some
preliminary points about the virtues of `{\em pointlessness}' and
`{\em algebraicity}' (as well as the related notions of `{\em
categoricity}' and `{\em functoriality}'), which figure
prominently in the said trilogy, in what we have said so far
herein, and in what follows.

\subsubsection{The point of pointlessness, algebraicity and of sheaf-theoretic localization (`gauging')
of structures and their concomitant `dynamicalization'}

We commence by first noting that the point-like character of the
events of the spacetime continuum is an ideal mathematical
assumption. In other words, the geometric manifold, as a
point-set, is an operationally unrealistic or non-pragmatic model
for `spacetime' \cite{rapzap1,rapzap2}. As a theoretical
alternative to the `pointed nature' of the manifold, one could
suggest to substitute its points by some kind of `open sets' about
them. Such a move is supposed to be of an operationally more
pragmatic flavor, since operationally realistic determinations or
`measurements' (localizations) of events---``{\em what we actually
do to produce spacetime by our measurements}''
\cite{sork1}---involve `coarse' (because instrumentally limited,
perturbing and indeterminate, as they either carry `experimental
error' or inflict uncontrollable changes onto the observable
fields) acts of `observation'. This operationally sound assumption
lies at the heart of the so-called `finitary' approach to
substituting the spacetime continuum by `discrete' spaces, as
originally pioneered by Sorkin well over a decade ago
\cite{sork0}.

`Operational pragmatism' aside for a moment, since also the
pointed character of the background spacetime manifold is arguably
{\em the} `geometrical reason' for (or origin of) its
singularities and their associated infinities,\footnote{For any
point in the spacetime manifold may be the host of a singularity
for a physically important (smooth) field on it---a {\it locus}
where the smooth field grows unbounded and its differentiability
(`smoothness') appears to break down.} the idea to substitute the
points of the manifold by coarse regions about them appears to be
in line with the general intuition that these potentially singular
points must somehow be `smeared out'---here for instance, be
substituted by `fatter', extended `open sets'.\footnote{In spirit
at least, this is akin to the by now bread-and-butter `blow-up'
idea and associated technique for resolving singularities in
algebraic geometry whereby one smears and, ultimately, resolves
the offensive geometrical point by erecting and spreading a bundle
(or better, sheaf!) of `directions' over it \cite{harts}. But more
about such ADG-based sheaf-theoretic resolutions of such
differential geometric `point-anomalies' shortly.}

On the other hand, from a physical point of view, there is another
glaring theoretical problem associated with Sorkin's so-called
finitary topological substitutes of continuous ($\cont$) point-set
manifolds in \cite{sork0}. To begin with, we mention that the
$T_{0}$-topological posets obtained from covering (a relatively
compact region $X$ of a) $\cont$-manifold $M$ by locally finite
open covers are based on the assumption that ``{\em the points of
$X$ are the carriers of its topology}'' \cite{sork0}. If at the
same time one wishes to carry theoretically this assumption to its
extreme, by suggesting for example that the spacetime topology
itself is an `observable' ({\it ie}, a dynamical and in principle
experimentally `measurable')  entity
\cite{grib,brezap,rapzap1,rapzap2}, one could infer that a theory
for `observable topology' is essentially a theory of spacetime
(point) measurements. However, again from an operational
viewpoint, {\em we never actually measure spacetime itself}---let
alone its ideal point events, but only the physical fields on
({\it ie}, triggering those events in) it
\cite{bohr1,bohr2,sork5}.\footnote{To be clear about this point,
what we usually measure is not the spacetime location of fields
and their particles (quanta)---{\it ie}, our measurements are
almost never `spacetime localizations' proper of fields,
particles, but only physical processes of energy-momentum transfer
between the observable and interacting fields (particles)
\cite{bohr1,bohr2}. One is tempted to say here that the (quantum)
dual (`complementary') picture of the space-time continuum, which
(locally) involves differential (co)tangent vector energy-momentum
quantities, is the experimentally significant one, not the
space-time (`position') picture itself. Loosely put, (dynamical)
changes not (static) locutions are physically important, because
measurable.} In effect, and this is one of the central didagmas of
ADG,

\bigskip \noindent (R6.3)\hskip 0.9in
\begin{minipage}{11cm}
\noindent the fields themselves are the `carriers' of (the
geometry of) `spacetime'---in particular, of its `topology'---and
the dynamical (and `observable'!) character of the former entails
the dynamical variability of the latter \cite{denisov,malrap3}. In
a Leibnizian-Machian sense, physical `space(time)' (geometry)
derives from the dynamical relations (which in turn can be
mathematically represented combinatory-algebraically) between the
`geometrical objects' ({\it ie}, the physical fields) themselves.
\end{minipage}

\vskip 0.1in

\noindent In other words, there is no {\it a priori} `spacetime'
as such, a fixed absolute ether-like entity, an `out there'
pre-existent container or carrier of the dynamical fields. Turning
the tables around, {\em the geometric properties of spacetime
({\it eg}, its topology) are inherent in the (dynamical) fields}.
Of course, the operationally ideal character of spacetime points,
and the more realistic idea of regions about them, can be retained
in this `field picture'---whether one assumes those fields to be
classical or quantum, as we read for instance from \cite{bohr2}:

\bigskip \noindent (Q6.4)\hskip 0.9in
\begin{minipage}{11cm}
\noindent ``{\small\em ...Classical electrodynamics operates with
the idealization of field components defined at every point of
space-time. Although in the quantum theory of fields these
concepts are formally upheld, it is essential to realize that only
averages of such field components over finite space-time regions
have a well defined meaning\footnote{Our emphasis
throughout.}...}''
\end{minipage}

\vskip 0.1in

\noindent with the notion of `average' highlighting the aforesaid
non-sharpness or `coarseness' and the smeared out character of our
field determinations.\footnote{At this, independently of whether
it is classical or quantum, level of field description, this
indeterminateness may be thought of as being either instrumental
(classical `fuzziness') or fundamentally epistemic/systemic
(quantum indeterminacy).}

Taking things one at a time, we note that all our ADG-based work
so far \cite{malrap1,malrap2,malrap3} has concentrated on
de-emphasizing the importance of a background spacetime---whether
it is `discrete' or a `continuum'---while placing emphasis on the
dynamical physical fields themselves, the `geometrical objects'
living on that background `surrogate'\footnote{The epithet
`surrogate' essentially means in our work that the sole role
played by the base (topological) space(time) in our ADG-based
scheme, is as a background (topological) space(time) enabling the
(sheaf-theoretic) soldering or localization (see step $(c)$ below)
of the said `geometrical objects'---the physical dynamical
fields---without actually participating at all in their
dynamics---the differential equations that they obey (or better,
that they {\em define}). In turn, this is a consequence of the
fact that, in ADG, unlike the manifold based CDG,
`differentiability' does not derive from a smooth geometrical base
space(time), but from the algebra inhabited stalks of the sheaves
engaged in the theory.} space(time). The conceptual and
methodological development of this `de-emphasization' of `static'
(pointed) background geometrical space(time) structures with the
concomitant concentration solely on the (algebraically
represented) dynamical fields {\it per se}, may be cast
progressively in the following steps:

\begin{equation}\label{eq8}
x{\stackrel{\mathrm{(a)}}{\mapto}}U{\stackrel{\mathrm{(b)}}{\mapto}}\sstruc(U)
{\stackrel{\mathrm{(c)}}{\mapto}}\struc_{U}
\end{equation}

\noindent which we briefly explain below:

\begin{itemize}

\item $(a)$ This is essentially Sorkin's step going {\em from points to
regions}, with a concomitant substitution of the continuum by a
`discrete', locally finite, directed graph---a $T_{0}$-topological
poset. However, as noted above, while this scheme appears to be
manifestly pointless and results in substituting the continuum by
a reticular, finitistic structure with an operational flavor,
points, in an ontological sense, are still tacitly assumed to be
the carriers of space(time) topology, and the aforesaid `finitary
substitutes' of continuous topology are regarded as `coarse
approximations' of (the $\cont$-topology of the) pointed
space(time) manifold, not as `autonomous' structures anyway.

\item $(b)$ This second step pertains essentially to the
`algebraic spatialization', based on a discrete version of
Gel'fand duality, procedure followed by Raptis and Zapatrin in
\cite{rapzap1,rapzap2}. In a nutshell, as also noted in some
detail in section 3, the topological information encoded in the
finitary $T_{0}$-posets mentioned above was transcribed to an
algebraic setting---the so-called incidence (Rota) algebras
associated with those posets. The basic idea was to substitute the
`spatial' posets (carrying the `discretized' $\cont$-manifold
topology) above by relational, combinatory-algebraic structures
effectively encoding the same (topological) information. This was
the preliminary, albeit important, step of the general idea of
{\em substituting} (`discrete') {\em space(time) by} (finite
dimensional and noncommutative) {\em algebras}, something that
enabled us to attain a finitistic and to a certain extent
`quantal' picture of spacetime (topology) \cite{rapzap1,rapzap2}.
The welcome bonus we got in going from pure `geometry' (`spatial
topology') to algebra in a finitistic setting was that the
corresponding incidence algebras were seen to be discrete {\em
differential} algebras (manifolds)---that is, they proved to
encode information not only about the topological structure proper
of discretized space(time), as their corresponding finitary posets
do, but also about its {\em differential} structure
\cite{rapzap1,rapzap2,zap1}.

\item (c) This last step is about making the combinatory-algebraic
structures of step $(b)$ {\em variable}, by {\em localizing} them
in a (finitary spacetime) sheaf-theoretic way (\cite{rap2}) over
their corresponding finitary topological posets of step $(a)$
\cite{malrap1}. This sheaf-theoretic localization procedure, when
the relevant locally finite posets (and their incidence algebras)
are not interpreted as finitary topological spaces proper, but as
{\em causal sets} or `causets' for short (and their incidence
algebras as {\em quantum causal sets} or `qausets' for short)
\cite{sork1,rap1}, has been physically interpreted as the process
of `{\em gauging qausets}', hence the aforesaid conception of the
relevant algebraic structures as being variable \cite{malrap1}. In
the latter paper, it was shown that this localization or gauging
procedure amounted to endowing the relevant finitary spacetime
sheaves (finsheaves) of qausets with a non-trivial (`discrete',
spin-Lorentzian) connection, which generalizes (in effect,
`curves') the flat Cartan-de Rham-K\"ahler differential (it too a
connection, albeit flat! \cite{mall1,mall2}) carried by the
incidence algebras (qausets) \cite{rapzap1,rapzap2,zap1}
`stalk-wise' in the respective finsheaves. This connection
variable sets the stage for the finitary gravitational dynamics of
qausets, as a finitary, causal and quantal version of the vacuum
Einstein equations of Lorentzian gravity is expressed solely in
terms of it (in fact, of its curvature)
\cite{malrap3}.\footnote{Indeed, on very general grounds, ``{\em
sheafification is localization}'' \cite{mall8}, which in turn
amounts to `gauging' and `dynamicalizing' the relevant (algebraic)
structures thus endow them with a non-trivial connection $\conn$
with respect to which the said dynamics is represented (as a
differential equation; for gravity in particular, the Einstein
equations are expressed via the curvature of the connection). All
this is the epitome of ADG.} Moreover, since the said connection
is categorically (in fact, functorially) defined {\it \`a la} ADG
as a (fin){\em sheaf morphism} \cite{malrap1,malrap2,malrap3}, the
base topological space plays absolutely no role in the said qauset
dynamics. This is in line with the general ADG-didactics that the
base topological space(time)---no matter what its character ({\it
ie}, whether it is assumed to be `discrete' or
`continuous')---does not influence at all the `inherently
algebraico-categorical' differential geometric mechanism in terms
of which the gravitational law ({\it ie}, the differential
equation obeyed by the gravitational connection field and its
curvature) is expressed (in ADG, as an equation between the
relevant sheaf morphisms).\footnote{See \cite{malrap3} for an
analytical discussion of this very important point, as well as
section 3 where we discuss the $\struc$-functoriality of the
gravitational dynamics in ADG. We will return to it in more detail
shortly, in 5.2.2 next.} Physically speaking, in ADG, the base
topological space(time), whether it is assumed to be a continuum
or a discretum, serves solely as a `surrogate' background stage
for the localization or `gauging' of the `geometrical objects'
(the dynamical fields) themselves, without contributing at all in
ADG's inherently algebraico-categorical differential geometric
mechanism in terms of which the said dynamical laws are expressed
as equations between the relevant sheaf morphisms.\footnote{See
footnote 82 earlier.}

\end{itemize}

\subsubsection{The interior Schwarzschild `dissolved' finitistically and algebraically in
an ADG-theoretic manner: a `static' (`spatial') point-resolution}

This is arguably the {\em neuralgic} part of the present paper and
it has been also presented (more laconically and in a slightly
different guise) in a recent paper by the second author
\cite{rap5}. In the ADG-theoretic `resolution' (or perhaps better,
evasion) to be presented below the interior Schwarzschild
singularity is regarded as a `static', `spatial', localized
point-singularity, and it is evaded in two different ways. The
first way, for reasons to be given below, may be coined {\em
direct} and {\em immediate}, while the second {\em indirect} and
{\em step-wise constructive}. So here is a brief outline of the
stages of a `syllogism' leading in a straightforward manner to the
finitistic-algebraic `resolution' of the inner Schwarzschild
singularity. In the course of this `syllogism' or `algorithm', all
the virtues of our ADG-approach to gravity highlighted earlier
throughout the paper, such as pointlessness, background spacetime
independence (`spacetimelessness'), $\struc$-functoriality and
synvariance, algebraicity, categorical bicompleteness of
differential triads {\it etc}, come to play their role in one way
or another.

\begin{itemize}

\item First we let $X$ be an open and bounded region of a spacetime
manifold $M$, from which initially, in the manner of Sorkin
\cite{sork0}, we consider only its topological ({\it ie},
$\cont$-continuous) structure.\footnote{That is, without
committing ourselves {\it ab initio} to its differential ({\it
ie}, $\smooth$-smooth) structure.}

\item We then let a point-particle of mass $m$ be situated at the `center' of
$X$,\footnote{In any case, we assume that $m$ is a point in $X$'s
interior (bulk) without evoking any boundary $\partial X$
construction.} as in \cite{df}.

\item Next, we cover $X$\footnote{Or in the jargon of ADG, `{\em locally
gauge}' $X$ \cite{mall1,mall2,malrap1,malrap2,malrap3}.} by a
locally finite open covering $\gauge_{i}$.\footnote{Indeed, the
$U$s in $\gauge_{i}$ are called `{\em open local gauges}' in ADG
\cite{mall1,mall2}.}

\item Subsequently, as we recalled in the last paragraph ({\bf v}) of 3.1.7,
we first discretize $X$ relative to $\gauge_{i}$ in the manner of
Sorkin, and then pass to the Gel'fand-dual representation of the
resulting finitary posets $P_{i}$ in terms of discrete
differential incidence algebras $\omg_{i}$.

\item Then we consider finsheaves \cite{rap2} of incidence algebras $\Omg_{i}$ in the manner first
studied in \cite{malrap1}.\footnote{Parenthetically, one may wish
to bring forth from \cite{rap1} the causet and qauset
interpretation that the $P_{i}$s and their associated $\omg_{i}$s
may be given, as well as the finsheaves thereof \cite{malrap1}.}

\item We then recall from 3.1.7 the fintriads
$\triad_{i}$ (of qausets) that the said finsheaves define.

\item We would like to digress a bit here and remind the reader of the
two different ways in which we can obtain $\triad_{i}$ from $X$:
the `roundabout', `{\em constructive}' one starting from $P_{i}$
and proceeding step-wise via the $\omg_{i}$s and the finsheaves
$\Omg_{i}$ thereof as described above, and the `{\em direct}' or
`{\em immediate}' one, via Papatriantafillou's categorical
push-out and pull-back (of differential triads)
results,\footnote{Especially `via her push-out result whereby the
classical differential triad supported by a manifold may be pushed
forward to the `moduli space' obtained from that manifold by an
(arbitrary) equivalence relation---in Sorkin's case in particular,
the quotient of $X$ by the $\stackrel{\gauge_{i}}{\sim}$-relation
we saw in (\ref{eqxx1}).} going directly, by what we called the
`Newtonian spark' in 3.1.7, from $X$ (now regarded not just as a
topological, but also as a differential manifold) and the
classical differential triad $\triad_{\infty}$ that it supports,
to the fintriad $\triad_{i}$ on the base $P_{i}$.\footnote{These
two different routes that one can follow in order to arrive at the
fintriad $\triad_{i}$ were coined at the beginning of this section
{\em indirect} (`constructive') and {\em direct} (`immediate'),
respectively.}

\item Having $\triad_{i}$ in hand either by direct or indirect means,
we then bring forth from \cite{malrap3} the result that, on the
said triads the vacuum Einstein equations of a ($f$)initary,
($c$)ausal and ($q$)uantal version of Lorentzian ($v$)acuum
Einstein gravity hold;\footnote{In \cite{malrap3} we abbreviated
this model by the acronym $fcqv$-Lorentzian gravity.} write:

\begin{equation}\label{eqfvg}
\ricci_{i}(\modl_{i})=0
\end{equation}

\item Then, from section 4 we recall that the said
finitary differential triads comprise an inverse/direct system
$\invtriad$ possessing, following Sorkin via Papatriantafillou's
categorical perspective on ADG, the CCDT
$\triad_{\infty}\equiv\striad$ as a projective/inductive limit
(\ref{eq20}).\footnote{Recall, inverse limit for the base $P_{i}$s
and direct limit for their (Gel'fand) dual $\omg_{i}$s inhabiting
the stalks of the $\Omg_{i}$s.}

\item Moreover, a plethora of finitary ADG-theoretic constructions, vital
for the formulation of a finitary version of Lorentzian gravity
regarded as a gauge theory, are based on those $\triad_{i}$s.
These include for example the aforementioned $fcqv$-Einstein
equations, the {\em $fcqv$-Einstein-Hilbert action functional}
$\eh_{i}$ from which these equations derive from variation with
respect to the Lorentzian gravitational $fcq$-connections
$\conn_{i}$, and the {\em $fcq$-moduli spaces}
$\sconn_{i}(\modl_{i})/\aut\modl_{i}$ of those gauge-equivalent
(self-dual) $fcq$-spin-Lorentzian connections---the
gauge-theoretic `configuration spaces' of our $fcqv$-version of
Lorentzian (vacuum) Einstein gravity. Thus, it is fitting at this
point to recall from \cite{malrap3}\footnote{Expression (150)
there.} the ``{\em 11-storeys' tower of $fcqv$-inverse and direct
systems}'' based on the $\triad_{i}$s in $\invtriad$:\footnote{In
the table below, the important notion of `{\em curvature space}'
is mentioned. The reader can refer to \cite{malrap3} (or of course
to the `originals' \cite{mall1,mall2,mall3}) for a more detailed
discussion of this notion, which however we will not be needing in
the sequel. Finally, the expression `{\em genuinely covariant}' at
the top level (7) is what we called earlier `{\em synvariant}' and
we will encounter it again later on.}

\newpage

\centerline{\doublebox{\bf `Standing on the shoulders of triads'}}

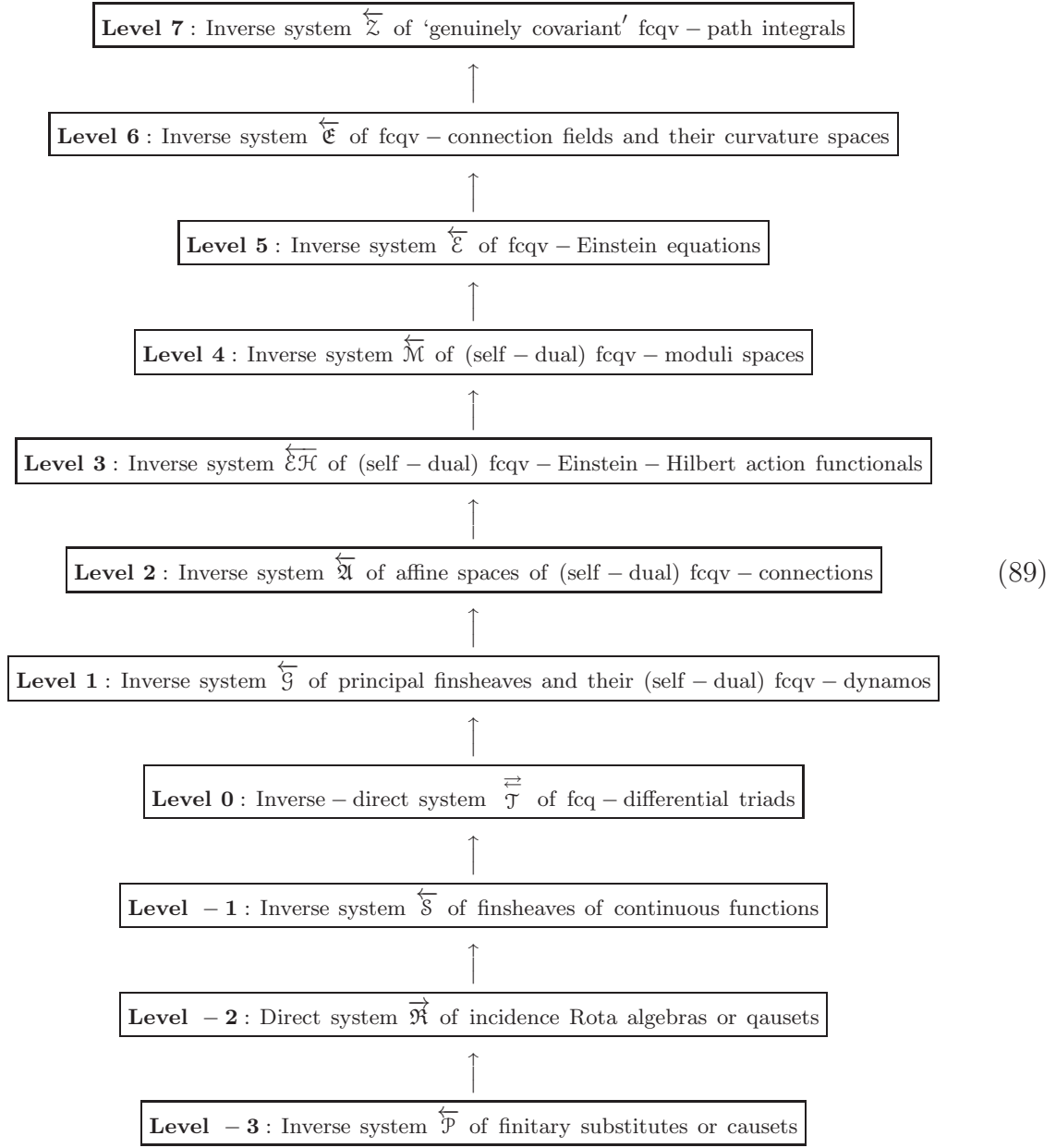
\medskip

\begin{equation}\label{eq27}
{\fontsize{0.13in}{0.13in}
\begin{CD}
\boxed{\mathbf{Level~7:}~\mathrm{Inverse~system~\invcq~of~`genuinely~covariant'~fcqv-path~integrals}}\\
@AAA\\
\boxed{\mathbf{Level~6:}~\mathrm{Inverse~system~\invel~of~fcqv-connection~fields~and~their~curvature~spaces}}\\
@AAA\\
\boxed{\mathbf{Level~5:}~\mathrm{Inverse~system~\inveinst~of~fcqv-Einstein~equations}}\\
@AAA\\
\boxed{\mathbf{Level~4:}~\mathrm{Inverse~system~\invmod~of~(self-dual)~fcqv-moduli~spaces}}\\
@AAA\\
\boxed{\mathbf{Level~3:}~\mathrm{Inverse~system~\inveh~of~(self-dual)~fcqv-Einstein-Hilbert~action~functionals}}\\
@AAA\\
\boxed{\mathbf{Level~2:}~\mathrm{Inverse~system~\invsconn~of~affine~spaces~of~(self-dual)~fcqv-connections}}\\
@AAA\\
\boxed{\mathbf{Level~1:}~\mathrm{Inverse~system~\finv~of~principal~finsheaves~and~their~(self-dual)~fcqv-dynamos}}\\
@AAA\\
\boxed{\mathbf{Level~0:}~\mathrm{Inverse-direct~system~\invtriad~of~fcq-differential~triads}}\\
@AAA\\
\boxed{\mathbf{Level~-1:}~\mathrm{Inverse~system~\invs~of~finsheaves~of~continuous~functions}}\\
@AAA\\
\boxed{\mathbf{Level~-2:}~\mathrm{Direct~system~\diromg~of~incidence~Rota~algebras~or~qausets}}\\
@AAA\\
\boxed{\mathbf{Level~-3:}~\mathrm{Inverse~system~\inv~of~finitary~substitutes~or~causets}}
\end{CD}}
\end{equation}

\noindent Papatriantafillou's results secure that all these
inverse-direct systems yield, like Sorkin's original projective
system $\inv$, their classical `continuum' counterparts at the
limit of infinite resolution of the (base) $P_{i}$s.\footnote{Or
what amounts to the same, at the limit of infinite (topological)
$\gauge_{i}$-refinement \cite{sork0} (or even, at the limit of
infinite sheaf-theoretic localization of qausets---inhabiting the
stalks of the respective finsheaves at the finitary level---over
$X$'s points).}

\item Of special interest to the proposed finitistic-algebraic `resolution' of the
inner Schwarzschild singularity here, is the inverse system
$\inveinst$ at level 5 in (\ref{eq27}) above. {\em The projective
limit of this system recovers the classical continuum vacuum
Einstein equations over the whole ({\it ie}, over all the points
of) $X$} (\ref{eqy23}). In particular, we wish to emphasize that

\begin{quotation}
\noindent {\em the (vacuum) Einstein equations hold over the,
offensive ({\it ie}, genuinely singular) from the CDG-theoretic
vantage, point-mass $m$ in the interior of $X$, and in no
sense\footnote{At least in the {\em differential geometric} sense
of the DGSs in which we are especially interested in the present
paper.} do they appear to break down there}.
\end{quotation}

\noindent In this sense we say that the interior Schwarzschild
singularity has been `resolved' by finitary-algebraic
ADG-theoretic means.

\end{itemize}

Below, we wish to make some further points in order to qualify
more this evasion:

\begin{itemize}

\item First, as noted in section 3.3, since in the ADG-theoretic
perspective on GR it is the algebraic $\struc$-connection $\conn$
and not the smooth metric $g$ (as in the original formulation of
the theory) that is the sole, fundamental (dynamical) variable,
and since moreover ADG is genuinely smooth background manifold
independent, the usual conception of the inner Schwarzschild
singularity as a DGS (with all the classical Analytic baggage that
the latter notion carries, such as analytic inextensibility,
incompleteness, topological boundary constructions to place that
singularity {\it etc}) is not valid in our theoresis, since {\em
neither the metric nor the $C^{k}$-extensibility\footnote{For any
order of differentiability $k=0,1,\ldots\infty ,\omega$.} of the
manifold supporting it are relevant, let alone important, issues
in the theory}.

\item Related to the point above is the fact that in ADG we replace the
usual CDG-based GR conception of a genuinely non-singular
spacetime `{\em the solution metric holds ({\it ie}, it is
non-singular) in the entire manifold $X$}' by the expression that
`{\em the field law ({\it ie}, the differential equation of
Einstein that $\conn$ defines via its curvature $\ricci$) is valid
throughout all the field's carrier (sheaf) space $\modl$ (over the
base topological space(time) $X$), which in turn can possibly host
sings}'. {\it Alias}, there is no breakdown whatsoever of
`differentiability', that is, of the differential equation that
$\conn$ defines, in our scheme. The field $(\conn ,\modl)$, and
the dynamical differential equations that it defines via its
curvature, $\ricci(\conn)(\modl)=0$\footnote{Read: `{\em the
curvature (gravitational field strength) $\ricci$ of the
gravitational connection field $\conn$ on the carrier,
representation (associated) sheaf space $\modl$ (over $X$)
vanishes identically (over the whole base topological space
$X$)}'.} is not impeded at all by any sings that the background
topological space $X$ might possess.

\item One should note that the particular finitary-algebraic inner Schwarzschild
singularity `resolution' presented above is closely akin to (or
one might even say that it `follows suit' from) the topological
resolution of $X$ (into its points) {\it \`a la} Sorkin
\cite{sork0}, in the following sense: as the ur-cell
$\Lambda(m)|_{\gauge_{i}}$ `smearing' the classically offensive
point $m\in X$ becomes `smaller' and
`smaller'\footnote{Equivalently, the topology $\tau_{i}$ generated
by the open sets in the $\gauge_{i}$s becomes finer and finer.}
with topological $\gauge_{i}$-refinement, the law of gravity holds
as close to the point-singularity $m$ as one wishes to get ({\it
ie}, at every level `$i$' of resolution or refinement of $X$ by
the open coverings $\gauge_{i}$); furthermore, at the (projective)
limit of infinite topological resolution (refinement) of $X$ into
its points,\footnote{Which, as noted earlier, in the spirit of
point-set topology are supposed to be the carriers of the
$\cont$-topology of the continuum $X$ \cite{sork0}.} one gets that
(\ref{eqy23}) actually holds on (over) $m$ itself.

\item In connection with the last remarks, it is also worth pointing out that the law
of vacuum Einstein gravity holds {\em both} at the `discrete',
$fcq$-level of the $P_{i}$s ($\forall i$) {\em and} at the
classical level limit corresponding to $X$,\footnote{Indeed, as
noted earlier, this inverse limit procedure may be interpreted as
the classical correspondence limit {\it \`a la} Bohr effecting the
`transition' from the $fcq$-level of the $\omg_{i}(P_{i})$s (and
their finsheaves $\Omg_{i}$), to the classical one of the
continuum $X$ \cite{rapzap1,rapzap2,malrap1,malrap2,malrap3}.}
which further supports our {\it motto} that {\em the ADG-picture
of (vacuum) GR, and the $fcq$-version of it, is genuinely
background independent}---{\it ie}, whether that background is a
continuum or a discretum. {\it In toto}, this emphasizes that our
ADG-perspective on classical or `quantal' gravity is manifestly
(base) spacetime free
\cite{malrap1,malrap2,malrap3,malrap4}.\footnote{As alluded to
towards the end of section 3, the ADG-theoretic formulation of GR
views gravity as a {\em pure gauge field theory} (or equivalently,
as a `{\em synvariant gauge theory of the third kind}')---one with
no allusion to (dependence on) an external continuous spacetime
manifold (continuum) or even a `discrete
$\stackrel{gauge_{i}}{\sim}$-orbifold' (`discontinuum'). The
reader should wait for the next section where numerous important
implications that this `gauge theory of the third kind' might have
for QG research are presented.} Furthermore, concerning the
CDG-problem of the inner Schwarzschild singularity and the usual
divergence of the gravitational field strength ($\ricci$) in its
immediate vicinity, this freedom may be interpreted as follows:
the vacuum Einstein equations hold {\em both} when a (locally)
finite and an uncountable continuous infinity of `degrees of
freedom' of the gravitational field are excited;\footnote{As it
were, when the gravitational field `occupies' and effectuates a
finite and an infinite number of point-events in the background
space(time) $X$.} moreover, unlike the CDG-based picture of inner
singularity, no infinity at all (in the analytical sense of
CDG)\footnote{For example, when $m$ is relegated to $X$'s boundary
$\partial X$ and a suitable topology is given to
$\overline{X}=X\cup\partial X$, as $\ricci\mapto m$, $\ricci$
diverges (as $1/r^{6}$).} for $\ricci$ appears as $m$ is
`approached' (in the categorical limit sense of $\infty\leftarrow
i$) by $\ricci_{i}(\conn_{i})$ upon (topological) refinement (of
$\Lambda(m)|_{\gauge_{i}}$). Plainly, there is no unphysical
infinity associated with this ADG-picture of the interior
Schwarzschild singularity, and in this sense the latter is
genuinely `resolved' (as it were, into locally finite `effects').

\item Of course, all this can be attributed to the fact that the base
topological space(time) $X$ (whether a continuum or a `discretum')
plays no role whatsoever in the inherently algebraic differential
geometric mechanism of ADG, which, as noted earlier, derives from
the algebra inhabited stalks of the (fin)sheaves involved and not
from the base space. Technically speaking, this is reflected by
the fact that the categorical in nature ADG-formulation of the
relevant differential equations (here, the Einstein equations)
involves (equations between) {\em sheaf morphisms}\footnote{And in
particular, $\struc$-morphisms such as $\ricci$.} which by
definition `see through' the generic base topological space $X$,
which in turn serves only as a surrogate scaffolding, without any
physical significance,\footnote{As it plays no role whatsoever in
the gravitational dynamics---the (vacuum) Einstein differential
equations (\ref{eqy23}).} used only for the localization of the
algebraic objects in the relevant (fin)sheaves.

\item Even more important than the remarks about the physical
insignificance of the base space $X$, but closely related to them,
is the issue of the $\struc$-functoriality of dynamics already
alluded to in section 3. Namely, the fact that the vacuum Einstein
equations (\ref{eqy23}) are (local) expressions of the curvature
$\ricci$ of the gravitational connection $\conn$, which curvature
is an $\struc$-morphism (or $\struc$-tensor)---a `geometrical
object' (an $\otimes_{\struc}$-tensor) in ADG jargon
\cite{malrap3}, means that our generalized coordinates (or
`measurements') in the structure sheaf $\struc$ (that {\em we}
assume to coordinatize and geometrically represent the
gravitational field $\conn$, and solder it on $\modl$, which is
anyway locally $\struc^{n}$) respect the gravitational field
(strength); or equivalently, it indicates that the field dynamics
`sees through' our (local) measurements in $\struc(U)$. As all the
singularities are inherent in $\struc$---the structure sheaf of
generalized algebras of `differentiable' coordinate
functions,\footnote{As noted earlier epithet `generalized'
pertains to the fact that in ADG one is free to use for structure
sheaf algebras different from the classical one $\smooth(M)$ of
smooth functions on a differential manifold. However, as it is the
case in the classical case too ($\struc\equiv\smooth_{M}$), it is
the structure sheaf that carries the singularities---{\it ie}, the
singularities are singularities of certain `differentiable'
functions in $\struc$, and they are `geometrically' localized
(situated) on some {\it locus} in the base space $X$
\cite{mall7,mall11}. In turn, ideally, $X$ itself derives from
$\struc$ (Gel'fand duality and spectral theory)
\cite{mall0,mall-1,mall1,mall8}.} it follows that the
$\struc$-functorial field dynamics `sees through' the
singularities built into $\struc$,\footnote{A $\struc$ that, we
emphasize it again, {\em we} assume anyway to coordinatize the
gravitational field $\conn$ and localize it on $\modl$, which by
definition is locally of the form $\struc^{n}$.} or equivalently,
but in a more philological sense, $\struc$ (and the singularities
that it carries) is `transparent' to the $\ricci(\conn)$ engaging
into the gravitational field dynamics---the differential equations
of Einstein (\ref{eqy23}). {\em In summa}, the field $(\modl
,\conn)$ (and the differential equation that it defines via its
curvature) does not stumble on or break down at any singularity
inherent in $\struc$ since it `passes through' them. In this
sense, the term `singularity-resolution' is not a very accurate
name to describe how ADG evades singularities; perhaps a better
term is `dissolution' or `absorption' in $\struc$.

\end{itemize}

\noindent A good example of the aforesaid singularity
`dissolution' in or `absorption' by $\struc$ is the ADG-theoretic
evasion of the inner Schwarzschild singularity regarded as a {\em
time-extended} or `{\em time-smeared}' (`temporally
distributional') {\em spacetime foam dense singularity} in the
sense of Mallios and Rosinger
\cite{malros1,malros2,mall3,malros3}. In the next sub-subsection
we present this distributional `dissolution'.

\subsubsection{The Schwarzschild singularity regarded as an extended spacetime foam dense singularity: a `dynamic'
(`temporal') time-line resolution}

Alternatively to the `static', `spatial', point-resolution of the
inner Schwarzschild singularity above, which was situated right at
the {\it locus} of the point-mass (`particle') source of the
(otherwise externally vacuum) spherically symmetric gravitational
field in $X$, as noted before we can also regard the Schwarzschild
region $X_{S}=X-L_{t}$ in (\ref{eqyy1}) as an `effective' regular
spacetime, where $L_{t}$ is the `wristwatch' time-line ($\R$) of
the particle at which the gravitational field is normally
considered to be singular \cite{df}. Of course, by excluding {\em
by hand} $L_{t}$ from $X$, one unfortunately appears to violate by
fiat Einstein's `disbelief' in (Q2.1) earlier that the
gravitational field ({\it ie}, the equations that it obeys) do not
hold on the line $L_{t}$, and also one appears to go against the
grain of Hawking's negative remarks about singularity-exclusion by
surgical excision or simply by elimination (omission) in (Q?.?).
Thus, in this subsection we will argue, based on ADG-grounds, that
the (vacuum) Einstein equations actually hold in full force over
$L_{t}$ ({\it ie}, over all $X$!) when that Euclidean time-axis is
thought of as being occupied by space{\em time} foam dense
singularities in the sense of Mallios-Rosinger
\cite{malros2,mall3,malros3} and 4.2 above. In fact, as promised
in the previous section, we go a bit further and combine the
methods of Mallios-Rosinger in \cite{malros2,malros3} and Sorkin
in \cite{sork0} effectively by bringing together the notions of
{\em singularity-refinement} and {\em topological-refinement}. All
in all, we shall regard the inner Schwarzschild singularity as a
distributional one in the sense of SFS, and `resolve' it in two
different ways---one `direct', the other `indirect'---similarly to
what we just did for the `static', point-resolution in 5.1.3:

\begin{enumerate}

\item First, we can directly view $L_{t}$ as being inhabited by
spacetime foam dense singularities in the manner of
\cite{malros2,malros3} and then straightforwardly borrow from
\cite{mall3} its main result, namely, that the vacuum Einstein
equations hold over all $L_{t}$ (hence, over all $X$!) when
sheaves of Rosinger's differential algebras of generalized
functions are used as structure sheaves of generalized arithmetics
(thus defining {\em spacetime foam differential triads}) in the
manner of ADG; and,

\item Second, we let the densely singular $L_{t}\simeq\R$ like in 1 above be covered by
suitably defined locally finite `{\em dense singularity open
coverings}',\footnote{Let us note here that it is perhaps more
sensible to consider a bounded region $X$ of the `wristwatch'
(proper) time-axis $L_{t}$ of the particle as befits a physically
realistic assumption about an actual {\em physical} particle
having a finite lifetime. This assumption is also suitable for
applying Sorkin's ideas from \cite{sork0} as he too considered a
bounded region of a $\cont$-spacetime manifold, as we saw earlier
in section 3.} thus combining the Sorkin and the Rosinger schemes.
We then naturally and constructively arrive at {\em spacetime foam
fintriads} by Papatriantafillou's push-out results, and have for
each such dense singularity-cover\footnote{With those coverings
assumed of course to comprise an inverse or projective system as
in \cite{sork0}.} the vacuum Einstein equations holding {\it \`a
la} \cite{mall3,malrap3} on the respective spacetime foam
fintriad. Finally, again via Papatriantafillou's inverse limit
results, we pass to the `continuum' projective limit of maximum
(infinite) {\em topological-{\it cum}-singularity refinement} to
show that the vacuum Einstein equations hold over the whole
(space)time---and in particular, over the densely singular
$X\subset L_{t}$), like in 1 above.\footnote{To be precise, as we
will see below in accordance with the remarks in footnote 150
about Hausdorff reflection, the vacuum Einstein equations hold
over the `large' projective limit non-Hausdorff space ({\it ie},
formally, the poset $P_{\infty}$ corresponding to the maximally
refined open cover $\gauge_{\infty}$ in Sorkin's scheme). Then,
plainly, the vacuum Einstein equations also hold over the densely
singular $X\subset P_{\infty}$, which as noted in footnote 150 can
be recovered by Hausdorff reflection \cite{kopperman,sork0} as a
{\em dense subset} of (closed points of) $P_{\infty}$, with each
point in turn hosting a singularity for some generalized function
in Rosinger's spacetime foam algebra.}

\end{enumerate}.

\paragraph{Direct distributional `resolution' without topological
discretization.} The direct evasion of the inner Schwarzschild
singularity in a distributional way begins by realizing that the
singular (locally) Euclidean `wristwatch time-axis' $L_{t}\simeq
|_{\mathrm{loc.}}\R$ above may as well be thought of as being
occupied by spacetime dense singularities in the sense of
\cite{malros2,malros3} as also exposed in section 4 earlier. This
effectively means that instead of assuming
$\struc\equiv\smooth_{X\equiv L_{t}}$ for structure sheaf in the
theory (as would befit the classical theoresis of the wristwatch
time-line of the point-particle as being a smooth or analytic
manifold \cite{df}), we let the fine and flabby sheaf
$\fa_{L,\ds,X}$ of Rosinger's differential algebras of generalized
functions (non-linear distributions) that we talked about in
section 4 to take its place. We then recall straight from
\cite{malros2,malros3} that the $\mathbf{K}$-algebraized space
$(X,\struc\equiv\fa_{L,\ds,X})$ supports {\it \`a la} ADG a
differential triad---the so-called {\em (s)pace-(t)ime (f)oam
differential triad}

\begin{equation}\label{eqstf1}
\triad_{stf}:=(X, \fa_{L,\ds ,X},\partial)
\end{equation}

\noindent with $\partial$ the usual flat connection we defined
abstractly in (\ref{eqy2}) having now a concrete realization as
the following $\mathbf{K}$-linear sheaf morphism

\begin{equation}\label{eqstf2}
\fa_{L,\ds}(U)\stackrel{\partial}{\mapto}\Omg^{1}(U)
\end{equation}

\noindent for a suitably defined $\fa_{L,\ds}(U)$-module sheaf
$\Omg^{1}(U)$, regarded along the basic ADG-lines as a free
$\fa_{L,\ds}(U)$-module of finite rank $n$ freely generated by the
usual coordinate differentials $dx_{1},\, dx_{2}\ldots dx_{n}$.
Point-wise in $\fa_{L,\ds}(U)$, that is, for one of its local
sections $\mathcal{F}$ (:a generalized function)

\begin{equation}\label{eqstf3}
\mathcal{F}\mapsto \sum_{i=1}^{n}(\partial_{i}\mathcal{F})dx_{i}
\end{equation}

\noindent with $\partial_{i}\equiv\frac{\partial}{\partial
x_{i}}$, as usual.

Then, again borrowing directly the result from
\cite{mall3},\footnote{Expression (5.7) there.} the vacuum
Einstein equations hold on $\triad_{stf}$; write

\begin{equation}\label{eqstf4}
\ricci(\modl_{stf})=0\footnote{The `$stf$' subscript of $\modl$
indicating that the relevant vector sheaf carrying (representing)
the gravitational connection field $\conn$ in this example is
locally a finite power of Rosinger's structure sheaf of spacetime
foam algebras: $\modl_{stf} |_{U}=(\fa_{L,\ds}(U))^{n}~(U\subset
X~\mathrm{open})$.}
\end{equation}

\noindent and, in particular, over the densely singular $L_{t}$.
Thus, once again the inner Schwarzschild singularity cannot be
thought of either as a DGS, or even in this example, as a SFS
\cite{clarke3}---both kinds of singularities meant here in an
abstract, generalized sense, since no base manifold is involved in
ADG,\footnote{That is, there is no issue here of smooth or
analytic extension of a background spacetime manifold as in the
CDG-based `definition' of DGSs and SFSs that we saw in section 2
\cite{clarke3,clarke4}.} while at the same time $\smooth(M)$ is
already included in Rosinger's spacetime foam algebras of
generalized functions (\ref{eqx10}).

\paragraph{Distributional `resolution' via topological
discretization and refinement.} As noted above, an alternative
distributional `resolution' of the interior Schwarzschild
singularity thought of as extending along $L_{t}$ proceeds by
combining Sorkin's topological with Rosinger's
singularity-refinement ideas in a constructive sort of way (making
explicit use also of Papatriantafillou's push-out and inverse
limits of differential triads results).

First, like we did above, we assume that the bounded region $X$ in
$L_{t}$ is densely packed with singularities, having
sheaf-theoretically localized on it Rosinger's differential
algebras of generalized functions ($\struc\equiv\fa_{L,\ds ,X}$).
Then we evoke Sorkin's locally finite open covers of $X$, {\it a
fortiori} assuming that the open sets that comprise them are also
densely singular in Rosinger's sense.\footnote{Refer again to
section 4.2.} This means that the open $U$s in Sorkin's
$\gauge_{i}$s are viewed here as being `{\em dense singularity
covering sets}'. Then we observe that topological refinement {\it
\`a la} Sorkin `induces' singularity refinement {\it \`a la}
Rosinger, which in turn effectively means that we have identified
the inverse systems (nets) $\inv$ (Sorkin)\footnote{This is
Sorkin's `{\em topological refinement net}'.} and $\invros$
(Rosinger)\footnote{This is Rosinger's `{\em singularity
refinement net}'.} that we encountered in sections 3 and 4,
respectively.\footnote{This rests on the fact about spacetime foam
dense singularities mentioned in 4.2, namely, that singularity
refinement $\sref$ (and its associated notion of quotient
regularity) of spacetime foam algebras of generalized functions is
`$\tref$-covariant'; where `$\tref$' is the relation of
topological refinement {\it \`a la} Sorkin in (\ref{eqx28}).}

Now, plainly, for every $\stackrel{\gauge_{i}}{\sim}$-partition
(quotienting) of $X$, by Papatriantafillou's push-out (of
differential triads on moduli spaces of arbitrary equivalence
relations) results, we obtain a finitary version of the spacetime
foam differential $\triad_{stf}$ in (\ref{eqstf1}) above

\begin{equation}\label{eqstf5}
\triad^{stf}_{i}:=(P_{i},
\fa_{L,\ds}(\Lambda(x)|_{\gauge_{i}}),\partial)\footnote{For
typographical clarity we have kept the finitarity index `$i$' as a
subscript as usual \cite{malrap1,malrap2,malrap3,rap5}, and have
promoted the `$stf$' acronym to a superscript.}
\end{equation}

\noindent on which, of course, a finitary version of the vacuum
Einstein equations in (\ref{eqstf4}) holds

\begin{equation}\label{eqstf6}
\ricci(\modl^{stf}_{i})=0
\end{equation}

\noindent Finally, and quite interestingly, the inverse limit of
the projective system of finitary spacetime foam differential
triads (\ref{eqstf5}) and, {\it in extenso}, of the vacuum
Einstein equations that hold on them (\ref{eqstf6}), is expected
to yield (via Ppatriantafillou's inverse limit results) at maximum
topological-{\it cum}-singularity refinement (of the covering
dense singularity open sets) (\ref{eqstf1}) and (\ref{eqstf4}),
respectively. Of course, like in Sorkin's case, this is true
modulo Hausdorff reflection \cite{kopperman,sork0}, which here
however enjoys a sound interpretation: much in the same way that
in Sorkin's case the original topological manifold $X$ is
recovered as a {\em dense} subset of the inverse limit space
obtained from $\inv$ upon infinite topological refinement, so here
the {\em densely singular} $X\subset L_{t}$, together with the
$\triad_{stf}$ (\ref{eqstf1}) and the vacuum Einstein gravity
(\ref{eqstf4}) that it supports, are obtained at the said
projective limit (of infinite singularity refinement). Evidently,
in this distributional `resolution' of the inner Schwarzschild
singularity a plethora of results from Sorkin, Papatriantafillou
and Rosinger were fruitfully combined under the roof of ADG.

\subsubsection{The  matter of the fact: the inclusion of matter
fields and gauge interactions into the vacuum Einstein equations}

\paragraph{Unitary fields with strong quantum undertones.} So far in this paper, and in the past trilogy
\cite{malrap1,malrap2,malrap3}, we have been talking in the light
of ADG about classical and potentially quantum aspects of {\em
vacuum} Einstein gravity---the dynamics of the `pure', `free'
gravitational field in empty space(time).\footnote{In fact, in our
ADG-theoresis here, {\em in the manifest absence of any `ambient',
background spacetime altogether}.} In the following couple of
paragraphs we would like to entertain for a little while the
possible inclusion of matter and, concomitantly, gauge fields, as
well as their dynamical interactions, into our ADG-perspective on
gravity. This inclusion, {\it prima facie} irrespective of
classical and quantum distinctions, seems to be mandated by the
fact that {\em gravity is a universal force} originating from and
coupling (applying) to all mass-energy-momentum manifestations of
matter in the world as well as interacting with the matter's
various gauge radiation fields, let alone of course that because
of its non-linearity, even in {\it vacuo} ({\it ie}, in the
absence of matter), gravity is also self-coupling.

Traditionally, the inclusion of matter (source) fields into
gravity is brought about by placing a non-vanishing stress-energy
tensor $T_{\mu\nu}$ on the right hand side of Einstein's
equations, while the incorporation of gauge (radiation) fields
emitted from the said sources is usually represented by a
so-called minimal coupling prescription---{\it ie}, basically by
augmenting the gravitational Christoffel connection with the
corresponding gauge potentials.\footnote{Accordingly, at the
Lagrangian (action) level, by adding to the Einstein-Hilbert
functional suitable expressions involving the curvatures (gauge
field strengths) of those potentials (plus possible interaction
cross-terms).}

Heuristically and tentatively speaking,\footnote{For a more
definitive and comprehensive ADG-theoretic treatment of gauge and
matter fields (possibly adjoined to gravity), the reader is
advised to wait for a forthcoming book \cite{mall4}.} the
ADG-theoretic `recipe' for including matter and gauge
(inter)actions into gravity is analogous to the standard one, but
at the same time quite different in basic concepts, technique and
theoretical scope, as well as in consequences and implications. To
begin with, in keeping with the basic field-axiomatics of ADG, as
noted earlier fields are ({\em by definition}) pairs $(\modl
,\conn)$, with $\conn$ a sheaf morphism defining the dynamical
(differential) equation of motion of the (free)
field,\footnote{For example, for a (free or `bare') fermion ({\it
eg}, electron or quark) source field, $\conn$ may be identified
with (a suitable ADG-version of) the Dirac differential operator
({\it viz.} connection), while for a bosonic radiation field ({\it
eg}, photon/Maxwell field or a general non-abelian Yang-Mills
field), $\conn$ may be identified with the usual gauge
connections.} while $\modl$ providing a local particle-states'
representation of the field.\footnote{From the ADG-theoretic
perspective on second and geometric prequantization noted earlier
\cite{sel,mall1,mall2,mall5,mall6,malrap2,malrap3,mall4}, (local)
particle (representation) states of {\em bosonic fields} are
identified with (local) sections of (associated) {\em line
sheaves} ({\it ie}, vector sheaves of rank $1$), while the (local)
particle states of {\em fermion fields} are represented by (local)
sections of {\em vector sheaves} of minimum rank $2$.} Then, as in
the case of the `free' (vacuum) gravitational Einstein equations,
the dynamical (differential) equations for free fermionic matter
(Dirac-like equations), or for their corresponding free bosonic
gauge fields (Maxwell's equations, Yang-Mills equations), are
again {\em equations between sheaf morphisms}. In addition, {\it
vis-\`a-vis} the well known problem (of the Minkowski manifold
based classical and quantum field theories of matter) of regarding
those matter-sources ({\it ie}, the particles or quanta of matter)
as singularities in their respective radiation gauge fields, the
$\struc$-functoriality of the said dynamical equations again
enables one, like in the case of vacuum Einstein gravity that we
saw before, to absorb whatever singularity is involved into the
judiciously chosen structure sheaf $\struc$ of generalized
arithmetics without perturbing the law itself ({\it viz.} the
differential equation defined by the corresponding connection
field $\conn$) the slightest bit
\cite{mall1,mall2,mall3,malrap3,mall4}.\footnote{Accordingly, like
in the case of vacuum gravity, the `self-invariances'
(`synvariances') or `auto-symmetries' of those bare field laws are
organized into the principal sheaves $\aut\modl$. Thus, these
field laws too are synvariant (`self-covariant'), without
reference to a background spacetime (manifold) \cite{malrap3}.}

Thus, given the $(\modl_{G},\conn_{G})$ representation of the
vacuum gravitational field satisfying (\ref{eqy23}),\footnote{The
capital-`g' ($`G'$) subscript indicating {\em gravitational} field
($\conn_{G}$) and its local particle-representation (associated)
sheaf ($\modl_{G}$).} the basic idea is to `adjoin' or `couple' to
it another ($g$)auge or ($m$)atter field
$(\modl_{g,m},\conn_{g,m})$---as it were, to combine the two
generally different kinds of fields into a joint, `integral',
`{\em unitary field}'\footnote{The word `unitary' here is meant in
the sense of Einstein (see below).}
$\conn^{'}=\conn_{\modl_{G}\otimes_{\struc}\modl_{g,m}}$ acting on
the {\em tensor product vector sheaf} $\Hom_{\struc}(\modl_{G}
,\modl_{g,m})=\modl_{G}\otimes_{\struc}\modl_{g,m}$, which
$\conn^{'}$ is in a sense `induced' by the individual (free or
just self-interacting)\footnote{Like gravity, non-abelian gauge
(Yang-Mills) fields are non-linear and hence self-coupling.}
$\struc$-connections $\conn_{G}$ and $\conn_{g,m}$ on $\modl_{G}$
and $\modl_{g,m}$, respectively.

We would like to slightly digress here and make a couple of
noteworthy observations about this gravity-{\it cum}-gauge/matter
$\otimes$-`entanglement':

\begin{itemize}

\item First, regarding the base topological spaces $X$ involved: it is tacitly assumed that the base topological
localization-spaces (say, $X_{G}$ and $X_{g,m}$) for the
individual (free) fields {\em combine by identification}---{\it
ie}, they are identified to one and the same base localization
space $X_{G}\equiv X_{g,m}=X$ on which the combined tensor product
vector sheaf $\modl_{G}\otimes_{\struc}\modl_{g,m}$ is then taken
to be soldered. That is, we assume that the two factors in the
$\otimes$-combined system `gravitational-{\em cum}-gauge/matter
field' have a common external localization `parameter space' ({\it
ie}, their tensor-entanglement $\conn^{'}$ `sees' a single, common
base topological space $X$).

\item Second, regarding the coordinate structure sheaves $\struc$ engaged in the said combination:
in principle, one can allow for the employment of different
structure sheaves of generalized arithmetics to coordinatize the
gravitational and the gauge/matter fields individually---{\it ie},
$\modl_{G}:\stackrel{\mathrm{loc.}}{\simeq}\struc^{n}_{G}$,
$\modl_{g,m}:\stackrel{\mathrm{loc.}}{\simeq}\struc^{l}_{g,m}$,
possibly with $\struc_{G}\not=\struc_{g,m}$.\footnote{This is in
keeping with the pragmatic or `ophelimistic' aspect of ADG noted
earlier in the context of the PARD, namely, that one can choose
freely how to `measure' (and localize or `geometrically
represent') the gravitational and the matter/gauge fields,
possibly by using different structure sheaves $\struc$ of
generalized arithmetics (coordinates) for each one. With respect
to the ADG singularity-evasion in particular, one may choose for
instance different structure sheaves of arithmetics---ones that
accommodate different kinds of sings---to cast the Einstein
equations for $\conn_{G}$ and (the Yang-Mills equations for)
$\conn_{g}$ (Yang-Mills fields). Such a flexibility is prominently
absent in the manifold based CDG treatment of both gravity and
Yang-Mills theories, whereby a common structure sheaf
$\struc\equiv\smooth_{X}$ is inevitably employed
(CDG-conservatism) so that all sings are branded as being
$\smooth$-smooth in one way or another (and one is implicitly
working within the category $\man$ of differential manifolds).}

\item And third, regarding the connection fields $\conn$
themselves: as noted above, the individual connections combine to
one---the induced `{\em $\otimes_{\struc}$-tensor product
connection}'
$\conn^{'}=\conn_{\modl_{G}\otimes_{\struc}\modl_{g,m}}$ on
$\modl_{G}\otimes_{\struc}\modl_{g,m}$. We furthermore assume
that, locally ({\it ie}, with respect to a local gauge $U\subset
X$), $\conn^{'}$ resolves into an analogue of the minimal coupling
expression of the usual theory

\begin{equation}\label{eqzzz1}
\conn^{'}|_{U\subset
X}=\partial+\aconn_{G}(\equiv\Gamma_{ij}^{k})+\aconn_{g}
\end{equation}

\noindent with $\partial$---the flat, `inertial' derivation
(connection)---common to all three kinds of evolution:
gravitational, gauge and matter.\footnote{This identification of
the $\partial$ is similar to the common `external parameter
localization/gauging topological space $X$' noted above.}

\item As a `bonus' remark, we note that the three observations above recall the way we actually
combine individual (particle) systems' states in conventional
mechanics (quantum or non-quantum), whereby, while we combine
(normally in quantum theory, Hilbert space) states by tensor
multiplication ({\it ie}, in quantum theory for example,
$\psi_{1}\in{\mathcal{H}}_{1},~\psi_{2}\in{\mathcal{H}}_{2}\mapto\psi_{1}\otimes\psi_{2}\in
{\mathcal{H}}_{1}\otimes{\mathcal{H}}_{2}$), {\em we retain
different spatial labels} (position coordinates), but at the same
time {\em we identify their (external) temporal evolution
parameters} (time coordinates).\footnote{Pun intended.} {\it In
summa},

\begin{equation}\label{eqzzz2}
\psi_{1}(x_{1},t_{1})\in{\mathcal{H}}_{1},~\psi_{2}(x_{2},t_{2})\in{\mathcal{H}}_{2}
\mapto\psi_{1}\otimes\psi_{2}(x_{1},x_{2},t)\in{\mathcal{H}}_{1}\otimes{\mathcal{H}}_{2}
\end{equation}

\noindent while, moreover, in relativistic quantum mechanics
(QFT), when we combine quantum field systems (living in individual
tensor product Hilbert=Fock spaces), we even merge the external
{\em space}time coordinates ({\it ie}, we identify even their
spatial coordinates), and suppose that they live in a joint tensor
product Hilbert (=Fock)\footnote{Strictly speaking, a Fock space
is a direct sum of tensor-multiplied Hilbert spaces.} space
(localized and fibered, as a bundle,\footnote{Or perhaps even
better, as a {\em sheaf}(!)---the associated, representation Fock
sheaf of nets (or better, sheaves) of algebras of quantum field
operators \cite{haag}.} over Minkowski space $\mathcal{M}$)

\begin{equation}\label{eqzzz3}
\phi_{1}(x_{1},t_{1}),~\phi_{2}(x_{2},t_{2}) \mapto
[\phi_{1}\otimes\phi_{2}](\mathbf{x})\in{\mathcal{F}}|_{\mathcal{M}}=
[{\mathcal{F}}_{1}\otimes{\mathcal{F}}_{2}]|_{\mathcal{M}},~(\mathbf{x}~\mathrm{a~point~in}~\mathcal{M})
\end{equation}

\noindent Finkelstein has described this `identification
phenomenon' in detail in \cite{df1}$^{*}$\footnote{See footnote
following the citation \cite{df1} for an explanation of the
asterisk.} when he talks about a similar identification of the
imaginary unit $(i^{2}=-1$ upon combination of systems in physical
theory, quantum or not:\footnote{By the way, the discussion in the
following quotation arises when Finkelstein remarks that real
($\R$) quantum mechanics ({\it ie}, quantum mechanics with real
coefficients/amplitudes) should be regarded as being more
fundamental than the complex ($\com$) one we use today. In what
follows, emphasis is ours. The first paragraph is really
parenthetical and is included only for continuity of the text. The
second and third paragraphs---the emphasized ones---are important
for our discussion here.}
\end{itemize}

\bigskip \noindent (Q6.5)\hskip 0.9in
\begin{minipage}{11cm}
\noindent ``{\small ...We expect that in nature the $\R$ theory
underlies the $\com$. The operator $i$ is a central or
superselection operator from the $\R$ point of view, and we are
familiar with the emergence of many such superselection operators
when a microscopic quantum theory condenses into a non-quantum
macroscopic one, as a result of random phases. Since $i$
transforms into $-i$ under time reversal $T$, we expect that the
superselection operator $i$ arises when an underlying quantum
spacetime structure condenses into the macroscopic non-quantum $t$
variable of elementary quantum mechanics and field theory.

{\em Imaginary units do not combine like symmetries under tensor
multiplication, however. If we look at two input spaces $V_{1}$
and $V_{2}$, in which some group symmetry $g$ is represented by
two operators $g_{1}$ and $g_{2}$, then in the tensor product
$V_{1}\otimes V_{2}$ the same symmetry $g$ acts as $g_{1}\otimes
g_{2}$. We say that under tensor multiplication, symmetries
multiply. It follows that under tensor multiplication,
infinitesimal symmetries add, unit factors $\mathbf{1}$
understood. However, if $i$ is represented by $i_{n}$ in $V_{n}$
($n=1,2$), then to to multiply a product vector
$\psi_{1}\otimes\psi_{2}$ by $i$, it suffices to multiply either
$\psi_{1}$ by $i_{1}$ or $\psi_{2}$ by $i_{2}$. Indeed, we have
the identification $i=i_{1}=i_{2}$. Under tensor multiplication,
the imaginary units $i$ of separate systems combine not by
multiplication or addition, but by identification.

They are not unique in this respect. When we combine systems in
non-quantum or quantum mechanics, the times $t$ also combine by
identification, and in field theories the spacetime coordinates
do. In general, when we combine systems we tensor multiply
symmetry transformations but identify their parameters, which in a
sense are dual. We understand this uniformly by regarding the
entries being combined not as independent factors in a tensor
product, but as subsystems of one embracing system, with unique
operators for time $t$ and imaginary $i$, among other group
parameters...}}''

\end{minipage}

\vskip 0.1in

\noindent And it must be further emphasized in view of the
quotation above that:

\begin{enumerate}

\item In ADG too, the dynamical `auto-symmetries' of the separate fields combine by tensor
multiplication in the joint, `unitary' gravity-{\it
cum}-matter-{\it cum}-gauge field\footnote{Here, the epithet `{\em
unitary}' pertains to what Einstein also referred to as the `{\em
total}' field (see quotations below). To avoid any confusion with
nomenclature used in previous work, in \cite{malrap3} the
adjective `unitary' for a (connection) field meant what we here
call `third gauged', namely, `autonomous', `external (base)
spacetime (continuum) independent'. Below, the adjunction of the
adverb `genuinely' to `unitary' is meant to capture precisely the
said `autonomous', `self-sustaining' quality of the fields
(combining into a unitary field) in ADG.}

\begin{equation}\label{eqzzz4}
\aut\modl_{G},\aut\modl_{g,m}\mapto\aut_{\modl_{G}\otimes_{\struc}\modl_{g,m}}=
\aut\modl_{G}\otimes_{\struc}\aut\modl_{g,m}\footnote{The same
$\struc$ for both gravity and matter/gauge fields is assumed
here.}
\end{equation}

\noindent while locally, in the same way that the fields'
connections `decouple', as a sum, into the the `minimal coupling'
expression (\ref{eqzzz1}), their `self-transmutations'
(auto-symmetries) too add ({\it eg}, they split into a direct sum
of local `auto-symmetries'). We write formally:

\begin{equation}\label{eq33}
\begin{array}{c}
\aut\modl_{G}|_{U}\equiv
{\mathcal{G}}{\mathcal{L}}(n,\struc)|_{U}:=gl(n,\struc);~\aut\modl_{g,m}|_{U}\equiv
{\mathcal{G}}{\mathcal{L}}(l,\struc)|_{U}:=gl(l,\struc)\mapto\cr
[{\mathcal{G}}{\mathcal{L}}(n,\struc)\otimes
{\mathcal{G}}{\mathcal{L}}(l,\struc)]_{U}=gl(n,\struc)\oplus
gl(l,\struc)
\end{array}
\end{equation}

\item Perhaps more important however is to observe, again in view of Finkelstein's
remarks above, that the constant sheaf $\cons\equiv\mathbf{K}$ of
complex ($\mathbb{K}=\com$), or even real ($\mathbb{K}=\R$),
scalars---which in ADG is by definition embedded into the (common,
in the example above) structure sheaf $\struc$---like the common
(identified) background parameter localization topological space
$X$ above, {\em does not partake into the field dynamics}---{\it
ie}, {\em it does not participate in the dynamical evolution of
the fields which is effectuated by the connections which are
constant $\mathbf{K}$-morphisms}. In contradistinction to
Finkelstein however, we maintain that {\em exactly because the
field dynamics `sees through' both the constant sheaf $\mathbf{K}$
and the background $X$ (which in turn both remain fixed), they are
`unobservable'}, for as we said throughout the present paper, {\em
the only `observables'\footnote{That is, `{\em $\struc$-measurable
dynamical quantities}'.} in ADG are the fields (connections)
themselves}.\footnote{To be more precise, the {\em curvatures} of
the fields, which, unlike the connections, are $\struc$-morphisms
(equivalently, $\otimes_{\struc}$-tensors---{\it viz.}
`geometrical objects' proper) \cite{malrap3}.} This of course
frees us from the `responsibility' to look for a quantum
description of spacetime structure in order to account for the
identifications of $X$ and $\mathbf{K}$ when we tensor-combine
gravitational, gauge and matter fields in ADG.\footnote{As we will
argue extensively in section 6, the quantization (or the quantum
structure) of (the background) spacetime itself should not be an
issue at all in a potential application of ADG-theoretic ideas to
QG.} For as stressed before, in ADG, unlike the usual spacetime
(continuum or discretum) based physical theories (particle or
field; relativistic or not; quantum or not), there is no external
(to the fields$\equiv$connections) variable spacetime (geometry)
as such, and the dynamical symmetries of the fields are their own
auto-transmutations in $\aut\modl$,\footnote{To say it again,
effectively these are the transformations of the fields' particles
(`local aspects of the fields'), whose states are represented by
local sections of $\modl$, which in turn is regarded as the
associated (representation) sheaf of the principal (group) sheaf
$\aut(\modl)$.} where no external (background) spacetime
`parameters' are actively involved.

\end{enumerate}

\paragraph{Matter singularities are incorporated into the `genuinely unitary
field' law.} After this small detour discussion about the
possibility of tensor-combining matter and gauge fields with
gravity (accompanied by some quantum undertones) into a `unitary'
ADG-field, we come to the `real' reason why, {\it vis-\`a-vis}
singularities, we wished in the first place to bring up the issue
of incorporating matter field actions into the Einstein field
equations by ADG-means. It has to do with our contention that ADG
may help us complete Einstein's unitary (or unified) field theory
programme and moreover possibly marry it with the apparently (for
Einstein at least) incompatible (with his spacetime continuum
based field theory of gravity) quantum theory.\footnote{This
contention of ours has been partly elaborated in \cite{malrap3},
and also speculatively, tentatively outlined in \cite{rap5}. It
will be developed in more detail in the last section.}

To begin with, as we contended throughout the past trilogy
\cite{malrap1,malrap2,malrap3},\footnote{Especially in the last
paper \cite{malrap3}.} and as we argue repeatedly in the course of
the present work, ADG offers the unique opportunity to develop a
genuinely unitary field theory, as Einstein had originally
envisioned that a genuine completion of his field theory program
should be based on the following three
accomplishments:\footnote{We will return to this issue, with
extensive relevant quotations of Einstein, in 7.5.3.}

To begin with, as we contended throughout the past trilogy
\cite{malrap1,malrap2,malrap3},\footnote{Especially in the last
paper \cite{malrap3}.} and as it was mentioned in various places
above, ADG offers us a unique opportunity to develop a
singularity-free, genuinely unitary field theory---one that is not
based at all on an external, background spacetime continuum with
its inherent singularities; moreover, one that, due to ADG's
purely algebraic character, accommodates quantum ideas from the
very start.\footnote{In this respect, the reader has to wait for
our remarks in section 6 about the `{\em inherently quantum}'
(self-quantum), `{\em third quantized}' character of the
(abstract) ADG-fields $(\modl ,\conn)$ (gravitational, or
Yang-Mills, or other).} Basically, Einstein had originally
envisioned that a genuine completion of his (albeit, spacetime
continuum based) field theory program should hinge on the
following three accomplishments:

\begin{enumerate}

\item First, to develop a unitary field dynamics with sole dynamical variable the
`{\em total}' (unitary) field itself, which satisfies certain
(partial) {\em differential} equations (:physical laws
differential geometrically represented).

\item Second, {\em the envisioned field dynamics (laws) to be free from singularities}.

\item And third, the material particles (`quanta') of the field to be `explained away' field-theoretically as `{\em
singularities in the total field}', while at the same time, {\em
their dynamical evolution} (in `time') {\em to be represented by
the differential equations for the unitary field itself and
nothing else, nothing more-nothing less}.

\vskip 0.1in

{\it In summa}, {\em a genuinely unified field dynamics should be
one of the unitary field `in-itself' that is singularity-free in
the sense that the particle-singularities of the field(s) should
be incorporated (or `absorbed') in the total field-law itself}.

\end{enumerate}

Below we give two Einstein quotations from \cite{einst9} which
corroborate the triptych above: the first expresses clearly his
anticipation that a field-theoretic completion of GR to a unitary
field theory should result in a singularity-free description even
of material point-particles, which act as sources of the various
radiation force-fields, but from GR's viewpoint they are genuine
or `true' singularities of the gravitational field.\footnote{Like
the inner Schwarzschild singularity located right at the
point-particle that we saw before.} In other words, for Einstein
one of the primary motivations for formulating a unitary field
theory is overcoming the problem of singularities troubling
primarily GR---arguably, the spacetime continuum based
relativistic field theory {\it par excellence}:

\bigskip \noindent (Q6.6)\hskip 0.9in
\begin{minipage}{11cm}
\noindent ``{\small ...The essence of this truly involved
situation {\small [{\it ie}, uniting gravity with the other forces
of matter]}\footnote{Our addition for completeness.} can be
visualized as follows: {\em A single material point at rest will
be represented by a gravitational field that is everywhere finite
and regular, except where the material point is located: there the
field has a singularity}\footnote{Our emphasis. For example, this
is the case with the inner Schwarzschild singularity at the fixed
point-mass source that we resolved earlier.}...Now it would of
course be possible to object: If singularities are permitted at
the locations of the material points, what justification is there
for forbidding the occurrence of singularities elsewhere? This
objection would be justified if the equations of gravitation were
to be considered as equations of the total field. [Since this is
not the case], however, one will have to say that the field of a
material particle will differ from a pure gravitational field the
closer one comes to the location of the particle. {\em If one had
the {\rm [unitary]}\footnote{Our addition for clarity.} field
equations for the total field, one would be compelled to demand
that the particles themselves could be represented as solutions of
the complete field equations that are free of irregularities
everywhere. Only then would the general theory of relativity be a
complete theory.}\footnote{Again, our emphasis.}...}''
\end{minipage}

\vskip 0.1in

\noindent The second quotation from \cite{einst9} expresses
Einstein's scepticism about quantum mechanics---especially, about
the apparent `pseudo-way' in which quantum theory purports to do
away with continuous structures when in fact {\em it still employs
the spacetime continuum in order to formulate the dynamics of
quantum wave amplitudes as differential equations proper}.
Moreover, in an indirect way, the words below put forward
Einstein's vision of a unitary field theory that may on the one
hand overcome the problem of singularities (in GR) and on the
other account the atomistic (quantum) structure of reality:

\bigskip \noindent (Q6.7)\hskip 0.9in
\begin{minipage}{11cm}
\noindent ``{\small ...It is my opinion that the contemporary
quantum theory represents an optimal formulation of the
relationships, given certain fixed basic concepts, which by and
large have been taken from classical mechanics. I believe,
however, that this theory offers no useful point of departure for
future development. This is the point at which my expectation
deviates most widely from that of contemporary physicists. {\em
They are convinced that it is impossible to account for the
essential aspects of quantum phenomena (apparently discontinuous
and temporally not determined changes of the state of a system,
simultaneously corpuscular and undulatory qualities of the
elementary carriers of energy) by means of a [field]\footnote{Our
addition for clarity.} theory that describes the real state of
things [objects] by continuous functions of space for which
differential equations are valid. They are also of the opinion
that in this way one cannot understand the atomic structure of
matter and radiation. They rather expect that systems of
differential equations, which might be considered for such a
theory, in any case would have no solutions that would be regular
(free from singularities) everywhere in four-dimensional
space}\footnote{Our emphasis}...}''
\end{minipage}

\vskip 0.1in

\noindent From the quotation above, and by `negation/exclusion' or
`elimination', one could say that Einstein, in contradistinction
to his contemporary quantum physicists:\footnote{And, it is fair
to say, in contrast also to the majority of current theoretical
physicists (quantum theorists, of course, included). Let it be
also noted here that for most (if not all) contemporary working
physicists, Einstein's unified field theory is regarded as being,
for all practical (research) purposes, a `closed and dead'
subject, or at best, a dated one of only historical value (except
perhaps, but only in a formal and `peripheral' sense, for
higher-dimensional scenarios such as Kaluza-Klein and string
theories, that regard themselves as `natural' continuations, or as
outcomes from `theory evolution', of Einstein's unified field
theory ideas \cite{witten}). However, in view of the basic
didactics of ADG, we feel compelled to look closer to such a
possible `unitary field theory revival', especially {\it
vis-\`a-vis} the current `hot' QG issues and problematics that we
will encounter in the next section. We feel that we are not just
anachronistically `digging up graves', while, anyway, in this one
we are in good company
\cite{stachel6,stachel7,stachel11,stachel9,stachel8}. Indeed, in
the last section we will argue strongly for a possible `revival'
(or even `fulfilment'!) of Einstein's unitary field theory in the
light of ADG.}

\vskip 0.1in

{\bf I.} Believed that a singularity-free field theory on the
spacetime continuum---whose laws are expressed differential
geometrically, {\it ie}, as differential equations---could still
be developed. This essentially implied his `unitary field theory'
vision,\footnote{See above.} albeit, one that still abides by the
background spacetime continuum (manifold) in which differential
equations can actually be formulated
(CDG-theoretically).\footnote{And in his commitment to the base
spacetime continuum {\it vis-\`a-vis} field theory Einstein was
quite adamant, in the sense that he did not believe that a field
theory (and the differential geometric methods effectuating it)
could stand on its own two feet without commitment to such a
background, as we will see in the last section. However, since, as
we have argued throughout the present work, singularities are
inherent in the manifold, Einstein's singularity-free unitary
field theory vision---one that is vitally based on CDG for its
differential geometric implementation---appears to us to be a
rather Quixotic attempt.}

\vskip 0.1in

{\bf II.} As also mentioned before, he also believed that such a
theory could account for the quantum structure of reality, in the
sense that the quantum particles of the source or radiation fields
will be described by everywhere (in the spacetime continuum)
singularity-free (regular) solutions to the total (unitary) field
equations.

\vskip 0.1in

{\bf III.} Finally, he criticized the fact that, in truth, quantum
theory, in spite of the apparent discontinuity (discreteness) of
quantum processes, still tacitly employs the continuum (as it
were, `in disguise') in the form of `continuous changes of' ({\it
ie}, again differential equations obeyed by) the probability
amplitudes for (states of) quantum systems, with those states
(wave functions) explicitly being defined on a spacetime continuum
({\it eg}, Schr\"odinger's non-relativistic or Dirac's
relativistic, or even the Klein-Gordon, wave equations).

\vskip 0.1in

\noindent Even more tantalizing are the following words taken from
three remarkable consecutive paragraphs in
\cite{einst10}\footnote{Which can be found on pages 92 and 93 in
article 13, titled `{\itshape Physics and Reality}' (reprinted
from the {\itshape Journal of the Franklin Institute}, {\bf 221},
313 (1936); see also \cite{malrap1}). The entire third paragraph
is written in {\em emphatic} script because of its relevance to
the present work.} which show, in order of appearance, an
`oscillation' or `undecidability' in Einstein's thought about
whether to opt for the classical geometrical spacetime continuum
and the continuous field theory based on it, or for an algebraic
and `discontinous' description of reality that quantum mechanics
appears to mandate,\footnote{To the knowledge of the present
authors, perhaps the best reference in which the sceptical,
ambivalent, almost `schizophrenic' attitude of Einstein on the one
hand towards continuous field theory and its differential
geometric constructions on the geometrical spacetime continuum,
and on the other, towards the finitistic-algebraic quantum theory,
is \cite{stachel5} (we would like to thank John Stachel for timely
sending us this paper). In the concluding section (section 7), we
argue in detail how ADG may serve as an appropriate (mathematical)
framework in which to bridge the gap between these two apparently
opposite (and irreconcilable to Einstein's mind!) modes of
description of physical reality---that is to say, how, with the
help of ADG, one can do field theory entirely algebraically ({\it
ie}, with finitistic-quantum methods and traits built into the
formalism from the very beginning), without at all the use of a
background spacetime continuum.} with the third paragraph, which
we emphasize due to its relevance here, showing clearly his
`wishful thinking' about a field theory---a total or unitary
gravity-{\it cum}-matter field theory---that could represent
particles (quanta) by singularity-free fields:

\bigskip \noindent (Q6.8)\hskip 0.9in
\begin{minipage}{11cm}
\noindent ``{\small ...{\em To be sure, it has been pointed out
that the introduction of a space-time continuum may be considered
as contrary to nature in view of the molecular structure of
everything which happens on a small scale. It is maintained that
perhaps the success of the Heisenberg method points to a purely
algebraic method of description of nature, that is to the
elimination of continuous functions from physics. Then, however,
we must also give up, by principle, the space-time continuum. It
is not unimaginable that human ingenuity will some day find
methods which make it possible to proceed along such a path. At
the present time, however, such a program looks like an attempt to
breathe in empty space}.\footnote{Our emphasis.}

There is no doubt that quantum mechanics has seized hold of a
beautiful element of truth, and that it will be a test stone for
any future theoretical basis, in that it must be deducible as a
limiting case from that basis, just as electrostatics is deducible
from the Maxwell equations of the electromagnetic field or as
thermodynamics is deducible from classical mechanics. However, I
do not believe that quantum mechanics will be the starting point
in the search for this basis, just as, vice versa, one could not
go from thermodynamics (resp. statistical mechanics) to the
foundations of mechanics.

{\small\em In view of this situation, it seems to be entirely
justifiable seriously to consider the question as to whether the
basis of field physics cannot by any means be put into harmony
with the facts of the quantum theory. Is this not the only basis
which, consistently with today's possibility of mathematical
expression, can be adapted to the requirements of the general
theory of relativity? The belief, prevailing among the physicists
of today, that such an attempt would be hopeless, may have its
root in the unjustifiable idea that such a theory should lead, as
a first approximation, to the equations of classical mechanics for
the motion of corpuscles, or at least to total differential
equations. As a matter of fact up to now we have never succeeded
in representing corpuscles theoretically by fields free of
singularities, and we can, a priori, say nothing about the
behavior of such entities. One thing, however, is certain: if a
field theory results in a representation of corpuscles free of
singularities, then the behavior of these corpuscles with time is
determined solely by the differential equations of the total
field.}\footnote{Again, emphasis is ours.}}''
\end{minipage}

\vskip 0.1in

\noindent The concluding apparently paradoxical `{\em hypothetical
certainty}' of Einstein in the last three lines of the quotation
above may be rephrased as follows: {\em should one be able some
day to represent field-theoretically particles (`quanta') in a
singularity-free manner, then the particle dynamics---as it were,
the evolution of those quanta `in time'---will be already inherent
in (or theoretically speaking, be the result of) the field
dynamics itself, and there would be no need to assume {\it a
priori} particles as fundamental theoretical entities on a par
with the field concept}.\footnote{As we will show and argue
explicitly in the next section, ADG puts particles ({\it viz.}
local sections of $\modl$) side-by-side the algebraic field
$\conn$ when it actually {\em defines} a field to be the pair
$(\modl ,\conn)$. However, it must be stressed here that, in
truth, the field $\conn$ is {\em the} fundamental notion in ADG,
with $\struc$, together with its inherent singularities,
introduced by us `observers' (`measurers' or `coordinators') in
order to localize, `geometrize', `coordinatize' the field, thus
extract its local, quantum-particle aspects
\cite{malrap2,malrap3,malrap4}. In other words, as we will see in
the next section when we expose the quantum aspects of
ADG-gravity, $\struc$ (and in effect $\modl$, which is locally
$\struc^{n}$) is introduced by us in order to particle-represent
the field ($\conn$), but this representation of the fields is
manifestly singularity-free, as Einstein in (Q?.?) above would
have liked it.} It is precisely in this sense that field
theory---Einstein's unitary field theory---aspired to `explain
away' particles and, accordingly, that ``{\em quantum theory could
be deducible from that future (unitary field) theoretical
basis}''. Indeed, some years earlier, in \cite{einst10}, Einstein,
upon concluding the section titled `{\itshape The Field Concept}'
in article 13, `{\itshape Physics and Reality}', reexpresses this
`wishful certainty' in a rather categorematic
fashion:\footnote{Again, the whole excerpt below is written in
{\em emphatic} script due to its significance in the present
work.}

\bigskip \noindent (Q6.9)\hskip 0.9in
\begin{minipage}{11cm}
\noindent ``{\small\em ...What appears certain to me, however, is
that, in the foundations of any consistent field theory, there
shall not be, in addition to the concept of field, any concept
concerning particles. The whole theory must be based solely on
partial differential equations and their singularity-free
solutions.}''
\end{minipage}

\vskip 0.1in

\noindent Moreover, in contradistinction to the commonly held idea
nowadays that Einstein envisioned a (unified) field-theoretic
scenario in which the particle-quanta of matter will be `explained
away' as singularities in the (total/unitary) field,\footnote{See
Eddington quotation below.} he in fact envisioned an even more
`pure', iconoclastic and particle/singularity-free field theory.
An excerpt from Clarke {\it et al.} \cite{clarke5} quotes
characteristically Einstein and Rosen from \cite{einst16} and
further adds (also about the Schwarzschild and the other
cosmological singularities in GR):

\bigskip \noindent (Q6.10)\hskip 0.9in
\begin{minipage}{11cm}
\noindent ``{\small ...Einstein was fundamentally unhappy with any
sort of discontinuity in general relativity. In a paper with Rosen
in 1935, Einstein posed the question: `{\em Is an atomistic theory
of matter and electricity conceivable which, while excluding
singularities in the field, makes use of no other field variables
than those of the gravitational field and those of the
electromagnetic field?}' He noted that: `{\em ...writers have
occasionally noted the possibility that material particles might
be considered as singularities of the field.\footnote{Again, see
Eddington quotation (Q6.11) next.} This point of view we cannot
accept at all. For a singularity rings so much arbitrariness into
the theory that it actually nullifies its own laws. Every field
theory, in our opinion, must therefore adhere to the fundamental
principle that singularities of the field are to be
excluded}'\footnote{This certainly foreshadows Einstein's later
maintaining (right at the end of his life!) that ``...{\em It is
my opinion that singularities must be excluded. It does not seem
reasonable to me to introduce into a continuum theory points (or
lines {\it etc.}) for which the field equations do not hold...}''
(Q2.1). It also indicates his basic conviction that the
`discontinuous' concept of a particle cannot be `married' to that
of the continuous field on the spacetime continuum---for him an
unbridgeable conceptual/theoretical divide on which we are going
to comment extensively under the prism of ADG-gravity and
ADG-field theory in general in subsection 8.5 in the sequel.}
\cite{einst16}. By `singularities of the field' Einstein meant not
only cosmological singularities but also the Schwarzschild
`singularity'...}''
\end{minipage}

\vskip 0.1in

\noindent We would like to leave aside for the time being (until
we pick the discussion up again in the next two sections {\it
vis-\`a-vis} current QG issues) the detailed arguments supporting
our view that ADG meets all the three requirements set by Einstein
above for the development of a genuinely unitary field theory,
together with quantum-particle traits and a manifest
non-commitment to a background spacetime (whether a continuum or a
discretum) built into it from the very start. For now we wish to
just quote Eddington from \cite{eddington2} about incorporating
`material particle singularities' in the (gravitational) field law
itself and then remark on how {\em ADG meets exactly his
words}:\footnote{Initially we thought that Eddington's words below
were the first to explicitly pitch the idea that material
particles could be viewed as `{\em singularities in the field}',
but recently the first author discovered that Michael Faraday had
actually pretty much the same idea significantly earlier, as we
quote Hermann Weyl from \cite{weyl3}: ``...not the field should
derive its meaning through its association with matter, but,
conversely, {\em particles of matter are ... singularities of the
field}'' (our emphasis). This quotation first appeared in the
first author's recent paper \cite{mall10}.}

\bigskip \noindent (Q6.11)\hskip 0.9in
\begin{minipage}{11cm}
\noindent ``{\small ...It is startling to find that the whole of
dynamics of material systems is contained in the law of
gravitation; at first gravitation seems scarcely relevant in much
of our dynamics. But there is a natural explanation. {\em\small A
particle of matter is a singularity in the gravitational
field},\footnote{Our emphasis.} and its mass is the pole-strength
of the singularity; consequently {\small\em the laws of motion of
the singularities must be contained in the
field-equations},\footnote{Again, our emphasis.} just as those of
electromagnetic singularities (electrons) are contained in the
electromagnetic field-equations...}''
\end{minipage}

\vskip 0.1in

\noindent Eddington's words above---especially those
emphasized---are tailor-cut for what ADG has accomplished here and
in the past {\it vis-\`a-vis} gravitational (as well as
Maxwellian/abelian and Yang-Mills/non-abelian gauge) singularities
\cite{mall1,mall2,mall3,malros1,malros2,malros3,malrap3,mall9,mall7,mall11,mall10,mall4}.
Indeed,

\bigskip \noindent (R6.4)\hskip 0.9in
\begin{minipage}{11cm}
\noindent ADG shows that one can actually `absorb' or `engulf'
singularities of any kind in the (judiciously and suitably chosen)
structure algebra sheaf $\struc$ of generalized
arithmetics,\footnote{`Judiciously and suitably chosen' pertains
here to two things as it has already been argued in the present
paper: first, the function that contains the singularity to belong
to the sheaf $\struc$ of generalized coordinate functions that one
chooses to employ in the theory, and second, that this $\struc$
employed is able to provide one with the basic differential
geometric substrate---the `domain of differentiability' on which
one can actually define $\partial$ (and {\it in extenso} $\conn$)
in the manner of (\ref{eqy2}) and (\ref{eqy4}); furthermore,
possibly without a base manifold ($\struc\not=\smooth_{M}$).}
while the field (generalized differential=connection) and the law
(differential equation) that it defines, still applies ({\it ie},
holds) in their very presence. Moreover, as we shall see in the
next section, from a quantum (particle) vantage, the (local)
particle states of the field are represented by the (local)
sections of the sheaf $\modl$ associated with ({\it ie}, carrying)
the field $\conn$.\footnote{Again, technically speaking, and from
a Kleinian perspective, $\modl$ is the sheaf associated with the
self-transmutations (geometrical auto-symmetries) of the field,
which comprise the principal sheaf $\aut\modl$. {\em $\modl$ is
the carrier space of $\conn$} \cite{malrap3}.} Thus, in view of
the fact that $\modl$ is by definition locally isomorphic to
$\struc^{n}$, the `particle-singularities' are built into the
`unitary field' $(\modl ,\conn)$ of ADG. In a very strong and
ostensive sense, {\em in ADG particles of matter (as well as gauge
forces) are singularities in their corresponding fields, with
these singularities in no way inhibiting the application ({\it
ie}, the holding) of the physical field-laws that those fields
define as differential equations proper}.\footnote{That is, in
contradistinction to Eddington's remarks and to the way
gravitational DGSs are normally perceived from the viewpoint of
the smooth manifold based CDG, singularities are in no sense
obstacles (let alone break-down points) of the fields, no matter
what their `pole strength' (for DGSs for example, no matter what
their `order of differentiability').}
\end{minipage}

\subsection{What's in a solution?}

It is now widely appreciated that, in view of the PGC, in GR
coordinates---the $\smooth$-smooth ones of the underlying pointed
manifold $M$---are just `{\em labels}', or `{\em names}' of $M$'s
points, without any physical significance since they do not
actively participate into the actual gravitational
dynamics.\footnote{See section 2. To be sure, underlying the
`labelness' of points by (smooth) coordinates is the
`monoamphidromous' or `tautosemous' relation (isomorphism) of
Gel'fand duality (spectral theory) identifying $M$'s points with
their smooth coordinates in $\smooth(M)$ (\ref{eq1}).} This is
supposed to be the central meaning of the
$\mathrm{Diff}(M)$-represented PGC, or equivalently, that the
physically `observable' (dynamical) quantities are
$\otimes_{\smooth(M)}$-{\em tensors}---what we called here `{\em
geometrical objects}', `$\struc\smooth_{M}$-{\em functorial}'
entities expressed via the homological tensor product functor
$\otimes_{\struc\equiv\smooth_{M}}$---like for example, in the
original formulation of GR, the sole dynamical variable is taken
to be the $\smooth$-smooth metric $g_{\mu\nu}$.\footnote{Thus, in
view of the last footnote, even GR, in an implicit way, is
fundamentally `{\em pointless}'. In the next section, in view of
Stachel's unfolding of the deeper significance of Einstein's hole
argument (and in effect, of the PGC), we will argue that this
`pointlessness' of GR is equivalent to the {\em dynamical}---as
opposed to merely {\em kinematical}---{\em individuation of
events}; albeit, an individuation not effectuated via the (smooth)
spacetime $\struc\equiv\smooth_{M}$-metric $g$ as in the usual
theory, but via the dynamical field $\conn$, which need not {\it a
priori} be {\em smooth} ({\it ie}, $\struc\not=\smooth_{M}$).}
Yet, when a new, particular solution $g_{\mu\nu}$ of the Einstein
equations is discovered, or even when an old well known one is
invoked for further study or application, it apparently becomes
physically important to investigate or make use of its
regularity-singularity behaviour on its domain of definition---the
differential spacetime manifold $M$. That is to say, in what
region of $M$ the said solution `holds'---the realm where the
gravitational field law appears to be valid.

On the other hand, given the `fine-tuning-sensitivity' of the
non-linear Einstein equations on boundary (`initial') conditions,
namely, that given a plethora of various such `external'
prescriptions\footnote{We stress it again that we tacitly assume
that boundary/initial conditions, like the various gauge choices
in gauge field theory, invariably lie with the `exosystem' (or
`episystem')---{\it ie}, with the experimenter (or at least with
the theoretician!)---and never with the physical `endosystem'
under experimental/theoretical scrutiny (here, the gravitational
field and the law that it obeys) \cite{df1}.} any point of $M$ can
be the {\it locus} of a singularity for a smooth solution (metric)
of the Einstein equations (Q2.?). Moreover, granted that the said
singularity will be of a $\smooth$-smooth function belonging to
the structure sheaf $\struc\equiv\smooth_{M}$\footnote{For after
all, to stress it once again, $g_{\mu\nu}$ is supposed to be
$\smooth(M)$-valued when referred to a (local) $\smooth$-chart.}
(that anyway was {\it a priori} assumed by the theoritician!), the
said occurrence of singularities would appear to indicate that
{\em we should actually undermine the physical significance of
solutions to the field equations}. In other words, it is the
conspiring of the differential manifold assumption to model {\em
physical} spacetime, coupled to the particular boundary conditions
for the field partaking in the law of gravity---both of which are
`contingent' features of the theory, externally chosen and
prescribed by the theorist)\footnote{That is, they have nothing to
do with the `physically objective' field of gravity (and the
differential law that it obeys), but with our subjective
theoretico-experimental tampering with (conditioning of and
measuring) it.}---that lead to limiting in one way or another, in
the form of singularities, the (entire)\footnote{That is, over all
the base space(time) $X\equiv M$.} physical validity of the law of
gravity. As a result, such limitations appear to be physically
unreasonable (Q2.1), unbelievable and, ultimately, unacceptable
(Q2.?).

{\it In summa}, on the one hand the PGC appears to be asking
(rhetorically):

\vskip 0.1in

\centerline{physically speaking, `{\em what's in a name?}' (PGC);}

\vskip 0.1in

\noindent and then the general practice and consensus in GR
appears to be asserting

\vskip 0.1in

\centerline{\em the physical significance---or anyway, of the
physical existence} (Q2.?)---{\em of $\smooth$-singularities,}

\vskip 0.1in

\noindent an oxymoron which reflects the deeply disturbing
conceptual and technical conflict between the
$\mathrm{Diff}(M)$-effectuated PGC of GR and the existence of
$\smooth$-singularities in GR mentioned in the first section,
which in turn results in the (apparently insuperable) difficulty
of defining precisely what is a singularity in the classical
spacetime manifold based relativistic field theory of gravity (GR)
\cite{geroch}---a difficulty, of course, that is pronounced within
the Analytic confines of the differential manifold based CDG
(Analysis) \cite{clarke4}.\footnote{That is to say, the difficulty
simply betrays the shortcomings and ineptness of the concepts and
methods of CDG in dealing with smooth singularities---of course,
quite an understandable inability considering the `{\em vicious
circularity}' of the problem at the bottom, namely, that one
cannot actually do (C)DG without a base $M$ (manifold monopoly and
conservatism), while at the same time singularities are inherent
in $\smooth_{M}$; when, as matter of fact, $M$ is nothing but
$\smooth_{M}$ (Gel'fand isomorphism).}

In the light of ADG however, in view of the basic didactics
emerging from the complete evasion of the interior Schwarzschild
singularity presented above---or better, from the incorporation of
that singularity into a suitably chosen structure sheaf of
generalized arithmetics (coordinates) while still retaining the
essentially algebraic differential geometric mechanism and the
physical law of gravity expressed as a differential equation based
on that mechanism in the very presence of that singularity---we
contend that:

$\bullet$ On the one hand, {\em the appearance of singularities
does not at all reduce the manifold of solutions}\footnote{That
is, the realm in which the gravitational field law holds.}
(Q2.?)---or to quote the first author from \cite{mall9}:

\bigskip \noindent (Q6.12)\hskip 0.9in
\begin{minipage}{11cm}
\noindent ``{\small the exclusion of singularities does not
actually reduce the manifold of solutions, since now the same
`singularities' can be incorporated into the solutions...}''

\end{minipage}

\vskip 0.1in

$\bullet$ While on the other hand, since we have witnessed the
total ADG-theoretic evasion of the base spacetime manifold and the
singularities inherent in it ({\it ie}, in
$\struc\equiv\smooth_{M}$), we come to ask

\vskip 0.1in

\centerline{\em what's in a solution?}

\vskip 0.1in

\noindent especially since the gravitational law ({\it ie}, the
field $\conn$ defining it) holds in all `space(time)'---the
`space(time)' which is anyway inherent in the field, in point of
fact, in the $\struc$ that {\em we} choose to represent it, which
not only does not actually partake into the law itself
(synvariance), but it may also carry singularities of the `worst'
kind.\footnote{Like for example when we choose for structure sheaf
Rosinger's algebras of generalized functions hosting spacetime
foam dense singularities of all pathological sorts.}

{\it In toto}, we contend that finding a particular solution to
the Einstein equations (and subsequently studying its
regularity-singularity properties for physics' sake) loses its
primary significance in ADG, as we quote the following concluding
paragraph from \cite{mall3}:

\bigskip \noindent (Q6.13)\hskip 0.9in
\begin{minipage}{11cm}
\noindent ``{\small ...On the other hand, it is in effect the
manifold that creates the `anomalies', or else `singularities', of
the equations involved, which presumably describe the physical
laws [{\it viz.} fields], the latter being inherent in the world
we perceive.\footnote{This is the PFR in ADG.} So, in agreement
with the `pointless' point of view, we now concentrate our
attention on the equation itself, the `manifold' being just the
space (or even `{\em solution space}') on which the equation has
meaning [or holds]. This aspect enables us to consider {\em
equations presenting `singularities'}, in the ordinary sense of
the latter term, insofar as the functions by means of which the
equations are expressed (or what amounts to the same, our
`structure algebra sheaf' $\struc$), provide us with the abstract
machinery by which we can still follow the classical procedure
within the present abstract framework. As a result, {\em in spite
of the presence of `singularities', the meaning and the
consequences and/or predictions of the equations at issue are all
in force, as if no singularities were present at
all!}\footnote{Our emphasis.} What amounts to the same, to
disregard the singularities just means that both the {\em vector
sheaf} (:bundle) and the {\em field} (:gauge potential) extend
across the `singularity', a fact that classically requires extra,
and not always easy(!), technical innovations.\footnote{For
example, as we saw in section 2, analytic extension, geodesic
completion, Sobolev smearings, topological boundary constructions
for the placement of singularities {\it etc}.} So our task would
be to give now the appropriate explanation to the predictions that
occasionally emerge from the equations considered. {\em Yet, for
this job, to know a particular solution of the equation under
consideration is, of course, not always that
important!}\footnote{Our emphasis.}...}''

\end{minipage}

\vskip 0.1in

\noindent These last remarks gain even more weight and
significance when one recalls that in ADG the basic---in fact, the
sole---gravitational variable is the connection $\conn$ and not
the metric $g_{\mu\nu}$. Thus apart from the fact that since in
ADG there is no background (spacetime) manifold whatsoever its
$\anal$- or $\smooth$-extensibility are `{\em non-issues}' in
ADG-gravity, at the same time the $\anal$- or $\smooth$-extension
of a solution-metric across its singularities also becomes a `{\em
non-issue}'. Which brings us to what we think is a subtle point
about singularities---especially when DGSs are concerned.

\paragraph{The basic theoretical fallacy: `judging and blaming the cause by its
effects'.} We noted earlier when we were discussing Finkelstein's
`resolution' of the exterior singularity of the Schwarzschild
solution-metric the apparent violation of the principle of
sufficient reason ({\it ie}, `time-symmetric laws $\Rightarrow$
time-asymmetric solutions'). Due to it, we there branded
singularities as `{\em differential geometric solution-metric
anomalies}'. There is more to this denomination than initially
meets the eye. The idea here is that by choosing a particular
coordinate system (or `gauge')---the Eddington-Finkelstein
frame---Finkelstein resolves the exterior singularity, but at the
cost of lifting (`breaking') the $\mathrm{Diff}(M)$-`symmetry' of
the dynamical equations. That is, the solution appears to mandate
the choice and the use (by the theorist!) of a special coordinate
system in order to manifest the non-pathological character of the
outer singularity, and as a result of this choice the PGC appears
to be sacrificed. Of course, this `symmetry-breaking' phenomenon
is not unusual, as Jackiw for example points out in \cite{jackiw}
when he discusses the significance of anomalies (and of various
other symmetry-breaking phenomena) in QFT:

\bigskip \noindent (Q6.14)\hskip 0.9in
\begin{minipage}{11cm}
\noindent ``{\small ...{\em In classical physics, the principal
mechanism for symmetry breaking, realized already within Newtonian
mechanics, is through boundary and initial conditions on dynamical
equations of motion}.\footnote{Our emphasis for what we want to
point out in the sequel.} For example, radially symmetric dynamics
for planetary motion allows radially nonsymmetric, noncircular
orbits with appropriate initial conditions...}''

\end{minipage}

\vskip 0.1in

\noindent And of course one cannot refrain from recalling a
similar situation in the context of GR, namely, Penrose's Weyl
curvature hypothesis in order to account for the `observed'
cosmological time-asymmetry: while the Einstein equations are
time-symmetric, the `phenomenological' time-asymmetry could be
accounted for by very special time-asymmetric initial conditions
posed at the Big Bang state of the universe ({\it ie}, the Weyl
curvature tensor---the part of the gravitational field strength
which measures the entropic degrees of freedom of the
gravitational field---was initially zero).

Similarly, our branding singularities as {\em differential
geometric solution-metric anomalies} may also be attributed,
generally speaking, to `initial conditions', in the following
sense: the singularities, which are responsible for lifting the
$\mathrm{Diff}(M)$-symmetry, are exactly due to {\em our} {\it a
priori} (initial) `kinematical' assumption (or model) of spacetime
as a differential manifold. The operative words here are `{\em our
initial}' in the sense that, as it was also explained earlier in
section 3, the assumption by us---the theorists---of spacetime as
a smooth manifold is akin to `initial conditions' in that while it
has nothing to do with the field-law itself (being externally
prescribed and fixed, like gauge-fixing choices or
initial/boundary conditions are), it affects the solutions
(`effects') of those laws (`causes').

Moreover, what is lurking in the above is what we regard as a
grave `theoretical fallacy' on the part of the physicist: the
anomalies (singularities) of the `effects' (solutions) of the
`causes' (differential equations representing the field-laws) are
often attributed back to the `causes' themselves---{\it eg}, to
perceive (as one indeed perceives!) of singularities as DGSs (or
even as SFSs) and infer that the gravitational field (law) breaks
down at them. Plainly, the mistake committed here is that one
blames the law (which anyway in ADG is {\em defined}, as a
differential equation, by the field $\conn$ itself---the original
`gravitational cause'), instead of the initial, {\it a priori}
posited and externally prescribed (by the theorist) spacetime
manifold assumption, for the shortcomings and failings (in the
guise of singularities) of the `effects' (solutions) of these
`causes' (laws). Ultimately, as repeatedly emphasized in the first
three sections, one is often tempted to blame the physics (the
dynamical field laws) instead of the mathematics (the manifold
based CDG), and what is really committed here is an instance of
what Finkelstein has recently coined (to the second author) `{\em
the mathetic fallacy}'\footnote{This fallacy roughly corresponds
to the nowadays commonplace positivist position of theoretical
physicists that physics {\em is} ({\it ie}, it is identified with)
the mathematical model used to describe it.} To stress it again,
singularities are faults and blemishes of the manifold based CDG,
not of the gravitational field (law) itself. Thus to qualify a bit
further the aforesaid denomination, singularities better be
thought of as {\em CDG-based solution-metric anomalies}; they are
{\em mathe(ma)tic fallacies}. All in all, {\em the manifold is to
`blame' (for the singularities), not the field}, and it is
needless to stress it again that that it is {\em we} that identify
physical spacetime with a background differential manifold
(arguably, for differentiability's sake).\footnote{The
CDG-conservatism and monopoly mentioned earlier. This is
understandable: how else can one apply differential geometry to
theoretical physics and gravity in particular?}

\subsection{Section's R\'esum\'e}

In this section we `resolved' (better, completely evaded) in an
ADG-theoretic manner the interior Schwarzschild singularity of the
gravitational field of a point-particle in two different ways.
First, we bypassed it in a finitistic-algebraic fashion issuing
from the finitary (:locally finite), causal and quantal version of
Lorentzian vacuum Einstein gravity presented in \cite{malrap3}. We
regarded the classical pathological point-mass {\it locus} as a
localized, static, spatial point-singularity and used the
categorical bicompleteness of $\ctriad$ to show that the vacuum
Einstein equations (\ref{eqy23}) of CDG-gravity hold both at the
`discrete' level of finsheaves of qausets as shown in
\cite{malrap3}, {\em and} at the `classical' (`smooth') continuum
limit of an inverse/direct system thereof, while no `infinity'
(divergence) of the gravitational field arises at either level
whatsoever. The inner Schwarzschild singularity is not a DGS, as
the manifold based Analysis (CDG) has so far forced us to think.
Second, we bypassed it also in distributional, `time-extended'
sense, by smearing it to a family of spacetime foam dense
singularities in the manner of
\cite{malros1,malros2,mall3,malros3} extending along the
wristwatch time-axis of the point-particle. Thus, the singularity
should not be thought of as an SFS either, as again the manifold
based Analysis has hitherto made us believe. Furthermore, in
either evasions no topological boundary constructions were evoked,
and the singularity was situated in the manifold's bulk; while
also, no geometrical spacetime configurations like a closed
trapped surface or a horizon were evoked to delimit the
applicability of Einstein's law in the vicinity of the
singularity. Finally, we combined the above finitistic evasion
{\it \`a la} Sorkin, with the `ultra-infinitistic' distributional
one {\it \`a la} Rosinger, we defined dense singularity coverings
of the singular spacetime, and argued that (\ref{eqy23}) again
holds both at the `discretum' and at the `continuum' levels. This
only goes to show what we have time and again emphasized in the
present paper, namely, that ADG-gravity is genuinely background
spacetimeless---whether that background is a `discontinuum' or a
`continuum'---and manifestly singularities' uninhibited.

\section{Evading Singularities ADG-Theoretically: What's in for Quantum Gravity?}

As we have repeatedly emphasized in the past trilogy
\cite{malrap1,malrap2,malrap3}, in the recent work \cite{rap5},
and in the present paper so far, spacetime singularities are
supposed to be a problem of {\em classical} gravity (GR) {\it per
se}, long before the quantization of the gravitational field
becomes an issue the theorist has to reckon with. However, in the
past many theoretical physicists have argued and have aired the
hope that a cogent quantum theory of gravity will shed more light
on, perhaps it will even resolve completely, the classical problem
of gravitational singularities \cite{pen5}. Recently for example,
Perry remarked in \cite{perry}:

\bigskip \noindent (Q7.1)\hskip 0.9in
\begin{minipage}{11cm}
\noindent ``{\small ...The existence of a singularity [in
GR]\footnote{Our addition.} is guaranteed by the singularity
theorems of Hawking and Penrose. At a singularity one typically
finds that the curvature tensor blows up, and in the context of
classical general relativity, one should regard this as being the
boundary of spacetime. {\em In classical theory, one generally
regards singularities as being unphysical and an indication that
the theory has broken down. It is hoped that a quantum theory of
gravity would enable us to describe black holes in a way that
would avoid such difficulties}\footnote{Our emphasis.}...}''
\end{minipage}

\vskip 0.1in

\noindent while, very recently \cite{husain}, Husain and Winkler
regard it as imperative that QG resolves in the end the
singularities of the classical theory (GR):

\bigskip \noindent (Q7.2)\hskip 0.9in
\begin{minipage}{11cm}
\noindent ``{\small ...It is well known that many solutions of
classical general relativity have curvature singularities. The
most commonly encountered are the initial and final singularities
in cosmological models, and the singularities inside a black hole.
{\em Just as the classical singularity of the Coulomb potential is
`resolved' by quantum theory, it is believed that any candidate
theory of quantum gravity must supply a mechanism for curvature
singularity resolution}\footnote{Our emphasis.}...}''
\end{minipage}

\vskip 0.1in

\noindent At any rate, generally speaking, if QG is approached
along QFTheoretic lines, the hope is that in much the same way QED
alleviated, or even resolved completely, the unphysical infinities
of the classical Maxwellian electrodynamics, so QG should do away
with the singularities and their associated infinities of the
classical relativistic field theory of gravity (GR). In fact, one
can argue that, independently of other motivations for quantizing
gravity, one must arrive at QG if anything in order to complete
the incomplete, because ridden by singularities, GR. In this
regard, from Gibbons' and Shellard's introductory remarks
\cite{gibbons} in the volume where Perry's quotation above was
found, one can isolate and highlight the following words:

\bigskip \noindent (Q7.3)\hskip 0.9in
\begin{minipage}{11cm}
\noindent ``{\small ...{\em It is interesting to note that the
singularity theorems gave a powerful impetus to the search for a
quantum theory of gravity because they show that classical general
relativity is an incomplete theory. The obviously attractive
direction in which to complete it is by passing to some quantum
version}.\footnote{Our emphasis.} This argument is quite
independent of the many other motivations for quantizing gravity
such as providing a consistent marriage between quantum theory and
spacetime physics or unifying gravity with the other forces of
nature...}''
\end{minipage}

\vskip 0.1in

\noindent In the present section we will discuss how one can
incorporate quantum features in ADG-gravity, side-by-side the
ADG-theoretic evasion of $\smooth$-smooth gravitational
singularities, thus be able to address certain currently important
issues in QG research.

\subsection{Infusing Quantum Ideas into `ADG-Gravity': Prequantization of Gravity via `Third Quantization'}

Our starting point is to mention the accomplishment along
ADG-theoretic lines of a {\em geometric prequantization of
gravity} \cite{malrap2,malrap3}. To this end, one first observes
that since the ADG-notion of field $(\modl ,\conn)$ is a
primitive, fundamental one, the infusion of quantum ideas into ADG
must somehow involve {\em field}, otherwise known as {\em second},
{\em quantization}---albeit, a novel, `non-standard' kind of field
quantization, with an ADG-twist.

But let us take things one at a time. We argued in 3.3.2 earlier
that the {\em background spacetimeless} (whether a continuum or a
discretum) and {\em solely connection-based} (half-order, `pure
gauge' formalism) ADG-theoretic formulation of GR points to the
ADG-conception of `{\em gravity as a gauge field theory of the
third kind}'---a `pure gauge', `external spacetimeless' theoresis
of the autonomous gravitational field `in-itself'. Closely akin to
this `third conception' of gravity as a gauge theory is the notion
of `{\em third quantization}', which we now come to discuss in
more detail.\footnote{The matters below have also been discussed
recently, but more laconically, in \cite{rap5}.}

\subsubsection{Bypassing first quantization and abstracting second quantization:
third, self, or spacetimeless field quantization}

To explain the novel notion of `{\em third quantization}', as
briefly noted above, since it has been amply appreciated ever its
inception \cite{mall1} that ADG refers directly and exclusively to
(the algebraic relations between) the `geometrical objects'---{\it
ie}, physically speaking, the {\em fields} themselves---that live
on `space(time)', without that external surrogate topological
localization background $X$ playing any role whatsoever in its
(differential geometric) concepts and constructions, {\em second}
or {\em field quantization} would {\it prima facie} appear to be
{\em the} appropriate vehicle via which to infuse quantum ideas in
our ADG-gravity.

Indeed, the idea to bypass altogether {\em first quantization} and
tackle directly issues of second quantization by ADG-means has
been worked out in the past \cite{mall6}, especially approached
via so-called {\em geometric (pre)quantization} concepts and
techniques \cite{mall1,mall2,mall5}.\footnote{Let it be noted here
that in the past it has been argued numerous times (and variously
motivated) by many workers in the field of (second) quantization
that in order to arrive at field quantization one need not first
pass through first quantization \cite{bohm,simms,wood,zeh}, in
spite of the traditional view to the opposite \cite{goldstein}.
For us, as we shall argue below, first quantization, even the
continuum based second quantization ideas (QFT), do not seem safe
and sound routes to arrive at a quantum theoresis of ADG-gravity,
because, generally speaking, we believe that one cannot arrive at
a genuinely quantum theoresis of gravity by quantizing `blindly'
the classical, manifold based field theory (GR), much in the same
way that one arrives at quantum mechanics by formally
(canonically, say) quantizing classical mechanics. A genuinely
quantum theoresis of gravity, as we will contend in the sequel,
cannot start from classical---especially from spacetime continuum
based---paradigms, and merely apply a formal `quantization
algorithm' to them. Rather, the description we are seeking for
better have quantum traits already built into it from the very
start.}

The upshot of these applications, that we are going to review
briefly below, is that ADG not only manages to model key ideas
from second and geometric (pre)quantization, such as the quantum
particle representation/interpretation of its fields and their
(sheaf cohomological) classification (into bosons and fermions)
according to their spin \cite{mall6},\footnote{An ADG-application
that was originally motivated by the, albeit explicitly manifold
based, bundle-theoretic musings of Selesnick in \cite{sel} (for
more on these results, see below).} but unlike the conventional
(Minkowski) manifold based (fiber bundle-theoretic) ideas and
techniques of second (QFT) and geometric (pre)quantization
\cite{sel,bandy}, it manages to do this without reference to a
background `space(time)', whether the latter may {\it ab initio}
assumed to be a `continuum' or a `discretum'. Precisely for this
reason we refer to the ADG-based field quantization as {\em third
quantization}. In other words, {\em third quantization is field
quantization without dependence on, reference or commitment to a
background ({\it ie}, external to the fields) spacetime, whether
discrete or continuous}.\footnote{As we will see shortly, it is
perhaps better and more accurate {\em not} to call the field
theory of ADG `third {\em quantized}'---and accordingly, the
theoretical scheme underlying it `third {\em quantization}'---as
if there is an underlying `{\em classical}' counterpart (theory)
that is being quantized. Below we will argue that the dynamically
autonomous (`autodynamical') ADG-fields $(\modl ,\conn)$ are in a
deep and subtle sense `{\em self-quantum}'. That is to say,
quantum ideas are built into the ADG-field formalism from the very
start, so that ADG-gravity better be called `{\em third quantum}'
or `{\em self-quantum third gauge field theory}'.}

Parenthetically we note at this point that this freedom that ADG
gives us in doing quantum field theory in a background spacetime
(manifold)-free way cannot be overemphasized. Especially {\it
vis-\`a-vis} the failure (so far) of attempts to arrive at a
cogent ({\it ie}, a conceptually sound and calculationally finite)
QG by applying QFTheoretic ideas and techniques to GR ({\it eg},
the non-renormalizability of gravity when regarded as a
perturbative quantum gauge field theory like the other three
fundamental forces, or even various deep technical and conceptual
problems when the theory is formulated as a non-perturbative,
background metric independent canonical loop QG), {\em third
quantum gravity as a gauge theory of the third kind} may prove to
be a fruitful route to the `true' QG. For instance, in a possible
`third quantum' scenario for gravity, and in striking
contradistinction to the other external (to the gravitational
field itself) spacetime manifold based scenaria for QG (such as
loop QG), {\em we do not expect `spacetime' (geometry) to be also
quantized} ({\it ie}, we do not regard the `problem' of the
quantum structure of spacetime as being `important', that is, as
being inextricably enmeshed with the problem of QG) simply because
{\em from the start there is no background spacetime in our
ADG-theoresis of gravity}, whether this theoresis may be
ultimately branded `classical' or `quantum'.\footnote{In fact, the
{\em classical}--{\em quantum} distinction loses its meaning in
ADG-gravity. See remarks in the sequel about spacetime
quantization, the questioning in the light of ADG-gravity of the
significance of the Planck length, and the {\em genuinely}
background (manifold) independent character of `third quantum
ADG-gravity'.} Let it be also noted {\em en passant} here that for
other approaches to QG, the attainment of a quantum description of
spacetime structure ({\it eg}, quantum set theory and quantum
topology) is supposed to be prior to---in fact, a necessary
stepping stone to---a genuinely quantum theoresis of gravity
\cite{df0,df1,df2,df3,ish0,ish,ish5,ish6,ish7,sel4,df4}. Still in
further contrast to our spacetimeless `third gauge' and `third
quantum' ADG-musings on gravity, there are certain theoretical
schemes that focus solely on a finitistic and quantum theoresis of
spacetime (or anyway, of space) itself
\cite{pen-1,pen,bostroem,brody}. These remarks will be of
significance a bit later when we comment on a recent `resolution'
of the interior Schwarzschild singularity by loop QG techniques
and results \cite{modesto,husain}, and we compare it with ours
herein.

But let us return to the ADG-based geometric (pre)quantization (of
gravity). In the general theory,\footnote{That is, not
particularly applied to a physical theory such as Maxwelliam
electrodynamics, Yang-Mills theory or gravity.} the three most
important results (for our exposition here) from the application
of the ADG-technology to geometric (pre)quantization and to second
quantization \cite{mall1,mall2,mall5,mall6,mall4} are:

\begin{enumerate}

\item The representation of (local) quantum-particle states of the
ADG-fields by (local) sections of the (associated) vector sheaves
$\modl$ that enter into the definition of the said fields as pairs
$(\modl ,\conn)$. In other words, the quantum-particle aspect of
the ADG-fields is already built into their very definition ({\it
viz}. they are sections of the $\modl$ involved in the $(\modl
,\conn)$). Thus, when one talks about an ADG-field, defined as the
pair $(\modl ,\conn)$, one quite literally understands a `{\em
particle($\modl$)-field($\conn$) pair}',\footnote{What in 6.2 next
we will coin, from a quantum-theoretic viewpoint, the `{\em
quantum self-dual unitary field}'.} with the first synthetic
(`particle') understood as the local aspect (section) of the
vector (sheaf $\modl$) representation of the connection field
$\conn$---the (local) `carrier' of the field. In a
nutshell,\footnote{Recalling also the diagram in 3.2.2.}

\begin{equation}\label{eq01}
\begin{array}{c}
(\mathrm{particle,~field})\equiv(\modl,\conn)
\Longleftrightarrow\cr
\mathrm{local~particle~representation~states~of}~\conn\longleftrightarrow
\mathrm{local~sections}\longleftrightarrow
\mathrm{vector~sheaf}~\modl
\end{array}
\end{equation}

\noindent with the last equivalence being a theorem in sheaf
theory.\footnote{Namely, {\em a sheaf is its (local) sections}.}

Moreover, by virtue of the fact that $\modl$ itself is by
definition locally (a power of) $\struc$, which structure sheaf in
turn is the host of all the singularities that may be present in
the theory,\footnote{Recall, in ADG all singularities are inherent
in the generalized arithmetics (coordinates) in $\struc$.} {\em in
ADG the particles may be regarded as `singularities in their
respective fields'}.

\item The (sheaf-cohomological) classification of the aforesaid
representation sheaves $\modl$ according to the spin of the
fields' particles. In short, (local) states of free (`bare') {\em
bosons} are identified with (local) sections of {\em line
sheaves},\footnote{By definition, {\em a line sheaf is a vector
sheaf of rank} $n=1$.} while (local) states of free {\em fermions}
are represented by (local) sections of {\em vector sheaves of
rank} $n\geq 1$.\footnote{The motivating `forerunners' of this
result, albeit ones that are not expressed in abstract
sheaf-theoretic terms, but rather in `classical' (vector)
bundle-theoretic ones (over the spacetime manifold), are
Selesnick's paper \cite{sel} (in fact, the above sheaf-theoretic
classification of bare elementary particles according to their
spin {\it \`a la} ADG has been coined `{\em the Selesnick
Correspondence}' in \cite{mall1,mall2,mall6,mall4}) and Manin's
`definition' of a {\em Maxwell field} as `{\em a connection on a
line bundle}' \cite{manin,manin1}.}

\item The upshot of 1 and 2 is a basic result from the application
of ADG-theoretic ideas to geometric
prequantization,\footnote{Especially in conjunction with the
ADG-version of {\em Weil's integrality theorem}
\cite{mall,mall1,mall2}.} and it reads: `{\em every (free)
elementary particle is prequantizable; that is, it entails by
itself a prequantizing vector sheaf}'
\cite{mall1,mall2,mall6,mall4}.

\end{enumerate}

\noindent We then complement the result stated above with the
following words taken from the first author's \cite{mall2}, which
highlight the implications and value of that result {\it
vis-\`a-vis} geometric, first and second quantization:\footnote{In
what follows emphasis is the first author's own.}

\bigskip \noindent (Q7.4)\hskip 0.9in
\begin{minipage}{11cm}
\noindent ``{\small ...One thus avoids, by the preceding
conclusion,\footnote{That is, that every free elementary particle
is prequantizable.} the so-called `{\em correspondence
principle}', namely, to pass first from the {\em 1st quantization}
of the physical system involved, a fact that lies, for that
matter, at the basis of {\em geometric quantization}. Indeed, the
latter theory aims basically to {\em bypass the Hamiltonian
mechanics altogether}.\footnote{Here the author cites Selesnick
\cite{sel} and \cite{mall1}.} Yet, within the same vain of ideas,
we also note that {\em there exist quantum mechanical observables
without known classical counterparts}\footnote{Here the author
cites \cite{bohm}.}...}''
\end{minipage}

\vskip 0.1in

\noindent `Third quantum vacuum ADG-gravity' has been
preliminarily investigated along geometric prequantization lines
and in a finitary, causal and quantal setting in
\cite{malrap2,malrap3}. In view also of the inner Schwarzschild
singularity `resolution' accomplished here both by
finitary-algebraic and distributional means, the main relevant
results are:

\begin{itemize}

\item The quantum-particle interpretation, as the {\em quanta of dynamical causality}---otherwise coined `{\em causons}',
of the $fcqv$-fields $\conn$ defining the finitary, causal and
quantal version (\ref{eqfvg}) of the vacuum Einstein equations for
Lorentzian gravity. The third epithet `{\em quantal}' ($q$) in the
denomination of $\conn$ rests precisely on this fact and it
suggests that {\em the dynamical vacuum Einstein equations}
(\ref{eqfvg}) {\em are}, to a certain extent, {\em already
quantum}.\footnote{We will return to discuss this `{\em already
quantum}' (or even, `{\em self-quantum}') trait of ADG-gravity
(\ref{eqy23}) in the sequel. The `{\em fully quantum}' ({\it ie},
not just `to a certain extent' as described above) ADG-gravity may
be formulated along a sum-over-connection-histories, that is, via
a path integral over connection space-type of dynamical scenario,
as we will loosely suggest in the sequel. More on this shortly.}

\item The sheaf-cohomological classification of the vector finsheaves of
qausets involved in $fcqv$-Einstein gravity \cite{malrap2,malrap3}
suggests that {\em causons are fermionic},\footnote{This recalls
the fermionic (Grassmannian) {\em chronons} involved in
Finkelstein's quantum (causal) net approach to quantum spacetime
and gravity \cite{df2,df6,df1,sel1,sel2,sel3,sel4}.} but with the
said vector sheaves one may also $\otimes_{\struc}$-associate a
line sheaf of {\em bosonic graviton-like states} mediating, by
carrying the gravitational force, between causons.\footnote{One
could speculate here on a possible `{\em supersymmetric}' scenario
that transmutes chronons to gravitons, and {\it vice versa}. One
thing is certain however, with every fermionic causon
(representation) state space $\modl$, there is always (implicitly)
a bosonic (graviton?) line sheaf $\mathcal{L}$
$\otimes_{\struc}$-entangled with it, since the following
`absorption' is implicit:
$\mathcal{L}\otimes_{\struc}\modl\simeq\modl$.}

\item Regarding `singularities' and other `discontinuous'
(`non-smooth') phenomena, the ADG-placing of particle-quanta ({\it
viz}. sections of)
$\modl\stackrel{\mathrm{loc.}}{\simeq}\struc^{n}$ side-by-side the
field $\conn$, amounts to integrating (or incorporating)
`particle-singularities'\footnote{Recall Einstein's intuition of
particles as `singularities in the field'.} into the Einstein law
(\ref{eqy23}) that the connection field defines (via its
curvature). Moreover, exactly because the said law is
$\struc$-functorial, we materialize Eddington's remarks in (Q?.?),
namely: {\em the laws of motion of the particle-singularities is
contained in the field-equations}.\footnote{However, unlike
Eddington's words that `{\em the particle's mass is the
pole-strength of the singularity}', which seem to indicate that
particles are `stumbling blocks' in the dynamical laws of motion
defined by the fields, in our ADG-scheme the $\struc$-absorption
of the particle-singularities (like we did for the point-mass at
the inner Schwarzschild singularity), exactly due to the
$\struc$-functoriality of dynamics, testifies to the opposite:
{\em particle-singularities are not obstacles in the dynamical
evolution of the field, and the latter `sees through' them}.}

\end{itemize}

\noindent This last remark, coupled to our earlier observation
that by geometric prequantization we totally bypass first
quantization and go directly to second, field
quantization,\footnote{In fact, as emphasized before, by
ADG-theoretic means we bypass 2nd quantization as well, and have a
background spacetimeless, `purely gauge' (third quantum) field
theory of gravity.} enables us to address some subtle
`categorical' issues regarding 3rd field quantization {\it \`a la}
ADG in comparison to 1st and even to 2nd quantization.

\subsubsection{The issue of `dynamical functoriality' versus `kinematical functoriality'}

To get things started, we read from Baez \cite{baez5}:\footnote{We
are grateful to Ms Kari Kelly for bringing this pre-print to our
attention.}

\bigskip \noindent (Q7.5)\hskip 0.9in
\begin{minipage}{11cm}
\noindent ``{\small ...There is a famous saying about quantization
due to Edward Nelson: `{\em First quantization is a mystery, but
second quantization is a functor!}'\footnote{Our
emphasis.}...

First quantization is a mystery. It is the attempt to get from a
classical description of a physical system to a quantum
description of the `same' system. Now it doesn't seem to be true
that God created a classical universe on the first day and then
quantized it on the second day. So it's unnatural to try to get
from classical to quantum mechanics.\footnote{We wholly concur.
Moreover, we believe that this is so even when one tries to apply
ideas from second quantized (field) theory (QFT) to GR, with the
latter regarded of course as a classical field theory. Read on.}
Nonetheless we are inclined to do so since we understand classical
mechanics better. So we'd like to start with a classical mechanics
problem---that is, a phase space and Hamiltonian function on
it---and cook up a quantum mechanics problem---that is, a Hilbert
space with a Hamiltonian operator on it. It has become clear that
there is no utterly general systematic procedure for doing so.

Mathematically, if quantization were `natural' it would be a {\em
functor} from the category whose objects are symplectic manifolds
(=phase spaces) and whose morphisms are symplectic maps
(=canonical transformations) to the category whose objects are
Hilbert spaces and whose morphisms are unitary operators. {\em
Alas, there is no such nice functor. So quantization is always an
ad hoc and problematic thing to attempt}.\footnote{Our emphasis.}
A lot is known about it, and more isn't. That's why first
quantization is a mystery...

...Note that there {\em is}\footnote{Baez's emphasis.} a functor
from the symplectic category to the Hilbert category, namely one
assigns to each symplectic manifold $X$ the Hilbert space
$L^{2}(X)$, where one takes $L^{2}$ with respect to the Liouville
measure. Every symplectic map yields a unitary operator in an
obvious way. This is called {\bf prequantization}.\footnote{Baez's
own {\em boldface} typescript.} The problem with it physically is
that a one-parameter group of symplectic transformation generated
by a positive Hamiltonian is not mapped to a one-parameter group
of unitaries with a {\em positive}\footnote{Again, Baez's
emphasis.} generator. So my conjecture is that there is no
`positivity preserving' functor from the symplectic category to
the Hilbert category.

Second quantization is the attempt to get from a quantum
description of a single-particle system to a quantum description
of a many-particle system. Starting from a Hilbert space
$\mathcal{H}$ for the single particle system, one forms the
symmetric (or antisymmetric) tensor algebra over $\mathcal{H}$ and
completes it to form a Hilbert space $\mathcal{K}$, called the
bosonic (or fermionic) {\bf Fock space}\footnote{Again, Baez's
{\bf boldface}.} over $\mathcal{H}$. Any unitary operator on
$\mathcal{H}$ gives a unitary operator on $\mathcal{K}$ in an
obvious way. {\em More generally, one has a functor called `second
quantization' from the Hilbert category to itself, which sends
each Hilbert space to its Fock space, and each unitary map to an
obvious unitary map. This functor \underline{is}\footnote{Baez's
emphasis.} positivity-preserving}\footnote{Our emphasis.}...}''
\end{minipage}

\vskip 0.1in

\noindent The first thing to highlight from the words above is
that {\em (geometric) prequantization is a functorial quantization
recipe after all}, but it fails when it comes to preserving the
`positivity of dynamical evolution' in the quantum regime ({\it
ie}, a positive Hamiltonian in phase space is not mapped to a
positive Hamiltonian operator on $\mathcal{H}$ generating a
continuous one-parameter unitary dynamical evolution in it). Thus
first quantization fails to be functorial. On the other hand,
second quantization is functorial as positive unitaries in the
single particle Hilbert space `lift' rather straightforwardly to
positive unitaries in the many quanta Fock space.

From our ADG-perspective, as also mentioned in (Q?.?) above, on
the one hand we totally bypass the {\em non-functorial first
quantization} procedure, and at the same time we apply the {\em
functorial (geometric) prequantization} directly to the dynamical
ADG-gravity field $(\modl ,\conn)$ itself. And, as a result,
although {\it prima facie} we do not abide by the by now standard
{\em categorical-kinematical structures of either classical or
quantum mechanics} ({\it eg}, symplectic
manifolds/symplectomorphisms; Hilbert spaces/unitary
maps),\footnote{In any case, in ADG we manifestly do not deal with
differential manifold phase spaces of classical systems.} in ADG
the functoriality of prequantization is directly reflected in the
dynamical equations (\ref{eqy23}) for vacuum Einstein gravity,
which are $\otimes_{\struc}$-functorial expressions, while, as we
will argue further below, they are in a deep sense `{\em already
quantum}' (self-quantum; `quantum
in-themselves').\footnote{Needless to mention in this respect that
the ADG-theoretic geometric prequantization of gravity (in effect,
what we earlier called `third quantization'), not only does not
start from a classical single-particle theory and then first
quantizes it, it does not start from GR and then applies
manifold-based field (second) quantization techniques either. To
stress it again, in ADG-gravity we do not start from the classical
theory (GR) and then quantize it, but rather we refer directly to
the dynamical field of gravity and expose its inherently
(pre)quantum character---and hence of the (abstract) dynamical
equations (\ref{eqy23}) that it defines. Let us also note here
that the (third quantum) field theory {\it \`a la} ADG does not
draw any distinction, as it is normally done like in (Q?.?) above,
between single-particle (finitely many degrees of freedom) first
quantized theory, and many-particle (infinitely many degrees of
freedom) second quantized theory (QFT). This is of course closely
akin to the fact that the ADG-field theory does not distinguish
between a background discretum or continuum spacetime, or {\it in
extenso}, state (configuration or phase) space.} {\it In toto}, we
could express this by saying that while in the usual continuum
based\footnote{Whether that continuum is the differential
spacetime manifold, or the differential configuration or phase
space manifold.} first and second quantization theories
functoriality pertains essentially to the `preservation' of the
kinematical structure as one progressively goes from
single-particle classical mechanics, to quantum mechanics (1st
quantization), to QFT (2nd quantization), in the {\it ab initio}
field-based and background spacetimeless ADG `3rd quantization'
(of gravity) pertains to the functoriality of the `already
quantum' dynamical equations (\ref{eqy23}). That is why we titled
the present sub-subsection `{\em dynamical versus kinematical
functoriality}'.\footnote{This is another manifestation of the
theoretical `phenomenon', already mentioned in section 3 and to be
further discussed in the next section in view of Einstein's hole
argument testing the PGC of GR, that {\em in ADG}, contrary to the
usual {\it aufbau} of physical theories so far, {\em dynamics is
prior to kinematics}.}

But now we are ready to discuss in greater detail below the `{\em
already quantum}' or `{\em self-quantum}' nature of the (3rd
quantum) ADG-fields mentioned above. Before we do this however, as
a general motivational question for developing the external
spacetimeless, 3rd or self-quantum (gauge) field theory (of
gravity)\footnote{And for the other Yang-Mills forces, including
Maxwellian electrodynamics \cite{mall1,mall2,mall4}.} we recall
Isham's words from the introduction of \cite{ish8}:

\bigskip \noindent (Q7.6)\hskip 0.9in
\begin{minipage}{11cm}
\noindent ``{\small Attempts to construct a quantum theory of
gravity provide many challenges for quantum theory itself. Some of
these involve conceptual issues...For example: {\em is it
meaningful to talk about quantum theory in the absence of any
background spatio-temporal structure?}\footnote{Our
emphasis.}...}''
\end{minipage}

\vskip 0.1in

\noindent In view of ADG and the 3rd quantum (gauge) field theory
that it supports, our answer to the question above is an
emphatically affirmative one: it is meaningful indeed to talk
about a background spacetimeless {\em quantum} (`pure gauge') {\em
field} theory (of gravity). Let us discuss more this `{\em
self-quantum}' character of its autonomous-dynamical
(`auto-dynamical') connection (gauge) fields in the manifest
absence of an external, background spacetime (whether a continuum
or a discretum) .

\subsection{A Genuine Field-Particle Duality: Heisenberg's Algebraic Indeterminacy and Bohr's
Complementarity Reformulated `Self-Referentially' in ADG-Theoretic
Terms}

On very general grounds, and apart from the distinctions between
first and second quantization noted above, one may view {\em
quantization as representation}. Wigner and Mackey's seminal work
on (unitary) `group imprimitivity' and `group quantization' theory
testifies to that \cite{wigner,mackey}. For example, it is well
established now that in the Minkowski manifold based (2nd
quantized) theory of fields (QFT), an elementary particle---the
quantum of the corresponding field---can be completely
characterized by the {\em irreducible representations} (`irreps')
of the Poincar\'e group of the external spacetime symmetries.
Indeed, irreps are {\em group characters} and represent the
particle's mass and spin (quantum) numbers, which are conserved
during the fields' dynamics.\footnote{General Noether's theorem
for continuous dynamical symmetries.} By the very definition of
irreps, particle states belong to minimal invariant subspaces of
the representation (Hilbert) space of the corresponding group.
Moreover, apart from the external spacetime symmetries, as it is
well known there are internal gauge symmetries whose characters
are the so-called `gauge charges'.\footnote{The usual Standard
Model ones being the $U(1)$-electric charge of the abelian
electrodynamics, as well as the flavor and colour internal gauge
charges conserved in the weak ($SU(2)$) and strong ($SU(3)$)
interactions, respectively.}

On the other hand, we have repeatedly emphasized thus far that the
notion of representation figures prominently in the very
definition of the ADG-fields $(\modl ,\conn)$, and in their
`inherently' (pre)quantized character;\footnote{Indeed, in view of
the aforementioned geometric prequantization results,
quantum-particle traits are built into the formalism from the very
start---in fact, from the very definition of ADG-fields as pairs
of the sort $(\modl , \conn)$.} for as noted above, the vector
sheaf $\modl$ is the representation, associated sheaf of (quantum)
particle states of (the principal sheaf $\aut\modl$ of the
dynamical self-transmutations) of the dynamically autonomous
connection field $\conn$. Albeit, this `dynamical autonomy'
pertains precisely to ADG's referring directly and solely to the
dynamical fields `in-themselves'---{\it ie}, without reference to
a background spacetime structure. As a result, the `pure gauge',
`esoteric' Kleinian geometry of the field's particle-quanta is not
described by extraneous ({\it ie}, externally prescribed and
fixed) spacetime attributes---{\it eg}, by the mass and spin
characters that quantum fields `inherit' from the external
Minkowski spacetime symmetries in the usual theory (QFT). In other
words, the particle-geometry (representation) of the ADG-fields is
not effectuated and expressed as usually, via the symmetries of
the external spacetime (manifold) on which the relevant fields are
normally thought of as being soldered. Keeping in mind this
remark, let us discuss further the aforesaid `{\em self-}' or
`{\em already quantum}' character of the auto-dynamical
ADG-fields.

On the face of all this, there is the following `syllogism': we
have the classic and perennial philosophical duality between {\em
state} (`being') and {\em change of state} (`becoming'). In
classical point-particle mechanics, operationally speaking, we
have mutually compatible {\em position-determinations}
(determinations of `stasis' or `position states') and {\em
momentum-determinations} (determinations of `kinesis' or
`change-of-position/motion states') in the classical phase space
continuum (point manifold) of the particle. In standard
(non-relativistic) quantum point-particle mechanics
however,\footnote{For example, in the case of a non-relativistic
particle moving on the real line $\R$.} these two kinds of acts of
determination are fundamentally supposed to be complementary or
conjugate (mutually exclusive), with their sharpness (accuracy of
determination) being limited by the quantum of action $\hbar$, an
operational complementarity that (in natural units $\hbar=1$) is
expressed algebraically by the so-called canonical (equal-time)
Heisenberg commutation relations (in the position or coordinate,
Schr\"{o}dinger picture)

\begin{equation}\label{eqz1}
[\hat{x},\hat{p}]=\imath 1;~\hat{x}=x,~\hat{p}=\frac{d}{dx}
\end{equation}

\noindent in which $\hat{x}$ and $\hat{p}$ are (self-adjoint)
operators acting on a certain Hilbert space $\mathcal{H}$ of
states ($\com$-valued functions) $\psi(x)$ of the quantum
particle---the so-called `carrier' or representation space of
these complementary acts of determination.\footnote{Let it be
noted here that in view of the Stone-Von Neumann theorem, all
irreducible unitary representations of the (kinematical)
Heisenberg algebra generated by the commutation relations in
(\ref{eqz1}) above are unitarily equivalent.} Operationally
speaking, the algebraic indeterminacy relations above can be read
as follows: a sharp act of determination of the position of the
quantum results in an uncontrollable, dynamical change
(perturbation) in its momentum---as it were, its momentum becomes
`fuzzy' and indeterminate.\footnote{And {\em vice versa} about
determining momentum sharply and losing information about the
quantum-particle's position. In quantum mechanics, `{\em you win
some, you lose some}'.}

In the standard Born-interpretation $\psi$ is then interpreted as
a probability amplitude wave-like `field' obeying in the
Schr\"odinger picture the well known differential (dynamical) and
unitary in $\mathcal{H}$ eigenvalue wave equation

\begin{equation}\label{eqz2}
-\imath\frac{d\psi(x,t)}{dt}=\hat{H}\psi(x,t)=E\psi(x,t)
\end{equation}

\noindent where $\hat{H}$ is the (hermitian) Hamiltonian operator
(total energy) of the particle, and $t$ an external time-parameter
by (or with respect to) which evolution is `temporally
coordinatized' and `geometrically ordered' (pictured).\footnote{In
a similar fashion, and informally speaking, in the QFT of matter,
where quantum fields are quantum systems with an infinite number
of degrees of freedom propagating in flat Minkowski space, one
follows suit and posits CCRs for the field $\phi(x)$ and its
conjugate momentum $\pi(x)$---$x$ now being the $4$-coordinates of
a point in Minkowski space. The particle interpretation is then
established by Fourier analysis of the field into its infinite
constituent modes and subsequent interpretation of the (positive
and negative frequency) coefficients of that series as creation
and annihilation operators of particles (quanta) acting on the
fundamental zero-mode state representing the (particle or quantum)
vacuum. Moreover, in quantum field theory there is what one
usually refers to as the Schwinger duality between the quantum
matter sources and their so-called radiation fields ({\it eg}, in
QED one has the electron source of the electromagnetic field)---a
duality that is the conceptual analogue of the wave-particle one
of non-relativistic quantum mechanics.}

All this is well known and fine. Now, in ADG we have the following
abstract or generalized analogue of the standard situation above:
we have the {\em unitary}, background spacetime (manifold)
independent---what we refer to as `spacetimeless'---field
$\field$\footnote{Hereafter to be referred to as `{\em the
$\U$-field}'.} being represented by the pair

\begin{equation}\label{eqz3}
\field :=(\modl ,\conn)
\end{equation}

\noindent $\field$ is what we would like to coin `{\em quantum
self-dual}' (or perhaps more suggestively, `{\em intrinsically
field-and-particle dual}') in view of the following remarks about
its very `definitional' traits:

\begin{itemize}

\item On the one hand we have $\modl$ built into $\field$---the
`carrier' or `representation' space' for the connection $\conn$
part of $\U$-field in (\ref{eqz3}) above.\footnote{As noted
numerous times before, in technical jargon, $\modl$ may be viewed
as the {\em associated sheaf} of the {\em `auto-symmetry' group
sheaf} $\aut\modl$ of its `auto-transformations' or
`self-transmutations'.} From a geometric prequantization vantage,
local sections of $\modl$ are interpreted as particle states of
the $\U$-field.\footnote{In fact, as noted before, from a
sheaf-theoretic vantage the basic result is that the sheaf $\modl$
{\em is nothing but its sections}
\cite{mall1,malrap1,malrap2,malrap3}.} At the same time, since by
definition $\modl$ is locally $\struc^{n}$, and since, as we noted
throughout this paper, $\struc$ represents our generalized
coordinate localizations (:local coordinate
determinations=`coordinatizations') of the $\U$-field, the local
sections of $\modl$ are nothing else but generalized `local
position' or `coordinate determinations' of the $\U$-field, which
are fittingly interpreted in a particle-kind of
way.\footnote{Arguably, a (point-)particle is an ideal,
`ultra-local' spatial (and static!) characterization of a
system---as it were, particles are completely localized [on
space(time)] aspects of a (quantum) system (see Einstein's
quotation next). In a quantum context, one may think for instance
of the localized point-like traces that the quantum leaves on the
silvered photographic plate in the double-slit experiment.}

For expository completeness, one could add here that the remarks
above about the local `coordinate' or `particle-position'
character of the $\modl$ part of the $\U$-field $\field$ are
supported very precisely in ADG by the following
(sheaf-)cohomological result \cite{mall1,mall2}:

\bigskip \noindent (R7.1)\hskip 0.9in
\begin{minipage}{11cm}
\noindent every vector sheaf $\modl$, which is locally, by
definition, a finite power of $\struc$, can be completely
characterized by a so-called {\em coordinate 1-cocycle}
$(g_{ab})\in Z^{1}(\gauge
,\mathcal{G}\mathcal{L}(n,\struc))$.\footnote{Where
$\gauge=(U_{a})_{a\in I}$ is a local frame (gauge) of $\modl$, as
usual.}
\end{minipage}

\vskip 0.1in

\item On the other hand, the $\conn$ component of $\field$ in
(\ref{eqz3}) above, being a generalized derivative (differential
$\partial$), acts locally on $\modl$'s (local) sections to change
them---a process representing dynamics in our theory.\footnote{In
much the same way that momentum represents a process of change of
position-states. In our scheme, and categorically speaking,
$\conn$ is a sheaf-morphism---a map or operator acting on and
changing the relevant $\modl$'s (local) sections.} Of course, in
our scheme there is no external agent\footnote{Or to the same
effect, a base spacetime (manifold)---a `background geometry' so
to speak---to which the $\U$-field is referred and relative to
which the dynamics that its $\conn$-component defines is
`coordinatized', `arithmetized', `quantified' and `geometrically
ordered' (pictured).} other than the $\U$-field itself to effect
these changes. Rather, when {\em we} externally try to
coordinatize or localize the $\U$-field, thus so to speak
`extract' its particle-like attributes\footnote{Which, as noted
above, are represented by the (local) sections of the $\modl$ part
of $\field$, which $\modl$, in turn, is completely characterized
sheaf-cohomologically by the coordinate $1$-cocycle $g_{ab}$
(R6.1).} by employing $\struc$,\footnote{The reader should notice
here that it is {\em us}---the external agents (observers,
experimenters, or `geometers/measurers')---that attempt to
coordinatize $\field$ via $\struc$.} the potential part $\conn$ of
the $\U$-field `kicks back (non-linearly) on itself' ({\it ie}, in
a sense, it `self-interacts', or `acts within $\field$') to
dynamically change these `sharp' (localized) position/particle
states---$\modl$'s (local) sections. Mathematically, one may think
of this as being reflected by the fact that $\conn$ is not an
$\struc$-morphism, which in turn means that our (external)
geometrical acts of localization (in $\struc$) or soldering of the
$\U$-field on $\modl$ cannot `arrest' its $\conn$ part\footnote{As
it were, our external attempts to `dissect' $\field$ into its
constituent parts $\modl$ and $\conn$, thus restrict it sharply
and exclusively to $\modl$---$\field${\em 's `particle
picture'}.}---our local gauge acts of pin-pointing, `arresting' or
`freezing' the $\U$-field (on $\modl$), and, as a result, $\conn$
eludes them---{\it ie}, it effects dynamical changes of particle
states (local sections of $\modl$).

At this point, and in connection with the foregoing discussion
about `the indeterminacy within $\field$', it is interesting to
bring forth from \cite{einst11} Einstein's remarks about the
essence of Heisenberg's uncertainty relations
(\ref{eqz1}):\footnote{in the quotation below emphasis is ours.}

\bigskip \noindent (Q7.7)\hskip 0.9in
\begin{minipage}{11cm}
\noindent ``{\small ...On the other hand, it seems to me certain
that {\small\em we must give up the idea of a complete
localization of the particles in a theoretical model.} This seems
to me to be the permanent upshot of Heisenberg's principle of
uncertainty...}''\footnote{And in a quite famous joking remark of
his, he said: ``{\small\em The more one chases quanta, the better
they hide.}'' \cite{einst2}.}
\end{minipage}

\vskip 0.1in

Of course, and this is another manifestation of the `{\em inherent
quantumness}'\footnote{See paragraph after the next.} of the
$\U$-field $\field$ (and {\it in extenso} of our ADG-theoretic
scheme), in a (generalized) sense analogous to Heisenberg's
so-called (noncommutative) `matrix mechanics', almost invariably
we effectively work not with $\modl$ itself, but with the sheaf
$\modl nd\modl=\mathcal{H}om_{\struc}(\modl,\modl)$ of
endomorphisms of the particle or quantum (states) itself which, by
virtue of the fact that $\modl$ is locally $\struc^{n}$,
corresponds to the {\em non-abelian matrix algebra sheaf}
$M_{n}(\struc)(U)$---the sheaf of (local sections of) algebras of
(in general) non-commuting $n\times n$ matrices with entries from
$\struc$. Accordingly, with respect to the $\conn$ aspect of
$\field$, we work with the connection
$\conn_{\modl\otimes_{\struc}\modl^{*}}$ on the tensor product
sheaf $\Hom_{\struc}(\modl
,\modl^{*})=(\modl\otimes_{\struc}\modl)^{*}=\modl^{*}\otimes_{\struc}\modl^{*}$,
a connection which is induced by the $\struc$-connection $\conn$
on $\modl$ \cite{mall1,mall2,malrap3}. In this way, the dynamical
changes that $\conn$ effects on the local particle (quantum)
states (sections) are `covariant' with the latter's (in general)
noncommuting self-transmutations in $\modl nd\modl$,\footnote{To
be precise, the said covariance-proper is captured by the group
sheaf $\aut\modl=(\modl nd\modl)^{^{\bull}}$ of {\em invertible}
endomorphisms of the particle $\modl$---the automorphisms of
$\modl$---which self-transmutations locally correspond to sections
living in $(M_{n}(\struc))^{^{\bull}}(U)$.} and the algebraic
$\conn$ lies at the quantum side of the quantum/classical
divide.\footnote{See the generalized ADG-theoretic analogue of
Bohr's correspondence principle expressed in the next
sub-subsection.}

For expository completeness, one could add here that, although the
algebraic connection $\conn$ manifestly eludes our local
$\struc$-coordinatizations (equivalently, `position
determinations' or `quantum particle-localizations') of the
$\U$-field $\field$ by effecting dynamical changes of the local
particle/position states (sections) of $\modl$ (something that as
noted above is implied by the statement that $\conn$ is not an
$\struc$-morphism), its measurable manifestation---the curvature
$\curv(\conn)$---is an $\struc$-morphism, which in turn, being a
closed $2$-form, defines a characteristic class $[\modl]$ of
$\modl$s (or equivalently, of $1$-cocycles $(g_{ab})$ determining
$\modl$) sheaf-cohomologically \cite{mall1}, as follows:

\begin{equation}\label{eqz4}
\frac{[\curv]}{2i}=[(g_{ab})]\equiv [\modl]
\end{equation}

\noindent {\it En passant} we note that, precisely because of this
expression, the $\U$-field $\field$, represented by the pair
$(\modl ,\conn)$ as in (\ref{eqz3}), can be equivalently
represented as:

\begin{equation}\label{eqz5}
\begin{array}{c}
\field :=(\conn ,\curv(\conn)),~\mathrm{or:}\cr (\conn
,\modl)\Leftrightarrow (\conn ,\curv(\conn))
\end{array}
\end{equation}

\noindent an equivalence which effectively rests on the {\em Chern
isomorphism}, the following abelian group isomorphism
\cite{mall1,mall2}:

\begin{equation}\label{eqz6}
\modl\leftrightarrow g_{ab}\in H^{1}(X,\struc^{\bull})\simeq
H^{2}(X,\Z)\ni\curv\leftrightarrow\modl
\end{equation}

\item All in all, the $\U$-field $\field$ has built into it the
particles (quanta)---represented by (the local sections of)
$\modl$, {\em and} the `gauge potential'\footnote{To use a popular
gauge-theoretic synonym for the connection.} for their dynamical
changes---represented by $\conn$.\footnote{In analogy with the
standard Heisenbergian quantum $x$--$p$ dichotomy, we could coin
the $\modl$ part of $\field$ `{\em the generalized
position/particle picture of the $\U$-field}', while the $\conn$
part of $\field$, `{\em the generalized momentum/field picture of
the $\U$-field}'.} The unitary field $\field$ is
self-indeterminate (self-dual or self-complementary\footnote{In
the sense that the generalized field/momentum aspect $\conn$ of
the $\U$-field $\field$ is complementary to its generalized
particle/position aspect $\modl$, so that one speaks of
$\field:=(\modl ,\conn)$ literally as `{\em particle-field
pair}'.}) and auto-dynamical.\footnote{In quantum jargon, one may
use the epithet `{\em coherent}' as a synonym to `{\em unitary}'
for the $\U$-field $\field$ in the sense that its dual character
is {\em inseparable} and `{\em holistic}': that is to say, {\em
one cannot think of the generalized field/momentum aspect $\conn$
of $\field$ apart from its generalized particle/position aspect
$\modl$}, and {\it vice versa}. $\field$ is not `fragmentable' or
`dissociable' into its `constituent' particle ($\modl$) and field
($\conn$) parts (aspects). Moreover, these two mutually
complementary aspects of $\field$ are autonomous---{\it ie}, in no
way dependent for the `subsistence' on a pre-existent and external
(background) geometrical space(time), and in particular, on a
continuum (3rd quantum character of the $\U$-field $\field$).} One
could then argue that in this sense {\em the unitary field
$\field$ is already `inherently quantum'},\footnote{Since the
epithet `{\em quantum}' is already preempted by the usual theory
and supported by its technical paraphernalia, we prefer the
adjective `{\em quantal}' \cite{malrap1,malrap2,malrap3}, also in
order to separate our position from the standard theory. `Quantal'
may be thought of as a synonym to the denomination `3rd quantum'
that we discussed earlier.} thus in no need of the (formal)
procedure of `quantization' to be exercised (in a forced, {\it ad
hoc} fashion by the theorist!) on it. Thus, in a subtle sense, our
theory has no `classical correspondents', and no formal statements
of the sort usually encountered in conventional quantum scenarios
about pre-existing classical theories, as for example `{\em the
classical theory is recovered at the limit as $\hbar\mapto 0$}',
are made.\footnote{Although we will make loose remarks in this
direction when we discuss the (spacetime) continuum-bound
appearance of fundamental constants (de)limiting (the range of
validity of) spacetime continuum and Calculus-based theories such
as SR ($c$), GR ($G$) and QM ($\hbar$), and their `conspiring' to
positing a fundamental length in Nature---the so-called Planck
length $\ell_{P}$, in section 7.2.}

The last remarks bring us to discuss a generalized ADG-theoretic
version of Bohr's {\em Correspondence Principle (or Limit)} (CP).

\end{itemize}

\subsubsection{A generalized version of the Correspondence Principle in ADG-theoretic terms:
the abstract algebra-vs-geometry `schnitt'}

A general expression of Bohr's CP is the following: {\em while the
`observable' properties of quantum mechanical systems can be
represented by non-commuting operators}\footnote{Acting on the
aforesaid (complex) Hilbert space $\mathcal{H}$ of states $\psi$
of the quantum system.} (the so-called `$q$-numbers'), {\em our
measurements of them always `yield' classical, commuting
numbers}\footnote{And the physics describing the measuring
apparatus is {\em classical physics}.} (the so-called
`$c$-numbers'), which are also fundamentally assumed to be {\em
real} ($\R$).\footnote{In turn, this expression of the CP is
intimately related to the act of {\em measurement} and its
associated {\em spontaneous collapse of the wave function} in the
usual Copenhagen interpretation of quantum mechanics, whereby,
upon measurement of a particular quantum mechanical observable
(represented by a certain self-adjoint operator $\mathcal{O}$ in
the aforementioned $\mathcal{H}$), the wave function (state)
$\psi\in\mathcal{H}$ of the quantum system is supposed to
`collapse' abruptly to one of $\mathcal{O}$'s eigenstates
$\psi^{\mathcal{O}}_{n}$ (with probability $|<\mathcal{O}\psi
|\psi^{\mathcal{O}}_{n}>|^{2}$), `yielding' in the process the
(real) number $o_{n}$---the eigenvalue of $\mathcal{O}$ at
$\psi^{\mathcal{O}}_{n}$ (projection postulate). In principle,
quantum theory, in its Copenhagen interpretation at least, says
nothing about where or when the collapse of the state vector
happens, hence the `{\em quantum divide}'---the so-called {\it
Heisenberg schnitt}---between $q$- and $c$-numbers (or quantum and
classical descriptions of the {\em quantum endosystem} and the
{\em quantum exo- or episystem} respectively \cite{df1}) is
indeterminate in the theory. In other words, the theory says
nothing about the `emergence of classicality' or `where one draws
the line between classical and quantum mechanical behavior', and
the separation between the classical observer (`experimenter') and
her measuring apparatus, and the quantum system (`experimentee')
is rather arbitrary. In the mathematical formalism, this
arbitrariness and the fuzziness of the transition from the quantum
to the classical realm (or {\it vice versa}) is reflected by such
`loose' and {\it ad hoc} pseudo-formal mathematical expressions
such as `{\em (canonical) quantization is deformation}' or `{\em
(canonical) quantization corresponds to substituting Poisson
brackets by commutators}' ($c\rightarrow q$), and conversely, that
{\em classical mechanics is recovered from quantum mechanics in
the `de-deformation' limit as $\hbar\mapto0$}' ($q\rightarrow c$).
All this is usually subsumed in what is generically known as the
{\em measurement problem} in quantum mechanics. Let it be also
noted here {\em en passant} that, for some researchers,
measurement---in its general conception as the process of ``{\em
transformation of possibilities {\rm [quantum potentialities
generically represented by $\psi$]} into facts {\rm [measurable
spacetime events]}}''---is the quintessential feature of the
quantum, capturing a {\em fundamental irreversibility or
`time-asymmetry'} supposedly inherent in quantum theory
\cite{haag}. Similarly, and in the context of QG, Penrose has not
only propounded a theory for quantum state reduction due to the
gravitational field \cite{pen1,pen3,pen4}, but also he has gone
even further to propose that ``{\em the true quantum gravity is a
time-asymmetric theory}''---a proposal basically resting on the
Weyl curvature hypothesis supporting time-asymmetric initial
conditions for the quantum Universe \cite{pen2}.}

As noted in the concluding lines of 6.2, in ADG the autonomous,
purely algebraic and background geometrical spacetime (manifold)
independent $\U$-field $\field$ is `{\em self-complementary}'---a
`{\em quantum self-duality}' which can be interpreted as an `{\em
inherent quantumness}'. Thus, $\field$ is {\em prima facie} in no
need of a formal procedure of quantization (and conversely, of a
CP), something which would presuppose an already existing
classical theory.\footnote{We will return to comment more on these
issues in the context of QG proper and, as a result, question
`quantization' altogether, in subsections 6.6 and 6.7. below.} To
be sure, on the mathematical side one could argue that there is a
sort of CP between ADG and CDG given by the formal substitution
(better, identification)

\begin{equation}\label{eqz7}
\struc\mapto\smooth_{M}\Leftrightarrow\struc\equiv\smooth_{M}
\end{equation}

\noindent whereby, {\em CDG is the particular instance of ADG in
case one uses $\smooth_{M}$ as the structure sheaf of generalized
coefficients (arithmetics) or coordinates}. In this special case
indeed one could maintain that {\em the smooth field} (with
respect to its differential geometric properties) {\em is
sustained by the smooth background locally Euclidean geometric
space(time), as in the classical theory (CDG)}.

However, this mathematical correspondence is just a particular
manifestation of a deeper physical `self-correspondence' which is
again inherent in the $\U$-field $\field$. In turn, this
correspondence may be `categorically' perceived as one between
{\em algebra} and {\em geometry}. To explain what we mean by the
latter, let us refer again to $\field$ as expressed by the pair
$(\modl ,\conn)$. $\field$, its part $\conn$ in particular, is a
purely algebraic, quantal entity, which in ADG is fundamentally
assumed to exist independently of an ambient (background)
geometrical space(time) (PFR). To be sure, {\em geometrization},
or what is the same, localization (of $\conn$ and {\it in extenso}
of $\field$), is achieved as soon as one introduces the {\em
commutative} algebra sheaf $\struc$ to coordinatize $\modl$---the
carrier space of $\conn$. The introduction of $\struc$ lies on the
classical side of the quantum divide: {\em it is us---the local
`observers' or `measurers'---who introduce $\struc$ in order to
localize, `pin-point', or solder $\conn$ on $\modl$, which is in
turn expressed as a finite local power of the $\struc$ employed}.
As far as geometry and space(time)\footnote{As noted earlier, in
the classical theory (CDG) for instance, the space(time) manifold
$M$ is extracted from $\struc\equiv\smooth_{M}$ by Gel'fand
duality and spatialization.} are concerned, they are completely
encoded in the abelian $\struc$---our measurements
(coordinatizations or localizations) of `it all'\footnote{And
recall that in ADG the `{\em it all}' above essentially
corresponds to $\conn$, and $\struc$ merely represents our actions
towards locating, measuring it---{\it ie}, the algebraic field
$\conn$.} yielding commutative `numbers', the local sections of
the arithmetics' sheaf $\struc$, in the process. At the same time,
the `arbitrary' choice\footnote{The epithet `arbitrary' here is
closely analogous to the completely arbitrary (gauge) choices (and
conditions) that the external macroscopic observer makes (and
imposes) about what property of the quantum system to measure.}
and introduction of $\struc$, and the concomitant (back re)action
of the (quantal) field $\conn$ to change the local particle states
of the quantum represented by the local sections of
$\modl=\struc^{n}$, is similar to how, upon the perturbing
measurement that the classical (exo)system exercises on the
quantum (endo)system, the latter's (local) states dynamically
change.\footnote{In turn, this is similar to how upon
determination of the position (particle aspect) of a quantum
system, the act of measurement dynamically disturbs (changes) its
momentum, which thus becomes indeterminate or `quantum fuzzy'.}

\paragraph{A quantum Zeno-type of paradox for the 3rd quantum
ADG-fields.} Note here the following apparent paradox: while the
algebraic $\struc$-connection field $\conn$ `derives' from
$\struc$,\footnote{After all, as discussed in section 3, all DG
boils down to $\struc$!} it ultimately elude them ({\it ie}, our
generalized measurements do not `respect' $\conn$ and the latter
is not an $\struc$-tensor---an $\struc$-sheaf morphism). In a
sense,  $\struc$ is $\conn$'s `source' (producer), but not its
`sink' (register). Once a $\struc$ `gives rise' to a
$\conn$,\footnote{That is, once {\em we} use an $\struc$ to
geometrically `capture' (represent) the field $\conn$.} $\conn$
ultimately eludes it. Equivalently, when we introduce (employ) a
specific $\struc$ to extract the particle/position-coordinate
aspect of the $\U$-unitary ADG-field $\field:=(\modl ,\conn)$---as
it were, to localize and geometrically represent $\conn$---the
latter reacts and effects dynamical changes of local
particle/position-states---the local sections of $\modl$.

\subsection{No Background Geometry to Work with: An Impediment or a Boon to
QG?}

As alluded to above, the dynamically autonomous (`synvariant'),
third gauge (`purely gauge'), third quantum (`self-quantum') and
background spacetimeless field $(\modl ,\conn)$ of ADG-gravity may
prove to be a suitable notion via which to address and tackle
certain caustic both structural and conceptual issues in current
QG research.

One of these major issues, especially in non-perturbative QGR in
its connection based loop QG version \cite{rovsm1,thiem3,smolin},
is to formulate the theory ({\it ie}, the quantum dynamics) in a
genuinely background independent fashion \cite{alvarez}. Roughly,
by `{\em background independence}' it is meant `{\em background
metric independence}'---{\it ie}, unlike in the usual (mainly
perturbative) approaches to QG where one fixes a (usually flat,
Minkowski) background metric in order to formulate the quantum
dynamics (and expand the relevant quantities about it, as well as
to impose meaningful commutation relations among
them),\footnote{The reader should note that this (flat) background
metric dependence is also a feature of the (perturbative)
string-theoretic approach to QG.} here there is no such desire
since, anyway, it appears to be begging the question to fix {\it a
priori} (and by fiat!), and moreover to duplicate, the one and
only dynamical variable of GR in its original formulation---{\it
ie}, the spacetime geometry (metric).\footnote{Let alone that, by
fixing the said background metric, one runs the risk of lifting
the manifest diffeomorphism invariance of the classical theory
(GR). (See below.)} Indeed, Geroch had noticed very early
\cite{geroch} this characteristic feature of GR when he tried, in
order to get a better idea of what is a singularity in GR, to
compare gravitational singularities with the infinities assailing
QFT:

\bigskip\noindent (Q7.8)\hskip 0.9in
\begin{minipage}{11cm}
\noindent ``{\small ...Our intuitive idea of what a singularity
should be in Einstein's theory comes from the relatively
well-understood infinities which arise in classical field
theories, e.g., electrodynamics and hydrodynamics. {\em
Unfortunately, general relativity differs from these theories in
one important respect: whereas in other field theories one has a
background (Minkowskian) metric to which the field quantities can
be referred, in general relativity the `background metric' is the
very {\rm [dynamical]}\footnote{Our addition.} field whose
singularities one wishes to describe}\footnote{Our
emphasis.}...}''
\end{minipage}

\vskip 0.1in

\noindent In the context of QG proper, Baez too, for example, is
similarly categorematic in \cite{baez0}:

\bigskip \noindent (Q7.9)\hskip 0.9in
\begin{minipage}{11cm}
\noindent ``{\small\em The main problem in quantum gravity is that
there is no background geometry to work with\footnote{Our
emphasis.}...}''
\end{minipage}

\vskip 0.1in

\noindent Let it be stressed here that Ashtekar and coworkers have
succeeded over the years in formulating loop QG in an
authentically fixed background metric independent way \cite{ash5},
albeit, {\em a smooth spacetime manifold is still retained in the
background} \cite{ash4,ash6}---or else, how could one still use
differential geometric ideas and constructions \cite{ashlew2} in
QG research?\footnote{This is another manifestation, now in QG
proper, of the CDG-conservatism and monopoly mentioned in section
2 in the context of the singularities of classical gravity (GR).
For example, the new connection variables \cite{ash} employed in
the loop approach to canonical QGR are {\em smooth}
(spin-Lorentzian) connections based on a differential spacetime
manifold, let alone that, as noted earlier, the smooth metric is
still implicitly present in the theory as it is carried by the
smooth tetrad ({\it vierbein}) variables (1st-order formalism). On
the other hand, the (canonical) quantum commutation relations
imposed in the theory are genuinely covariant and no fixed
(Minkowski) metric is evoked to effectuate them.} We thus read
from \cite{ash4}:

\bigskip \noindent (Q7.10)\hskip 0.9in
\begin{minipage}{11cm}
\noindent ``...{\small In this approach, one takes the central
lesson of general relativity seriously: {\em gravity is geometry
whence, in a fundamental theory, there should be no background
metric.}\footnote{Our emphasis.} In quantum gravity, geometry and
matter should both be `born quantum mechanically'. Thus, in
contrast to approaches developed by particle physicists, one does
not begin with quantum matter on a background geometry and use
perturbation theory to incorporate quantum effects of gravity.
{\em There is a {\rm [background/base]}\footnote{Our addition to
make a point.} manifold but no metric},\footnote{Our emphasis.} or
indeed any other physical fields, in the background...}''
\end{minipage}

\vskip 0.1in

\noindent and Ashtekar goes on to stress further:

\bigskip \noindent (Q7.11)\hskip 0.9in
\begin{minipage}{11cm}
\noindent ``{\small ...Although there is no natural unification of
dynamics of all interactions in loop quantum gravity, it does
provide a kinematical unification. More precisely, in this
approach one begins by formulating general relativity in the
mathematical language of connections, the basic variables of gauge
theories of electro-weak and strong interactions. Thus, now the
configuration variables are not metrics as in Wheeler's
geometrodynamics, but certain spin connections; the emphasis is
shifted from distances and geodesics to holonomies and Wilson
loops. Consequently, the basic kinematical structures are the same
as those used in gauge theories. {\small\em A key difference,
however, is that while a background metric is available and
crucially used in gauge theories, there are no background fields
whatsoever now. This absence is forced on us by the requirement of
diffeomorphism invariance (or `general covariance')}.\footnote{Our
emphasis.}

This is a key difference and it causes a host of conceptual as
well as technical difficulties in the passage to quantum theory.
For most of the techniques used in the familiar Minkowskian
quantum theories are deeply rooted in the availability of a flat
background metric. It is this structure that enables one to single
out the vacuum state, perform Fourier transforms to decompose
fields canonically into creation and annihilation parts, define
masses and spins of particles and carry out regularizations of
products of operators. Already when one passes to quantum field
theory in curved space-times, extra work is needed to construct
mathematical structures that can adequately capture underlying
physics. {\em In our case, the situation is much more drastic:
there is no background metric whatsoever!}\footnote{Our emphasis.}
...}''
\end{minipage}

\vskip 0.1in

\paragraph{ADG-gravity is genuinely background independent.} By
contrast, in ADG the theory\footnote{That is, the third gauge,
third quantum (vacuum) gravitational dynamics (\ref{eqy23}).} is
not only formulated solely in terns of the gravitational
$\struc$-connection $\conn$ without at all the presence of a
metric (`purely gauge, half-order formalism'),\footnote{To be
sure, as noted earlier in section 3, one ({\it ie}, {\em we}, the
experimenters/measurers/geometers) can externally ({\it ie}, by
hand) impose an $\struc$-metric $\rho$, and then require that {\em
it} be compatible with the gravitational $\struc$-connection field
$\conn$, but this $\rho$ has nothing to do with the {\em physical}
gravitational field $\conn$ itself. One may think of $\rho$ as an
optional, auxiliary extra structure without immediate physical
meaning in ADG-gravity.} but also, {\it a fortiori}, no base
differential spacetime manifold is used whatsoever in (the
differential geometric formulation of) the theory.\footnote{That
is, it is not necessary in the theory that one assumes up-front
$\struc\equiv\smooth_{X}$ as structure sheaf.} In this sense,
which is even more `drastic' than the (still manifold-bound)
situation in loop QG,\footnote{See last sentence in (Q?.?) above.}
{\em ADG-gravity is truly background independent}. Of course, it
goes without saying that, since singularities are inherent in
$\smooth_{M}$ ({\it ie}, in the base differential manifold $M$),
loop QG (and of course other continuum based approaches to
QG)\footnote{Including string theory.} still has to reckon with
them---that is to say, they persist being problems for the theory,
hence the theory still aims at `resolving' them in one way or
another.\footnote{We will comment on this shortly in connection
with a recent `resolution' of the inner Schwarzschild singularity
achieved by loop QG means in \cite{modesto,husain}.}

\subsection{Penrose's `Singularity Manifold': Not an Issue
Whether QG Would (let alone Should) Remove Singularities}

In the beginning of this section we saw that currently
\cite{perry,gibbons,pen5} (Q?.?, Q?.?) there is hope from
theoretical physicists that QG will (or perhaps {\em should})
remove the singularities of the classical manifold and CDG-based
theory (GR). Joshi in \cite{joshi}, for example, is expressly
optimistic (even though he is aware of the manifest absence of a
full-fledged QG theory):

\bigskip \noindent (Q7.12)\hskip 0.9in
\begin{minipage}{11cm}
\noindent ``{\small ...Now, if one takes the quantum fluctuations
of the spacetime geometry into account, the spacetime
singularities or the geodesic incompleteness property of the
spacetime is not invariant under such metric perturbations.
Quantum fluctuations in spacetime geometry will change the
geodesic completeness properties of the spacetime and {\em it may
be possible to envisage a scenario in which, in spite of our not
having a full quantum gravity theory as yet, the classical problem
of spacetime singularities might be resolved}\footnote{Our
emphasis.}...}''

\end{minipage}

\vskip 0.1in

\noindent and similarly, Ashtekar also expressed hope for the
potential `resolution' of singularities by QG in the following
words taken from \cite{ash0}:\footnote{All emphasis below is
ours.}

\bigskip \noindent (Q7.13)\hskip 0.9in
\begin{minipage}{11cm}
\noindent ``{\small\em ...Returning to the singularities of
general relativity, the hope is that, in quantum gravity, a
similar interference would occur between probability amplitudes
for various space-time geometries and prevent the occurrence of
infinities...}''

\end{minipage}

\vskip 0.1in

\noindent On the other hand, we may consider the following
exchange taken from the relatively recent debate between Hawking
and Penrose about the `nature of spacetime'
\cite{hawk2}:\footnote{Again, in Penrose's reply below, emphasis
is ours.}

\bigskip \noindent (Q7.14)\hskip 0.9in
\begin{minipage}{11cm}
\noindent ``{\small\bf Question:} {\small Do you think that
quantum gravity removes singularities?}

\vskip 0.1in

\noindent {\small\bf Penrose:} {\small\em I don't think it can be
quite like that...A true theory of quantum gravity should replace
our present concept of spacetime at a singularity. It should give
a clear-cut way of talking about what we call a singularity in
classical theory. It shouldn't be simply a nonsingular spacetime,
but something drastically different\footnote{Our emphasis.}...}''
\end{minipage}

\vskip 0.1in

\noindent We maintain that Penrose is talking about a drastically
different, from the usual notion, `{\em singularity manifold}' and
not merely about one from which singularities are just surgically
excised in an {\it ad hoc} fashion---in his own words, ``{\em
simply a nonsingular spacetime manifold}'' with singularities
removed by `theoretical fiat'.\footnote{Recall Hawking's words,
also taken from \cite{hawk2}, in (Q?.?).} In other words, a
genuine theory of QG should be clear and definitive about the
notion and role of singularities in GR---as it were, it should be
clear about their origin, their (physical) meaning and their
potential utility,\footnote{That is to say, their physical
significance and implications, if any.} while at the same time it
should point out precisely to the reasons for their `anomalous'
status in the smooth manifold and CDG-based GR. On the whole, one
could say that QG should delimit sharply the latter's domain of
consistency and applicability, thus make plain why singularities
are (differential geometric) anomalies and in what sense they are
physically significant, if they are at all.

Ultimately, and since the sought after QG is supposed to shed more
light on, or even remedy completely, the problem of singularities
of GR \cite{ash0,joshi}, we hope that by the ADG-theoretic means
employed herein to evade singularities we have on the one hand
{\em given a clear-cut way of talking about what we call a
singularity in the classical theory},\footnote{And it must be
emphasized once more that {\em there is no well defined notion of
singularities in the classical theory} (GR)
\cite{geroch,clarke4}!} and on the other, formulate (even if
indirectly) the notion of a `singularity manifold'. Indeed,
ADG-gravity has highlighted and clarified precisely these two
points, namely that,

\begin{enumerate}

\item Singularities, at least in their most clear-cut perception (and `definition') so far as DGSs, are
inherent in $\smooth_{M}$, and as a result, they are the intrinsic
`faults' of the base (spacetime) manifold-effectuated CDG. They
appear to be `incompleteness' or even `breakdown' points of
Einstein's gravitational field law of GR simply because the latter
is modelled differential geometrically ({\it ie}, as a
differential equation) by the smooth manifold based CDG-means,
which in turn carry (in the background $M$) the seeds of their own
inapplicability and destruction in the guise of
$\smooth$-singularities.

\item Concerning the notion of `singularity manifold', from the our ADG-theoretic perspective
this is understood as pertaining to a realm which is not just a
differential spacetime manifold with its singular points either
being surgically excised in an {\it ad hoc} fashion so as to
retain classical differentiability and regular (smooth) laws of
physics in the remaining `effective manifold',\footnote{Thus abide
by the aforesaid `manifold conservatism and monopoly'.} or just
being pushed to the boundary of its regular points again in order
to retain the usual laws in the smooth spacetime `interior' (bulk)
\cite{clarke3,clarke4}, but {\em one that replaces our present
concept of spacetime at a singularity and is drastically different
from simply a nonsingular spacetime} (manifold). Indeed, quite on
the contrary, in the fundamentally base manifoldless ADG-gravity a
`singularity manifold' can be as pathological-looking (always from
the CDG-viewpoint) as Rosinger's spacetime foam dense singularity
manifolds we encountered earlier, yet we saw that the
gravitational field law defined by $\conn$ still holds at their
very presence.

\item Thus, {\it in summa}, in the following way ADG delineates clearly what is a
singularity in the classical theory: {\em a singularity}---at
least, a DGS---{\em is an internal blemish of $M$, ultimately, a
genetic shortcoming of CDG, coming from our identifying `physical
spacetime' with a base differential manifold. All singularities
are in a strong sense `coordinate' ones, yet, by being absorbed
into $\struc$, while we are still able to retain (by ADG-means)
all the essentially algebraic differential geometric mechanism
that is not at all dependent on a geometrical background manifold
(`spacetime'), they present no obstruction or breakdown sites to
the gravitational dynamics (Einstein equations) defined by the
gravitational field $\conn$}.\footnote{Singularities are
`transparent' to $\conn$.} GR, as a {\em physical} theory
`defined' by the gravitational dynamics (Einstein equations), by
no means breaks down at singularities, only its mathematical
scaffolding---the manifold supported CDG---collapses due to its
own faults (singularities). The ``{\em present concept of
spacetime at a singularity}'' is replaced by the `holistic',
`unitary' concept of the ADG-field $(\modl ,\conn)$ having
integrated or incorporated singularities of any kind into $\struc$
(or equivalently, into the associated sheaf $\modl$ that {\em we}
employ to represent $\conn$). Moreover, {\it prima facie} no
(formal process of) `quantization' (of the classical theory) is
evoked to be able to deal with (the) singularities (of the
classical) theory, while at the same time ADG-gravity inherently
({\it ie}, `by construction' or `by virtue of its fundamental
building blocks'---the fields $(\modl ,\conn)$) possesses quantum
features. {\em ADG-gravity is quantum from the very start}.

\end{enumerate}

\noindent In view of this `self-quantumness' of ADG-gravity, in
the next subsection we question and criticize many attempts in the
past to arrive at a conceptually sound and calculationally
consistent (finite) QG not only as a straight-out quantization of
GR along QFTheoretic lines while still retaining a background
differential manifold (for DG's sake), but also by attempts to
quantize directly spacetime itself.\footnote{In general terms,
these two attempts may be coined in bundle-theoretic jargon, `{\em
quantization of the fiber}' (while still retaining a base
continuum) and `{\em quantization of the base}', respectively.}
Our basic criticism revolves about the following ADG-gravity
motivated question: {\em in view of the fact that an external
`spacetime'}---whether `continuous' or `discrete'---{\em plays
absolutely no role and therefore has no physical significance
whatsoever in the 3rd gauged, 3rd quantized ADG-gravity, which
relies solely on the purely algebraic, `auto-dynamical'
(`synvariant') gravitational $\struc$-connection field $\conn$,
what right do we have (or more mildly put, of what relevance is to
us) to attempt to quantize spacetime in order to formulate a
quantum theory of gravity?} The point here is that ADG-gravity
achieves a quantum theoresis of the gravitational field itself,
without the involvement of any external `{\em spacetime
paraphernalia}', which are thus not begging for any `quantization'
whatsoever.\footnote{Also, in ADG-gravity the traditional
classical/quantum distinction inevitably loses its meaning, and so
does the formal Correspondence Principle (CP) which typically is
thought of as `undoing quantization'. Accordingly, the Planck
length should also be subjected to criticism (see below).}

As a contrasting warmup to the next subsection, we bring forth
some recent words of Baez from \cite{baez4}:

\bigskip \noindent (Q7.15)\hskip 0.9in
\begin{minipage}{11cm}
\noindent ``{\small ...Here I would like to propose another
possibility, namely that {\em quantum theory will make more sense
when regarded as part of a theory of spacetime}\footnote{Baez's
emphasis.} ...}''
\end{minipage}

\vskip 0.1in

\subsection{Whence Quantization of Spacetime for QG? Trying to Fit an Elephant (GR) Into a Flea's Pyjamas (QFT=SR+QM)
in the Presence of a Chimera: the Background Spacetime Continuum}

We begin our critique by commenting first on various attempts at
`{\em quantizing the gravitational field-fiber while retaining a
base geometrical differential manifold}'.

\subsubsection{Attempts at quantizing the fiber: arriving at QG by emulating QFT?}

Let us first recall some general doubts that Einstein had (as
early as prewar times!) about the application of QFTheoretic ideas
to GR. By now it has been well established that, for Einstein, the
essential, `defining' so to speak, feature of his unitary field
theory research programme would be {\em some sort of
generalization of the PGC of GR}---in any case, the discovery of a
logically simple unifying principle underlying both matter and
gravity---and certainly not a `regress' to a fixed background
spacetime (geometry), even if quantum principles were to be
suitably accommodated, as is the case in the nowadays flat quantum
field theories of matter and, {\it in extenso}, in various
attempts at formulating QG abiding by principles and constructions
borrowed from QFT on the Minkowski space continuum.
Characteristically, Stachel remarks in \cite{stachel1}:

\bigskip \noindent (Q7.16)\hskip 0.9in
\begin{minipage}{11cm}
\noindent ``{\small ...Of course, one could just abandon the
dynamical view of the space-time structure, and return to the
pre-general-relativistic concept of this structure as a given,
non-dynamical one. Indeed, this is the route that
(special-)relativistic quantum field theory has taken. I need
hardly add that to Einstein such an abandonment represented not
progress but a singularly dangerous regression:\footnote{And here
Stachel quotes Einstein.} `{\small\em You consider the transition
to special relativity as the most essential thought of relativity.
I consider the reverse to be correct}'\footnote{Our
emphasis.}...}''
\end{minipage}

\vskip 0.1in

\noindent In particular, Einstein was rather critical and
suspicious towards attempts to {\em quantize} in one way or
another, by applying QFTheoretic concepts and methods (be it on a
flat or a curved background spacetime manifold),\footnote{On a
curved spacetime manifold the approach is usually referred to as
`{\em quantum field theory on a curved spacetime}' and is regarded
as a stepping stone---a `semi-classical' or `semi-quantum'
one(!)---to the full QG theory \cite{birrel,fulling}.} his
relativistic field theory of gravity (GR). Even more generally, he
was skeptical about any research endeavor that starts from a
classical theory and applies to it in a `blind', {\it ad hoc}
fashion a formal quantization procedure (`algorithm' or `recipe')
in order to arrive at a quantum description of Nature---as it
were, to obtain a formal quantum analogue (or simile!) of a
pre-established classical theory.\footnote{For example, the
nowadays so-called {\em Quantum General Relativity} (QGR) (or
perhaps more fittingly coined, {\em Quantized General Relativity})
approach to QG may be thought of as the attempt {\it par
excellence} to quantize GR by persistently retaining a classical
smooth geometrical background spacetime manifold. Of course, {\em
Canonical Quantum Gravity} is {\em the} `canonical' (pun intended)
example of such (classically based) endeavors (see Finkelstein
quotations below). Bold pioneers of such attempts to `quantum
field theorize' about GR (albeit, in particle physics sort of
ways) in order to arrive at QG are Feynman \cite{feyn2} and
Weinberg \cite{wein}.} For the attempts to quantize a pre-existing
classical field theory for example, we read from the concluding
lines of \cite{stachel8} about Einstein's gut-feeling of `{\em
no-go}' (regarding applications of QFT to GR):

\bigskip \noindent (Q7.17)\hskip 0.9in
\begin{minipage}{11cm}
\noindent ``{\small\em ...I do not believe that it will lead to
the {\rm [QG]}\footnote{Our addition.} goal if one sets up a
classical theory and then `quantizes' it}.\footnote{Our emphasis.}
{\small This way was indeed successful in connection with the
interpretation of classical mechanics and the interpretation of
the quantum facts by modification of the theory on a fundamentally
statistical basis}. {\small\em But I believe that, in attempts to
transfer this method to field theories, one will hit upon steadily
mounting complications and upon the necessity to multiply the
independent assumptions enormously}\footnote{Again our
emphasis.}...''
\end{minipage}

\vskip 0.1in

\noindent Parenthetically, independently of the project of
field-quantization (of gravity), and in view of the first sentence
in the quotation above, we recall from \cite{malrap3} Landau and
Lifshitz's description of the `paradoxical' character of quantum
theory {\it vis-\`a-vis} classical mechanics from which it was
originally derived (by means of `quantization'):\footnote{The
quotation below also appears in our last paper \cite{malrap3}.
Below, emphasis is ours.}

\bigskip \noindent (Q7.18)\hskip 0.9in
\begin{minipage}{11cm}
\noindent ``{\small\em ...Quantum mechanics occupies a very
unusual place among physical theories: it contains classical
mechanics as a limiting case, yet at the same time requires this
limiting case for its own formulation...}.''
\end{minipage}

\vskip 0.1in

\noindent To return to Einstein, his warnings for a `{\em
dangerous regression}' pertains precisely to our blind applying
the QFT-algorithm to GR in order to quantize gravity as a field
theory, albeit, persistently abiding by the spacetime continuum.
Arguably, the said `{\em dangerous regression}' may be attributed
to the following `{\em dangerous analogy}', which is based on the
paradigm of the classical (field) theory {\it par excellence}, GR:

\bigskip \noindent (R7.2)\hskip 0.9in
\begin{minipage}{11cm}
\noindent {\em Einstein's equations in the presence of matter},
$G_{\mu\nu}=\kappa T_{\mu\nu}$, {\em tempt one to infer that,
since the right hand side of the equations---which subsumes the
energy-momentum contribution of matter (sources) to (of) the
gravitational field---has been successfully quantized by using the
concepts and methods of QFT, by `symmetry' (or analogy) the same
must be done to the left hand side---as it were, to `quantize
(field theoretically the spacetime) geometry'}. For, if
matter-energy-momentum is quantum---in fact, `quantized' field
theoretically---why not try to quantize spacetime geometry as
well?\footnote{Generally speaking, this may be thought of as {\em
the} motivating analogy behind recent attempts at formulating a
so-called `quantum Riemannian geometry' \cite{ash4,vaas,ash6}.} To
apply a `quantum metaphor' to Wheeler's one-line {\it r\'esum\'e}
of GR ``{\em energy-matter here, tells spacetime geometry how to
curve there}'': {\em quantization of energy-matter here, entails
quantization of spacetime geometry there}.\footnote{In our
opinion, an even more misleading inference which is usually drawn
from the `dangerous analogy' to the classical theory above is that
{\em one should attempt to somehow quantize spacetime structure
itself}---{\it ie}, one must give a quantum description of the
smooth spacetime manifold of GR (not just of the gravitational
field that lives on it), or equivalently, and more mathematically
speaking, one could attempt to quantize the manifold based
Calculus (CDG) ({\it eg}, formulate a noncommutative kind of
differential geometry \cite{connes,connes1}). Implicit here is a
dissection or dissociation of the gravitational field-fiber from
its base spacetime, with a concomitant ascription of an
independent (physical?) reality to the latter. We will return to
comment on this `{\em dangerously misled}' above from the
perspective of ADG-gravity in the next sub-subsection as well as
in the concluding section.}
\end{minipage}

\vskip 0.1in

\noindent However, Einstein would not `bite the tempting bait'
that GR and QFT presented him. He was caustically critical
(albeit, in an uncertain, `agnostic' manner)\footnote{This
agnosticism of his, originating from his deep doubts that a
singularity-free (unitary) field-theoretic description of quantum
reality---one that {\it a fortiori} was free from the shackles of
the spacetime continuum, could actually be formulated. (In the
next section we will discuss in detail, in the light of ADG and
\cite{stachel1}, this `second', `anti-continuous field theory on
the spacetime continuum', facet of Einstein.)} of the attempts of
his contemporaries at quantum field theorizing `non-general
covariantly' about the generally covariant GR, as we read again
from \cite{stachel8}:

\bigskip \noindent (Q7.19)\hskip 0.9in
\begin{minipage}{11cm}
\noindent ``{\small\em ...Contemporary physicists do not see that
it is hopeless to take a theory that is based on an independent
rigid space (Lorentz-invariance) and later hope to make it
general-relativistic (in some natural way)...}''

{\normalsize\rm and}

``{\small\em ...I have not really studied quantum field theory.
This is because I cannot believe that special relativity theory
suffices as the basis for a theory of matter, and that one can
afterwards make a non-generally relativistic theory into a
generally relativistic one. But I am aware of the possibility that
this opinion may be erroneous\footnote{In both quotes above
emphasis is ours.}...}''

\end{minipage}

\vskip 0.1in

\noindent More recently, and in the context of canonical
quantization (of gravity), Finkelstein too has aired similar
doubts about quantizing (canonically) a classical theory (like
GR). In \cite{df1} he notes for instance:

\bigskip \noindent (Q7.20)\hskip 0.9in
\begin{minipage}{11cm}
\noindent ``{\small ...Canonical quantization works because the
Poisson bracket and its commutator replacement have the same
meaning...Indeed, commutator relations become Poisson bracket
relations as $\hbar\rightarrow
0$}...{\rm\small[However,]}\footnote{Our addition for
text-continuity.} {\small in relativity there are informal
understandings about how to correct a non-relativistic theory to
make a relativistic one that has more chance of working. After we
formulate the relativistic theory we do not need the
pre-relativistic one in principle, and can discard it and the
relativization rule. They are part of history, not part of the
theory.}

{\small\em Canonical quantization too is not part of the theory
but only a scaffolding to use while repairing the theory and to
discard after its work is done. There is no fundamental classical
system underlying any quantum system.}\footnote{Our emphasis.}
{\small The quantum commutator relations are more fundamental than
the classical Poisson brackets, which are approximate consequences
in a suitable classical limit...}''

\end{minipage}

\vskip 0.1in

\noindent and he subsequently gives {\em physical} reasons for
{\em not} canonically quantizing a classical theory (like gravity
{\it \`a la} GR), especially when the classical spacetime
structure ({\it ie}, the manifold) supporting the latter is
viewed---as it is viewed in {\em Quantum Relativity} (QR)---as a
{\em macroscopic quantum effect} akin to superconductivity or
superfluidity:

\bigskip \noindent (Q7.21)\hskip 0.9in
\begin{minipage}{11cm}
\noindent ``{\small\em ...We should not apply canonical
quantization to spacetime structure if spacetime is a
supercondensate. Canonical quantization is a reasonable way to
reconstruct a quantum theory from its classical behavior at high
quantum numbers, but it will not recover a quantum theory from the
behavior of a low temperature supercondensate.

For example, one could not discover the helium atom by canonically
quantizing the macroscopic two-fluid theory of superfluid helium
Nor could one could discover the electron theory of solids by
canonically quantizing the field theory of the Josephson potential
of a superconductor}.\footnote{Our emphasis. In gravity too,
``{\em one would expect to find the graviton by canonically
quantizing the Einstein equations of GR no more than one would
hope to discover the fine structure of the water molecule by
straight-out quantizing the Navier-Stokes equations of
hydrodynamics}'' (David Finkelstein loosely quoted from
\cite{df2}). Let it be noted here that to much the same
conclusion, but by significantly different motivations and
arguments (issuing mainly from thermodynamics), Ted Jacobson was
also led in \cite{jacob}.} {\small Despite their macroscopic
nature, such fields are themselves best understood as {\rm
[order]} parameters of coherent quantum modes, not as limits of
normal operators (observables) as $\hbar\rightarrow 0$}.

{\small Instead {\em we suppose that spacetime structure is
already quantized}.\footnote{Our emphasis.} Some of the
macroscopic variables with which we describe spacetime and gravity
today, such as the spacetime metric tensor, already have a quantum
nature, like order parameters of a superfluid...}''

\end{minipage}

\vskip 0.1in

\noindent The fundamental `supposition' (or `motto') of QR
emphasized in the last paragraph in (Q?.?) above---namely, that
{\em spacetime structure is already quantized}---is on the one
hand a {\em conceptual or intuitive meeting point} between our
ADG-theoretic perspective on QG and QR, and at the same time on
the other, a {\em technical point of departure} of ADG from QR.
This paradoxical situation arises from the fact that while the
conceptual meeting point is that both ADG and QR basically assume
that {\em all is quantum}\footnote{For example, we have argued
(and will argue more in the sequel) that the ADG-gravitational
field $(\modl ,\conn)$ is in a strong sense `{\em already
quantum}' (`3rd-quantized').}, their (technical) ways of
implementing this fundamental intuition differ significantly, in
the following sense:

\bigskip \noindent (R7.3)\hskip 0.9in
\begin{minipage}{11cm}
\noindent While ADG fundamentally assumes that `{\em all is
field}'---an inherently dynamical (autodynamical) and already
quantum (3rd-quantum and 3rd-gauge gravitational) field---without
reference at all to a background spacetime (be it a continuum or a
discretum),\footnote{In fact, {\em in the algebraic ADG there is
no space(time) at all}, or if one wishes to think in such
`geometrical' (`spatio-temporal') terms, {\em `spacetime geometry'
is already inherent in the dynamical gravitational field} (for
more on this, see below).} which thus does not exist at all (in a
physical sense)\footnote{That is to say, as a {\em dynamical}
entity.} in the theory, QR assumes that `{\em all is spacetime}'
structure ({\it eg}, topology)---an already quantized spacetime
structure ({\it eg}, quantum topology or quantum
geometry).\footnote{At this point we should stress again our
preference for the expression `{\em already quantum}' instead of
the `{\em already quantized}' used in (Q?.?) above, for if {\em
all is quantum ab initio}, presumably it has no need of being
`quantized' (even if the latter word is used `philologically' to
distinguish it all from `fictitious' non-quantum entities, which
anyway do not exist in Nature). All in all, the following question
is begging the question: {\em how does one quantize the quantum?}
In any case, in the next sub-subsection we will criticize head-on,
with ADG-gravity in hand, any attempt at quantizing spacetime
itself.}
\end{minipage}

\vskip 0.1in

\noindent In the next sub-subsection we criticize the aforesaid
`point of departure' ({\it ie}, {\em quantum spacetime structure
instead of quantum field}), and then we comment on the `meeting
point' ({\em all is quantum}), always from an ADG-theoretic
vantage.

But before we go on to critically comment on various attempts to
quantize the base spacetime itself, let us close the present
`quantization of the field-fiber' sub-subsection by mentioning the
currently most plausible and potentially successful of all
attempts (so far) at quantizing GR as a field theory; albeit,
still manifestly abiding by a base differential manifold. This is
the so-called {\em Quantum General Relativity} (QGR) scenario, and
in particular, its approach via {\em Loop Quantum Gravity} (LQG)
and its recent offspring, {\em Quantum (Riemannian) Geometry}
(QRG)
\cite{rovsm,ash1,thiem2,thiem3,smolin,ash2,ashlew3,ashlew4,ash4,ash6,ash5,rovelli}.\footnote{This
is not meant to be a comprehensive array of references on QGR, LQG
and QRG, but the citations in \cite{thiem2,thiem3,smolin,rovelli}
for example constitute pretty much a complete list.} Of course, it
goes without saying that this is not a place for us to review this
approach to QG, but we would like to highlight certain basic
conceptual features and results of it so as to set a platform for
a brief comparison with ADG-gravity:\footnote{Many of the items in
the following list have already been mentioned and discussed in
the present paper, as well as in our past trilogy
\cite{malrap1,malrap2,malrap3} and the more recent paper
\cite{rap5}.}

\begin{itemize}

\item LQG approaches QG as a quantum {\em gauge theory}, since the
basic gravitational variables involved in it are Ashtekar's `new
variables': spin-Lorentzian ($SL(2,\com)$-valued) connections
\cite{ash}.\footnote{Arguably, the original interest in Ashtekar's
new gravitational variables was due to the fact that the latter
simplify significantly the (Dirac) constraints when gravity (GR)
is treated canonically.}

\item At the same time, the {\it vierbein} (comoving tetrad frame)
field is also another set of fundamental variables (alongside the
connection) so that the formalism is closely akin to Palatini's
so-called first-order formalism whereby, as noted earlier,
variation of the corresponding action functional with respect to
the tetrad field yields the Einstein equations, while variation
with respect to the connection field yields the metric
compatibility (torsionless) condition for the latter.

\item As also stressed earlier, although LQG has managed to arrive at a genuinely
background metric independent formulation of (canonical) QGR,
still heavily relies on a base differential manifold for its
(differential geometric) formulation,\footnote{This means that the
aforesaid connection and tetrad fields are {\em smooth}.} with all
the well known technical and conceptual problems that come along
with this dependence ({\it eg}, the inner product/quantum measure
problem, the problem of time, the problem of observables {\it
etc}).

\item On the other hand, since LQG still employs a base differential manifold, the
problem of singularities in the classical theory of gravity (GR)
still persists\footnote{For after all, as we argued extensively in
this paper, the culprit for the $\smooth$-smooth gravitational
singularities in the manifold based GR is our assumption of
spacetime as a differential manifold $M$. Singularities are
inherent in $M$---as it were, they are innate in $\smooth_{M}$.}
and has to be dealt with in one way or another. Indeed, as we
shall see shortly, LQG `resolves' singularities by quantizing the
base spacetime itself, via its offspring theory coined QRG. Thus,
broadly speaking, LQG belongs to that general category of
approaches to QG that maintains (or better, hopes) that a cogent
quantum theoresis of gravity will resolve (or even ultimately, do
away with) singularities (Q?.?); moreover, this could be achieved
by a quantization of spacetime itself---as it were, as we shall
see soon, the quantization of the (base) spacetime, which is {\it
a priori} assumed to be a classical smooth continuum, is an
`outcome' (or `byproduct') of the quantization of the
gravitational field (fiber).

\end{itemize}

\noindent The last remarks hint at to the next sub-subsection
where we comment on attempts to quantize (or arrive at a quantum
theory of) the base spacetime itself. The general philosophy there
is that since gravity is usually associated (geometrically) with
the structure of spacetime, quantum (or a quantization of) gravity
could not be thought of apart from quantum (or a quantization of)
spacetime structure.

\subsubsection{Attempts at quantizing the base: the spacetime
continuum stands in the way towards quantizing gravity as a field
theory, so why not quantize it first?}

If a `direct' quantization---canonical (Hamiltonian), covariant
(Lagrangian, action based) or other---of GR, regarded as a
continuous field theory on the smooth spacetime continuum, fails
for one reason or another, the next tenable position would be to
attempt to quantize spacetime structure itself. That is to say,
one could suppose for instance that since the `second quantization
algorithm' fails to yield when applied to the field-inhabited
fibers themselves, one should perhaps try to search for (or
better, ascribe!) quantum traits such as discreteness,
noncommutativity {\it etc} in (to) the classical base spacetime
manifold itself---the domain of definition of the said (continuum)
fields. In other words:

\bigskip \noindent (R7.4)\hskip 0.9in
\begin{minipage}{11cm}
\noindent {\em If the (gravitational) field-fiber (over the
spacetime continuum)\footnote{For instance, in bundle-theoretic
terms, one think of the spacetime metric representing the
gravitational field as a (local) cross-section of the smooth
symmetric tensor bundle over the spacetime manifold $M$, acting
(as a real symmetric bilinear form) on pairs of local sections of
the tangent bundle on $M$ (vector fields).} appears to
persistently resist quantization, then why not try to quantize
directly the base spacetime manifold itself?}

\end{minipage}

\vskip 0.1in

\noindent Finkelstein {\it et al.} for example have recently
stated this apparent need for quantizing the spacetime base
explicitly and succinctly in \cite{dfetal}:

\bigskip \noindent (Q7.22)\hskip 0.9in
\begin{minipage}{11cm}
\noindent ``{\small ...{\em Quantizing gravity without quantizing
space-time introduces yet a further fragility}.\footnote{Our
emphasis. `Fragility' in \cite{dfetal} is a term with technical
group-theoretic connotations that we should not be bothered about
here. The reader is referred to the paper for its explanation.}
The canonical group of Einstein gravity is much bigger than the
classical diffeomorphism group $\mathrm{Diff}$, for it also
includes transformations that couple field variables into the
space-time coordinates, like de Donder or harmonic coordinates,
and $\mathrm{Diff}$ does not. In the quantum theory the
corresponding space-time coordinates will inherit the
non-commutativity  of the field variables. It is fragile to
postulate that a frame exists in which the space-time coordinates
commute with the field variables...}''
\end{minipage}

\vskip 0.1in

\noindent The quantization of spacetime itself may also be
alternatively motivated as well as be implemented in many
different ways issuing from ideas grounded in the so far three
main approaches to QG, namely, the Euclidean QG approach, the
Ashtekar (LOG and QRG) programme and superstring theory
\cite{ish10}. In the last reference, Isham discusses the issue of
whether the basic, essentially classical spacetime continuum
based, formalism of quantum mechanics can be suitably modified in
a way so as to suit the theoretical possibility that spacetime
itself be quantized. Characteristically, we quote him saying:

\bigskip \noindent (Q7.23)\hskip 0.9in
\begin{minipage}{11cm}
\noindent ``{\small ...The Euclidean programme, the Ashtekar
scheme, and superstring theory all use what are, broadly speaking,
standard ideas in quantum theory. In particular, they work with an
essentially classical view of space and time---something that,
arguably, is a {\em prerequisite}\footnote{Isham's emphasis.} of
the standard quantum formalism. {\em This raises the important
question of whether quantum theory can be adapted to accommodate
the idea that spacetime itself ({\it i.e.}, not just the metric
tensor) is subject to quantum effects: surely one of the most
intriguing challenges to those working in quantum
gravity}.\footnote{Our emphasis.}

(New Section 5.2) {\bf Quantising Space-time.} {\em Some of the
many issues that arise can be seen by contemplating how one might
try to quantise spacetime `itself' by analogy with what is done
for---say---the simple harmonic oscillator, or the hydrogen atom.
Of course, this may be fundamentally misguided... Nevertheless, it
is instructive to think about the types of problem that occur if
one does try to actively quantise spacetime itself---if nothing
else, it reveals the rather shaky basis of the whole idea of
`quantising' a given classical structure}\footnote{Again, our
emphasis.}...}''
\end{minipage}

\vskip 0.1in

For an instance of a reason why one might attempt to quantize
spacetime itself, the appearance of a fundamental, `minimum'
space(time) length---the so-called Planck length $\ell_{p}$ (and
correspondingly, the Planck time $t_{P}$)---when (the physical
constants of) SR ($c$), GR ($G$) and QM ($\hbar$) are envisioned
to be combined in the (still uncharted) QG domain, could be taken
as suggesting a natural regularization spacetime continuum cut-off
in a covariant, path-integral quantization of gravity for example.
Similarly, in view of (R6.?) above, one could also suppose that
some sort of {\em quantum spacetime geometry}, being involved in
the left hand side of the `quantized' Einstein equations, should
be formulated
\cite{ashrovsm,ash2,ash3,ashlew3,ashlew4,ash4,ash6}.\footnote{The
last two references give a thorough account of the (primarily
conceptual) development of this LQG-offshoot approach to QG, QRG.}
With respect to the LQG approach to non-perturbative (canonical)
QGR for instance \cite{rovsm,thiem2,thiem3,smolin}, a head-on
quantization of spacetime has led to discrete expressions
(`eigenvalues') for the spatial length, area and volume quantum
operators \cite{rovsm1,ashlew3,ashlew4,loll1,thiem1}, while the
quantization of volume also entails indirectly a quantization of
time, since the Hamiltonian (constraint), which is supposed to
generate the dynamical time-evolution in canonical QGR, can be
expressed in terms of the volume operator having a discrete
eigen-spectrum \cite{modesto,husain}.\footnote{We will comment
shortly on how this spacetime quantization result enables one to
dynamically extend spacetime past the inner Schwarzschild without
encountering any infinity (divergence) of the Ricci tensor. In
this {\em indirect} way---{\it ie}, via spacetime
quantization---LQG claims to `resolve' the interior Schwarzschild
singularity \cite{modesto,husain}, much like we achieved
herein---albeit, {\em directly}, without having to quantize (an
{\it a priori} assumed background differential manifold)
spacetime, which anyway does not exist {\it ab initio} (`by
definition') in ADG-gravity.} Vaas \cite{vaas} has wrapped-up the
basic achievement of LQG and QRG as follows:

\bigskip \noindent (Q7.24)\hskip 0.9in
\begin{minipage}{11cm}
\noindent ``{\small ...This is how quantum geometry revolutionizes
our world-view: {\em Space is quantized like matter!}\footnote{Our
emphasis. For some recent ideas towards a head-on quantization of
time, the reader can refer to \cite{bostroem}.}...}''
\end{minipage}

\vskip 0.1in

\noindent and, remarkably, he then continues and stresses the
fundamentally relational and {\it a priori} spacetimeless
underpinnings of LQG, which are very close (at least in spirit) to
the dynamically `autonomous', algebraic and background
spacetime-free ADG-gravitational fields in which `spacetime' is
inherent:\footnote{And its spin-network/spin-foam relatives---see
for instance \cite{baez1,perez,barrett}.}

\bigskip \noindent (Q7.25)\hskip 0.9in
\begin{minipage}{11cm}
\noindent ``{\small ...The question why nothing can be made to fit
into half the volume of the smallest unit of space is meaningless
in view of such `space atoms'. It is based on the incorrect
assumption of an absolute space in which things fit in... Space
and time are not at all fundamental, but rather built up from more
basic structures [graph-like entities called `spin
networks']\footnote{Our addition for continuity.}...There is
`nothing' between these graphs. Those entities rest only on
themselves, so to speak. `{\em The spin networks do not exist in
the space. Their structures produce the space,}' Smolin stresses.
`{\em And they are nothing but abstractly defined
relations...}'\footnote{Our emphasis of Smolin's quotation.}...}''
\end{minipage}

\vskip 0.1in

\noindent while, very recently, Smolin \cite{smolin} highlighted
the main achievement of LQG as follows:

\bigskip \noindent (Q7.26)\hskip 0.9in
\begin{minipage}{11cm}
\noindent ``{\small ...The result [from the past 17-years' work on
loop QG] is a language and dynamical framework for studying the
physics of quantum spacetimes, which is completely consistent with
the principles of both general relativity and quantum field
theory. The picture of quantum spacetime geometry which emerges is
to many compelling, independently of the fact that it has been
derived from a rigorous quantization of general relativity. The
basic structure that emerges is of a new class of quantum gauge
field theories, which are background independent, in that no fixed
spacetime metric is needed to describe their quantum dynamics.
Instead, the geometry of space and time is coded in the degrees of
freedom of a gauge field\footnote{Like ADG-gravity, LQG regards QG
as a quantum gauge theory---one that fundamentally hinges on the
notion of connection. However, as stressed earlier, in LQG and
QRG, unlike in ADG-gravity, although a fixed background metric is
not used, a background smooth manifold that supports the said
gravitational connection-fields {\em is}.}...}''
\end{minipage}

\vskip 0.1in

\noindent Similarly, in Finkelstein's QR there appears to be a
fundamental fermionic `{\em quantum of time}'---the so-called {\em
chronon}. A chronon represents an elementary quantum causal
(dynamical) process. An ensemble of chronons and their mutual
interactions (interrelations) is then supposed to weave (define) a
finitistic (`discrete'), causal and quantal network (relativized
and quantized directed graph topology) which replaces the fixed or
`frozen' smooth Lorentzian spacetime manifold of the classical
theory (GR) and recovers it as a superconducting phase (`coherent
vacuum state'). In turn, the latter corresponds to a macroscopic
quantum effect resulting from a coherent Bose-Einstein type of
quantum condensation of the said elementary quantum causal
processes (chronon Cooper pair-like coherent quantum
agglomerations) \cite{df2,df6,df5,df1,df3,sel1,sel2,sel3,sel4}.

There is also another `discrete' approach to Lorentzian QG, the
{\em causal set (causet) theory} due to Sorkin {\it et al.}
\cite{bomb87,sork4,sork-1,sork1,sork2,sork3,sork10,sork9}. Causets
are locally finite models of spacetime in which what is solely
retained from the relativistic Lorentzian continuum of events is
the causal or chronological precedence relation between them,
while in turn the classical spacetime manifold of GR is regarded
as a coarse (`statistical') approximation of the causet, which is
thus viewed as the fundamental structure in the quantum deep.
Causets as such are not quantum. A classical sequential growth
(stochastic) dynamics has already been proposed for them in
\cite{ridesork}, while a quantum (stochastic) dynamics is
currently pursued along quantum measure-theoretic lines
\cite{sork6,sork7,sork8}.\footnote{Loosely speaking, here one is
looking for a decoherence functional-type of object on the
kinematical space of `kinematically consistent' ({\it ie},
discrete general covariance and Bell causality obeying) causet
histories (and their `observables').}

More on the mathematical side, one could also envision an indirect
quantization of the classical smooth base spacetime manifold by
quantizing not the continuum itself, but, as it were, the
(differential) Calculus (CDG) which is based on it. This was
essentially the aim of Connes' celebrated {\em Noncommutative
(Differential) Geometry} (NDG), which, as noted earlier, was
achieved primarily by functional-analytic means, while arguably a
differential manifold is still somehow retained in the background
\cite{connes,kastler,connes1}.

On the other hand, in view of all that has been said and achieved
in the context of ADG-gravity, we come to ask:

\bigskip \noindent (R7.5)\hskip 0.9in
\begin{minipage}{11cm}
\noindent {\em If (the) spacetime (continuum) does not physically
exist in the first place, already at the classical level of
theoresis of gravity (GR)}---that is to say, like the fathers of
GR had intuited and warned, `{\sl it corresponds to nothing {\rm
[physically]} real, being a mode by which we think, not a
condition in which we live---our mental creation so to speak}'
(Einstein Q?.?, Eddington Q?.?), and certainly not an element of
Physis---{\em isn't it begging the question to {\it ad hoc} (as it
were, by `mathematical fiat') posit its existence\footnote{In
order to actually apply CDG-ideas to Physics.} and then, moreover,
try to somehow quantize it on top?

All in all, aren't we creating (mathematical) models, subsequently
raising them to idols if they prove to be successful,\footnote{And
the (spacetime) continuum, with the Calculus that comes hand in
hand with it, has undoubtedly been the most successful (regarding
physical applications) mathematical structure that we have
invented so far.} while at the same time we forget their
`terrestrial origin' ({\it ie}, the fact that they are our own
`fictions' in the first place) and regard them as `necessary
truths',\footnote{See the Einstein quotation next.} only to put an
inordinate amount of effort (mental and material resources!) into
trying to shoot them down when they actually prove to be
problematic and pathological ({\it eg}, singularities in QG and
the general miscarrying of the spacetime continuum in the quantum
deep); and moreover, due to our aforementioned `forgetfulness', we
tend to attribute these pathologies to Physis when the problems
simply expose the fact that our creations are of limited
applicability and validity, certainly not Nature's own
shortcomings and inadequacies?}
\end{minipage}

\vskip 0.1in

\noindent Einstein's words come once again to wake us up to this
reality---the `{\em Golem unreality of spacetime}':\footnote{The
following quotation concluded our last paper \cite{malrap3}.}

\bigskip \noindent (Q7.27)\hskip 0.9in
\begin{minipage}{11cm}
\noindent ``{\small\em ...Concepts {\rm [like the spacetime
continuum]}\footnote{Our addition for making our point clearer.}
which have proved useful for ordering things easily assume so
great an authority over us, that we forget their terrestrial
origin and accept them as unalterable facts. They then become
labelled as `conceptual necessities', `a priori situations', etc.
The road of scientific progress is frequently blocked for long
periods by such errors. It is therefore not just an idle game to
exercise our ability to analyse familiar concepts, and to
demonstrate the conditions on which their justification and
usefulness depend, and the way in which these developed, little by
little\footnote{Our emphasis throughout.}...}'' (1916)
\cite{einst7}
\end{minipage}

\vskip 0.1in

\noindent In view of the third gauge (`pure gauge'), third quantum
(`self-quantum') field-theoretic underpinnings of ADG-gravity that
we have in hand, the critical remarks above about quantizing the
base spacetime structure itself acquire further significance. The
bottom-line is that ADG not only manages to model key ideas from
second and geometric (pre)quantization, such as the quantum
particle interpretation/geometric representation ($\modl$) of the
gravitational field ($\conn$) and its (sheaf cohomological)
classification as a boson (`graviton': particle mediating the
quantum gravitational force; line sheaf of representation) and/or
a fermion (`causon': the `matter'-like quantum of causality acting
as a `source' for the said `graviton'; vector sheaf of
representation), but also unlike the conventional manifold based
(fiber bundle-theoretic) ideas and techniques of second (QFT) and
geometric (pre)quantization \cite{simms,wood,sel,bandy}, it
manages to do this without reference to a background
`space(time)', whether the latter external, `ambient realm' is
taken to be a `continuum' or a `discretum'.

This freedom from a base spacetime that ADG gives us about
infusing QFTheoretic ideas to QG research cannot be
overemphasized. Especially due to the aforesaid faring poorly (so
far) of attempts to arrive at a cogent QG by applying QFTheoretic
ideas and techniques to GR by retaining a base differential
manifold ({\it eg}, the non-renormalizability of gravity when
regarded as a perturbative quantum gauge theory like the other
three fundamental forces, or the $\mathrm{Diff}(M)$-associated
problems in non-perturbative quantum gauge-theoretic scenaria for
canonical QGR such as LQG), the base spacetimeless third
quantized, third gauged ADG-gravity may prove to be a fruitful
route indeed to QG---a route bypassing directly all the problems
that the said approaches encounter due to their assuming a smooth
base spacetime continuum. Moreover, and this should be emphasized
here,

\bigskip \noindent (R7.6)\hskip 0.9in
\begin{minipage}{11cm}
\noindent  in glaring contradistinction to the other external (to
the gravitational field itself) spacetime manifold based (quantum
gauge-theoretic) scenaria for QG (such as LQG), {\em we do not
expect `spacetime' to be also quantized---that is, we do not
regard the `problem' of the quantum structure of spacetime as
being `important', that is, as being inextricably entwined with
the problem of QG---simply because from the start there is no
background spacetime in our ADG-theoresis of gravity and the
ADG-gravitational field is `already quantum'}.
\end{minipage}

\vskip 0.1in

\noindent {\it En passant}, let it be also noted here that for
other approaches to QG, the attainment of a quantum description of
spacetime structure {\it per se} is supposed to be prior to---in
fact, an apparently necessary stepping stone to---a genuinely
quantum theoresis of gravity
\cite{df0,df2,df6,df5,df3,df1,sel1,sel2,sel3,sel4,ish10,ish5,ish6,ish7,ish8,df4,dfetal}.
Still in further contrast to our spacetimeless ADG-musings on
gravity, there are certain theoretical schemes that focus solely
on a finitistic and quantum theoresis of spacetime (or anyway, of
`space') itself \cite{pen-1,pen,brody}.

These critical, `contrasting' remarks will be of significance in
the next sub-subsection when we comment on a recent `resolution'
of the interior Schwarzschild singularity by LQG techniques and
results \cite{modesto,husain}, and we compare it with our
fundamentally base spacetimeless ADG-theoretic evasion presented
above.

\subsubsection{A recent `resolution' of the inner Schwarzschild
singularity and its comparison with our ADG-evasion}

As noted earlier, all the differential manifold and {\it in
extenso} CDG-based approaches to QG will inevitably have to reckon
with the problem of smooth spacetime singularities in the
classical theory (GR), simply because singularities are `inherent'
in the background differential manifold employed in GR. Thus,
arguably, all these approaches hope (or even expect!) that a
cogent quantum theory of gravity will (eventually) `resolve' (or
even ultimately remove!) singularities in some way.

Among these QG approaches, and to that a very promising one(!), is
LQG\footnote{Together with string theory, LQG has proven to be so
far the most successful approach to (non-perturbative) QG.}
\cite{thiem2,thiem3,smolin}. Notably, within the past half decade,
in the context of loop quantum cosmology---the application of LQG
ideas and results to quantum cosmology, it has been shown that the
initial (`Big Bang') singularity `predicted' by GR can be indeed
`resolved' or `bypassed'
\cite{boj1,boj2,ash3}.\footnote{Coincidentally, at about the same
time as the loop quantum cosmology result above, in the context of
the string theory approach to QG, the so-called `ekpyrotic
scenario' also claimed a successful `passing through' the original
cosmological singularity \cite{turok,turok1,turok2}.} However,
even more significant to the present paper is the following very
recent result of Modesto \cite{modesto}, which was also arrived at
by LQG means: in a single sentence, {\em the Schwarzschild black
hole singularity of the classical theory (GR) `disappears' in
(L)QG}.\footnote{See also \cite{husain} for the wider context of
black hole singularity resolution by LQG means.} In the present
sub-subsection we would like to describe briefly this
`disappearance', comment on it and juxtapose it against the
`resolution' (better, `complete evasion') of the same singularity
({\it ie}, the interior Schwarzschild one) that was accomplished
above by ADG-theoretic means.\footnote{As noted in \cite{rap5},
this comparison should by no means be taken as an attempt to
`downplay' the remarkable indeed result of Modesto, let alone to
undermine the significant import and value of LQG as a whole. The
aim of the comparison below is simply to highlight the basic
differences in general approach, principle and `attitude' of the
two `resolutions' (and {\it in extenso} of the two theories,
ADG-gravity and LQG), leaving the final word of judgment, critique
or praise to the reader and, perhaps more appropriately, to time
({\it ie}, to future developments and applications of the two
theoretical schemata).}

Let us first point out again that since, as mentioned before, LQG,
although manifestly background metric independent, still expressly
employs a base differential (spacetime) manifold for its basic
concepts and constructions, the problem of singularities in the
classical theory (GR) persists and has to be somehow dealt with in
the quantum theory. Without going into any (technical) detail, and
quite synoptically, in \cite{modesto} the inner Schwarzschild
singularity is resolved in the following steps:\footnote{The
reader is referred to \cite{modesto} for detailed arguments,
calculations and pertinent citations. However, in addition to some
of the references therein, we also provide some more relevant
references. In this respect, refer also to \cite{husain}.}

\begin{enumerate}

\item To begin with, one expresses the Ricci scalar curvature,
which as noted earlier diverges as $1/r^{6}$ near the interior
($r=0$) Schwarzschild singularity, in terms of the spacetime
volume.

\item Then, one evokes {\em the} major result in loop QG, namely,
that the said volume is quantized---{\it ie}, it is promoted to a
volume operator having a discrete eigen-spectrum.\footnote{This
volume-quantization \cite{ashlew4} is just one of a series of
significant results in Ashtekar's QRG programme, which is an
offspring of LQG \cite{ash4,ash6,thiem2,thiem3,smolin}, along with
quantization of length \cite{thiem1} and area \cite{ashlew3} (see
also \cite{rovsm1,rovelli}).} Thus, near the Schwarzschild black
hole, $\ricci$ is rendered finite and the classical continuum
infinities are controlled (in a sense, `regularized') by quantum
theory.

\item Moreover, one can show that the said `regularization' is not
`kinematical' ({\it ie}, one that is not {\it a priori} fixed by
hand),\footnote{Like for example the spacetime discretizations in
lattice QCD.} but it is a dynamical one, as the Hamiltonian
(constraint), which regulates the dynamical, time-evolution in the
canonical approach to QGR originally underlying LQG
\cite{thiem2,thiem3}, can also be expressed in terms of the volume
operator. Thus, as Modesto shows and argues in detail, {\em the
spacetime can be dynamically extended past the inner Schwarzschild
singularity, with no infinity involved at all in the process}.

\item On the other hand, from a differential geometric viewpoint,
the upshot of all this is that the said dynamical evolution, which
is classically represented by a {\em differential equation} on the
underlying spacetime continuum,\footnote{After all, the
Hamiltonian (constraint) in the classical canonical theory (GR) is
the generator of temporal diffeomorphisms of the underlying
spacetime manifold.} is now substituted, in view of the said
quantization of spacetime (geometry) in LQG, by a reticular, {\em
difference equation} (discretely parametrized by the coefficients
of the physical quantum eigenstates of the volume operator). {\it
In summa}, one can say that {\em the inner Schwarzschild
singularity is resolved thanks to the quantization-{\it
cum}-discretization of the base spacetime continuum itself}.

\end{enumerate}

\noindent Based on the brief description above, our comparison of
the two `resolutions' of the interior Schwarzschild singularity
({\it ie}, the ADG-gravity one and the LQG one) hinges on two
fundamental in our opinion differences:

\begin{itemize}

\item {\bf I.} Unlike in the LQG `resolution' where a
quantization of spacetime appears to be necessary, in the ADG
`resolution' this is not so, for the theory is `inherently
background spacetimeless' ({\it ie}, the theory is indifferent as
to whether that background is a `classical continuum' or a
`quantal discretum' \cite{malrap3,malrap4});\footnote{To stress it
once again, in ADG-gravity the quest for a quantization of the
base spacetime {\it per se} is essentially `begging the question':
{\em in the first place, in ADG, what `spacetime' is one talking
about?} Another way to say this, perhaps even more
iconoclastically, is that, {\em from the ADG-viewpoint, gravity}
({\it ie}, the dynamically `autonomous' third gauge and third
quantum gravitational field) {\em has nothing to do with
`spacetime', so that a possible quantum theoresis of the former is
in no need of a quantum description of the latter}. The
ADG-gravitational field is in no need of a background spacetime
for its (dynamical) sustenance.} and as a consequence of this
difference,

\item {\bf II.} Unlike the situation in the LQG `resolution' where the said base spacetime
quantization and concomitant discretization appears to mandate the
abandoning of the picture of `gravitational dynamical evolution'
as a differential equation proper (and, as a result, the
abandonment of differential geometric ideas in the quantum
regime), in the ADG `resolution' all the (essentially algebraic)
differential geometric machinery (of the geometrical background
spacetime continuum) is retained in full effect (manifestly
independently of that background, and {\it a fortiori}, even if
that background is a `discretum' where differential geometric
ideas would traditionally---{\it ie}, from the CDG-viewpoint of
the base differential manifold---seem to fail to apply).

\end{itemize}

\noindent This apparently necessary involvement of the process of
quantization-{\it cum}-discretization of the classical base
spacetime continuum in order to render a physical quantity (here
for example, the Ricci curvature scalar) finite (thus physically
meaningful!), is characteristic of the general intuitive
anticipation (and at times `firm expectation'!) of current
theoretical physicists working on QG that, in view of the fact
that there is a space-time scale---the so called Planck
length-time---arising from combining the fundamental constants of
relativity ($c$ from SR and $G$ from GR) and quantum ($\hbar$)
theory which are supposed to be consistently merged into the
elusive QG, below that scale the classical `infinitistic'
continuum picture of spacetime (and as a result, of the
singularities' and infinities' assailed continuous field theory,
whether classical or quantum, based on it) should be replaced by
something more `finitistic' and `quantal'---the `{\em true quantum
spacetime geometry of the genuine QG}', so to speak. Based on the
{\em theoretical} paradigm of ADG-gravity, we would like to
challenge in the next subsection this, so popular nowadays, {\em
theoretical} expectation.

But first, as a warmup to the following subsection, let us invoke
from \cite{pen6} some words of Penrose about, on the one hand the
invaluable role that the spacetime continuum (:smooth manifold)
has played in our formulation of physical theories hitherto, and
on the other, the apparent `need' to scrap it off below Planck
length ({\it ie}, in the QG regime):\footnote{In the quotation
below all emphasis is Penrose's, unless noted.}

\bigskip \noindent (Q7.28)\hskip 0.9in
\begin{minipage}{11cm}
\noindent ``{\small According to present-day theory, all the
phenomena of physics take place within the framework of a certain
differentiable manifold referred to as the {\em space-time
continuum}. Our familiarity with this idea is such that it is now
regarded as almost `obvious' that space and time should constitute
such a structure. However, before discussing the nature of this
structure, it is worth examining something of what lies behind
this belief. Indeed, there is the definite possibility that some
future theory may be found which describes nature more accurately
than present theory, but for which the differentiable manifold
picture of space-time would not be appropriate. We should not
close our minds to such a possibility, but also we should keep in
mind the extraordinary range over which the present-day view is
such an excellent approximation.

The very accurately `locally Euclidean' nature of space, and the
continuity of time, would, indeed, seem to have supplied the prime
motivation, in the first instance, for the rigorous development of
the continuum concept. At the time of Zeno, no such rigorous
concept of continuum existed, so that the idea of a limit, in
space or time, seemed puzzling. It does not seem puzzling to us
today, but perhaps we are wrong not to be puzzled! The standard
resolution of Zeno's paradoxes refers more to the {\em
mathematical} continuum concept than to the nature of space-time
itself. The view of space-time as forming a continuum would imply
that a continuous nature would persist, no matter how much a
system is magnified. But it is not at all clear that continuous
descriptions are really appropriate on a scale small enough that
quantum phenomena become important. For example, at a scale of
$10^{-33}$cm (approximately the radius of an elementary particle),
the mere attempt at localization of the position of a particle to
that accuracy will, as a consequence of the uncertainty principle,
imply the probable occurrence of a very large momentum, with the
implication that {\em new} particles are created, some of which
may be indistinguishable from the original particle. Thus the
concept of `position' for the original particle becomes obscured.
More alarming, moreover, is the picture presented if we allow
ourselves to discuss phenomena at a dimension of the order of
$10^{-33}$cm. At such a dimension, the quantum fluctuations in the
curvature of space-time (if both present-day quantum theory and
gravitation theory can be accurately extrapolated to this degree)
would be large enough to produce alterations in topology. Thus the
view of space-time at this dimension would be some kind of chaotic
linear superposition of different topologies---a picture in no way
resembling a smooth manifold.

Whether or not it is meaningful to talk about the nature of
space-time at such dimensions is not at all clear. But if it is
{\em not} meaningful, then we {\em certainly} cannot refer to
space-time as accurately consisting a smooth manifold. On the
other hand, it may be argued that the smooth manifold is {\em
adequate} for the discussion of all relevant physical processes.
It is my personal view that this cannot {\em ultimately} be the
case. {\em I do not believe that a real understanding of the
nature of elementary particles can ever be achieved without a
simultaneous deeper understanding of the nature of space-time
itself}.\footnote{Our emphasis.} But if we are concerned with a
level of phenomena for which such an understanding is {\em not}
necessary---and this will cover almost all present-day
physics---then the smooth manifold picture presents an
(unreasonably!) excellent framework for the discussion of
phenomena...}''
\end{minipage}

\vskip 0.1in

\noindent In contradistinction to Penrose, our relevant comment
here to the emphasized text above, apart from questioning the
Planck length (which we will do next), is that, in our ADG-based
view, {\em a better understanding of the nature of elementary
particles (:`field quanta') cannot be achieved without the
development of a `better' field theory---one that is not at all
dependent on a background spacetime (continuum)}.\footnote{In
anticipation of 8.5 in the sequel.} Of course, we agree with
Penrose insofar as one abides by the idea that {\em spacetime is
inherent in the dynamical fields} (and their quanta), so that a
deeper (or perhaps, a different from the current) understanding of
field theory is needed.

\subsection{Whence the Planck Length?}

To get straight to the point, we wish to open this subsection
emphatically, with an apparently `agnostic' and, as it happens,
`rhetorical' from the viewpoint of ADG-gravity question:

\bigskip \noindent (R7.7)\hskip 0.9in
\begin{minipage}{11cm}
\noindent In a purely algebraic (`relational') theory, like
ADG-gravity, where no background `geometrical' spacetime---whether
a continuum or a discretum---is involved at all in our
(differential geometric) calculations and constructions ({\it in
toto}, in our Calculus), what is the significance (and the use) of
having a supposedly fundamental space-time length like
$\ell_{P}$-$t_{P}$ (the Planck length-time)?
\end{minipage}

\vskip 0.1in

\noindent To be sure, in the spacetime continuum based QFTs of
matter, as well as in various manifold based approaches to QG, the
`utilitarian' or `pragmatic' attitude towards $\ell_{P}$-$t_{P}$
is to use it as a {\em cut-off scale} (with respect to which the
continuum is then effectively replaced by a granular structure) in
order to regularize the corresponding (ultraviolet divergent) path
integrals (for quantum matter and gauge dynamics),\footnote{Think
for instance of lattice QCD.} or even to come up with a finite
value for the black hole horizon's entropy (in QG). The result of
course, like in Modesto's LQG-based resolution of the inner
Schwarzschild singularity revisited above, is that one is `forced'
to scrap-off the picture of spacetime as a smooth continuum and,
inevitably, with it also abandon the applicability of CDG-ideas to
QG. We corroborate this by reading from \cite{sork0}:

\bigskip \noindent (Q7.29)\hskip 0.9in
\begin{minipage}{11cm}
\noindent ``{\small ...That matter on the smallest scales sheds
its continuous nature is indicated by several features of
present-day physics. In particular, {\em the short-distance
`cut-offs' required (apparently) by both quantum field theory (to
`regularize' the functional integral) and `quantum gravity' (to
render black hole entropy finite) seem ultimately foreign to the
notion of differentiable manifold embodied in classical general
relativity. Their stubborn presence suggests, rather, that there
is a discrete substratum underlying spacetime and accounting
naturally for the appearance of a minimal length in the effective
theories we now possess}\footnote{Our emphasis.}...}''
\end{minipage}

\vskip 0.1in

\noindent Moreover, in more-or-less the same line of thought, and
exclusively in the context of QG, we read from \cite{ash0}:

\bigskip \noindent (Q7.30)\hskip 0.9in
\begin{minipage}{11cm}
\noindent ``{\small ...In quantum gravity, there appear three
fundamental constants of nature: Planck's constant, $\hbar$, comes
from quantum mechanics; Newton's constant, $G$, from gravity; and,
the velocity of light, $c$, comes from special relativity. The
three constants have physical dimensions. There is a combination
of them---{\rm [the Planck length $\ell_{P}$]}---with the
dimension of length. {\em No other physical theory has a
fundamental length built into it...The planck length, on the other
hand,\footnote{As opposed, for example, to the Bohr radius of
atomic physics, which involves the mass $m$ and charge $e$ of the
electron, as well as the Planck constant $\hbar$, in its
expression, and which is thus `contingent' (due to the particular
parameters $m$ and $e$) only in the realm of specific systems,
namely, the atoms.} refers only to universal constants and not
parameters specific to a sub-class of physical systems. Therefore,
quantum effects of gravity are expected to be significant to all
physics around and below the planck-length. The common belief is
that our usual picture of space-time as a four dimensional
continuum would cease to be a good approximation at the planck
length and that the `microscopic' picture of space-time may be
very complicated}.\footnote{Our emphasis.} This would have a
profound effect on both general relativity and quantum
theory...}''
\end{minipage}

\vskip 0.1in

\noindent while Ashtekar further adds that:

\bigskip \noindent (Q7.31)\hskip 0.9in
\begin{minipage}{11cm}
\noindent ``{\small ...Unification of the principles of quantum
mechanics and special relativity has given rise to the quantum
theory of fields. An outstanding example of such theories is
quantum electrodynamics, the quantum theory of charged particles
and electromagnetic fields. In the development of this theory, the
focus was on calculations leading to predictions which can be
tested against experiments, rather than on issues of mathematical
rigor.\footnote{See Dirac's remarks in (Q?.?) and (Q?.?).} And the
theory has had brilliant successes with experiments...However, in
the calculations leading to these predictions, one has to
integrate certain expressions over energies of virtual photons
which mediate the interaction, and these integrals diverge because
the range of integration extends to infinite energies. (The
resulting infinities are called ultra-violet divergences...) A
systematic procedure---called renormalization---has been invented
to subtract out these infinities and to obtain finite answers, and
it is these answers that have had experimental success. Thus, the
procedure `works', but seems rather ad-hoc. Now, {\em integration
up to infinitely high energies and momenta corresponds, in the
physical-space language, to integration down to infinitely small
time and space intervals. Thus, as was emphasized by the founders
of renormalization theory, infinities arise because one assumes
that the smooth-continuum picture of space-time is valid to
arbitrarily small distances. The true structure of space-time is
presumably very complicated and the renormalization procedure may
be only a convenient trick to get the correct answer without
bothering about the details of these complications. Thus, it is
only when we have a reasonably good picture of the quantum
structure of space-time that we can really understand why the
renormalization procedure works}\footnote{Our emphasis. A
`critique' of renormalization in the context of ADG-gravity
follows in the next sub-subsection.}...}''
\end{minipage}

\vskip 0.1in

\noindent Indeed, the Planck length is often invoked to counter
various arguments by many researchers, especially those with a
bent towards particle physics,\footnote{That is, those that
predominantly believe that one could (or even should!) arrive at a
quantum theory of gravity by applying concepts and techniques from
the continuum based QFT of matter to GR---as it were, like we said
earlier, to quantize the field-fiber while leaving the continuous
spacetime base intact.} who regard gravity's non-renormalizability
as a `fatal blow' to any attempt at arriving at a conceptually
consistent, but perhaps more importantly from a practical
perspective, a calculationally finite(!), QG. In particular, they
doubt a possible formulation of a consistent and finite {\em
non-perturbative} QG, as Ashtekar's words below, taken from
\cite{ash4} which expounds his new QRG theory (in the general
context of non-perturbative canonical LQG), corroborate:

\bigskip \noindent (Q7.32)\hskip 0.9in
\begin{minipage}{11cm}
\noindent ``{\small ...In classical gravity, Riemannian geometry
provides the appropriate mathematical language to formulate the
physical, kinematical notions as well as the final dynamical
equations. This role is now taken by quantum Riemannian
geometry... In the classical domain, general relativity stands out
as the best available theory of gravity, some of whose predictions
have been tested to an amazing accuracy, surpassing even the
legendary tests of quantum electrodynamics. However, if one
applies to general relativity the standard perturbative techniques
of quantum field theory, one obtains a `non-renormalizable'
theory, i.e., a theory with uncontrollable
infinities..}\footnote{Our emphasis.}

...{\small In the particle physics circles, the answer\footnote{To
the question posed a bit earlier in that paper by Ashtekar: ``{\em
Does quantum general relativity, coupled to suitable matter, exist
as a consistent theory non-perturbatively?}''} is often assumed to
be in the negative, not because there is concrete evidence against
non-perturbative quantum gravity, but because of an analogy to the
theory of weak interactions, where non-renormalizability of the
initial `Fermi theory' forced one to replace it by the
renormalizable Glashow-Weinberg-Salam theory. {\em However this
analogy overlooks the crucial fact that, in the case of general
relativity, there is a qualitatively new element. Perturbative
treatments pre-suppose that the spacetime can be assumed to be a
continuum at all scales of interest to physics under
consideration. Since this is a safe assumption for weak
interactions, non-renormalizability was a genuine problem.
However, in the gravitational case, the scale of interest is given
by the Planck length $\ell_{\mathrm{Pl}}$ and there is no physical
basis to pre-suppose that the continuum picture should be valid
down to that scale.}\footnote{Our emphasis.} The failure of the
standard perturbative treatments may simply be due to this grossly
incorrect assumption and a non-perturbative treatment which
correctly incorporates the physical micro-structure of geometry
may well be free of these inconsistencies...}''
\end{minipage}

\vskip 0.1in

On the other hand, in striking contradistinction to the above, in
the two so far quite  successful applications of ADG to classical
(GR) and QG, spanning the entire `background spacetime
spectrum'---from the `{\em ultra-continuum}', and thus in a strong
sense `{\em ultra-singular}' and `{\em ultra-infinite}'
\cite{mall2,malros1,malros2,mall3,malros3,mall9,mall7}, to the
`{\em ultra-discretum}', and hence `{\em non-smooth}' and `{\em
ultra-finite}' \cite{malrap1,malrap2,malrap3}---no issue of a
forced, by hand as it were, discretization (in order to manage our
Calculus, and to secure `analyticity'---good analytic
behavior---of the usual perturbation series expansion) arose
whatsoever.\footnote{Thus, Ashtekar's words above, that ``{\em
perturbative treatments pre-suppose that the spacetime can be
assumed to be a continuum at all scales of interest to physics
under consideration}'', in a strong sense force one to identify
the terms `{\em perturbative techniques}' with `{\em analytic
techniques}'---in other words, {\em `perturbability' vitally
depends on analyticity}, and of course the latter on the
assumption that the base spacetime is a smooth (even more so, an
{\em analytic}!) manifold. Then, the crux of Ashtekar's argument
above is that while perturbability and renormalizability are
legitimate requirements for the electroweak forces since there is
no `natural' minimum spacetime scale (so that in principle all
energy-momenta, no matter how large, should be integrated over in
the perturbation expansion), for the gravitational force this is
not so due to the appearance of the Planck length-time.} In other
words, unlike (Q7.?) above, {\em no fundamental scale such as
$\ell_{P}$ is invoked at all in order to enable us to apply our
concepts and carry out our calculations, since our
`ADG-calculus'---that is ADG's essentially algebraic (relational)
differential geometric machinery subsuming the dynamical relations
between the fields `in themselves'---does not depend at all on a
background (spacetime) structure, be it `continuous'
(`infinitistic') or `discrete' (`finitistic')}. Moreover, in a
finitistic realm, where the continuum based CDG would appear to
fail, ADG applies unhindered \cite{malrap3}.

{\it In summa}, and in view of the genuinely unitary, pure (3rd
gauge and 3rd quantum) field-theoretic viewpoint\footnote{As noted
repeatedly earlier, a theoretical scheme dealing solely and
exclusively with the fields themselves, without reference to an
external, ambient spacetime no matter what its `nature' ({\it ie},
`discrete' or `continuous').} that ADG enables us to maintain and
practice, we answer to our opening question (R7.?) above with a
another `rhetorical' question:

\vskip 0.07in

\centerline{\em Whence $\ell_{P}$?}

\vskip 0.07in

\noindent We firmly believe that this question, far from trying to
be `eccentric' or `smug', touches on a very subtle and important
(for the comprehension and appreciation at least of the spirit of
the potential application of ADG to QG) point, namely, that while
`{\em big}' questions and associated arguments, such as whether QG
exists non-perturbatively or not, vitally hinge in one way or
another on the `nature' (character) of (a background) spacetime
structure ({\it ie}, whether it is a `continuum' or a `discretum',
a `classical' or a `quantal' structure), {\em in ADG there is no
(background) spacetime at all}, so in a sense such questions
become `irrelevant' and the issues they purport to address are
`non-problems' for ADG-gravity. At the same time, the Planck
length becomes an `obsolete chimera' in our theory, simply
reflecting that we have been asking perturbability (analyticity)
and renormalizability questions from within ({\it ie}, from the
perspective of) the `wrong' (mathematical) framework ({\it ie},
the framework of CDG or Analysis)---which is `wrong' in the sense
that it {\it a priori} regards as important the `nature' of the
underlying spacetime (manifold) and views the fields involved as
being inextricably tied to (or dependent on) it.\footnote{Shortly,
in 7.5.2, we will argue further in support of this point.}

\paragraph{The `unity' and `universality' of physical law.} The remarks above prompt us, in complete analogy to
Einstein's mode of expression (and doubt!) in (Q2.1) about
singularities and the `incompleteness of determination' of field
theory that they entail, to maintain in view of ADG's
field-realism (or perhaps better, field-solipsism) (PFR):

\bigskip \noindent (R7.8)\hskip 0.9in
\begin{minipage}{11cm}
\noindent  It is our opinion that the (external) spacetime
scale-dependence of physical law must be excluded from a
fundamental candidate theory for QG. It does not seem (physically)
reasonable to us to `dissect' the unitary field by introducing
(artificial) `effective' scales (for an external to the field
itself continuum spacetime structure) in order to delimit the
range of validity of the law that it obeys (in fact, the law that
the field {\em defines}, as a differential equation, in the first
place). Ultimately, scale (like `spacetime')\footnote{These two
words always go together: we talk about {\em spacetime scales}. At
any rate, even if one is talking about the operationally more
realistic sounding term like {\em energy-momentum scale}, quantum
theory can always translate the latter (via quantum duality or
complementarity) back to time-space scale (in this operational
respect, see the Einstein quotation (Q?.?) below). On precisely
this quantum duality rests Ashtekar's argument in (Q?.?) above:
the higher the energy-momenta (one wishes to integrate over), the
smaller the time-space scales one has to dig deeper into
(ultraviolet infinities are quantum dual/complementary to
infinitesimal scales that the spacetime continuum allows one to
reach, even if just theoretically).} is introduced and fixed by us
observers or experimenters,\footnote{From an ADG-perspective,
scale, like spacetime, is carried by (is inherent in) $\struc$.}
thus it is a contingent not a fundamental feature of Nature---or
anyway, something that could not possibly influence or determine
the validity of field dynamics in any way. If anything, such a
dependance of the law of gravity on a fundamental spacetime scale
shows an {\em antinomy} of the very term `physical {\em law}' (pun
intended!) and its supposed universality---{\it ie}, that the law
holds in the whole `{\em World}', when the latter word however is
not understood as `spacetime' as usual, but as `field'. The World
{\em is} the dynamical fields that comprise it, nothing else
(PFR).
\end{minipage}

\vskip 0.1in

\noindent In other words, we believe in the `unity' and
`universality' of the (gravitational) field and thus cannot accept
that it obeys (defines!) one gravitational law above, say,
$\ell_{P}$ (where spacetime is a supposed to be a continuum)---the
so-called classical gravitational Einstein equations of (GR), and
a different one below $\ell_{P}$ (where spacetime is intuited to
be `reticular' and `quantal')---the elusive law of QG, this
difference being in turn reflected by the external (to the
dynamically autonomous, `unitary' field) continuum/discretum
`divide' resulting from the `{\it ad hoc}' (as it were, forced by
theoretical fiat) introduction (by us!) of the Planck length. This
remark

\begin{itemize}

\item on the one hand puts into perspective the nowadays
popular anticipation-{\it cum}-imperative that a cogent quantum
theory of gravity {\em should} have GR as an `{\em effective
theory}' \cite{burgess} or, what is essentially the same, as a
so-called `{\em low energy limit}'---arguably, the main
`prognosis' of the string-theoretic approach to QG
\cite{veneziano,manin1}.

\item and in pretty much the same gist, that QG {\em should} yield GR at
some `{\em classical correspondence limit}', much in the same way
that GR reduces to (non-relativistic) Newtonian gravity as
$c\mapto\infty$, while quantum mechanics (or field theory) yields
its classical correspondent when $\hbar\mapto
0$,\footnote{Actually, in the case of QFT---the standard
denomination of relativistic quantum mechanics of systems with an
infinite number of degrees of freedom, the correspondence
principle (yielding a non-relativistic classical field theory) is
supposed to be represented by the joint application of the formal
limits $c\mapto\infty$ and $\hbar\mapto 0$.}

\item and on the other hand, it exposes what we have time and
again highlighted in the present paper and throughout our past
trilogy \cite{malrap1,malrap2,malrap3}, namely, that {\em the}
culprit for all the difficulties and anomalies encountered in our
differential geometric models and calculations (Calculus!) in
physics is {\em our} assumption that physical space(time) is an
external (to the fields), background continuum---a manifold
mediating our calculations and at the same time facilitating the
geometrical interpretation of our CDG-based constructions and of
the physical inferences that we draw based on them.

\end{itemize}

Of course, it must be made clear here, in order to avoid any
misunderstanding, that all these `{\em fundamental physical
constants}' and their combination to $\ell_{P}$ are in every sense
`real', as they delimit the range of validity of their
corresponding theories---albeit, {\em our spacetime continuum
based and thus CDG-theoretically modelled theories and their
defining dynamical laws!}\footnote{That is, theories that both
conceptually and technically regard the base spacetime continuum
as being on a par with the dynamical laws (defining these
theories). In point of fact, the spacetime continuum is viewed as
being deeper than and prior to the laws themselves, when one
considers that the latter are being modelled after differential
equations and that the continuum is assumed {\it a priori}
precisely in order to secure those vital `conditions of
differentiability' (and variation!) of the physically observable
quantities engaging in the laws---the fields and their quanta
(particles).}---like SR ($c$), GR ($G$), QM ($\hbar$) and QFT ($c$
and $\hbar$), but {\em it would be a blatantly arbitrary, {\it
prima facie} uncalled for (and perhaps prove to be `false' in the
long run of QG research) inference, just because our mathematical
model of spacetime as a manifold and the CDG-panoply based on it
suffer or appear to be problematic and of limited validity in the
quantum domain, to posit that Nature herself has a fundamental
length or time duration}.\footnote{Compare this with our basic
motto that `{\em Nature has no singularities}' in sections 1 and
2. This is why we emulated in (Q?.?) above Einstein's words in
(Q2.1).} To stress it again, since in ADG-gravity (as well as in
the ADG-formulation of the dynamics of the other fundamental
forces, such as the Yang-Mills theories) there is no geometrical
background spacetime---be it a discretum or a continuum---but only
the fields and their algebraically represented dynamical
interrelations ({\it ie}, the by differential equations modelled
physical laws), such {\it a priori} and arguably {\it ad hoc}
assumptions and resulting practices are foreign to its very
essence (or even, mathematically speaking, to its very
`definition'!).

{\it En passant}, it must be mentioned here that the supposedly
physical importance of the Planck length (and the potential
development of a new `perspective on gravity' at scales below
it)\footnote{A quantum theory of gravity like the one envisioned
today?} was, to the knowledge of these authors, first anticipated
by Eddington, who remarks prophetically in \cite{eddington2}:

\bigskip \noindent (Q7.33)\hskip 0.9in
\begin{minipage}{11cm}
\noindent ``{\small ...There are three fundamental constants of
nature which stand out pre-eminently. The velocity of light,
$3.00\times 10^{10}$ c.g.s. units; dimensions $LT^{-1}$. The
quantum, $6.55\times 10^{-27}$ c.g.s. units; dimensions
$ML^{2}T^{-1}$. The constant of gravitation, $6.66\times 10^{-8}$
c.g.s. units; dimensions $M^{-1}L^{3}T^{-3}$.

From these we can construct a fundamental unit of length whose
value is $4\times 10^{-33}$cms. There are other natural units of
length---the radii of the positive and negative unit electric
charges---but these are of an altogether higher order of
magnitude.

With the exception of Osborne Reynold's theory of matter,
{\small\em no theory has attempted to reach such fine-grainedness.
But it is evident that this length must be the key to some
essential structure. It may not be an unattainable hope that some
day a clearer knowledge of the processes of gravitation may be
reached; and the extreme generality and detachment of the
relativity theory [from the other theories of matter] may be
illuminated by the particular study of a precise mechanism}...}''
\end{minipage}

\vskip 0.1in

\noindent Also, from an operational viewpoint, equally suggestive
are the following remarks of Einstein found in \cite{stachel8}:

\bigskip\noindent (Q7.34)\hskip 0.9in
\begin{minipage}{11cm}
\noindent ``{\small ...If one does {\em not}\footnote{Einstein's
emphasis.} want to introduce rods and clocks as independent
objects into the theory, {\em then one must have a structural
theory in which a fundamental length enters, which then leads to
the existence of a solution in which this length occurs, so that
there no longer exists a continuous sequence of `similar'
solutions. This is indeed the case in the present quantum theory,
but has nothing to do with its basic
characteristics}.\footnote{Our emphasis.} Any theory which has a
universal length in its foundations, and, on the basis of this
circumstance, qualitatively distinguished solutions of definite
extent, would offer the same thing with respect to the question
envisioned here...}''

\end{minipage}

\vskip 0.1in

\noindent However, {\em both Einstein and Eddington} had on the
one hand criticized the background spacetime continuum of GR
(Q?.?), and on the other they had intuited that a completion of GR
(to a unitary field theory so as to include the other matter
sources and their radiation force-fields) should consist of the
field-dynamics alone, and what's more, they intuited and
`demanded' that the singularities (of matter; the matter
particles/quanta) be included in that total field dynamics (Q?.?).
To a certain degree, {\em ADG has achieved both}, without
involving any background spacetime at all and, as a result,
fundamental spacetime scale either.

We may address and question the issue of a fundamental spacetime
scale from another point of view relevant to ADG-gravity. For
instance, we may begin with an `aphorism' and claim that {\em the
notion of `scale' (or `measurement gauge'!) has meaning only in an
operational context, in fact, it is closely associated with our
measurement actions/acts of measurement}. In our scheme, the
notion of `scale' is inextricably tied to the generalized
arithmetics $\struc$---the realm where our acts of measurement
`take values'. A generalized `gauge' or `scale' principle is the
following:

\bigskip \noindent (R7.9)\hskip 0.9in
\begin{minipage}{11cm}
\noindent {\em The laws of Nature are scale (gauge) independent.}
\end{minipage}

\vskip 0.1in

\noindent which in turn, in the case of the spacetime continuum
({\it ie}, when we take $\struc\equiv\smooth_{M}$) would appear to
contradict the nowadays widespread view:

\begin{itemize}

\item that the law of gravity does not hold (or breaks down!) at
singularities (or below Planck length-time scale), or more-or-less
equivalently in the continuum based QFTs of matter,

\item that there is a natural length-duration spacetime
scale\footnote{The epithet `natural' pertaining to the fact that
these scales can be expressed in terms of fundamental {\em
physical} constants.}---the so-called Planck scale
($\ell_{P}\approx 10^{-35}m$-$t_{P}\approx 10^{-40}s$)---below
which the classical spacetime continuum gives way to a `discrete'
and `quantal' space(time) structure.

\end{itemize}

\noindent On the other hand, if one wished to retain these
numerical physical constants in ADG-gravity, one could indeed do
so by regarding them as {\em global sections of the constant
sheaf} $\cons\equiv\mathbf{K}:=\mathbf{R}$ (of reals), {\em which
is by definition embedded into} $\struc$. However, {\em being
elements of $\cons$, physical constants do not contribute at all
to the ADG-gravitational field-dynamics which is effectuated via
the ADG-theoretic, essentially algebraic and background
independent, differential geometric mechanism}. Equivalently, the
ADG-gravitational field $\conn$, by virtue of being (by
definition) a $\cons$-morphism, `sees through' (via the dynamics
that it defines) the fundamental constants. Another way to say
this, the physical constants, being {\em constant}, do not partake
into the field dynamics. The ADG-gravitational field dynamics,
being also completely background (spacetime) geometry free (which
is inherent in $\struc$, while the gravitational dynamics is
$\struc$-functorial as repeatedly emphasized earlier), is also
`blind' or `indifferent' to various scales that we might employ to
gradate or `tessellate' this fiducial base spacetime.\footnote{See
Riemann's quotation below.}

To stress it once again in order to `digest' it, physical
constants (and their combinations), with values in the real
numbers' continuum ($\R$), are `physical enough' to the extent
that they set `gauges' or `scales', or even `natural units of
measurement', for delimiting energies (speeds or momenta, actions,
energies, or dually, time-durations {\it etc}) at which various
interactions between fields and their quanta (particles) become
significant (`observable') and the corresponding (spacetime
continuum based field) theories about them are valid (as it were,
as effective (field) theories \cite{cliff}), but what `right' do
we have to first combine them to $\ell_{P}$ (or $t_{P}$)

\begin{equation}\label{eq100}
\ell_{P}=\sqrt{\frac{G\hbar}{c^{3}}}=1.6\times 10^{-33}cm;~
t_{P}=\sqrt{\frac{G\hbar}{c^{5}}}=5.3\times 10^{-44}s
\end{equation}

\noindent and then infer that the latter combination singles out a
`preferred', `natural', `universal' length/time-scale below which
(classical continuum spacetime based) field physics fails to
yield? Could it instead be that

\bigskip\noindent (R7.10)\hskip 0.9in
\begin{minipage}{11cm}
\noindent \noindent {\em it is precisely because of the way we
formulate these continuum field theories differential
geometrically, by means of (or by mediation in the guise of
coordinates---in effect, of our `measurements' and `smooth field
localizations' on) a background spacetime continuum, that it
appears to be necessary to introduce a fundamental length scale in
terms of which (and what's more, in order for) our analytic
(Calculus based) calculations (to) make sense ({\it eg}, to be
rendered finite)?}
\end{minipage}

\vskip 0.1in

For one can recall here Riemann's words in \cite{riemann}, which
we quote {\it verbatim} from \cite{mall7}:

\bigskip\noindent (Q7.35)\hskip 0.9in
\begin{minipage}{11cm}
\noindent \noindent ``{\small Ma\ss~bestimungen erfordern eine
Unabh\"{a}ngichkeit der Gr\"{o}\ss en vom Ort, die in mehr als
einer Weise stattfinden kann.}'' : ``{\small\em Specifications
{\rm [:\;measurements]} of mass require an independence of
quantity from position, which can happen in more than one way.}''
\end{minipage}

\vskip 0.1in

\noindent By analogy with ADG's singularity `absorptions' or
`dissolutions' in $\struc$, one may conceive here, at least in a
metaphorical sense, of the so-called `perturbation expansion' (of
gravitational interactions in the context of perturbative QGR) and
the associated `renormalization technology' inherited from the
continuum based QFT, as attempts to `{\em extend analyticity}' at
the {\it loci} of interaction by {\em absorbing infinities} in the
physical parameters (constants) involved. However, these
attempts---all of them based of course on the analytic means of
Calculus (CDG)---are doomed to failure, and this failure is
usually attributed, physically speaking, to the dimensionality of
Newton's constant $G$ involved in the `minimal length' $\ell_{P}$
in (\ref{eq100}) above. In turn, this minimal length is supposed
to represent a natural `regularization cut-off' beyond which this
`perturbative analyticity' (renormalization series) cannot be
extended further so that the smooth spacetime continuum is
supposed to give way to something (which one might still call
`spacetime') more `discrete', `quantal' and `inherently
cut-off'---something, of course, that can `regulate' the
uncontrollable infinities of the spacetime continuum.

But GR is simply the relativistic field-theoretic extension of
Newtonian gravity, {\em one that is vitally based on the smooth
base continuum} (at least for its differential geometric
representation). Thus, the infinite field-redefinitions involved
in attempts to renormalize `field-quantized gravity' are closely
analogous to the infinities associated with
$\smooth$-singularities, with the important difference that in
gravity---at least in its original second order formulation in
terms of the metric {\it \`a la} Einstein (or even in its
subsequent tetrad-connection first order formulation {\it \`a la}
Palatini-Ashtekar), in striking contradistinction to the other
(classical or quantum gauge) field theories of matter, {\em the
background geometry (metric) is itself the dynamical variable, so
the relevant field quantities involved cannot be referred or
`expanded' with respect to it}. We saw for instance earlier in
(Q?.?) how Geroch notes in the introduction of \cite{geroch} the
similarities between the infinities associated with singularities
of the gravitational field and the infinities of the other
(classical, continuum based) field theories of matter; however,
with the important difference that with the former there is no
fixed background geometry with to which the gravitational field
({\it ie}, the metric itself!, in the standard theoresis of GR)
can be referred and `perturbatively' expanded (to remove those
infinities). Parenthetically, it must be further noted here, that
Geroch goes a bit further and attributes exactly to that
fundamental difference between gravity and other classical field
theories the difficulty in arriving at a precise definition of
singularities in GR:

\bigskip\noindent (Q7.36)\hskip 0.9in
\begin{minipage}{11cm}
\noindent ``{\small\em ...In view of the faulty analogy with which
we must work,\footnote{The analogy referred to here being the one
between the singularities in GR and the infinities in the
(classical and quantum) matter field theories.} it is not
surprising that (a) there is no widely accepted definition of a
singularity in general relativity, and (b) each of the proposed
definitions is subject to some inadequacy...}''
\end{minipage}

\vskip 0.1in

This whole rationale is pregnant to a `{\em general conjecture}'
(or `hunch') that emerges in the light of ADG-gravity: Calculus
(Analysis), which is manifestly dependent on a background
$\smooth$-manifold, cannot be carried through to the (`true')
quantum gravity regime,\footnote{See quotations of Einstein,
Feynman and Isham in the sequel (Q?.?).} while the incurable
infinities that arise are precisely due to our inappropriate
application of differential geometric
ideas\footnote{`Inappropriate' in the sense that for our
differential geometric `aufbau' we always refer to a background or
`intervening' spacetime (geometry).} in the quantum deep. In this
line of thought, we read from \cite{mall7} for example:

\bigskip \noindent (Q7.37)\hskip 0.9in
\begin{minipage}{11cm}
\noindent ``{\small ...So we can think here of the famous {\em
`Planck scale'}, that, of course, it is not a matter of
analysis/algebra, but rather of the particular manner ({\it viz.}
still classical differential geometry), that it is here undertaken
to exploit `{\em geometry}' in the quantum
domain\footnote{`Geometry' here refers to what one might call
`{\em physical geometry}'---the (structural) analysis directly of
the dynamical, essentially algebraic (relational), attributes of
fields and their particles (quanta)---and not to what one could
call `{\em mathematical geometry}', which is usually understood as
the (structural) analysis of a given, fixed and `inert'
`space(time)'. Of course, the more subtle distinction is between
(differential) geometry understood in a Cartesian
(analytic-arithmetic) way---what we will call `{\em mathematical
geometry}' in 7.?, and (differential) geometry seen from a
Leibnizian (algebraic-relational) perspective---what we will call
`{\em physical geometry}' in 7.?.}...}''
\end{minipage}

\vskip 0.1in

As briefly alluded to above, and in the spirit of the present
paper, the following rather `standard' intuition appears to be
{\it prima facie} `natural': {\em at sub-Planckian scales the
smooth spacetime continuum gives way to something more finitistic,
something `inherently cut-off'}. Moreover, it is widely hoped that
the desired successful `quantization' of the gravitational field
(to a conceptually consistent and calculationally finite quantum
theory of gravity) will `cure', or even evade altogether, not only
the non-renormalizable quantum field theoretic infinities of QGR,
but the more robust ones associated with the $\smooth$-smooth
singularities. {\em Isn't such an anticipation `begging the
question', though?} For if we suppose (as many researchers
currently do suppose!) that GR and the smooth spacetime manifold
supporting it will arise from the `true' quantum gravity theory at
a Bohrean `macroscopic', `correspondence' or `low energies' limit
({\it ie}, at large scales/weak gravitational field strengths),
would not the emergent GR on the differential spacetime manifold
automatically carry with it the entire differential geometric
anomalies (`singularities') baggage---the burden that the
purported quantization of GR so painstakingly tried to get rid of
while still retaining the CDG picture of a base spacetime manifold
and the smooth fields on it?\footnote{This in turn seems to place
serious doubts on the QGR (`quantum', or better, `{\em quantized}'
GR) program. See next remark.}

\bigskip \noindent (R7.11)\hskip 0.9in
\begin{minipage}{11cm}
\noindent On the other hand, if `space(time)' (and its `geometry')
is inherent in the dynamical fields and their dynamical
interrelations, and if we suppose (as we actually do suppose in
ADG-gravity) that {\em everything is  field, and that the field is
by itself already quantum, as well as that the field represents
the actions of (its) quanta without the intervention of any
`space(time) as such' whatsoever}, {\em why attempt at all to
`save' space(time) by speculating that, if it is not `continuous',
it must then be `discrete'?} ({\it ie}, take $\ell_{P}$ at `face
value'); moreover, {\em why at all insist that the `true' quantum
gravity is (somehow) `quantized GR' and at the same time expect
that a successful quantization of GR will alleviate, ultimately
remove, singularities and their nonsensical
infinities?}\footnote{As it recently claims to have achieved in
LQG as we discussed earlier, where the quantization of the
gravitational field results in the quantization and concomitant
discretization of the background continuum, which `quantum
discretization' is in turn employed to `resolve' the interior
Schwarzschild singularity and other cosmological ones
\cite{boj1,modesto,husain}.}
\end{minipage}

\vskip 0.1in

\noindent With respect to the first question, {\em why depend at
all on a background `spacetime' arena---be it a continuous or a
discrete realm---when all that there is are fields representing
the dynamical interrelations of quanta?}\footnote{Previously, in
the context of a $\smooth$-smooth spacetime manifold, we referred
to this tendency as `{\em CDG or Calculus conservative}'. Here we
extend this to a general `{\em background spacetime
conservatism}'.} Presumably, finitistic and quantal are the
(dynamical) (inter)actions of field-quanta `in themselves'; {\em
quanta `see' no ambient spacetime as such}---rather, it is their
dynamics---their field-dynamics---that `{\em defines}' (`creates')
physical spacetime (geometry).\footnote{In this sense we will
distinguish between `physical' and `mathematical' (spacetime)
geometry in subsection 7.? later.} Antonio Machado's verse from
his poem `{\itshape The Road}' \cite{machado} comes to mind:

\bigskip \noindent (Q7.38)\hskip 0.9in
\begin{minipage}{11cm}
\noindent ``{\small Traveller there are no paths; paths are made
by walking.}''
\end{minipage}

\vskip 0.1in

\noindent as well as Wheatley and Kellner-Rogers' remarks from
their inspired book \cite{kellner}:\footnote{In the quotation
below, emphasis is ours.}

\bigskip \noindent (Q7.39)\hskip 0.9in
\begin{minipage}{11cm}
\noindent ``{\small\em The future cannot be determined; it can
only be experienced as it is occurring. Nature\footnote{Here, the
authors say `{\em Life}' instead of `{\em Nature}'.} doesn't know
what it will be until it notices what it has just become.}''
\end{minipage}

\vskip 0.1in

\noindent {\it In toto}, all there is are these field-quanta,
their existence and `sustenance' is not conditional on a
background space(time), and it is high time, following Leibniz, to
devise a Calculus that refers directly to them; while, perhaps
more importantly for Physics, base our (differential geometric)
calculations on such a (generalized or abstract) Differential
Calculus.

\vskip 0.1in

{\it In toto}, we are tempted to say, in a Leibnizian way, that in
ADG fields are `{\em differential geometric entelechies}' or
`monads', in the sense that they are in no need of reference to an
ambient (external or background) space(time) to support
themselves;\footnote{And recall Leibniz
\cite{leibniz2,leibniz,leibniz1}: ``{\em Monads are
windowless.}''.} they are irreducible (atomic or `{\it
ur}'),\footnote{And recall Einstein: ``{\em The notion of field is
a primary, not further reducible one.}'' \cite{einst3}.}
autonomous (differential) geometric (in fact, {\em purely
algebraic}) entities, with the aforesaid `autonomy' being
physically interpreted as `{\em autodynamicity}': the
ADG-gravitational field alone {\em defines} the (vacuum)
gravitational dynamics, without that dynamics being in any way
dependent on---or perhaps better, conditioned and constrained
by---an external (background) spacetime manifold.\footnote{We have
in mind here the usual spacetime diffeomorphism
($\mathrm{Diff}(M)$) constraints that create a host of problems in
regarding gravity as a quantum gauge theory
\cite{ivanenko,weinstein0,weinstein}. In the next section we
devote a whole sub-subsection (7.?.?) to this
Aristotelian-Leibnizian character of the third gauge, third
quantum ADG-gravitational field.}

On the face of the above, we come to ask:

\bigskip \noindent (R7.12)\hskip 0.9in
\begin{minipage}{11cm}
\noindent  To what extent is the dimensional `$G$' involved in
$\ell_{P}$ `bound' to the continuum picture of spacetime on which
both the non-relativistic Newtonian gravity and GR fundamentally
depend? Then, {\it mutatis mutandis} for $\hbar$.\footnote{And one
must recall that Planck introduced $\hbar$ in order to regularize
the continuous black body radiation spectrum!} Isn't that
dependence of `$G$' on $M$ analogous to the dependence of Bohr's
radius on the specific physical systems in focus---{\it ie}, the
material atoms?\footnote{We implicitly suppose for a moment, by
assuming a `Devil's Advocate' stance against the problem of
quantum gravity, that at sub-Planckian scales the system in focus
is spacetime itself.} On the other hand, {\em what reasons have we
got to assume that in the `true' quantum gravity theory the basic
systemic (quantum) variable is `spacetime', rather than the
gravitational field alone---`in-itself', so to speak?}\footnote{Of
course, the gravitational field conceived now not
CDG-theoretically, but in some other way---{\it eg}, solely as an
`algebraico-categorical' connection $\conn$ like in ADG---still
retaining though its fundamental `differentiability' and
`observability' ({\it ie}, its `dynamical variability') through
$\struc$ ({\it ie}, effectively through $\ric(\conn)$, which is an
$\struc$-morphism) attribute.} For, {\em who gave us `spacetime'
in the first place?} Arguably, it is {\em we} that assumed it in
the first place.\footnote{And arguably, we assumed it in the first
place in order to be able to do Calculus based on it (manifold and
CDG-conservatism and monopoly).} So, {\em isn't it possible that
the commonly intuited $\ell_{P}$ `cut-off', below which the
spacetime manifold---the so-called continuum, is supposed to give
way to a `discrete' spacetime structure---a
`discretum',\footnote{See quotation of Sorkin in (Q7.?) next.} is
due to our inappropriate employment of CDG-theoretic ideas to
address the problem of the (presumably inherently\footnote{As we
presume in the present paper. See also \cite{malrap3}.}) quantum
gravitational field?}
\end{minipage}

\vskip 0.1in

For example, we read from \cite{booth}:

\bigskip \noindent (Q7.40)\hskip 0.9in
\begin{minipage}{11cm}
\noindent ``{\small ...Einstein's General Theory of Relativity has
proved to be one of the most successful and enduring theories in
physics, and its predictions have been verified in numerous
experiments. However, it stands alone amongst field theories in
that it is not scale invariant. For example, the differential form
of Maxwell's equations, which elegantly describe the
electromagnetic field, do not define any intrinsic scale.
Conversely, Einstein's field equations, which describe the way
that matter curves spacetime, are linked to an apparently
arbitrary scale determined by the Newtonian gravitational
constant, $G$...{\em Numerous attempts have been made to develop a
theory of gravitation that is scale invariant, and yet retains the
key properties of General Relativity, such as the principle of
general covariance}\footnote{Our emphasis.}...The concept of a
uniquely defined Planck scale is one of the two principal
motivations for the pursuit of a quantum theory of gravity, the
other being the need for the curvature terms in the gravitational
field equation to be quantized in order to be equivalent to the
quantized matter fields. {\em If the Planck factor is removed, we
need to ask whether there is still a need for a theory of Quantum
Gravity, at least in the form currently being sought. There is no
fundamental requirement that the gravitational field should have
an inherent quantum structure, and it may well be more reasonable
to think of the gravitational field as being quantised as a
consequence of the matter fields with which it
interacts}.\footnote{Again, emphasis is ours.}

...Finally, one can speculate that by more clearly defining the
distinction between the realms of General Relativity and Quantum
Mechanics, we can actually move closer towards constructing a
paradigm that unifies the two theories...}''
\end{minipage}

\vskip 0.1in

\noindent Booth's words above are in accord and at the same time
in discord with ADG-gravity, in the following sense:

\begin{itemize}

\item The background spacetime manifoldless (in fact,
spacetimeless altogether) ADG-gravity is `by definition' (or
construction) scale-free, since no external (to the gravitational
field $\conn$ itself) spacetime is involved whatsoever. Moreover,
the PGC of GR is now transferred to the field `in-itself', thus it
has been substituted by synvariance, which is implemented via
$\aut\modl$, not $\mathrm{Diff}(M)$. {\em No spacetime, no
fundamental spacetime scale either}.

\item Thus, from the ADG-gravitational perspective, a uniquely defined Planck scale is {\em not} a
principal motivation for QG, and a quantization of the
gravitational field like one quantizes the matter fields (on the
right hand side of Einstein's equations), is not our principal
concern.

\item Hence, by removing the Planck factor we genuinely question
the ``{\em need for a theory of Quantum Gravity, at least in the
form currently being sought}'', that is to say, with a
differential spacetime manifold still present at the background
and with a fundamental `cut-off' scale such as Planck's which is
introduced in order to `tame' our Calculus (perturbatively) at
infinitely small scales.

\item On the other hand, we do believe that ``{\em the gravitational field should have
an inherent quantum structure}'' like the other quantum matter
fields, but we cannot accept that its quantization is ``{\em a
consequence of the matter fields with which it interacts}''. For
after all, one should still be able to account in a quantum
mechanical way for the gravitational field {\it in vacuo}, while
apparently insuperable problematics, both conceptual and
technical, arise in various endeavors to do that while maintaining
a base manifold \cite{torre1,torre2,weinstein0,weinstein}

\end{itemize}

\noindent At any rate, it must also be emphasized here that this
`algebraic-discretum' alternative to the smooth background
geometrical spacetime continuum is supposed to be a `{\em real}'
and `{\em radical}' alternative, in the sense that it is expected
to retain few features (if any at all) from continuum
physics---especially, it is doubtful that it will retain the
background geometrical space-time interpretation. Here is a
quotation of Einstein, found in \cite{einst12}, that supports
precisely this:\footnote{See the last subsection 8.?, and
especially 8.?.?, for this continuum-discretum `debate' {\it
vis-\`a-vis} Einstein.}

\bigskip \noindent (Q7.41)\hskip 0.9in
\begin{minipage}{11cm}
\noindent ``{\small ...The alternative continuum-discontinuum
seems to me to be a real alternative; i.e., there is no
compromise. {\em By discontinuum theory I understand one in which
there are no differential quotients. In such a theory space and
time cannot occur, but only numbers and number-fields and rules
for the formation of such on the basis of algebraic rules with
exclusion of limiting processes}.\footnote{Our emphasis.} Which
way will prove itself, only success can teach us...}''
\end{minipage}

\vskip 0.1in

{\it In summa}, our thesis {\it contra} $\ell_{P}$-$t_{P}$ can be
distilled to the following:

\bigskip \noindent (R7.13)\hskip 0.9in
\begin{minipage}{11cm}
\noindent Both the classical field theory of gravity (GR) and the
relativistic quantum theory of matter (QFT) depend on a smooth
spacetime manifold---curved for the former, flat for the latter.
In attempts to quantize gravity by applying QFTheoretic rules to
GR---as it were, to arrive at a `Quantum or Quantized General
Relativity' theory, either canonically or covariantly (via path
integrals), perturbatively or not----the continuum gets in the way
in one way or another.\footnote{For example, the so-called
`problem of time' or the `inner product problem' ({\it ie}, the
metric on the physical Hilbert space of states for the quantum
gravitational field) in (non-perturbative) canonical QGR, or the
functional quantum measure in a sum-over-histories (path integral)
type of scenario, or even the non-renormalizability of
(perturbative quantum) gravity which simply shows that one cannot
continue analyticity down to the infinitely small (when, quite
paradoxically, {\em analyticity is defined by the infinitely
small}---{\it ie}, by the locally Euclidean smooth base
continuum!).} To circumvent these continuum-problematics, one
evokes the Planck length and argues that in the QG regime there is
a fundamental spacetime scale below which the continuum picture
ceases being a faithful representation of the `true' spacetime
geometry. In contradistinction, ADG-gravity is a pure gauge
field-theoresis of gravity that is fundamentally background
spacetimeless---whether this background is `continuous' or
`discrete'---while at the same time the third gauge, third quantum
ADG-gravitational field $\conn$, which defines (via its curvature)
the (vacuum) Einstein equations (\ref{eqy23}) is {\it ab initio},
by the very ADG-formalism, already or self-quantum, as we argue
further in the next sub-subsection.
\end{minipage}

\subsubsection{The distinctions (epithets) `classical' and
`quantum' lose their meaning in ADG-gravity: in what sense the
ADG-formulated Einstein equations are not already quantum?}

The crux of the foregoing arguments is that in (vacuum)
ADG-gravity the traditional denominations of gravity as
`classical' and `quantum' lose their standard meaning in the
theory. We maintain that such distinctions are essentially due to
the presence of a background spacetime manifold which on the
classical side of the quantum divide remains intact, while on the
quantum side, although it is initially assumed in order to set up
differential geometrically any field-theoretic approach to QG
({\it eg}, canonical QGR, LQG and QRG) it is eventually supposed
to be discretized and quantized, mainly by evoking the minimal
Planck scale. By contrast, since in the `pure field' theoresis of
gravity {\it \`a la} ADG a base spacetime, continuum or discretum,
is not involved at all, a fundamental space-time length-duration
has to go, thus, {\it in extenso}, those two standard epithets to
`gravity' become insignificant. Moreover, since the
ADG-gravitational field $(\modl ,\conn)$, at least from a
geometric prequantization vantage, is `inherently quantum' (as we
argued extensively earlier), in the ADG-approach to QG no issue
arises whatsoever of quantizing spacetime itself. {\it In toto},
{\em the nowadays popular quest for a quantization of spacetime
structure is begging the question in ADG-gravity}.

This is already reflected in the quantum interpretation carried by
the ADG-gravitational vacuum Einstein equations (\ref{eqy23}),
whereby the connection field $\conn$ acts, as a sheaf morphism,
via its (Ricci) curvature scalar on the local sections of the
associated (representation) vector sheaf $\modl$, which in turn
from a geometric (pre)quantization vantage represents the local
quantum particle states of the field. This is the non-linear,
`self-(inter)acting', character of the ADG-gravitational field.
Let it be noted here that in a `fuller' quantum theoresis of the
ADG-gravitational dynamics, a non-smooth manifold based path
integral over connection space scenario is envisaged to capture
the `complete' dynamics for QG from the purely gauge-theoretic
ADG-perspective.\footnote{See 7.9.1 in the sequel.}

\subsubsection{The twilight of a theoretical paradigm-shift in QG}

So far we have time and again emphasized that ADG-gravity is
fundamentally (background) spacetimeless---in particular, it is
background manifoldless. Moreover, the very formulation of
ADG-gravity is fundamentally different from the manifold and
smooth metric-based one originally due to Einstein. It certainly
resembles the recent Ashtekar formulation (AF) essentially
regarding (quantum) gravity as a (quantum) gauge theory, but it
differs from it as well in at least three, closely related to each
other, important ways:

\begin{itemize}

\item Like the AF, the basic dynamical variable is the
gravitational connection field $\conn$, but unlike it, no metric
in the guise of the smooth tetrad field of the first-order
formalism is involved (ADG-gravity$\equiv$half-order formalism).

\item In addition, unlike the AF, no external (to the connection) smooth background
spacetime manifold is involved: only the algebraic
$\struc$-connection field $\conn$ is present. ADG-gravity is
genuinely background independent ({\it ie}, not only background
metric, but also differential manifold independent) and {\em what
has been retained from the original formulation is the
gravito-inertial interpretation of the theory, not its
chrono-geometric one}.

\item As a consequence, the theory recognizes no fundamental (minimal)
space-time scale in the guise of the Planck length-duration
($\ell_{P}$-$t_{P}$) and also it does not regard `spacetime
quantization' as a physically meaningful quest(ion). This puts
into perspective the whole QG enterprize (perturbative or not) as
no infinity or singularity is present in the theory so that
neither $\ell_{P}$ would have to be invoked as a regularization
`cut-off' scale in order to render physical quantities finite, nor
the `classical continuum'-to-`quantal discretum' (below
$\ell_{p}$) transition is expected to be effectuated by spacetime
quantization and concomitant discretization as it is currently
maintained in non-perturbative QG scenarios such as LQG
\cite{boj1,modesto,husain}. The bottom line is that ADG-gravity is
a third gauge, third quantum theory of the gravitational field
`in-itself', with the notion of a geometrical spacetime playing no
role in the theory whatsoever.

\end{itemize}

\noindent These three glaring differences are in our opinion
pregnant to a significant Kuhnian paradigm-shift \cite{kuhn} in QG
research issuing from ADG-gravity, which we itemize in three parts
below:

\begin{itemize}

\item The first fundamental notion in GR that must be reconsidered and revised
in the light of ADG-gravity is that of {\em spacetime event}. In
the manifold based GR, the basic aftermath of Einstein's hole
argument as we will extensively describe and comment on in 8.5 is
that, basically, a spacetime event is a point $p$ of the base
differential manifold $M$ {\em together} with the smooth
gravitational field $g_{\mu\nu}$ defined on it and satisfying the
Einstein equations.\footnote{{\it A fortiori}, in view of the hole
argument, any (smooth) `coordinate-dressing' $x(p)$ of $p$ has no
physical significance in view of the general covariance of the
Einstein equations ($\mathrm{Diff}(M)$ implementing the PGC in
GR).} There is no {\em spacetime point event-set} interpretation
of $M$ without the dynamical gravitational field on
it.\footnote{This is the Einstein-Stachel `{\em no field, no
spacetime event}' {\it motto}, as we will see in the next section.
Here already one witnesses the germ of the idea that dynamics is
essentially prior to kinematics if one recalls that the basic
kinematical structure in GR is that of a base differential
manifold and a smooth Lorentzian metric on it.} In
contradistinction, in ADG-gravity the dynamics is formulated
solely in terms of the gravitational connection field $\conn$, not
the metric, without any reference to a(n {\it a priori} posited
and fixed, kinematical) background `geometrical spacetime'
structure (arena). Thus, the notion of {\em space-time event} and
of the chrono-geometrical interpretation of the gravitational
field $g_{\mu\nu}$ that goes hand in hand with it cannot survive
in the theory.

\item Related to the above is the issue of causality. Since
$g_{\mu\nu}(x)$ represents the local causal structure of spacetime
as at every $x\in M$ it stands for the field of {\em
infinitesimal} or {\em differential} spacetime locality (local
causality) \cite{malrap1} which thus becomes a dynamical variable
in GR, in ADG-gravity causality too must be revised. In other
words, the causal nexus between events, which in the manifold
based GR is represented by the smooth variable $g_{\mu\nu}$
obeying the differential equations of Einstein, is replaced by the
single notion of connection-field $\conn$. This is a Jungian
synchronicity-type of coincidence in `chronological names' (pun
intended): the chronological {\em causal connection} is subsumed
under the dynamical {\em connection field} $\conn$. To stress it
once again, {\em in ADG-gravity there are no (dynamical) causal
connections between spacetime events, only the background
spacetimeless (dynamical) gravitational connection field}
$\conn$.\footnote{This puts into perspective the `discrete' causet
approach to QG in which the causal connection between `events' is
the sole notion, while dynamics is sought after a functional on
the kinematical space of causet histories.} In ADG-gravity, `time'
({\it ie}, chronology and causality) derive from ({\it ie}, they
are inherent in) $\conn$ and the dynamics that it defines.

\item Now that we have scrapped-off (the background) {\em spacetime} (manifold) and opted
for a choro-chrono-geometrical interpretation-free and purely
algebraic gravitational field $\conn$, what happens to the by now
commonly accepted ideas about {\em space-time measurements}? One
immediately anticipates that measurements of `spacetime position'
(locution) have no longer any place or meaning in the theory. To
appreciate how formidable this revision of the classical theoresis
of spacetime structure and gravity, and especially of its
potentially quantal albeit persistently differential manifold
based versions, one may bring forth Weinstein's words from
\cite{weinstein} about space-time observables (:measurable
dynamical quantities) in non-perturbative canonical QG and how the
presence of the $\mathrm{Diff}(M)$-constraint group associated
with the background smooth manifold $M$ make the whole enterprize
of regarding and treating (canonical) QGR as a quantum gauge
theory proper an almost impossible task:

\bigskip \noindent (Q7.42)\hskip 0.9in
\begin{minipage}{11cm}
\noindent ``{\small ...In this paper, I show that general
relativity is not a gauge theory at all, in the specific sense
that gauge theory has in elementary particle physics. This issue
is of crucial importance in attempts to quantize general
relativity, because {\em in quantum theory, the generators of
gauge transformations are emphatically not treated as observables,
while the generators of spatiotemporal (e.g., Lorentz)
transformations are in fact the canonical
observables}.\footnote{Our emphasis.} Thus the discussion in this
paper sheds light on the origin of some of the deep and
longstanding difficulties in quantum gravity, including the
problem of time, a familiar form of which arises from treating the
parametrized time-evolution of canonical general relativity as a
gauge transformation\footnote{The `parametrized time-evolution'
referred to above is nothing else but the time-diffeomorphisms
generated by the Hamiltonian (constraint) in canonical GR regarded
as a constrained gauge system like the other three gauge forces of
matter. The main point that Weinstein wishes to make in
\cite{weinstein} is precisely that (quantum canonical) GR should
not (in fact, it {\em cannot}) be viewed as a (quantum) gauge
theory exactly because the $\mathrm{Diff}(M)$ group of the
background manifold $M$ is {\em not} a local gauge group (the
structure group of a principal fiber bundle) like in
electrodynamics or the Yang-Mills theories of matter.}...}''

\end{minipage}

\vskip 0.1in

\noindent Thus in the base spacetime manifoldless ADG-gravity
there are no spacetime measurements as such and, concomitantly, no
spacetime observables either. That we actually do {\em not}
measure the space and time locution of quantum fields and their
elementary particles has been convincingly argued in the classic
pair of papers by Bohr and Rosenfeld \cite{bohr1,bohr2}; moreover,
here, in the pure gauge field-theoretic context of ADG, we {\it a
fortiori} posit that {\em there is no space-time `observables' to
measure in the first place}. Having said this, there actually are
abstract acts of measurement and localization of the ADG-fields:
these are effected precisely when one introduces into the theory
the abstract sheaf $\struc$ of generalized arithmetics or
`coordinates', which in turn locally `analyzes' the carrier
(representation sheaf) space $\modl$ of the field $(\modl ,\conn)$
into $\struc^{n}$ and identifies the field's quantum particle
states with its local sections. Furthermore, we assume that if
there is any (geometrical) space(time) at all in the theory, then
it is effectively encoded in $\struc$ (Gel'fand
duality).\footnote{In the same way that in the classical theory
(CDG) $M$ is recovered from $\struc\equiv\smooth_{M}$ by Gel'fand
representation theory. In other words, it is the, external to the
gravitational field $\conn$, `measurer' or `geometer' (`observer')
who geometrizes (localizes and measures in `spacetime' terms) the
gravitational field by bringing in her own $\struc$.} On the other
hand, since we have only fields and no spacetime (field
solipsism), we have {\em field observables} in the theory. As we
have repeatedly emphasized in the present work, as field
observables we regard the `geometrical objects' in the theory.
These are the $\otimes_{\struc}$-tensors ({\it ie}, the
$\struc$-sheaf morphisms), with the curvature $\curv(\conn)$ of
the connection being the archetypical one. In turn, since the
gravitational field dynamics is expressed via the curvature field
observable, the dynamics is $\struc$-functorial (`synvariant'),
and the field `sees through' our generalized measurements in
$\struc$ (and {\it in extenso}, through the `geometrical
spacetime' encoded in $\struc$)---in other words, our measurements
`respect' the field dynamics (PFR). As a result, neither our
generalized measurement acts are field-perturbing

\end{itemize}

\paragraph{The Kuhnian gist of ADG: the physical unreality of spacetime.} One could say that the central thesis in
Kuhn's seminal work \cite{kuhn} is that scientific revolutions
happen when a different (from the traditional, `standard' ones)
theoretical paradigm---essentially, a different theory---is
adopted by scientists to `look at the World'. Indeed, theoretical
paradigms or `theories' are thought of by Kuhn as frameworks
within the boundaries and from the perspective of which we view
what's `out there'.\footnote{In this respect the term `theory' is
in line with the original Greek
`$\theta\epsilon\omega\rho\acute{\iota}\alpha$'---way of looking
at (and thinking about!) the World.} There is no physical reality
(and objective Truth!) out there that we are trying to capture (or
even approximate) with our theories. Whatever we can utter about
the world we can do so from within the confines of the theoretical
paradigm we have adopted. In a deep sense, a theory really
determines what we see and the means we devise in order to see
`it'.\footnote{That is, experiments are planned and experimental
apparatuses are designed according to a theory. In a deep sense,
theory comes before experiment.}

In the particular case of ADG and its application to ADG-gravity,
all differential geometry and its application to gravity boils
down to our choice of $\struc$. There is no {\it a priori} posited
(smooth) space(time) geometry `out there'; only the one that {\em
we} carry in the algebraic baggage $\struc$ we have adopted. The
significant paradigm shift that ADG and ADG-gravity brings about
is that no external smooth space(time) is required in order to do
differential geometry (and apply it to gravity); one simply adopts
an $\struc$ (suitable for tackling the particular physical problem
one has in mind, and providing one with the basic $d$ with which
one can actually do DG) via which then one `looks at the world'
with `geometrical eyes'. Of course, the physical objectivity
(reality) of the world ({\it ie}, of the field) is secured in
precisely that the physical laws that the fields ({\it viz.}
connections) define are $\struc$-functorial ({\it ie}, `invariant'
under changes of `point of view' $\struc$), which in turn entails
the PARD and its expression as a natural transformation, as well
as the PFR (field solipsism).

Also quote Feynman from \cite{feyn3}:

\bigskip \noindent (Q7.43)\hskip 0.9in
\begin{minipage}{11cm}
\noindent ``{\small {\bf Q:} So these infinities have plagued
quantum field theory for over a generation. Do you think that a
fundamental theory of different particle interactions can still
contain these infinities? Or do you think that Dirac was right to
say he couldn't believe any theory that contained these
infinities?

\noindent {\bf A:} Well, obviously there are no infinities in
observation---the mass of the electron is not infinite... Now,
[renormalization technology aside,] {\em it should be possible one
day for someone to work out more carefully in a different way a
set of equations in which there aren't any infinities and which
have the same consequences. I don't mean by inventing a new
physics, but rather by reorganizing the statement of what it is
you do to make the calculations less awkwardly written. So it's
just a matter of mathematical technology in that case...It
therefore must be possible to say what the result is without going
through the infinities. So I think that those infinities are
somehow technical. We're formulating the theories incorrectly when
we first write them down}\footnote{Our emphasis.}...}''

\noindent ``{\small {\bf Q:} Of course, the really tough problem
as far as the infinities are concerned is gravity...How to solve
the problem of divergences?

\noindent {\bf A:} ...The question is whether gravity has to be a
quantum mechanical theory, like the other quantum mechanical
phenomena associated with the other particles. It doesn't seem
possible to have the world partly classical and partly quantum
mechanical. Therefore, for example, the fact that you can't
observe a position and a momentum at the same time with arbitrary
accuracy---which is what we know from quantum mechanics---should
apply to gravity also. We shouldn't be able to use gravitational
forces to determine the position and momentum of a particle beyond
a certain accuracy, because we'd run into an inconsistency. In
trying to modify gravity theory to make it into a quantum theory
we discover infinities just like we did in electrodynamics, but
which are much more difficult to sweep under the rug. They're much
more serious. I don't know how gravity fits in these things, but
it has to fit in. It presents a very large number of problems
beside the infinities.

In the quantum field theories, there is an energy associated with
what we call the vacuum in which everything has settled to the
lowest energy; that energy is not zero---according to the theory.
Now gravity is supposed to interact with every form of energy and
should interact then with this vacuum energy. And therefore, so to
speak, a vacuum would have a weight---an equivalent mass
energy---and produce a gravitational field. Well it doesn't! The
gravitational field produced by the energy in the electromagnetic
field in a vacuum---where there's no light, just quiet,
nothing---should be enormous, so enormous, it would be obvious.
The fact is, it's zero! Or so small that it's completely in
disagreement with what we'd expect from the field theory. This
problem is sometimes called the cosmological constant problem. It
suggests that we're missing something in our formulation of the
theory of gravity.}'' (PTO$\mapto$)

\end{minipage}

\vskip 0.1in

\bigskip \noindent (Q7.43---cont'd)\hskip 0.9in
\begin{minipage}{11cm}
\noindent ``{\small It's even possible that the cause of the
trouble---the infinities---arises from gravity interacting with
its own energy in a vacuum. And we started off wrong because we
already know there's something wrong with the idea that gravity
should interact with the energy of a vacuum. So I think the first
thing we should understand is how to formulate gravity so that it
doesn't interact with the energy in a vacuum. Or maybe we need to
formulate the field theories so there isn't any energy in a vacuum
in the first place. {\em In other words, there are some mysteries
associated with the problem of quantizing gravity which go beyond
the infinities. They have to do with the formulation of the theory
in the first place}.\footnote{Again, emphasis is ours.}

\noindent ``{\small {\bf Q:} There are also some conceptual
issues. If you're applying quantum mechanics to gravity, then in a
sense you're applying quantum mechanics to space and time...}''}

\end{minipage}

\vskip 0.1in

\noindent There is a plethora of issues raised by Feynman above
that we could address under the prism of ADG and ADG-gravity:

\begin{enumerate}

\item First thing to note is that Feynman, in view of the
unphysical infinities that plague (quantum) field theories of
matter and especially (in attempts to quantize) GR, explicitly
maintains that for tackling them {\em no new physics is needed,
but rather new mathematical technology}. In his own words, it is
as if ``{\em we are formulating the theories incorrectly when we
first write them down}''. Indeed, that's completely in line with
the gist of ADG-gravity: we hold that we do not propose some new
physics---as it were, new physical laws---for after all Einstein's
equations (\ref{eqy23}) formally remain the same in our theory and
one could say, as Feynman insists, that {\em the ADG-gravitational
equations have the same `consequences' as in the original,
manifold based theory} (GR).\footnote{For the dynamical law
(Einstein equations) remains formally the same, while the
classical theory can be simply recovered by assuming
$\struc\equiv\smooth_{X}$.} However, what radically changes is the
theoretical-mathematical framework and technology via which we
view and {\em physically interpret} them `same'
equations.\footnote{Which from a semantic point of view they are
not the same anymore!} What really changes is the way we arrive
at, write and interpret the law of gravity in the first place.

\item The said way on the one hand forces no background geometrical
smooth spacetime interpretation of the equations and the
underlying mathematical formalism (CDG), so that the theory {\it
ab initio} encounters no problem of unphysical infinities, and on
the other, it allows us to interpret them quantum mechanically
from the very start. That is, we suggest that when Feynman says
that we are formulating gravity incorrectly in the first place,
and apart from the `physical reason' he gives ({\it ie}, the
gravitational interaction with the vacuum energy)
infinities-aside, from the perspective of ADG-gravity this
corresponds to the `classical' or traditional way we formulate GR
by the CDG-technology, in effect, by assuming a smooth spacetime
manifold in the first place. Then our Calculus (mathematical
technology) glaringly miscarries with quantum theory ({\it eg},
the appearance of singularities and meaningless infinities)---{\it
ie}, when one tries to apply quantum ideas to the base manifold
and CDG-based relativistic field theory of gravity (GR).

\item On the other hand, the inherently spacetime manifoldless
ADG-gravity goes against Feynman's `pessimistic anticipation'
maintaining that any attempt to localize ({\it ie}, measure the
position and momentum) of a particle beyond certain accuracy by
using gravitational fields will run into an inconsistency ({\it
eg}, an infinity for a physically measurable physical quantity).
That is to say, we do not regard such inconsistencies as {\em
physical} problems, but as shortcomings of the very
theoretical-mathematical framework and formalism (the manifold
based CDG) within which we formulate the physical theory in the
first place. Our disagreement with Feynman becomes even more
prominent when he says at the end of the quotation above that
``{\em if you're applying quantum mechanics to gravity, then in a
sense you're applying quantum mechanics to space and time}''.
Under the prism of ADG-gravity, not only we do not apply quantum
mechanics to ({\it ie}, quantize) gravity (as a field theory), but
also in the first place there is no space-time in the theory to
quantize. ADG-gravity is a background spacetimeless, third gauge
and third quantum field theory in which no singularity or infinity
is involved at all.

\end{enumerate}

\subsubsection{Questions of Renormalizability Questioned and
`Renormalized'}

Let us first consider some remarks by Feynman about
renormalization found in \cite{feyn0}:

\bigskip \noindent (Q7.44)\hskip 0.9in
\begin{minipage}{11cm}
\noindent ``{\small\em ...{\small\rm [Due to nonsensical
infinities],}\footnote{Our addition for textual continuity.} I
don't think we have a completely satisfactory relativistic
quantum-mechanical model...Therefore, {\small\em I think that the
renormalization theory is simply a way to sweep the difficulties
of the divergences of electrodynamics under the rug}\footnote{Our
emphasis.}...}''

\end{minipage}

\vskip 0.1in

\noindent and attune it with Dirac's well documented not accepting
the QFT's infinities, as well as for his optimistic vision that a
time will come when a truly relativistic and genuinely
infinities'-free quantum mechanics will be developed, as quoted
below from \cite{dirac1}:

\bigskip \noindent (Q7.45)\hskip 0.9in
\begin{minipage}{11cm}
\noindent ``{\small ...This was all satisfactory so long as one
considered only a single particle. There remained, of course, the
problem of two or more particles interacting with each other. Then
one soon found that there were serious difficulties. Applying the
standard rules, all one could say was that the theory did not
work. The theory allowed one to set up definite equations.}
{\small\em When one tried to interpret those equations, one found
that certain quantities were infinite according to the theory,
when according to common sense they should be
finite}.\footnote{Our emphasis.} {\small That was a very serious
difficulty in the theory, a difficulty that still has not been
completely resolved.}

{\small\em Physicists have been very clever in finding ways of
turning a blind eye to terms they prefer not to see in an
equation}.\footnote{Our emphasis.} {\small They may go on to get
useful results, but this procedure is of course very far from the
way in which Einstein thought that nature should work.}

{\small\em It seems clear that the present quantum mechanics is
not in its final form}.\footnote{Again, our emphasis.} {\small
Some further changes will be needed, just about as drastic as the
changes made in passing from Bohr's orbit theory to quantum
mechanics}. {\small\em Some day a new quantum mechanics, a
relativistic one, will be discovered, in which we will not have
these infinities occurring at all\footnote{Once again, our
emphasis.}...}''

\end{minipage}

One may start with the observation that singularities are already
insuperable problems within the classical theory of gravity (GR
based on CDG), as it were, long before the `quantization' of that
theory becomes an issue. Yet, some people, including Ashtekar,
Hawking, Joshi and Penrose to name a few, have hoped that a
genuine QG will not only shed more light on the problem of
singularities, but also it may ultimately resolve
them.\footnote{Well, the exchange between Hawking and Penrose in
(Q7.?) below shows rather that while SWH starts from singularities
as being `given'---hopefully to be `abolished' by a quantization
of the classical theory (GR), RP intuits a drastic revision of
``{\em spacetime at a singularity}''---a {\em singularity
manifold} so to speak---according to which, so to say, ``{\em a
true quantum gravity does not quite remove singularities}''; or in
other words, {\em he does not expect that a head-on, `blind'
quantization of the classical theory will resolve the singularity
problem} ({\it ie}, in a sense SWH's question is begging the
question!). On the other hand, RP expects such a revision of the
concept of spacetime at a singularity to lay bare and clear what
is a singularity in GR, thus resolving the puzzle that Geroch puts
forth in \cite{geroch}.} To that we have the testimony of recent
expectations and remarkable results(!) in Loop Quantum Gravity and
Cosmology \cite{boj1,modesto,husain}. {\it Prima facie}, such
hopes could derive for instance from the paradigm of the weaker
(than gravitational singularities) but still troublesome
infinities assailing QFT, since the quantization of classical
field theories enlightened us significantly about the nature
of---in QED, for example---the electromagnetic (photon) field
right at the electron source, as well as about the (mathematical)
nature of quantum fields in general ({\it eg}, their smeared out,
distributional character). Admittedly, if it was not for the QFT
infinities, the entire renormalization (group) program, the
appearance of anomalies and other physically important issues
({\it eg}, spontaneous symmetry breaking and the Higgs mechanism,
confinement and asymptotic freedom, topological aspects of gauge
theories) that were of enormous conceptual, technical and
calculational import to our current understanding of the
fundamental gauge forces of matter, would simply have not come
about \cite{jackiw}.

\paragraph{'t Hooft's vision following Feynman: to replace `Analysis' in QFT by perturbation diagrams and
emphasize the importance of (renormalizable) gauge theory
(especially in QG).} Relevant to the gist of the discussion above
would be to recall Feynman's mistrust of (differential) geometric
ideas and associated `calculational technology' (:``{\em fancy
schmanzy differential geometry}'') in (Q4.??), and instead his
preference to tackle quantum gravity issues expressly
perturbatively, along QFTheoretic lines \cite{feyn,feyn2}. That
is, he chose to first write down the relational-finitistic
(combinatorial) Feynman diagrams for gravity, and then look for a
possible geometrical (`spacetime') interpretation of the
calculational rules and algebra accompanying them; moreover, he
insisted that emphasis should be placed on the {\em
gauge-theoretic} character of gravity,\footnote{Especially for the
quantization of gravity, the formalism that appears be more
suitable---albeit more heuristic---(than the canonical) is the
path integral one. By the way, this is exactly what we espouse in
the present paper-book {\it vis-\`a-vis} ADG-gravity and its
quantum theoresis (see 7.9.1 below).} rather than on its original
(and `accidental'!) `spacetime-(geo)metric' one due to
Einstein.\footnote{Especially for the quantization of gravity, the
canonical formalism, which relies heavily on differential
geometric (CDG) ideas, appears to be the more geometrically bent
(and mathematically/analytically more rigorous than the functional
integral) approach.} This attitude, strategy and method
foreshadowed and influenced significantly subsequent developments
in quantum gauge field theory research in general, as we witness
for example in 't Hooft and Veltman's introductory remarks in
their celebrated paper \cite{thooft6}:

\bigskip \noindent (Q7.46)\hskip 0.9in
\begin{minipage}{11cm}
\noindent ``{\small With the advent of gauge theories it became
necessary to reconsider many well established ideas in quantum
field theories. The canonical formalism, formerly regarded as the
most conventional and rigorous approach, has been abandoned by
many authors. The path-integral concept cannot replace the
canonical formalism in defining a theory, since path integrals in
four dimensions are meaningless without additional and rather ad
hoc renormalization prescriptions.

Whatever approach is used, the result is always that the S-matrix
is expressed in terms of a certain set of Feynman diagrams. {\em
Few physicists object nowadays to the idea that diagrams contain
more truth than the underlying formalism, and it seems only
rational to take the final step and abandon operator formalism and
path integrals as instruments of analysis}.\footnote{Our
emphasis.}

Yet it would be very shortsighted to turn away completely from
these methods. Many useful relations have been derived, and many
more may be in the future. What must be done is to put them on a
solid footing. {\em The situation must be reversed: diagrams form
the basis from which everything must be derived. They define the
operational rules},\footnote{Again, our emphasis.} and tell us
when to worry about Schwinger terms, subtractions, and whatever
other mythological objects needs to be introduced.

The development of gauge theories owes much to path integrals and
it is tempting to attach more than a heuristic value to path
integral derivations. Although we do not rely on path integrals in
this paper, one may think of expanding the exponent of the
interaction Lagrangian in a Taylor series, so that {\em the
algebra of the Gaussian integrals becomes identical to the scheme
of manipulations with Feynman diagrams}.\footnote{Emphasis is
ours.} That would leave us with the problems of giving the correct
$i\epsilon$ prescription in the propagators, and to find a decent
renormalization scheme.

There is another aspect that needs emphasis. {\em From the outset
the canonical operator formalism is not a perturbation theory,
while diagrams certainly are perturbative objects. Using diagrams
as a starting point seems therefore to be a capitulation in the
struggle to go beyond perturbation theory. It is unthinkable to
accept as a final goal a perturbation theory, and it is not our
purpose to forward such a notion}.\footnote{Once again, our
emphasis.} On the contrary, it becomes more and more clear that
perturbation theory is a very useful device to discover equations
and properties that may hold true even if the perturbation
expansion fails...}''

\end{minipage}

\noindent These words should really be coupled (in order to gain
more weight) to some other remarks that 't Hooft made about QG
proper a little bit later, in \cite{thooft}:

\bigskip \noindent (Q7.47)\hskip 0.9in
\begin{minipage}{11cm}
\noindent ``{\small ...[In the beginning of the paper, after 't
Hooft mentions that in all attempts to quantize gravity one
encounters fundamental natural units of length, time and mass
(energy) in the form of $\ell_{P}$, $t_{P}$ and $E_{P}$, he
continues:]\footnote{These are the Planck scales for length, time
and energy (mass), respectively. Our addition for textual
continuity and completeness.} But then the theory contains a
number of obstacles. First there are the conceptual difficulties:
the meaning of space and time in Einstein's general relativity as
arbitrary coordinates, is very different from that of space and
time in quantum mechanics. {\em The metric tensor $g_{\mu\nu}$,
which used to be always fixed and flat in quantum field theory,
now becomes a local dynamical variable}.\footnote{Our emphasis.}

Advances have been made, from different directions,\footnote{Here
the author gives a couple of references, which we omit.} to devise
a language to formulate quantum gravity, but then the next problem
arises: {\em the theory contains essential infinities such that a
field theorist would say: it is not renormalizable. This problem
may be very serious. It may very well imply that there exists no
well determined, logical, way to combine gravity with quantum
mechanics from first principles. And then one is led to the
question: should gravity be quantized at all?}\footnote{Our
emphasis again.}...[to the conclusion of the paper]

...Even so, a renormalized perturbation expansion [for
gravity]\footnote{Our addition for clarity.} would only be a small
step forward. At very small distances the gravitational effects
must be large, because of the dimension of the gravitational
constant, so the expansion would break down at small distances
anyhow. {\em We have the impression that not only a better
mathematical analysis is needed, but also new physics. What we
learned is that in such a theory the metric tensor might not at
all be such a fundamental concept}.\footnote{Again, emphasis is
ours.} In any case, its definition is not unambiguous.}''

\end{minipage}

Yet, in spite of all this, we still have Dirac's haunting
criticism of this {\it ad hoc} removal of infinities in QFT by the
procedure of renormalization:

\bigskip \noindent (Q7.48)\hskip 0.9in
\begin{minipage}{11cm}
\noindent ``{\small\em Sensible mathematics involves neglecting a
quantity when it turns out to be small---not neglecting it just
because it is infinitely great and you do not want it.}''
\cite{dirac}
\end{minipage}

\vskip 0.1in

\noindent even more so when we try to apply conventional quantum
ideas---`quantize' as it were---not just to matter as in QFT, but
also to spacetime and gravity itself:

\bigskip \noindent (Q7.49)\hskip 0.9in
\begin{minipage}{11cm}
\noindent ``{\small\em Our present quantum theory is very good
provided we do not try to push it too far...We do not try to apply
it to particles with very high energies and we do not try to apply
it to very small distances. When we do try to push it in these
directions we get equations which do not have sensible solutions.
We get interactions always leading to infinities...It is because
of these difficulties that I feel the foundations of quantum
mechanics have not yet been correctly established. Working with
the present foundation people have done an awful lot of work in
making application in which they can find rules for discarding the
infinities but these rules, even though they may lead to results
in agreement with observations, are artificial rules, and I just
cannot accept that the present foundations are
correct.\footnote{All emphasis is ours.}}'' \cite{goldman}
\end{minipage}

\vskip 0.1in

\noindent Quite, for as we argued earlier in this subsection in
connection with our background spacetimeless ADG-gravity based
doubts about the importance of the fundamental physical constants
and their conspiring to setting a supposedly natural scale below
which classical spacetime and gravity break down signifying at the
same time the onset of quantum gravitational effects whose law
still eludes us, from a quantum gauge field-theoretic viewpoint
the issue of renormalization (and renormalizability of a gauge
field theory) is intimately tied to the assumption of a spacetime
continuum, hence it too must be questioned in the light of ADG.
But let us first quote again 't Hooft on this from the
introduction to \cite{thooft7}:

\bigskip \noindent (Q7.50)\hskip 0.9in
\begin{minipage}{11cm}
\noindent ``{\small ...Space and time are continuous. This is how
it has to be in all our theories, because it is the only way known
to implement the experimentally established fact that we have
exact Lorentz invariance. {\em It is also the reason why we must
restrict ourselves to renormalizable quantum field theories for
elementary particles}.\footnote{Our emphasis.} As a consequence,
we can consider unlimited scale transformations and study the
behavior of our theories at all scales. This behavior is important
and turns out to be highly nontrivial. {\em The fundamental
physical parameters such as masses and coupling constants undergo
an effective change if we study a theory at a different length and
time scale, even the ones that had been introduced as being
dimensionless. The reason for this is that the renormalization
procedure that relates these constants to physically observable
particle properties depends explicitly on the mass and length
scale used}.\footnote{Again, our emphasis.}}''
\end{minipage}

\vskip 0.1in

\noindent We may summarize and highlight the main points in all
the quotations above regarding how a quantum gauge field theoresis
of gravity fares {\it vis-\`a-vis} renormalization in the light of
our ADG-gravity paradigm, as follows:

\begin{itemize}

\item One, like Feynman originally did, may abandon Analysis
up-front and attempt to head-on quantize gravity diagrammatically
(perturbatively). Then, especially viewing gravity as a gauge
theory (like the other renormalizable fundamental quantum gauge
forces of matter), sooner or later the procedure of
renormalization must be evoked.

\item Renormalizability is intimately tied to the assumption of a
spacetime continuum \cite{thooft8,collins}, which of course shows
no {\it a priori} `preference' for a fundamental scale
(regularization `cut-off'), while at the same time it is
`responsible' for the singularities and associated infinities in
field theory---the {\it raison d'\^etre et de faire} of
renormalization in the first place. Indirectly, in a strong sense
Analysis is still to be blamed.

\item As a result, a theoretical physicist with a particle physics
bent (:QFTheorist) would be entirely content with such a situation
if gravity did not prove to be perturbatively non-renormalizable
(for one thing, due to the dimensionful Newton constant),
something that may be turned around to a `non-problem' if one
wishes to concede that there actually is a minimal space-time
length-duration in Nature---an inherent regularization `cut-off'
scale in the elusive QG theory one is after.\footnote{See again
the Ashtekar quotes (Q7.?) and (Q7.?) above.}

\item Then one is in principle bound to recognize that QG `exists'
non-perturbatively after all (and the theoretical physicist ceases
being a die-hard particle theorist!). Of course, such a
`conversion' of a `perturbative particle physicist' to a
`non-perturbative quantum gravitist' is not without its technical
cost and conceptual concessions: for one thing, the
`renormalization rug' the particle physicist used to sweep the
matter gauge field-infinities under is now, in the context of
gravity, replaced by the Planck scale, with the concomitant
relegation of the elusive `true' QG law below
it.\footnote{Metaphorically speaking, the renormalization rug is
now replaced by the black hole horizon-membrane, and only a
quantum theory of gravity is supposed to tell us what is going on
in its interior, as it were below the Planck scale
\cite{thooft2,thooft4,thooft5,thooft7}.} This means that the
spacetime continuum, which is invaluable for renormalization, is
given up, and also, inevitably, the standard Analysis (CDG)
supported by it.

\item We appear to have gone in circles: in the beginning we
abandoned Analysis to work diagrammatically
(relationally-combinatorially) {\it \`a la} Feynman. When we
realized that the consistency of the rules of playing with these
diagrams and the physical sensibility of the interpretation of the
results of the diagrammatic method are dependent in one way or
another on the spacetime continuum and its (differential/integral)
Calculus---in fact, they fundamentally presuppose it\footnote{For
example, in the continuum one can in principle integrate over
infinite momenta-energies, or dually, over arbitrarily short
distances-time durations.}---we decided to go `non-perturbative'
and we in a strong sense reinstated the background differential
spacetime manifold and its CDG, only to get rid of it by the
process of quantization.\footnote{This is what has happened in LQG
(and its affine LQC and QRG theories) for example where, although
QG is approached non-perturbatively as a quantum gauge theory in a
manifestly background metric independent way, a differential
manifold is still retained as a base, only to be quantized in the
later stages of the development of the theory, while at the same
time the spacetime quantization is used to show how to bypass and
`resolve' the initial cosmological and (Schwarzschild) black hole
singularities \cite{boj1,modesto,husein}.}

\end{itemize}

\noindent In view of all this, we are confident to say that the
background spacetimeless ADG may be the ``better'' (than the
continuum based CDG) ``mathematical analysis needed'' (Q7.?) in
QG. In the new mathematical ADG-framework, not only (quantum)
gravity is regarded as a (quantum) gauge theory (of the third
kind), but also exactly because no external (to the gauge
gravitational field itself), background spacetime manifold is
involved at all, the (non-)renormalizability of gravity along
QFTheoretic lines, as well as the appearance of a minimal
spacetime scale are really `non-problems'/`non-issues' in the
theory. Let alone that no singularities and associated unphysical
infinities plague ADG-gravity; hence, like we asked about the
Planck length above: `{\em whence renormalization of gravity?}'

We would like to conclude this subsection by paralleling the way
we view ADG's role in QG against 't Hooft's critique of (the
mathematics of) (super)string {\it vis-\`a-vis} QG from the
introduction to \cite{thooft7}---a critique that is similar to his
words quoted before (Q7.?) about a better analysis and a new
physics in the Planck regime:

\bigskip \noindent (Q7.51)\hskip 0.9in
\begin{minipage}{11cm}
\noindent ``{\small ...The same objection [about whether general
relativity is genuinely united with quantum
mechanics]\footnote{Our addition for textual clarity and
continuity.} can be brought against the `superstring' approach to
quantizing gravity. The space-time in which the superstring moves
is a continuous space-time, and yet we have a distance scale at
which a smooth metric becomes meaningless. On the other hand a
flat background metric is usually required at ultrashort
distances, even in string theories. {\em It is my conviction that
a much more drastic approach is inevitable. Space-time ceases to
make sense at distances shorter than the Planck length. Here again
I reject a purely mathematical attack, particularly when the math
is impressive for its stunning complexity, yet too straightforward
to be credible. The point here is also that our problem is not
only a mathematical one but more essentially physical as well:
what is it precisely that we want to know, and what do we know
already?}\footnote{Our emphasis.}}''
\end{minipage}

\vskip 0.1in

\noindent In certain ways we agree with 't Hooft, while in others
we disagree. For starters, it is true that (perturbative) string
theory still effectively uses a base  spacetime continuum and a
smooth background metric, thus we agree with his dissatisfaction
about that. But then we disagree with both his invoking of a
fundamental scale as a counter-argument against the smooth
manifold, and with his doubts that a different mathematical
framework and approach to QG will enlighten the problem. At the
same time, we share his conviction that the problems of QG are
deeply conceptual (not just technical) ones (Q?.?), but also think
that a fundamental revision and scrutiny of the basic conceptual
aspects of the mathematics (in particular, DG) that we use to
formulate (and concomitantly {\em interpret}!) our physical
theories is what is really needed in current QG research: a new
mathematical-theoretical framework is pregnant to new ideas and
opens new routes of interpretation.

\subsection{All is Quantum: `ADG-Gravity' is `Inherently Quantum'---a Summary of our Credo (so far)}

We would like to use this subsection to summarize in an
`aphoristic' way our theses about ADG-gravity {\it vis-\`a-vis}
QG. Let us itemize them:

\begin{itemize}

\item From the ADG-viewpoint {\em all is field, and the field is third gauge and third quantum}.

\item {\em The ADG-gravitational field}, represented by a
dynamically autonomous, purely algebraic connection $\conn$, {\em
is `already' or `inherently' quantum, and therefore in no need of
a formal procedure of `quantization'}.

\item It follows that {\em there is no formal distinction in the
theory between a `classical' and a `quantum' domain}. Strictly
speaking, there is no analogue of Bohr's correspondence principle
(and the issue of the emergence of classicality) in our
ADG-theoresis of gravity.\footnote{No `GR as low energy limit of
QG'.}

\item Since the theory is fundamentally background spacetimeless
(whether the latter may be taken to be a classical continuum or a
quantal discretum), {\em no need arises to quantize spacetime
itself}.

\item It follows that, from our vantage, {\em the problem of QG is not fundamentally
related to the problem of the quantum structure of space and
time}.

\item The finitary, causal and quantal gravitational field, as
represented \`a la ADG by a connection on a finsheaf of qausets
\cite{malrap1,malrap2,malrap3,rap5}, is already geometrically
(pre)quantized. (\ref{eqy23}), in our view, {\em represents the
quantum Einstein equations}.

\item {\em The so-called classical gravity (GR) pathologies ({\it eg},
singularities) as well as the unphysical infinities of the
manifold and hence CDG-based quantum gauge field theories of
matter are not problems and shortcomings of the physics ({\it ie},
of the physical laws), but of the mathematics used to represent
and calculate the outcomes (thus interpret as well as draw
consequences from) those physical laws---{\it ie}, they are
anomalies of our background manifold mediated and effectuated
Calculus (CDG)}.

\end{itemize}

\subsection{Sheaf Theory, the `Narrow' and `Delicate' Passage from-Local-to-Global-and-Back, and
its Potential Import to QG Research}

\subsubsection{Brief technical generalities}

First let us make it clear up-front that the notion of {\em sheaf}
has {\em by definition} (and by construction!) both {\em local}
and {\em global} characteristics built in itself. To begin with,
and technically speaking for a little while, a sheaf is a {\em
local-topological} structure (or construction), being defined as
we saw earlier as a {\em local homeomorphism} between a base
topological space $X$ and the sheaf space $\Phi$
\cite{bredon,mall1}. In more detail, and for functional sheaves in
particular,\footnote{That is, sheaves of functions $\phi$ (on
$X$), such as those to which all physical fields are supposed to
belong.} one begins by gathering field-information (`field-data')
$\phi$ locally in $X$---{\it ie}, `field-values' for every open
subset $U$ in $X$. This collection of local `data' about $\phi$ is
then thought of as constituting a so-called  {\em presheaf} (of
functions $\phi$ subjected to restriction maps relative to the
lattice of open subsets $U$ of $X$) \cite{bredon,mall1,rap2}.
Subsequently, a sheaf is constructed from the said presheaf, by a
procedure called {\em sheafification}, essentially by {\em
collating, gluing, or stitching together}, the aforesaid local
information about $\phi$\footnote{In particular, {\em compatible}
local information about the field---{\it ie}, $\phi$-data or
`values' which agree on the overlap ($U\cap V$) of the open
regions of $X$ ($U,V\subset X$).} into the {\em global sheaf
space} $\Phi$ in such a way that one can show that $X$ and $\Phi$
are {\em topologically equivalent, at least locally} ({\it ie},
they are locally homeomorphic to each other).\footnote{As a matter
of fact, functional presheaves `{\em locally characterized}' are
{\em complete}, which is tantamount to their being sheaves
\cite{mall1,mall2,mall8}.} In fact, one can go as far as to say
that

\vskip 0.1in

\centerline{{\em ``sheafification is localization''}
\cite{mall8}.\footnote{A very powerful statement indeed, at least
{\it vis-\`a-vis} physics, since we have argued throughout our
past trilogy \cite{malrap1,malrap2,malrap3} that, in turn, ``{\em
localization is gauging}'' and concomitantly ``{\em gauging
entails dynamical variability}''. Thus, {\it in toto}, {\em sheaf
theory appears to be the appropriate language in which to
formulate dynamical physical theories of a gauge character}---{\it
ergo} ADG(!), especially if one is faithful to one's intuition (or
simply theoretical choice) that the dynamical laws of physics
should be represented by {\em differential} equations proper ({\it
ie}, that some sort of differential geometry should be formulated
along sheaf-theoretic lines).}}

\vskip 0.1in

\noindent Concomitantly, a very instrumental notion in sheaf
theory, and one that plays a crucial role in the aforementioned
stitching up $\Phi$ from $\phi|_{U}$, is that of the {\em
continuous (local) sections} of a sheaf---maps $s$ from a
$U\subset X$ to $\Phi$ that are themselves local homeomorphisms
\cite{mall1}. In fact, that the notions of a sheaf and of (the
collection of) its (local) sections (for all $U\subset X$) are
tautosemous is captured by the well known `slogan-result' in sheaf
theory that {\em a sheaf is its (continuous local)
sections}---that is to say, that the (compatible) local
information about the field (as encoded in the local sections) is
glued together in such a `{\em coherent}' way so as to comprise
the total or global sheaf space, which conversely can be analyzed
or `decomposed' at will into the local sections that make it up
\cite{mall1,mall8}.\footnote{Alternatively, a sheaf may be viewed
as a {\em fibered space}, with the total space $\Phi$ being the
set-theoretic (disjoint) union (over all the points $x$ of the
base topological space $X$) of its {\em fibers} or {\em stalks}
$\Phi_{x\in X}$ which, as we noted earlier in this paper, are
inhabited by the so-called {\em germs} of its continuous local
sections---the `ultralocal' elements of a sheaf
\cite{mall1,rap2}.} In this (mathematical) sense, {\em a sheaf
models effectively the transition from local to global, and back}.

\subsubsection{From local (`micro'/particle) to global
(`macro'/field) and back: the essence of sheaf theory vis-\`a-vis
field theory \`a la ADG}

After these brief technical generalities about how the
mathematical concept and structure of a sheaf is suitable to
accommodate the transition from local to global and
back,\footnote{To stress it again, the direction `{\em
from-local-to-global}' is the process of gluing (or composition)
of local function (field) data that sheafification achieves, while
the opposite direction, `{\em from-global-to-local}' refers to the
(dis)section (or analysis) of the total sheaf space into its local
(continuous) sections that ultimately constitute it.} we wish to
bring forth and discuss recent tendencies of (mathematical)
physicists to apply sheaf theory to different research areas of
theoretical physics---from QFT and its infinities, to GR and its
singularities, and hopefully to QG---exactly thanks to the
aforesaid virtue of the notion of sheaf in helping one attain both
a worm's (local/particle/`micro') and a bird's
(global/field/`macro') eye-view of field theory.

In particular, we will focus on how ADG's conception and
definition of a field, as being represented by a pair $(\modl
,\conn)$, is tailor-cut for implementing that desired transition
from local to global, and conversely, especially in the context of
QG. In the process, we will juxtapose, as a {\it contrapunctus} as
it were, certain aspects of Feynman's `global' ({\it ie}, entire
spacetime histories of a) particle viewpoint of QED {\it versus}
the more `commonplace' perspective on QED as a local QFT based on
local field interactions and differential equations modelling
(local) dynamical propagations (in the Minkowski continuum), as
concisely described in his Nobel prize address \cite{feyn0}.

So for starters, in Feynman's recollection of how he came to
develop the spacetime (histories) view of QED \cite{feyn0}, we
witness two motivations that led him to that development:

\begin{itemize}

\item i) the infinite self-energy of the electron: ``{\em The first
{\rm [source of difficulties]} was an infinite energy of
interaction of the electron with itself}'', and

\item ii) the infinities associated with the infinite number of degrees
of freedom of fields on the Minkowski spacetime continuum: ``{\em
The other difficulty came from some infinities which had to do
with the infinite numbers of degrees of freedom in the field}''.

\end{itemize}

\noindent In view of these difficulties, Feynman was willing to
drop the notion of `local field' (interactions) altogether:

\bigskip \noindent (Q7.52)\hskip 0.9in
\begin{minipage}{11cm}
\noindent ``{\small ...And, so I suggested to myself, that
electrons cannot act on themselves, they can only act on other
electrons.} {\small\em That means that there is no field at all.
You see, if all charges contribute to making a single common
field, and if that common field acts back on all the charges, then
each charge must act back on itself. Well, that was where the
mistake was, there was no field.}\footnote{Our emphasis.} {\small
It is just that when you shook one charge, another would shake
later...}''
\end{minipage}

\vskip 0.1in

\noindent Thus, in a nutshell, Feynman envisaged that direct
interactions solely in terms of (a finite number of) quanta
(particles), and without the notion of a `global' field (which,
anyway, in a Machian sense, is the quanta that produce/comprise
it) would solve the problem of infinities:

\bigskip \noindent (Q7.53)\hskip 0.9in
\begin{minipage}{11cm}
\noindent ``{\small\em ...There is no field at all; or if you
insist on thinking in terms of ideas like that of a field, this
field is always completely determined by the action of the
particles that produce it}.\footnote{Our emphasis.} {\small You
shake this particle, it shakes that one, but if you want to think
in a field way, the field, if it's there, would be entirely
determined by the matter which generates it, and therefore, the
field does not have any independent degrees of freedom and the
infinities from the degrees of freedom would be removed...}''
\end{minipage}

\vskip 0.1in

\noindent Furthermore, and perhaps more importantly for our
ADG-perspective here, he was ever ready to `dump' the local/{\em
differential}/Hamiltonian method (of field-theoretic quantization)
for a more global/{\em integral}/Lagrangian one (a covariant
action-based quantization---what is commonly known as a `sum over
histories' dynamical scenario)\footnote{A global dynamics that
subsumes under an integral---a global object---the contribution of
the local dynamical history---the local `differential behavior' as
it were---of every (local) particle comprising the `global'
field.}---albeit, a scenario that, almost by
definition,\footnote{For, arguably, one cannot speak of `history'
without an {\it a priori} notion of `(space)time' (paths or
trajectories).} still abides by a background spacetime as

\bigskip \noindent (Q7.54)\hskip 0.9in
\begin{minipage}{11cm}
\noindent ``{\small ...I would like to emphasize that by this time
I was becoming used to a physical point of view different from the
more customary one. {\em In the customary point of view, things
are discussed as a function of time in very great detail. For
example, you have the field at this moment, a differential
equation gives you the field at the next moment and so on; a
method, which I shall call the Hamilton method, the time
differential method. We have instead a thing that describes the
character of the path throughout all space and time. The behavior
of nature is determined by saying her whole spacetime path has a
certain character...}\footnote{Our emphasis.} If you wish to use
as variables only the coordinates of particles, then you can talk
about the property of the paths---but the path of one particle at
a given time is affected by the path of another at a different
time. If you try to describe, therefore, things differentially,
telling what the present conditions of the particles are, and how
these present conditions will affect the future---you see, it is
impossible with particles alone, because something the particle
did in the past is going to affect the future. Therefore, you need
a lot of book-keeping variables to keep track of what the particle
did in the past. {\small\em These are called field
variables.}\footnote{Our emphasis.} You will, also, have to tell
what the field is at this present moment, if you are to be able to
see later what is going to happen. From the overall space-time
view of the least action principle, the field disappears as
nothing but book-keeping variables insisted on by the Hamiltonian
method...}''
\end{minipage}

\vskip 0.1in

\noindent The parallel of the remarks above with ADG's conception
of a particle-field pair $(\modl ,\conn)$, as well as the
potential purely covariant path-integral quantization of such
fields, is remarkable indeed, as we briefly explain below:

\begin{enumerate}

\item First, let it be stressed that Feynman, one could say, opted for
an `integral' (global) rather than a `differential' (local)
method. This `global-{\it versus}-local' dichotomy is {\em the}
quintessential difference between the `covariant' (path integral,
sum-over-histories, Lagrangian action based) and the `canonical'
(differential Hamiltonian based) approaches to quantization in
general. Choosing the global method is supposed to capture more
naturally the `non-local' feature of quantum mechanics whereby
(quantum field) actions (`wave functions') at a certain spacetime
point (or region) can influence others located at some other
region.\footnote{Always under the condition---an axiom in QFT---of
{\em Einstein locality} (local causality) requiring that the two
regions must not be spacelike to each other.} This is the
quintessential aspect of coherent quantum superposition---{\em
the} defining property of quantum systems ({\it eg}, quantum
fields). In the global method all field histories must be taken
into account and contribute---in a proportion weighed by the
action---to the total path integral quantum field amplitude. On
the other hand, the truly local aspects of fields are the (point-)
particles, which are localized in spacetime by use of coordinate
labels---the basic `local field variables'. Of course, as Feynman
notes above, if one wishes to describe things (:the dynamics)
`{\em differentially}' (:differential geometrically) the
coordinates used must be `{\em differentiable}' (`smooth')
functions, so that the presence (and use) of a background
spacetime manifold appears to be almost mandatory.

\item Second, the ADG particle-field pair $(\modl ,\conn)$ appears
to suit well Feynman's description above and in the penultimate
quotation, as on the one hand the field (:$\conn$) is completely
determined by its particles (after all, recall that the
carrier/respresentation sheaf $\modl$ of $\conn$ is completely
determined by its local sections, which represent local quantum
particle states of the field), and on the other, the local
particle information (:sections) is coherently sheaf-theoretically
stitched up to comprise the entire sheaf space $\modl$ on which
$\conn$ acts. $\conn$, as befits a differential operator, acts
locally on $\modl$'s sections, but at the same time `sees' the
entire carrier `particle-space' $\modl$. In turn, the quantum
dynamics of the field is envisaged to be modelled after a
`global', path integral-type of scenario over the affine space of
connection fields (on $\modl$), which is the appropriate
kinematical space in our ADG-theoresis of gravity.\footnote{See
7.9.1 below.} This is how the unitary ADG particle-field pair
$(\modl ,\conn)$ has both local (particle:=:$\modl$) and global
(field:=:$\conn$) aspects built into it. Of course, as Feynman
says, if one wishes to describe the field histories locally and
differentially (:differential geometrically) one introduces the
coordinate structure sheaf $\modl$ of differentiable functions
(variables)\footnote{And since Feynman arguably had in mind
$\struc\equiv\smooth_{X}$, a geometrical spacetime manifold $X$
necessarily got engaged, even if implicitly, into his scheme.} and
the field is relegated solely to a book-keeping device, only
keeping track of the (differential) local particle actions.

\end{enumerate}

\paragraph{Sheaf theory in QFT: emphasis on locality.} Interestingly enough,
in the flat Minkowski manifold based context of
algebraic\footnote{A scheme essentially based on the theory of
(noncommutative) Von Neumann or $C^{*}$-algebras in which the
operator-valued distributions modelling quantum fields are
supposed to take their values.} or axiomatic QFT, where the notion
of {\em locality} is of paramount importance,\footnote{For
example, one of the basic axioms of axiomatic QFT is that of {\em
micro-causality}, otherwise known as {\em Einstein locality} or
{\em the axiom of local causality}.} Rudolph Haag in \cite{haag1}
stresses the potential importance of sheaf theory especially
concerning its essentially local character. Characteristically, we
quote him from that book:\footnote{The reader should note that
(s)he will not be able to find this quotation in the 1st 1992
edition of Haag's book. In the 2nd `expanded' edition, these words
can be found on page 326. In this quotation all emphasis is ours.}

\bigskip \noindent (Q7.55)\hskip 0.9in
\begin{minipage}{11cm}
\noindent ``{\small {\bf Germs.} {\small\em We may take it as the
central message of Quantum Field Theory that all information
characterizing the theory is strictly local i.e. expressed in the
structure of the theory in an arbitrarily small neighborhood of a
point}.\footnote{Our emphasis.} For instance in the traditional
approach the theory is characterized by a Lagrangean density.
{\small\em Since the quantities associated with a point are very
singular objects, it is advisable to consider neighborhoods. This
means that instead of a fiber bundle one has to work with a sheaf.
The needed information consists then of two parts: first the
description of the germs, secondly the rules for joining the germs
to obtain the theory in a finite region}\footnote{Again, emphasis
is ours.}...}''

\end{minipage}

\vskip 0.1in

\noindent These remarks of Haag are tailor-cut {\it vis-\`a-vis}
the sheaf-theoretic ADG-field theory and in particular ADG-gravity
at least in its finitary-algebraic guise developed in
\cite{malrap1,malrap2,malrap3}, as we explain below:

\begin{itemize}

\item First we highlight the remark about the singular character of the
pointed spacetime continuum and its `smearing' or `blowing-up' by
open sets about them. In a sheaf-theoretic setting, one gathers
field-information relative to such `local open gauges' and
formulates the field-dynamics locally, as a differential equation,
with respect to them.\footnote{One should also highlight here
Haag's remark that the mathematical structure suitable for doing
this is {\em not} a fiber bundle, but a sheaf.}

\item The field-law is seen to hold at every such open set and
then one stitches up or `collates' those local field data and
build the total field (representation) sheaf space $\modl$ on
which then one can show that the field law---{\it eg}, the law of
gravity (\ref{eqy23})---holds entirely.

\item Moreover, in the ADG-setting we can use the bicompleteness of $\ctriad$ and
actually show that the field-law holds at the limit of infinite
refinement (point-localization) of the said open sets covering the
`experimentally actuated' spacetime region $X$ under focus. That
is, the field-law of gravity holds on (or better, over) each and
every point of the underlying background and in its entirety.

\end{itemize}

\noindent This discussion brings us to some other remarks of
Stachel \cite{stachel3} about `the actual practice in
GR-research'. Like Haag above, but for different reasons, Stachel
insists that the `appropriate' mathematics, which faithfully
represents what general relativists actually do in solving locally
the Einstein equations and then extend the solution globally so
that the spacetime (manifold) topology is not fixed globally {\it
a priori}, but it rather `emerges' from stitching up the local
gravitational solution-field data, is {\em not} fiber bundle
theory, but {\em sheaf theory} and in particular {\em sheaf
cohomology}.\footnote{We will encounter Stachel's pertinent
quotation (Q8.?) in the sequel, when we discuss his deeper
interpretation and aftermath of Einstein's hole argument.}

Moreover, regarding QG, there is another telling quotation of
Stachel, which complements nicely (Q8.?) mentioned in the last
footnote and emphasizes the need for developing an approach to QG
which takes into account both the locality of GR and the
`globality' of quantum mechanics. What follows are the concluding
remarks in \cite{stachel}:

\bigskip \noindent (Q7.56)\hskip 0.9in
\begin{minipage}{11cm}
\noindent ``{\small ...The moral I want to draw from this final
point may be related to something that Chris Isham talked about.
Suppose you take seriously the point of view that there is
something fundamentally local about the way general relativity
approaches a problem.\footnote{Indeed, GR is after all a local
field theory---the local relativistic field theory {\it par
excellence}, so that Haag's words from the last quotation apply
{\it a fortiori} here.} Then there is another fundamental tension
between the basic approaches of general relativity and of quantum
mechanics since quantum mechanics, in a deep sense, is
fundamentally global in its approach to problems. It doesn't make
much sense to talk about the wave function on one patch of
space-time, or the sum over all paths on one patch of space-time.
In solving a quantum-mechanical problem you have to consider the
{\small\em whole}\footnote{Stachel's emphasis.} manifold from the
beginning. The conventional mathematical approach to general
relativity, which starts with a manifold, masks this tension.
{\small\em If we develop a mathematical formulation of general
relativity that emphasizes the element of locality from the
beginning, it would emphasize this contrast more sharply. Such an
emphasis on the tension may be a necessary stage in finding its
ultimate solution}\footnote{Our emphasis.}...}''
\end{minipage}

\vskip 0.1in

\noindent Regarding singularities {\it per se}, we would like to
quote Heller from \cite{hell1} who also notes that in order to
cope effectively with all types of gravitational singularities,
Geroch's algebraic formulation of GR \cite{geroch4} should be cast
sheaf-theoretically:

\bigskip \noindent (Q7.57)\hskip 0.9in
\begin{minipage}{11cm}
\noindent ``{\small ...To include other type of singularities
(also curvature singularities) into the theoretical scheme,}
{\small\em one must change from Einstein algebras to sheaves of
Einstein algebras...To allow stronger type of singularities to
become part of the theory, we should change from Einstein algebras
to sheaves of Einstein algebras...}''\footnote{Our emphasis.}
\end{minipage}

\vskip 0.1in

\noindent And also with Sasin from \cite{hell}:

\bigskip \noindent (Q7.58)\hskip 0.9in
\begin{minipage}{11cm}
\noindent ``{\small\em ...To cover all types of singularities, one
must change to a sheaf of Einstein algebras...}''\footnote{Again,
our emphasis.}
\end{minipage}

\vskip 0.1in

\noindent {\it En passant}, and in view of these quotes, we would
like to ask: physically speaking, what does it mean for `{\em
singularities to be part of the theoretical scheme}'? From the
ADG-viewpoint, since the (physical) theory {\em is} the
(dynamical) laws that define it, we could interpret the first
remark above as intuiting that {\em singularities must be
integrated or `absorbed' into the theory; ultimately, into the
physical laws (differential equations) that define it}. But this
is precisely what ADG-gravity achieves: it `integrates', `engulfs'
or `absorbs' the singularity into the $\struc$-connection field
$\conn$ which, in turn, `sees' or `passes through' them as the law
(differential equation) that it defines does not stumble, let
alone halt (or break down), at all on them ($\struc$-functoriality
of the ADG-gravitational dynamics).\footnote{See Eddington
quotation above about integrating singularities into the dynamical
equations.}

\paragraph{Sheaf theory in QG: the transition from `micro' to
`macro'.} From what we have said so far about developing a
theoretical scheme for QG that goes naturally from
local-to-global:

\begin{itemize}

\item From Stachel's remarks, it appears that, formally, the tension between
GR and QM is the one between local (differential) and global
(integral) Calculus methods.

\item The upshot of his thoughts is his hunch about the potential import of
sheaf theory and sheaf cohomology in order to relieve that
tension.

\item Then comes Feynman's opting for a global, (path) {\em
integral}-type of quantization of any field theory, instead of a
local (Hamiltonian) {\em differential} one; {\em as well as his
questioning the local field concept altogether}---recall, in his
view, the field is relegated to a local book-keeping device for
the entire history-actions of particle-quanta.

\item Then also along came Haag who, in glaring contradistinction to Feynman, insisted that the quintessence of (Q)FT is that
the theory is strictly local, thus, like Stachel, he propounded a
sheaf-theoretic formulation of algebraic QFT.

\item Finally, Heller {\it et al.} too pitched the idea of casting sheaf-theoretically the algebraic formulation of GR
{\it \`a la} Geroch in order to deal effectively with all sorts of
gravitational singularities.

\end{itemize}

\noindent All in all, the idea of using sheaf-theoretic methods in
GR, QFT and, ultimately, QG has been brewing in the mind of
theoretical physicists for quite some time lately, but let it be
stressed here that this idea still revolves in one way or another
about a base (spacetime) manifold---{\it ie}, that the sheaves
envisaged are still soldered on a differential manifold; they are
{\em smooth sheaves}. Stachel in particular has regarded this
manifestation of the `manifold monopoly' in sheaf theory as an
obstacle to the further application of sheaf theory to QG, as we
recall from (Q?.?) his words that ``{\em ...As far as I know, no
one has followed up on this suggestion,\footnote{That is, to use
sheaf theory and sheaf cohomology in GR and QG.} and my own recent
efforts have been stymied by the circumstance that all treatments
of sheaf theory that I know assume an underlying manifold...}''.

However, to the mind of these authors, Selesnick's remarks below,
from a private communication with the second author \cite{sel5},
emphasize in the best way the passing from local to global with
the help from sheaf (and topos) theory {\it vis-\`a-vis}
QG:\footnote{Below, all emphasis is ours.}

\bigskip \noindent (Q7.59)\hskip 0.9in
\begin{minipage}{11cm}
\noindent ``{\small\em ...One of the primary technical hurdles
which must be overcome by any theory that purports to account, on
the basis of microscopic quantum principles, for macroscopic
effects (such as the large-scale structure of what appears to us
as space-time, {\it ie}, gravity) is the handling of the
transition from `localness' to `globalness'. In the `classical'
world this kind of maneuver has been traditionally effected either
measure-theoretically---by evaluating largely mythical integrals,
for instance---or geometrically, through the use of sheaf theory,
which, surprisingly, has a close relation to topos theory. The
failure of integration methods in traditional approaches to
quantum gravity may be ascribed in large measure to the
inappropriateness of maintaining a manifold---a `classical'
object---as a model for space-time, while performing quantum
operations everywhere else. If we give up this classical manifold
and replace it by a quantal structure, then the already
considerable problem of mediating between local and global (or
micro and macro) is compounded with problems arising from the
appearance of subtle effects like quantum entanglement, and more
generally by the problems arising from the non-objective nature of
quantum `reality'. Although there is a rich and now highly
developed mathematical theory of `noncommutative geometry' (which
has had considerable success in application to traditional quantum
field theories), a concomitant noncommutative sheaf theory seems
to have been slow in coming...}''

\end{minipage}

\vskip 0.1in

\noindent We agree with most of Selesnick's remarks above,
especially with the one about the currently mathematically
ill-defined path integrals abounding in theoretical
physics\footnote{For it is true that we do not have yet a well
developed integration theory over infinite-dimensional function
spaces as envisaged by Feynman.}, as well as with his remarks that
we perform all sorts of quantization procedures on the
object-fibers ({\it eg}, fields) living on a spacetime manifold
while leaving this classical continuum background intact. On the
other hand, as has been repeatedly commented on throughout the
present paper-book, we would not go as far as Selesnick goes and
maintain that what is actually needed is a quantization of that
classical background, let alone that one has to resort to
noncommutative mathematics ({\it eg}, noncommutative geometry,
sheaf and topos theory) in order to tackle the problem of QG.
Rather, we hold that {\em no background geometrical spacetime, and
in particular a base manifold, is needed at all in an entirely
algebraic (sheaf-theoretic) approach to QG regarded as a quantum
gauge theory}. Of course, we have ADG-gravity in mind.

But let us now proceed and address from the perspective of ADG the
nowadays caustic QG issue of black hole information-loss. The
following discussion will again bring forth the broad character,
versatility and potential import of ADG-ideas in QG research.

\subsubsection{Field information lost in a black hole and found in
ADG: the brighter side of black holes}

By now it must have become clear that in ADG, and especially in
ADG-gravity, singularities are not regarded as being sites past
which the dynamical gravitational field law (Einstein equations)
cannot be continued, let alone as {\it loci} in (or at the
boundary of) `physical spacetime' (whatever that means in
ADG-gravity before the field dynamics is set up) where the
Einstein equations break down like the spacetime manifold and,
{\it in extenso}, the CDG-based approaches to classical and
quantum gravity have hitherto made us believe. Expressly, for
example, we cannot (and we do not!) think of background
geometrical spacetime configurations, such as a closed trapped
surface and its ensuing black hole horizon, to be formed by the
{\em physical} process of gravitational collapse for instance, for
one thing simply because there is no background spacetime
whatsoever ({\it ie}, a `continuous' or `discrete' base/medium
on/in which fields dynamically propagate) in the theory to begin
with. This puts into perspective, question, and to a large extent
doubt, the viability in our theoresis of the by now standard and
thought of as being basic concepts, technical tools and general
working philosophy of black hole theory, in which every black hole
is normally supposed to conceal in its core, as it were hidden
behind its horizon, a spacetime singularity.

In view of the above, in this sub-subsection we wish to comment in
particular on the nowadays caustic and to a certain degree
controversial issue of `{\em black hole information, or quantum
coherence, loss}'. In a nutshell, very roughly and simplified, the
original scenario due to Hawking \cite{hawk3} had it that a
quantum (particle), initially prepared by an external (to the
black hole) experimenter (:`observer') in a pure state,
dynamically evolves past the black hole horizon (:`falls into the
black hole') and is eventually thermally radiated (:`spat out') in
a mixed state (described by the external observer by a density
matrix)---an account that appears to violate the linear evolution
law of quantum mechanics thus corresponding to a loss of quantum
coherence (a sort of `quantum superposition breakdown'). An
obvious `explanation', viewing the black hole interior as a `black
box'-type of `region of no pure return', would be an epistemic
uncertainty on the part of the external observer\footnote{For
instance, she does not have a clue about the dynamical evolution
laws past the horizon's opaque veil, so that any sort of
predictability by her is lifted.} to the effect that she, in a
hidden variables kind of way, averages over the unobservable
particle states beyond the horizon---{\it ie}, she effectively
`thermalizes' (or `thermodynamicizes') the system. In other words,
the quantum, once past the horizon, is regarded as being coupled
to a heat bath in the black hole interior, from which it is
subsequently emitted in a thermal (mixed) state
\cite{thooft4,thooft5}, with the concomitant breaking of quantum
coherence and the loss of information.

Of course, one could immediately observe that Hawking's arguments,
which are closely akin to his black hole instability and
associated evaporation results, are glaringly semi-classical (:QFT
on a black hole background environment), so that again all our
shortcomings should be attributed to our not having (as yet!) a
cogent QG to describe the physical laws in the black hole
interior.\footnote{As it happens, we do not have a quantum theory
of black holes, or what essentially amounts to the same, a theory
of quantum (microscopic, of Planck size and mass) black holes
where quantum gravitational effects are expected to reign supreme
\cite{thooft4}; however, a QFTheoretic (scattering matrix)
`approximation' to such a theory, describing what the external
observer---the one outside the horizon---sees as in-going (to the
back hole) and out-coming (from the black hole) states of quanta,
has already been worked out \cite{thooft5}.} Metaphorically
speaking, information loss is essentially attributed, once again,
to our epistemic uncertainty and theoretical `inadequacy' in the
realm past the black hole horizon. In other words, currently only
a quantum theory of gravity is expected to describe the
gravitational field and what happens to matter and their radiation
fields right at a black hole singularity.\footnote{And let it be
noted here, as Susskind remarks in \cite{susskind}, that while for
the external observer quantum linearity and coherence appears to
be violated, for an in-falling observer (and particle-quanta), one
might argue, there could be a ``{\em brick wall}'' at the horizon
that ping-pongs back outside the in-falling quantum matter in a
pure state. This alternative however would violate the EP of GR,
since we know that the horizon is a `soft', flat, semi-permeable
membrane. Thus there's no way out of the impasse: for the external
observer QM is violated, while for an in-falling one, GR. It has
been suggested by numerous people in the past that to arrive at
the `true' QG (some of) the fundamental principles of both GR and
QM should be somehow modified. Penrose, for example, remarks in
\cite{pen5}: ``{\em Is this {\rm [{\it ie}, the QFTheoretic]} the
right way to think about quantum gravity, or should we be looking
for some more even-handed marriage, with some give on both
sides?}''.} In a similar vain, it also follows that QG is expected
to unveil the quantum (microscopic) origin of black hole
thermodynamics, whose laws (especially the famous second,
Bekenstein-Hawking entropy one) essentially subsume our ignorance
(arising from the coarse thermodynamic description) about the
black hole's `true' microscopic degrees of freedom and their
quantum microstates.

From all this heat (and debate) that the issue of black hole
information loss has generated over the years, we wish to isolate
an apofthegma and its corollary from a recent paper by Leonard
Susskind \cite{susskind} in which he tries to come to grips and
resolve this apparently paradoxical---or in his own words, ``{\em
weird}''---scenario:

\bigskip \noindent (Q7.60)\hskip 0.9in
\begin{minipage}{11cm}
\noindent \ovalbox{\bf Apofthegma:} ``{\small ...I can now state
the principle of Horizon Complementarity. All it says is that {\em
no observer ever sees a violation of the laws of
nature}.\footnote{Our emphasis.} More specifically it says that to
an observer that never crosses the horizon, the horizon behaves
like a complex system which can absorb, thermalize, and re-emit
all information that falls on it. No information is ever lost. In
essence, the world on the outside of the horizon is a closed
system...}''

\vskip 0.05in

\noindent \ovalbox{\bf Corollary:} ``{\small ...It is clear from
Horizon Complementarity that {\em a revision is needed in the way
we think about information being localized in
spacetime}.\footnote{Our emphasis.} In both classical relativity
and quantum field theory the spacetime location of an event is
invariant, that is, independent of the observer. Nothing in the
theory prepares us for the kind of weirdness I described
above...}''

\end{minipage}

\vskip 0.1in

\noindent The ADG-gravity answer to black hole coherence loss
`paradox' involves again a sort of `cutting the Gordian knot'-type
of reasoning similar in spirit to the interpretation of the
ADG-theoretic resolution of the inner Schwarzschild singularity we
gave above:

\begin{enumerate}

\item For starters, on general grounds and as noted earlier,
on the face of the background spacetime manifoldless ADG-gravity
we do not accept (because it never arises in our theoresis!) that
there is a black hole event horizon surrounding a spacetime
singularity beyond which, and in the immediate vicinity of the
singularity, the so-called `classical' Einstein equations cannot
be continued and, ultimately, do not hold (`break down'). It
follows that we do not view the horizon as demarcating not only a
physical, but also a `theoretical' boundary outside of which one
law of gravity---the so-called `classical' one of Einstein
(GR)---holds, but that another one---the still to be discovered QG
law---is in force in its interior.\footnote{This is completely
analogous to our earlier critique of the Planck length/time as
denoting a fundamental scale in Nature below which the Einstein
field law of gravity (and the spacetime continuum supporting it at
least differential geometrically) breaks down, and especially it
applies to quantum black holes of radius of around the Planck
length.} As a `counter-scene' to our comments here, let us bring
forth from the introduction of \cite{thooft5} the following words:

\bigskip \noindent (Q7.61)\hskip 0.9in
\begin{minipage}{11cm}
\noindent ``{\small There is quite a bit of controversy (and
confusion) regarding the nature of physical law governing a black
hole. Some of the difficulties have their origin in the
deceptively clean picture given by the `classical' (here this
means `non-quantum mechanical') solutions of Einstein's equations
of gravity in the case of gravitational collapse. The metric
tensor describing the fabric of space-time appears to be smooth
and well-behaved in the vicinity of a region we call the
`horizon', a surface beyond which there are space-time points from
which no information can reach the outside world...}''

\end{minipage}

\vskip 0.1in

\noindent Here we once again promulgate the unity and universality
of the physical law of gravity---{\it ie}, Einstein's law {\it \`a
la} ADG-gravity (\ref{eqy23}) holds everywhere and the
gravitational field `sees' neither a background spacetime or
horizon, nor even a spacetime singularity to that matter (since it
is still in force over it, as we saw in Schwarzschild's case). In
a peripheral, slanted sense, this unity accords with Susskind's
apofthegma above that `{\em no observer ever sees a violation of
the laws of nature}'.\footnote{The sense is slanted, because on
the one hand the relativity of the ADG-gravitational field,
effectuated via $\aut\modl$ concerns it and it alone, without
reference to an external observer/measurer/experimenter localized
(with coordinate origin at a point) in an external (to the
ADG-gravitational field) spacetime, and on the other, even when an
external observer brings along with her an $\struc$ to
coordinatize (or `measure') and geometrically represent the field,
the $\struc$-functoriality of the law that the latter defines
secures that the field `sees' no background spacetime (horizon),
or even singularity (all of which are built into $\struc$).}

\item About the closely related to the above `field-information conservation' and `spacetime
localization' mentioned in Susskind's corollary, the
spacetimeless, dynamically autonomous, unitary ADG-gravitational
field $(\modl ,\conn)$ defining (\ref{eqy23}), is not `leaking'
information to its `environment',\footnote{If one can call that
the background, surrogate localization (topological) space $X$,
which does not partake into the gravitational dynamics.} while, as
we emphasized earlier, the sheaf-theoretic character of the
ADG-gravitational law secures that local field information is
coherently stitched up to comprise the global field law. To stress
it once again, (\ref{eqy23}) holds on the total sheaf space
$\modl$ over the whole surrogate localization background $X$,
without the latter, in contradistinction to Susskind's corollary
above, being interpreted as `spacetime' (so that one can speak of
`spacetime localization' proper). Again, in a peripheral and
slanted sense, the unitary and autonomous $(\modl ,\conn)$ may be
regarded as a(n information-wise) `hermetically' closed (and
`sealed' from an external/background to it `spacetime
environment') dynamical system.

\item In continuation of the point made above, the geometrically (pre)quantized
ADG-gravitational field $(\modl ,\conn)$---with $\modl$ now
regarded as the associated/representation (Hilbert) sheaf of local
quantum (particle) states of the (principal sheaf $\aut\modl$ of
dynamical auto-transformations of the) field---may be viewed as a
closed {\em quantum} system, and (\ref{eqy23}) as its quantum
evolution law. Again, there is no information leaking to the
ambient spacetime environment, since {\it ab initio} there is no
such milieu in the theory. The (pre)quantized ADG-gravitational
dynamics (\ref{eqy23}), although manifestly non-linear (as befits
gravity), it still respects locally coherent quantum
superpositions (between the local quantum particle states of the
field---{\it ie}, the local sections of $\modl$), reminiscent of
the EP in GR whereby locally the (curved) spacetime continuum is
isomorphic to the (flat) Minkowski vector space \cite{malrap1}.
{\it A fortiori}, here, since no external (to the field) smooth
spacetime continuum is involved, thus no $\mathrm{Diff}(M)$
either, there is no ``{\em possible breakdown of the superposition
principle}'' \cite{weinstein0,weinstein} and the concomitant
lifting of quantum coherence. {\it In toto}, the ADG-gravitational
field is a self-quantum, coherent auto-dynamical system.

\item The last thing that we would like to comment on, which also
nicely wraps up the whole discussion above about black hole
information loss, is 't Hooft's proposal to regard all ultra-small
(of localization radius smaller than the Planck length) and
super-heavy (with mass larger than the Planck mass) elementary
particles as microscopic (quantum) black holes
\cite{thooft4,thooft5}. To this end, let us first quote him
extensively once again from \cite{thooft5}:

\bigskip \noindent (Q7.62)\hskip 0.9in
\begin{minipage}{11cm}
\noindent ``{\small ...But what if the hole is small, so that
classical observers are too bulky to enter? Or let us ask a
question that is probably equivalent to this: suppose one keeps
track of {\em all}\footnote{'t Hooft's emphasis.} possible states
a black hole can be in, is it then possible to describe the hole
in terms of pure quantum states alone? Will the very tiny black
holes evolve according to conventional evolution equations in
quantum physics or is the loss of information a fundamental new
feature, even for them?

{\em There is a big problem with any theory in which the loss of
information is accepted as a fundamental item. This is the fact
that all effective laws become fuzzy}\footnote{Our emphasis.}...

[Then 't Hooft gives an example in quantum mechanics where a
system dynamically evolves from pure to mixed states. He then
continues...]

In our example we clearly see what the remedy is. {\em The extra
uncertainty had nothing to do with quantum mechanics; the
Hamiltonian was not yet known because of our incomplete knowledge
of the laws of physics... Returning with this wisdom to the black
hole, what knowledge was incomplete? Here I think one has a
situation that is common to all macroscopic systems: because of
large number of quantum mechanical states it was hopelessly
difficult to follow the evolution of just one such state
precisely. One was forced to apply thermodynamics. The outcome of
our calculations with black holes got the form of thermodynamic
expressions because of the impossibility, in practice, to follow
in detail the evolution of any particular quantum
state}.\footnote{Again, our emphasis. Here, let us mention {\it en
passant} that a philosopher would say that the uncertainty (and
the probabilities accompanying it) involved in quantum mechanics
is fundamentally different from the statistical mechanical
uncertainty of thermodynamics. The first is an `ontological' ({\it
ie}, `inherent' in the theory/in the quantum mechanical system
under scrutiny) uncertainty, while the second one `epistemic' (it
reflects more our---{\it ie}, the external
observer's/experimenter's ignorance about the system). In the
formalism this is reflected by the fact that quantum mechanical
states are represented by probability amplitudes, as opposed to
the probability distributions of classical probability theory. (To
be sure, the latter are obtained by squaring the amplitude, but
the square is altogether a different animal than the square
root...)}

But this does mean that our basic understanding of black holes at
present is incomplete...If we want to understand how a black hole
behaves when it reaches the Planck mass, we expect thermodynamic
expressions to break down.

The importance of a good quantum mechanical description is that it
would enable us to link black holes with ordinary particles. The
Planck region may well be populated by a lot of different types of
fundamental particles. Their `high energy limit' will probably
consist of particles small and heavy enough to possess a horizon
and thus be indistinguishable from black holes...

The reason why black holes should be used as a starting point in a
theory of elementary particles is that anything that is tiny and
heavy enough to be considered an entry in the spectrum of ultra
heavy elementary particles (beyond the Planck mass) must be
essentially a black hole.

Black holes are defined as solutions of the classical, i.e.
unquantized, Einstein equations of General relativity. This
implies that we only know how to describe them reliably when they
are considerably bigger than the Planck length and heavier than
the Planck mass. What was discovered by Hawking in 1975 is that
these objects radiate and therefore must decrease in size. It is
obvious that they will sooner or later enter the domain that we
presently do not understand.\footnote{That is, the QG domain below
Planck length. (Our footnote.)}

...[For the study of microscopic (quantum) black
holes,]\footnote{Our addition for noematic and textual
continuity.} what we require is first of all some quantum
mechanically pure evolution operator, and secondly that this
operator be consistent with all we know of large scale physics, in
particular general relativity.

At first sight these requirements are in conflict with each other.
General relativity predicts unequivocally that gravitational
collapse is possible, and this produces a horizon with all its
difficulties...Anyway, we know that the amendments needed at the
horizon all refer to Planck scale physics...}''

\end{minipage}

\vskip 0.1in

\noindent Again, the whole issue revolves in one way or another
about the problem of `spacetime localization' as also Susskind
mentioned above: the thing is that the general opinion nowadays is
that one cannot localize in space-time a quantum particle with
accuracy higher (:uncertainty smaller) than the Planck length-time
without creating a black hole (singularity). Equivalently, in the
Planck regime, where quantum gravitational effects are expected to
be significant, all elementary particles can be modelled after
black holes; hence a quantum theory for black holes appears to be
mandated and come hand in hand with a quantum theory of gravity.
Moreover, the theoretical threshold or divide (alongside the
Planck length) appears to be the horizon, outside of which the
non-quantum GR (and the picture of spacetime as a classical smooth
continuum) holds galore, but inside of which it appears to break
down.

Concerning ADG-gravity, (\ref{eqy23}) may be regarded as a
(pre)quantum evolution equation for the ADG-gravitational field;
while a fully quantum gravitational propagator, as noted earlier,
is currently expected to be modelled after a path integral-type of
dynamics over the $\sconn /\aut\modl$-moduli space of
ADG-gravitational connections---a dynamics which in turn will not
encounter any hindrance from a spacetime horizon or a singularity
concealed behind it.\footnote{See sub-subsection 7.9.1 in the
sequel.} {\it In summa}, let it be emphasized here that in a
strong sense the entire present `paper-book' comes precisely as an
`antithesis' to 't Hooft's thesis above.

\end{enumerate}

\subsection{Genuinely Quantum and Purely Gauge Vacuum Einstein
Gravity}

As noted in section 4, the idea to view Einstein gravity (GR) as a
gauge theory, based predominantly on the notion of (affine,
non-metric) connection and less on the metric as in the original
pseudo-Riemannian geometry based theory of Einstein, dates back to
the pioneering work of Weyl, Eddington and Cartan.\footnote{The
first two also wanted to unite gravity with electromagnetism---the
first attempts at a unified field theory.} {\it Vis-\`a-vis} QG,
this idea seems to be quite a natural one if one wishes to (as
many theoretical physicists do indeed aspire to) `unify' gravity
with the other three fundamental forces of Nature, since the
latter have proven to be gauge forces {\it par excellence}.

Just three years before Ashtekar's ground-breaking discovery of
the new (spin-Lorentzian connection) variables for classical and
quantum gravity \cite{sen,ash}---a discovery that imparted
tremendous impetus to the quest for formulating GR as a gauge
theory proper, hence it revived the interest in approaching QG
like one approaches the other three quantum gauge theories of
matter---Ivanenko and Sardanashvily made some very telling, almost
prophetic (at least for Ashtekar's work), remarks in their well
known Physics Report \cite{ivanenko}:

\bigskip \noindent (Q7.63)\hskip 0.9in
\begin{minipage}{11cm}
\noindent ``{\small ...These and some other
difficulties\footnote{The authors here have just mentioned a few
problems encountered in GR, such as the weak field formulation of
the EP of GR, its associated (in)famous problem of gravitational
energy positivity, and of course, more important to our treatise
here, the problem of gravitational singularities.} of the GR
picture of gravity motivate one's attempts to reformulate
gravitational theory from non-conventional standpoints extending
the framework of Einstein's GR.

But why gauge gravity? Can the gauge treatment of gravity really
solve the above-mentioned problems? Beforehand nobody knows. But
today many of these problems seem to be put in the shadow of the
urgent goal of the gravity unification with the elementary
particle world. Just this goal stimulated by the grand unification
program in contemporary particle physics puts the gauge version in
the forefront of modern gravitation research.

Today, gauge theory provides the theoretical base of all modern
unification attempts in particle physics. It has become clear that
weak and electromagnetic interactions can be successfully unified
by the Weinberg-Salam gauge model, and there is strong evidence
that strong interaction is also mediated by gauge particles or
gluons within the framework of chromodynamics. In field theory
gauge potentials become a standard tool for describing
interactions with very different symmetries. And apparently the
single gap in the modern gauge picture still remains gauging the
external or space-time symmetries of fields and particles, that
includes the gauge gravity also.

Moreover, gauge theory using the mathematical formalism of fiber
bundles realizes bin fact the known program of the 1920s to build
the geometric unified picture of various interactions. And it is
strange enough that gravitation theory, being the first example of
field geometrization, has still not any recognized gauge version.
Although the first gauge treatment of gravity was suggested
immediately after the gauge theory birth itself
[...]\footnote{Here the authors provide some references which are
omitted in the present paper.}

The main dilemma which during 25 years has been confronting the
establishment of the gauge gravitation theory, is that gauge
potentials represent connections on fiber bundles, while
gravitational fields in GR are only metric or tetrad (vierbein)
fields.

Connections as fundamental quantities appeared together with the
metric in Weyl's and Eddington's generalizations of GR on gravity
with nonmetricity and torsion,\footnote{That is, as noted earlier,
when one does not require the metric compatibility of the
connection (metric or torsionless connection). One should
certainly mention in this respect Cartan's work in what is now
known as Einstein-Cartan theory (`torsionful spin-gravity')
\cite{gosch}.} and in this quality were again recognized by
Einstein in his last scientific paper [...].\footnote{Again, a
reference originally cited by the authors at this point, is
omitted here.} But even in the gauge gravitation theory
connections cannot at all substitute the metric, because there are
no groups, whose gauging would lead to the purely gravitational
part of space-time connections (Christoffel symbols or
Fock-Ivanenko spinorial coefficients). To separate such
gravitational components from gauge fields, e.g., of the Lorentz
group, metric or tetrad fields have to be introduced.

At the present time the gravitation theory is viewed actually as
the affine-metric theory possessing two independent potentials,
namely, metric and connection, and just this constitutes the
peculiarity of the gauge approach to the gravitation theory in
comparison with the gauge models of the Yang-Mills type for
internal symmetries...}''
\end{minipage}

\vskip 0.1in

\noindent It is fair to say that to this day, even after
Ashtekar's new connection variables formulation of gravity three
years after the quotation above, all the obstacles and problems
stressed above regarding the (quantum) gauge theoresis of gravity,
although alleviated, still remain in effect. Of course, as noted
earlier, the new spin-connection variables, by simplifying the
gravitational constraints, tremendously facilitated the canonical
quantization of gravity regarded as a constrained gauge
system\footnote{For the general theory of quantization of gauge
systems with constraints, see \cite{henneaux}.} ({\it \`a la}
Dirac-Bergmann), and it achieved the formulation of canonical QGR
in a genuinely background metric independent fashion. However,
Ashtekar's first order formalism

\begin{itemize}

\item on the one hand it still is in essence an affine-metric theory in the sense of Ivanenko and
Sardanashvily, as it regards the spacetime metric (in the guise of
the {\it vierbein} field) as a gravitational dynamical variable on
a par with the affine connection. That is, the Ashtekar scheme is
{\em not} a purely gauge ({\it ie}, solely connection based)
formulation of gravity like the half-order (third gauge) formalism
of ADG-gravity is,

\item and on the other hand, it still employs a background
(spacetime) differential manifold, thus also, inevitably, it
involves the four (primary or first-class)
$\mathrm{Diff}(M)$-external spacetime constraints of GR (which
anyway it aims to simplify in the first place).

\end{itemize}

\noindent Moreover, as noted earlier and in \cite{rap5}, it is
precisely the {\em non-gauge} character of the $\mathrm{Diff}(M)$
group of the external spacetime manifold that prevents one from
formulating gravity as a gauge theory like the other three
fundamental forces \cite{weinstein0,weinstein}. This `non-gauge
character' of $\mathrm{Diff}(M)$ means that it is `wrong' (and
certainly misleading) to think of the
$\mathrm{Diff}(M)$-invariances of the (external spacetime manifold
based) theory of gravity (GR) as gravitational gauge symmetries
proper. For plainly, technically speaking at least, the
diffeomorphisms in $\mathrm{Diff}(M)$ are (by definition) the
(external) `symmetries' (automorphisms) of the (external)
$\smooth$-smooth base spacetime manifold ({\it ie},
$\mathrm{Aut}(M)\equiv\mathrm{Diff}(M)$), {\em not elements of the
automorphism group of a principal fiber bundle} as in the other
three gauge theories of matter. In other words, diffeomorphisms
are {\em external} not {\em internal} symmetries. Indeed, as
repeatedly mentioned throughout this paper, especially {\it
vis-\`a-vis} the problem of (canonical/Hamiltonian or
covariant/Lagrangian) quantization of gravity, regarding
$\mathrm{Diff}(M)$ as gravity's gauge symmetry group proper leads
to a number of formidable problems, such as the so-called {\em
problem of time} and the {\em inner product/quantum measure
problem}.\footnote{Let alone that even in the classical theory
(GR) $\mathrm{Diff}(M)$ causes problems in defining precisely what
is a singularity in the theory (recall that in the present paper
we `defined' singularities as solution-metric differential
geometric anomalies) as well as, by underlying the PGC, it
apparently halted for a couple of years (till its significance was
further appreciated) the development of the theory at its very
birth. This refers to Einstein's hole argument, which we are going
to tackle under the prism of ADG-gravity in the next section.}

Of particular interest to us here is {\em the problem of defining
physically meaningful observables} in a possible (canonical)
quantization of (vacuum) Einstein gravity. This problem too is due
to the external spacetime $\mathrm{Diff}(M)$-invariances of the
classical theory (GR), as the following excerpt from
\cite{weinstein} corroborates:

\bigskip \noindent (Q7.64)\hskip 0.9in
\begin{minipage}{11cm}
\noindent ``{\small ...The distinction between the two sorts of
invariance\footnote{Here Weinstein refers to the distinction noted
above, namely, that while the gauge symmetries are internal and
can be organized into a principal fiber bundle, the
$\mathrm{Diff}(M)$-invariances are external and cannot, thus they
should not be regarded as gauge symmetries proper.} is absolutely
vital in the context of quantum theory. We understand how to
quantize gauge theories. A primary aspect of this is to represent
some subset of the classical gauge-invariant quantities (classical
observables) as self-adjoint operators (quantum observables)
obeying relevant commutation relations. However, {\em it is not at
all clear what it means to quantize a diffeomorphism-invariant
theory. One can attempt to turn the diffeomorphism-invariant
quantities into observables---this is effectively what one is
doing when one formally treats the diffeomorphism group as a gauge
group}.\footnote{Our emphasis.} However, in addition to leading to
the problem of time in canonical gravity [...]\footnote{Here
Weinstein gives the following references:
\cite{ish2,kuchar,weinstein0}, as well as another one of his not
included at the back.} and possibly a breakdown of the
superposition principle [...],\footnote{Reference given here in
the original is omitted.} {\em this approach leaves one with
virtually no known observables whatsoever for vacuum gravity in
the compact case}\footnote{Our emphasis.} [...]\footnote{Here
Weinstein refers to the papers by Torre
\cite{torre1,torre2}.}...}''
\end{minipage}

\vskip 0.1in

\noindent Indeed, similarly, in the `bottom-up', discrete approach
to Lorentzian QG called causet theory \cite{sork10}, Sorkin has
for a long time been categorematic about and rather critical of
any attempt at defining physically meaningful observables in the
presence of the $\mathrm{Diff}(M)$ implementing the PGC in the
classical theory (GR) as it is especially done in the canonical
QGR scenario, unless of course one is willing to sacrifice general
covariance altogether:\footnote{Let it be noted here that in
causet theory there is still a `discrete' analogue of the
$\mathrm{Diff}(M)$-modelled PGC of the smooth continuum based
`top-down' approaches to QG (such as the canonical QGR and its
associated LQG scenario), which is called `{\em causet-vertex
labelling independence}'---much in the same way that in GR the
gravitational dynamics (Einstein equations) is indifferent to our
smooth coordinate labelling of the points of the underlying
manifold (see quotation after the one next).}

\bigskip \noindent (Q7.65)\hskip 0.9in
\begin{minipage}{11cm}
\noindent ``{\small ...{\bf What are the `observables' of quantum
gravity?}

\hskip 0.05in

{\em Just as in the continuum the demand of diffeomorphism
invariance makes it harder to formulate meaningful
statements},\footnote{Think, for example, of the statement that
light slows down when passing near the sun. (Sorkin's own
footnote.) Emphasis above is ours.} so also for causets the demand
of discrete general covariance has the same consequence, bringing
with it the risk that, even if we succeeded in characterizing the
covariant questions in abstract formal terms, we might never know
what they meant in a physically useful way. I believe that a
similar issue will arise in every approach to quantum gravity,
discrete or continuous ({\em unless of course general covariance
is renounced})\footnote{In the context of canonical quantum
gravity, this issue is called `the problem of time'. There,
covariance means commuting with the constraints, and the problem
is how to interpret quantities which do so in any recognizable
spacetime language. (Again, Sorkin's own footnote. The last
sentence in this footnote is omitted here.) Emphasis in the
parenthesis above is ours.}...}'' \cite{sork10}
\end{minipage}

\vskip 0.1in

\noindent In the following quotation, taken from
\cite{sork6},\footnote{We will provide a more complete quotation
in (Q8.?) in the sequel, when, in the context of our analysis of
Einstein's hole argument from the perspective of ADG-gravity, we
will criticize not only the usual smooth continuum based
$\mathrm{Diff}(M)$-implemented general covariance, but also
`discrete' general covariance like the one proposed in Sorkin {\it
et al.}'s causet scenario.} Sorkin {\it et al.} go a bit further
and note:

\bigskip \noindent (Q7.66)\hskip 0.9in
\begin{minipage}{12cm}
\noindent ``{\small ...After all, labels in this discrete setting
are the analogs of coordinates in the continuum, and the first
lesson of general relativity is precisely that such arbitrary
identifiers must be regarded as physically meaningless: the
elements of spacetime---or of the causet---have individuality only
to the extent that they acquire it from the pattern of their
relations to the other elements. It is therefore natural to
introduce a principle of `discrete general covariance' according
to which `the labels are physically meaningless'.

But why have labels at all then? For causets, the reason is that
we don't know otherwise how to formulate the idea of sequential
growth, or the condition thereon of Bell causality, which plays a
crucial role in deriving the dynamics. {\em Ideally perhaps, one
would formulate the theory so that labels never entered, but so
far no one knows how to do this---anymore than one knows how to
formulate general relativity without introducing extra gauge
degrees of freedom that then have to be cancelled against the
diffeomorphism invariance}.\footnote{Our emphasis.}}

\end{minipage}

\vskip 0.1in

\noindent In striking contrast to all of the points raised in the
quotes above, in ADG-vacuum Einstein gravity there is no external
(to the gravitational field) background spacetime manifold, and
hence no $\mathrm{Diff}(M)$ either. All there is is the
ADG-gravitational field $(\modl ,\conn)$ `in-itself', and the
vacuum gravitational dynamics that it defines (\ref{eqy23}). Since
the connection is the sole dynamical variable in our theoresis of
gravity (half-order formalism), without any
`metric-commitment',\footnote{That is, without the smooth
spacetime metric $g$ (second-order formalism), or the tetrad $e$
(first-order formalism) being involved at all in the theory.}
ADG-gravity is a purely gauge theory; moreover, with quantum
traits built into the formalism from the very start---{\it ie},
from the very definition of the ADG-gravitational field as the
pair $(\modl ,\conn)$. However, like Sorkin posits above, in
ADG-gravity {\em covariance is inevitably replaced by
synvariance}---{\it ie}, one considers the (principal) group sheaf
$\aut\modl$ of dynamical self-transmutations
(`auto-transformations') of the ADG-gravitational field $(\modl
,\conn)$ `in-itself', so that no need to `gauge' (localize) the
external (Lorentz) spacetime symmetries, which anyway do not exist
(since the external spacetime does not exist in the first place!),
arises whatsoever. In this light, on the one hand Ivanenko and
Sardanashvily's remarks that ``{\em ...apparently the single gap
in the modern gauge picture still remains gauging the external or
space-time symmetries of fields and particles, that includes the
gauge gravity also...but even in the gauge gravitation theory
connections cannot at all substitute the metric, because there are
no groups, whose gauging would lead to the purely gravitational
part of space-time connections (Christoffel symbols or
Fock-Ivanenko spinorial coefficients).\footnote{The ADG
$\struc$-connections are {\em not} (interpreted as) `{\em
space-time connections}', as they are purely gauge fields and no
external to them background spacetime (manifold) is involved.} To
separate such gravitational components from gauge fields, e.g., of
the Lorentz group, metric or tetrad fields have to be
introduced...}'' are rendered obsolete, and on the other, Sorkin's
position about general covariance that ``{\em ...Ideally perhaps,
one would formulate the theory so that labels never entered, but
so far no one knows how to do this---anymore than one knows how to
formulate general relativity without introducing extra gauge
degrees of freedom that then have to be cancelled against the
diffeomorphism invariance...}'' is simply overcome since, apart
from the ADG-gravitational connection $\conn$, there is no other
dynamical (gauge) degree of freedom involved that has to vanish
when operated on by the diffeomorphism constraints (which anyway
do not exist in ADG-gravity, because {\it ab initio} there is no
background manifold).

In particular, regarding the issue of `observables' (especially in
the context of QG), since in the inherently (third) quantum vacuum
ADG-gravity the sole dynamical variable is the connection $\conn$,
the only `{\em measurable} dynamical variable'---what is commonly
understood by the term `{\em observable}'--- is the curvature of
the connection $\curv(\conn)$. The latter, on the one hand is
measurable because it is $\struc$-valued,\footnote{With $\struc$
the structure sheaf of generalized coordinates or arithmetics in
the theory.} and on the other, exactly because it is an
$\struc$-morphism by the very definition of the ADG-gravitational
field as the pair $(\modl ,\conn)$, it remains (gauge) invariant
under the action of the principal group sheaf $\aut\modl$ (of
dynamical self-transmutations) of the field on the associated
(quantum particle states' representation) sheaf $\modl$ of the
field. Equivalently, and categorically speaking, the dynamical
(differential) equations (\ref{eqy23}) in which $\ricci(\conn)$
partakes (in fact, it defines!), are $\struc$-functorial
(`synvariance'). All in all, in ADG-gravity the sole observable
proper is the most important `{\em geometrical object}' ({\it ie},
an $\struc$-tensor) concomitant of the sole dynamical variable in
the theory---the connection field $\conn$: $\curv(\conn)$.

From a more general `theory-{\it aufbau}' perspective, this
accords with Einstein's advice to Heisenberg\footnote{When
Heisenberg asked for his opinion whether quantum theory, much in
the way that Einstein had developed the theory of relativity
(especially the operational SR), should be based solely on
observable entities.} that, roughly, {\em not only a physical
theory cannot be built solely on observable entities, but also,
quite on the contrary, that it should determine what is observable
in the first place} \cite{heisenberg}. In ADG-gravity too the
theory fundamentally hinges on the notion of connection (:field)
$\conn$, which is an unobservable entity since it is not a
`geometrical object' ({\it ie}, it is not an $\struc$-morphism or
$\struc$-tensor). However, the curvature field $\curv(\conn)$ that
the algebraic connection field $\conn$ determines, via which the
vacuum gravitational dynamics is in turn expressed (\ref{eqy23}),
is an observable proper.

\subsubsection{Path integral quantization of ADG-gravity}

We envisage a genuinely covariant, background manifold independent
and inherently singularities' and infinities' free quantum
gravitational dynamics. Since the relevant physical kinematical
configuration space in ADG-gravity is the moduli space $\sconn
/\aut\modl$ of the affine space $\sconn$ of gravitational
connections modulo the `internal' gauge transformations in
$\aut\modl$, the said quantum dynamics can be modelled after an
abstract functional (`path') integral involving a generalized,
{\em Radon-type of measure} on $\sconn /\aut\modl$. The abstract
{\em differential} and {\em integral} calculus on $\sconn$, that
has been developed in gory detail in \cite{mall1} and \cite{mall4}
respectively, will be of great import here.

Since the ADG-technology is expressly base manifold-free, the main
feature of this quantum dynamics is that {\em it is manifestly
background independent}, with `background independence' here
meaning, as emphasized numerous times before, not only `background
{\em metric} independence' as for example in the Loop Quantum
Gravity (LQG) approach to non-perturbative canonical QGR
\cite{ash5,thiem2,thiem3,smolin,alvarez}, but also `base {\em
differential spacetime manifold} $M$ independence'
\cite{malrap1,malrap2,malrap3,rap5}. By the background $M$
independence of ADG-gravity we can totally bypass formidable
obstacles in coming up with a suitable generalized measure to
model the envisaged dynamics, which are due to $\mathrm{Diff}(M)$
\cite{baez2,baez3,baez6}. For as we repeatedly stressed above, in
the purely gauge ADG-gravity, $\mathrm{Diff}(M)$ is replaced by
$\aut\modl$ (and covariance by synvariance!) \cite{malrap3,rap5}.

\subsection{Section's R\'esum\'e}

Again, we summarize this long section by itemizing its basic
tenets:

\begin{enumerate}

\item Much in the same way that the external, background spacetimeless ADG-gravity may be regarded as a
gauge theory of the third kind---a genuinely, purely gauge theory,
with gauge group of the field's auto-symmetries organized in the
principal sheaf $\aut\modl$---it can also be regarded as a third
quantized (or better, third or self-quantum) field theory. This is
the basic result from the geometric prequantization of gravity:
$\modl$ is the associated (representation) sheaf of $\aut\modl$
and its (local) sections may be thought of as the
quantum-particle/generalized position states of the field $\conn$.
In turn, $\conn$ acts on them to (dynamically) change them, as
(\ref{eqy23}) depicts, and in this action we witness the quantum
field-particle duality, which reflects a generalized version of
Heisenberg's indeterminacy relation: by employing $\struc$ to
localize (geometrically represent) the field $\conn$ (as acting on
its associated representation sheaf $\modl$, which by definition
is locally isomorphic to $\struc^{n}$), the latter acts on the
carrier space's ($\modl$'s) sections to change them. In other
words, the ADG-gravitational field $(\modl ,\conn)$ is `quantum
self-dual' (self-quantized), something that can be easily proven
sheaf-cohomologically.

\item With the self-quantum ADG-gravity we have in our hands a
genuinely {\em background independent} and quantal theoresis of
the gravitational field, with background independence meaning here
not only background metric independence (since anyway no metric is
involved in our purely gauge theory), but also background
differential manifold independence. In this vain, no issue arises
of quantizing the external (to the gravitational field) base
spacetime manifold itself and, for example, some recent
`resolutions' of the inner Schwarzschild singularity using LQG
means in which the singularity is resolved by evoking some
discretization-quantization of the space-time continuum results
from QRG, are quite irrelevant/redundant from our ADG-gravity
viewpoint. In a nutshell, we do not expect a quantum theory of
gravity to remove singularities, in the same way that we do not
believe that a conceptually cogent and calculationally finite QG
will be arrived at as a quantization (along QFTheoretic lines) of
the classical theory (GR).

\item From the point made above it sort of follows that we cannot
believe on the (physical existence of the) Planck space-time
length either: that is, we do not believe in it either technically
(calculationally)---as a cut-off length introduced by fiat in
order to regularize divergent analytic expressions ({\it eg},
perturbation series expansions) and manage/control our continuum
based `field-Calculus'---or more importantly, conceptually as
supposedly demarcating a `theoretical boundary' above which the
classical field theory (law) of gravity on the spacetime continuum
holds (GR), but below which another (set of elusive) law(s) holds
in the so-called QG regime, and all this because our continuum
based Analysis miscarries (or anyway, it proves to be problematic,
of limited applicability and validity in the quantum domain). {\it
In toto}, ADG-gravity recognizes no fundamental (space-time) scale
(or length-duration) as limiting the dynamical field-law of
gravity, since there is no spacetime (continuum) in the theory to
begin with (let alone that ADG-gravity does not recognize as
conceptually sound any formal procedure of quantization of a
pre-existing so-called `classical' theory as GR).

\item A fully quantum dynamics for gravity is currently anticipated
to be modelled after a genuinely covariant, functional or `path'
integral-type of scenario over the affine moduli space $\sconn
/\aut\modl$ of purely gauge-equivalent gravitational
$\struc$-connections $\conn$ \cite{mall4}, with the epithets
`genuinely' and `purely' indicating our authentically gauge
version of the PGC of GR being effectuated by the dynamical
self-transmutations (`auto-symmetries') of the gravitational field
dwelling in $\aut\modl$, and expressly not in $\mathrm{Diff}(M)$
since no external (to the field itself), base spacetime manifold
$M$ is involved in the theory. This absence will significantly
enhance and facilitate our search for a suitable Radon-type of
measure on the said moduli space to model our
`sum-over-connection-histories' dynamics, and it will not stumble
and falter on $M$ and its `symmetry' group $\mathrm{Diff}(M)$
\cite{ashlew2,ashlew5,baez2,baez3,baez6}.

\end{enumerate}

\section{Physico-Philosophical Finale: `ADG-Gravity', GR, Singularities, QG and the Euclidean
vs the Cartesian Conception of Differential Geometry}

In this concluding section of the present `paper-book' we give a
physico-philosophical account of the potential import and
aftermath of ADG-gravity {\it vis-\`a-vis} classical gravity (GR),
its $\smooth$-smooth spacetime singularities and QG, as well as we
discuss the wider significance of ADG for (future) developments in
applied (predominantly to the problem of QG) differential
geometry.

\subsection{The Einstein-Feynman-Isham `No-Go' of CDG in the Quantum
Deep}

Below there are three quotations, in chronological order, by
Einstein, Feynman and Isham respectively, about the miscarrying of
the spacetime continuum (manifold) and, {\it in extenso}, about
the inadequacy of the usual differential calculus (CDG) based on
it in the quantum regime, and in particular, in QG research. Our
comments following each quotation will depict a progressive
refinement in the meaning of the words (and attitude!) of those
people about the inapplicability of the manifold and its Calculus
(Analysis) in our attempts to capture the elusive `{\em true
quantum theory of gravity}'.

But before we present these, let us recall from \cite{riemann}
certain `prophetic' remarks of Riemann about (the status and
validity of Euclidean) `geometry in the infinitely small':

\bigskip \noindent (Q8.1)\hskip 0.9in
\begin{minipage}{11cm}

\centerline{\underline{\bf Riemann}}

\noindent ``{\small ...{\em It is upon the exactness with which we
follow phenomena into the infinitely small that our knowledge of
their causal relations essentially depends... It seems that the
empirical notions on which the metrical determinations of space
are founded, the notions of a solid body and of a ray of light,
cease to be valid for the infinitely small. We are therefore quite
at liberty to suppose that the metric relations of space in the
infinitely small do not conform to the hypotheses of (Euclidean)
geometry...The question of the validity of the hypotheses of
geometry in the infinitely small is bound up with the question of
the ground of the metric relations of space}\footnote{Our emphasis
throughout.}...}''
\end{minipage}

\vskip 0.1in

\noindent \underline{\bf Brief Comments:} Without reading too much
in them, there is both quantum and relativity anticipations in
Riemann's words above. The first sentence has more of a quantum
flavor, as the exactness with which we can follow (the causal
relations between) phenomena---{\it ie}, the precision of our
determination (localization or measurement) of events (and the
establishment of their causal nexus)---in the very small is (we
would say nowadays) limited by the quantum of action (quantum
indeterminacy). The rest of the quotation reminds one of
relativity ideas, since the empirical notions (and implicitly, the
operationalist or instrumentalist means) used to found (and
determine) the `metric relations of space'---in Euclidean
geometry, these are the `optical' straight lines of light (and
their incidence relations), as well as the (relations between)
rigid (imponderable) bodies---are intuited to perhaps `miscarry'
in the infinitely small. {\it In toto}, it is speculated that
Euclidean geometry is not the `true' theory of space in the very
small;\footnote{These ideas are not that surprising bearing in
mind that Riemann was Gauss' student so that he must have been
exposed to (the novel back then) non-Euclidean geometry ideas.
Also, let it be mentioned here that, apart from his mentioning
`causal relations' in the first sentence, Riemann is primarily
concerned with a `static' theory of `pure space' (`being'),
without bringing in arguments about `time' (`dynamics' or
`becoming') {\it per se}, so that the geometry that he envisions
is not a physical (dynamical) `space-{\em time}' one as in
Einstein's theory of relativity ({\it eg}, no clocks are
mentioned). One could say that for Riemann `space' is an
objective, `real substance' out there, whose metrical relations
(attributes) {\em we} determine.} moreover, in retrospect, by
identifying the term `{\em geometry in the infinitely small}' with
`{\em infinitesimal geometry}'---what we would nowadays call {\em
differential geometry} or {\em infinitesimal calculus}---by a
small stretch of the imagination we can maintain that {\em Riemann
questioned the validity of the differential calculus in the very
small} (`quantum deep'?). Which brings us to the main triptych of
quotations by Einstein, Feynman and Isham.\footnote{These three
quotations can be also found right before the introduction to
\cite{malrap3}, but there we made no significant comments on
them.}

First comes a quotation of Einstein taken from
\cite{stachel},\footnote{This is a more extended (Q?.?).} which at
the end is also augmented by the conclusion of a later quotation
of his found in \cite{einst3}:\footnote{These are the concluding
words in (Q?.?). As usual, {\em emphasized} text in quotations is
written in {\em emphasis} script, while in square brackets and
{\rm roman script} are our additions for clarity and completeness
of expression.}

\bigskip \noindent (Q8.2)\hskip 0.9in
\begin{minipage}{11cm}

\centerline{\underline{\bf Einstein}}

\noindent ``{\small `...you have correctly grasped the drawback
that the continuum brings. If the molecular view of matter is the
correct (appropriate) one; {\it ie}, if a part of the universe is
to be represented by a finite number of points, then the continuum
of the present theory contains too great a manifold of
possibilities. I also believe that this `too great' is responsible
for the fact that our present means of description miscarry with
quantum theory. {\em The problem seems to me how one can formulate
statements about a discontinuum without calling upon a continuum
space-time as an aid; the latter should be banned from theory as a
supplementary construction not justified by the essence of the
problem---a construction which corresponds to nothing real. But we
still lack the mathematical structure unfortunately}.\footnote{Our
emphasis.} How much have I already plagued myself in this way of
the manifold!...}'', and...

\noindent {\small\em ``...This {\rm [`discrete' (`discretum'
based) or `discontinuous' scenario]} does not seem to be in
accordance with a continuum theory, {\em and must lead to an
attempt to find a purely algebraic theory for the description of
reality. But nobody knows how to obtain the basis of such a
theory}.''}

\end{minipage}

\vskip 0.1in

\noindent \underline{\bf Brief Comments:} Einstein here appears to
be in favor of a `discontinuous' (`discrete' or finitistic)
description\footnote{That is, one based on a `discontinuum' or a
`discretum'.} of physical reality in the quantum deep, instead of
a manifold based, continuous (field) one, since the spacetime
continuum (and the field theory based on it) ``{\em miscarries
with quantum theory and...it corresponds to nothing real}''.
Moreover, he intuits that {\em the desired `discontinuum' based
theory should be formulated by purely algebraic means}, but
unfortunately, at that time, theorizing along such
finitistic-algebraic lines was just `wishful thinking', lacking a
sound (mathematical) basis (foundation).\footnote{In 8.5.? we will
return to comment further, in the light of ADG, about Einstein
propounding a `genuinely discrete', ``{\em purely algebraic
physics}'' \cite{stachel1} {\em contra} the traditional (and more
commonly associated with the name of Einstein) geometric spacetime
continuum (manifold) based field-physics.}

We now come to some similar remarks by Feynman in \cite{feyn1}
{\it contra} the continuum which, furthermore, mention the words
{\em geometry} and {\em infinities} (coming from singularities).

\vskip 0.1in

\bigskip \noindent (Q8.3)\hskip 0.9in
\begin{minipage}{11cm}

\centerline{\underline{\bf Feynman}}

\noindent ``{\small{\em ...the theory that space is continuous is
wrong, because we get...infinities} {\small\rm [ viz.
`singularities']} {\em\small and other similar difficulties}
...{\small\rm [ while]} {\small\em the simple ideas of geometry,
extended down to infinitely small, are wrong\footnote{Our
emphasis.}...}}''
\end{minipage}

\vskip 0.1in

\noindent \underline{\bf Brief Comments:} Feynman, like Einstein,
is clearly against the continuum, which miscarries with quantum
theory (Q?.?)---the description of physical reality at the very
small. For him, this miscarrying is exemplified by the infertility
of ``{\em the simple ideas of geometry}'' when applied to the {\em
infinitely} small,\footnote{One could say, at infinitely small
`scales' or `distances'.} this `wrongness of geometrical concepts'
being manifested by singularities and their associated physically
nonsensical infinities.\footnote{That is, for Feynman, what ``{\em
corresponds to nothing real}'' about the geometrical spacetime
continuum that Einstein mentioned in (Q8.?), is the latter's
singularities and their unphysical infinities---precisely our
thesis here.} Of course, one can subsume this `{\em application of
ideas of geometry to infinitely small}' by the more widely used
term `{\em infinitesimal geometry}', or better, `{\em
infinitesimal calculus}', or even, `{\em infinitesimal
analysis}'\footnote{And it is fitting to recall here Leibniz's
perception of a {\em Geometric Calculus} as a relational
(algebraic) {\it Ars Combinatoria}---a {\em combinatorial art}, as
communicated to de l'H\^{o}pital: ``{\em the secret of Analysis
lies} [precisely] {\it in an apt combination of symbols''}
\cite{mall7}. On the one hand, this appears to corroborate the
aforesaid `purely algebraic physics' propounded by Einstein, and
on the other, of course, it was subsequently vindicated by
Feynman's own combinatory-diagrammatic method of describing
quantum dynamical propagations and interactions.}---all synonyms
of {\em CDG}.

But {\em explicitly}, the terms `classical {\em differential}
geometry', the `{\em Planck scale}' and `{\em quantum gravity}'
appear in the following remarkable quotation of Isham \cite{ish}:

\vskip 0.1in

\bigskip \noindent (Q8.4)\hskip 0.9in
\begin{minipage}{11cm}

\centerline{\underline{\bf Isham}}

\noindent ``{\small{\em ...at the Planck-length scale,
differential geometry is simply incompatible with quantum
theory}...{\small [so that]} {\small\em one will not be able to
use differential geometry in the true quantum-gravity
theory\footnote{Our emphasis.}...}}''

\end{minipage}

\vskip 0.1in

\noindent \underline{\bf Brief Comments:} Isham is straight-out
categorematic: {\em CDG (and the classical $\smooth$-smooth
manifold model of spacetime supporting its constructions) does not
go hand in hand with the quantum, and it will therefore be of no
import to QG research}. On the other hand, and this is one of the
basic theses of the present paper, from an ADG-theoretic point of
view it is not exactly that differential geometric ideas cannot be
used in the quantum regime---as if the intrinsic differential
geometric mechanism (which in its essence is of an algebraic
nature as we have amply argued throughout this paper) fails in one
way or another when applied to the realm of QG---but rather that
when that mechanism is geometrically effectuated or implemented
(represented) by the (mediation in the guise of the smooth
coordinates of the) background $\smooth$-smooth spacetime manifold
as in CDG, then all the said problems crop up and are
insurmountable (within the confines of, {\it ie}, with the
concepts and the methods of, the manifold based CDG). Thus, to
pronounce this subtle but crucial from the ADG-perspective
difference, we maintain that the second part of Isham's quotation
above should also carry the adjective `{\em classical}' in front
of `{\em differential geometry}', and read: `{\em one will not be
able to use differential geometry}' (or equivalently, a
geometrical base differential spacetime manifold) `{\em in the
true quantum-gravity theory}'. {\it In toto}, the aforesaid subtle
distinction hinges on the physical non-existence of a background
smooth spacetime manifold, not of the inapplicability of the
(essentially algebraic) mechanism of differential geometry.

\vskip 0.1in

\subsection{Euclid and Leibniz versus Descartes and Newton vis-\`a-vis DG}

In this subsection we would like to comment briefly on a common
`misconception' of the term `{\em Euclidean geometry}' and the
misleading view of DG in general that this misconception
subsequently led to.

It is fair to say that {\em geometry in Ancient Greece}, which
found its apotheosis in Euclid's axiomatics \cite{euclid}, {\em
was of a purely relational character}. That is, for Euclid for
example, geometry pertained to the study of `incidence',
`parallel', `congruence', `contiguity', `tangent', `inscription',
`circumscription' {\it etc} {\em \underline{relations} between the
various `geometrical objects'} such as points, lines, triangles,
circles {\it etc}. {\it In toto}, Euclid did geometry entirely
{\em relationally}, by using solely the `geometrical objects'
themselves, without the `intervention' of any `space' ({\it eg},
without considering the `space' in which those objects were
`embedded' so to speak).\footnote{The second author recalls a
telling exchange with David Finkelstein, over breakfast, during
the 5th Quantum Structures conference in Cesena (Italy), in which
David told him (as a mild critique of the fact that in ancient
Greece {\em geometry} not {\em algebra} prospered, while it
appears that for the description of the quantum deep nowadays,
algebra is needed not geometry), quite philologically and
picturesquely, that ``{\em for the ancient Greeks, geometry was
drawing and comparing circles and triangles on sandy
beaches}''---the operative word here being `{\em comparing}', not
of course `{\em drawing on sandy beaches}'.}

It was Descartes', revolutionary for his time, idea to introduce
numbers (coordinates) to represent (`coordinatize') the said
`geometrical objects' that `{\em arithmetized}' Euclid's
relational geometry. Objects thus became parts of the
coordinatized `Cartesian space' ({\it eg}, the Cartesian plane to
which the aforementioned circles, triangles {\it etc} belonged)
and their relations became `algebraic' equations between their
coordinates.\footnote{In fact, the objects themselves became
algebraic equations---{\it eg}, the planar circle of radius $r$ is
nothing but the set of point-{\it loci} in the Cartesian plane
whose coordinates $(x,y)$ satisfy the polynomial equation
$x^{2}+y^{2}=r^{2}$.} This marked the birth of the so-called {\em
Analytic Geometry} in which Euclidean geometry was in a sense
algebraicized.\footnote{Thus, analytic geometry may be understood
as the `procedure' towards an `algebraic analysis of geometry'.
Later on, by a process of abstraction, this procedure gave rise to
{\em Algebraic Geometry} where, very loosely, one extracts
`geometrical information' ({\it eg}, `space') from algebra(ic
structures)---{\it eg}, spectral theory of polynomial rings and
other algebraic varieties---and one focuses on solutions of {\em
algebraic} equations ({\it eg}, the circle mentioned above is the
`geometrical object'---{\it ie}, the `geometry'---corresponding to
the solution of the {\em algebraic} equation
$x^{2}+y^{2}-r^{2}=0$).} {\it En passant}, let it be noted here
that, more than two centuries after the cartesian
coordinatization-{\it cum}-arithmetization-{\it
cum}-algebraicization of geometry, Descartes' idea was still
considered to be a bold and radical one, as we read for example
from Hermann Weyl's \cite{weyl}:

\bigskip \noindent (Q8.5)\hskip 0.9in
\begin{minipage}{11cm}
\noindent ``{\small\em ...The introduction of numbers as
coordinates {\rm [in the Cartesian fashion]}\footnote{Our
addition.} was an act of violence\footnote{Our emphasis. This Weyl
quote appears also in a quotation of Shiing-Shen Chern given
next.}...}''
\end{minipage}

\vskip 0.1in

\noindent while Cassirer \cite{cassirer}, antithetically, praises
on epistemological grounds the revolutionary Cartesian
arithmetization of geometry:

\bigskip \noindent (Q8.6)\hskip 0.9in
\begin{minipage}{11cm}
\noindent ``{\small ...[First, Cassirer praises Descartes' method
of analytic geometry, quoting him:] `{\em The sciences in their
present condition are masked and will only appear in full beauty
when we remove their masks; whoever surveys the chain of the
sciences will find them no more difficult to hold in mind than the
series of numbers}'\footnote{Cassirer's own emphasis.} [Then, he
comments on this quotation:] This is the goal of the philosophical
method: to conceive all its objects with the same strictness of
systematic connection as the system of numbers. From the
standpoint of the exact sciences in the time of Descartes, this is
the only manifold which is built up from a self-created beginning
according to immanent laws, and thus can conceal within itself no
question in principle insoluble for thought. The demand that
spatial forms be represented as forms of number and be wholly
expressed in the latter, may appear strange when regarded from the
standpoint of the Cartesian ontology; for in this, `extension'
signifies the true substance of the external objects and is thus
an original and irreducible condition of being. But here the
analysis of being must be subordinated to the analysis of
knowledge. {\em We can only bring space to exact intelligibility
by giving it the same logical character as hitherto belonged only
to number. Number is not understood here as a mere technical
instrument of measurement. Its deeper value consists in that it
alone is completely fulfilled the supreme methodological
postulate, which first makes knowledge knowledge. The conversion
of spatial concepts into numerical concepts thus raises all
geometrical enquiry to a new intellectual level}\footnote{Our
emphasis.}...}''
\end{minipage}

\vskip 0.1in

Now, in our view, when it comes to {\em differential} geometry,
there is an analogous dichotomy of views about Calculus or
Infinitesimal Analysis\footnote{Or, to emulate the term Analytic
Geometry above, `{\em Infinitesimal Analytic Geometry}'.} that may
be traced back to the very creators of differential calculus,
namely, Leibniz and Newton. The first, like we saw in (Q2.?),
envisioned a {\em purely relational} (`combinatorial') Geometric
Calculus in which ``{\em one operates directly on the geometrical
elements,\footnote{What we call here `geometrical objects'.}
without the mediation of coordinates}''; or what amounts to the
same, {\em one refers to the (algebraic) relations between the
`geometrical objects' {\it per se}, without the mediation (or in
the presence) of any (background) `space'}. By contrast, Newton
developed differential calculus in the manifest presence of a
background `space' (assisting the geometrical
picturization/representation of his differential calculus'
concepts and constructions)\footnote{Again, think of the
derivative (or differential) of a function at a point in the
latter's domain, as the slope of the tangent line to the
graph-curve of the function at that point.} to what is presently
identified---via modern technical terms and constructions---with
the manifold based CDG. In complete analogy to the Euclidean-{\it
vs}-Cartesian conception of geometry described above, we have here
the Leibnizian-{\it vs}-Newtonian conception of {\em differential}
geometry.

In fact, this is more than just an analogy if one considers that
the `space' in, on and via which we nowadays develop CDG
(Calculus) {\it \`a la} Newton is a {\em locally Euclidean} ({\it
ie}, locally $\R^{n}$) one---a (differential) manifold $M$, which
is in turn identical to the algebra $\smooth(M)$ of its points'
coordinates (Gel'fand duality). This is a (local)
`arithmetization' (`coordinatization') of the points of that
ambient `space' identical to Descartes' embedding and
coordinatization of Euclid's geometrical objects in the cartesian
plane ($\R^{2}$).

\bigskip \noindent (R8.1)\hskip 0.9in
\begin{minipage}{11cm}
\noindent Alas, we have been misled by the term `locally {\em
Euclidean}', which should better be called `locally {\em
Cartesian}' and hence {\em the usual manifold based CDG should
better be thought of as a Cartesian-Newtonian conception of DG}.
In contradistinction, {\em ADG is a Euclidean-Leibnizian
conception and practice of DG}, as there is no background, locally
Cartesian `space' (:manifold) that intervenes in our calculations
(in our Calculus!) in the guise of coordinates, while the
differential geometric machinery of the theory refers to and
derives from the (algebraic) relations between the `geometrical
objects in-themselves'---the connection fields.
\end{minipage}

\subsubsection{The Promethean `haute couture' of modern
differential geometry}

At this point, keeping the remarks made above in mind, it would be
nice to dwell a bit on the following telling quotation of
Shiing-Shen Chern taken from \cite{chern1}:\footnote{The quotation
that follows is broken up into five `paragraphs'---marked
(I)--(V)---on which we comment separately afterwards.}

\bigskip \noindent (Q8.7)\hskip 0.9in
\begin{minipage}{11cm}
\noindent ``{\small ...It was Descartes who in the seventeenth
century  revolutionized geometry by using coordinates. Quoting
Hermann Weyl, `{\em The introduction of numbers as coordinates was
an act of violence}' \cite{weyl}.\footnote{Quotation (Q8.?)
above.} From now on, paraphrasing Weyl, figure and number, like
angel and devil, fight for the soul of every geometer {\bf (I)}...

...General coordinates need only the property that they can be
identified with points; i.e., there is a one-to-one correspondence
between points and their coordinates---their origin and meaning
are inessential. {\bf (II)}

If you find it difficult to accept general coordinates, you will
be in good company. It took Einstein seven years to pass from his
special relativity in 1908 to his general relativity in 1915. He
explained the long delay in the following words: `{\sl Why were
another seven years required for the construction of the general
theory of relativity? The main reason lies in the fact that it is
not so easy to free oneself from the idea that coordinates must
have an immediate metrical meaning}'.\footnote{Quote (Q?.?) given
earlier in section 3, when we were exploring the deeper meaning of
(the smooth) coordinates (of the base manifold) in GR.} {\bf
(III)}

After being served by coordinates in the study of geometry, we now
wish to be free from their bond. This leads us to the fundamental
notion of a {\em manifold}.\footnote{Chern's own emphasis.} A
manifold is described locally by coordinates, but the latter are
subject to arbitrary transformations. In other words, it is a
space with transient or relative coordinates (principle of
relativity). {\bf (IV)}

I would compare the concept with the introduction of clothing to
human life. It was a historical event of the utmost importance
that human beings began to clothe themselves. No less significant
was the ability of human beings to change their clothing. If
geometry is the human body and coordinates are clothing, then the
evolution of geometry has the following comparison.}

\vskip 0.05in

\centerline{\small Synthetic geometry~~~~Naked man}

\centerline{\small Coordinate geometry~~~~Primitive man}

\centerline{\small Manifolds~~~~Modern man {\bf (V)}...''}

\end{minipage}

\vskip 0.1in

\begin{itemize}

\item {\bf (I)} It is an interesting philological remark that Weyl made, that is, to identify
the Euclidean figure---what we called earlier the `{\em inherently
geometrical object}' (or equivalently, the `geometrical object
in-itself')---with an {\em angel}, while the Cartesian
number---the coordinates introduced `by hand' or `by fiat' ({\it
ie}, by ``{\em an act of violence}'') by Descartes to represent
arithmetically those genuinely geometrical figures of
Euclid---with the {\em devil}. Like the {\em Promethean fire}, the
Cartesian arithmetization (and concomitant algebraicization) of
the purely relational Euclidean geometry, was a double-edged
sword, ``{\em a boon and at the same time a curse brought by a
man, from the Gods, to Man}'' \cite{aeschylus}. Descartes brought
Euclid's geometry `down to earth'---as it were, from Olympus to
Athens---for after all, as Aeschylus says in {\itshape Prometheus
Bound} \cite{aeschylus}:

\vskip 0.1in

\centerline{(Q8.8)\hskip 0.3in ``{\small\em ...Number {\rm [like
fire]} is the most remarkable of human
inventions...}''\footnote{``$A\rho\iota\theta\mu
\acute{o}\nu~\acute{\epsilon}\xi o\chi o\nu~\sigma
o\phi\iota\sigma\mu\acute{\alpha}\tau\omega\nu$''. Our
[...]-addition and emphasis.}}

\vskip 0.1in

\item {\bf (II)} Like we emphasized in sections 2 and 3, in the case of the (locally) Cartesian---what is
`misleadingly' called today (locally) Euclidean---coordinate
patches of a differential manifold $M$,\footnote{When we speak of
(locally) Cartesian coordinates, we do not just mean {\em
orthocanonical} coordinate systems as the epithet {\em Cartesian}
has come to signify in modern (analytic) geometry. We mean general
coordinate (numerical) identifiers (labels) of the (Euclidean)
points of a manifold ({\it ie}, more in line with the general
sense of `coordinates' due to Gauss and his student Riemann). But
the idea is the Cartesian identification of a point of an
$n$-dimensional (real) manifold, with an (arbitrary) $n$-tuple of
(real) numbers.} this 1--1 correspondence (`identification')
between the geometrical (Euclidean) points of $M$ with their
smooth coordinates in $\smooth(M)$ is what we referred to as
Gel'fand duality (spectral theory) in section 2, and represented
it by the amphidromous arrows in (\ref{eq1}). To something more
significant, in ADG the `{\em origin and meaning}' of a structure
sheaf $\struc$ chosen to deal with a differential geometric
situation (problem) `{\em is inessential}', insofar that $\struc$
furnishes one with the essentially algebraic differential
mechanism---{\it ie}, it effectively provides one with a
$d$---with which one can do {\em differential} geometry in the
first place.\footnote{See also (III) and (IV) next.}

\item {\bf (III)} We commented extensively on this in section 3. The
upshot was that although (the real smooth) coordinates (of the
manifold) do not have an immediate metrical meaning, on the one
hand they are the elements where the components of the metric take
values (thus the metric qualifies as being a smooth
$\otimes_{\smooth_{M}}$-tensor field in GR), and more importantly
on the other, they are the very vital (`kinematical')
preconditions for setting up (formulating) GR differential
geometrically, since a differential manifold is nothing else but
$\smooth_{M}$.

\item {\bf (IV)} That coordinates do not have an immediate metrical
meaning in GR essentially means that they do not have any
dynamical (physical) significance in the theory, as they do not
partake into the gravitational dynamics (Einstein equations). This
is the meaning of the PGC of GR---{\it ie}, that the Einstein
equations are generally covariant (form-invariant under general
coordinate transformations). Thus, the biunique (1--1
correspondence) smooth coordinate identifiers (labels) of the
points of $M$ are not physically significant, and hence GR is
fundamentally pointless---or to emulate Chern, coordinates (and
the manifold's points that they correspond to) are `{\em
dynamically transient}'. Of course, as emphasized in sections 2
and 3, the PGC and, what Chern also refers to above, the principle
of relativity (PR), are essentially supported by the PFR, namely,
to paraphrase again Einstein from (Q2.?), `{\em what has the
dynamical field of gravity to do with the coordinate systems that
\underline{we} propose to describe it?}'. Recall again Aeschylus
from above: number,\footnote{Here, the locally Cartesian
coordinate-identifiers of the points of a manifold.} a remarkable
invention as it may be, it is still \underline{our} concept, and
no matter how successful our descriptions of Nature have been
hitherto based on it,\footnote{And in general, regardless of the
enormous success in both pure and applied mathematics (theoretical
physics) that the concept of a locally Cartesian space (:a
manifold) and the CDG-framework based on it has enjoyed last
century.} soon its `human limitations' are reached.\footnote{In
the case of the background manifold, the singularities of GR, and
the general miscarrying of the manifold based CDG in the quantum
(especially in the QG) domain.} Perhaps it is in this particular
light, namely, that we impose our (undoubtedly ingenious
mathematical) inventions, concepts and constructions to
Nature---or equivalently, that we ``{\em forget their terrestrial
origin and accept them as unalterable facts, which then become
labelled as `conceptual necessities', `a priori situations',
etc}''\footnote{In view of Aeschylus' quotation (Q8.?) above, our
comments about the locally Cartesian character of a manifold, as
well as Einstein's words here about the man-made nature of {\em
our} concepts, it is a quite remarkable coincidence that a
manifold was called a {\em Man}-ifold (pun intended).}
(Q?.?)---that Weyl's remark in (Q8.?) should be understood: {\em
it was a violent act\footnote{Albeit, a {\em revolutionary} one,
as Chern says in paragraph I of the quotation above.} against
Nature to introduce numbers as coordinates and, {\it in extenso},
to identify in physics physical spacetime (geometry) with that of
a locally Cartesian space(time)}. However, in line with the
quintessential didactics of ancient Greek tragedy, sooner or later
such `{\em anthropic hubris}' is `punished' by Nature\footnote{By
the Gods in ancient Greek tragedy.}---{\it ie}, at least
\underline{our} (mathematical) theories of Her crash (or
breakdown) at the limit of their applicability and
validity.\footnote{Like the CDG-based GR {\it contra}
singularities.}

\vskip 0.08in

\centerline{(R8.2)\hskip 0.3in {\em It is always relieving to see
Nature `outsmarting' us in the end.}}

\vskip 0.08in

\item {\bf (V)} Chern's account of the historical development of
geometry and its comparison with the `progress' that humans made
in clothing themselves, is quite interesting in its own right: at
the first, more `primitive' so to say level, he puts synthetic
geometry, which is more-or-less as we described the (Euclidean)
geometry in Ancient Greece above---{\it ie},  relational and
coordinate-free. The synthetic (or the Euclidean) geometer is
compared to `naked man', as he lays bare (he studies) his naked
body (:geometry=the `geometrical objects'---points, lines,
surfaces---and their relations) as it is (`in-themselves'). At the
next level he puts the Cartesian introduction of coordinates,
which marks the commencement of Analytic Geometry---the
arithmetization of geometry. Indeed, coordinates may be likened to
the clothes in which we dress ourselves: the (points of the)
`geometrical objects' of Euclid are `labelled' by numbers and they
satisfy algebraic equations. At the last level we encounter the
modern (differential) geometer studying the (differential)
geometry of manifolds---(smooth) `spaces' which are locally
labelled by coordinates in the Cartesian fashion, but with these
coordinates having, as Chern says, only `relative' or `transient'
significance since they are `variable' ({\it ie}, liable to
general/arbitrary coordinate transformations). Of course, it would
hardly be an exaggeration to claim that geometers arrived at the
notion of {\em manifold} in order to be able to do {\em
differential} geometry (CDG or Analysis).

Below is {\em our} (slightly and suitably modified) version of
Chern's table in (Q8.?) above:

\vskip 0.05in

\centerline{\small Euclidean-Synthetic Geometry~~~~Naked man}

\centerline{\small Cartesian Analytic Geometry~~~~Primitive man}

\centerline{\small Locally Cartesian (Differential) Geometry on
Manifolds~~~~Modern man}

\centerline{\small Background Manifoldless (Differential) Geometry
(ADG)~~~~Future man: `back to nakedness'}

\vskip 0.05in

\noindent with the last entry indicating the current tendency (in
differential geometry via ADG) to free ourselves from a background
`space(time)' (:manifold). It should be noted here that as
theoretical physics research in GR gave tremendous impetus to
mathematicians (geometers) to develop the {\em differential}
geometry of smooth manifolds (CDG), so nowadays QG research has
given us numerous reasons and motivations to `throw away' the
background (spacetime) manifold and, if possible, still develop
differential geometry, entirely algebraically (categorically), in
its absence---{\it eg}, like in ADG.
\end{itemize}

\noindent In this line of thought, we cannot refrain from quoting
the following pertinent words from the prologue to the Russian
edition of the first author's book \cite{mall1}:

\bigskip \noindent (Q8.9)\hskip 0.9in
\begin{minipage}{11cm}
\noindent ``{\small\em .....This is an unexpected help from ADG
now when the necessity has grown to study manifolds with
singularities and even to remove the underlying space (for
example, `spacetime') and proceed to a direct description of the
structures on this manifold, which may be important for many
branches of contemporary mathematical and theoretical
physics\footnote{Our emphasis.}...}''
\end{minipage}

\vskip 0.1in

\noindent Indeed, ADG scraps off the underlying manifold and
proceeds directly to study (purely algebraico-categorically, {\it
ie}, sheaf-theoretically) the (relations between the)
`(differential) geometrical objects' that live on that `space',
which `space', in turn, does not contribute at all to ADG's
inherently algebraic differential geometric mechanism. Thus
viewed,

\vskip 0.08in

\centerline{(R8.3)\hskip 0.3in {\em ADG is a Euclidean-Leibnizian
way of doing DG}}

\vskip 0.08in

\noindent and the current (differential) geometer, after trying
all his clothes on, frees himself completely from coordinates and
`space', and goes back to nakedness---as it were, `back to the
future man'. But in the next subsubsection, let us dwell a bit on
how and in what sense the `ADGeometer' actually manages not to be
flashed and dazed by the glittering and luxurious clothes of
Analysis, and goes back to the bare basics (ie, without his
fashionable background geometrical manifold clothes).

In this gist, let us close this discussion with some remarkable,
from the Leibnizian ADG-vantage, words of Ernst Cassirer from
\cite{cassirer} about `infinitesimal analysis' ({\it ie}, DG or
Calculus), which follow immediately after his appraisal of
Descartes' method of arithmetizing geometry quoted above(Q8.?):

\bigskip \noindent (Q8.10)\hskip 0.9in
\begin{minipage}{11cm}
\noindent ``{\small ...Criticism of these methods [of Descartes'
in the field of infinitesimal geometry]\footnote{Our addition for
textual clarity and continuity.} begins with Leibniz and is
brought to a first conclusion with his founding the analysis of
position. It is charged that analysis is not able to establish the
universal principle of order, upon which it prides itself, within
the field to be ordered, but that it is obliged to have recourse
to a point of view external to the object considered. The
reference of a spatial figure to arbitrarily chosen coordinates
introduces an element of subjective caprice into the
determination; the conceptual character of the form is not
established on the basis of properties purely within itself, but
is expressed by an accidental relation, which may be different
according to the choice of the assumed system of reference.
Whether from among all the various equations which can be applied,
according to this process, in the expression of a spatial figure,
the relatively simplest is chosen depends upon the individual
skills of the calculator, and thus upon an element which the
strict progress of method seeks to exclude. {\em If this defect is
to be avoided, a procedure must be found which is equal to the
analytic methods in conceptual rigor but which accomplishes the
rationalization wholly within the field of infinitesimal geometry
and of pure space. The fundamental spatial forms are to be grasped
as they are `in themselves' and understood in their own laws
without translation into abstract numerical
relations}\footnote{Our emphasis.}...}''
\end{minipage}

\vskip 0.1in

\noindent Cassirer hit the nail on the head {\it vis-\`a-vis} the
Euclidean-Leibnizian ADG: what he refers to as `{\em fundamental
spatial forms}' are the Leibnizian `{\em geometrical objects}' we
saw in the Bourbakis' quotation (Q2.?), that, in turn, correspond
to the `algebraic' (:relational) and Euclidean (in the sense we
gave to this epithet above) ADG-fields ({\it viz}, connections
$\conn$),\footnote{And their `purely geometrical' concomitants
({\it eg}, curvatures $\curv(\conn)$).} which are, to paraphrase
Cassirer, ``{\em grasped} (by the ADG theory) {\em `in themselves'
and understood in their own laws without translation into abstract
numerical relations}''. That is to say, to stress it once more,
{\em the ADG-fields are autonomous, self-sustaining (differential)
geometric entities in no need of Cartesian arithmetization; or
equivalently, in no need of the intervention of `space' in the
guise of (numerical) coordinates}.

\subsubsection{Freeing DG from the glittering trappings and
the golden shackles of Analysis}

As we have been arguing back in section 4, CDG---the differential
Calculus or Analysis on smooth manifolds---is a
topologico-algebraic affair, since basically, in order to
differentiate ({\it ie}, to define a differential $d$), one must
be able to take differences and form products (algebraic
structure), as well as take limits thereof (topological
structure). In the guise of the topological algebra $\smooth(M)$
(or equivalently, the structure sheaf $\smooth_{M}$), the
differential manifold $M$ underlying CDG is a rich enough
structure to accommodate the usual notion of differentiability
({\it ie}, furnish us with a $d$ with which we can actually do DG
via $M$) and in this sense the underlying `space(time)' mediates
(or intervenes in) our usual differential geometric
calculations---our Differential Calculus.

On the other hand, as it was also highlighted in section 4, {\em
ADG's principal virtue is that it provides a purely algebraic
framework for doing DG, without commitment to or dependence on an
underlying `space'} (and at that, a differential manifold
one!)---as it were, its principal achievement is `{\em the
complete algebraicization of Calculus}'. In other words, in view
of the topologico-algebraic character of $d$ noted above, in ADG
we underplay and `atrophize' the `spatial', topological
(:geometric) aspect of Analysis---{\it ie}, the fact that CDG has
its basis on a base\footnote{Pun intended.} (`smooth') space
(:manifold)---and we pronounce its purely algebraic aspect.
Indeed, as we argued many times in the present paper, the
arbitrary base topological space $X$ on which the algebra and
differential module sheaves of interest are soldered plays no role
in the {\em `inherently' algebraic} differential geometric
mechanism of ADG, which, as it happens, derives from the stalk of
the said sheaves---{\it ie}, from the algebraic structures
dwelling in the relevant sheaf spaces, not in the base space. For
anyway, the basic differential operators in ADG ({\it ie}, the
connections $\partial$ and $\conn$) are defined as {\em sheaf
morphisms}, so {\em by definition} the base topological space $X$
plays essentially no role in the ADG-notion of
`differentiability'.\footnote{Another way to say this is that, as
{\em Algebraic} Geometry is defined by {\em algebraic} equations
(as mentioned a couple of footnotes before), so {\em Differential}
Geometry is defined by {\em differential} equations, which on the
one hand in CDG are made possible by the intervention of a
background `space' (:geometrical representation of the `evolution'
or the `change-relations' that the differential equations stand
for), but on the other, in ADG they are equations between sheaf
morphisms---{\it ie}, they are relations between the geometrical
objects themselves---without a mediating (geometrical
representation) `space' whatsoever. Shortly we will come to
comment on this `end of geometrical picturization' in ADG, and its
potential significance in theoretical physics.}

In the light of the remarks above, the usual Cartesian (Analytic)
conception of (differential) geometry,\footnote{Recall from a
couple of footnotes above our calling the usual Differential
Calculus `{\em Infinitesimal Analytic Geometry}'.} brings to mind
George Birkhoff's words in \cite{birkhoff}:

\bigskip \noindent (Q8.11)\hskip 0.9in
\begin{minipage}{11cm}
\noindent ``{\small\em ...{\rm [There is this]} disturbing secret
fear that geometry may ultimately turn out to be no more than the
glittering intuitional trappings of analysis\footnote{Our
emphasis.}...}''
\end{minipage}

\vskip 0.1in

\noindent which can be suitably modified a bit here to the
following nightmarish feeling---a `Kafka Castle' \cite{kafka} type
of foreboding---about DG:

\bigskip \noindent (R8.4)\hskip 0.9in
\begin{minipage}{11cm}
\noindent  There is this {\em disturbing secret fear} that {\em
Differential} Geometry may ultimately turn out to be no more than
{\em the glittering intuitional trappings of Infinitesimal
Analysis}.
\end{minipage}

\vskip 0.1in

\noindent But what are we really trapped by in our yearning to do
DG? The answer we suggest is: {\em by} (the underlying) {\em
`space'}.

\bigskip \noindent (R8.5)\hskip 0.9in
\begin{minipage}{11cm}
\noindent  Insofar as Analysis (:the usual Differential Calculus)
is understood as the (apparently inextricable, at least from the
classical, manifold based viewpoint) enmeshing of Algebra
(`relations') and Topology (`space'), what has so far `trapped'
our differential geometric endeavors is (the base) `space'
(:manifold), and {\em the fear is that DG will turn out to be no
more than CDG} ({\it ie}, Infinitesimal Analysis on Manifolds).
\end{minipage}

\vskip 0.1in

\noindent In contradistinction, {\em ADG, by suppressing the
underlying topological (`spatial') and at the same time by
pronouncing the overlying algebraic (`relational') aspect of
DG,\footnote{The terms `underlying' and `overlying' here are meant
to refer to the base (topological) and the (algebra inhabited)
sheaf spaces involved in ADG, respectively. ADG `dissects' the
topological from the algebraic aspect of differentiability, it
`downplays' the former and `overplays' the latter.} comes to
dispel that fear that Birkhoff mentions above and free us} ({\it
ie}, our DG) {\em from}, to use a phrase by Chris Isham
\cite{ish}, {\em `the golden shackles' of the geometrical manifold
based CDG}.

The reader should not underestimate here the significance of the
epithets `{\em glittering}' (to trappings) and `{\em golden}' (to
shackles) used by Birkhoff and Isham, respectively. For there is
no doubt that the CDG of smooth manifolds has been of great import
to both pure and applied mathematics (in particular, in
theoretical and applied physics); nevertheless, its `negative
spell' and shortcomings has been felt in both disciplines. In
particular, it has been felt very early on by Einstein {\it
vis-\`a-vis} the singularities of GR (which are due to the base
spacetime continuum), and the `inherently finitistic-algebraic'
nature of the quantum. We thus recall again, from (Q?.?),
Einstein's words about a `discontinuum based theory' as opposed to
the (spacetime) continuum based CDG of GR:

\bigskip \noindent (Q8.12)\hskip 0.9in
\begin{minipage}{11cm}
\noindent ``{\small\em ...By discontinuum theory I understand one
in which there are no differential quotients. In such a theory
space and time cannot occur, but only numbers and number-fields
and rules for the formation of such on the basis of algebraic
rules with exclusion of limiting processes\footnote{Our
emphasis.}...}''
\end{minipage}

\vskip 0.1in

\noindent These words are remarkable indeed, for they show that
Einstein had not only deep physical intuition, but also
penetrating mathematical vision as well. He advocated ``{\em a
purely algebraic theory for the description of reality}'' (Q?.?)
not based on the spacetime continuum, thus inevitably a theory not
founded on CDG. To his mind, `differential quotients', essentially
the basic process of taking derivatives (:the basic structural
procedure in doing DG!), could not occur since limiting processes
had to be excluded---in effect, since topology ({\it ie}, the
background `space') had to be excluded.\footnote{For there is no
notion of limit (:convergence) without topology.} Thus Einstein
separated (in his mind at least) the topological from the
algebraic aspects of DG, scrapped the first and opted for the
second; however, since in the CDG of manifolds the algebra is
apparently inextricably entwined with the (continuum) topology,
dropping the latter meant for him the complete abandonment of
differential geometry\footnote{And of the spacetime interpretation
that goes hand in hand with the background geometrical continuum.}
(in the quantum regime). By contrast, ADG does not at all drop
differential geometry: it simply abandons the base
continuum\footnote{And its geometrical spacetime interpretation.}
with its infinitary limit processes (and retains a general, in
principle arbitrary, topological space on the background), while
at the same time it emphasizes the algebraic character of
DG---what we called earlier DG's inherently algebraic differential
geometric mechanism.

\bigskip \noindent (R8.6)\hskip 0.9in
\begin{minipage}{11cm}
\noindent All in all, we can now say, to parallel Leibniz's
remarks to de l'H\^opital,\footnote{See footnote ? above.} that
`{\em the secrets} (or perhaps better, the virtues!) {\em of
Analysis lie in the algebra that it employs, while its pitfalls}
(or better, shortcomings, especially in the quantum deep) {\em in
the intervening `geometry' (or topology, or even `space(time)')
that it also employs, with its infinitistic, limiting
procedures'...}''
\end{minipage}

\vskip 0.1in

In the light of the remarks above, it is fitting to make the
following important, {\it vis-\`a-vis} the singularities and their
associated unphysical infinities of the spacetime manifold
grounded physical theories,\footnote{GR and QFTs of matter
included.} general observation in the form of an `aphorism':

\bigskip \noindent (R8.7)\hskip 0.9in
\begin{minipage}{11cm}
\noindent {\em In algebra} ({\it ie}, in the `relational' aspect
of DG) {\em there is no `infinity' involved whatsoever.
Infinities} (in DG) {\em arise when the background topological
continuum} ({\it ie}, the base `spatial' aspect of DG), {\em with
its infinitary limit processes, gets involved} (in our
differential geometric calculations---our Differential Calculus),
{\em like in the spacetime continuum based CDG}.\footnote{Which
manifold founded CDG is {\em the} mathematical framework within
(or {\em the} mathematical language in) which both GR and the QFTs
of matter are formulated.}
\end{minipage}

\subsection{What is Differential Geometry?---ADG's Unifying Platform (based on a theme by Shing-Shen Chern)}

Motivated by the discussion in the previous subsection, in the
present one we would like to recapitulate the basic tenets of ADG
under the prism of the main stages in the historical development
of geometry---placing particular emphasis on issues concerning
{\em differential} geometry {\it per se}---as perceived and (in
the order) presented by Shing-Shen Chern in his paper ``{\itshape
What is geometry?}'' \cite{chern}. What will emerge from this
summary-via-comparison is that ADG judiciously selects and
combines, for the (axiomatic/abstract) development of an entirely
algebraic (:sheaf-theoretic) and base $\smooth$-smooth
manifold-free {\em differential} geometry, certain key ideas and
central notions in the developmental history of geometry, as well
as it revises or totally evades others pertaining to {\em
differential} geometry proper. To this end, the discussion in the
previous subsection will come in handy.

First of all, it goes without saying of course that this is no
place and no time\footnote{For there is no {\em space-time} in the
first place! (pun intended)} for us to attempt at a `definition'
(or concise description) of such a broad mathematical subject as
DG. We will rather present certain ideas, mainly motivated and
supported by lessons we have learned from developing and applying
ADG to theoretical/mathematical physics---especially to classical
and QG---which could help one outline and sketch, albeit coarsely,
quintessential in our opinion features that a predominantly
physically motivated (mathematical) theory, such as ADG, ought to
possess in order to qualify as being `differential geometric'
proper. In this endeavor we will also set the stage for certain
important distinctions between `physical' and `mathematical' space
and geometry {\it \`a la} Peter Bergmann that we are going to draw
shortly.\footnote{In subsection 8.4 next.} What is going to emerge
from these distinctions is that what ought to qualify as (D)G
proper is in our view fundamentally conditioned by---in fact,
inextricably tied to---{\em physical} considerations. For if we
have learned anything at all from Einstein's GR is that geometry,
far from being a crystalline-rigid structure {\it a priori} fixed
by the theoretician (better, the mathematician!) and being
detached from natural processes (of dynamical propagation and
interaction}, is a flexible, conditioned by and subject to {\em
physical dynamics}, enterprize \cite{df7,df8,putnam}.

After these anticipatory remarks about `{\em the physicality of
geometry}', the starting point of our present discussion about
what is or what `ought' to qualify in our ADG-based view as
differential geometry is the rather basic observation, already
touched on briefly in footnote ? above, that as {\em algebraic}
geometry essentially began as the study of the (solution)
properties and associated (solution) spaces of {\em algebraic}
({\it eg}, polynomial) equations, so similarly {\em differential}
geometry may be generally perceived as being primarily concerned
with the study of {\em differential equations} (DEs) and their
{\em solution spaces}. One can gain more insight into and
understand the `naturalness' of this broad, `mock definition' of
DG by dissecting the term into its two constituent words---{\em
differential} and {\em geometry}---in the following way:

\paragraph{Differential $d$: the condition for setting up the DE; the origin of the entire DG set-up.}
We commence with an `aphorism', namely that

\bigskip \noindent (R8.8)\hskip 0.9in
\begin{minipage}{12cm}
\noindent {\em there can be conceived no differential equation and
{\it in extenso}, by our pseudo-definition above, no DG, without
first having some sort of `differential' (operator) $\kd$.}
\end{minipage}

\vskip 0.1in

\noindent The question that follows then is {\em where does this
`differential' come from}? Since $\kd$---in effect, the concept of
derivative---is a {\em topologico-algebraic} notion as repeatedly
stressed before, while, {\it a fortiori}, the operation of
differentiation is a {\em local} process, the usual CDG of
(finite-dimensional) manifolds $M$ secures a (mathematical)
milieu---a `space'---in which a $\kd$ can be snugly accommodated,
as follows:

\begin{itemize}

\item Since by definition an $n$-manifold is {\em locally}
isomorphic to $\R^{n}$, with the latter coming equipped with the
usual topological and differential structure of the real line $\R$
(and the Cartesian products thereof), $\kd$ is well secured; but
perhaps more importantly, from a functional (and from our
ADG-theoretic) viewpoint,

\item Since (a real smooth manifold) $M$ itself is equivalent to $^{(\R)}\smooth(M)$ by
Gel'fand duality and its associated spectral theory (\ref{eq1}),
the latter, which is {\em the} archetypical example of a
(non-normable) {\em topological algebra}, secures a place for the
topologico-algebraic $\kd$, with the {\em local} nature of $\kd$
being in turn satisfied by the sheaf-theoretic localization of
$^{(\R)}\smooth(M)$---as the structure {\em sheaf}
$^{(\R)}\smooth_{M}$---over $M$ (simply regarded as a topological
space).\footnote{For, after all, the notion of sheaf is, by
definition, a {\em local-topological} one, being defined as a {\em
local homeomorphism}.}

\item All in all, we recall from \cite{malrap3}, albeit slightly
modified, the following diagram summarizing the `{\it raison
d'\^etre et de faire}' of CDG:

\[
\begin{array}{c}
\boxed{\mathbf{Background~Geometry~(BG)}~M}\cr\cr
\mathrm{CDG\equiv
\smooth\!\!-\!\!Manifolds}~M\equiv\smooth_{M}(\equiv\mathrm{BG})\stackrel{(a)}{\mapto}\mathrm{Tangent~
bundles}\stackrel{(b)}{\mapto}\cr \mathrm{Smooth~Vector~
Fields}(\equiv\mathrm{Derivations})\stackrel{(c)}{\mapto}\mathrm{Differential~
Equations}
\end{array}
\]

\noindent in the following sense: {\em the smooth manifold was
`made' (or `invented') for the tangent bundle $(a)$, which in turn
was made for the vector fields $(b)$, which were finally made for
the differential equations} $(c)$; hence, by `arrow-transitivity',
the diagram above vindicates our `mock definition' of DG, that is
to say, that {\em the manifold supported CDG was ultimately
`invented' for the DEs}. To summarize, in CDG the base manifold
$M$, as a background geometrical space, `mediates' our
Calculus---{\it ie}, the smooth coordinates $\smooth(M)$ play a
crucial role in our setting up DEs---as it is the vital
pre-condition for securing a $\kd$ in terms of which DEs can be
formulated in the first place.

\end{itemize}

In glaring contrast, as it was also observed in \cite{malrap3}, in
the purely algebraic (:sheaf-theoretic) ADG, no smooth base
manifold is used at all in order to `mediate our calculations',
ultimately, to set up DEs. In fact, {\em ADG refers directly to
the algebraic fields} ({\it viz.} connections) {\em themselves,
which define directly DEs without dependence on a background
geometrical manifold $M$ to provide us with a differential, which
differential (or the generalized differential---{\it ie}, the
connection $\conn$) in turn comes from (structure sheaves of)
algebra(s)}:

\[
\begin{array}{c}
\boxed{\mathbf{No~Background~Geometry~(BG)}~M}\cr\cr
\mathrm{ADG}\equiv\mathrm{`Algebra'}\stackrel{(a^{'})}{\mapto}\mathrm{Fields}~(\modl
,\conn)\stackrel{(b^{'})}{\mapto}\mathrm{Differential~Equations}
\end{array}
\]

\noindent which, again by arrow-transitivity, can be read as
follows: {\em ADG refers in an algebraico-categorical way directly
to the dynamical fields---represented by pairs such as `$(\modl
,\conn)$'---and to the differential equations (physical laws) that
they define, without the intervention (neither conceptually nor
technically) of any notion of (background geometric manifold)
space(time) $M$, or equivalently, independently of any intervening
smooth coordinates} $\smooth(M)$. In other words, {\em ADG deals
directly with the differential equations (the laws of physics),
which now are `categorical equations' between sheaf
morphisms---the $\struc$-connections $\conn$ acting on the (local)
sections of the associated vector (representation) sheaves $\modl$
under consideration.}

\paragraph{`Geometry' and `space': the result from solving the
DE; the `secondary effect' of the DE; `solution space'.} The
foregoing discussion can be distilled to the following: the entire
DG enterprize revolves in one way or another about the notion of
differential $\kd$ which one wants to possess in order to actually
do DG---{\it ie}, write down (and hopefully solve!) DEs in the
first place. In the usual theory (CDG), $\kd$ is secured by the
character of the underlying (background) `space' (:manifold)
employed at the very basis of the theory. In this sense, in CDG
(or Analysis) `geometry' (the base geometrical $M$) precedes $\kd$
as well as the DEs that can be formulated with it. Indeed, Chern
in \cite{chern} has pondered on the `mysterious' origin of $\kd$:

\bigskip \noindent (Q8.13)\hskip 0.9in
\begin{minipage}{11cm}
\noindent ``{\small\em ...A mystery is the role of
differentiation. The analytic method is most effective when the
functions involved are smooth. Hence I wish to quote a
philosophical question posed by Clifford Taubes: Do humans really
take derivatives? Can they tell the difference?}\footnote{We are
going to answer to this question, from a physical point of view,
in the next subsection when we talk about `{\em physical
(differential) geometry}'.}...''
\end{minipage}

\vskip 0.1in

\noindent which is of course settled ({\it ie}, $\kd$ is not that
mysterious after all!) within the realm of the smooth manifold
based CDG (Analysis).

In striking contradistinction however, in ADG no base $M$ is
employed to furnish us with a differential (or a generalized one,
$\conn$).\footnote{Which in turn means, {\it \`a la} Chern, that
{\em differentiation still remains a `mystery'}.} The latter is
obtained in the theory abstractly (categorically) from algebra,
not (point set-theoretically) from a background geometrical
`space'. The DEs in ADG are equations between sheaf morphisms (the
$\kd$s involved in the theory). In ADG, at the origin (or basis)
of DG is algebra (loosely speaking, a `relational' structure),
which furnishes us with an algebraico-categorical $\kd$, in terms
of which DEs can then be written down. The terms `geometry' or
`space' then pertain to the `{\em solution space}'---the realm
where the relevant DEs actually hold, much like we said some
footnotes before about {\em algebraic} geometry and we gave the
example of the circle as the `solution space' of the {\em
algebraic} equation $x^{2}+y^{2}=r^{2}$. This is exactly how we
think ADG-theoretically, for instance, about the differential
(vacuum) Einstein equations (\ref{eqy23}): there, the `geometrical
object' in the theory---namely, the curvature $\ricci(\conn)$ of
the algebraic gravitational connection field $\conn$---satisfies
(\ref{eqy23}); moreover, it does so on the entire `carrier' (or
representation) sheaf space $\modl$.\footnote{Which itself comes
from algebra, as by definition it is (locally) of the form
$\struc^{n}$.} This observation lies at the heart of the
ADG-conception of DG and more importantly of its general viewing
`geometry' (or `space') as the result (`solution space') of
algebra---{\it ie}, of the differential equation defined by the
algebraic $\kd$ (or equivalently, $\conn$). In turn, in the domain
of physics, this reflects our general attitude about the priority
of dynamics over kinematics and about the higher significance of
the field law itself than of its solution that we saw earlier, as
well as about what actually constitutes the `{\em physical
geometry}' (or `{\em physical space(time)}') that we will see in
the next subsection.

But before we proceed to the next subsection where we argue,
inspired by Peter Bergmann, in favor of a `physical geometry'
instead of a `geometrical physics', let us borrow from the
conclusion of \cite{chern} the six ``{\em major developments in
the history of geometry}'' and their corresponding protagonists,
and juxtapose them against the basic features of ADG, which we
claim combines basic features from all the basic `epochs' in the
development of geometry and especially of {\em differential}
geometry. The following table, summarizing this juxtaposition, is
`self-explanatory' in view of what has been said above:

\begin{tabular}{p{0.65\textwidth}p{0.65\textwidth}}
\hspace{0.5\textwidth} & \hspace{0.5\textwidth} \cr \hskip 0.4in
\underline{\large\bf Chern} &  \hskip 0.7in\underline{\large\bf
ADG}\cr & \cr {\sl Axioms (Euclid)} & ADG is an {\em axiomatic},
Euclidean ({\em relational}) conception of {\em differential}
geometry. \cr & \cr {\sl Coordinates (Descartes, Fermat)} &
Structure algebra sheaf $\struc$ of generalized ari-\\ & thmetics
(`coordinates'); all differential geo-\\ & metry boils down to
it---{\it ie}, to {\em algebra}. \cr & \cr {\sl Calculus (Newton,
Leibniz)} &
 ADG is more {\em Leibnizian} ({\em relational-algebraic})\\ & than {\em Newtonian}
 ({\em geometrical-analytic}) $\Rightarrow$ do\\ &
DG with the geometrical objects (fields) themselves, without at
all the intervention of\\ & Cartesian
$\smooth$-coordinates$\equiv$spacetime manifold $M$.
 \cr & \cr {\sl Groups (Klein, Lie)} & Geometrical objects (fields and their quanta)\\ & are
 totally characterized by their transformation (symmetry) groups
$\Rightarrow$ {\em Synvariance}$\equiv${\em Covariance} {\em with
respect to the field itself, without reference} {\em to a
background smooth spacetime manifold} $M$; work with $\aut\modl$,
not with $\mathrm{Aut}(M)\equiv\mathrm{Diff}(M)$.\cr & \cr {\sl
Manifolds (Riemann)} & A $\smooth$-manifold (CDG) is a particular
instance of ADG; when $\struc\equiv\smooth_{M}$, and it is not a
basic structure for doing DG, but merely the classical one. \cr &
\cr {\sl Fiber bundles (Elie Cartan, Whitney)} & Vector and
algebra (associated or representation) sheaves (of $\aut\modl$).
\cr
\end{tabular}

\vskip 0.1in

\subsubsection{Modern comparisons: ADG versus Noncommutative and Synthetic Differential
Geometry}

With the table above in mind, and since we have briefly alluded
earlier to algebraic geometry, perhaps it would be fruitful at
this point to throw in some fleeting remarks about ADG {\it
vis-\`a-vis} the general notion of a sheaf and its historical
development, as well as about the potential relationship between
ADG and the other two more `popular' nowadays---because they have
been around and worked out longer, as well as because so far they
have been more widely applied (to theoretical physics---in
particular, to spacetime and gravity)---theories of DG: Connes'
{\em Noncommutative Differential Geometry} (NDG)
\cite{connes,kastler} and Kock-Lawvere's {\em Synthetic
Differential Geometry} (SDG) \cite{kock,laven}.

\paragraph{Sheaves in algebraic geometry: the notion of `space',
`symmetry' and their quantum descendants.} The origin of the
notion of {\em sheaf} goes as far back as Weierstraas
\cite{grarem}, when Analysis was taking its first `toddler' steps.
Indeed, sheaf theory became the `bread and butter' of subsequent
investigations in the field of Complex Analysis.\footnote{One has
just to recall the great import of sheaf theory into such central
Complex Analysis' subjects as Riemann surfaces, K\"ahler-Stein
manifolds, holomorphic (vector) bundles {\it etc}, with which one
usually associates the names of Oka, Stein, Henri Cartan, Grauert,
Remmert, Leray and many others (no references provided).} Two of
the most commonly used definitions of a sheaf are due to Leray, as
{\em a complete} (functional) {\em presheaf}, and subsequently due
to Lazard, as {\em a local homeomorphism} \cite{mall1}.

Sheaf theory underlies virtually all modern developments in the
field of {\em algebraic} geometry \cite{harts,shaf,eisenbud} whose
historical development goes hand in hand with the evolution of the
notion of `space', its `geometry' ({\it eg}, topology), and their
`extraction' or `derivation' from various algebraic structures and
varieties \cite{cartier}.\footnote{Here one has just to recall
such basic algebraic geometric notions as ringed spaces, schemes
and their various spectral topologies, as well as the names of the
pioneers in this field such as the `Bourbakis'
\cite{cartier1}---the group of French mathematicians which
includes among others Schwartz, Weil, Dieudonn\'e, Chevalley,
Cartan, Serre, Grothendieck---as well as Gel'fand, Zariski, Swan,
Shafarevich, Hartshorne, Manin, and many more.}

Undoubtedly, one of the turning and high points in the development
of sheaf theory is its `categorification'---{\it ie}, the infusion
of homological algebra (category-theoretic) ideas (and jargon!)
into sheaf theory. Eilenberg and MacLane's inspired category
theory provided the basic platform on which to found fundamental
sheaf-theoretic notions\footnote{Here one has just to recall how
basic algebraic geometric notions were formulated categorically,
as for example presheaves as contravariant functors, the process
of sheafification as a functor, (global) sections as functors,
spectral functoriality, geometric morphisms, and many more.}
Perhaps, the apotheosis of the category-theoretic perspective on
sheaf theory {\it vis\`a-vis} algebraic geometry is the notion of
a topos \cite{macmo}, originally due to
Grothendieck.\footnote{Just recall Grothendieck's revolutionary
abstraction, via categorification, of topology ({\it eg},
Grothendieck sites), the categorical definition of sheaves and
general fibered spaces over such abstract `pointless' topologies
(suitable for generalized sheaf-cohomology theories), and the
concomitant organization of those sheaves into `larger'
categorical universes---the celebrated Grothendieck topoi.}
Subsequently, the notion of topos was further refined, distilled
and abstracted by Lawvere and Tierney so that it almost detached
itself from algebraic geometry proper and made contact with logic
\cite{gold,lambek}\footnote{In fact, it is more fair to say that
topos theory `unified' (the sheaf theory based algebraic) geometry
with logic, especially after it was realized that every topos has
an internal `language' (or logic), which, unlike the standard
Boolean (classical) one of the topos $\mathbf{Set}$ of `constant
sets', it is intuitionistic (in this sense one speaks for example
of the topos $\mathbf{Shv}(X)$ of sheaves of sets over a
topological space $X$ as being `a universe of variable sets')
\cite{lawvere,macmo}.}

Of great importance in algebraic geometry are {\em noncommutative}
algebraic varieties (rings, modules, algebras)---the object of
study of the so-called noncommutative algebraic geometry
\cite{voyst1,voyst2}. Here the purely mathematical interest
focuses around defining `noncommutative spaces' and their
`noncommutative geometry' ({\it eg}, `noncommutative topology'
\cite{bor3}).\footnote{In the case of topology for example,
interest lies in the possible definition of a spectrum for a
noncommutative algebra \cite{mul}.} This is supposed to make
contact with quantum physics (and possibly/hopefully with QG), in
the sense that the proposed noncommutative spaces could possibly
represent `quantum space' (or even, `quantum spacetime')
\cite{mulpel1,mulpel2,voyst3}. In noncommutative (algebraic)
geometry too categorical language and techniques abound
\cite{voyst4}, while of crucial importance in this area is
arriving at some `natural' notion of `noncommutative sheaves'
\cite{bor2}.\footnote{The epithet `natural' here meaning, for
example, that the construction of noncommutative sheaves over the
aforesaid `noncommutative (spectral) topologies' should be
functorial.} Noncommutative sheaves then are also envisaged to
inherit the epithet `quantum', the structures supposedly having
quantum physical significance \cite{groote}. Finally, the interest
naturally arises of grouping these noncommutative sheaves into a
`noncommutative topos' structure \cite{bor1}, whose quantum
interpretation (as a `quantum topos') could provide us with a
mathematical universe in which the logical and the geometrical
properties of quantum `systems' (such as the spatial topology, the
space{\em time} topology, or the causal structure) are intimately
entwined and mutually affecting.

\paragraph{Sheaves in differential geometry.} Although algebraic
geometry is teeming with sheaves and in its domain sheaf theory
prospers, sheaves have not figured `seriously' in {\em
differential} geometry proper, at least not until the advent of
ADG; or before that, they have been indirectly and to a lesser
extent involved, via topos theory, in SDG.\footnote{Here one could
perhaps add the categorical, Grothendieck-style of perspective on
abstract differential structures, such as the so-called {\em
differential modules} ($D$-modules) \cite{kato-1} as well as the
versatile (and exotic) cohomology theories based on them
\cite{kato2,kato0} (the second author wishes to thank Goro Kato
for introducing him to this beautiful mathematics).} This may be
partly explained by the `dominance of Calculus'---the enormous
success that smooth manifolds and fiber bundles over them (the
basic entities of CDG) have enjoyed in both pure and applied
mathematics (theoretical physics) \cite{hermann}.\footnote{Another
instance of the `differential manifold-conservatism' and the
`CDG-monopoly' we talked about in the early sections of this
paper.} However, as we have repeatedly noted in the present paper,
the manifold based CDG ``{\em simply miscarries with quantum
theory}'' (Einstein), so that one ``{\em will not be able to use
it in the true QG theory}'' (CDG).

Concerning QG, one could say that the road of applying sheaf
theory `bifurcates': on the one hand we encounter vigorous
research activity focusing on noncommutative (quantum)
space(time)s and their sheaves, being essentially motivated by
(noncommutative) {\em algebraic} geometry as noted above; while on
the other, we have ADG, which is a direct application of sheaf
theory to {\em differential} geometry {\it per se}, and {\it
vis-\`a-vis} QG applications, it does not address at all the
structure of (the background) `space(time)' ({\it eg}, whether it
is `discrete' or `continuous'; `commutative'/`classical' or
`noncommutative'/`quantal'), but rather it refers directly to the
`geometrical objects' that live on that (in principle arbitrary
and not participating to the DG mechanism itself) surrogate
(sheaf-theoretic localization) background.\footnote{In a nutshell,
as repeatedly noted before, from the ADG-theoretic perspective on
gravity, the issue of a quantum description (`quantization') of
space(time) structure itself is a `non-issue', since `spacetime'
(especially the smooth continuum of GR) does not (physically)
exist.}

However, there is on the one hand Connes' NDG, where
noncommutative algebras abound in a differential geometric context
(but no sheaf theory {\it per se} appears to have been seriously
used so far), and Kock-Lawvere's SDG, which revises CDG at a
`logical level', by employing topos theory (which as noted above,
has a close relation to sheaf theory, since for one thing, the
`canonical' example of a topos having a non-classical,
non-Boolean, intuitionistic-type of internal logic, is the topos
$\mathbf{Shv}(X)$ of sheaves of structureless sets over an in
principle arbitrary topological space $X$). Which brings us to the
comparison between ADG {\it contra} NDG and SDG that we initially
set out to comment on in this sub-subsection.

\begin{itemize}

\item\underline{\bf ADG versus NDG:} The crucial observation which may enable
one to relate straightforwardly ADG with NDG is that, in a
nutshell, the latter may be `reduced' to ({\it ie}, it may be
viewed as a particular `restriction'/case of) the former when one
assumes a {\em non-abelian algebra sheaf} as {\em structure sheaf}
$\struc$ of generalized coordinates or `arithmetics' in the
theory, as well as (sheaves of) modules of `differential
form'-like objects over it. Moreover, as we already emphasized
back in 4.2.2, in ADG there are `naturally' occurring NDG-related
ideas and structures, principally in the guise of what we called
`noncommutative Kleinian (esoteric) geometry' of the ADG-fields
$(\modl ,\conn)$.\footnote{See remarks in footnote ?? before.}

Regarding spacetime singularities however, and following some
critical remarks about NDG in \cite{malros1,malros2}, it must be
noted that, while with the passage to noncommutative structure
algebra sheaves enables one to deal with stronger singularities
than in the abelian case---especially in the classical (CDG) case
corresponding to assuming $\smooth_{X}$ for structure sheaf ({\it
ie}, the base space(time) $X$ is a differential manifold), NDG
still falls far short of encompassing the complete range of
singularities encountered even in Schwartz's linear distribution
theory. Moreover, the only type of derivation (differential
operator) defined in NDG is a commutator with a fixed operator,
hence it is quite a restricted notion of differential operation
even within the category of Banach algebras. Thus, it is even less
capable of dealing with with singularities on arbitrary closed
nowhere dense sets (of finite-dimensional Euclidean or locally
Euclidean---{\it ie}, manifold-type of---spaces) \cite{malros1},
let alone with the so-called `spacetime foam dense singularities'
like the ones teeming Rosinger's abelian differential algebras of
generalized functions, which have been successfully treated
ADG-theoretically in \cite{malros2} and used earlier here in
connection with the ADG-evasion of the inner Schwarzschild
singularity.

On the whole, one could say that NDG, {\it \`a la} Connes, may be
perceived as an attempt to `quantize Calculus' by functional
analytic means, while the notion of manifold is still manifestly
retained at the background. On the other hand, the whole
algebraico-categorical, being sheaf-theoretic, machinery of ADG
has been developed independently of any notion of manifold, a
notion which, {\it vis-\`a-vis} the problem of QG, has been
repeatedly criticized by many workers in the field of QG.
Moreover, ADG, being free of any manifold concept, is able to cope
with problems pertaining to `singularities' by applying methods of
non-linear PDEs ({\it eg}, Rosinger's `differential algebras of
generalized functions' \cite{ros1,ros2,ros}). Finally, it is also
worth noting here the conceptual simplicity of the machinery of
ADG.\footnote{That is, passing to noncommutative structure sheaves
may only complicate (mathematically) issues ({\it ie},
pragmatically, if we can do all DG using commutative algebras, why
resort to noncommutative ones? \cite{vinogradov})---let alone that
it is not at all clear, physically speaking, that the use of
noncommutative coordinates has anything to do with the problem of
QG {\it per se} (again, what does `quantum spacetime' have to do
with QG?). Of course, it would be just short-sighted not to
acknowledge the manifold applications that NDG has enjoyed in
fundamental physics in recent times---from the study of gauge
theories of the standard model as well as spacetime and gravity
\cite{connes2,connes3,connes1}, to the so-called $D$-branes in
open string theory in the presence of a background $B$-field (no
references given). Noncommutative geometry has also been suggested
as a UV cut-off in QFT and it has been recently appreciated that
noncommutative Yang-Mills theory is exactly solvable (again, no
references provided).}

\item\underline{\bf ADG versus SDG:} From a technical perspective,
a preliminary step one could take in order to initiate a fruitful
comparison between ADG and SDG, both abstractly (mathematically)
and with respect to (physical) applications to QG, is to show
somehow that the category of differential triads is in fact an
(elementary) {\em topos} \cite{macmo}---a so-called `cartesian
closed category'. This is only `natural' and feasible since
$\ctriad$ is {\em bicomplete}---{\it ie}, it closes under (finite)
categorical (inverse) limits and (direct) colimits, it has {\em
canonical subobjects}---{\it ie}, a canonical subobject classifier
$\mathbf{\Omega}$ can be defined in $\ctriad$, it possesses
canonical {\em products and coproducts}, as well as an {\em
exponential structure} \cite{pap1,pap2,pap3,pap4,pap5}. In
particular, in connection with finitary, causal and quantal
ADG-gravity, the immediate vision is to organize the finsheaves of
qausets in \cite{malrap3}, on which a locally finite, causal,
quantum and singularity-free\footnote{This paper.} version of
(vacuum) Lorentzian gravity holds, into a topos in which to
address deep logico-geometrical issues in QG \cite{macmo}. This
potential application of ADG has been variously anticipated in the
past \cite{rap3,rap9,rap4,rap6,rap8,rap5}, and it is currently
being under intense development \cite{rap7}.

From a purely mathematical perspective, under the prism of topos
theory, it would be particularly interesting to see how can one
carry out the basic ADG-theoretic constructions {\em internally}
in the $\ctriad$-topos by using the internal, intuitionistic-type
of language (logic) of the topos \cite{lambek,macmo}.\footnote{For
example, by stepping into the constructive world of the topos, one
could bypass the `problem' (because
$\struc$-functoriality-violating) of defining {\em derivations} in
ADG (Chris Mulvey in private communication with the second
author). {\it En passant}, the reader will have already noticed
that no notion of `{\em tangent vector field}' is involved in
ADG---{\it ie}, no maps in $Der:~\struc\mapto\struc$ are defined,
as in the classical geometrical manifold based theory (CDG).
Loosely, this can be justified by the fact that in the purely
algebraic ADG the (classical) geometrical notion of `tangent
space' to the (arbitrary) simply topological base space $X$
involved in the theory, has essentially no meaning, but more
importantly, no physical significance since $X$ itself plays no
role in the (gravitational) equations defined as differential
equations proper via the derivation-free ADG-machinery.}

Ever more importantly for a purely mathematical comparison between
ADG and SDG, having delimited the topos-theoretic (logical)
background underlying both ADG and SDG, one can then compare the
notion of {\em connection}---arguably, the key concept that
actually qualifies both theories as being {\em differential}
geometries proper---as this concept appears in a categorical guise
in both theories
\cite{kock1,kock,laven,mall1,mall2,mall4,vas1,vas2,vas4}.

Finally, one may try for an ambitious, `synthetic juxtaposition'
of the so far various and numerous physical applications of ADG
\cite{mall1,mall3,malros1,malros2,malrap1,malrap2,malrap3,mall5,mall6,mall7,rap7,mall4,mall9,mall11,mall10}
and SDG
\cite{guts0,guts1,guts2,guts,guts3,grink,buttish1,ish4,buttish2,buttish3,buttish4,ish3}
(as well as topos theory in general \cite{trifonov,kato,kato1}),
to GR, quantum theory, and their desirable `unison', QG. Even with
regard to this modern potential `unification' of the purely
algebraico-categorical, sheaf-theoretic ADG and the similar
topos-based SDG, we have (even if indirectly, but quite
prophetically) Einstein's `approval', for we recall from the
conclusion of \cite{stachel1}\footnote{With the original citation
being \cite{fraenkel}.} the following telling Fraenkel--Einstein
exchange:

\bigskip \noindent (Q8.14)\hskip 0.9in
\begin{minipage}{11cm}
\noindent ``{\small ...In December 1951 I had the privilege of
talking to Professor Einstein and describing the recent
controversies between the (neo-)intuitionists and their
`formalistic' and `logicistic' antagonists; {\em I pointed out
that the first attitude would mean a kind of atomistic theory of
functions, comparable to the atomistic structure of matter and
energy. Einstein showed a lively interest in the subject and
pointed out that to the physicist such a theory would seem by far
preferable to the classical theory of continuity. I objected by
stressing the main difficulty, namely, the fact that the
procedures of mathematical analysis, e.g., of differential
equations, are based on the assumption of mathematical continuity,
while a modification sufficient to cover an
intuitionistic-discrete medium cannot easily be imagined. Einstein
did not share this pessimism and urged mathematicians to try to
develop suitable new methods not based on continuity}\footnote{Our
emphasis.}...}''
\end{minipage}

\vskip 0.1in

\item {\it In toto}, pictorially, but sketchily, the relation which we envision between ADG {\it
contra} SDG and NDG can be cast as follows:

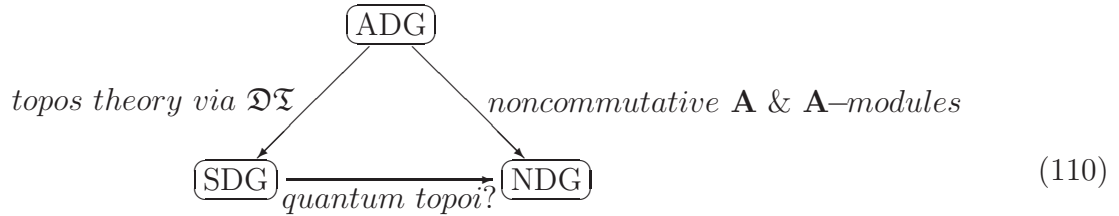
\begin{equation}\label{eqo}
\Atriangle[\ovalbox{ADG}`\ovalbox{SDG}`\ovalbox{NDG};topos~theory~via~\ctriad`noncommutative~\struc~\&~\struc\!\!-\!\!
modules` quantum~topoi?]
\end{equation}

\noindent with the potential relation between SDG and NDG being
possibly effectuated, via ideas issuing from noncommutative
algebraic geometry (as well as its related noncommutative operator
algebra and functional analysis), by the aforementioned notion of
noncommutative (`quantum') topology, the noncommutative sheaves
based on it, and the (`quantum') topoi thereof. For example, the
quest for a quantum topos may follow the route via
$C^{*}$-quantales---structures that may well qualify as
noncommutative (`quantum') topologies proper
\cite{mul,mulpel1,mulpel2},\footnote{`Originating', one could say,
by a desire to find a noncommutative version of the
Gel'fand-Naimark representation theorem for abelian
$C^{*}$-algebras, thus arrive at the noncommutative analogue of
the {\em spectrum} (of a non-abelia $C^{*}$-algebra)---a
topological space `inherent' in the algebras involved, which is a
generalized, classical one (a so-called {\em locale} \cite{macmo})
in the case of commutative $C^{*}$-algebras \cite{mulpel1,mul1}.}
as the following `analogy' depicts:

\begin{equation}\label{eqoo}
\frac{\rm (sheaves~over)~locales}{\rm
(sheaves~over)~quantales}=\frac{\rm topoi}{\rm ?}
\end{equation}

\noindent with the elusive (abstract) noncommutative (`quantum')
topos structure being expected to replace the question-mark in
equation (\ref{eqoo}) above, as well as in the diagram (\ref{eqo})
just before it.\footnote{Jim Lambek, Chris Mulvey, Steve Selesnick
and Freddy Van Oystaeyen in private communication with the second
author.} And one has to recall again here the central role that
(noncommutative) $C^{*}$-algebras play in NDG \cite{connes,block}.

\end{itemize}

\subsubsection{ADG versus `Quantizing on a Category'}

The discussion above about the potential links and close
affinities between the categorico-algebraic ADG with NDG and SDG
in the general light of homological algebra (category and
topos-theoretic) ideas, brings us to another mathematical approach
to QG to which we would like to relate ADG. This is Isham's very
recent `{\em Quantizing on a Category}' (QC) general mathematical
scheme \cite{ish5,ish6,ish7,ish8}. The principal goal of QC is to
quantize systems with configuration (or more generally, history)
spaces consisting of `points' having `internal structure'. The
main motivation behind this approach is the apparent failure of
applying the conventional quantization concepts and techniques to
`systems', such as causets or space(time) topologies for instance,
whose configuration (or general history) spaces are far from being
point-like differential manifolds consisting of structureless
points. Isham's approach basically hinges on two innovations:
first to regard the relevant entities as objects in a category,
and then view the categorical arrows as analogues of momentum
(`derivation maps') in the usual (manifold based) theories.
Although this approach includes the standard manifold based
quantization recipes, it goes much further by making possible the
quantization of systems whose `state' spaces are not smooth
continua ({\it eg}, causets or finitary topological spaces).

There appear to be close connections, both conceptually and
technically, between QC and ADG---affinities which one would be
tempted to explore further. {\it Prima facie}, and generally
speaking, both schemes concentrate on evading in one way or
another the pointed differential manifold---be it the
configuration space of some physical system, or the background
spacetime arena of classical or quantum (field) physics, and they
both employ `pointless', categorico-algebraic methods. Both focus
on an abstract (categorical) representation of the notion of
derivative (or derivation)---QC, by abstracting from the usual
continuum based notion of vector field (derivation), arrives at
the categorical notion of an `arrow field' which may be thought of
as a map or morphism (:section of certain non-functional
presheaves and their associated representation Hilbert presheaves
involved in QC) which respects the internal structure of the
system-objects that one focuses on ({\it eg}, topological spaces
or causets); while in ADG, the notion of derivative (:differential
$d$) is abstracted and generalized to that of a connection field
$\conn$, defined categorically as a sheaf morphism, on a sheaf of
suitably algebraized structures ({\it eg}, causets or finitary
topologies
\cite{rapzap1,rapzap2,rap1,rap2,malrap1,malrap2,malrap3,rap5}).

On the other hand however, in `philosophical' aims and general
outlook, ADG and QC differ: the second aims at quantizing ({\it
ie}, developing quantum theories of) non-continuum models of
`space' ({\it eg}, finitary topologies) or `space{\em time}' ({\it
eg}, causets), while ADG is indifferent to the character of any
background space(time) and it develops a field theory with quantum
traits built into the formalism from the very start, hence in no
need of (a formal process of) quantization. Nevertheless, the
technical and conceptual affinities between ADG and QC are so
strong that it is well worth attempting a marriage of the two.

\subsection{`Physical Geometry' or `Geometrical Physics'? (ADG-Theoretic Variations on a Theme by
Peter Bergmann)}

We would like to remark under the prism of ADG, and ADG-gravity in
particular, on a fundamental difference in our opinion between
what Peter Bergmann called in \cite{berg0} `{\em physical
geometry}' and `{\em geometrical physics}'. Let us open these
remarks with the following `duality' between algebra and geometry
originally intuited by Sophie Germain already back in the late
17th century \cite{germ}:

\bigskip \noindent (Q8.15)\hskip 0.9in
\begin{minipage}{11cm}
\noindent ``{\small\em L' alg\`{e}bre n' est qu' une
g\'{e}om\'{e}trie \'{e}crite, la g\'{e}om\'{e}trie n' est qu' une
alg\`{e}bre figur\'{e}e.}''\footnote{``{\small\em Algebra is
nothing but a `written geometry', {\rm [while]} geometry is
nothing but a `figured algebra'}''.}
\end{minipage}

\vskip 0.1in

\noindent Loosely and intuitively speaking, from the vantage of
the applied ADG to theoretical physics, `{\em physical
Geometry}'---what Sophie Germain calls `{\em alg\`{e}bre
figur\'{e}e}'---is the geometry `deriving' from the (`dynamical')
relations between the geometric fields themselves, which relations
can in turn conveniently be (mathematically) grounded in algebra.
In this sense, physical geometry is Euclidean\footnote{See
comments earlier on our ADG-based perception of what is commonly
referred to as Euclidean geometry.} and ADG appears to be well
suited to represent it mathematically. The word `deriving' above
is meant to suggest that {\em as `physical geometry' we regard the
`geometry' or `structure' of the `solution space' of the dynamical
field equations, which are mathematically expressed algebraically}
({\it ie}, `relationally'---here by sheaf-theoretic means). This
is related to what we posited earlier, namely, that `{\em
kinematics (geometry) comes after or is the result (outcome) of
dynamics}'.\footnote{On a par is the statement that `{\em
geometrical spacetime is the effect of, or is inherent in, the
algebraic field}'.} In other words, much in the same way of our
Wheeler-type of `principle'-{\it motto} that {\em no theory is a
physical theory unless it is a dynamical theory}, we here hold
that {\em no geometry is a physical geometry unless it is the
result of a dynamics}. Most of this has already been anticipated
by the first author in the recent paper \cite{mall10}.

On the other hand, at least insofar as differential geometric
considerations are concerned, thus far `{\em geometrical physics}'
has been exercised (implemented) by CDG-theoretic means. CDG is a
theoretical framework which, almost by definition, gives a central
operative role and, as a result, a physical significance, to an
{\it a priori} fixed, kinematical-geometrical background smooth
spacetime manifold, as well as to the analytic,
Cartesian-Newtonian way in which (smooth) coordinates (`point-like
space(time) locations') mediate or intervene in our calculations
(Differential Calculus) so as to mislead one into thinking that
the most important `supplier' of the differential geometric
mechanism and technotropy is the {\em underlying} pointed
spacetime manifold, rather than, in a Euclidean-Leibnizian
fashion, the (algebraic relations between the) `{\em overlying}'
geometro-physical objects themselves ({\it ie}, the fields and
their particle-quanta). In this sense, `geometrical physics' is
part, or an aspect, of the more general term `mathematical
physics'.\footnote{Which should be contrasted against the term
`{\em physical mathematics}' that we prefer (see last subsection
8.6).}

Below, Peter Bergmann \cite{berg1} makes it precise what a `{\em
Physical Geometry}', in contradistinction to a `{\em Geometric
Physics}', really ought to be:

\bigskip \noindent (Q8.16)\hskip 0.9in
\begin{minipage}{11cm}
\noindent ``{\small ...In concluding, permit me to come back to my
original question: to what extent do geometric considerations play
a role in these theories? I believe that the answer is largely
heuristic. {\em Any physical theory derives its justification from
physical motives, and how well that theory helps us understand the
physical universe}.\footnote{Our emphasis.} The judgment as to
desirable aspects of the invariance group of a theory must
ultimately rest on the physical decision as to the equivalence of
apparently distinct descriptions of a physical situation (choice
of coordinate frame, choice of gauge frame, etc). But given a
conceptual framework, geometric realizations, and geometric
imagery will often prove a powerful tool in elaborating the
theory...

...Such transformations might point in a direction in which {\em
fields get unified under an invariance group in which the
space-time manifold no longer plays a pre-eminent role. A world
point would derive its identity from its dynamic environment, or
it might possess no identity at all.}\footnote{Again, our
emphasis.}

Whether such a theory can still be called geometric is a question
of terminology, not of principle. {\em Certainly, physicists will
go in directions suggested to them by physical considerations, not
by a mathematical discipline, and that is what geometry
is.}\footnote{Emphasis is ours.} Considerations of an aesthetic
nature are important, to be sure; for that we have the testimony
of such giants as Einstein and Dirac. {\em But physicists must not
lose touch with its foundation, soil as it were: we must above all
remain sensitive to what Nature in its subtlety attempts to tell
us\footnote{Our emphasis.}.}''}

\end{minipage}

\vskip 0.1in

In the penultimate paragraph of the quotation above, and in
connection with the dynamical (not kinematical) character of
(smooth) coordinates---the `dynamicalization' of
coordinates---propounded in \cite{stachel2}, we should mention
here ADG's thesis that {\em `space(time)' is inherent in the
field}, which appears to turn upside-down the hitherto conception
in physics that the kinematics---the `possibility space'---of a
theory should be delimited (defined) prior to the dynamics---the
`actuality space'.\footnote{This is our basic thesis, multiply
emphasized earlier in this paper-book, that `{\em dynamics
(algebra) comes before kinematics (geometry)}'.} And let it be
stressed here that for us, in contradistinction to Bergmann, {\em
this} ({\it ie}, that algebra/dynamical relations comes before
geometry/kinematical spacetime) {\em is indeed a question of
fundamental principle, not just of terminology!} All this is
already subsumed under the basic ADG-field pair $(\modl ,\conn)$:
the dynamical Law is (defined by) $\conn$, while the (geometrical)
expression (:geometrical representation) of the Law is
$\curv$---one that in turn delimits its `solution space' ({\it
ie}, the entire sheaf space $\modl$ on which the law holds, which
is completely described by its sections---the quantum particle
states of the field).\footnote{Recall again that,
sheaf-cohomologically speaking, $\modl$ is completely
characterized by $\curv(\conn)$ ({\it eg}, Weil's integrality).}

Of course, in line with Bergmann's words above, in ADG-gravity not
only ``{\em the space-time manifold no longer plays a pre-eminent
role}'', but in fact {\em it plays no role at all}. Furthermore,
the categorico-algebraic ADG-formalism is fundamentally pointless,
thus no question arises as to {\em where from does a world point
derive its identity}. Spacetime itself (and at that, not
necessarily a continuum, or even a discretum!) derives (its
identity) from the dynamical fields: in this sense we have claimed
throughout this paper that {\em `background geometrical spacetime'
is inherent in the `geometrical objects' (:the dynamical physical
fields) that comprise `it'}.\footnote{As repeatedly noted before,
in ADG-gravity and ADG-field theory in general, we cease talking
about `{\em world points}', or equivalently about `{\em spacetime
events}', and we speak solely of `{\em world fields}'. {\it In
extenso}, we do not talk about `{\em spacetime geometry}', but
exclusively about `{\em field geometry}', with the noun `geometry'
pertaining to the (structure of the) `solution space' of the
dynamical field-law---the realm where the field law holds.
Fittingly, ADG was originally coined the (differential) `{\em
geometry of vector sheaves}' \cite{mall1}, it being understood in
a generic way that that the vector sheaves $\modl$ involved---the
representation or carrier spaces for the connection fields
$\conn$---are the universes where the field laws such as
(\ref{eqy23}) hold.}

And finally, about the last paragraph in (Q8.16) above: we totally
agree with Bergmann, and in a wider sense, that a physical theory
is {\em not} its underlying mathematical formalism,\footnote{This
accords with our earlier maintaining that `physical spacetime'
(whatever that means) is traditionally modelled after a smooth
manifold (:locally Euclidean space) just because the original
mathematical formalism underlying GR happened to be the CDG-based
(pseudo)-Riemannian geometry.} needless to say that a physical
theory does {\em not} progress solely from input(s) from an
advancing mathematical discipline.\footnote{However, see also
counterpoints to this position {\it \`a la} Dirac and Faddeev
raised in 8.6 below.} In other words, the primary considerations
of a physicist are {\em physical} ones, not mathematical. On the
other hand, again if we abide by the Wheelerian-type of principle
that {\em a physical theory is defined by, and is nothing but, the
dynamical laws that it purports to describe and aspires to
interpret}, coupled to the initial choice-{\it cum}-`assumption'
(of the theoretical physicist) to mathematically model those laws
differential geometrically ({\it ie}, as differential equations)
for whatever intuitive reason she may be inspired and motivated
by,\footnote{For example, for the sake of (infinitesimal)
locality.} the mathematical apparatus (:theory of DG) appears to
be intimately entwined with the physical theory. Choosing the
standard CDG for that purpose inevitably brings into play the
`metaphysical' geometrical background manifold, which is almost
inevitably interpreted physically as `spacetime'. In other words,
we are inclined to hold that the differential spacetime manifold
(:a background locally Euclidean space) was forced on physicists
by a mathematical discipline (CDG), {\em not} by physical
considerations proper. However, if the physicist chooses ADG, one
leaves out {\it ab initio} the metaphysical\footnote{And in view
of singularities, `{\em paraphysical}'!} mathematical
extra-baggage of the geometrical base manifold (and, inevitably,
its pseudo-physical interpretation as `spacetime'), and she
directly refers to/focuses on the dynamical fields (:connections)
and the laws (:differential equations) that they define. And, to
emulate Bergmann's expression above, {\em this is what `physical
(differential) geometry'} (:the geometry defined by the laws
defining a physical theory) {\em is all about}.

\subsubsection{The end of `geometrical picturization' in theoretical physics:
breaking Sophie Germain's `symmetry-duality' between algebra and
geometry in favor of the first}

By now it must have become clear to the reader that the
`iconoclastic' stance against (differential) geometry that we
assume in view of the ADG-theoretical paradigm and its physical
applications---ADG-field theory in general and ADG-gravity in
particular---is the following: the principal `icon' that has been
`religiously' venerated by theoretical physicists is that of a
background geometrical space-time---be it a continuum or a
discretum.\footnote{In Riemann's words \cite{riemann}, a
continuous or a discrete manifold. Arguably, at least in field
physics, the continuum has dominated the discretum, although
traditionally the concept of a (material) particle is supposed to
carry within it some fundamental `discreteness' and `pointedness'
(one usually speaks of `point-particles'). On the other hand, the
(dynamical) evolution of point-particles is normally taken to be
`continuous' ({\it ie}, differential equations in a space-time
continuum).}---and it is this `icon' (or better, `{\em idol}')
that we would like to `cut'\footnote{In Greek,
`$\epsilon\iota\kappa o\nu
o\kappa\lambda\acute{\alpha}\zeta\omega$' (:verb), from which the
noun `iconoclasm' derives, is a compound word consisting of the
noun `$\epsilon\kappa \acute{o}\nu\alpha$' (:icon, picture) and
the verb `$\kappa\lambda\acute{\alpha}\zeta\omega$' (:ancient
Greek meaning `to cut' or `to separate', and then `to compare and
classify' things---think of the English word `class' and the Greek
word `$\kappa\lambda\acute{\alpha}\sigma\mu\alpha$' meaning `cut
part' or `section', and in mathematics `fraction' or
`proportion'). According to Webster's Encyclopedic Unabridged
Dictionary of the English Language, an `{\em iconoclast}' (I
kon$^{'}$a klast$^{'}$, noun) is: (i) a breaker or destroyer of
images, especially those set up for religious veneration, and/or
(ii) one who attacks cherished beliefs, traditional institutions,
{\it etc.}, as being based on error or superstition. Historically,
in Byzantium (723-843AC), `{\em iconoclasm}' or `{\em iconomachy}'
was the polemic current against `{\em iconolatry}'---the
worshipping of Christian icons (predominantly in churches). The
`canonical' example of an iconoclast scientist is Galileo and
Darwin.} (or better, tear down) in fundamental physics.
Undoubtedly, geometrical picturization in physics, through the
medium and `lens' of an ambient spacetime environment intervening
(via coordinates/Cartesian arithmetics) in our physical models and
calculations with them, has proven to be invaluable in the past,
but we feel it is high-time we left it behind, as a dated
intuitive/heuristic method of tackling issues especially in the
quantum deep where the usual differential geometric means of the
manifold based CDG appear to be inadequate and at best
problematic.

In this stance, again we are not alone. Einstein too, although he
relied heavily on geometrical picturization
(heuristically/intuitively) and on the spacetime continuum based
Riemannian geometry (technically) in developing GR and the theory
was a great success---an enormous single-handed achievement, had
subsequently expressed scepticism about `geometric reasoning' and
the concomitant `geometrization of physics' (especially {\it
vis-\`a-vis} the {\it aufbau} of a unified field theory!), as
Bergmann recalls below:

\bigskip \noindent (Q8.17)\hskip 0.9in
\begin{minipage}{11cm}
\noindent ``{\small ...{\small\em What is geometry?}\footnote{Our
emphasis. Bergmann's question recalls Chern's `{\em What is
Geometry?}' in \cite{chern,chern1} that we commented on in the
previous subsection. Let it be stressed here however that the
first was rather trying to address the question of what we wish to
coin `{\em physical geometry}' (and {\it in extenso} of `{\em
physical space(time)}'), while the second attempted to answer to
that question by giving an outline of the development of what we
would like to call `{\em mathematical geometry}' (and {\it in
extenso} of `{\em mathematical space}'). What we wish to draw here
is a clear distinction between physical and mathematical geometry
and, {\it in extenso}, space(time). The `geometry' that the
mathematician is interested in---the `mathematical geometry' and
`space', is very different from the `physical geometry' that the
physicist is interested in, although as it has been repeatedly
noted before, in the theoretical physicist's mind---which has been
greatly influenced by mathematics---there is a conflation (and
resulting confusion) of the two terms, at least insofar as the
background differential manifold is interpreted as `physical
spacetime' with the concomitant use of CDG-concepts and methods in
field theory, classical or quantum.} I suspect that there is no
answer to this question that will satisfy
everybody...\footnote{Then Bergmann attempts to loosely define
(mathematical) geometry as ``{\em any kind of mathematical
structure that begins with the construction of a set of points
that satisfies the minimal properties of continuity that
justifying one in speaking of a space...}'' (something that
recalls the mathematical notion of a {\em topological space}---the
minimal structure that a space can have apart of course from being
a bare/structureless set), and he then gives examples of
properties of and structures on such geometric spaces ``{\em that
have been used by physicists in their pursuits and endeavors to
understand nature}''.} Depending on the properties ascribed to a
new model for space-time, its structures might lend themselves to
interpretations that are reminiscent of fields known to
physicists. {\small\em How does such `geometrization' contribute
to unification? Einstein stated repeatedly that he did not
consider geometrization of physics a foremost or even a meaningful
objective, and I believe that his comments remain valid today.
What really counts is not a geometric formulation or picturization
but a real fusing of the mathematical structures intended to
represent physical fields.\footnote{Our emphasis again.} }''}

\end{minipage}

\vskip 0.1in

\noindent {\it In toto}, we believe that ADG-field theory, by
down-playing and atrophizing geometrical picturization (by means
of a geometrical background spacetime) while at the same by
concentrating on the algebraic (dynamical) relations between the
physical fields themselves (without recourse to a metaphysical
background), has contributed significantly towards materializing
Einstein's meaningful objective above, namely, {\em to fuse the
mathematical structures intended to represent physical fields
under the general and unifying notion of connection $\conn$}
(:generalized differential)---the principal notion with which one
can actually do (differential) geometry after all.

\subsection{Einstein's `Organic Theory' Approached via a Rhetorical (for ADG) Question: Can One Do Field
Theory Without the Spacetime Continuum?}

\begin{quotation}
\noindent {\small\em ``...Adhering to the continuum originates
with me not in a prejudice, but arises out of the fact that I have
been unable to think up anything organic to take its place...''}
\end{quotation}

The starting point for the following discussion is Einstein's
confessionary words in (Q2.?), together with his wish to find a
purely algebraic theory for the description of reality (Q2.?,
Q?.?). In (Q?.?), `{\em organic}' is understood here in the
context of ADG as meaning, on the one hand simply on pragmatic
grounds, {\em a theory that works as well as the spacetime
continuum (manifold) based field theory}, and on the other, as
{\em a `self-sustaining' theory, which is `autonomous', without
its `paracytic' dependence on extraneous, auxiliary structures}
({\it eg}, `spacetime'\footnote{This external `spacetime' be it a
continuum or a discretum. (See the 7.5.1 next for
this.})---structures that are vital for its (mathematical)
subsistence, consistency and operativeness. Such autonomous
theoretical entities were the Leibnizian monads, which
(``$\epsilon\nu\tau\epsilon\lambda\epsilon\chi\epsilon\iota\alpha\nu~\tau\epsilon~\kappa\alpha\acute{\iota}~\alpha\upsilon
\tau\alpha\rho\kappa\epsilon\iota\alpha\nu~\epsilon\chi
o\upsilon\sigma\iota\nu$'')\footnote{Translation from the original
Greek: ``{\em Monads have entelechy ({\it ie}, an `inner
perfection', a `built-in end') and are self-sufficient/sustaining
({\it ie}, they are `autonomous').}''}
\cite{leibniz2,leibniz,leibniz1}.

\subsubsection{Reconciling the ``better known Einstein'' with Stachel's ``other Einstein''}

First, as it has already been mentioned and analyzed to some
extent in \cite{malrap3} and as it has been transparent in various
quotations above, we shall state it up-front that Einstein's
`dissatisfaction' with `continuous' field theory on the spacetime
continuum was essentially due to two reasons:

\begin{enumerate}

\item The singularities and associated unphysical infinities
plaguing his GR on the smooth spacetime manifold---the spacetime
continuum.

\item The (successes of the) finitistic and algebraic quantum
theory (of matter).

\end{enumerate}

The basic `thesis' in \cite{stachel1} is that there were two
(facets of) Einstein(s):

\begin{enumerate}

\item The more well known Einstein, advocating a `continuous'
field theory on the spacetime continuum (SR, GR, unitary field
theories for gravity and electromagnetism)---\texttt{What does the
`completion of GR' consist in?}\footnote{From the Einstein's
autobiography \cite{einst9}: a completion consists in i) {\em
evading singularities}, and ii) {\em describing the
atomistic/quantum structure of reality}.} Let us call that facet
of Einstein `{\em Geometric Continuum Einstein}' (GCE).

The basic argument of Einstein is that {\em if the spacetime
continuum is to be renounced as being fundamental, then a field
theory (based on it) must also go}. We borrow two quotations of
his from \cite{stachel1})\footnote{See there for exact
references.} that corroborate this inextricable dependence of
field theory on the spacetime continuum and the equivalent
`implication-by-negation' `{\em no field theory}$\Rightarrow${\em
no spacetime continuum}':

\bigskip \noindent (Q8.18)\hskip 0.9in
\begin{minipage}{11cm}
\noindent ``{\small ...I consider it entirely possible that
physics cannot be based upon the field concept, that is on
continuous structures. Then {\small\em nothing}\footnote{Our
emphasis.} will remain of my whole castle in the air including the
theory of gravitation, but also nothing of contemporary
physics\footnote{Implying at the same time that theoretical
physics till his time was essentially `{\em continuous field
physics on the spacetime continuum}'.}...}''

\end{minipage}

\vskip 0.1in

\noindent Especially for GR, as Einstein notes above, which is the
field theory {\it par excellence}, field theory and the spacetime
continuum appear to go hand in hand, and it is almost unimaginable
that one can separate them ({\it ie}, retain one and abandon the
other). Stachel puts it succinctly in \cite{stachel1} as the
apparently inevitable shortcoming in implementing the PGC of GR if
one throws away the spacetime continuum and, as a result, the
field-theoretic outlook as well:

\bigskip \noindent (Q8.19)\hskip 0.9in
\begin{minipage}{11cm}
\noindent ``{\small ...Suppose one believes that future physics
will continue to employ the space-time continuum as a fundamental
element, as Einstein often did. {\small\em Then, if further
progress in physics is to incorporate the dynamical view of
space-time inherent in the general theory, it is hard to see how
it can escape from the field viewpoint. How else could one express
general covariance?}\footnote{Our emphasis.}...}

...{\small The important point to emphasize is that this was
undoubtedly Einstein's final viewpoint about the progress of
physics.}{\small\em If the space-time continuum is to remain a
fundamental element of future physics, some sort of
generally-covariant field-theoretic generalization of the general
theory of relativity is required. As mentioned above, he devoted
almost forty years of his life to the search for such a
generalized theory.}\footnote{Our emphasis.}

{\small I am sure you caught the caveat in my last remark:
`{\small\em If the space-time continuum is to remain a fundamental
part of future physics}'\footnote{Again, our emphasis.} This
caveat brings me, finally, to the subject of this paper: `The
Other Einstein'}.\footnote{What we coin below, `{\em Algebraic
Finitistic Einstein}' (AFE).}''

\end{minipage}

\vskip 0.1in

\noindent But before we go on to describe Einstein's other facet,
the AFE, let us again borrow from \cite{stachel1} an Einstein
quote that further supports the aforesaid `{\em no spacetime
manifold, no field theory}' motto of the GCE:

\bigskip \noindent (Q8.20)\hskip 0.9in
\begin{minipage}{11cm}
\noindent ``{\small ...I must confess that I was not able to find
a way to explain the atomistic character of nature. My opinion is
that if the objective description through the field as an
elementary concept is not possible, then one has to find the
possibility to avoid the continuum (together with space and time)
altogether. {\small\em But I have not the slightest idea what kind
of elementary concepts could be used in such a
theory}\footnote{Our emphasis.}...}''

\end{minipage}

\vskip 0.1in

\noindent We also encounter in Einstein's autobiographical sketch,
as recalled in \cite{pais}:

\bigskip \noindent (Q8.21)\hskip 0.9in
\begin{minipage}{11cm}
\noindent ``{\small ...It appears dubious whether a field theory
can account for the atomistic structure of matter and radiation as
well as of quantum phenomena. Most physicists will reply with a
convinced `{\em No}',\footnote{Einstein's emphasis.} since they
believe that the quantum problem has been solved in principle by
other means. However that may be, Lessing's comforting words stay
with us:} `{\small The aspiration to truth is more precious than
its possession'...}''
\end{minipage}

\vskip 0.1in

\noindent {\it Ergo}, Einstein did not have a (mathematical)
theory in his hands that would enable him to do field theory
without using the background spacetime manifold, a theory which,
consequently, would do away or evade singularities as well as
address `finitistic quantum questions'.\footnote{Of course, we
have ADG \cite{mall1}, and its various applications so far in
classical and quantum gravity
\cite{malros1,malros2,malros3,mall3,malrap1,malrap2,malrap3}, and
herein. Remarkably for the concluding sentence in (Q8.20) above,
{\em ADG's elementary (fundamental) concept is again that of a
field ({\it viz.} connection); albeit, one that is not at all
based on a spacetime continuum}, while the basic means employed
are {\em entirely algebraic} (:sheaf-theoretic).}

\item The lesser known (and much overlooked in our opinion!) Einstein, advocating
a ``{\em purely algebraic physics}'' \cite{stachel1}, possibly
based on a `discretum'---anyway, one without ties to a base
spacetime continuum. Let us call that facet of Einstein `{\em
Algebraic Finitistic Einstein}' (AFE).

\item For Einstein, this continuum/discretum' schism is a
fundamental, {\em non-bridgeable} one---a strictly mutually
exclusive theoretical `binary alternative' which basically rests
on the fact that, for him, it was almost unimaginable that one
could do field theory---in fact, do {\em differential geometry}
solely with those fields!---on, say, a reticular (`discrete')
background space(time), let alone one teeming with singularities
and supporting quantum constructions which are {\it prima facie}
incompatible with continuous field descriptions on the spacetime
continuum. The unabridgeable continuum-discretum divide, and more
importantly, the essential difference between the two being that
while the continuum supports {\em differential geometric}
constructions, the discontinuum does not, is explicitly stated in
the following remarkable words by Einstein \cite{einst12}:

\bigskip \noindent (Q8.21)\hskip 0.9in
\begin{minipage}{11cm}
\noindent ``{\small\em ...The alternative continuum-discontinuum
seems to me to be a real alternative; i.e., there is no
compromise. By discontinuum theory I understand one in which there
are no differential quotients. In such a theory space and time
cannot occur, but only numbers and number-fields and rules for the
formation of such on the basis of algebraic rules with exclusion
of limiting processes}.\footnote{Our emphasis.} {\small Which way
will prove itself, only success can teach us...}''
\end{minipage}

\vskip 0.1in

\item ADG offers a sheaf-theoretic way of doing a differential
geometry based field theory independently of the `nature' or
character of the background space(time)---{\it ie}, whether it is
a continuum or a discretum---but one that deals directly with the
dynamical fields---what we have called throughout the present
paper-book, the `{\em geometrical objects}'---`in-themselves'
({\it ie}, autonomously).

\end{enumerate}

\noindent Stachel's observation (in accordance with GCE) is the
following: field (and even particle) theory cannot be thought of
independently of the (background) spacetime continuum, so that if
one does away with one concept, the other must go as well.
However, the continuum supports differential (and differentiable!)
physical quantities (fields), and laws expressed as differential
equations between them. Moreover (from the bottom of the first
page of \cite{stachel1}), in the context of GR, in order to be
able to formulate the PGC, the continuum postulate appears to be
necessary---in other words, one cannot express general covariance
without a smooth spacetime manifold. {\em Or can (s)he?}, we would
like to ask rhetorically in view of ADG-field theory and
ADG-gravity in particular.

At the same time, he argues (again in accordance with Einstein,
but now with AFE), if we drop the continuum, the next tenable
position is for a `discontinuum'---what we here refer to as a
`discretum', not being able to support differential equations
though, but only relations between discontinuous, discrete
quantities.\footnote{Here, the reader should go back and read the
telling excerpt from a dialogue between Einstein and Fr\"ankel in
(Q8.14) about the apparently unbridgeable
`continuum'/`discontinuum' (also found at the end of
\cite{stachel1}). Especially telling there are Fr\"ankel's
position that ``{\em the procedures of mathematical analysis,
e.g., of differential equations, are based on the assumption of
mathematical continuity}''---in effect, on the assumption of a
background continuum (:manifold). In 8.3 earlier, we expressed
ADG's `counterpoint' to this point by Fr\"ankel, which is another
instance of what we have here coined `CDG and manifold
conservatism and monopoly'.}

\subsubsection{Not a question whether it is a continuum or a discretum;
for \underline{`it'} is not!}

Following our remarks in subsection 7.6 before, the Planck
space-time scale $\ell_{P}$-$t_{P}$ appears to be a `catchall' for
everyone working in QG.\footnote{We owe this graphic
characterization to Chris Isham (private e-correspondence with the
second author).} However, in a theoretical/mathematical scenario
for differential geometry, such as ADG, in which a base
space(time)---be it a `continuous' manifold/continuum or a
`discrete' space/discontinuum or `discretum') plays no role
whatsoever in the `intrinsic' or `inherent' so to speak, and
essentially algebraic, differential geometric mechanism, as well
as in our calculations (Calculus!) based on it, {\it prima facie}
{\em there appears to be no Planck length-time issue at all}.

That is to say, to stress it again, the operative role that
$\ell_{P}$-$t_{P}$ plays in the usual continuum based theories as
a `natural cut-off' below which the manifold is supposed to give
way to something fundamentally reticular or
granular/finitistic\footnote{That something still perceived though
as `space(time)', even if in some `oblique', `slanted' way. Such
for instance is the primitive conception of `spacetime' at
sub-Planckian scales advocated by causet theory
\cite{bomb87}.}---a `discretum' from the point of view of which
the usual spacetime continuum and the GR based on it are purported
to be coarse, `effective theories' (Q?.?), is in a strong sense
rendered obsolete, and there is no need at all to evoke it.

In other words, from a differential geometric (but not classical
analytic in the usual Cartesian--(locally) Euclidean sense of the
smooth manifold\footnote{See 7.? above.}) perspective, it is not a
question whether `{\em it}' ({\it ie}, `spacetime') is a continuum
or a `discretum', for `{\em it' is not}---{\it ie,} it plays no
role in our ADG-based `Calculus', whose mechanism/machinery
derives from the algebraic structure---{\it ie}, the (dynamical)
relations between the `geometrical objects' themselves that live
on `space(time)', without at all the intervention in the classical
Euclidean--Cartesian--analytic sense of the latter into the said
inherently algebraic differential geometric mechanism in the guise
of coordinates, as in the classical, $\smooth$-manifold-based
theory.

All in all, we maintain that $\ell_{P}$-$t_{P}$ enters our
physics, because of the way (the manifold way, which is the only
way we know so far of doing differential geometry!\footnote{Hence
the CDG-conservative attitude mentioned throughout the present
paper-book.}) we try to implement differential geometric ideas to
the physics of the very small, while, we hold, the very notions of
`very small' and `very large' (scale) lose their meaning when
viewed from the ADG-perspective. For, in any case, as Sorkin
mentions in \cite{sork1},

\bigskip \noindent (Q8.22)\hskip 0.9in
\begin{minipage}{11cm}
\noindent ``{\small\em ...Historically you could say that Quantum
Theory deals with the very small and General Relativity with the
very large, but the essence of the distinction is not really one
of size. Rather, `the quantum of action' is in general important
whenever no more than a few degrees of freedom are
excited,\footnote{This generally, but by no means always, means
that only a few particles are involved. (Sorkin's original
footnote.)} while gravity---or in other words General
Relativity---is important whenever a large enough amount of energy
is compressed into a small enough space. More specifically,
gravity is important when the ratio $Gm/rc^{2}$ is of order unity,
where $m$ is the total mass-energy, $r$ is the radius of the
region into which it has been compressed, and $G$ and $c$ are
respectively the gravitational constant and the speed of
light...}''
\end{minipage}

\vskip 0.1in

\noindent which, of course, behooves us to analyze the role the
physical constants play in these theories, and their eventual
contribution into the $\ell_{P}$-$t_{P}$ below which quantum
gravitational effects are supposed to become important.

To make the point stronger here, we would like quote a passage
from a four year-old now report of a referee concerning a research
project proposal of the second author whose main objective was to
apply the ADG-formalism so as to on the one hand develop a
finitary, causal and quantal Lorentzian gravity
\cite{malrap1,malrap2,malrap3}, and on the other tackle the
problem of spacetime singularities (in classical GR), as done in
\cite{rap5} and of course herein. (S)he first summarized the
project:

\bigskip \noindent (Q8.23)\hskip 0.9in
\begin{minipage}{11cm}
\noindent ``{\small ...The candidate proposes broadly three topics
of study, all following naturally out of his previous work. First
he aims to develop a finitary or `discrete' analysis of the
spin-connection formulation of general relativity, which would
blend elements from the approaches of Ashtekar, Finkelstein and
Sorkin. He would then apply this new formalism to the problem of
quantum gravity. In this work, he would use heavily the `ADG'
formalism of Mallios. His second project would apply the ADG
technology in an attempt to `resolve' the singularities of the
classical Einstein equations, such as that residing at the center
of a black hole...[and (s)he then wrote in parenthesis]...{\em
This is somewhat at odds with the first project, since, if
successful, it would to some extent remove the need for any
postulate of spacetime discreteness\footnote{Our emphasis.}...}}''
\end{minipage}

\vskip 0.1in

\noindent As if the `resolution' of the black hole singularity and
associated unphysical (curvature) infinity was actually the result
of quantization and spacetime discretization.\footnote{As it has
been repeatedly emphasized throughout the present paper-book, and
recently in \cite{rap5}, in ADG-gravity, at least as it concerns
QG, no {\it a priori} spacetime discretization and quantization is
evoked to evade the gravitational singularities, such as the
Schwarzschild one, normally thought of as lying at the center of a
black hole.}

For anyway, to stress it once again in the present work (as we
have also time and again done it throughout our past tetralogy
\cite{malrap1,malrap2,malrap3,rap5}), in ADG the base topological
space (on which the vector, algebra, and $\struc$-module sheaves
in focus are soldered), be it a continuum or a discretum, plays
absolutely no role in the inherently algebraic differential
geometric mechanism, which derives as it were `from the stalk'
(ie, from the `geometrical objects'---physically speaking, the
dynamical physical fields `in-themselves'). Ultimately, it is not
a question whether it is a continuum or a discontinuum, because it
plays no epicurical role in the said differential geometric
mechanism---or more importantly, in the physical dynamics
expressed as (differential) equations between the relevant sheaf
morphisms representing the fields ({\it viz.} connections) and
their curvatures ($\struc$-functoriality and $\struc$-natural
transformation PARD of the dynamics). In a philological sense, in
ADG the generic and in principle arbitrary base (topological)
space $X$ recalls a bit Archimedes' famous dictum ``{\em Give me
somewhere to stand, and I shall move the Earth}''\footnote{In the
original ancient Greek: ``$\mathrm{\Delta \acute{o}\varsigma}$
$\mathrm{\mu o\acute{\iota}}$ $\mathrm{\pi\tilde{\alpha}}$
    $\mathrm{\sigma\tau\tilde{\omega}}$
    $\mathrm{\kappa\alpha\acute{\iota}}$ $\mathrm{\tau\acute{\alpha}\nu}$ $\mathrm{\gamma\tilde{\alpha}\nu}$
    $\mathrm{\kappa\iota\nu\acute{\eta}\sigma\omega}$''.} \cite{archimedes},\footnote{See also Simplicious \cite{simple}.} in the sense that once $X$ has been chosen and
specified as a `surrogate' background in order to
(sheaf-theoretically) localize and solder the `geometrical
objects' ({\it ie}, the fields $\conn$, their auto-symmetries'
principal sheaves $\aut\modl$, and their associated representation
quantum-particle sheaf spaces $\modl$),\footnote{Archimedes'
``{\em somewhere to stand}''.} it is then for all
practical-calculational (:differential geometric-differential
Calculus) purposes effectively discarded, as it does not partake
at all in the sheaf-morphism expressed differential equations
modelling the field-dynamics.\footnote{Archimedes' ``{\em moving
the earth}''.} This has already been anticipated and discussed to
a great extent in the first author's \cite{mall7,mall10,mall11}.

Last but not least pertinent here is the remark, following from
Finkelstein's telling words in the introduction of \cite{df0}:

\bigskip \noindent (Q8.24)\hskip 0.9in
\begin{minipage}{11cm}
\noindent ``{\small ...Until we find a satisfactory theory of
space-time structure, we shall be beset by the dilemma of the
discrete versus the continuous, the idea already posed by Riemann,
in much the following terms:

(a) A discrete manifold has finite properties, whereas a
continuous manifold does not. {\em Natural quantities are to be
finite}.\footnote{Our emphasis.} The world must be discrete.

(b) A discrete manifold possesses natural internal metrical
structure, whereas a continuous manifold must have its metrical
structure imposed from without. {\em Natural law is to be
unified}.\footnote{Again, our emphasis.} The world must be
discrete.

(c) A continuous manifold has continuous symmetries, whereas a
discrete manifold does not. {\em Nature possesses continuous
symmetries}.\footnote{Emphasis is ours.} The world must be
continuous

The third argument is especially serious for rotational and
Lorentz symmetry, which are much more serious to counterfeit than
translational symmetry. Subgroups can be found as dense as desired
in the translation group that are not everywhere dense, but I do
not think that they exist for the rotation or Lorentz groups.

Since Riemann a new approach to this dilemma has become available.
The same question about matter asked for two millennia---Is it
continuous or is it discrete?---has at last been answered in this
century: No. Matter is made neither of discrete objects nor [of
continuous]\footnote{Our addition for clarity or completeness.}
waves but of quanta. In most familiar terms, a quantum is an
object whose coordinates form a noncommutative algebra...A quantum
manifold is a third possibility for space-time too. This
possibility would pass us cleanly between the horns of Riemann's
dilemma:

(a$^{'}$) A quantum manifold, like a discrete one, {\em has better
convergence than a continuous manifold}\footnote{Our
emphasis.}---remember Planck and the black body.

(b$^{'}$) A quantum manifold, like a discrete one, {\em is born
with internal structure, and is even more unified, being
coherent}.\footnote{Once again, our emphasis.}

(c$^{'}$) A quantum manifold, like a continuous one, {\em
possesses continuous symmetries}.\footnote{Our emphasis
again.}...}''
\end{minipage}

\vskip 0.1in

\noindent that

\bigskip \noindent (R8.9)\hskip 0.9in
\begin{minipage}{11cm}
\noindent {\em In the same way that the quantum passes through the
horns of Riemann's discretum-{\it vs}-continuum dilemma by being
neither (`continuous') wave (field) nor (`discrete')
particle---paradoxically, as it were, {\em neither and at the same
time both}!---so here the ADG-field is indifferent to whether the
background spacetime is `discrete' (particle-like) or `continuous'
(field-like)---it is a coherent combination of field ($\conn$) and
particle ($\modl$) qualities. In this sense too, as we argued
before, the ADG-field is `in-itself' already quantum---a `{\em
self-}', or `{\em third quantized}' entity and {\it a fortiori}
{\it prima facie} in need of quantization.}
\end{minipage}

\vskip 0.1in

\noindent However, there is a lot more in Finkelstein's remarks in
(Q8.24) above that we can comment on in the light of the
`autodynamical', `self-quantum' and `background spacetime
indifferent' ({\it ie}, whether the geometrical base spacetime is
`continuous' or `discrete') ADG-field theory that we hereby
propound. Let us itemize our comments:

\begin{itemize}

\item First of all, it goes with little saying that our
ADG-gravity efforts are {\em not} beset in any way by Riemann's
continuous-{\em versus}-discrete spacetime dilemma. {\em In the
background spacetimeless ADG-gravity, the question whether
spacetime is a discretum or a continuum is begging the question,
for there is no `spacetime' external to the gravitational field
itself}.

\item From (a), we isolate the remark, which accords with our
basic {\it credo} here, that {\em there are no infinities (or
singularities) in (the laws of) Nature}.

\item From (b), we highlight our basic thesis here that the
ADG-gravitational field is dynamically autonomous, in no need of
an external (background) spacetime structure (whether continuous
or discrete) to be prescribed from without. The ADG-gravitational
field $(\modl ,\conn)$ is `unified' (what we called `unitary') and
`quantum coherent' (what we coined `third quantum').

\item From (c) above and its ensuing remarks (about Lorentz and rotational Lie symmetries),
which appear to favor the argument for a spacetime continuum
picture, we fundamentally disagree with the position that ``{\em
Nature possesses continuous symmetries}''---in point of fact, with
the position that Nature possesses any symmetry at all. In the
same way that we do not hold that the notion of (an external to
the fields) geometrical spacetime---whether discrete or
continuous---is physically meaningful, we also maintain, now from
a Kleinian perspective, that its symmetry group (whether discrete
or continuous, respectively) is of no physical significance
either.\footnote{To put it in another, more general, way: {\em
`symmetry' goes hand in hand with `measurement', that is to say,
like `space(time)' and its `geometry', it lies with
\underline{us}---the external to the fields
themselves---`observers', `measurers' and `symmetry perceivers'}.}
Since we have repeatedly argued throughout this work that `{\em
all is field}' (to the extent that if there is any `geometrical
spacetime'---what is usually called `spacetime geometry'---in our
ADG-theoresis, it is inherent in the dynamical fields defining the
laws of Nature), if there is any `symmetry' at all in our scheme,
it is encoded in the principal sheaf $\aut\modl$ of dynamical
autosymmetries of the ADG particle-field pair $(\modl
,\conn)$.\footnote{The intimate connection between `symmetry' and
geometry can be seen in our scheme in the fact that as soon as one
prescribes (externally to the field $\conn$ of course!) a
structure sheaf $\struc$ of generalized arithmetics
(`coordinates'), which in turn may carry inherently in it a
`discrete' or a `continuous' spectral spacetime geometry (Gel'fand
duality) and represents $\modl$ locally as $\struc^{n}$, the
corresponding principal sheaf $\aut\modl$ of the dynamical field
autosymmetries becomes (again locally) the group sheaf
$M_{n}(\struc(U))^{\bullet}$ ($n$ being the rank of the associated
sheaf $\modl$).}

\item Then Finkelstein draws from the analogy between spacetime and
matter\footnote{An analogy already present in the usual
`action-reaction' way the (non-vacuum) Einstein field equations
are physically interpreted, namely, that matter (fields) on the
right hand side curve spacetime geometry on the left. This, as
noted earlier, `misleads' one into thinking that since the fields
on the right hand side are quantum or quantized (QFT), so must be
the spacetime geometry on the left---arguably, {\em the} analogy
motivating and guiding most of the current approaches to QG.} and
directs the quest towards the development of a `{\em quantum
manifold}'---one that is neither discrete nor continuous and at
the same time both ((a$^{'}$--c$^{'}$).\footnote{Interestingly
enough, with his remark that ``{\em a quantum is an object whose
coordinates form a noncommutative algebra}'', in a way he
prophetically anticipated Connes' noncommutative (differential)
geometry, and in footnote 2 he also refers to the pioneering work
of Snyder \cite{snyder} as (to his knowledge) the first to suggest
that space-time is a quantum space---a `quantum manifold' defined
as a space whose points are labelled by noncommuting coordinates.}

\item However, to stress it once more, since in ADG-gravity (the structure of) spacetime is not an
issue at all, the quest for its `true' quantum structure ({\it
eg}, arriving at it via some process of quantization, or even
developing from scratch a full-fledged `quantum theory of
spacetime') does not arise either.\footnote{At the same time, to
emphasize it once again, if there is any `noncommutative geometry'
involved in ADG-gravity, it must be sought in the non-abelian
Klein group $M_{n}(\struc(U))^{\bullet}$ of (local) dynamical
auto-symmetries of the coherent and autonomous ADG-gravitational
particle-field pair $(\modl ,\conn)$ as well as of the dynamical
equations (\ref{eqy23}) that it defines.}

\end{itemize}

\noindent The last remarks about the background geometrical
spacetime continuum-{\it contra}-discontinuum indifference or
`blindness' of ADG-field theory, as well as its associated unitary
`third quantum coherence' and `unitarity' {\it vis-\`a-vis} field
and particle attributes of the ADG-field $(\modl ,\conn)$, brings
us to comment on the later Einstein's\footnote{As later Einstein
we would consider Einstein's work after the accomplishment of
GR---at least as Einstein's ideas and work after 1920, which were
predominantly focused around his unitary (nowadays called unified)
field theory vision.} `schizophrenic' and ambivalent attitude
towards the geometrical spacetime continuum and the continuous
field theory based on it on the one hand, and on the other, the
possibility of developing a purely algebraic theory for the
description of the world in the quantum deep, one that is {\it a
fortiori} based on a fundamental discretum (with the concomitant
loss of the spacetime interpretation in the inherently reticular
quantum domain).

\subsubsection{ADG-theoretic marriage of the `two Einsteins':
a possible completion of GR to a genuinely unitary field theory}

Einstein's dissatisfaction with the spacetime continuum based
field theory granted, he apparently turned a blind eye to his
`second nature' urging him to look for a purely algebraic and
finitistic method for the description of reality due to the
singularities of GR and the finitistic-algebraic paradigm of
quantum mechanics, and insisted till the end of his life on
looking for a unified (better, `unitary') field theory (of the
electromagnetic with the gravitational forces)\footnote{Back then
the two known elementary/fundamental forces of Nature.} which
could on the one hand evade the problem of singularities, and on
the other `explain' the atomistic character of matter and
radiation, thus evading altogether quantum theory.

Below are two quotations from \cite{einst9}: the first expresses
clearly his anticipation that a field-theoretic completion of GR
to a unitary field theory should result in a singularity-free
description even of material point particles, which act as sources
of the various radiation force-fields, but from GR's viewpoint
they are genuine or `true' singularities of the (gravitational)
radiation field. In other words, {\em one of the primary
motivations for formulating a unitary field theory is overcoming
the problem of singularities troubling primarily GR}---arguably,
the background geometrical spacetime continuum based field theory
{\it par excellence}:

\bigskip \noindent (Q8.25)\hskip 0.9in
\begin{minipage}{11cm}
\noindent ``{\small ...The essence of this truly involved
situation [{\it ie}, uniting gravity with the other forces of
matter---{\it ie}, electromagnetism]\footnote{Our addition.} can
be visualized as follows: {\em A single material point at rest
will be represented by a gravitational field that is everywhere
finite and regular, except where the material point is located:
there the field has a singularity}\footnote{Our emphasis. This is
the Schwarzschild singularity at the fixed point mass that we
`resolved' earlier, and in \cite{rap5}, by ADG-means.}... Now it
would of course be possible to object: {\em If singularities are
permitted at the locations of the material points, what
justification is there for forbidding the occurrence of
singularities elsewhere?}\footnote{Our emphasis again.} This
objection would be justified if the equations of gravitation were
to be considered as equations of the total field. [Since this is
not the case], however, one will have to say that the field of a
material particle will differ from a pure gravitational field the
closer one comes to the location of the particle. {\em If one had
the {\rm [unitary]}\footnote{Again, our addition.} field equations
for the total field, one would be compelled to demand that the
particles themselves could be represented as solutions of the
complete field equations that are free of irregularities
everywhere. Only then would the general theory of relativity be a
complete theory.}...}''
\end{minipage}

\vskip 0.1in

\noindent The second quotation from \cite{einst9} expresses
Einstein's well documented scepticism about quantum
mechanics---especially about the `pseudo' way in which quantum
theory purports to do away with continuous structures when in fact
it still employs the spacetime continuum in order to formulate the
dynamics of quantum wave amplitudes ({\it eg}, quantum fields) as
{\em differential equations}---while, in an indirect way, it puts
forward his support of a unitary field theory that may on the one
hand overcome the problem of singularities in GR, and on the other
`explain' the atomistic structure of reality:

\bigskip \noindent (Q8.26)\hskip 0.9in
\begin{minipage}{12cm}
\noindent ``{\small ...It is my opinion that the contemporary
quantum theory represents an optimal formulation of the
relationships, given certain fixed basic concepts, which by and
large have been taken from classical mechanics. I believe,
however, that this theory offers no useful point of departure for
future development. This is the point at which my expectation
deviates most widely from that of contemporary physicists. {\em
They are convinced that it is impossible to account for the
essential aspects of quantum phenomena} (apparently discontinuous
and temporally not determined changes of the state of a system,
simultaneously corpuscular and undulatory qualities of the
elementary carriers of energy) {\em by means of a theory that
describes the real state of things [objects] by continuous
functions of space for which differential equations are
valid}.\footnote{Our emphasis.} They are also of the opinion that
in this way one cannot understand the atomic structure of matter
and radiation. They rather expect that systems of differential
equations, which might be considered for such a theory, in any
case would have no solutions that would be regular (free from
singularities) everywhere in four-dimensional space. {\em Above
everything else, however, they believe that the apparently
discontinuous character of elementary processes can be described
only by means of an essentially statistical theory, in which the
discontinuous changes of the systems are accounted for by
continuous changes of the probabilities of the possible
states}\footnote{Our emphasis again.}...}''
\end{minipage}

\vskip 0.1in

\noindent From the quotation above, and `by negation/exclusion',
one could say that Einstein, in contradistinction to his
contemporary quantum physicists:\footnote{And, it is fair to say,
in contrast also to the majority of current quantum theorists.}

\begin{itemize}

\item Believed that a singularity-free field theory on the
spacetime continuum---whose laws are expressed differential
geometrically, {\it ie}, as differential equations---could still
be developed. This essentially implied his `unitary field theory'
vision,\footnote{See above.} albeit, one that still abides by the
spacetime continuum (manifold) in which differential equations can
be formulated.

\item As also noted before, he also believed that such a theory
could account for the quantum structure of reality, in the sense
that the quanta of the source or radiation fields will be
described by everywhere (in the spacetime continuum)
singularity-free ({\it ie}, regular) solutions to the total ({\it
ie}, unitary) field equations (if we had them).

\item Finally, he maintained that, in truth, quantum theory, in
spite of the apparent discontinuity of quantum processes, still
tacitly employs the continuum (as it were, `in disguise') in the
form of `continuous changes of' ({\it ie}, again differential
equations obeyed by) the probability amplitudes for (states of)
quantum systems, with those `field-states' (wave functions) too
being defined on a spacetime continuum ({\it eg}, Schr\"odinger's
non-relativistic or Dirac's relativistic wave equations).

\end{itemize}

Even more forceful and telling are the following words taken from
three remarkable consecutive paragraphs in
\cite{einst10}\footnote{Which can be found on pages 92 and 93 in
article 13, titled `{\itshape Physics and Reality}' (reprinted
from the {\itshape Journal of the Franklin Institute}, {\bf 221},
313 (1936); see \cite{malrap1}). The entire third paragraph is
written in {\em emphatic} script for emphasis.} which show, in
order of appearance, an `oscillation', `ambivalence', or
`indecisiveness' in Einstein's thought about whether to opt for
the background geometrical spacetime continuum/field theory or for
an algebraic/discontinous description of reality {\it \`a la}
quantum mechanics, with the third paragraph showing clearly his
`wishful thinking' about a field theory that could represent
particles (quanta) by singularity-free fields:

\bigskip \noindent (Q8.27)\hskip 0.9in
\begin{minipage}{12cm}
\noindent ``{\small ...{\em To be sure, it has been pointed out
that the introduction of a space-time continuum may be considered
as contrary to nature in view of the molecular structure of
everything which happens on a small scale. It is maintained that
perhaps the success of the Heisenberg method points to a purely
algebraical method of description of nature, that is to the
elimination of continuous functions from physics. Then, however,
we must also give up, by principle, the space-time
continuum}.\footnote{Our emphasis.} It is not unimaginable that
human ingenuity will some day find methods which make it possible
to proceed along such a path. {\em At the present time, however,
such a program looks like an attempt to breathe in empty
space}.\footnote{Our emphasis again.}

There is no doubt that quantum mechanics has seized hold of a
beautiful element of truth, and that it will be a test stone for
any future theoretical basis, in that it must be deducible as a
limiting case from that basis, just as electrostatics is deducible
from the Maxwell equations of the electromagnetic field or as
thermodynamics is deducible from classical mechanics. However, I
do not believe that quantum mechanics will be the starting point
in the search for this basis, just as, vice versa, one could not
go from thermodynamics (resp. statistical mechanics) to the
foundations of mechanics.

{\em In view of this situation, it seems to be entirely
justifiable seriously to consider the question as to whether the
basis of field physics cannot by any means be put into harmony
with the facts of the quantum theory. Is this not the only basis
which, consistently with today's possibility of mathematical
expression, can be adapted to the requirements of the general
theory of relativity? The belief, prevailing among the physicists
of today, that such an attempt would be hopeless, may have its
root in the unjustifiable idea that such a theory should lead, as
a first approximation, to the equations of classical mechanics for
the motion of corpuscles, or at least to total differential
equations. As a matter of fact up to now we have never succeeded
in representing corpuscles theoretically by fields free of
singularities, and we can, a priori, say nothing about the
behavior of such entities. One thing, however, is certain: if a
field theory results in a representation of corpuscles free of
singularities, then the behavior of these corpuscles with time is
determined solely by the differential equations of the
field.}\footnote{All emphasis is ours.}}''
\end{minipage}

\vskip 0.1in

\noindent The concluding `hypothetical certainty' of Einstein in
the last three lines of the quotation above may be rephrased as
follows: {\em should one be able some day to represent
field-theoretically particles (`quanta') in a singularity-free
manner, then the particle dynamics---as it were, the evolution of
those quanta in time---will be already inherent in (or
theoretically speaking, be the result of) the field dynamics
itself, and there would be no need to assume {\it a priori}
particles as fundamental theoretical entities side-by-side the
field concept}. It is precisely in this sense that field
theory---Einstein's unitary field theory---aspired to `explain
away' particles and, accordingly, that ``{\em quantum theory could
be deducible from that future (unitary field) theoretical
basis}''. Indeed, earlier in \cite{einst10}, Einstein, upon
concluding the section titled `{\itshape The Field Concept}' in
article 13, `{\itshape Physics and Reality}', reexpresses this
certainty in quite a categorematic fashion:\footnote{Again, the
whole excerpt below is written in {\em emphasis} script for
emphasis(!)}

\bigskip \noindent (Q8.28)\hskip 0.9in
\begin{minipage}{12cm}
\noindent ``{\small\em ...What appears certain to me, however, is
that, in the foundations of any consistent field theory, there
shall not be, in addition to the concept of field, any concept
concerning particles. The whole theory must be based solely on
partial differential equations {\rm [for the fields
alone]}\footnote{Our addition.} and their singularity-free
solutions.}''
\end{minipage}

\subsubsection{What (mathematical) theory was Einstein looking for in order to materialize
his unitary field theory vision?}

We have seen in many quotations before Einstein's `agnosticism'
and `scepticism' about a possible future theory that can on the
one hand accommodate both his unitary field theory life-long grand
project and the purely algebraic and finitistic quantum. Those
doubts undoubtedly originate from his conviction that a field
theory necessarily depended on the geometrical background
spacetime continuum, while the latter apparently gravely
miscarried with the combinatory-algebraic quantum theory.

Especially {\it vis-\`a-vis} differential geometry, it is worth
recalling here Pais' remarks in the concluding paragraph of
\cite{pais}:

\bigskip \noindent (Q8.29)\hskip 0.9in
\begin{minipage}{12cm}
\noindent ``{\small ...Yet as his life drew to a close, occasional
doubts on his [unitary field theory]\footnote{Our addition.}
vision arose in his mind. In the early fifties he once said to me,
in essence: `{\em I am not sure that \underline{differential
geometry} is the framework for further progress, but if it is, I
believe I am on the right track'}\footnote{Our emphasis and
underlining.}...}''
\end{minipage}

\vskip 0.1in

\noindent And one should immediately contrast this against
Einstein's own remarks in (Q8.?), but especially, against
Feynman's and Isham's quotations in (Q8.?) and (Q8.?),
respectively.

\paragraph{A hypothetical theoretical scenario for a singularity-free, purely algebraic field theory in
the quantum domain and its potential ADG-theoretic realization.}
In view of the remarks above, we would like to suggest that ADG,
ADG-field theory, and ADG-gravity in particular (as well as the
ADG-theoretic formulation of the other gauge, Yang-Mills forces
\cite{mall1,mall2,mall4}), is `tailor-cut' for realizing his
unitary field vision, for the reasons we itemize below:

\begin{itemize}

\item First of all, ADG is a purely
{\em algebraic} and essentially {\em categorical} theory, as it
relies on sheaf theory and sheaf cohomology.

\item The central notion in ADG is that of a {\em field} ({\it viz.}
connection), hence the theory is (mathematically speaking) {\em
differential geometric}; albeit, in contradistinction to the CDG
and thus background spacetime manifold based current classical and
quantum field theories, it does not base its concepts and
technotropy on a base differential manifold---in fact, on any
background `space(time)' whatsoever, be it a `classical continuum'
or a `quantal discretum'.

\item The ADG connection fields allow one to write total
differential field equations for the dynamical laws of Nature,
equations that {\it a fortiori} are {\em free from singularities}.

\item In contradistinction to Einstein's expectations however, the
ADG particle-field pairs $(\modl ,\conn)$ are not forced in any
way to sacrifice the important particle aspect of
fields.\footnote{After all, in the quantum domain the
field-particle duality (also commonly known as Bohr's
Complementarity) is an important aspect of the (physical
interpretation of the) theory.} On the contrary, it places it
({\it ie}, $\modl$, whose local sections represent particle states
of the field) side-by-side the field (:$\conn$) concept, without
any singularity arising in the theory.

\item Finally, in contrast to Einstein's reservations about the
applicability of differential geometric ideas in the quantum
domain, as repeatedly explained earlier, the ADG-fields are
`inherently quantum' (:third quantized).

\end{itemize}

\subsubsection{No potholes in ADG's path: evading the whole of Einstein's
hole argument (based on a theme by John Stachel)}

We have emphasized numerous times throughout this paper, as well
as in the past trilogy \cite{malrap1,malrap2,malrap3}, the totally
base spacetime manifold free and purely algebraic (categorical)
way in which ADG is able to formulate classical gravity (GR) and
how it goes some way in capturing quantum features of that elusive
QG theory. With respect to the latter aim in particular, we have
argued, primarily in \cite{malrap3}, how this background manifold
independence can evade directly and entirely certain caustic
problems that the smooth spacetime manifold presents in various
attempts to quantize, either canonically (`Hamiltonianly') or
covariantly (`Lagrangianly'), GR; namely, {\em the inner product
problem} and {\em the problem of time}, both of which essentially
involve the diffeomorphism group $\mathrm{Diff}(M)$ of the base
spacetime manifold $M$, which group in turn implements
mathematically the PGC of GR.

In this sub-subsection we will see, in the classical context of
GR, how the said manifold independence of ADG enables us to evade
directly and completely a gedanken scenario originally proposed in
1913 by Einstein and Grossmann in order to put to a test the PGC
of GR, usually referred to as {\em Einstein's hole argument} (EHA)
\cite{einst13}. We do not intend to give here a detailed account
of EHA since many thorough expositions of it exist in the
literature---\cite{hoffmann,norton,stachel3,stachel5,stachel0,stachel6,stachel9},
to mention a few---but we will rather discuss its deeper meaning,
its far reaching (conceptual) consequences and aftermath as
elaborated by Stachel in \cite{stachel5,stachel0,stachel9}. We
will show how that deeper significance of the EHA is captured
precisely by the `{\em spacetime manifold free and solely
gravitational field and its automorphisms
(synvariance)\footnote{As represented by the algebraic connection
variable $\conn$.}}' picture of GR that ADG affords. Thus, in the
process of our elaborations on the EHA and its consequences under
the prism of ADG, we will not just stop at the straightforward and
quite obvious cutting the Gordian knot-type evasion-result `{\em
no background (smooth) spacetime manifold and consequently no
$\mathrm{Diff}(M)$, no EHA}' that ADG achieves,\footnote{We
provided arguments of such a kind for the direct ADG-theoretic
evasion of the inner product problem and the problem of time in
the manifold based approaches to QG, whether canonical or
covariant, in \cite{malrap3}---{\it ie}, `{\em no base spacetime
manifold and, as a result, no $\mathrm{Diff}(M)$, no inner product
problem or problem of time}'.} but we will show how the EHA's
deeper significance, as unfolded by Stachel, is already included
in (as it were, it is a particular instance of) the general
reversal of the traditional priority of kinematics (spacetime
geometry) over dynamics (algebra) that ADG enables us to maintain
not only in the context of (classical) gravity (GR), but also with
respect to the other (quantum) gauge forces (Yang-Mills theory).
All in all, the arguments about the EHA to be given below
vindicate the basic consequence of the ADG-theoretic perspective
on gravity (and, {\it in extenso}, on the other gauge
theories),\footnote{For anyway, from an ADG-theoretic perspective,
GR is another kind of gauge theory
\cite{malrap1,malrap2,malrap3}.} namely, that {\em the
algebraic/dynamical gravitational field} (as represented by a
connection $\conn$ on a suitable vector sheaf $\modl$) {\em is
(physically) prior to the geometrical/kinematical spacetime
manifold}, or more laconically put, that {\em dynamics (algebra)
is prior to kinematics (geometry)}.

To recapitulate, the way in which the EHA is (physically)
interpreted by Stachel allows us, as we shall see below,

\begin{itemize}

\item (a) not only to argue that the background differential
spacetime manifold free ADG, the `synvariant' notion of
gravitational field dynamics that the latter supports, as well as
the (smooth) metric-free (purely gauge-theoretic) formulation of
GR solely in terms of the algebraic $\struc$-connection $\conn$
that goes hand in hand with `synvariance', are able to bypass in
one-go the whole hole argument,\footnote{Pun intended.} but also,
and perhaps more importantly,

\item (b) to present a concrete example of the radical `{\em
dynamics before kinematics}' conceptual reversal that ADG is
pregnant to, as we discussed earlier in 3.2.3.

\end{itemize}

\noindent The bottom-line of the following presentation of the EHA
as expounded by Stachel will be that the smooth coordinates of the
points of $M$ are not what distinguishes and individuates them as
{\em physical events} proper, but that the dynamical field
$g_{\mu\nu}$ of GR, as a {\em solution} to the dynamical field
equations (Einstein equations), assumes that individuating role.
Alas, in ADG not only there is no {\it a priori} posited
background geometrical spacetime manifold, let alone a smooth
Lorentzian metric on it (which is normally assumed to represent
the gravitational field), but also the sole dynamical
gravitational variable is the entirely algebraic,
$\smooth$-smoothness unrelated $\struc$-connection $\conn$ on the
vector sheaf $\modl$ one chooses to employ as the carrier (or
particle state) space of the gravitational field $\conn$. But
without further ado or thinking ahead, let us first expose the EHA
and summarize Stachel's physical interpretation of it.

\paragraph{Brief exposition of the EHA \`a la Stachel.} Thus, let us first recapitulate the EHA as distilled and
expressed in modern (differential geometric) language by Stachel
\cite{stachel5}. So first, in accordance with the original
formulation of GR by Einstein, the `structural conditions' for the
expression of the HA are those set by the basic kinematics of GR
which involves a four-dimensional differential manifold $M$ (with
its points being coordinatized by $\smooth(M)$) and a smooth
metric tensor field $g_{\mu\nu}$ of Lorentzian
signature.\footnote{Actually, as we shall see below, the {\it a
priori} fixation of the Lorentzian manifold kinematics of
GR---`{\it a priori}' here essentially means `{\em before the
dynamical equations for the gravitational metric are solved}'---is
exactly what the EHA, as interpreted and generalized by Stachel,
comes to question.}

The next assumption of the EHA is to consider a subset $\hole$ of
$M$ which is devoid of any matter---an empty region in $M$
representing the `{\em hole}'---while matter fields are supposed
to be present outside (and at the boundary of) this hole, that is,
in its complement $M-\hole$. Then, Einstein assumed that (the
points of) $M$ (are) is initially charted by a coordinate system
${\mathcal{X}}=(x_{\mu})$ relative to which the energy-momentum
stress tensor $T$ assumes values $T_{\mu\nu}^{M-{\hole}}$ in
$M-{\hole}$ and it vanishes identically in the empty hole
($T_{\mu\nu}^{{\hole}}=0$), while the metric $g$ takes on values
$g_{\mu\nu}$ and satisfies the generally covariant Einstein field
equations on the whole $M$.

Now, apply a smooth coordinate transformation
$f:~{\mathcal{X}}\mapto{\mathcal{X}}^{'}$ (or diffeomorphism
$f:~M\mapto M$) of the following kind:

\begin{itemize}

\item $f$ is the identity on $M-{\hole}$
(${\mathcal{X}}_{M-{\hole}}={\mathcal{X}}^{'}_{M-{\hole}}$),

\item while inside the hole $\hole$,
${\mathcal{X}}_{{\hole}}\not={\mathcal{X}}^{'}_{{\hole}}$.

\end{itemize}

\noindent It follows that with respect to the new coordinates
${\mathcal{X}}^{'}$, $T_{\mu\nu}^{'{\hole}}=0$ and
$T_{\mu\nu}^{'M-{\hole}}=T_{\mu\nu}^{M-{\hole}}$---{\it ie},
$T_{\mu\nu}=T_{\mu\nu}^{'}$ everywhere on $M$. On the other hand,
while $g^{'}_{\mu\nu}=g_{\mu\nu}$ identically on $M-{\hole}$,
$g^{'{\hole}}_{\mu\nu}\not=g^{{\hole}}_{\mu\nu}$ in general,
although $g^{'}$ too is supposed to satisfy the same set of
generally covariant field equations. Moreover, if we include the
coordinated point arguments of these fields---{\it ie}, write
$g_{\mu\nu}(x)$ relative to $\mathcal{X}$ and
$g_{\mu\nu}^{'}(x^{'})$ relative to ${\mathcal{X}}^{'}$---Einstein
intuited the following apparent contradiction:

\begin{itemize}

\item  while $g_{\mu\nu}^{'}(x^{'})$ and $g_{\mu\nu}(x)$
correspond to the {\em same} gravitational field,\footnote{After
all, the Einstein equations are generally covariant.}

\item the $f$-pushed forward $g^{'}=f^{*}(g)$,\footnote{Stachel
calls $g^{'}$ `{\em the metric $g$ dragged along $f$}'.} {\em but
still referred to the initial coordinate system} $\mathcal{X}$
({\it ie}, write $g^{'}(x)$ by suppressing indices), also
satisfies the Einstein equations on $M$ {\em but does not
correspond to the same gravitational field as} $g(x)$ (or
equivalently, as $g^{'}(x^{'})$) {\em even though the two fields
agree on the matter-filled $M-\hole$ and on the boundary
$\partial{\hole}$ of $\hole$}.

\end{itemize}

\noindent Einstein went ahead and interpreted this contradiction
as the following apparent {\em violation of causality} coming from
a mathematical non-uniqueness of solutions on the entire spacetime
$M$ of the field equations: {\em generally covariant field
equations for gravity determine two distinct gravitational fields}
({\it ie}, solutions $g(x)$ or $g^{'}(x^{'})$, and $g^{'}(x)$)
{\em in the presence of the same matter sources}. Thus he
abandoned the assumption of generally covariant field equations
for gravity and required that further constraints on the
coordinates and their transformations should be imposed within the
hole $\hole$ as well as in going to it from the `matter-plenum'
region $M-\hole$.

At this point it must be emphasized that there have been rather
simplistic analyses and interpretations of the EHA, one of them,
\cite{hoffmann}, regarding it as an `anomaly' in Einstein's
thought and work in the sense that Einstein, `blinded' by his
life-long commitment to causality, overlooked the rather trivial
fact that although the values of the components of a tensor such
as the metric $g$ change under a coordinate transformation ({\it
ie}, $g^{'}_{\mu\nu}\not= g_{\mu\nu}$), the tensor itself remains
the same---{\it ie}, it represents the same gravitational field.
Characteristically, we quote Bannesh Hoffmann:

\bigskip \noindent (Q8.30)\hskip 0.9in
\begin{minipage}{12cm}
\noindent ``{\small\em ...It is clear from both the argument
pertaining to the unchanging $T_{ab}$ and the argument pertaining
to the changing $g_{ab}$ that Einstein was saying that if the
components of a tensor change under a coordinate transformation,
the tensor itself changes. But in fact, just the opposite is the
case. Precisely because $g_{ab}$ is a tensor, the $g_{ab}$ and
$\bar{g}_{ab}$ actually represent the same field even though
$\bar{g}_{ab}\not= g_{ab}$. In general, we should expect the
components of a tensor to change when the coordinates are changed.
That is the nature of tensors. Indeed, it is the first thing one
learns in studying the tensor calculus...How could Einstein have
made so elementary an error?...I suggest that it was because
Einstein had long been profoundly concerned about causality and
determinism...}''
\end{minipage}

\vskip 0.1in

\noindent Stachel, however, in \cite{stachel5} would not settle
for such simplistic interpretations of the EHA:

\bigskip \noindent (Q8.31)\hskip 0.9in
\begin{minipage}{12cm}
\noindent ``{\small...In trying to interpret these
passages,\footnote{Immediately preceding this quotation were two
translated passages from \cite{einst13,einst14}.} then, I have
proceeded on the assumption that he was trying to express
something nontrivial...Thus, I am led to reject one common
interpretation of the `hole' argument, which assumes that Einstein
did not realize that the transformation of the components of the
metric tensor under a coordinate transformation results in a
re-description of the {\em same}\footnote{Stachel's emphasis.}
gravitational field in a {\small\em different}\footnote{Stachel's
emphasis.} coordinate system...So Einstein's point is {\small\em
not}\footnote{Stachel's emphasis.} that $G^{'}(x^{'})$ and $G(x)$
are different.\footnote{Stachel, following Einstein and Grossmann
in \cite{einst14}, uses the symbol `$G$' for the totality of the
components $g_{\mu\nu}$ of the metric tensor $g$.} His point is
that $G^{'}(x)$ and $G(x)$---same $x$---are different} {\rm
[quoting then Einstein and Grossmann from \cite{einst14}]}:
`{\small\em $G^{'}(x)$ also describes a gravitational field with
respect to $K$\footnote{Our coordinate frame ${\mathcal{X}}$
above.} which however does not correspond to the actual (i.e.,
originally given) gravitational field}'\footnote{Our
emphasis.}...''
\end{minipage}

\vskip 0.1in

\noindent For, after all, Einstein well understood that $g(x)$ and
$g^{'}(x^{'})$ represent the same gravitational field, but was
baffled that $g^{'}(x)$ and $g(x)$---{\em both referred to}
$\mathcal{X}$---appeared to be different.

Stachel, after a careful semantic analysis of the EHA in
\cite{stachel5}, was led to the following subtle `conclusion': the
question whether $g(x)$ and $g^{'}(x)$ are {\em physically}
equivalent ({\it ie}, whether they represent the same
gravitational field) or not is not a purely mathematical one as
long as one does not regard coordinates as {\em physically
individuating} the points of $M$---that is, as long as one does
not regard coordinates as the structure that qualifies the points
of $M$ to {\em physical events} proper. Stated in a positive way,
if one maintains {\it a priori} (like Einstein did in the HA) that
coordinates physically individuate the points of $M$---{\it ie},
that the assignment-labelling $M\ni p\mapsto x(p)$ physically
distinguishes, as events, the points of $M$, then one can indeed
maintain that $g^{'}(x)$ is physically distinct from (inequivalent
to) $g(x)$. However, if one does not think of coordinates as
physically individuating structures on the spacetime manifold, the
original $g(x)$ and the $f^{*}$-dragged metric $g^{'}(x)$---again,
{\em both referred to the same coordinate system}---are physically
indistinguishable (equivalent). In other words, {\em the hole in
Einstein's hole argument lies with the argument}\footnote{Double
pun intended.}---{\it ie}, with the coordinate $x$ in the argument
of $g$.

Einstein, Stachel argues convincingly in \cite{stachel5}, perhaps
because he, still as late as 1913, had not shed completely the
idea that coordinates have a direct metrical significance (Q2.7),
abandoned ``{\em with a heavy heart}'' the PGC of GR precisely
because he thought of coordinates as physically individuating the
points of $M$ as events proper. Furthermore, Stachel asserts that

\bigskip \noindent (Q8.32)\hskip 0.9in
\begin{minipage}{12cm}
\noindent ``{\small ...the physical equivalence or nonequivalence
of these two fields is {\small\em not}\footnote{Stachel's
emphasis.} a purely mathematical question. Or, more accurately, it
does not {\small\em become}\footnote{Stachel's emphasis again.}
one unless and until one introduces the additional, {\small\em
nonmathematical}\footnote{Or better, `{\em physically minded}'.
Again, Stachel's emphasis.} assumption or postulate that, in
regions where no matter is present, the points of a manifold are
physically individuated only by the properties that they inherit
from the metric field...}''
\end{minipage}

\vskip 0.1in

\noindent In other words, in {\it vacuo} ({\it ie}, in the absence
of matter), {\em it is the dynamical field itself}---in GR, the
metric $g_{\mu\nu}$ {\em satisfying the Einstein equations}---{\em
and not the {\it a priori} prescribed (kinematical) coordinates,
that is physically individuating the points of $M$ as spacetime
events proper}.

At the same time, Stachel makes it clear that a mathematician
would in a sense be `justified' in regarding the coordinate
labelling of the of the points of $M$ as {\em mathematically}
individuating structures since for example, as we have repeatedly
noted throughout the present paper, $M$ (initially regarded as a
structureless set) inherits its topological and differential
structure from $\smooth(M)$. Indicatively, we quote him from
\cite{stachel5}:

\bigskip \noindent (Q8.33)\hskip 0.9in
\begin{minipage}{12cm}
\noindent ``{\small ...One might be tempted to assert that
coordinate-labelling systems provide such individuating fields for
the points of the manifold. And so they do for {\small\em
mathematical}\footnote{Stachel's emphasis.} properties, such as
the differential-topological structure of the manifold. But the
point is (no pun intended!) that no mathematical coordinate system
is {\small\em physically}\footnote{Stachel's emphasis.}
distinguished per se; and without such distinction there is no
justification for physically identifying the points of a
manifold---which are homogeneous anyway---as physical events in
space-time. Thus, the mathematician will always correctly regard
the original and the dragged-along fields as distinct from each
other. But the physicist (or indeed anyone applying differentiable
manifolds) must examine the question in a different light; and the
answer will depend upon the means available for making
extra-mathematical distinctions between the points of the
manifold---or, more concisely upon the presence of individuating
structures...}''
\end{minipage}

\vskip 0.1in

\noindent Let it be added here that this {\em mathematically
individuating} role that coordinates play for the points of a
smooth manifold has also been advocated by Auyang. In
\cite{auyang}, in accounting about Gauss and Riemann's
ground-breaking work in infinitesimal (differential) geometry, she
remarks about the role of coordinates (in the former's theory of
surfaces):

\bigskip \noindent (Q8.34)\hskip 0.9in
\begin{minipage}{12cm}
\noindent ``{\small ...The Gaussian coordinates [arbitrary
curvilinear coordinates]\footnote{Our addition.} consistently
assign to each point on a surface a unique pair of ordered
numbers, and refrain from going further. {\em The Gaussian
coordinates individuate, but neither relate nor
measure}}.\footnote{Auyang's emphasis.} {\small\em This is the
idea utilized in differential geometry, where the bare
differentiable manifold is just a system of identifiable
points\footnote{Our emphasis.}...}''
\end{minipage}

\vskip 0.1in

Of course, in a strong sense this is the very physical essence of
the PGC of GR; namely, that {\em in GR, points are not sacrosanct,
that no coordinate system is physically preferred, and that
coordinates are not physically distinguishable (`observable' so to
say entities partaking into the gravitational dynamics)}. Quite on
the contrary, as also mentioned in the first section of the
present paper in the categorical language of ADG, the `observable'
geometrical object (or the `measurable dynamical entity') in GR is
an $\otimes_{\struc}$-tensor (or equivalently, an
$\struc$-morphism),\footnote{In the case of a base differential
manifold $M$, $\struc\equiv\smooth_{M}$.} such as the metric or
the Riemann curvature tensors---objects that `see through' the
base spacetime manifold's points and our $\smooth$-smooth
coordinatizations of them. Plainly then, Stachel's `hunch' that
Einstein's support of the HA and his tentative abandonment of the
PGC of GR until 1915 is `evidence' that as late as 1913 he was
still not convinced of the metrical insignificance of coordinates.
For since the metric is the sole dynamical variable in his
formulation of GR, one could say that still by 1913--14 he was not
sure about the aforesaid {\em dynamical insignificance of
coordinates}---the essence of the PGC\footnote{Which, as he notes
in \cite{einst14} ``{\em is the principle of relativity in its
most far reaching sense}''---see our elaborations about this point
in section 1.} maintaining that ``{\em systems of differential
equations {\small\rm [about the gravitational field $g_{\mu\nu}$]}
are generally covariant---i.e., they remain the same with respect
to arbitrary substitutions of the $x_{\nu}$}''
\cite{einst14}.\footnote{See Stachel's translation of the original
paper written in German in \cite{stachel5}.} But Einstein, already
in 1916 \cite{einst15}, reinstated the PGC in GR `at the expense
of' the EHA and its apparent implication that the smooth
coordinates of the spacetime continuum have a direct metrical,
hence dynamical, meaning:

\bigskip \noindent (Q8.35)\hskip 0.9in
\begin{minipage}{12cm}
\noindent ``{\small\em ...The laws of nature are to be expressed
as equations which hold good for all systems of coordinates, that
is, are co-variant with respect to arbitrary susbstitutions
whatever (generally co-variant)...}\footnote{Einstein's own
emphasis.}''
\end{minipage}

\vskip 0.1in

\noindent All in all, since the smooth coordinates of (the points
of) the spacetime manifold $M$ are not dynamical entities in GR,
and since as we mentioned in section 1 {\it \`a la} Wheeler, no
theory is a {\em physical} theory unless it is a dynamical
theory,\footnote{That is to say, a physical theory {\em is}
(defined by) its dynamics.} in GR the points of $M$, or their
smooth coordinate labels in $\smooth(M)$, are non-physical
(mathematical) structures, and the physical-{\it
versus}-mathematical individuation distinction that Stachel draws
as the deep aftermath of the EHA highlight precisely this.

Furthermore, in a remarkable passage from \cite{stachel5}, Stachel
gives the crux of his arguments about the whole of the EHA:

\bigskip \noindent (Q8.36)\hskip 0.9in
\begin{minipage}{12cm}
\noindent ``{\small ...More generally, spatio-temporal
individuation of the points of the manifold in a
general-relativistic model is possible only after specification of
a particular metric field, i.e., {\small\em only after the field
equations of the theory (which constitute its dynamical problem)
have been solved}.\footnote{Our emphasis.} Once this is done, the
points of the manifold-with-metric become full-fledged physical
events, endowed with gravitational as well as spatio-temporal
properties...}''
\end{minipage}

\vskip 0.1in

\noindent as well as its `negative' statement taken from
\cite{stachel3}:

\bigskip \noindent (Q8.37)\hskip 0.9in
\begin{minipage}{12cm}
\noindent ``{\small ...In any theory in which the metrical
structure is given {\small\it a priori}, the physical identity of
points of the space-time manifold is indeed established
independently of any dynamical considerations. But, in general
relativity the metrical structure forms part of the set of
dynamical variables, which must be determined {\small\rm [that is
to say, {\em solved}!]} before the points of spacetime have any
physical properties...}''
\end{minipage}

\vskip 0.1in

\noindent These words strike a very sensitive chord in the present
paper. To begin with, they perfectly accord with what Peter
Bergmann---the advocate {\it par excellence} for the `{\em
pointlessness of GR}'---says in (Q?.?) above:

\bigskip \noindent (Q8.38)\hskip 0.9in
\begin{minipage}{12cm}
\noindent ``{\small\em ...A world point would derive its identity
from its dynamic environment, or it might possess no identity at
all\footnote{Our emphasis.}...}''
\end{minipage}

\vskip 0.1in

\noindent Moreover, they corroborate---in fact, they are a
particular instance of---the central didagma of ADG, namely, that,
physically speaking, {\em dynamics is prior to kinematics}. Here,
in the context of GR, the (mathematical) kinematics of GR positing
up-front $M$ as a pointed differential manifold {\it a priori}
assumed to support a smooth Lorentzian metric\footnote{See (Q?.?)
from \cite{sork2} above.} does not physically qualify (cannot be
physically interpreted) as a gravitational {\em spacetime of
events} unless first its dynamical equations (for the dynamical
metric field) have been prescribed (and possibly, solved!).

Stachel succinctly resumes all this to the following `aphorism',
which he calls ``{\em Einstein's criterion}''---generally
speaking, the aftermath of the EHA-{\it contra}-the PGC of GR:

\bigskip \noindent (Q8.39)\hskip 0.9in
\begin{minipage}{12cm}
\noindent $\bullet$ ``{\small\em No metric, no
anything.\footnote{Stachel's own emphasis.}}''
\end{minipage}

\vskip 0.1in

\noindent a motto which we can readily generalize
ADG-theoretically to:

\bigskip \noindent (R8.10)\hskip 0.9in
\begin{minipage}{12cm}
\noindent $\bullet$ {\em No (dynamical) field, no anything}, or
equivalently, to

\noindent $\bullet$ {\em No dynamics, no spacetime}, or even more
abstractly, to

\noindent $\bullet$ {\em No dynamics, no kinematics; (spacetime)
geometry is the result of algebra ({\it ie}, of the gravitational
dynamics)---as it were, the `solution space(time)' of the physical
law corresponding to the dynamics, the realm where the field
equations hold ({\it ie}, where they are valid).}\footnote{In this
respect, we may also recall Stachel's own generalization in
\cite{stachel2} (and in \cite{stachel7}) of the ``{\em no metric,
no anything}'' motto above the ``{\em dynamic individuation of
fundamental entities}''.}
\end{minipage}

\vskip 0.1in

\noindent Of course, this ADG-generalization of the main aftermath
or deeper significance of the EHA as uncovered by Stachel rests
precisely on its direct and complete evasion, as described in the
following steps:

\begin{enumerate}

\item For the ADG-theoretic formulation of the (vacuum) Einstein
equations of GR no base differential ({\it ie}, $\smooth$-smooth)
spacetime manifold is used whatsoever \cite{mall3,malrap3};
therefore, the question (or problem in the original manifold based
GR-formulation of the EHA) of implementing the PGC via
$\mathrm{Diff}(M)$ becomes immediately a `non-issue'
(`non-problem').

\item Moreover, since in the ADG-theoretic formulation of (vacuum)
Einstein gravity the sole dynamical variable is not a smooth
metric $g_{\mu\nu}$ (on a locally Euclidean background manifold),
but an algebraic $\struc$-connection $\conn$ (on a suitable vector
sheaf $\modl$, which is locally isomorphic to $\struc^{n}$), the
original Einstein-Grossmann formulation of the EHA in terms of
$g_{\mu\nu}$ is rendered practically irrelevant or obsolete.

\item All in all, since for ADG, physically speaking, `{\em all
(and there) is (only the) field}' $(\conn ,\modl)$, without the
({\it a priori}, even mathematically speaking!) existence of an
external, base spacetime manifold, on the one hand the two points
above, from a Leibnitzian-Kleinian (relational/algebraic) sense,
point to the generalization of the PGC of GR involving
$\mathrm{Diff}(M)$ to the {\em Principle of Field Synvariance}
(PFS) involving solely $\aut(\modl)$, and on the other, they
automatically entail that the gravitational field ({\it viz.} the
$\struc$-connection $\conn$) and the algebraically
(:sheaf-theoretically) implemented dynamical law (differential
equation---here, the vacuum Einstein equations) that it defines is
manifestly prior the physical (`spacetime') geometry---the
solution space of the field equations that the field itself is
pregnant to. To emphasize it again, {\em first comes dynamics
(algebra), then kinematics (geometry); dynamics (algebra) is the
`cause' of kinematics (spacetime geometry)}; hence, {\em no
dynamics, no spacetime}.

\end{enumerate}

\paragraph{Even `discrete general covariance' questioned
ADG-theoretically.} In 7.5.2 above we argued that, from an
ADG-theoretic perspective, it makes no difference whether the base
`space(time)' is assumed to be a continuum or a discretum, for
{\em the theory is intrinsically background spacetimeless}. To
further support this claim, we wish to comment in the light of ADG
on certain remarks of Brightwell {\it et al.} in \cite{sork6}
about the search for a `discrete' analogue of the PGC of GR in the
finitistic-relational theoretical setting of the causet approach
to QG, as well as about how the so-called problem of time---which
is closely related to the PGC of GR via
$\mathrm{Diff}(M)$---appears from a discrete vantage. If anything,
these comments will be of value to the exposition here, because we
have already worked out a finitary, causal and quantal version of
vacuum Einstein gravity by ADG-theoretic means \cite{malrap3}.
Moreover, in the latter paper we maintained, as we did above, that
our genuinely background independent theory evades in one-go hard,
both conceptually and technically, problems in (the still manifold
based, both canonical or covariant, approaches to) QG such as {\em
the inner product problem} and {\em the problem of time}. Here, we
argue on top that this evasion is total, in the sense that the
said problems are bypassed not (only) because ADG is base
spacetime manifold independent, {\em but because it is background
spacetimeless period}, whether this background is continuous or
reticular.

Thus, we first quote Brightwell {\it et al.} In \cite{sork6}, upon
discussing how to implement the PGC of GR in the discrete realm of
causets, the authors remark:\footnote{The quotation below is
sectioned into three parts. Our comments following it are itemized
according to this three-fold partition.}

\bigskip \noindent (Q8.40)\hskip 0.9in
\begin{minipage}{12cm}
\noindent ``{\small ...{\bf (1)} After all, labels in this
discrete setting are the analogs of coordinates in the continuum,
and the first lesson of general relativity is precisely that such
arbitrary identifiers must be regarded as physically meaningless:
the elements of spacetime---or of the causet---have individuality
only to the extent that they acquire it from the pattern of their
relations to the other elements. It is therefore natural to
introduce a principle of `discrete general covariance' according
to which `the labels are physically meaningless'.

{\bf (2)} But why have labels at all then? For causets, the reason
is that we don't know otherwise how to formulate the idea of
sequential growth, or the condition thereon of Bell causality,
which plays a crucial role in deriving the dynamics. {\em Ideally
perhaps, one would formulate the theory so that labels never
entered, but so far no one knows how to do this---anymore than one
knows how to formulate general relativity without introducing
extra gauge degrees of freedom that then have to be cancelled
against the diffeomorphism invariance}.\footnote{Our emphasis.}

{\bf (3)} Given the dynamics as we {\em  can}\footnote{Brightwell
{\it et al.}'s emphasis.} formulate it, {\em discrete general
covariance plays a double role}.\footnote{Our emphasis.} On one
hand it serves to limit the possible choices of the {\em
transition probabilities}\footnote{Brightwell {\it et al.}'s
emphasis.} in such a way that the labels drop out of certain `net
probabilities', a condition made precise in
[...].\footnote{Reference omitted. The reader is referred to
\cite{sork6} for this.} {\em This is meant to be the analog of
requiring the gravitational action-integral $S$ to be invariant
under diffeomorphisms (whence, in virtue of the further assumption
of locality, it must be the integral of a local scalar concomitant
of the metric)}. On the other hand, general covariance limits the
{\em questions}\footnote{Their emphasis.} one can meaningfully ask
about the causet ({\it cf.} Einstein's `hole argument'). {\em It
is this second limitation that is related to the `problem of
time', and it is only this aspect of discrete general covariance
that I am addressing in the present talk}.\footnote{Our emphasis.}
{\bf (3)}}''
\end{minipage}

\vskip 0.1in

\noindent {\bf (1)} First, in our ADG-theoretic perspective on
gravity, coordinates (or coordinate-labelling of spacetime points)
play no role in the actual gravitational dynamics.\footnote{As
noted many times before, in this sense ($\smooth$-)coordinates (or
equivalently, the points of $M$), are physically meaningless.} In
our ADG-based scheme, the gravitational dynamics is effectuated
(represented) via $\struc$-sheaf morphisms---in particular, via
the curvature of the connection \cite{malrap3}. As a result, the
supposed `coordinate $\struc$-individuation of the elements of the
background structure (space)'---in ADG, the base topological space
$X$---plays no role in the dynamics, which is in turn expressed
purely relationally (algebraically) solely in terms of the
$\struc$-connection (field) $\conn$ (on the `carrier sheaf space'
$\modl$). Thus, as we saw earlier, the PGC in our theory,
implemented via $\aut\modl$, concerns only the field $(\modl
,\conn)$ `in-itself', while the generalized coordinate
$\struc$-labelling (or what amounts to the same, the background
space(time) dependence\footnote{We tacitly abide by the assumption
that space(time) is inherent in $\struc$ ({\it eg}, the way that
$M$ is the spectrum of $\smooth(M)$ by Gel'fand duality).})
practically disappears ({\it ie}, it is not involved at all in the
gravitational dynamics---Einstein's equations).

\vskip 0.05in

\noindent {\bf (2)} Second, to the question ``{\em why have labels
at all then?}'', our ADG-based reply is simply that {\em we assume
$\struc$ in order to get $\conn$;\footnote{After all, {\em all DG
boils down to $\struc$} (3.2), and $\conn$ is viewed
ADG-theoretically as nothing else but a generalized derivative
operator $\partial$.} but once we have been supplied with a
connection, $\struc$ disappears from our calculations---the
differential equations\footnote{Which are equations between
$\struc$-morphisms ($\otimes_{\struc}$-tensors).} for gravity that
we can set up ADG-theoretically}. Plainly then, we have a
formulation of GR as a {\em pure gauge theory} ({\it ie}, solely
in terms of the gravitational connection $\conn$) manifestly
without introducing any ``{\em extra gauge degrees of
freedom\footnote{Again, the only dynamical gauge variable in our
scheme is $\conn$ \cite{malrap3}.} which have to be cancelled
against the diffeomorphism invariance}'', which anyway does not
exist in our manifold-free theory.\footnote{Again, no background
$M$, no $\mathrm{Diff}(M)$ implementing the PGC in the classical
theory \cite{malrap3}.}

\vskip 0.05in

\noindent {\bf (3)} And third, about the double role that GC plays
in the gravitational dynamics ``{\em as we can formulate it}''
ADG-theoretically:

\begin{itemize}

\item (i) The gravitational action integral $S$---in our theory, a
functional only of the connection $\conn$ \cite{malrap3}---is
manifestly $\aut\modl$-invariant, since $\conn$ itself is a
manifestly local entity (variable), and $S$ is thus the integral
of the Ricci scalar curvature $\otimes_{\struc}$-tensor
$\ricci(\conn)$ (equivalently, $\struc$-sheaf morphism) of the
connection $\conn$, {\em not of the metric}.\footnote{Recall
again, in our manifold-free formulation of GR, the gravitational
variable is the connection, not the metric \cite{malrap3}.}

\item (ii) On the other hand, GC, as formulated purely
algebraically by ADG-theoretic means, and in a background
spacetime (whether `continuous' or `discontinuous') independent
way, solely in terms of the self-transmutations of the
gravitational field $(\modl ,\conn)$ itself in $\aut\modl$, {\em
does not limit at all the meaningful questions one can ask about
this background} (whether the latter is a continuum/manifold or a
discretum/causet, say), {\em simply because the latter does not
physically exist at all in our theory}---{\it ie}, {\em it plays
no role whatsoever in the gravitational dynamics as formulated
ADG-theoretically (in fact, algebraico-categorically) as equations
between sheaf morphisms}. In other words, and as a critique of the
continuum (manifold) viewpoint, as it was emphasized in
\cite{malrap3} and repeatedly noted in connection with the EHA
above,

\vskip 0.1in

\centerline{{\em no $M$, no} $\mathrm{Diff}(M)$, {\em no EHA},
{\em and no problem of time}.}

\vskip 0.1in

ADG simply cuts the `Gordian knot' of the background spacetime
manifold, thus it evades automatically all the problems in
classical ({\it eg}, PGC effectuated via $\mathrm{Diff}(M)$, EHA)
and quantum gravity (problem of time, inner product problem) that
go hand in hand with the assumption of a background smooth
spacetime continuum.

\end{itemize}

\paragraph{Bridging ``the outstanding gap between our creed and our
deed'' in GR.} Stachel \cite{stachel3}, based on the
aforedescribed aftermath of the deeper meaning of the EHA, went
even further and noted a discrepancy or discord in the way we
mathematically lay out up-front the Lorentzian manifold kinematics
of GR---interpreting it {\it a priori} as a spacetime of physical
events---and the actual way that physicists go about and solve the
Einstein equations. Characteristically, we quote him from
\cite{stachel3}:

\bigskip \noindent (Q8.41)\hskip 0.9in
\begin{minipage}{12cm}
\noindent ``{\small ...The standard mathematical formulation of
the general theory {\small\rm [of relativity]} starts with a bare
differentiable manifold, the points of which are interpreted as
physical events, and then imposes various fields on this manifold.
In particular, a second-rank symmetric covariant tensor field is
interpreted as the space-time metrical structure interrelating
these events, and as the gravitational potentials. While this way
of introducing the space-time structure is quite adequate in the
case of non-generally covariant theories, in which this structure
is given {\small\it a priori}, {\small\em it does not do full
justice to Einstein's relational concept of space-time, or to the
actual practice of relativists in formulating and solving
problems}\footnote{Our emphasis.}... The standard treatment does
not correctly describe the actual practice of relativists in
solving the field equations...I need hardly remind the audience
that we do not start out with a fixed manifold topology, but solve
the field equations on a generic patch.\footnote{That is to say,
{\em initially, we determine the field locally}.} Once such a
solution has been found, the problem of finding its maximal
{\small\em global}\footnote{Our emphasis.} extension is a major
one. It involves a lot of hard work, which generally includes the
formulation of criteria for an acceptable extension in order to
arrive at a unique answer. Questions such as maximal analytic
extension, geodesic completeness (timelike and/or null), and the
adjunction of ideal boundary points at infinity (null and/or
spatial) have generated some of the most interesting research on
general relativity in recent years...\footnote{Indeed, in
particular in research on the analysis of $\smooth$-smooth
spacetime singularities \cite{clarke4}, as we saw in section 2.}
Yet we persist in formalizing what we do as if the choice of
manifold topology were made at the outset. It appears to me unwise
to allow this gap between official ideology and daily practice.
Not only does it mislead students and researchers in other fields
attempting to understand the nature of our work; {\small\em the
gap may conceal important new insights into the nature of the
theory. The correct mathematical formulation of a problem often
suggests further and unexpected avenues of progress\footnote{Our
emphasis.}...}}''
\end{minipage}

\vskip 0.1in

\noindent Then, quite remarkably, right at the end of
\cite{stachel3} Stachel goes on and drops a very telling for us
hint regarding the use of the `right' mathematics for dealing with
the aforesaid problem of going from local-to-global solutions of
the dynamical Einstein equations for gravity rather than freezing
{\it ab initio} ({\it ie}, before actually solving the field
equations) the (global) spacetime manifold topology:\footnote{The
reader should refer to our earlier discussion in 6.8 about the
issue of going sheaf-theoretically from-local-to-global
(solutions) in GR {\it \`a la} ADG.}

\bigskip \noindent (Q8.42)\hskip 0.9in
\begin{minipage}{12cm}
\noindent ``{\small ...However, the fibre bundle approach clearly
does not solve the second problem discussed in the previous
section.\footnote{That is, not fixing up-front the global topology
of the manifold, and globalizing a local solution to the Einstein
equations---{\it in toto}, (analytically) extending a local
solution (a local region where the law holds) to a global one (the
total spacetime manifold where the law is valid).} The topology of
the base manifold is given {\small\it a priori}, so that a
different fibre bundle must be introduced {\small\it a posteriori}
for each topologically distinct class of solutions. The process of
finding the global topology cannot be formalized within the fibre
bundle approach. {\small\em It appears that sheaf cohomology
theory is the appropriate mathematical theory for dealing with the
problem of going from local to global solutions}\footnote{Our
emphasis.}... Let me close by appealing to the mathematically more
sophisticated members of our group to work on closing this
outstanding gap between our creed and our deed.}''
\end{minipage}

\vskip 0.1in

\noindent {\rm Needless to say that {\em sheaf cohomology lies at
the very heart of ADG} \cite{mall1,mall2,malrap2} and the issue of
the law being valid both locally and globally is settled by the
very nature of the sheaf-theoretic methods of ADG---in effect,
{\em this is precisely what the vacuum Einstein equations that
$\conn$ satisfies on $\modl$ means}\footnote{Again, see
6.8.}---but what's perhaps even more important, in ADG all this is
achieved manifestly without an external, base spacetime manifold.}

\paragraph{Individuation is distinction/discrimination; ultimately, difference:
dynamical individuation of `spacetime events'.} {\rm From the
foregoing ADG-theoretic generalization of the deeper significance
of the EHA it follows that, in a subtle sense, it is the dynamical
field $\conn$ alone and `in itself' that physically individuates
spacetime events.\footnote{If there are any such things at all,
for in ADG the field itself substitutes the spacetime (events) of
GR, being the sole dynamical (thus physically significant!) entity
in the theory.} In other words, the gravitational connection field
$\conn$ is the sole physically individuating structure in the
`world'---a `world' physically featureless (and meaningless) when
devoid of $\conn$---ultimately, when devoid of dynamics (that
$\conn$ defines). To emphasize this (dynamical) {\em field
solipsism} once more:}

\bigskip \noindent (R8.11)\hskip 0.9in
\begin{minipage}{12cm}
\noindent {\em All {\rm ({\it ie}, the `world')} is {\rm (the
dynamical)} field.}
\end{minipage}

\vskip 0.1in

\noindent Furthermore, in connection with our remarks earlier in
section 3 (3.3) about the connection, rather than the
traditional/original metric, ADG-theoretic formulation of GR and
therefore the essentially gauge-theoretic character of gravity, we
would like to stress here on the one hand the {\em mathematical}
significance of the notion of connection and its fundamental role
in setting up a theory of {\em differential}
geometry,\footnote{Let us stress it once again, ADG's original
motivation was to formulate an entirely algberaic notion of
connection, and its starting point so to speak was the first
author's observation that the usual differential $\partial$ is
another sort of connection \cite{mall1,mall2}.} and on the other,
its role as a {\em physically} individuating structure.

With respect to the first aspect of $\conn$, we bring forth from
\cite{weyl1} Weyl's words about the importance of the notion of
(affine) connection (and of the parallel transport of vectors that
it effects) in Riemannian---more generally, in
differential---geometry:\footnote{As noted in section 3, Weyl was
the founder of gauge theory in attempts to unite gravitodynamics
(Einstein's equations) with electrodynamics (Maxwell's
equations).}

\bigskip \noindent (Q8.43)\hskip 0.9in
\begin{minipage}{12cm}
\noindent ``{\small ...The later work of Levi-Civita, Hessenberg,
and the author, shows quite plainly that the fundamental
conception on which the development of Riemann's geometry must be
based if it is to be in agreement with nature, is that of the
infinitesimal parallel displacement of a vector...{\rm [Hence,]}
{\small\em a truly infinitesimal geometry must recognize only the
principle of the transference of a length from one point to
another infinitely near to the first.}}\footnote{Weyl's own
emphasis.}''
\end{minipage}

\vskip 0.1in

\noindent While, for the important role that the connection plays
as a physically individuating structure, we bring forth from
(Q?.?} Stachel's remarks about Eddington's following up Weyl's
affine connection based (unified field) theory:
\footnote{Eddington who, as noted in section 3, attempted to
further Weyl's (affine) connection based gauge theory.}

\bigskip \noindent (Q8.44)\hskip 0.9in
\begin{minipage}{12cm}
\noindent ``{\small ...Noting the very special nature of Weyl's
generalization, Eddington started by assuming that} {\small\em
there was no a priori connection between the metric and initially
arbitrary symmetric connection}.\footnote{Our emphasis. Just like
in ADG, where $\conn$ is primary, but $g_{\mu\nu}$ of secondary
importance.} {\small He observed that the curvature or Riemann
tensor and its contraction, usually referred to today as the Ricci
tensor, may be formed from the affine connection without any
metric. This Ricci tensor, however, will in general not be
symmetric, even though the connection is.} {\small\em What has
this to do with physics?}\footnote{Again, our emphasis.} {\small
Eddington noted that an affine connection enables us to compare
tensors at neighboring points (in particular, to say when two
neighboring vectors are parallel). He regarded the possibility of
such a comparison between quantities at neighboring points in
space-time as the minimum element necessary to do any physics:}
`{\small\em For if there were no comparability of relations, even
the most closely adjacent, the continuum would be divested of even
the rudiments of structure and nothing in nature would resemble
anything else'\footnote{Our emphasis.}...}''\footnote{Quotation
from \cite{eddington}.}
\end{minipage}

\vskip 0.1in

\noindent In other words, {\em no individuation means no
difference---no $\conn$---ultimately, no dynamics}.

\paragraph{The vicious circle of the smooth coordinates: an important ADG-theoretic clarification.}

\subsubsection{``I have an answer; can somebody please tell me the
question?''---Woody Allen's joke: a `principle' for QG?} Elaborate
a bit on the following telling remarks of Stachel from
\cite{stachel}:

\bigskip \noindent (Q8.45)\hskip 0.9in
\begin{minipage}{12cm}
\noindent ``{\small ...But the moral [of the hole story] is that
if you are trying to formulate a quantum theory of general
relativity (or of any generally covariant theory), you must always
bear in mind that you do {\small\em not}\footnote{Stachel's
emphasis.} have a manifold of events to start with, before finding
a solution to the quantized equations for the metric field. This
raises the following question: suppose one {\small\em
had}\footnote{Stachel's emphasis.}  a consistent quantum formalism
for general relativity (we should only be so lucky!), and had
found a solution to the quantized field equations; what could one
do with it? If you have a wave function of the universe, for
example, how do you interpret it physically? Usually, one
interprets the quantum formalism in a background space-time, so
that such a question may be answered as follows: You put an
apparatus {\small\em here} to create a particle {\small\em now},
and then you put an apparatus {\small\em there} to detect it
{\small\em then}; one can then use quantum probability amplitudes
(such as wave functions) to compute transition probabilities
between two such events. But in general relativity you have no a
priori {\small\em here} and {\small\em now} or {\small\em there}
and {\small\em then}.\footnote{All Stachel's emphasis.} What `here
and now' and `there and then' signify are among other things that
{\em one must solve the equations in order to
determine}.\footnote{Our emphasis.} I am reminded of Gertrude
Stein's last words. She roused herself from a coma to ask
worriedly: `{\small\em What is the answer?}' Then she smiled and
asked: `{\small\em What is the question?}' and lapsed into her
final coma. {\small\em In a sense, in matters of quantum gravity
one must know the answer before one knows the
question.}\footnote{Our emphasis.}}''
\end{minipage}

\vskip 0.1in

\noindent {\it Ergo}, Woody Allen's joke at the heading of this
sub-subsection. Physical propositions involving geometric
spacetime locutions ({\it eg}, this or that event: `here and now'
or `there and then') are essentially about the answer
(solution/geometry/kinematics), not about the question (dynamical
equations/algebra/dynamics). Gertrude Stein's, or more strikingly,
Woody Allen's apparently oxymoronic joke is very fitting here,
because the way we attempt to dress a generally covariant theory
(such as GR) with the standard formalism and interpretation of QM
is paradoxical---literally speaking, {\em we are begging the
question}. We are posing kinematical questions to an essentially
(and solely) dynamical theory!

\subsubsection{Field axiomatics and field realism: field over
spacetime (point events) and an abstract conception of `causal
nexus'}

This is one aftermath of the EHA and especially of its
ADG-theoretic generalization, which is represented by the
statement `{\em dynamics over kinematics}'. A point manifold $M$
does not have the `right' to qualify physically as spacetime (and
its points as events) unless the dynamics (Einstein equations) is
prescribed and solved on it.

In ADG, and in particular in the ADG-theoretic perspective on
gravity, the sole fundamental, irreducible, {\it ur} as it were,
notion is that of a {\em field}, which is represented by the pair
$\modl ,\conn)$. The field determines---in fact, {\em
defines}---the dynamics (differential equation). There is no {\it
a priori} spacetime in ADG. If there is any spacetime (geometry)
at all, then that is inherent in the field---it is the realm where
the field holds, where the field equations are valid (satisfied).

\subsubsection{Field over matter: generalizing Mach's principle in
the light of ADG}

In the first section we referred to ADG as being a {\em
Leibnizian-Machian} theory (R2.??). Throughout the paper we
`justified' the first epithet, `{\em Leibnizian}', on various
physico-mathematical grounds---primarily, in that ADG is a purely
relational, algebraic Calculus ({\it ie}, theory of DG), in the
development and expression of which a background point-manifold
({\it ie}, a locally Euclidean space)---or what amounts to the
same, the smooth coordinates thereof---plays no role whatsoever.
Rather, ADG is concerned solely with the `geometrical objects
in-themselves' and with their algebraic (inter)relations, without
an external, `intervening' space(time) being needed at all for the
mediation and (differential) geometrical realization
(picturization) of these relations (Q?.?). But how about the
second epithet, `{\em Machian}'? As seen in (Q?.?), this epithet
bears almost the same meaning as the one we gave in the light of
ADG to the term `Leibnizian'; albeit, unlike Penrose's remarks
pertaining to the theory of spin-networks, our ADG-based
interpretation of that term is in the context of DG proper.

Here, however, we would like to give a slightly `slanted'
interpretation to the epithet `Machian'---one that is more in line
with the `traditional' meaning of Mach's principle, as it was
originally conceived and used by Einstein in the {\it aufbau} of
GR. Furthermore, under the general philosophical prism of ADG, we
are going to provide a generalization of Mach's principle to the
effect that we will contend that `{\em field is prior to matter}',
thus slightly distorting both the {\em geometry-matter symmetry}
of GR (Einstein's equations), as well as the {\em field-particle
duality} of quantum theory. The ultimate apofthegma will be that
Mach's principle, viewed from an ADG-theoretic vantage, is
completely analogous to the ADG-theoretic generalization of the
aftermath of EHA, which, as we argued above, can be concisely
stated as `{\em dynamics (algebra, the field) over kinematics
(geometry, spacetime)}', with the only difference being that,
here, the statement concerns matter (particles): `{\em field over
matter}'.

Usually, Mach's principle in GR is thought of as a `heuristic
explanation' of the origin of inertia: loosely speaking, the
inertial properties of a material body are determined
`immediately, at a distance' by all the other masses in the
Universe---by the `global', total distribution of matter in the
Universe, so to say.\footnote{See Pauli's quotation (Q?.?) in
7.5.9 next.} We read, for example, from \cite{einst-1}:

\bigskip \noindent (Q8.46)\hskip 0.9in
\begin{minipage}{11cm}
\noindent ``{\small ...[Since Galileo and Newton] no one thought
of giving up the concept of space, for it appeared indispensable
in the eminently satisfactory whole system of natural science.
{\em Mach}, in the nineteenth century, {\em was the only one who
thought seriously of an elimination of the concept of space, in
that he sought to replace it by the notion of the totality of the
instantaneous distances between all material bodies. He made this
attempt in order to arrive at a satisfactory understanding of
inertia}\footnote{Our emphasis.}...}''
\end{minipage}

\vskip 0.1in

\noindent At the same time on the other hand, Mach's principle may
be perceived as a `relational' principle originally adopted by
Einstein in order to `balance' or `causalize' the dynamical
equations for gravity: matter is the source (`cause') of gravity,
and the spacetime-geometrical left hand side of Einstein's
equations is determined (`balanced') by the matter-energy
distribution on the right hand side. As Wheeler has time and again
put this `action-reaction'-type of `symmetry' between gravity and
matter in GR \cite{mtw}:

\bigskip \noindent (Q8.47)\hskip 0.9in
\begin{minipage}{11cm}
\noindent ``{\small Space acts on matter, telling it how to move.
In turn, matter reacts back on space, telling it how to
curve...}''
\end{minipage}

\noindent In fact, foreshadowed by Einstein's interpretation of
Mach's principle as {\em an attempt to eliminate space}, an
`extreme' expression of Mach's principle may be taken as holding
that {\em it is the (relations between) material bodies that
determine the (geometrical) structure of spacetime (gravity);
spacetime (structure) devoid of matter is a purely (mathematical)
artifact}. Characteristically, we quote Torretti from
\cite{torretti} regarding the crucial role that Mach's principle
(MP) played in building GR:

\bigskip \noindent (Q8.48)\hskip 0.9in
\begin{minipage}{12cm}
\noindent ``{\small ...The philosophical component of GR lies
mainly with Mach's Principle...The insight which Mach's Principle
is supposed to express certainly guided Einstein during the long
strenuous search that led to the formulation of General
Relativity. In traditional philosophical language that insight can
be stated thus: {\em Absolute space and absolute time---as well as
absolute spacetime, without matter, are physically inviable
mathematical contraptions}\footnote{Our emphasis.}... On the other
hand, [in Newtonian gravity]\footnote{Our addition.} {\em the
presence of matter makes no difference in the structure of
spacetime. According to Einstein, such lopsidedness in the mutual
relationship between two physical entities is utterly at variance
with everything we know of nature}.\footnote{Our emphasis.} Such
was the insight that Einstein sought to embody in the Mach
Principle of 1918. This can be rephrased as follows: {\em The
linear connection, or rather, the spacetime metric on which it
depends in a relativistic theory, is fully determined by the
distribution of matter...However, unless one understands the
`distribution of matter' in a Pickwickian sense, the field
equations do not agree with the above version of Mach's Principle,
since they also have solutions if the stress-energy tensor is
identically zero.}\footnote{Our emphasis again.} Later in life
Einstein rejected the 1918 Mach Principle. In 1954 he wrote to
Felix Pirani: `{\em One shouldn't talk any longer of Mach's
principle, in my opinion. It arose at a time when one thought that
ponderable bodies were the only physical reality and that in a
theory all elements that are not fully determined by them should
be conscientiously avoided. I am quite aware of the fact that for
a long time, I, too, was influenced by this fixed
idea.}'\footnote{Once again, our emphasis.}...}''
\end{minipage}

\vskip 0.1in

\noindent Indeed, one can contend that Einstein, although
originally influenced by Mach's principle in raising up GR, he
eventually abandoned it in view of the existence of Einstein
equations in {\it vacuo}---in other words, {\em the existence of a
gravitational field in the absence of matter}.\footnote{Again, see
Pauli's quotation (Q?.?) in 8.?.? next.}

Recall Pauli \cite{pauli}:

\bigskip \noindent (Q8.49)\hskip 0.9in
\begin{minipage}{12cm}
\noindent ``{\small ...In the subsequent development of the
general theory of relativity a problem cropped up which could not
be finally settled. Ernst Mach had suggested that inertia might be
traced back entirely to the action of distant masses. If this
principle of Mach's were correct, Einstein's G-field would have to
vanish if all matter were removed. In setting up his theory
Einstein was probably guided by this principle and regarded it as
correct. But it has not been possible to deduce it from the
equations of the theory. {\em It seems to be inherent in the
nature of the field concept that while the field is influenced by
the distribution of mass, it nevertheless continues to exist as an
independent reality even when all masses are removed. What the
ultimate solution will be we do not know}\footnote{Our
emphasis.}...}''
\end{minipage}

\vskip 0.1in

\subsubsection{Field solipsism versus structural realism vis-\`a-vis QG}

According to Stachel \cite{stachel7}, there are three broad
interpretations of {\em structural realism}:

\begin{enumerate}

\item The Pythagorean interpretation according to which {\em only
structures exist}, while our perceptions of the World are
illusory, together with a concomitant straight-out renunciation of
the reality ({\it ie}, the existence) of matter.

\item The Platonic scenario according to which there is an
unstructured reality (void) out there, including non-dynamical
(inert) matter, on which {\em structures are imposed from
without}---as it were, to give the World shape, form and dynamism.

\item The Aristotelian theoresis, that Stachel espouses, which
constitutes the philosophical background against which his ideas
on QG are presented in \cite{stachel7},\footnote{See also below.}
according to which---and we quote Stachel:

\end{enumerate}

\bigskip \noindent (Q8.50)\hskip 0.9in
\begin{minipage}{12cm}
\noindent ``{\small ...The world is composed of entities of
various kinds, and {\em it is inherent in their nature to be
structured in various ways. It is impossible to separate these
entities from their essential structural properties and relations.
Matter is inherently propertied, structured and dynamic
(Aristotle)}\footnote{Our emphasis.}...}''
\end{minipage}

\vskip 0.1in

\noindent We, on the other hand, supported by ADG, have taken in
this paper the philosophical stance of {\em field solipsism} or
{\em pure field realism}, which can be expressed in a way that
combines (and at the same time renounces) traits of all the three
positions above,\footnote{Although, on the whole, like Stachel, we
tend to abide more by the Aristotelian interpretation of
structural realism rather than by its Pythagorean or Platonic
antagonists.} especially once the primitive notion of {\em field}
(in its ADG-theoretic guise) substitutes that of {\em structure},
as follows:

\bigskip \noindent (R8.12)\hskip 0.9in
\begin{minipage}{12cm}
\noindent Fields are real---nothing else really exists in the
(physical) World.\footnote{This is in accordance with the
Pythagorean position 1 above \cite{stachel7}.} Nothing physically
real (significant) would remain if all fields were removed from
the World.\footnote{This is in accordance with the Platonic
position 2 above \cite{stachel7}. {\it Vis-\`a-vis} the notion of
`spacetime' in particular, no `spacetime' whatsoever remains if
all fields are removed (`switched off') from the world. For
`spacetime', if such a `thing' (`concept') exists even
theoretically speaking (not actually, {\it ie}, physically!), is
inherent in the fields.} The World is composed solely and entirely
of fields\footnote{To use an Aristotelian term, {\em the
(physical) World is a `field-plenum'} \cite{arist}.} which are
inherently dynamic and structured, with their essential dynamism
and structural relations being inextricably entwined ({\it ie},
inseparable from each other).\footnote{In our scheme, dynamics
({\it ie}, the field) `derives' from structural-algebraic
relations. This is in accordance with the Aristotelian position 3
above \cite{stachel7}.} Moreover, as noted before, spacetime
(geometry) and matter are inherent in the (dynamical) fields---to
use an Einsteinian expression (Q2.?): {\em both spacetime
(geometry) and matter are structural qualities of the dynamical
fields, not having an independent existence of their own}.

At the same time, we are inclined to depart from Stachel's main
argument and position in \cite{stachel7} that ``{\em the
fundamental physical entities,}''---in our theoresis, the
ADG-fields---``{\em while they have `quiddity' (a basic nature),
they lack `haeccity' (inherent individuality)}''. While we agree
with Stachel that the ADG-fields cannot (and should not) be
thought of `individually'---{\it ie}, apart from the dynamical
relations and processes that they engage into, as it were, apart
from the dynamical laws that they define (possibly also engaging
into dynamical interactions with other fields),\footnote{Here, the
word `other' surely assumes already some individuation!} we do not
go as far as to claim that ``{\em all properties that do not
inhere in their nature are relational}''. For this is exactly the
point, namely, that the ADG-fields are `autochthonous',
`auto-dynamical', `dynamically holistic' (`unitary' or `integral')
algebraic entities, with all dynamical relations (physical laws)
inherent in them (since the fields define them in the first
place). {\em There is no `predicative relations' (properties)
other than the dynamical ones defined by the fields}. In other
words, there is no structure (or structural realism) apart from
dynamical one.\footnote{This is part and parcel of our fundamental
{\it motto} that dynamics comes before kinematics. Structure,
understood as kinematical structure ({\it eg}, ambient, {\it a
priori} posited and fixed, spacetime geometry), is a consequence
(result) of the (field) dynamics. In this respect, our
philosophical stance again `structural (field) realism' is even
more Aristotelian (and, from the point of view of Differential
Calculus, Leibnizian) than Stachel's, as we shall also elaborate
in further extent in the last section (8.5.10).}
\end{minipage}

\vskip 0.1in

\noindent Of course, {\em our perceptions} of the fields' inherent
or innate structure and dynamism\footnote{What {\it in toto} one
might refer to as `{\em dynamical structure}' and, {\it in
extenso}, the philosophical position underlying it as `{\em
dynamical structural realism}' (or even as `{\em real structural
dynamism}').} are effectuated (triggered) by our
measurements/observations\footnote{What one might call `{\em
quantified perceptions}'.} of them---that is to say, their
intrinsic structure is brought out by our
measurements/coordinatizations of them (conveniently encoded in
the $\struc$ that we employ); it is expressed or represented (by
us) as `{\em geometry}'.\footnote{`{\em Geometry}' here being
broadly understood (`defined') as `{\em the structural analysis of
some space}' \cite{malrap3,stachel2}.} Coordinatizations ({\it
ie}, the introduction of generalized/abstract coordinates $\struc$
by us `measurers' or `geometers' external to those real fields)
may be thought of as Platonic (in Stachel's sense described above)
attempts at individuating, dissecting the fields and imparting (or
forcing) structure ({\it eg}, `spacetime geometry') on them from
without.

In particular, the inherent {\em dynamical structure} of the
fields is expressed as {\em differential geometry}, with the
fields ({\it viz.} connections) being represented
ADG-theoretically, via (our measurements in) $\struc$, as pairs
$(\modl ,\conn)$. The introduction of $\struc$ (by
us)\footnote{And, as a result, the whole `differential geometric
enterprize' (:theory of differential geometry) that {\em we}
engage into (:develop) in order to study and describe those
physically real and external to us fields. For after all, {\em
differential geometry is our activity, not Nature's. There is no
differential geometry in Nature}.} may be regarded as an attempt
to `individuate' the holistic fields---as it were, they are
attempts to localize (locally measure in our `local
laboratories'/local gauges) and extract from their `global',
all-pervading nature, their local (particle/`position') aspects
(properties). However, the structure (geometry) encoded in
$\modl$\footnote{Recall, $\modl$ is (locally) of the form
$\struc^{n}$ ($n$ a finite positive integer).} is inseparable---in
a quantum sense, {\em coherent}---from the dynamical change
represented by $\conn$, and this coherence (or `unitarity') may be
interpreted quantum mechanically as complementarity (and
operationally, as self-indeterminacy).\footnote{See subsection 6.2
above about the innate (global) field-(local) particle duality of
the ADG-fields $(\modl ,\conn)$. The ADG-fields are coherent
(inseparable), self-quantum entities, and thus `{\em individual}'
({\it ie}, in a literal sense, they cannot be divided into their
field $\conn$ and particle $\modl$ aspects).} All in all, for us,
{\em differential} ($\conn$) {\em geometry} ($\modl$) is
synonymous to {\em dynamical} ($\conn$) {\em structure}
($\modl$)---a structure analyzed by means of the $\struc$ that
{\em we employ} to `coordinatize' or `measure' the dynamical
fields, to quantify (`arithmetize', in a Cartesian sense) our
perceptions of them so to speak, and thus extract their local
particle traits (`properties').\footnote{Recall again, the local
sections of $\modl$ represent, from a geometric (pre)quantization
standpoint, the local quantum-particle states of the field.} Only
in such a `slanted' quantum-theoretic sense, {\em our field
structural realism is `observer' (and observation)
dependent}.\footnote{And exactly because of these `extraneous
interventions' by us `observers' or `measurers of the fields'
(geometers!), such a realism is not Platonic ({\it ie}, not
strictly detached), but it is in accord with the observer
dependent physical reality that quantum theory advocates.} So, on
the whole, there are perceptions,\footnote{In contradistinction to
the Pythagorean position 1 above \cite{stachel7}.} and the
dynamical structure inherent in the fields is perceived/expressed
(`quantified'/`Cartesianly arithmetized' or `geometrized', and
brought out) from without---in point of fact, from us
`observers'.\footnote{Partly, as the Platonic perspective 2 on
structural realism maintains \cite{stachel7}, but unlike it, these
interventions are not as invasive, as the conventional quantum
theory has it, so as to maintain a strict `observer dependent
reality'. For one thing, the dynamics that the fields define are
$\struc$-covariant or $\struc$-functorial (:`synvariant'), so that
the dynamical laws see through our attempts of coordinatization
and individuation of the fields. {\it In summa}, there are
dynamically autochthonous fields out there, and we simply
(differential) geometrize them in order to represent them
mathematically. Mathematics is ours, physics Nature's.}

\paragraph{What GR structure(s) to quantize?} Then, against the philosophical background of the Aristotelian
interpretation of structural realism, Stachel goes on and
addresses the question of what structures of GR one `should'
quantize as an attempt to arrive at a quantum theory of gravity.
In what follows we quote the whole of section 6 in \cite{stachel7}
where this question is posed, as we wish to respond to it from our
ADG-theoretic `{\em field solipsistic}' vantage.\footnote{In the
following excerpt we have Greek letter-ordered the paragraphs of
the original text so as to facilitate our analysis of them in the
sequel.}

\bigskip \noindent (Q8.51)\hskip 0.9in
\begin{minipage}{12cm}
\noindent {\small $\alpha$. There are a number of space-time
structures that occur in the general theory of relativity. The
{\em chrono-geometry} is represented mathematically by a {\em
pseudo-metric tensor field} on a four-dimensional manifold. The
{\em inertio-gravitational field} is represented by a {\em
symmetric affine connection} on this manifold. Then there are {\em
compatibility conditions} between the previous two structures (the
covariant derivative of the metric with respect to the connection
must vanish).

$\beta$. It is possible to start from the metric field and derive
from it the unique connection (the Christoffel symbols) that
automatically satisfies the compatibility conditions. This is what
was first done historically, and still done in most textbooks.

$\gamma$. It is also possible to treat {\em metric and connection}
as {\em initially independent}, and then allow the compatibility
conditions to emerge, along with the field equations, from a
Palatini-type variational principle.

$\delta$. Each of these methods may be combined with a {\em tetrad
formalism} for the metric, combined with various mathematical
representations of the connection, e.g., connection forms, tetrad
components of the connection.

$\epsilon$. But it is possible to abstract from the (four-)
volume-defining property of the metric, resulting in a {\em
conformal structure} on the manifold. This is all that is needed
to represent mathematically the {\em causal structure} of
space-time. It is also possible to abstract from the preferred
parametrization of geodesics associated with the connection,
resulting in a {\em projective structure} on the manifold. This
represents mathematically the class of {\em preferred (time-like)
paths} in space-time. Compatibility conditions between the causal
and projective structures can be defined, which guarantee the
existence of a corresponding compatible metric and connection
(Ehlers, Pirani and Schild).\footnote{Here Stachel most probably
refers to the celebrated paper \cite{ehlpirsch} and subsequent
follow-up work \cite{ehlsch}.}

$\zeta$. {\em Just as it was so important in the historical
development of quantum mechanics to choose an appropriate
formulation of classical mechanics, to which to apply some
quantization technique, it may well be the case that one or the
other of these formulations of general relativity will be more
helpful in solving the quantum gravity puzzle in one or more of
the various ways in which it has been--or will
be--posed.}\footnote{Our emphasis.}}
\end{minipage}

\vskip 0.1in

\noindent In what follows we itemize our reply, to the aforesaid
question `{\em What Structures} (from GR) {\em to Quantize?}' that
Stachel puts as heading to the last section (6) of
\cite{stachel7}, according to our letter-ordering of its
paragraphs above. Our reply below will recall various important
issues that have been discussed in many places throughout the
present paper-book, and may be regarded as a {\it r\'esum\'e} of
key conceptual issues in classical and quantum gravity that have
been addressed, highlighted and resolved under the prism of ADG
herein:

\begin{itemize}

\item $\alpha$ First of all, {\em in the ADG-theoretic formulation
of GR, there is no (background) spacetime structure at all}---{\it
ie}, a base realm on which $g$ is soldered to measure for instance
spacetime intervals, or to characterize smooth paths (supposedly
followed by material particles) as being timelike, null, or
spacelike (chrono-geometry). That is, in ADG's perspective on
gravity there is, physically speaking,\footnote{And by `{\em
physically speaking}', we are talking about a structure that has
absolutely no {\em dynamical} role in the theory---{\it ie}, it is
not a dynamical variable (`observable') in the theory.} no
chrono-geometry whatsoever, represented by a pseudo-metric tensor
field on a four dimensional spacetime manifold, since, to be
blatant, {\em there is no spacetime manifold `out there' to begin
with}. To emphasize it again, {\em GR {\it \`a la} ADG is
fundamentally spacetimeless, while the traditional geometrical
picturization and associated `geometrical reasoning'\footnote{In
terms of points, lines, surfaces {\it etc}, as usual.} by means
(or with the help) of a base spacetime manifold is of little
(conceptual) import in our purely algebraic (relational) theory}.
On the other hand, {\em in ADG the inertio-gravitational
field---the only physical (dynamical) variable in the theory---is
represented by the $\struc$-connection
$\conn$}.\footnote{Moreover, exactly because $\conn$ (unlike the
$\struc$-metric $\rho$, or its curvature $R(\conn)$) is not a
tensor (in our language, an $\struc$-morphism), `resists'
decomposition into its inertial (flat) $\partial$ and
gravitational $\aconn$ parts, unless of course we---the external
`episystem' in the whole scenario---evoke a local (coordinate)
gauge and `unnaturally' split it into $\partial+\aconn$.} Then,
only when $\struc$ is introduced (by us---the external `episystem'
of `observers' or `measurers') and concomitantly the locally
$\struc^{n}$ vector sheaf $\modl$ is adjoined to the purely
algebraic gravitational field $\conn$ as its {\em carrier} (or
representation/geometrization) {\em space},\footnote{Recall, in
ADG, by the term {\em field} we refer to the pair $(\modl
,\conn)$.} can the $\struc$-metric $\rho$ be employed on $\modl$
and (voluntarily) be made compatible with $\conn$ (if we wish)
\cite{malrap3}. Otherwise, it ({\it ie}, the metric) has no {\it a
priori} status (interpretation) as a measurement device on (a)
spacetime (manifold), since {\it a fortiori} there is no spacetime
manifold to start with.\footnote{To say it once more, ADG-gravity
is a purely (third) gauge formulation of gravity, based solely on
the notion of connection---``{\em the gravito-inertial field}''
itself according to Stachel---without {\it a priori} reference to
a smooth background spacetime manifold or to a smooth metric based
on it. Thus, also no ``{\em chrono-geometric}'' structure and
interpretation is {\it ab initio} evoked in ADG-gravity.} To
stress it once more, the gravitational field $\conn$ alone exists
`out there' independently of us (field solipsism), while we
capture it and realize/represent it (differential) geometrically
by our generalized arithmetics (`measurements') in $\struc$,
bringing along the (our) $\struc$-metric $\rho$, which may be made
compatible with it. $\conn$ {\em is primary and fundamental},
$\rho$ {\em secondary and contingent}.\footnote{Depending on {\em
our} choice of $\struc$.}

\item $\beta$ That the metric determines a unique connection (and
vice versa) is in our view a historical (mathematical) accident of
the manifold based Riemannian geometry---however, it is an
identification or equivalence (of the metric with the connection)
that masks the physical significance of the latter ($\conn$) and
the insignificance of the former ($g$). That is to say, that the
gravitational field came to be identified with $g$ is also in our
view a theoretical accident, with the mathematics ({\it ie}, the
CDG-based Riemannian geometry) to blame for this
`misrepresentation' and `misinterpretation'.\footnote{In our
theoresis, we are talking about {\em the connection compatibility
of the metric}, not about {\em the metric compatibility of the
connection}, as usually ({\it ie}, in the smooth pseudo-Riemannian
geometry based original formulation of GR by Einstein, as well as
what is ``{\em still done in most textbooks}'' nowadays.} Rather,
the metric-represented gravitational field conceals the
fundamental gauge-theoretic character of gravity.\footnote{Recall
Feynman's words in (Q?.?).} The recent tendencies to pronounce the
importance of the notion of connection in GR, thus view gravity as
another gauge theory \cite{sen,ash,ash1,ash2,ash4} are certainly
important, albeit they are still bound by the ``{\em golden
shackles of the manifold}'' and of the CDG based on it \cite{ish},
while at the same time the (smooth) metric is still present in the
theory in the guise of the (smooth) comoving tetrad ({\it
vierbein}) field.\footnote{This is to the effect, as emphasized
earlier, that while Ashtekar's formalism and its associated
quantization scenario (LQG and canonical QGR, as well as its
offshoot QRG) treat non-perturbative quantum gravity as a quantum
gauge theory in a manifestly fixed background {\em metric}
(geometry) independent way, they are still not able to formulate
QG in a background {\em differential manifold} independent way.
For after all, the connections, tetrad fields {\it etc} used there
are {\em smooth}, and the differential geometric methods are those
of the manifold based CDG.}

\item $\gamma$ In ADG it is not an issue whether connection and
metric are independent or not: $\conn$ {\em is primary, while} $g$
{\em secondary}, not the other way round. However, it is
interesting indeed that the metric compatibility condition for the
connection\footnote{That is, a torsionless connection.} derives,
along with the field equations, from a Palatini-type of
variational principle \cite{malrap3}.

\item $\delta$ The tetrad ({\it vierbein}) formalism for the
metric, together with the $sl(2,\com)$-valued $1$-form
representation of the gauge potential part $\aconn$ of the
self-dual (spin-Lorentzian) connection, is what is essentially
involved in the Ashtekar `new variables' approach to classical and
quantum gravity \cite{malrap3}.

\item $\epsilon$ Here Stachel notes another metric-{\it
versus}-connection `duality'. Since the spacetime metric field, at
least locally, delimits the causal structure of the
world,\footnote{That is to say, locally at every spacetime point,
according to the EP, $g_{\mu\nu}$ reduces to the Minkowskian
$\eta_{\mu\nu}$, which delimits the local causal structure at each
event, {\it ie}, which events in the immediate neighborhood of
each event causally influence (causal past) and are influenced by
(causal future) it. In turn, GR may be thought of as the dynamical
theory of the field of local causality \cite{malrap1}.} one may
base one's theoresis of gravity on causality alone (which itself
determines the conformal structure of the world).\footnote{Since
conversely, as it is well known, causality (mathematically
represented by a partial order relation) determines nine out of
ten components of the Lorentzian metric of GR, the tenth being the
spacetime volume element ({\it ie}, the conformal determinant of
$g_{\mu\nu}$). Indeed, some `bottom-up', `discrete' approaches to
QG, such as Sorkin {\it et al.}'s causet theory that Stachel also
mentions in \cite{stachel7}, capitalize on causality as a
fundamental notion.} On the other hand, one could work directly
with the causal (null and timelike) geodesics determined by the
connection (resulting in what Stachel calls above the `projective'
structure of the world), and then try to bring together the
conformal and the projective structures. In glaring
contradistinction, such an endeavor would be meaningless in the
purely algebraic ADG-gravity, since the theory is fundamentally
base spacetimeless, while such `classical' (Aristotelian and
Kantian) categories of (physical) thought, such as space(-time),
events and causality, are all replaced by the notion of field
({\it viz}. algebraic connection), which is a fundamental, ``{\em
primary, not further reducible}'' \cite{einst3} notion in the
theory.

\item $\zeta$ In this respect we fundamentally and expressly
diverge in opinion from Stachel: as noted numerously in the
present paper-book, we do not expect QG to arise as the result of
quantizing the (structures of the) classical theory (GR), so that
the question of choosing the most appropriate (for quantization)
formulation of GR in order to arrive at a cogent QG is from the
ADG-viewpoint `begging the question'. This basic difference in
`working philosophy' conceals a host of novel ideas in ADG-gravity
not encountered in any QG approach so far; and to name a few:
background smooth manifold independence,\footnote{In fact,
background spacetime independence altogether: whether this
background is `continuous' or `discrete'.} QG as a dynamically
autonomous third-quantum gauge field theory of the third kind,
absence of a fundamental space-time length in Nature, no
(perturbative non-)renormalizability, no (black hole) information
loss, no spacetime quantization, and no Correspondence
Principle.\footnote{For example, no `emergence of classicality',
or viewing GR as an `effective theory'.}

\end{itemize}

\paragraph{Field solipsism over structural realism.} Field-solipsism: \texttt{All is field,
and the field is (inherently) quantum}
$\Longrightarrow$ \texttt{All is quantum} (Finkelstein
\cite{df1}).

$\bullet$ In Wittgenstein's Tractarian sense \cite{witt}, `{\em
solipsism is pure realism}', as follows:\footnote{In the quotation
below, all emphasis is Wittgenstein's unless noted otherwise.}

\bigskip \noindent (Q8.52)\hskip 0.9in
\begin{minipage}{11cm}
\noindent ``{\small ...{\em The limits of my language} means the
limits of my world...Logic fills the world: the limits of the
world are also its limits...What solipsism {\em means} is quite
correct, only it cannot be {\em said}, but it shows itself...The
world and life are one. I am my world (the microcosm). The
thinking, presenting subject; there is no such thing...The subject
does not belong to the world, but it is a limit of the
world...From nothing in {\em the field of sight} can it be
concluded that it is seen from an eye...This is connected with the
fact that no part of our experience is also a priori. Everything
we see could also be otherwise. Everything we can describe at all
could also be otherwise. There is no order of things a priori.
{\em Here we see that solipsism strictly carried out coincides
with pure realism}.\footnote{Our emphasis.} The I in solipsism
shrinks to an extensionless point and there remains the reality
co-ordinated with it...}''
\end{minipage}

\subsubsection{The ADG-fields as
Aristotelian-Leibnizian `entelechies': `autodynamical monads'
(with windows)}

Recall Aristotle's conception of potentia and `autodynamics' from
\cite{popper} (compare it with our ADG-conception of `autonomous',
`unitary fields'):

\bigskip \noindent (Q8.53)\hskip 0.9in
\begin{minipage}{11cm}
\noindent ``{\small ...[Aristotle says that] the [Platonic] Form
or essence is {\em in}\footnote{Popper's emphasis.} the thing,
not, as Plato said, prior and external to it. {\em For Aristotle,
all movement or change means the realization (or `actualization')
of some of the potentialities inherent in the essence of a
thing}\footnote{Our emphasis.}...Accordingly, the essence, which
embraces all the potentialities of a thing,, is something like its
internal source of change or motion...`{\em Nature}', he writes in
the Metaphysics,\footnote{See \cite{aristotle}.} {\em belongs also
to the same class as potentiality; for it is a principle of
movement inherent in the thing itself}'\footnote{Again, our
emphasis.}...}''
\end{minipage}

\subsection{Bird's-Eye-View: Mathematical Physics or Physical
Mathematics?, and What `It' Should Be (ADG-Theoretic Musings and
`Post-Anticipations' of Paul Dirac and Ludwig Faddeev)}

We argued above, based on the main didactics of ADG-gravity in the
light of some `directives' given by Peter Bergmann (inspired by
Einstein), in favor of a `physical geometry' than of a
`geometrical physics'. In a nutshell, we posited that
`geometry'---in particular {\em differential} geometry---and
`space(time)' is of interest to the physicist, not {\it a
priori}---as it were, as a pure mathematical discipline offering
the convenience of `geometrical picturization' (:representation)
of physical situations and processes, but insofar as it is the
`result' or the `outcome' of physical laws (:dynamical field
laws)---processes which, in turn, {\it a priori} know no
`geometry' (:`spacetime'), but quite on the contrary, they `give
birth to it'.\footnote{That is, the laws (:the fields defining
them) are the causes of `space(time)', and conversely, `physical
geometry' the effect of these dynamical field laws. On this hinges
the aforenoted reversal of the traditional priority of kinematics
over dynamics: indeed, we posited in the light of ADG that, in a
deep sense, {\em dynamics comes before kinematics}.} Accordingly,
we argued for an algebraization (:`relationization'), rather of a
geometrization, of physics, which finds its `natural habitat' (at
least regarding the application of differential geometric ideas to
theoretical physics) in ADG.

Here we would like to extend this `philosophy' in favor of a
`physical mathematics' rather than the most commonly used term
`mathematical physics'. This extension finds us in accord with the
thesis taken recently by Ludwig Faddeev in \cite{faddeev} {\it
vis-\`a-vis} what mathematical physics should (or anyway, ought
to) be. Briefly, Faddeev maintains that we should finally break
away from the classical developmental route followed so far by
theoretical physics, which can be resumed by the cycle:
`experiments$\mapto$predictions$\mapto$mathematical
elaborations/formulations$\mapto$further experiments {\it etc.}',
and implore all our mathematical resources to plough deeper into
`physical reality', leaving experiments (and experimentalists!) to
`catch up' with the new mathematics (and with theoreticians!), not
the other way round.

Especially, we would like to borrow from \cite{faddeev} some
prophetic remarks in this line of thought by Dirac
\cite{dirac3}:\footnote{Once again, we split the quote below into
two parts and we comment on them separately after.}

\bigskip \noindent (Q8.54)\hskip 0.9in
\begin{minipage}{11cm}
\noindent ``{\small ...The steady progress of physics requires for
its theoretical foundation a mathematics that gets continually
more advanced. This is only natural and to be expected. What,
however, was not expected by the scientific workers of the last
century was the particular form that the line of advancement of
the mathematics would take, namely, it was expected that the
mathematics would get more complicated, but would rest on a
permanent basis of axioms and definitions, {\em while actually the
modern physical developments have required a mathematics that
continually shifts its foundation and gets more abstract...It
seems likely that this process of increasing abstraction will
continue in the future and that advance in physics is to be
associated with a continual modification and generalization of the
axioms at the base of mathematics rather than with logical
development of any one mathematical scheme on a fixed
foundation.}\footnote{Our emphasis.} {\bf (I)}

There are at present fundamental problems in theoretical physics
awaiting solution [...]\footnote{Dirac here mentions a couple of
outstanding mathematical physics problems of his times. We have
omitted them.}the solution of which problems will presumably
require a more drastic revision of our fundamental concepts than
any that have gone before. Quite likely these changes will be so
great that it will be beyond the power of human intelligence to
get the necessary new ideas by direct attempt to formulate the
experimental data in mathematical terms. The theoretical worker in
the future will therefore have to proceed in a more indirect way.
{\em The most powerful method of advance that can be suggested at
present is to employ all the resources of pure mathematics in
attempts to perfect and generalise the mathematical formalism that
forms the existing basis of theoretical physics, and {\sl
after}\footnote{Dirac's own emphasis.} each success in this
direction, to try to interpret the new mathematical features in
terms of physical entities}\footnote{Our emphasis throughout.}...}
{\bf (II)}''
\end{minipage}

\vskip 0.1in

\begin{itemize}

\item {\bf (I)} The words from this paragraph to be highlighted with ADG in mind are:
`{\em a mathematics that gets more abstract}' and `{\em advance in
physics is to be associated with a continual process of
abstraction {\rm [leading to a]} modification and generalization
of the axioms at the base of mathematics}'. Indeed, the axiomatic
ADG essentially involves an abstraction of the fundamental notions
of modern differential geometry ({\it eg}, connection), resulting
in a modification and generalization of the latter's basic axioms
\cite{mall1,mall2,mall4}. And it is precisely this abstract and
generalized character of ADG that makes us hope that its
application could advance significantly (theoretical) physics (and
in particular, classical and quantum gravity research).

\item {\bf (II)} In this paragraph, apart from breaking from the
traditional circle mentioned above ({\it ie}, Dirac's anticipation
that `{\em new ideas {\rm [won't come]} by direct attempts to
formulate the experimental data in mathematical terms}'), what
should be highlighted is on the one hand Dirac's prompting us
`{\em to generalise the mathematical formalism that forms the
existing basis of theoretical physics}', and on the other, `{\em
to try to interpret the new mathematical features in terms of
physical entities}'. Again, ADG comes to fulfill Dirac's vision,
since {\em the} (or at least the bigger part of the) mathematics
that lies at the heart of current theoretical physics---namely,
(the formalism of) {\em differential geometry} ({\it ie}, CDG on
smooth manifolds)---is abstracted and generalized, while after
this generalization has been achieved, the physical application
and interpretation (of ADG's novel concepts and features) has been
carried out (especially in the theoretical physics field of
classical and quantum gravity research). We believe that this is
`{\em a powerful method of advance}' indeed.

\end{itemize}

\noindent But perhaps it is more important to conclude this
`paper-book' by stressing once again that ADG is not so much a
{\em new} theory of DG---the main `{\em mathematical formalism
that forms the existing basis of theoretical physics}'---but a
theoretical framework that abstracts, generalizes, revises and
recasts the existing CDG by isolating and capitalizing on its
fundamental, {\em essentially algebraic} (`relational' in a
Leibnizian sense) features, which are not dependent at all on a
background smooth geometrical `space(time)' (:manifold). In a way,
from the novel viewpoint of ADG, we see `old' and `stale' problems
({\it eg}, the $\smooth$-singularities of the manifold and CDG
based GR) with `new' and `fresh' eyes. Schopenhauer's words from
\cite{schopenhauer} immediately spring to mind:

\bigskip \noindent (Q8.55)\hskip 0.9in
\begin{minipage}{11cm}
\noindent ``{\small\em ...Thus, the task is not so much to see
what no one has yet seen, but to think what nobody yet has thought
about that which everybody sees\footnote{All emphasis is
ours.}...}''
\end{minipage}

\subsection{Section's R\'esum\'e}

As we did with the previous sections, we may wrap-up this final
section by itemizing and highlighting its main theses:

\begin{enumerate}

\item First, we recall {\it \`a la} Isham (Q8.?) that the general consensus until today is
that one could not use CDG, and {\it in extenso} a smooth base
manifold, in attempts to arrive at the true QG theory---{\it eg},
by applying the continuum based QFTheoretic technology (and the
smooth fiber bundle mathematical panoply of modern gauge theories)
to GR. Arguably, the strongest reasons for this `no-go' attitude
is on the one hand the singularities of GR, which are supposed
`classical' issues one has to reckon with before the actual
quantization of the gravitational is evoked,\footnote{Although it
is characteristically expected nowadays that QG will relieve GR
from its pathological singularities.} and on the other the
irremovable (non-renormalizable) infinities that (perturbative)
QFTheoretic approaches to QG are assailed by.

\item By virtue of the above, we argued, motivated by ADG and ADG-gravity,
that such a drastic measure as a complete abandonment of
differential geometric ideas in the quantum regime is not really
necessary, as long, of course, as a base (spacetime)
manifold---{\em the} culprit for the aforesaid (differential
geometric and physical) diseases---is truly abandoned. {\it Ergo},
scrap CDG and invite ADG in the quantum deep. In this line of
thought, we argued that in ADG no infinitary (limit) processes
occur as in the standard Analysis (Differential Calculus) on
manifolds, because the role of a background space(time) is
atrophized (and topology is `suppressed'), while algebra is being
emphasized and capitalized on. And there is no infinity in
algebra! Furthermore, we argued {\it \`a la} Einstein and
Bergmann, that from the purely algebraico-categorical
(relational-pointless) perspective of ADG, one is not inclined
anymore to `geometrize physics' and `picturize' (via some
pre-existing, mathematically posited by fiat, underlying
`spacetime') one's constructions. Rather, `physical spacetime
geometry', if anyone still wishes to think in such `classical
picturesque' terms, is inherent in the fields---or better, it is
the result (`solution space') of the dynamical laws that the
algebraic (relational) fields define (differential geometrically
speaking, as differential equations). Here we have another example
of the novel ADG-gravity {\it motto} that `{\em dynamics (algebra)
comes before kinematics (geometry)}'.

\item We then suggested that ADG could qualify as a strong candidate for the (mathematical)
`organic' theory that Einstein was searching for in order to
materialize his unitary field theory {\it vis-\`a-vis} either the
singularities of his geometrical manifold based GR and the
quantum-particle aspects of the other field theories of matter.
Here we encountered a plethora of issues and questions that
Einstein raised and we answered ADG-theoretically. Most
characteristic of all was his `ambivalence' and associated
agnosticism about a `discrete-algebraic' non-field-theoretic
physics which is not based on a (spacetime) continuum {\it versus}
a `continuous-(differential) geometric' one that is manifestly
based on a background geometrical (spacetime) continuum. We argued
that ADG-gravity (and the general viewpoint of ADG facing
Yang-Mills and the other gauge forces of matter) passes through
the horns of Einstein's dilemma most characteristically by
formulating a background spacetimeless field theory---one that is
indifferent as to whether an external (to the fields themselves)
spacetime is a `quantal-discretum' or a `classical continuum'. A
characteristic example of such an ADG-theoretic bypass of a major
problem that Einstein encountered during the {\it aufbau} of GR,
regarded as a relativistic field theory of the gravitational field
on the spacetime continuum, is his (and Grossmann's) hole argument
which was originally proposed in order to test the viability of
the PGC, which is usually modelled after $\mathrm{Diff}(M)$.
Precisely due to the background $M$ independence of ADG-gravity,
this becomes a `non-argument' in our theoresis, while Stachel's
deep interpretation of the conceptual consequences of the EHA was
essentially seen to be another manifestation of one of the
pillar-aftermaths of ADG-gravity noted above, namely that dynamics
(the field) is prior to kinematics (`spacetime geometry'), or
equivalently, that spacetime is inherent in the dynamical field.

\item We then arrived at a host of arguments and theses based on this `field-over-spacetime'
priority, such as `field over spacetime events and the causal
nexus between them', `field over matter' in a Machean sense, as
well as `field solipsism and autodynamicity' in an
Aristotelian-Leibnizian `windowfull' sense, to name a few.

\item Finally, as an extension and generalization of Bergmann's
`physicalization of geometry' (as opposed to `geometrization of
physics') mentioned above, we maintained that there is no
`mathematical physics' as such, but only `physical mathematics';
while, based on some telling remarks of Dirac as recalled by
Faddeev, we put forth that theoretical physics research---and
especially QG---is in need of more abstract and general(ized)
mathematical concepts and structures, so that when it comes to
applications of differential geometric ideas in QG, ADG is well
qualified to be identified with such a framework.

\end{enumerate}

\section{Appendix: Glossary for ADG-Gravity}

Below is a list, in alphabetical order, of certain {\em novel}
basic concepts, ideas and principles, that have been mentioned in
this work, in the past trilogy \cite{malrap1,malrap2,malrap3}, as
well as in the recent paper \cite{rap5}. These central notions,
some of them carrying rather standard meaning and
connotation,\footnote{Indeed, some of the notions appearing in the
list below have well established meaning and connotations in
theoretical physics' nomenclature, jargon and general literature.
Their new `definitions' given here in no way aim at confusing
things (and if they do so, we sincerely apologize in advance for
our inability to come up with different names/terms). Rather, they
are presented on the one hand just to challenge some of those
traditional terms, and on the other simply for the reader to have
ready at his disposal a concise explanation of certain new and/or
`exotic'-sounding concepts that she encounters during the reading
of this work. Since these terms are randomly dispersed and occur
repeatedly throughout this voluminous work, we believe that the
compact concentration of them into a glossary will significantly
facilitate her reading. Finally, let us mention that some of the
`definitions' in the list are interdependent, so the acronym `{\em
sdb}' (followed by a number $x$ from 1 to 42) written in
parenthesis after the appearance of a not-yet-defined term means
`{\em see definition (of that yet undefined term in $x$) below}'.}
were born during the development of {\em ADG-gravity} and we feel
that they are important for understanding it better.

\begin{enumerate}

\item {\bf ADG-gravitational field:} This pertains to the pair $(\modl
,\conn)$, where $\conn$ is the gravitational connection field
proper ({\em sdb} 6), and $\modl$ its (local quantum-particle)
associated (representation) vector sheaf ({\em sdb} 4).

\item {\bf Algebraicity---relationalism:} This pertains to the
Leibnizian-Machean trait of ADG, and {\it in extenso} of
ADG-gravity, that all differential geometry, and as a result the
ADG-gravitational dynamics (Einstein equations) that are
represented differential geometrically as differential equations
proper, refers directly to (in fact, it derives solely from) the
algebraic relations between the `geometrical objects' ({\it ie},
the {\em physical fields}; {\em sdb} 18) that live on a surrogate
base (spacetime) scaffolding, without that background playing any
role whatsoever (in the guise of coordinates) in that differential
geometric mechanism ({\em sdb} 10) nor in the said autonomous
field-dynamics ({\em sdb} 5), which in turn is formulated just via
that essentially algebraic (:sheaf-theoretic) mechanism.

\item {\bf Analytic (smooth and continuous) manifold extension:}
From an ADG-vantage, the usual notion of analytic ($\anal$),
smooth ($\smooth$), or just continuous ($\cont$) extensibility (of
a manifold $M$) simply pertains to changing (`enlarging') the
structure sheaf $\struc$ ({\em sdb} 17) of analytic, smooth, or
continuous functions that one assumes up-front to chart (label or
coordinatize) $M$'s points, it being implicitly assumed here that
one invariably starts with $M$ as a structureless point-set and
then one `dresses' it with a topological ($\cont$), differential
($C^{i};~i=1\ldots\infty$), or analytic ($\anal$) structure
(sheaf) thus qualify it as a topological, differential, or
analytic manifold, respectively. From a differential geometric
viewpoint, this hinges on the fact that a differential manifold
$M$ is nothing but the (algebra or structure sheaf of)
differentiable functions (in $\smooth(M)$ or $\smooth_{M}$) on it
(Gel'fand duality; {\em sdb} 23), so that when one wishes to
extend it, one should simply enlarge one's (`space' or algebra of)
differentiable functions.

\item {\bf Associated sheaf:} This is the vector sheaf $\modl$ of
representation of the connection field $\conn$ {\em sdb} 6). It is
the carrier (or action) space of $\conn$ and its (local) sections
represent (local) particle quantum states of the said field ({\em
sdb} 12). Technically speaking, $\modl$ is the sheaf associated
with the principal sheaf $\aut\modl$ ({\em sdb} 26)  of
self-transmutations (dynamical Kleinian `auto'-symmetries; {\em
sdb} 5, 21) of the (quantum-particle states of the) field.

\item {\bf Autodynamics---dynamical autonomy:} This pertains to the fact that the
dynamical (vacuum) Einstein equations in ADG-gravity is expressed
solely in terms of the (curvature of the) connection field $\conn$
`in-itself' ({\em sdb} 12, 23), without reference to a background
spacetume (whether `discrete' or `continuous').

\item {\bf Background independence:} The currently fashionable in QG research notion of `background
independence' acquires a new, deeper meaning in ADG-gravity.
Unlike in various perturbative and non-perturbative approaches to
QG where this term means `background {\em metric} independence'
({\it ie}, no background geometry), while a background manifold is
still employed in order to be able to do CDG (CDG-conservatism and
monopoly; {\em sdb} 8), in ADG-gravity not even a smooth base
(spacetime) manifold, let alone a smooth metric, is used in the
theory. In this sense, ADG-gravity is genuinely and completely
background independent. 

\item {\bf Categoricity of (gravitational) dynamics:} The (vacuum) gravitational
dynamics is expressed via sheaf morphisms. In particular, via the
curvature of the connection, which is an $\struc$-morphism---a
fact which bears on the functoriality of the dynamics in
ADG-gravity ({\em sdb} 20).

\item {\bf CDG and manifold conservatism and monopoly in physics:} By this we
mean that all physical theories that have been formulated (so far)
in the language of differential geometry (basically, the dynamics
that define them as physical theories proper are modelled after
differential equations) employ in one way or another a base
differential manifold---a (locally) Euclidean space, thus
effectively they use the conceptual and technical means of
Classical Differential Geometry (CDG). Even in the case of
singularities, which are arguably `internal blemishes' or `faults'
of the background manifold employed, we have so far used
persistently CDG (Analysis) to deal with them ({\em sdb} 8).

\item {\bf CDG-vicious circle vis-\`a-vis singularities:} This
refers to the fact that since spacetime singularities are `innate'
anomalies and pathologies of the background spacetime manifold,
the employment of the manifold based CDG (Analysis) to deal with
({\it eg}, resolve) them is essentially an `oxymoron'. In other
words, there are genuine singularities---{\it ie}, singularities
that CDG cannot cope with---in the manifold based GR.

\item {\bf Conflict between the PGC and singularities:} Since
singularities are `inherent' in the manifold, the Principle of
General Covariance (PGC) of GR, which is modelled in the usual
theory via the diffeomorphism group $\mathrm{Diff}(M)$ of the
underlying differential spacetime manifold $M$, appears to come in
conflict with them and moreover it makes a precise definition of
them, within the base manifold confines of CDG, an impossible
task.

\item {\bf Connection field:} This is the `gauge field proper' part $\conn$ of
the ADG-field $(\modl ,\conn)$. Technically, it is a
$\cons\equiv\mathbf{K}$-morphism, not an $\struc$-morphism. This
property qualifies $\conn$ as a purely algebraic entity, and not
as a `geometrical object' proper ({\em sdb} 25). From a
gauge-theoretic point of view, the connection may be identified
with the gauge (potential) field, although by the latter term
theoretical physicists usually understand only the local part
$\aconn$ of $\conn :=\partial +\aconn$ .

\item {\bf Coordinate (virtual) singularities:} In the context of
the manifold and {\it in extenso} CDG-based GR, the notion
`coordinate singularities' pertains to certain {\it loci} of the
spacetime continuum that appear to host singularities for some of
the $\smooth(M)$-components of the smooth spacetime
metric-solution of Einstein's equations, simply because those {\it
loci} have been charted by (referred to) an `inappropriate' system
of coordinates (frame). It is therefore understood that by a
suitable change of system of (smooth) coordinates ({\it eg}, by an
appropriate smooth extension), coordinate singularities disappear,
as for instance when one changes coordinates from the
Cartesian-Schwarzschild frame to the Eddington-Finkelstein one in
order to `resolve' the exterior Schwarzschild singularity.

\item {\bf Curvature (of a connection):} In ADG, the curvature of
a connection, $\curv(\conn)$, is also a sheaf morphism; albeit, an
$\struc$-morphism. This property qualifies $\curv$ as a
`geometrical object' proper ({\em sdb} 25). From a gauge-theoretic
vantage, $\curv$ may be identified with the gauge field strength.

\item {\bf DGSs, SFSs and VESs:} So far, the Analytic (CDG-based) taxonomy of
spacetime singularities splits into three general classes: (i)
differential geometric singularities (DGSs)---points at a suitably
defined (topological) boundary $\partial M$ of the spacetime
manifold $M$ for which there is no $C^{k}$-differential extension
of (the solution-metric on) $M$ so as to incorporate them with the
other regular points in $M$'s interior; (ii) solution field
singularities (SFSs)---again, boundary points of $M$ for which
there is no (analytic) extension of the (metric on the) latter
that removes them and at the same time it is a distributional
solution of the Einstein field equations; and (iii), various
energy singularities (VESs)---again, boundary points for which
there is no (analytic) extension of (the metric on) $M$ that
removes them satisfying at the same time various energy
conditions.

\item {\bf Differentiability, differential (geometric) mechanism:}
By differentiability, in a certain theoretical framework (for
doing differential geometry), we mean the possibility of defining
a differential operator $d\equiv\partial$---{\it ie}, the ability
to differentiate mathematical entities within that framework.
Since differentiation is usually perceived as a (local)
topologico-algebraic (analytic) notion, in the usual theory (CDG)
it is the (locally) Euclidean character of the base space
(manifold) that secures the definition of $d$ and, {\it in
extenso}, differentiability. In CDG it is the background space
(manifold) that provides one with `differentiability'
(differential structure). In contradistinction, in ADG, which is
base manifold free, the topological (`spatial background') aspect
of differentiability becomes atrophic and what is pronounced is
$d$'s algebraic character. Indeed, the basic recognition of ADG is
that $d$ is a particular instance of the general notion of
connection $\conn$ (`generalized derivative') and to secure the
(definition of the) latter, all that one needs is a structure
algebra sheaf $\struc$ of generalized arithmetics---abstract
`differentiable functions' (`coordinates')---on an in principle
arbitrary topological space $X$. In turn, the classical theory
(CDG) is obtained when one assumes $\struc\equiv\smooth_{X}$ for
structure sheaf of generalized coordinates (or what amounts to the
same by Gel'fand duality, a base differential manifold $X\equiv
M$). {\it In toto}, we maintain that in ADG the essentially
algebraic differential geometric mechanism derives from the stalk
of the sheaves involved and not from the base space, which is
merely a topological space. ADG is an algebraization of Analysis
emphasizing the latter's algebraic (relational) qualities and at
the same time deemphasizing the latter's topological
(spatial-geometrical) `dependencies'.

\item {\bf Differential geometric monads (with `windows'):}
Differential triads are the `{\it ur}'-elements, the basic
building units of ADG. On them one can erect the entire
differential geometric edifice ({\it eg}, modules of higher-order
differential forms, connection, curvature {\it etc}.), while the
base space employed plays absolutely no role in the purely
algebraic (:sheaf-theoretic) differential geometric mechanism
`encoded' in (or `carried' by) those units. In particular, for the
most important ADG-notion of connection (field) $\conn$, this
background independence has made us qualify the ADG-fields $(\modl
,\conn)$ as (dynamically) autonomous, `unitary'
entities---entities in no need of a background space(time) for
their (differential geometric) subsistence. Due to their
(differential geometric) autonomy and self-sufficiency, triads may
be coined, in honor of Leibniz (and of the purely relational
character of Calculus that he had envisaged), `differential
geometric monads'. Granted that an external space does not
influence at all the purely algebraic differential mechanism
inherent in the triads, it does not mean that the latter, like the
Leibnizian monads, are `windowless': on the contrary, they form a
category $\ctriad$ and the arrows in it simply show that triads
`communicate' with each other ({\it ie}, they have windows). For
the connections (fields) in particular, which are sheaf morphisms,
the fact that the laws of physics (as differential equations) are
represented by equations between sheaf morphisms, shows that
fields actually `communicate' ({\it eg}, they interact).

\item {\bf Dynamics as algebra (`cause'):} This pertains to the
fact that in ADG-gravity the algebraic $\struc$-connection field
$\conn$ defines the gravitational dynamics---the actual law of
motion of the gravitational field---without an {\it ab initio}
commitment to an {\it a priori} fixed, external (ambient),
kinematical (possibility) space(time). Antonio Machado's verses
come to mind: ``{\em Traveller there are no paths; paths are made
by walking}'' \cite{machado}, and it is the precisely the fields
that `do the walking' by their dynamics, in the manifest absence
of an ambient/external, pre-existing, geometrical space-time.)
This is consistent with the aforesaid `dynamical autonomy' of the
ADG-fields and it suggests that in ADG-gravity the notion of the
causal nexus (between events) in the external, geometrical
space(time) is replaced by the autonomous, `bootstrapping' actions
of the spacetimeless ADG-fields (and their inherent
particle-quanta). Moreover, the geometrical physical space(time)
is in a sense `inherent' in those algebraic (relational)
fields---it is the `result' (`solution space') of the dynamics
that the fields define.

\item {\bf Field---generalized causality:} This, as also briefly
alluded to above, refers to the observation that in ADG-gravity,
causality and the `geometrical pattern' of causal ties between
spacetime events is replaced by the (gravitational) field and the
dynamics that this defines---nothing more, nothing less.

\item {\bf Field-particle duality:} This is closely related to the
notion of third quantization ({\em sdb} ) and it refers to the
self-dual character of the autonomous, unitary ADG-fields $\field
:=(\modl ,\conn)$: on the one hand we have the `proper' field
aspect $\conn$ of $\field$, and on the other, from a geometric
prequantization vantage, the local sections of $\modl$ represent
the local quantum-particle states of the field---$\field$'s
particle aspect.

\item {\bf Functional sheaves:} This refers to sheaves of
functions that can be used as structure sheaves in ADG, although
ADG in principle admits also non-functional sheaves, as long as
the latter can accommodate (provide one with) a differential
operator $d$ (with which one can actually do differential
geometry).

\item {\bf Functoriality (kinematical and dynamical):} In general
terms, functoriality of a construction (in a categorical setting)
pertains to the quality of the construction that all entities
involved in it respect the relevant categories concerned. In the
standard kinematical sense, functoriality means that in the
process of the construction certain key kinematical (structural)
features are preserved---{\it eg}, quantization perceived as a
structural-functorial procedure: geometric prequantization being
functorial, first quantization being non-functorial, and second
quantization being functorial. By contrast, in ADG-gravity
functoriality is not of a kinematical, but of a dynamical kind.
This pertains to the fact the dynamical equations of Einstein are
functorial with respect to the structure sheaf $\struc$ of
generalized coordinates, as the curvature (of the gravitational
connection field) involved in them is an $\struc$-morphism, a
$\otimes_{\struc}$-tensor, with the latter being the homological
tensor product functor. On this functoriality hinges the
ADG-theoretic generalization of the PGC of the manifold based GR
(namely, the principle of Synvariance) as well as ADG-gravity's
evasion of $\smooth$-singularities by first absorbing (or
integrating) them into $\struc$, and then by making the dynamics
manifestly $\struc$-independent ({\it ie}, $\struc$-functorial).

\item {\bf Gauge theory of the third kind:} In contradistinction to the
so-called gauge theories of the first (global gauge symmetries)
and second (local gauge symmetries) kind, ADG-gravity---in fact,
the ADG-theoresis also of Maxwellian electrodynamics (abelian
gauge symmetries) and Yang-Mills theories (non-abelian gauge
symmetries)---is formulated solely in terms of the (gravitational)
connection field $\conn$ (half-order formalism); moreover and more
importantly, no external, background spacetime (manifold) is
involved in the theory. Due to these two features, the
ADG-theoresis of gravity (and of the other fundamental gauge
forces) is coined gauge theory of the third kind.

\item {\bf Generalized arithmetics (coordinates)---structure
sheaf:} In ADG, where the principal motto is that `{\em
differentiability is independent of smoothness}' \cite{malrap2},
one can use a structure sheaf $\struc$ of (algebras of)
`differentiable coordinates' different from the usual (classical)
one $\smooth_{M}$ of smooth ones on a differential manifold $M$.
Such $\struc$s are called generalized arithmetics and they may be
far from smooth ({\it eg}, `discrete', or distributional and
`ultra-singular'). From an ADG-theoretic viewpoint, all
differential geometry boils down to $\struc$, since, for one
thing, it is the `domain' of (or `source space' for) the basic
differential $\partial$ and effectively of the connection field
$\conn$ (since by definition $\modl$ is locally a finite power of
$\struc$). From a physical point of view, the elements of $\struc$
model our acts of coordinatization, representation, measurement
and concomitant `geometrization' of the fields involved.

\item {\bf `Geometrical' objects, `algebraic' objects:} By
`geometrical' objects we mean fields that our generalized
arithmetics or measurements in $\struc$ respect. Geometrical
objects are mathematically modelled in ADG by $\struc$-sheaf
morphisms, or perhaps better, by $\otimes_{\struc}$-tensors. The
canonical example of a geometrical object is the curvature (field)
of the connection (field), $\curv(\conn)$. By contrast,
`algebraic' objects are those fields not respected by our
generalized coordinates---fields that elude our attempts (in
$\struc$) at localizing, representing and `geometrizing'
(measuring) them. Algebraic objects are not
$\otimes_{\struc}$-tensors and they evade our acts at sharply
localizing (measuring) them by using $\struc$. The canonical
example of an algebraic object is the connection field $\conn$
itself, which is not an $\struc$-sheaf morphism, but only a
$\cons\equiv\mathbf{K}$-morphism. That $\conn$ cannot be sharply
determined by $\struc$ ({\it ie}, the field cannot be localized
and be quantum particle-represented by $\modl$, which is locally
isomorphic to $\struc^{n}$) is reflected by the fact that it acts
(as a `derivation') on the local particle (generalized position)
states ({\it ie}, the local sections) of $\modl$ and changes them
as a momentum-like operator. This of course bears on the fact that
the unitary ADG-field $(\modl ,\conn)$ is third- or self-quantum.

\item {\bf Half-order formalism:} The formulation of ADG-gravity
solely in terms of the connection field $\conn$, as opposed to the
second order formalism---the original formulation of GR in terms
of the metric (Einstein)---as well as to the so-called first order
formalism whose basic variables are the (smooth) connection and
the (smooth) vierbein/tetrad frame (Palatini, Ashtekar).
Furthermore, in contrast to both the first and the second order
formalism, in the third order ADG-formalism for gravity no
$\smooth$-smooth base spacetime manifold is involved at all.

\item {\bf Internal, `auto-symmetries':} This refers to the `symmetries' of
the ADG-gravitational field $(\modl ,\conn)$, and concomitantly,
to the `invariances' of the dynamical law (Einstein equations)
that the field defines. Because the field is independent of an
external, background spacetime (manifold) and it is of a purely
gauge (third gauge) character, these symmetries may be coined
`internal', or `esoteric'---they are the invariances of the field
(law) `in-itself', what one could call `auto' or
`self-symmetries'. Mathematically, they are organized into the
principal group sheaf $\aut\modl$ of (dynamical)
`auto-transmutations' of the field ($\conn$) and its
particle-quanta ($\modl$).

\item {\bf Kinematics as geometry (`effect'):} This pertains to the
fact that in ADG-gravity the role of kinematics and dynamics is
reversed, in the sense that one does not think of a geometrical
kinematical (possibility) space as being {\it a priori}
fixed---{\it ie}, prescribed before the `algebraic' dynamics is
given---by the theoretician. Rather, kinematics (geometry) is the
result of dynamics (algebra), and in this respect one may think of
the traditional term `physical spacetime geometry' as the
`solution space(time)'---the `place' where the dynamical field law
of gravity holds. In terms of the algebraic and geometrical
objects distinction made above and of our perception of $\conn$ as
an abstract causal nexus (causality/causal connection), one may
formally write this `geometry is the effect/result of algebra' (or
equivalently, that `algebra is the cause of geometry') by
$\conn\Rightarrow\curv(\conn)$.

\item {\bf Kleinian field-particle geometry:} This pertains to
identifying {\it \` a la} Felix Klein the internal, `esoteric'
geometry of the particle-field pair $(\modl ,\conn)$ with its
group (sheaf) $\aut\modl$ of dynamical self-transmutations---the
field's automorphisms.

\item {\bf Local (open) gauges, local frames, local measurements:}
The commonly used term in local gauge theory (of the second kind)
`local gauge', in the context of ADG-gravity refers simply to an
open subset $U$ of the base topological (localization) space $X$
(of the sheaves) involved. $U$ may be thought of as an abstract
`local laboratory', a `local reference frame', or even better, a
`local measuring device' ({\it ie}, a local gauge!), to which our
local measurements (field-coordinatizations) in $\struc
|_{U}\equiv\struc(U)$ belong. Accordingly, given a `coordinatizing
open cover' $\gauge=\{ U_{\alpha}\}_{\alpha\in I}$ of $X$, $e^{U}$
defines a local frame or a local choice of basis (or gauge!) for
the vector sheaf $\modl$. The $e_{i}$s in $e^{U}$ are local
sections of $\modl$ ({\it ie}, they are elements of $\modl
|_{U}\equiv\modl(U)\equiv\Gamma(U,\modl)$) constituting a basis of
$\modl(U)$---{\it ie}, they span uniquely, with linear
coefficients in $\struc(U)$, every local section of $\modl$ living
in $\modl(U)$.

\item {\bf Localizing-sheafifying-gauging-curving-relativizing-dynamicalizing:} From an
ADG-theoretic viewpoint, these six terms are essentially synonyms.
Localization pertains to soldering the various algebraic
structures involved on the base topological space $X$.
Technically, and in the sheaf-theoretic context of ADG, this is
accomplished by sheafification. From a physical viewpoint
localization of the various (algebraic) structures involved and
their symmetries is tautosemous to `gauging'---{\it ie}, endowing
the (locally independent) structures with a non-trivial
connection---whose geometrical expression is the curvature field.
Finally, the assignment of a $\conn$ means essentially that the
`relativized' structures involved ({\it ie}, structures referred
to and expressed in a particular local gauge) are variable---in
fact, that they are dynamically variable, and their dynamical
variation is expressed differential geometrically by the
differential equations that the said connection field defines
(here, the Einstein equations, which are geometrically expressed
not directly in terms of the connection which anyway is not a
geometrical, but an algebraic, object, but indirectly via the
curvature of the connection, which is the geometrical object {\it
par excellence}).

\item {\bf Natural transformation theory:} This is ADG's categorical version
of the general notion of `transformation theory' ({\it eg}, in the
sense of Dirac) underlying any process of relativization. To
explain this, in GR, relativization---{\it ie}, the Principle of
Relativity (PR)---can be expressed by saying that the description
of the gravitational dynamics is `independent' of reference to any
particular system of (smooth) coordinates, which culminates in the
standard $\mathrm{Diff}(M)$-representation of the PGC of the
differential manifold based CDG underlying GR. In the background
manifoldless ADG-gravity however, the PGC is generalized to the
notion of Synvariance, which is in turn equivalent to the
dynamical $\struc$-functoriality of the ADG-formulated Einstein
equations (with $\struc$ not necessarily restricted to be
$\smooth_{M}$ as in the manifold based CDG). This functoriality
then amounts to a relativization of $\struc$ in ADG-gravity, which
is expressed via the Principle of Algebraic Relativity of
Differentiability (PARD). Finally, the latter is mathematically
represented by a pair of `second order' functors (natural
transformations) mapping the $\struc_{1}$-functorial
ADG-gravitational dynamics expressed in a certain (chosen)
structure sheaf $\struc_{1}$ of generalized coordinates, to the
$\struc_{2}$-functorial ADG-Einstein equations expressed in
another (different) structure sheaf $\struc_{2}$. The physical
upshot of PARD is the Principle of Field Realism (PFR) in
ADG-gravity.

\item {\bf Newtonian spark---background space(time)
forgetfulness:} This pertains to the general `phenomenon' in ADG
whereby once one has `extracted' the inherently algebraic
differential geometric mechanism from the stalk of the algebra and
vector sheaves involved, or even possibly from a locally Euclidean
base space (manifold) as in the classical theory (CDG), we develop
all our differential geometric concepts and constructions
independently of that surrogate background (sheaf-theoretic)
localization space(time). As it were, we totally forget about the
base space(time) and work exclusively in the (algebra inhabited)
sheaf space occupied by the really (physically) significant
objects---the fields and their algebraic interrelations.

\item {\bf Physical (spacetime) geometry:} By `physical (spacetime) geometry' we understand the
`solution space' of the dynamical law, or more generally, the
`space' where the dynamical law holds. In ADG-gravity, this is the
representation (carrier) sheaf (space) $\modl$ for (of) the
gravitational field $\conn$---the `space(time)' (defined and
occupied by the quantum particle states of the field) where the
(vacuum) Einstein equations of ADG-gravity hold.

\item {\bf Principal sheaf:} In ADG-gravity, this is the group sheaf
$\aut\modl$ of (dynamical) `auto-symmetries' of the (dynamically)
autonomous ADG-gravitational field $(\modl ,\conn)$. In turn, it
is understood that $\aut\modl$ has $\modl$ as its associated
(representation) vector sheaf.

\item {\bf Principle of Algebraic Relativity of
Differentiability:} Since one of the basic morals of ADG is that
all differential geometry boils down to $\struc$, the idea is that
by changing $\struc$ (with these changes dictated by actual
physical problems---{\it eg}, when one needs to change structure
sheaf of differentiable functions in order to include in the new
structure sheaf an apparently singular, from the point of view of
the old structure sheaf, function; {\it alias}, when one wishes to
incorporate or `absorb' the singular function into the structure
sheaf of generalized arithmetics),

\item {\bf Principle of Field Realism:} The dynamical laws of
physics which are (mathematically) defined (as differential
equations) by the ADG-fields $(\modl ,\conn)$ are independent of
our `generalized measurements'. That is, they are independent of
the structure sheaves $\struc$ that we---the external (to the
fields) `observers', `measurers' (`geometers'), or
`coordinators'---employ to represent (`geometrize', or
`coordinatize on some geometrical spacetime') those fields. This
is reflected in the aforesaid functoriality of the ADG-field
dynamics, and {\it in extenso} in the PARD, with its natural
transformations' representation.

\item {\bf Real (genuine) singularities:} Let it be stressed up-front that from the perspective of
ADG-gravity, this notion is both conceptually and technically a
chimera. In the usual CDG-based GR, real (true or genuine)
singularities are {\it loci} in the spacetime continuum that
`resist' analytic extension of that manifold so as to include them
with the other regular points of $M$ (in other words, one cannot
further enlarge one's algebra of differentiable coordinate
functions, always staying within the realm of $\smooth_{M}$, so as
to include the singular ones with the other regular, smooth ones).
As a result, true singularities are thought of as being situated
at the boundary of a maximally (analytically) extended, albeit
incomplete, spacetime manifold. They are `defined' by exclusion
(they are sites where all Analysis fails!), so that there is no
precise and straightforward definition of real singularities in
the manifold and CDG-based GR. In contradistinction, in the base
smooth manifoldless ADG-gravity where all singularities are
integrated or `absorbed' into (a suitably chosen) $\struc$ while
the Einstein law that the ADG-gravitational field defines is
$\struc$-functorial (synvariant) hence not at all impeded by any
singularity, there is no real (genuine) singularity (at least in
the DGS and SFS sense above---{\it ie}, as sites where the
differential equation represented field law breaks down, or where
even a generalized, distributional solution fails to be one). From
an ADG-theoretic vantage, in a deep sense all singularities are
coordinate (virtual) ones as they are embodied into (a suitably
chosen) $\struc$ and the $\struc$-functorial gravitational
dynamics is not at all impeded by their presence (Einstein's
equations hold in their very presence).

\item {\bf Singularities as differential geometric solution-metric
anomalies:} In view of the fundamental conflict between the PGC of
GR and the existence of singularities, the latter may be thought
of as `differential geometric solution-metric anomalies' in the
sense that while the field law (Einstein equations) are
$\mathrm{Diff}(M)$-invariant, its solution-metric may possess
singularities where `the smoothness of the law' appears to be
broken.

\item {\bf Singularity absorption and evasion:} This refers to the
basic singularity-evasion `strategy' of ADG: upon encountering a
singularity, one should look for a structure sheaf $\struc$ that
contains it while at the same time it provides one with the
(essentially algebraic) differential geometric mechanism with
which one can write the Einstein equations (as differential
equations proper) in the very presence of the said singularity.

\item {\bf Synvariance:} Due to the manifest absence of an external
(to the gravitational field itself) background spacetime manifold
in ADG-gravity, the $\mathrm{Diff}(M)$-modelled PGC of the
manifold based GR is replaced by the group (sheaf) $\aut\modl$ of
dynamical self-transmutations of the ADG-gravitational field
$(\modl ,\conn)$ `in-itself' (to be precise, $\aut\modl$ is the
symmetry group sheaf of the local quantum-particle states of the
field, which, at least from a geometric prequantization vantage,
are modelled after the local sections of $\modl$). There is no
external spacetime structure dynamically varying with (`covarying'
with) the gravitational field. All there is `out there' is the
gravitational field $\conn$ alone and the dynamics (the vacuum
Einstein equations) that it defines via the action of its Ricci
curvature on its own local particle states ({\it ie}, the sections
of) $\modl$: $\ricci(\modl)=0$. At the same time, thanks to the
$\struc$-functoriality of the ADG-gravitational dynamics, no
matter what $\struc$ we employ to coordinatize or geometrize ({\it
ie}, localize and particle-represent) the gravitational connection
field $\conn$, we do not `perturb' it (and the dynamics that it
defines via its curvature) at all, no matter what singularities
that (the function-like objects in) $\struc$ may be assumed to
carry.

\item {\bf Third Quantization:} In contradistinction to both first
(non-relativistic quantum particle mechanics) and second
(relativistic quantum field mechanics) quantization, for the
formulation of which an external space-time manifold is invariably
involved in one way or another, third quantization (better, third-
or self-quantum field theory) pertains to the ADG-scheme of doing
field theory solely in terms of the fields $(\modl ,\conn)$
`in-themselves' without a base space-time manifold; and moreover,
with quantum traits built into these autonomous fields from the
very start ({\it ie}, virtually `from their very definition').

\item {\bf Unitary (`monadic') field theory:} In the context of ADG, the epithet `unitary'
to `field theory' pertains less to Einstein's original vision of
`one single field for all forces' and more to (also Einstein's,
less popular however, vision of) a `total field' obeying
(defining) total dynamical (differential) equations that are
`free' from ({\it ie}, in no way impeded by, let alone breaking
down in the presence of) singularities, it incorporates its
particle-quanta as singularities in the law (differential
equation) that it defines, and (as a bonus to Einstein's
vision---and something that Einstein could not have possibly
envisioned in the CDG-based field theory that he had in mind and
was advocating) the background spacetime continuum plays no role
whatsoever in the field's autonomous dynamics (`autodynamics').
{\it In summa}, a genuinely unitary field theory in our
ADG-theoretic sense is concerned only with the field, the whole
field and nothing but the field.

\end{enumerate}

\end{document}